%% file: EFT-2HDM+MLRSM.tex
\documentclass[11pt,a4paper,oneside]{article}

\pdfoutput=1
\usepackage{jheppub}

\usepackage{amsmath}
\usepackage{verbatim}
\usepackage{graphicx}
\usepackage{mathrsfs}
\usepackage{appendix}
\usepackage{caption}
\usepackage{float}
\usepackage{subfig}
\usepackage{array,multirow}

\usepackage{graphicx}
\usepackage{amssymb}
\usepackage{slashed}
\usepackage{amsmath}
\usepackage{setspace}
\usepackage{lscape}
\usepackage{xcolor}
\newcommand{\1}{\Phi}
\newcommand{\2}{\Phi^{\dagger}}
\newcommand{\3}{\tilde{\Phi}}
\newcommand{\4}{\tilde{\Phi}^{\dagger}}
\renewcommand{\l}{\Delta_{L}}
\renewcommand{\L}{\Delta_{L}^{\dagger}}
\renewcommand{\r}{\Delta_{R}}
\newcommand{\R}{\Delta_{R}^{\dagger}}
\newcommand{\at}[2][]{#1|_{#2}}

\title{Hilbert Series and Plethystics: Paving the\\ path towards 2HDM- and MLRSM-EFT}

\author{Anisha,} \author{Supratim Das Bakshi,} \author{Joydeep Chakrabortty,} \author{and Suraj Prakash} \author{\\}

\affiliation{Indian Institute of Technology Kanpur, Kalyanpur, Kanpur 208016. INDIA. \\}

\emailAdd{anisha, sdbakshi, joydeep, surajprk@iitk.ac.in}

\abstract
{Effective Field Theory (EFT) technique is one of the most elegant ways to capture the impact of high scale theory, if any, at some low energy by incorporating higher mass dimensional ($\geq 5$) effective operators ($\mathcal{O}_i$). The low energy EFT is described in terms of only light degrees of freedom (DOF) which can appear on-shell. An essential task while developing the EFT framework is to compute these  $\mathcal{O}_i$'s. Hilbert Series (HS) is a novel and mathematically robust method to construct the complete set of gauge invariant independent, effective operators. The HS requires the knowledge of the transformation properties of the light DOF and the covariant derivatives under the internal gauge symmetries and conformal groups. The Hilbert Series method, by its virtue, automatically takes care of the redundancies in the operator set due to the Equations of Motion (EOMs) of fields and Integration by Parts (IBPs) with impeccable accuracy.
	
	\noindent
	In this paper, we have adopted this methodology to construct the complete set of independent operators up to dimension-6 in the ``Warsaw"-like basis for two different Beyond Standard Model scenarios -- Two Higgs Doublet Model (2HDM) and Minimal Left-Right Symmetric Model (MLRSM). For both these cases, we have calculated the corrections to the scalar, gauge boson and fermion mass spectra due to the dimension-6 operators. The additional contributions to all the Feynman vertices are computed and their impact on different observables, namely Weak mixing angle, Fermi constant, $\rho$ and oblique $(S,T,U)$ parameters. We have further discussed how the magnetic moments of charged leptons and production and decay of the massive BSM particles, e.g., charged scalar and different rare processes are affected in the presence of effective operators. We have also constructed the effective scalar four-point interactions and commented on the possible reinvestigation of the theoretical constraints, e.g., unitarity and vacuum stability within these frameworks. 
}

\begin{document}
	
	\maketitle

	\newpage

\section{Introduction}

The Standard Model (SM) of particle physics is the most successful theory to describe the dynamics of the sub-atomic particles. Most of the predictions of the SM are verified experimentally with impeccable accuracy. Nevertheless, the SM fails to explain many experimental observations, e.g., neutrino mass, the anomalous magnetic moment of the muon, dark matter, etc. Thus we are confident about the necessary existence of some theory Beyond Standard Model (BSM). The actual form of the BSM is yet to be unveiled, but it is expected that low energy physics, e.g., SM has emerged from some more fundamental theory lying at some higher scale. If we believe in such a proposal, then one needs to derive the low energy lagrangian from that relatively complete theory. To do so, the massive modes of that complete theory, i.e., the heavier particles having mass $>$ scale of low energy theory must be integrated out. 
In the process,  higher mass dimensional operators, suppressed by the mass  of the integrated out particles,  are generated alongside the renormalizable lagrangian describing relatively low energy physics. Then the $new$ lagrangian at low energy will be written as $\mathcal{L}=\mathcal{L}^4+\sum\limits_{i\in \mathbb{N}} \mathcal{C}_{i+4} \frac{\mathcal{O}_{i+4}}{\Lambda^{i}}$, where $\mathcal{L}^4$ contains all renormalizable terms up to mass dimension-4, and $\mathcal{O}_{i+4}$'s are the effective operators of mass dimension $[M]^{i+4}$. The $\mathcal{C}_{i+4}$'s are the associated Wilson Coefficients (WCs).  These effective operators and their respective  WCs are expected to capture the footprints of the high scale theory. This idea is the basic anthem of the so-called ``Effective Field Theory" (EFT) which provides a platform to realise the high scale physics in terms of low energy effective lagrangian.
One can consult Refs.~\cite{Georgi:1991ch,Appelquist:1974tg,Ross:1978wt,Weinberg:1980wa,Gaillard:1985uh, Cheyette:1985ue, Henning:2014wua,Henning:2016lyp,Henning:2015alf,Chiang:2015ura,Huo:2015exa,Huo:2015nka,Drozd:2015rsp,delAguila:2016zcb,Fuentes-Martin:2016uol,Ellis:2016enq,Gripaios:2018zrz,Falkowski:2015wza,Celis:2017hod,Criado:2017khh,Aebischer:2017ugx,Aebischer:2018bkb,Bakshi:2018ics,Criado:2019ugp,Gavela:2016bzc,Criado:2018sdb,Misiak:2018gvl} where different aspects of effective field theory have been discussed.

But can we construct a low energy effective field theory even if we do not know the exact nature of the high scale physics? Yes, that is indeed possible.  One can always add gauge invariant operators of different mass dimensions to  the renormalizable low energy lagrangian \cite{Arzt:1993gz,Wudka:1994ny,Georgi:1994qn,Kaplan:1995uv,Kaplan:2005es,Manohar:1996cq,Burgess:2007pt,Rothstein:2003mp,Gaillard:1985uh, Cheyette:1985ue, Henning:2014wua,Henning:2016lyp,Henning:2015alf,Lehman:2015via,Chiang:2015ura,Huo:2015exa,Huo:2015nka,Lehman:2015coa,Wells:2015uba,Drozd:2015rsp,delAguila:2016zcb,Fuentes-Martin:2016uol,Ellis:2016enq,Falkowski:2015wza,Misiak:2018gvl,Passarino:2019yjx,Cohen:2019wxr,Zhang:2016pja,Fuentes-Martin:2016uol,Ellis:2016enq,Willenbrock:2014bja,delAguila:2016zcb,Jenkins:2013fya}. 
Based on this idea, over the years, EFT has become one of the most sophisticated tools to sense the presence of new physics, even without knowing it exactly. This version of EFT relies only on the light degrees of freedom (DOF). The  new physics is captured in the WCs, which are free parameters and blind to the exact form of the BSM theory. Here, one must include the ``complete set of gauge invariant independent operators of particular mass dimension" such that the operators can form a basis  \cite{Buchmuller:1985jz,Grzadkowski:2010es}. This idea was first implemented in the context of the SM in \cite{Buchmuller:1985jz,Grzadkowski:2010es}. It has been noted in Refs.~\cite{Grzadkowski:2010es,Jenkins:2013zja, Jenkins:2013wua,Alonso:2013hga}  that based on the SM  interactions and the particle content one can construct 59 gauge invariant independent dimension-6 operators in a specific basis, known as the ``Warsaw" basis.
Here, the lepton, as well as baryon number symmetries, are assumed to be respected. While constructing these operators, the most subtle aspect is the identification of independent operators. Since we are dealing with the dynamical field variables, the derivatives (more specifically covariant derivative) appear in the operator structures almost in the same footing as the quantum fields. Thus it is very much possible that two differently structured operators may be related to each other by Equations of Motion (EOMs) of the fields or they may differ by a total derivative. The appearance of the total derivative is related to one of the properties of the derivative operator -- Integration by Parts (IBPs) \cite{GrosseKnetter:1993td,Buchmuller:1985jz,Grzadkowski:2010es,Barzinji:2018xvu}. These redundancies can be avoided by incorporating the EOMs and IBPs explicitly. One can look into the Refs. \cite{Buchmuller:1985jz,Grzadkowski:2010es} where the SM-Effective Field Theory (SMEFT) with dimension-6 operators was constructed based on this principle.  In \cite{Lehman:2014jma} the complete set of dimension-7 operators has been computed. The methodology discussed in these papers was adequate, but can be tedious to implement for BSM models with extended symmetries and(or) more particles.

Recently a new approach has been developed to construct the higher mass dimensional ($\geq 5$) operators based on the Hilbert Series (HS) method within the SM framework \cite{Henning:2015daa,Kobach:2017xkw,Henning:2015alf,Lehman:2015coa,Trautner:2018ipq,Hays:2018zze,Trautner:2018ipq,Henning:2017fpj,Hanany:2010vu,Feng:2007ur,Lehman:2015via}. This provides a mathematically elegant, and rigorous platform to reveal the algebraic structures of the SMEFT. In this construction,  information of the underlying global, gauge, and also space-time symmetries is utilised in the form of ``characters" of representations of the field variables. We would like to specifically mention here that the derivative operators are treated in the same footing as the quantum fields in this construction. The derivatives transform under the conformal group, which is $SO(4,2)$ in our case. 
This method automatically takes care of the redundancy in the operator set and generates all independent operators, which is its best merit.
Thus explicit implementations of the EOMs and IBPs can be avoided which may be cumbersome for general gauge structures involving multiple field representations.  This algebraic method certainly widens the reach of the EFT program beyond the SM.

Now, coming back to the utility of the EFT framework, we have to remember that a finite number of WCs are capable of representing almost all possible BSM scenarios which can affect the low energy physics.
In general, effective field theory provides a better handle to deal with the unknown BSM physics, if it is there and interferes with the SM interactions. Most of the BSM models contain too many free parameters and suffer from a lack of predictability. Therefore, from that perspective, the EFT approach is very economical and captures the new physics in a much more concise form. For example, if there is a deviation between the experimental data and the SM predictions, then it can be explained in terms of a few effective operators. The corresponding WCs can be constrained very acutely as the DOF of the analysis will be less as compared to a complete BSM theory. The fitted values of these WCs are obtained using experimental data and based on the low energy effective lagrangian.

An essential aspect of phenomenological analyses is identifying the parameter space for different BSM models allowed by the experimental data till date.  In the SMEFT framework, the BSM effects appear in the form of effective operators along with the renormalizable SM lagrangian.
For decades, the  precision low energy observables, e.g., Weak mixing angle, Fermi constant, $\rho$ and oblique ($S,T,U$) parameters  have been used to determine the precise room left for the physics beyond the Standard Model \cite{Gates:1991uu,Dugan:1991ck,Blondel:1989ev,Holdom:1990tc,Kennedy:1990ib,Peskin:1990zt,Peskin:1991sw,Kennedy:1991sn,Altarelli:1990zd,Gori:2013mia,Cacciapaglia:2006pk,Englert:2019zmt,LlewellynSmith:1983tzz,Veltman:1977kh,Kim:1980sa,Wheater:1982yk,Gluck:1981mt,Renken:1982ap,Hagiwara:1994pw,Bjorn:2016zlr,Voloshin:1992sn,Ellis:2018gqa}. The impact of these effective operators on the  low energy precision observables, magnetic moments, and theoretical constraints in the scalar sector  are discussed in \cite{Georgi:1991ci,Buchmuller:1985jz,Kennedy:1988sn,Kennedy:1990fj,Kennedy:1988sn,Altarelli:1990wt,Altarelli:1990zd,Holdom:1990tc,Einhorn:1981cy,Grinstein:1991cd,Low:2009di,Almeida:2018cld,Contino:2016jqw,Trott:2014dma,Falkowski:2014tna,Falkowski:2017pss,Carpentier:2010ue,Falkowski:2018dmy,Ellis:2018gqa}. 
Their role  in the context of  collider phenomenology, neutrino masses, and rare processes are addressed in  \cite{Manohar:2013rga,Han:2004az,Bonnet:2011yx,Bonnet:2012nm,delAguila:2011zs,Grojean:2013kd,Contino:2016jqw,Brehmer:2015rna,Gorbahn:2015gxa,Berthier:2015oma,Skiba:2010xn,Englert:2013wga,Banerjee:2012xc,Banerjee:2013apa,Almeida:2018cld,Dekens:2018pbu,Dawson:2018jlg,Hays:2018zze,Dawson:2018liq,Karmakar:2019vnq,Crivellin:2016ihg,DiazCruz:2001tn,Contino:2014aaa,Contino:2013kra,Bar-Shalom:2018ure,Gomez-Ambrosio:2018pnl,Graf:2018ozy,delAguila:2012nu,Bhattacharya:2015vja,Ghosh:2017coz,Corbett:2017qgl,King:2014uha,Crivellin:2018ahj,Geng:2016auy,Elgaard-Clausen:2017xkq,Cepedello:2017eqf,Liao:2016qyd,Alte:2018xgc,Khandker:2012zu,Gori:2013mia,Elias-Miro:2013eta,Wells:2015uba,Ellis:2014jta,Trott:2014dma,Englert:2014cva,Biekoetter:2014jwa,Belusca-Maito:2014dpa,Ellis:2014dva,Einhorn:2013tja,Elias-Miro:2013mua,Mebane:2013zga,Gupta:2014rxa,Berthier:2015gja,Efrati:2015eaa,Chen:2013kfa,Dedes:2018seb,Davidson:2018kud,Vryonidou:2018eyv,Hesari:2018ssq,Baglio:2018bkm,Silvestrini:2018dos}.

In the process of adjudging the BSM parameter space under the light of present and future proposed colliders, and other low energy experiments we consider mostly the renormalizable  BSM lagrangian. But these TeV scale BSM scenarios (NP1) could be effective theories which are generated from more complete theories (NP2) lying at an energy scale just beyond the access of present experiments. Then considering only the renormalizable BSM lagrangian may be misleading while drawing the exclusion limits in BSM parameter space. It has also been noted that the final remarks on the parameters of NP1 may alter significantly once we include the effective operators \cite{Contino:2013kra,Bar-Shalom:2018ure,Gomez-Ambrosio:2018pnl,Hays:2018zze,Graf:2018ozy,Karmakar:2019vnq}. 
Also, in the light of future colliders, like FCC \cite{Golling:2016gvc,Contino:2016spe,Blondel:2018mad}, the realisation of the effective TeV scale BSM scenarios could be significant. Recent works have already introduced the novelty of effective BSM (BSM-EFT) scenario in the context of 2HDM-EFT, see Refs.~\cite{Belusca-Maito:2016dqe,Crivellin:2016ihg,Karmakar:2017yek, Karmakar:2018scg, Trautner:2018ipq,Karmakar:2019vnq}.

In this paper, we have borrowed the concept and methodology of Hilbert Series (HS) that has been used for SMEFT construction in \cite{Henning:2015daa,Kobach:2017xkw,Henning:2015alf,Lehman:2015coa} and implemented to understand BSM-EFT. This paper, to our knowledge, is one of the earlier attempts to construct the complete BSM-EFT lagrangian up to dimension-6 effective operators using the HS method. To start with, we have briefly discussed the underlying principle of effective operator construction using Hilbert Series. Around the TeV scale, many BSM theories have been developed. Among them, we have adopted the following scenarios: Two Higgs doublet (2HDM) and  Minimal Left-Right Symmetric  (MLRSM) models. We have computed the complete set of independent dimension-6 operators in the ``Warsaw"-like basis for these two BSM scenarios using the HS method. We have considered the most generic Two Higgs Doublet Model, i.e., no $Z_2$ symmetry has been imposed.
In \cite{Crivellin:2016ihg, Karmakar:2017yek} the dimension-6 operators are computed for $Z_2$ symmetric 2HDM based on the method discussed in \cite{Buchmuller:1985jz,Grzadkowski:2010es}. We have noted a few erroneous operators in \cite{Crivellin:2016ihg} and corrected them.  We have further revisited the spectrum computation performed in \cite{Crivellin:2016ihg, Karmakar:2017yek} but with the most generic operator set. We have carried a similar task for the effective MLRSM case. For both the models, we have discussed the complete effective lagrangian exhaustively. We have further computed all the modified Feynman vertices after the inclusion of  dimension-6 operators.
We have rigorously estimated the impacts of those effective operators on different low energy observables, e.g., Weak mixing angle, Fermi constant, $\rho$ and oblique $(S,T,U)$ parameters. Then we have mentioned how the production, and decay of the massive BSM particles can be affected within this EFT framework in the presence of the modified interactions. In this connection, we have  outlined the possible sources of additional contributions to the magnetic moment interactions and rare lepton flavor, and number violating processes, e.g., neutrinoless double beta decay, radiative and three-body decays of charged leptons.  We have also derived the modified amplitude for the scalar four-point interactions and reasoned the necessity of further investigation of the theoretical constraints, e.g., unitarity, and vacuum stability. Then we have concluded with a note mentioning the possible future directions one can think of, based on this work.




\input{hilbert-observable.tex}

\pagebreak


\input{2HDM.tex}

\pagebreak


\input{MLRSM.tex}

\section{Impact of Effective operators on Observables}
\label{sec:impact-HDO}

\subsection{Outline to grab the impact of effective operators}
\label{subsec:outline-impact-HDO}

\begin{figure}[h!]
	\centering
	{
		\includegraphics[trim={2.5cm 0 2.5cm 0},scale=0.35]{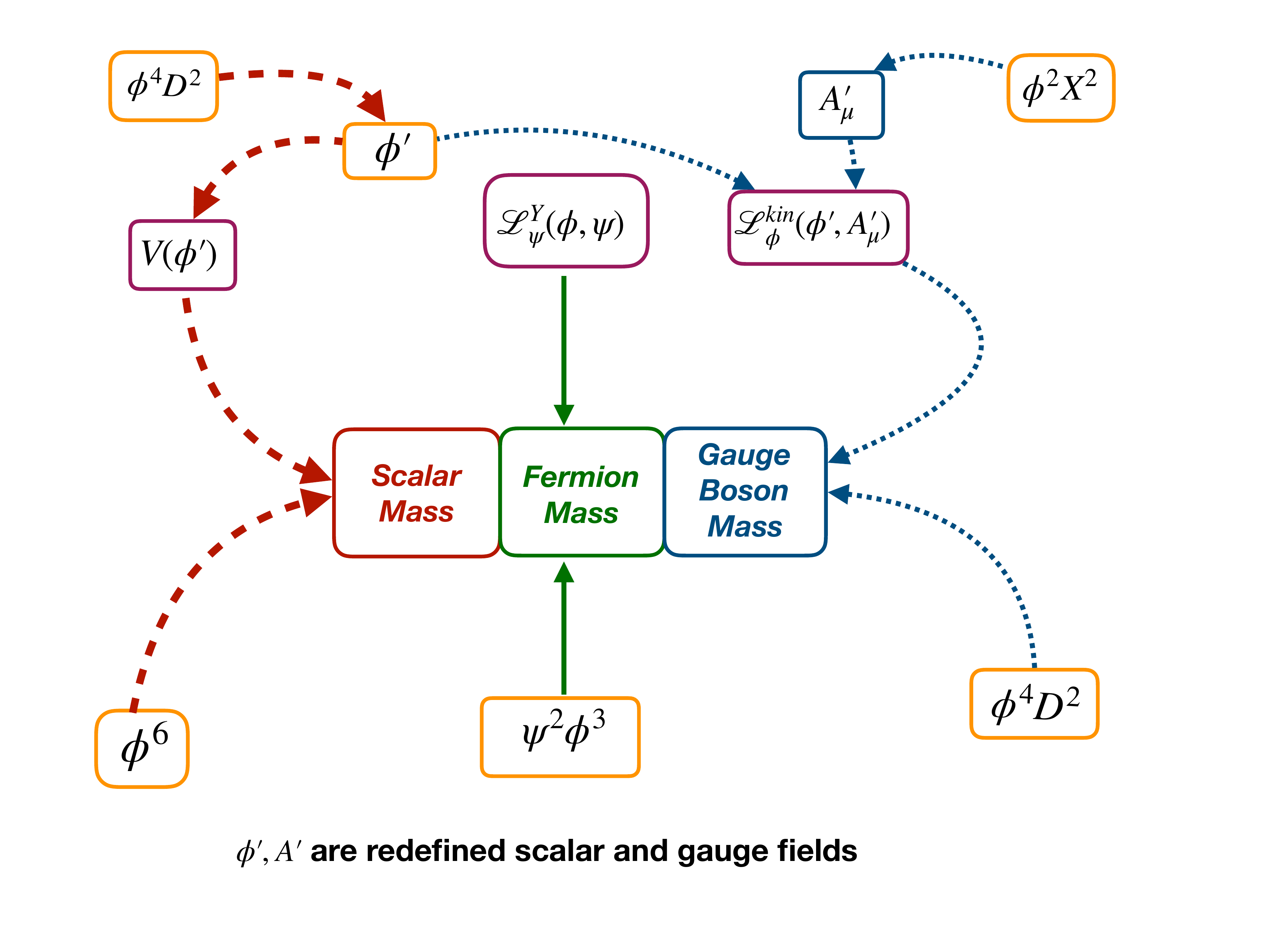}
	}
	\caption{Trajectories outlining the impact of effective operators on the spectrum. }\label{fig:spectrum-flowchart}
\end{figure}

In this paper, the impact of the effective dimension-6 operators is catalogued in a two-fold way: 
(i) shift in the mass spectrum and (ii) modification in the Feynman vertices. In the earlier sections, we have discussed how the mass spectrum gets modified in the presence of higher dimensional operators and computed the complete spectra for two scenarios -- 2HDM- and MLRSM-EFT in detail. We have prepared a schematic drawing,  Fig.~\ref{fig:spectrum-flowchart}, to summarise the earlier sections. 

In the presence of redefined fields ($A^{\prime}_{\mu}$, $\phi^{\prime}$), the effective lagrangian can be written as:
\small{\begin{eqnarray}\label{eq:lagrangian-ren-dim6}
		\mathcal{L}^{(4)+(6)}&=&\mathcal{L}^\text{kin}_{A}(A^{\prime}_{\mu})+\mathcal{L}^\text{kin}_{\phi}(\phi^{\prime},A^{\prime}_{\mu})+\mathcal{L}^\text{kin}_{\psi}(\psi,A^{\prime}_{\mu})+V(\phi^{\prime})\nonumber \\
		& & +\mathcal{L}^\text{Y}_\text{Dirac}(\phi^{\prime},\psi)+\mathcal{L}^\text{Y}_\text{Majorana}(\phi^{\prime},\psi) + \frac{1}{\Lambda^2} \sum_{j}\mathcal{C}_{j} \mathcal{O}_{j}.
\end{eqnarray}}
But this does not tell the full story. In the following section, we have systematically discussed how the interaction vertices get modified in the presence of these effective operators, and how they affect the low energy observables. To get an idea of how different observables are connected to different parts of the lagrangian (which is a function of shifted fields), we have prepared  Table~\ref{tab:observables}. We have organised the next part of the paper as follows: First, we have computed all the relevant modified Feynman vertices using the effective lagrangian in Eq.~\ref{eq:lagrangian-ren-dim6}. Then we have estimated the low energy observables, e.g., Weak mixing angle, Fermi constant, $\rho$ and oblique parameters. We have also outlined the possible sources of anomalous magnetic moment and theoretical constraints on the scalar potential to capture the impact of the effective operators. We have further highlighted how different collider processes, e.g., production and decay of heavy BSM particles, get modified within this effective theory framework. Instead of going for a detailed exploration, we have chosen selective channels relevant to the considered models. We have concluded this section by paving the path towards the possible phenomenological explorations of these interesting BSM scenarios in colliders and other experiments. In passing, we would like to mention that Baryon Number Violating operators listed in Tables \ref{tab:2HDM-psi4-BNV} and \ref{tab:MLRSM-psi4-b-viol}, are not being used in the further analysis.

\begin{table}[h!]
	\centering 
	\renewcommand{\arraystretch}{2.0}
	{\small\begin{tabular}{|c|c|}
			\hline
			{\textbf{Observables}}&{\textbf{Effective lagrangian terms}}\\ 
			\hline
			{Weak mixing angle} ($\theta_W$) &  \\ 
			\cline{1-1}
			Fermi constant ($G_F$)  & $\mathcal{L}^{kin}_{\psi}(\psi,A^{\prime}_{\mu})+  \frac{1}{\Lambda^2} [\psi^2 \phi^2 D +\psi^4]$    \\
			\cline{1-1}
			
			$\rho$ parameter &  \\
			\hline
			Oblique parameters ($S,T,U$)&$\mathcal{L}^{kin}_{A}(A^{\prime}_{\mu})+\mathcal{L}^{kin}_{\phi}(\phi^{\prime},A^{\prime}_{\mu})+\frac{1}{\Lambda^2} [\phi^4 D^2 +\phi^2 X^2]$\\  
			\hline
			Magnetic moment & $\mathcal{L}^{Y}_{Majorana}(\phi^{\prime},\psi)+\frac{1}{\Lambda^2} [\psi^2 \phi X]$\\ 
			\hline
			Lepton Flavor Violating (LFV) and&
			\multirow{2}{*}{$\mathcal{L}^{Y}_{Majorana}(\phi^{\prime},\psi) +\mathcal{L}^{kin}_{\phi}(\phi^{\prime},A^{\prime}_{\mu})+\frac{1}{\Lambda^2} [\phi^4 D^2 +\psi^2\phi^3]$}\\ 
			Lepton Number Violating (LNV) processes &
			\\
			\hline
			Scalar Quartic couplings &$V(\phi^{\prime})+\frac{1}{\Lambda^2} [\phi^6]$\\ 
			
			\hline
	\end{tabular}}
	\caption{Low energy observables and their origin in effective lagrangian.}
	\label{tab:observables}
\end{table}




\input{Weak_mixing_angle.tex}


\input{Oblique.tex}


\input{Magnetic-moment.tex}


\input{LFV-LNV.tex}


\input{Theoretical-constraints.tex}






\section{Conclusions and Remarks}

In this paper, we have implemented the Hilbert Series (HS) method to construct the complete set of independent effective  operators up to dimension-6 for two BSM scenarios: 2HDM and MLRSM. To start with, we have demonstrated the building blocks for the HS: Plethysthic Exponentials (PEs) and Haar measure. The PEs are constructed for bosonic and fermionic fields using the transformation properties of the fields under the internal gauge symmetry and Lorentz group. In this connection, we have also mentioned how this information is encrypted in the characters of those representations. Once the PEs are constructed, they are integrated on the group space using the Haar measure. These allow us to obtain the full Hilbert Series in a polynomial form. In our case, the index parameter for this polynomial expansion is the mass dimension. Thus we have been able to truncate the series based on the mass dimensionality of the operators. We have restricted ourselves up to dimension-6 operators. This method also takes care of two significant constraints related to the derivative operators: Equations of Motion (EOMs) and Integration by Parts (IBPs). Thus all the redundancies are removed and a complete set of independent operators for a given mass dimension is computed.

After constructing the operators, we have analysed how they can affect the predictions of the renormalizable version of the adopted scenarios: 2HDM and MLRSM. First, we have calculated the scalar and gauge field redefinitions due to $\phi^4 D^2$ and $\phi^2 X^2$ class of operators respectively and have estimated the corresponding modifications in the spectra. We have also focused on other sets of operators, which directly affect the spectrum, and included their contributions. Noting down all such contributions, we have provided the full mass spectra for scalar and gauge fields within the effective theory frameworks of 2HDM and MLRSM. We have further estimated the additional contributions to the charged and neutral fermion masses from the relevant operators. After completing the spectrum analysis, we have looked into the low energy observables which are precisely measured in the experiments, namely Weak mixing angle, Fermi constant, $\rho$ and oblique ($S,T,U$) parameters. For both the scenarios, we have analytically computed the contributions to these observables. We have also listed all the relevant modified Feynman vertices which affect the above mentioned low energy processes, and are also involved in the production and decay modes of the BSM particles leading to interesting phenomenological signatures at the LHC, FCC. We have concluded with a discussion on the impacts of these effective operators on the magnetic moments of charged fermions, LNV and LFV processes, and  the theoretical constraints, like tree unitarity using four-point scalar interactions and vacuum stability, i.e., boundedness criteria of the scalar potential.

Our future plan is to embed the 2HDM- and MLRSM-EFT frameworks in FeynRules using Refs.~\cite{Dedes:2019uzs,Alwall:2014hca,Degrande:2011ua,Alloul:2013bka,Dedes:2017zog} to adjudge them in the light of collider and low energy experiments and precision observable tests.

\section{Acknowledgments}

This work is supported by the Department of Science and
Technology, Government of India, under the Grant IFA12/PH/34 (INSPIRE Faculty Award);  the Science and Engineering Research Board, Government of India, under the agreement
SERB/PHY/2016348 (Early Career Research Award), and Initiation Research Grant, agreement no. IITK/PHY/2015077, by IIT Kanpur. The Feynman diagrams have been generated using JaxoDraw \cite{Binosi:2003yf}.

\section*{}

\providecommand{\href}[2]{#2}
\addcontentsline*{toc}{section}{}
\bibliographystyle{JHEP}
\bibliography{EFT-2HDM+MLRSM}

\end{document}

%% file: hilbert-observable.tex
\section{Hilbert Series and Effective Operators}

In this section, we have briefly sketched the Hilbert Series (HS) method  \cite{Henning:2015daa,Kobach:2017xkw,Henning:2015alf,Lehman:2015coa,Trautner:2018ipq,Trautner:2018ipq,Henning:2017fpj,Hanany:2010vu,Feng:2007ur,Lehman:2015via} based on which our BSM-EFT construction has been developed.  There are two important ingredients to define the HS: (i) Plethystic Exponentials (PEs) and (ii) the Haar measure.
The generic form of the Hilbert Series can be given as \cite{Hanany:2010vu,Feng:2007ur,Henning:2017fpj,Henning:2015alf}:
{\small\begin{equation}\label{eq:HS}
		\mathcal{H}[\phi] = \prod^{n}_{j=1} \int_{\mathcal{G}_j} \hspace{0.1cm}\underbrace{d\mu_j}_{\text{Haar}\hspace{0.1cm}\text{Measure}} \hspace{0.1cm}\underbrace{PE[\varphi,  R]}_{\hspace{0.1cm}\text{Plethystic}\hspace{0.1cm}\text{Exponential}},
\end{equation}}
where, $\varphi$ is a  spurion variable which can be either a scalar ($\phi$), a fermion ($\psi$) or a gauge field ($A_{\mu}$).  Here, the PEs are integrated on the symmetry (group) space using the Haar measure $d\mu$. The Plethystic Exponentials for bosonic and fermionic fields are written as \cite{Hanany:2010vu,Feng:2007ur,Henning:2017fpj,Henning:2015alf}:
{\small\begin{eqnarray}\label{eq:PE}
		PE[\phi, R] &=& \exp\left[\sum_{r=1}^{\infty}\frac{\phi^r\chi_R(z^r_j)}{r}\right],\\
		PE[\psi, R] &=& \exp\left[\sum_{r=1}^{\infty}(-1)^{r+1}\frac{\psi^r\chi_R(z^r_j)}{r}\right],
\end{eqnarray}}
respectively.
Here, $\chi_R(z^r_j)$ is the ``Weyl" character corresponding to the representation $R$ of the  spurion fields ($\phi,\psi$) under the symmetry group $\mathcal{G}_j$.

For a given group $\mathcal{G}$, the form of the Haar measure can be expressed as \cite{Weyl,Dieck,Gray:2008yu}:
{\small\begin{equation}\label{eq:Haar}
		\int_{\mathcal{G}} d\mu_{\mathcal{G}} = \frac{1}{(2\pi i)^k}\oint_{|z_1|=1}....\oint_{|z_{\sigma}|=1}\frac{dz_1}{z_1}....\frac{dz_{\sigma}}{z_{\sigma}}\prod_{\alpha^+}\left[1-\prod_{m=1}^{k}(z_m)^{\alpha_m^+}\right],
\end{equation}}
where $\alpha_m^+$,  and $k$ are the positive roots and rank of the group $\mathcal{G}$ respectively.

Thus it is evident that the HS construction relies on two pieces of information: (a) the DOF, i.e., particle content and (b) the transformation properties of the particles under the given symmetries. If we restrict ourselves to the quantum fields, i.e., particles then the information we have gathered so far is sufficient to compute the HS. But we are interested in a complete set of operators, and that includes operators involving {\it {derivatives}}. Since the derivatives transform under space-time symmetry, we need knowledge of the conformal group along with the internal symmetries\footnote{The transformation properties of the fields, having non-zero spin, under the space-time symmetry group must be taken into account. The forthcoming discussion addresses this as well.}. Due to the intrinsic properties of derivative, additional constraints might appear in the form of Equations of Motion (EOMs), and Integration by Parts (IBPs). It leads to the overcounting in the operators. Thus we should carefully identify the redundancy in the operator set to construct only independent operators so that they can form a basis. Now we will discuss how the HS construction can automatically take care of this issue, which is one of the most significant merits of this program.

To do so, we need to construct the characters of the derivative operator $\mathcal{D}$ under the conformal group $SO(4,2)$\footnote{As we are working in 3+1 dimensional space-time.} \cite{Dolan:2005wy,Gruber:1975sn,Dobrev:2004tk,Bourget:2017kik,Mack:1975je,Ferrara:2000nu,Cardy:1991kr, Henning:2015alf,Henning:2015daa,Henning:2017fpj}. The extra two dimensions ($+2$) increase the rank of the conformal group by one unit which appears as a scaling dimension ($\Delta$) of the representation. In our analysis, $\Delta$ is the mass dimension of the operator/fields, e.g., 1 for boson, 3/2 for fermions, etc. The remaining part of the conformal group is the Lorentz group. But as the Lorentz group (LG) is not compact, its unitary representation is infinite dimensional \cite{DiFrancesco:1997nk}. Thus we will realize Lorentz group as $SO(4,\mathbb{C}) \simeq SU(2)_L \otimes SU(2)_R$. The Verma module characters \cite{DiFrancesco:1997nk} of  $SO(4,\mathbb{C})$ with highest weight $(l_1, l_2)$ $\big[ \chi^{(4)}_{(l_1,l_2)}(x_1,x_2) \big]$ can be written as the  product of the Verma module characters of $SU(2)_L$ with highest weight $j_1$ $\big[\chi_{j_1}(x)\big]$, and 
$SU(2)_R$ with highest weight $j_2$ $\big(\chi_{j_2}(y)\big)$ as \cite{Dolan:2005wy,Siegel:1988gd,Gruber:1975sn,Dobrev:2004tk,Bourget:2017kik,Henning:2015alf,Henning:2015daa,Henning:2017fpj}:
\small\begin{equation}
	\chi^{(4)}_{(l_1,l_2)}(x_1,x_2) = \chi_{j_1}(y_1)*\chi_{j_2}(y_2),
\end{equation}
where $l_1$ = $j_1+j_2$, $l_2$ = $j_1-j_2$, $x_1$ = $y_1^{1/2}y_2^{1/2}$, and $x_2=y_1^{1/2}y_2^{-1/2}$.
Using this generic notion as discussed by F. A. Dolan in \cite{Dolan:2005wy}, 
the \textit{characters of unitary irreducible representations} of $SO(4,2)$ can be given as \cite{Dolan:2005wy,Henning:2015alf,Henning:2015daa,Minwalla:1997ka,Vasiliev:2004cm,Barabanschikov:2005ri,Henning:2017fpj}:
\small\begin{eqnarray}
	\chi^{(4)}_{[j_1+j_2+2;j_1,j_2]}(s,y_1,y_2) &=& s^{j_1+j_2+2}\left(\chi_{j_1}(y_1)\chi_{j_2}(y_2)-s\chi_{j_1-\frac{1}{2}}(y_1)\chi_{j_2-\frac{1}{2}}(y_2)\right)P^{(4)}(s,y_1,y_2),\nonumber\\
	\chi^{(4)}_{[j_1+1;j_1]}(s,y_1,y_2) &=& s^{j_1+1}\left(\chi_{j_1}(y_1)-s\chi_{j_1-\frac{1}{2}}(y_1)\chi_{\frac{1}{2}}(y_2)+s^2\chi_{j_1-1}(y_1)\right)P^{(4)}(s,y_1,y_2),\nonumber\\
	\chi^{(4)}_{[j_2+1;j_2]}(s,y_1,y_2) &=& s^{j_2+1}\left(\chi_{j_2}(y_2)-s\chi_{j_2-\frac{1}{2}}(y_2)\chi_{\frac{1}{2}}(y_1)+s^2\chi_{j_2-1}(y_2)\right)P^{(4)}(s,y_1,y_2), \nonumber
\end{eqnarray}
where $P^{(4)}(s,y_1,y_2)$ is the momentum generating function for  $SO(4,\mathbb{C})$.
Now using this construction, we can write down the necessary and relevant characters for our analysis \cite{Dolan:2005wy,Henning:2015alf,Henning:2015daa,Henning:2017fpj}:
{\small\begin{align}\label{eq:lorentz_char}
		\chi^{(4)}_{[1;(0,0)]}(\mathcal{D}, \alpha, \beta) &=\mathcal{D} P^{(4)}(\mathcal{D}, \alpha, \beta) \times \Big[ 1 - \mathcal{D}^2 \Big],\\
		\chi^{(4)}_{[\frac{3}{2},(\frac{1}{2},0)]}(\mathcal{D}, \alpha, \beta) &= \mathcal{D}^{\frac{3}{2}} P^{(4)}(\mathcal{D}, \alpha, \beta) \times \Big[\alpha + \frac{1}{\alpha} -\mathcal{D}\Big(\beta + \frac{1}{\beta}\Big)\Big],\\
		\chi^{(4)}_{[\frac{3}{2};(0,\frac{1}{2})]}(\mathcal{D}, \alpha, \beta) &= \mathcal{D}^{\frac{3}{2}} P^{(4)}(D, \alpha, \beta) \times \Big[\beta + \frac{1}{\beta} -\mathcal{D}\Big(\alpha + \frac{1}{\alpha}\Big)\Big],\\
		\chi^{(4)}_{[2;(1,0)]}(\mathcal{D}, \alpha, \beta) &= \mathcal{D}^2 P^{(4)}(\mathcal{D},\alpha ,\beta) \times \Big[\alpha^2+\frac{1}{\alpha ^2}+1 - \mathcal{D}\Big(\alpha +\frac{1}{\alpha }\Big) \Big(\beta +\frac{1}{\beta }\Big) + \mathcal{D}^2\Big],\\
		\chi^{(4)}_{[2;(0,1)]}(\mathcal{D}, \alpha, \beta) &= \mathcal{D}^2 P^{(4)}(\mathcal{D},\alpha ,\beta) \times \Big[ \beta^2+\frac{1}{\beta ^2}+1 - \mathcal{D}\Big(\alpha +\frac{1}{\alpha }\Big) \Big(\beta +\frac{1}{\beta }\Big) + \mathcal{D}^2\Big],    
\end{align}}
where the momentum generating function $P^{(4)}(\mathcal{D}, \alpha, \beta)$ can be written as \cite{Henning:2015alf,Henning:2015daa,Henning:2017fpj},
{\small\begin{equation}
		P^{(4)}(\mathcal{D}, \alpha, \beta) = \Bigg[\left(1-\mathcal{D}\alpha  \beta\right) \left(1-\frac{\mathcal{D}}{\alpha  \beta }\right) \left(1-\frac{\mathcal{D}\alpha}{\beta}\right) \left(1-\frac{\mathcal{D}\beta}{\alpha}\right)\Bigg]^{-1}.
\end{equation}}

In the presence of derivative operators which transform under the Lorentz group, the Hilbert Series in Eq.~\ref{eq:HS} is redefined in the following form
\cite{Henning:2017fpj,Henning:2015alf,Henning:2015daa}:
{\small\begin{equation}\label{eq:HS-modified}
		\mathcal{H}[\phi] =\prod^{n}_{j=1} \int_{LG} \hspace{0.05cm} {d\mu_{_{LG}}}  \int_{\mathcal{G}_j} \hspace{0.1cm}{d\mu_j} \hspace{0.1cm}{PE[\varphi, \mathcal{D}, R]}_{\hspace{0.1cm}} \frac{1}{    P^{(4)}(\mathcal{D}, \alpha, \beta)},
\end{equation}}
where, $\mathcal{D}$ is the spurion variable corresponding to the covariant derivative operator. 
The Plethystics in Eq.~\ref{eq:PE} also get modified as \cite{Henning:2017fpj,Henning:2015alf,Henning:2015daa}:
{\small\begin{eqnarray}\label{eq:PE-modified}
		PE[\phi,\mathcal{D}, R] &=& \exp\left[\sum_{r=1}^{\infty}  \left(\frac{\phi}{\mathcal{D}^{\Delta_{\phi}} }\right)^r \frac{\chi_R(z^r_j,\alpha^r,\beta^r)}{r}\right],\\
		PE[\psi, \mathcal{D},R] &=& \exp\left[  \sum_{r=1}^{\infty}(-1)^{r+1}  \left(\frac{\psi}{\mathcal{D}^{\Delta_{\psi}} }\right)^r \frac{\chi_R(z^r_j,\alpha^r,\beta^r)}{ r}\right].
\end{eqnarray}}

The Haar measure for Lorentz group is given as \cite{Henning:2017fpj,Henning:2015alf,Henning:2015daa}:
{\small\begin{equation}\label{eq:LG}
		d\mu_{_{LG}} = \Big[\frac{1}{2 \alpha}\left(1-\alpha^2\right) \left(1-\frac{1}{\alpha^2}\right)\Big] \Big[\frac{1}{2 \beta}\left(1-\beta^2\right) \left(1-\frac{1}{\beta^2}\right)\Big].
\end{equation}}

In this paper, we have worked with the following groups: $U(1), SU(2), SU(3)$. The Haar measure for these groups are given as \cite{Weyl,Dieck,Gray:2008yu}:
{\small\begin{eqnarray}\label{eq:Haar-ex}
				d\mu_{U(1)} &=& \frac{1}{x},\\
				d\mu_{SU(2)} &=& \frac{1}{2 y}\left(1-y^2\right) \left(1-\frac{1}{y^2}\right),\nonumber \\
				d\mu_{SU(3)} &=& \frac{1}{6z_1z_2}(1-z_1z_2)\left(1-\frac{z_1^2}{z_2}\right)\left(1-\frac{z_2^2}{z_1}\right) \left(1-\frac{1}{z_1 z_2}\right)\left(1-\frac{z_1}{z_2^2}\right)\left(1-\frac{z_2}{z_1^2}\right). \nonumber
\end{eqnarray}}\label{eqn:haar_measure}

We have worked with few specific representations under these above mentioned internal symmetry groups. We have enlisted the characters of those representations \cite{Hanany:2008sb} below :
\begin{itemize}
	\item $U(1)$
		{\small\begin{equation}
			z^q,\;\;\;  q \text{\; is\; the\; $U(1)$\; charge\; of\; the\; representation}, 
	\end{equation}}\label{eqn:U1_char}
		\item $SU(2)$ 
	\begin{eqnarray}
		\left[1\right] & \equiv &  2 = \bar{2} = z_1 + \frac{1}{z_1}, \\
	  \left[2\right] & \equiv & 3 \equiv 2 \otimes 2 - 1  \equiv  \left(z_1 + \frac{1}{z_1}\right)^2 - 1 = z_1^2 + \frac{1}{z_1^2} + 1, 
	\end{eqnarray}\label{eqn:su2_char}
	\item $SU(3)$
	{\small\begin{eqnarray}
		\left[1,0\right] &\equiv& 3 = z_1 + \frac{1}{z_2} + \frac{z_2}{z_1},\\
		\left[0,1\right] &\equiv& \bar{3} = \bar{2} = z_2 + \frac{1}{z_1} + \frac{z_1}{z_2},\\
		\left[1,1\right] &\equiv&
		8 \equiv 3 \otimes \bar{3} - 1 = \left(z_1 + \frac{1}{z_2} + \frac{z_2}{z_1}\right)\left(z_2 + \frac{1}{z_1} + \frac{z_1}{z_2}\right)-1\\
		&=&\frac{z_1^2}{z_2}+\frac{z_1}{z_2^2}+z_1z_2+\frac{1}{z_1z_2}+\frac{z_2}{z_1^2}+\frac{z_2^2}{z_1}+2.
	\end{eqnarray}}\label{eqn:su3_char}
\end{itemize}
\noindent
In the above equations, the Dynkin indices  of the respective representations are depicted as $[..]$.

\begin{figure}[h!]
	\centering
	{
		\includegraphics[trim={2cm 0 1.2cm 0},scale=0.4]{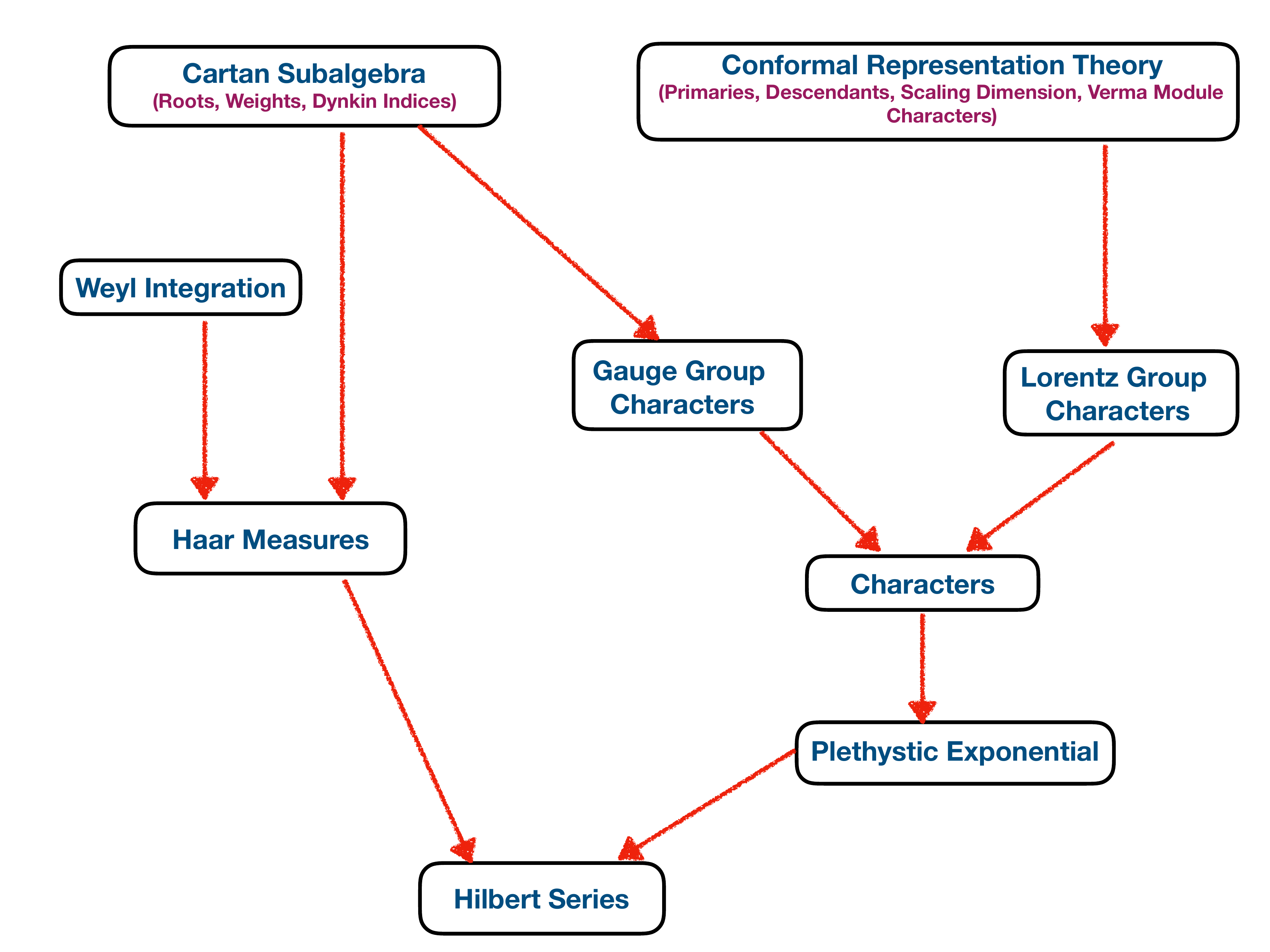}
	}
	\caption{Flow chart depicting the Hilbert Series construction. }\label{fig:HS}
\end{figure}

Based on the HS method, using the information above the complete and independent set of effective operators up to dimension-6 has been constructed for 2HDM- and MLRSM-EFT frameworks mimicking the ``Warsaw" basis. First we have constructed the raw HS\footnote{ In Ref.~\cite{Henning:2015daa} the derivative operator construction for specific cases are performed using ``{\it Macaulay2}" \cite{Grayson}, an ``{\it algorithm based on  Commutative Algebra and Algebraic Geometry}" \cite{Schenck,Cox,Sturmfels}} \cite{Henning:2015alf,Henning:2015daa,Henning:2017fpj}, then we have refined those operators and presented them in a workable format. In Fig.~\ref{fig:HS}, we have sketched the integrated steps that lead to the Hilbert Series. 
\clearpage

%% file: 2HDM.tex
\section{Two Higgs Doublet Model (2HDM)}

The Two Higgs Doublet Model (2HDM) extends the Standard Model (SM) in the most elementary way through the addition of an identical second Higgs doublet \cite{Gunion:1989we,Gunion:2002zf,Branco:2011iw,Carena:2013ooa,Haber:2013mia,Chen:2013jvg,Mrazek:2011iu,Dev:2014yca,Bhattacharyya:2015nca,Crivellin:2015hha,Crivellin:2016ihg}. In 2HDM the internal gauge symmetry is the same as the SM: $SU(3)_C\otimes SU(2)_L\otimes U(1)_Y$. In Table~\ref{tab:2HDM-fields}, we have listed the particle content of this model, and have depicted their representations and charges. In the following section, we have revisited the work of \cite{Crivellin:2016ihg} using our set of operators (which are not the same as in Ref.~\cite{Crivellin:2016ihg}). Before going into the EFT description of the 2HDM scenario, we have provided a brief discussion on the spectrum computation of this model following the path paved in \cite{Crivellin:2016ihg}.

\begin{table}[h!]
	\centering
	\renewcommand{\arraystretch}{1.75}
	\begin{tabular}{|c|c|c|c|c|c|}
		\hline
		\textbf{Field} & \textbf{$SU(3)_C$} & \textbf{$SU(2)_{L}$}&\textbf{$U(1)_{Y}$}&Spin&Lorentz Group\\
		\hline
		$\phi_1$    &1&2&1/2&0&Scalar\\
		$\phi_2$    &1&2&1/2&0&Scalar\\
		\hline
		$Q_L$       &3&2&1/6&1/2&Spinor\\
		$(u^c)_L$     &$\bar{3}$&1&2/3&1/2&Spinor\\
		$(d^c)_L$     &$\bar{3}$&1&-1/3&1/2&Spinor\\
		$L_L$       &1&2&-1/2&1/2&Spinor\\
		$(e^c)_L$     &1&1&-1&1/2&Spinor\\
		\hline
		$G^A_{\mu}$ &8&1&0&1&Vector\\
		$W^I_{\mu}$ &1&3&0&1&Vector\\
		$B_{\mu}$   &1&1&0&1&Vector\\
		\hline
	\end{tabular}
	\caption{2HDM: Quantum numbers of the fields.} \label{tab:2HDM-fields}
\end{table}

\subsubsection*{The Renormalizable Lagrangian}

Based on the transformation properties of the particles, the renormalizable lagrangian for 2HDM can be written as \cite{Gunion:1989we,Gunion:2002zf,Crivellin:2016ihg}:
{\small\begin{eqnarray}\label{eq:2HDM-ren-lag}
	\mathcal{L}^{(4)}_{2HDM}& = &-\frac{1}{4} Tr[G_{\mu\nu}G^{\mu\nu}] -\frac{1}{4} Tr[W_{\mu\nu}W^{\mu\nu}] -\frac{1}{4}B_{\mu\nu}B^{\mu\nu}\nonumber \\
	&&+\; (D_{\mu}\phi_1)^{\dagger}(D_{\mu}\phi_1) + (D_{\mu}\phi_2)^{\dagger}(D_{\mu}\phi_2) - V(\phi_1 , \phi_2)\nonumber \\
	&&+\; i(\bar{L}\gamma^\mu D_\mu L + \bar{Q}\gamma^\mu D_\mu Q + \bar{e}\gamma^\mu D_\mu e + \bar{u}\gamma^\mu D_\mu u + \bar{d}\gamma^\mu D_\mu d) + L_Y+h.c.,
	\end{eqnarray}}
where the components of field strength tensors are given as \cite{Gunion:1989we,Gunion:2002zf,Crivellin:2016ihg}:
{\small\begin{eqnarray}\label{eq:2HDM-field-tensor}
	G^A_{\mu\nu} &=& \partial_\mu G^A_\nu - \partial_\nu G^A_\mu + g_3           f^{ABC}G^B_\mu G^C_\nu,\nonumber\\
	W^I_{\mu\nu} &=& \partial_\mu W^I_\nu - \partial_\nu W^I_\mu + g \epsilon^{IJK}W^J_\mu W^K_\nu,\\
	B_{\mu\nu} &=& \partial_\mu B_\nu - \partial_\nu B_\mu \nonumber.
\end{eqnarray}}
for $SU(3)_C, SU(2)_L, U(1)_Y$ respectively with $\{ A,B,C\} \in (1,..,8)$,  $\{ I,J,K\} \in (1,2,3)$. The general form of the covariant derivative can be given as \cite{Gunion:1989we,Gunion:2002zf,Crivellin:2016ihg}:
{\small\begin{equation}\label{eq:2HDM-cov-der}
	D_\mu   \equiv (\partial_\mu + ig_3 \frac{T^A}{2}G^A_\mu + ig  \frac{\tau^I}{2}W^I_\mu + ig'Y B_\mu).
	\end{equation}}
Let us consider that $D_\mu$ acts on a quantum field ($\varphi$) which is charged under some gauge quantum number, e.g.,  $SU(3)_C$. Then the covariant derivative $D_\mu$ will contain a term proportional to the respective gauge field, e.g.,  $ ig_3 \frac{T^A}{2}G^A_\mu $. 
Here, $g_3,g,g'$ are the gauge couplings and $G^A_\mu, W^I_\mu, B_\mu$ are gauge fields corresponding to $SU(3)_C, SU(2)_L, U(1)_Y$ respectively. $Y$ is the hypercharge of the field on which this operator is acting.  Here, $\frac{T^A}{2}$ and $\frac{\tau^I}{2}$ are the normalized generators of $SU(3)_C$ and $SU(2)_L$ groups respectively. 
\\
The Yukawa coupling is given as \cite{Gunion:1989we,Gunion:2002zf,Crivellin:2016ihg}:
{\small\begin{eqnarray}\label{eq:2HDM-Yukawa}
	\mathcal{L}_Y &=&- y^e_{1}\bar{L}\phi_1 e - y^e_{2}\bar{L}\phi_2 e - y^d_{1}\bar{Q}\phi_1 d- y^d_{2}\bar{Q}\phi_2 d - y^u_{1}\bar{Q}\tilde{\phi}_1 u - y^u_{2}\bar{Q}\tilde{\phi}_2 u + h.c.,
\end{eqnarray}}
where $\tilde{\phi} = i\tau_2\phi^{*}$, and $y_{1,2}^{e,u,d}$ are the respective Yukawa coupling matrices.
\\
The scalar potential for generic 2HDM is given as \cite{Gunion:1989we,Gunion:2002zf}:
\begin{eqnarray}\label{eq:2HDM-scalar-pot}
	V(\phi_1, \phi_2) &= & m^2_{11}(\phi_1^{\dagger}\phi_1) + m^2_{22}(\phi_2^{\dagger}\phi_2) - m^2_{12}(\phi_1^{\dagger}\phi_2 + \phi_2^{\dagger}\phi_1) + \lambda_1(\phi_1^{\dagger}\phi_1)^2 + \lambda_2(\phi_2^{\dagger}\phi_2)^2  \nonumber \\
	& & +  \lambda_3(\phi_1^{\dagger}\phi_1)(\phi_2^{\dagger}\phi_2) + \lambda_4(\phi_1^{\dagger}\phi_2)(\phi_2^{\dagger}\phi_1) +
	\frac{1}{2}\lambda_5[(\phi_1^{\dagger}\phi_2)^2 + (\phi_2^{\dagger}\phi_1)^2]  \\
	& & +  \textcolor{purple}{\big(\lambda_6(\phi_1^{\dagger}\phi_1) + \lambda_7(\phi_2^{\dagger}\phi_2)\big)(\phi_1^{\dagger}\phi_2 + \phi_2^{\dagger}\phi_1)}.\nonumber
	\end{eqnarray}
 Here, the colored parts signify the $Z_2$ symmetry ($\phi_1  \to -\phi_1, \phi_2 \to \phi_2$) violation. These  $Z_2$ violating terms in the above equation are not considered in \cite{Crivellin:2016ihg}.

In this model the spontaneous symmetry breaking (SSB) of SM gauge symmetry is induced by the vacuum expectation values (VEVs) of both the Higgs doublets. After SSB, the Higgs fields are given as \cite{Gunion:1989we,Gunion:2002zf,Crivellin:2016ihg}:
\begin{equation}\label{eq:2HDM-scalar-ssb}
\phi_1 = \begin{pmatrix}
\phi^+_1\\
(v_1 + h_1 + ia_1)/\sqrt{2}\\
\end{pmatrix}, \hspace{2cm} 
\phi_2 = \begin{pmatrix}
\phi^+_2\\
(v_2 + h_2 + ia_2)/\sqrt{2}\\
\end{pmatrix},
\end{equation}
where $v_1$ and $v_2$ are the VEVs of $\phi_1$ and $\phi_2$ respectively. The electro-weak VEV is given as $v=\sqrt{v_1^2+v_2^2}$. Among these unphysical scalar fields, we have two CP-odd scalars $(a_1,a_2)$, two CP-even scalars $(h_1,h_2)$ and two charged scalars $(\phi_1^+,\phi_2^+)$.

After  SSB, the scalar mass terms in the lagrangian is given as \cite{Gunion:1989we,Gunion:2002zf,Crivellin:2016ihg}:
{\small\begin{equation}\label{eq:2HDM-scalar-mass-matrix}
	\mathcal{L}^{(4)}_{M_H} = \frac{1}{2} \begin{pmatrix}
	a_{1} \ \ a_{2}
	\end{pmatrix}m^2_{i}\begin{pmatrix}
	\hspace{0.2cm}a_{1}\hspace{0.2cm} \\ \hspace{0.2cm}a_{2}\hspace{0.2cm}\end{pmatrix} 
	+ 
	\begin{pmatrix}
	\phi^-_{1} \ \ \phi^-_{2}
	\end{pmatrix}m^2_{+}\begin{pmatrix}
	\hspace{0.2cm}\phi^+_{1}\hspace{0.2cm} \\ \hspace{0.2cm}\phi^+_{2}\hspace{0.2cm}
	\end{pmatrix} 
	+ \frac{1}{2} \begin{pmatrix}
	h_{1} \ \ h_{2}
	\end{pmatrix}m^2_{r}\begin{pmatrix}
	\hspace{0.2cm}h_{1}\hspace{0.2cm} \\ \hspace{0.2cm}h_{2}\hspace{0.2cm}
	\end{pmatrix}.
	\end{equation}}
The scalar mass matrices are explicitly written as \cite{Crivellin:2016ihg}:
{\small\begin{equation}\label{eq:2HDM-scalar-mass-CP-odd-tree}
	m^2_{i} =\left(-m_{12}^2+v_1 v_2 \lambda _5+\frac{1}{2} v_1^2 \lambda _6+\frac{1}{2} v_2^2 \lambda _7\right)
	\begin{pmatrix}
	-\frac{v_{2}}{v_{1}} & 1  \\
	1  &  -\frac{v_{1}}{v_{2}}
	\end{pmatrix}, \nonumber
	\end{equation}}

{\small\begin{equation}\label{eq:2HDM-scalar-mass-CP-even-tree}
	m^2_{r} =\begin{pmatrix}
	-\frac{\lambda _7 v_2^3}{2 v_1}+\frac{3}{2} v_1 \lambda _6 v_2+\frac{m_{12}^2 v_2}{v_1}+2 v_1^2 \lambda _1 & -m_{12}^2+v_1 v_2 \left(\lambda _3+\lambda _4+\lambda _5\right)+\frac{3}{2} v_1^2 \lambda _6+\frac{3}{2} v_2^2 \lambda _7 \\
	-m_{12}^2+v_1 v_2 \left(\lambda _3+\lambda _4+\lambda _5\right)+\frac{3}{2} v_1^2 \lambda _6+\frac{3}{2} v_2^2 \lambda _7 & -\frac{\lambda _6 v_1^3}{2 v_2}+\frac{3}{2} v_2 \lambda _7 v_1+\frac{m_{12}^2 v_1}{v_2}+2 v_2^2 \lambda _2 \\
	\end{pmatrix}, \nonumber
	\end{equation}}

{\small\begin{equation}\label{eq:2HDM-scalar-mass-charge-tree}
	m^2_{+} =\left(-m_{12}^2+\frac{1}{2} v_1 v_2 \left(\lambda _4+\lambda _5\right)+\frac{1}{2} v_1^2 \lambda _6+\frac{1}{2} v_2^2 \lambda _7\right)\begin{pmatrix}
	-\frac{v_{2}}{v_{1}} & 1  \\
	1  &  -\frac{v_{1}}{v_{2}}
	\end{pmatrix}.
	\end{equation}}

Using the potential minimization conditions \cite{Gunion:1989we,Gunion:2002zf,Crivellin:2016ihg}
\small{\begin{eqnarray}\label{eq:2HDM-scalar-pot-min-tree}
	\frac{\partial V(\phi_{1},\phi_{2})}{\partial v_{1}}&=&m_{11}^2 v_1-m_{12}^2 v_2+\lambda _1 v_1^3+\frac{1}{2} \lambda _3 v_2^2 v_1+\frac{1}{2} \lambda _4 v_2^2 v_1+\frac{1}{2} \lambda _5 v_2^2 v_1\textcolor{purple}{+\frac{3}{2} \lambda _6 v_2 v_1^2+\frac{1}{2} \lambda _7 v_2^3} \nonumber \\
	&=&0, \nonumber\\ 
	\frac{\partial V(\phi_{1},\phi_{2})}{\partial v_{2}}&=&-m_{12}^2 v_1+m_{22}^2 v_2+\frac{1}{2} \lambda _3 v_2 v_1^2+\frac{1}{2} \lambda _4 v_2 v_1^2+\frac{1}{2} \lambda _5 v_2 v_1^2+\lambda _2 v_2^3\textcolor{purple}{+\frac{1}{2} \lambda _6 v_1^3+\frac{3}{2} \lambda _7 v_2^2 v_1}\nonumber \\
	&=&0,
	\end{eqnarray}}
we can recast the mass-square parameters of the potential as \cite{Gunion:1989we,Gunion:2002zf,Crivellin:2016ihg}:
\small{\begin{eqnarray}\label{eq:2HDM-min-relations-tree}
	m_{11}^2  &=& \frac{m_{12}^2 v_2}{v_1}-\frac{1}{2} \lambda _3 v_2^2-\frac{1}{2} \lambda _4 v_2^2-\frac{1}{2} \lambda _5
	v_2^2-\lambda _1 v_1^2\textcolor{purple}{-\frac{3}{2} \lambda _6 v_1 v_2-\frac{\lambda _7 v_2^3}{2 v_1}} , \nonumber\\ 
	m_{22}^2 &=& \frac{m_{12}^2 v_1}{v_2}-\frac{1}{2} \lambda _3 v_1^2-\frac{1}{2}
	\lambda _4 v_1^2-\frac{1}{2} \lambda _5 v_1^2-\lambda _2 v_2^2\textcolor{purple}{-\frac{\lambda _6 v_1^3}{2 v_2}-\frac{3}{2} \lambda _7 v_2 v_1}.
	\end{eqnarray}}

The physical scalar fields ($	\Phi^{\prime}$) are related to the unphysical ones ($	\Phi$) through the relations below, which further can be used to obtain physical scalar mass matrices \cite{Gunion:1989we,Gunion:2002zf,Crivellin:2016ihg}:
{\small\begin{eqnarray}\label{eq:2HDM-scalar-field-rotation}
	\begin{pmatrix}
	\Phi^{\prime}
	\end{pmatrix}=\begin{pmatrix}
	\cos\beta_{j} & \sin\beta_{j} \\
	-\sin\beta_{j} & \cos\beta_{j}
	\end{pmatrix}
	\begin{pmatrix}
	\Phi
	\end{pmatrix}, \hspace{0.5cm}
	\beta_{j} \equiv \{r,i,+\}.
	\end{eqnarray}}
Here,
\begin{eqnarray}\label{eq:2HDM-scalar-mass-basis}
\Phi &\equiv& \Bigg\{\begin{pmatrix}
h_{1} \\ h_{2}
\end{pmatrix},\begin{pmatrix}
a_{1} \\ a_{2}
\end{pmatrix},\begin{pmatrix}
\phi^{+}_{1} \\ \phi^{+}_{2}
\end{pmatrix}\Bigg\}, \hspace{0.2cm}
\Phi^{\prime} \equiv \Bigg\{\begin{pmatrix}
H_{1} \\ H_{2}
\end{pmatrix},\begin{pmatrix}
A \\ \tilde{G}
\end{pmatrix},\begin{pmatrix}
H^{\pm} \\ \tilde{G}^{\pm}
\end{pmatrix}\Bigg\}.
\end{eqnarray}
While computing the physical states, we find two Goldstone bosons (CP-odd: $\tilde{G}$,  charged: $\tilde{G}^{\pm}$) which get eaten up through the Higgs mechanism and generate masses for $Z$ and $W^{\pm}$ gauge bosons. 
The CP-odd, charged scalar and CP-even mass matrices are diagonalized by rotation angles \cite{Gunion:1989we,Gunion:2002zf,Crivellin:2016ihg}:
{\small\begin{eqnarray}\label{eq:2HDM-scalar-rotation-angle-tree}
	\tan{2\beta}_{+} & = & \tan{2\beta}_{i} =  \frac{2 v_{2} {v_{1}}}{v_{2}^2-v_{1}^2},\\
		\tan{2\beta}_{r} &= &\frac{2 v_1 v_2 \left(-2 m_{12}^2\textcolor{purple}{+3 \lambda _6 v_1^2}+v_2 \left(2 \lambda _3 v_1+2 \lambda _4 v_1+2 \lambda _5 v_1\textcolor{purple}{+3 \lambda _7 v_2}\right)\right)}{2 m_{12}^2 \left(v_1^2-v_2^2\right)\textcolor{purple}{-\lambda _6 \left(v_1^4+3 v_2^2 v_1^2\right)}+v_2 \left(-4 \lambda _1 v_1^3\textcolor{purple}{+3 \lambda _7 v_2 v_1^2}+4 \lambda _2 v_2^2 v_1\textcolor{purple}{+\lambda _7 v_2^3}\right)}\;, \nonumber
	\end{eqnarray}}
leading to physical masses for one CP-odd scalar $(A)$, one charged scalar $(H^{\pm})$, and two CP-even scalars $(H_1,H_2)$ respectively:
{\small\begin{eqnarray}\label{eq:2HDM-scalar-physical-tree}
	m^2_{A}&=&-v^2 \left(\lambda _5-\frac{m_{12}^2}{v_1 v_2}\textcolor{purple}{+\frac{\lambda _6 v_1}{2 v_2}+\frac{\lambda _7 v_2}{2 v_1}}\right),\\
	m^2_{H^{\pm}}&=& -v^2 \left(\frac{\lambda _4}{2}+\frac{\lambda _5}{2}-\frac{m_{12}^2}{v_1 v_2}\textcolor{purple}{+\frac{\lambda _6 v_1}{2 v_2}+\frac{\lambda _7 v_2}{2 v_1}}\right),\\
	m^{2}_{H_{1}}&=&\frac{1}{2 v_1 v_2}\Big[v_1 \sin ^2\beta _{h } \left(2 m_{12}^2 v_1\textcolor{purple}{-\lambda _6 v_1^3}+v_2^2 \left(4 \lambda _2 v_2\textcolor{purple}{+3 \lambda _7 v_1}\right)\right)\nonumber \\
	&-&v_1 v_2 \sin \left(2 \beta _{h }\right) \left(-2 m_{12}^2\textcolor{purple}{+3 \lambda _6 v_1^2}+v_2 \left(2 \lambda _3 v_1+2 \lambda _4 v_1+2 \lambda _5 v_1\textcolor{purple}{+3 \lambda _7 v_2}\right)\right)\nonumber \\
	&+&v_2 \cos^2\beta _{h } \left(2 m_{12}^2 v_2+4 \lambda _1 v_1^3\textcolor{purple}{+3 \lambda _6 v_2 v_1^2-\lambda _7 v_2^3}\right)\Big],\\
	m^{2}_{H_{2}}&=&\frac{1}{2 v_1 v_2}\Big[v_2 \sin ^2\beta _{h } \left(2 m_{12}^2 v_2+4 \lambda _1 v_1^3\textcolor{purple}{+3 \lambda _6 v_2 v_1^2-\lambda _7 v_2^3}\right)\nonumber \\
	&+&v_1 v_2 \sin \left(2 \beta _{h }\right) \left(-2 m_{12}^2\textcolor{purple}{+3 \lambda _6 v_1^2}+v_2 \left(2 \lambda _3 v_1+2 \lambda _4 v_1+2 \lambda _5 v_1\textcolor{purple}{+3 \lambda _7 v_2}\right)\right)\nonumber \\
	&+&v_1 \cos^2\beta _{h } \left(2 m_{12}^2 v_1\textcolor{purple}{-\lambda _6 v_1^3}+v_2^2 \left(4 \lambda _2 v_2\textcolor{purple}{+3 \lambda _7 v_1}\right)\right)\Big].
	\end{eqnarray}}
The Higgs kinetic terms $(D^{\mu}\phi_1)^{\dagger}(D_{\mu}\phi_1)$, $(D^{\mu}\phi_2)^{\dagger}(D_{\mu}\phi_2)$ contain the  informations regarding the gauge boson masses which acquire the following forms after SSB: \small{\begin{eqnarray}\label{eq:2HDM-GB-physical-tree}
	m_Z^2 &=& \frac{1}{4}(g^2 + g^{\prime 2})v^2,\\
	m_W^2 &=& \frac{1}{4}g^2v^2.
	\end{eqnarray}}
The physical gauge bosons $(W_\mu^{\pm},Z_\mu,A_\mu)$ are related to the unphysical ones 
$( W^1_{\mu},  W^2_{\mu},  W^3_{\mu}, B_{\mu})$ as:
\small{\begin{eqnarray}\label{eq:2HDM-GB-physical-basis-tree}
	W_{\mu}^{\pm} &=& \frac{1}{\sqrt{2}}(W^1_{\mu} \mp iW^2_{\mu}),\\
	W^3_{\mu} &=& Z_{\mu}\cos\theta_{\text{w}} + A_{\mu}\sin\theta_{\text{w}} ,\\
	B_{\mu} & =& -Z_{\mu}\sin\theta_{\text{w}} + A_{\mu}\cos\theta_{\text{w}} .
	\end{eqnarray}}
After SSB, the fermion masses are generated through the Yukawa terms as:
{\small\begin{equation}\label{eq:2HDM-fermion-mass-tree}
	m_f = \frac{1}{\sqrt{2}}\left(v_1y_1^f + v_2y_2^f\right), \hspace{1.0cm} f = e,u,d.
	\end{equation}}
where $y_{1,2}^f$ are the $(3\times 3)$ Yukawa matrices for three generations of fermions.



\section{Effective operators for 2HDM, their categorisation and impact on spectrum}

\begin{figure}[h!]
	{
		\includegraphics[trim={2.0cm 0 2.0cm 0},scale=0.43]{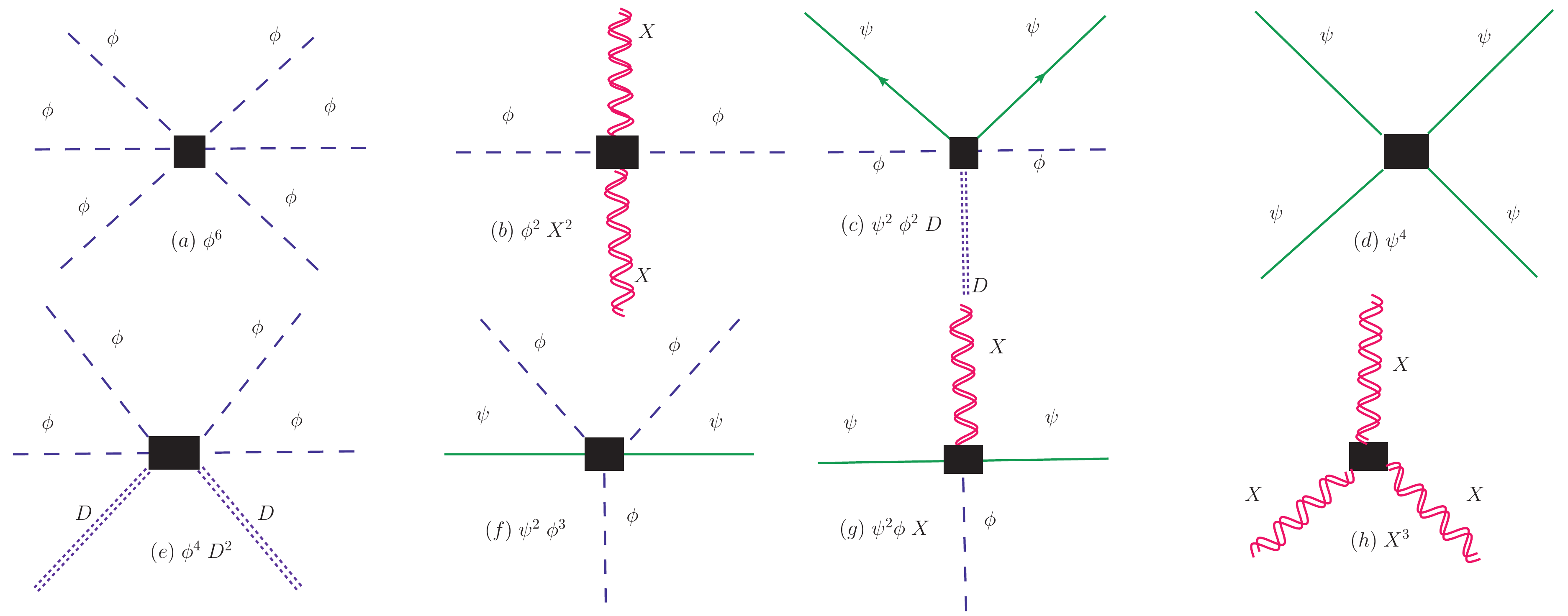}
	}
	\caption{Effective operators in ``Warsaw"-like  basis representing the following class of operators:
		(a) $\phi^6\;$ \textcolor{red}{[6 + 7$\times$2 = 20]}, (b) $\phi^2X^2\;$\textcolor{red}{[32]}, (c) $\psi^2\phi^2D\;$\textcolor{red}{[14 + 10$\times$2 = 34]}, (d) $\psi^4\;$\textcolor{red}{[20 + 5$\times$2 = 30]} (Baryon Number Conserving) + \textcolor{red}{[4$\times$2 = 8]} (Baryon Number Violating), (e) $\phi^4D^2\;$\textcolor{red}{[8 + 6$\times$2 = 20]}, (f) $ \psi^2\phi^3\;$\textcolor{red}{[24$\times$2 = 48]}, (g)  $\psi^2\phi X\;$\textcolor{red}{[16$\times$2 = 32]}, (h) $ X^3\;$\textcolor{red}{[4]}. }\label{fig:operator-class-2HDM}
\end{figure}

We have constructed the Hilbert Series for 2HDM using the  information from Table~\ref{tab:2HDM-fields} and following the thumb rules mentioned in the flowchart of Fig.~\ref{fig:HS}.  The complete set of dimension-6 operators are classified as: $\{ \phi^6, \phi^2X^2, \psi^2\phi^2D, \psi^4, \phi^4D^2,   \psi^2\phi^3, \psi^2\phi X, X^3\}$. These operators are depicted in Tables~\ref{tab:2HDM-phi6}-\ref{tab:2HDM-psi4-BNV}, and they are schematically shown in Fig.~\ref{fig:operator-class-2HDM}. The number of independent operators are also mentioned in the caption of Fig.~\ref{fig:operator-class-2HDM} as $\mathcal{O}\;[..]$. The effective operators in the context of 2HDM have been discussed in \cite{Contino:2013kra,Contino:2014aaa,DiazCruz:2001tn,Karmakar:2017yek,Crivellin:2016ihg,Kikuta:2011ew,Karmakar:2019vnq,Karmakar:2018scg}.  As we are working with the most generic 2HDM framework, we have also included the $Z_2$ violating operators, which are written in colored texts. Our method predicts the existence of $\mathcal{O}^{21(21)}_{\square}$ operator in $\phi^4 D^2$ class instead of $ Q_{\varphi D}^{12(12)}$ and $Q_{\varphi D}^{(1)21(2)} $ operators given in \cite{Crivellin:2016ihg}. For 2HDM, the Hilbert Series yields \textcolor{red}{228} independent operators of mass dimension-6. This includes the distinct Hermitian Conjugates (h.c.) as well. In our analysis, we have assumed  the Wilson Coefficients  to be real. While writing the explicit structures of these operators apart from the field variables we have used few tensors, e.g., $\delta_{\mu \nu}$ (Kronecker delta),  $\epsilon^{\mu \nu \rho}$ (Levi-Civita), etc. to form the invariants. 
As the operator set is enormous and their intricate structures are a bit involved, we have provided a suitable table where we have remarked on the nomenclature of the operators. We hope this would help the readers to follow the operators ($\mathcal{O}$), and their respective Wilson Coefficients ($\mathcal{C}$) which are also named following the same rules, see Table~\ref{tab:operator-nomenclature}.

\begin{table}[h!]
	\centering
	\renewcommand{\arraystretch}{1.80}
	{\scriptsize\begin{tabular}{|c|c|c|c|c|}
			\hline
			\multirow{2}{*}{\textbf{Operator}}&
			\multirow{2}{*}{\textbf{Symbol}}&
			\multirow{2}{*}{\textbf{Superscript}}&
			\multirow{2}{*}{\textbf{Subscript}}&
			\multirow{2}{*}{\textbf{Remarks}}\\
			
			\textbf{Class}&&&&\\
			
			\hline
			$\phi^6$&
			$\mathcal{O}^{i_1 i_2 ... i_6}_{6}$&
			$\{i_1,...i_6\}\rightarrow \{1,2\}$&
		    6 $\rightarrow$ mass dimension of $\phi^6$&
			1 $\rightarrow \{\phi_1,\phi_1^{\dagger}\}$; 2 $\rightarrow \{\phi_2,\phi_2^{\dagger}\}$\\
			\cline{1-4}
			
			$\phi^2X^2$&
			$\mathcal{O}^{ij}_{\phi X/\phi \tilde{X}}$&
			$\{i,j\} \rightarrow \{1,2\}$&
			$X \rightarrow \{W, B, G, WB\}$&\\
			\cline{1-4}
			
			$X^3$&
			$\mathcal{O}_{X/\tilde{X}}$&&
			$X \rightarrow \{W, G\}$&
			for $X\equiv WB$, $\tilde{X}\rightarrow \tilde{W}B$\\
			\cline{1-5}
			
			\multirow{4}{*}{$\psi^2\phi X$}&
			\multirow{4}{*}{$\mathcal{O}^{i}_{f X}$}&
			\multirow{4}{*}{$i\rightarrow \{1,2,\tilde{1},\tilde{2}\}$}&
			\multirow{4}{*}{$f\rightarrow \{u,d,e\}$}&
			\multirow{2}{*}{For ($u,d$), the operator} \\
			
			&&&& contains $Q$ and for ($e$), one has $L$\\ 
			
			&&&&\multirow{2}{*}{$u$ couples with $\tilde{\phi_1},\tilde{\phi_2}$}\\
			
			&&&&$e,d$ couple with $\phi_1,\phi_2$\\
			\cline{1-5}
			
			\multirow{4}{*}{$\psi^2\phi^3$}&
			\multirow{4}{*}{$\mathcal{O}^{i(jk)}_{f_1 f_2}$}&
			\multirow{4}{*}{$\{i,j,k\}\rightarrow \{1,2,\tilde{1},\tilde{2}\}$}&
			\multirow{4}{*}{$f_1\rightarrow \{Q, L\},\hspace{0.2cm}f_2\rightarrow \{u,d,e\}$}&
			\multirow{2}{*}{The field outside ``( )" is coupled }
			\\ 
			
			&&&& directly to the fermions,\\ 
			
			&&&&
			\multirow{2}{*}{while fields inside ``( )" are}\\
			
			&&&& 
			contracted with each other\\
			
			\cline{1-5}
			
			\multirow{2}{*}{$\psi^2\phi^2D$}&
			$\mathcal{O}^{ij}_{f\phi D}$&
			\multirow{2}{*}{$\{i,j\}\rightarrow \{1,2,\tilde{1},\tilde{2}\}$}&
			\multirow{2}{*}{$f\rightarrow \{Q, L, u, d, e\}$}&
			[1] $ \implies SU(2)$ Singlet\\
			
			&
			$\mathcal{O}^{ij[1]/ij[3]}_{f\phi D}$&
			&
			&
			[3] $ \implies SU(2)$ Triplet (involves $\tau^I$)\\ \cline{1-5}
			
			\multirow{3}{*}{$\phi^4D^2$}&
			$\mathcal{O}^{ij(kl)}_{\square}$&
			\multirow{3}{*}{$\{i,j,k,l\}\rightarrow \{1,2\}$}&
			\multirow{2}{*}{$\square$ for terms like }&
			\multirow{3}{*}{(i) $\implies$ $D$ acting on $\phi_i$}\\
			
			&
			$\mathcal{O}^{(i)jk(l)}_{\phi D}$&
			&
			$(\phi_{i}^{\dagger}\phi_{j})\square(\phi_{k}^{\dagger}\phi_{l})$, else $\phi D$&
			\\
			
			&
			$\mathcal{O}^{ij(k)(l)}_{\phi D}$&
			&&
			\\ \cline{1-5}
			
			\multirow{3}{*}{$\psi^4$}&
			\multirow{3}{*}{$\mathcal{O}^{[i]}_{f_1 f_2}$/$\mathcal{O}^{[i]}_{f_1 f_2 f_3 f_4}$}&
			\multirow{3}{*}{$i\rightarrow \{1,3,8\}$}&
			\multirow{3}{*}{$\{f_1,..f_4\}\rightarrow \{Q, L, u, d, e\}$}&
			[1]$\implies$ $SU(2)$, $SU(3)$ Singlet\\
			
			&
			&&
			&
			[3]$\implies$ $SU(2)$ Triplet (involves $\tau^I$)\\
			
			&
			&&
			&
			[8]$\implies  SU(3)$ Octet (involves $T^A$)\\
			
			\hline
	\end{tabular}}
	\caption{2HDM : Summary of the nomenclature of operators.}    \label{tab:operator-nomenclature}
\end{table}

We have started with mentioning the dimension-5 operators which are equivalent to the Weinberg operator.  These operators lead to the Majorana mass terms for the SM light neutrinos. Here, we would like to note that the dimension-5, and -6 operators are suppressed by a mass scale $\Lambda^{-1}$ and $\Lambda^{-2}$ respectively, but they are not explicitly mentioned in the tables to keep them tidy. But in the later sections where the explicit computation with these operators has been performed, the scales are mentioned appropriately.

\subsection*{{\Large$\bullet$} Dimension-5 and Dimension-6 Operators: Spectrum of 2HDM-EFT}

\begin{table}[h]
	\centering
	\renewcommand{\arraystretch}{2.5}
	{\small\begin{tabular}{|c|c|c|c|c|c|}
			\hline
			$\mathcal{O}_{\phi L}^{11}$&
			$(L^TC i \tau_2 \phi_1)(\tilde{\phi_1}^{\dagger} L) + h.c.$&
			$\mathcal{O}_{\phi L}^{22}$&
			$(L^TC i \tau_2 \phi_2)(\tilde{\phi_2}^{\dagger} L) + h.c.$&
			$\mathcal{O}_{\phi L}^{12}$&
			$\textcolor{purple}{(L^TC i \tau_2 \phi_1)(\tilde{\phi_2}^{\dagger} L) + h.c.}$\\
			\hline
	\end{tabular}}
	\caption{2HDM: $\phi^2L^2$ \textcolor{red}{[3$\times$2 = 6]} class of operators.}
	\label{table:2HDM-dim5}
\end{table}

\begin{table}[h!]
	\centering
	\renewcommand{\arraystretch}{1.5}
	{\small\begin{tabular}{|c|c|c|c|}
			\hline
			$\mathcal{O}^{111111}_{6}$&
			$(\phi_1^{\dagger}\phi_1)^3$ & 
			$\mathcal{O}^{222222}_{6}$&
			$(\phi_2^{\dagger}\phi_2)^3$ \\
			
			$\mathcal{O}^{111122}_{6}$&
			$(\phi_1^{\dagger}\phi_1)^2(\phi_2^{\dagger}\phi_2)$&
			$\mathcal{O}^{112222}_{6}$&
			$(\phi_1^{\dagger}\phi_1)(\phi_2^{\dagger}\phi_2)^2$ \\
			
			$\mathcal{O}^{122111}_{6}$&
			$(\phi_1^{\dagger}\phi_2)(\phi_2^{\dagger}\phi_1)(\phi_1^{\dagger}\phi_1)$&
			$\mathcal{O}^{122122}_{6}$&
			$(\phi_1^{\dagger}\phi_2)(\phi_2^{\dagger}\phi_1)(\phi_2^{\dagger}\phi_2)$\\
			
			$\mathcal{O}^{121211}_{6}$&
			$(\phi_1^{\dagger}\phi_2)^2(\phi_1^{\dagger}\phi_1)$ + h.c. &
			$\mathcal{O}^{121222}_{6}$&
			$(\phi_1^{\dagger}\phi_2)^2(\phi_2^{\dagger}\phi_2)$ + h.c. \\
			
			$\mathcal{O}^{111112}_{6}$&    
			\textcolor{purple}{$(\phi_1^{\dagger}\phi_1)^2(\phi_1^{\dagger}\phi_2)$ + h.c.}&
			$\mathcal{O}^{122222}_{6}$&
			\textcolor{purple}{$(\phi_1^{\dagger}\phi_2)(\phi_2^{\dagger}\phi_2)^2$ + h.c.}\\
			
			$\mathcal{O}^{121221}_{6}$&
			\textcolor{purple}{$(\phi_1^{\dagger}\phi_2)^2(\phi_2^{\dagger}\phi_1)$ + h.c.}&
			$\mathcal{O}^{121212}_{6}$&
			\textcolor{purple}{$(\phi_1^{\dagger}\phi_2)^3$ + h.c.}
			\\
			
			$\mathcal{O}^{112212}_{6}$&
			\textcolor{purple}{$(\phi_1^{\dagger}\phi_1)(\phi_2^{\dagger}\phi_2)(\phi_1^{\dagger}\phi_2)$ + h.c.}&
			&\\

			\hline
	\end{tabular}}
	\caption{2HDM: $\phi^6$ \textcolor{red}{[6 + 7$\times$2 = 20]} class of operators.}\label{tab:2HDM-phi6}
\end{table}    

\begin{table}[h!]
	\centering
	\renewcommand{\arraystretch}{1.5}
	{\small\begin{tabular}{|c|c|c|c|c|c|}
			\hline        
			$\mathcal{O}^{11}_{\phi G}$&
			$(\phi_1^{\dagger}\phi_1)G^A_{\mu\nu}G^{A\mu\nu}$&
			$\mathcal{O}^{11}_{\phi \tilde{G}}$&
			$(\phi_1^{\dagger}\phi_1)G^A_{\mu\nu}\tilde{G}^{A\mu\nu}$&
			$\mathcal{O}^{22}_{\phi G}$&
			$(\phi_2^{\dagger}\phi_2)G^A_{\mu\nu}G^{A\mu\nu}$\\
			
			$\mathcal{O}^{11}_{\phi W}$&
			$(\phi_1^{\dagger}\phi_1)W^I_{\mu\nu}W^{I\mu\nu}$&
			$\mathcal{O}^{11}_{\phi \tilde{W}}$&
			$(\phi_1^{\dagger}\phi_1)W^I_{\mu\nu}\tilde{W}^{I\mu\nu}$&
			$\mathcal{O}^{22}_{\phi \tilde{G}}$&
			$(\phi_2^{\dagger}\phi_2)G^A_{\mu\nu}\tilde{G}^{A\mu\nu}$
			\\
			
			$\mathcal{O}^{11}_{\phi B}$&
			$(\phi_1^{\dagger}\phi_1)B_{\mu\nu}B^{\mu\nu}$&
			$\mathcal{O}^{11}_{\phi \tilde{B}}$&
			$(\phi_1^{\dagger}\phi_1)B_{\mu\nu}\tilde{B}^{\mu\nu}$&
			$\mathcal{O}^{22}_{\phi W}$&
			$(\phi_2^{\dagger}\phi_2)W^I_{\mu\nu}W^{I\mu\nu}$\\
			
			$\mathcal{O}^{22}_{\phi B}$&
			$(\phi_2^{\dagger}\phi_2)B_{\mu\nu}B^{\mu\nu}$&
			$\mathcal{O}^{22}_{\phi \tilde{B}}$&
			$(\phi_2^{\dagger}\phi_2)B_{\mu\nu}\tilde{B}^{\mu\nu}$&
			$\mathcal{O}^{22}_{\phi \tilde{W}}$&
			$(\phi_2^{\dagger}\phi_2)W^I_{\mu\nu}\tilde{W}^{I\mu\nu}$
			\\
			
			$\mathcal{O}^{11}_{\phi WB}$&
			$(\phi_1^{\dagger}\tau^I\phi_1)W^I_{\mu\nu}B^{\mu\nu}$&
			$\mathcal{O}^{11}_{\phi \tilde{W}B}$& 
			$(\phi_1^{\dagger}\tau^I\phi_1)\tilde{W}^I_{\mu\nu}B^{\mu\nu}$&
			$\mathcal{O}^{22}_{\phi WB}$& $(\phi_2^{\dagger}\tau^I\phi_2)W^I_{\mu\nu}B^{\mu\nu}$\\
			
			$\mathcal{O}^{22}_{\phi \tilde{W}B}$&
			$(\phi_2^{\dagger}\tau^I\phi_2)\tilde{W}^I_{\mu\nu}B^{\mu\nu}$&
			$\mathcal{O}^{12}_{\phi G}$&
			\textcolor{purple}{$(\phi_1^{\dagger}\phi_2)G^A_{\mu\nu}G^{A\mu\nu}$}&
			$\mathcal{O}^{12}_{\phi \tilde{G}}$&
			\textcolor{purple}{$(\phi_1^{\dagger}\phi_2)G^A_{\mu\nu}\tilde{G}^{A\mu\nu}$}\\
			
			$\mathcal{O}^{21}_{\phi G}$&
			\textcolor{purple}{$(\phi_2^{\dagger}\phi_1)G^A_{\mu\nu}G^{A\mu\nu}$}&
			$\mathcal{O}^{21}_{\phi \tilde{G}}$&
			\textcolor{purple}{$(\phi_2^{\dagger}\phi_1)G^A_{\mu\nu}\tilde{G}^{A\mu\nu}$}&
			$\mathcal{O}^{12}_{\phi W}$&
			\textcolor{purple}{$(\phi_1^{\dagger}\phi_2)W^I_{\mu\nu}W^{I\mu\nu}$}\\
			
			$\mathcal{O}^{12}_{\phi \tilde{W}}$&
			\textcolor{purple}{$(\phi_1^{\dagger}\phi_2)W^I_{\mu\nu}\tilde{W}^{I\mu\nu}$}&
			$\mathcal{O}^{21}_{\phi W}$&
			\textcolor{purple}{$(\phi_2^{\dagger}\phi_1)W^I_{\mu\nu}W^{I\mu\nu}$}&
			$\mathcal{O}^{21}_{\phi \tilde{W}}$&
			\textcolor{purple}{$(\phi_2^{\dagger}\phi_1)W^I_{\mu\nu}\tilde{W}^{I\mu\nu}$}
			\\
			
			$\mathcal{O}^{12}_{\phi B}$&
			\textcolor{purple}{$(\phi_1^{\dagger}\phi_2)B_{\mu\nu}B^{\mu\nu}$}&
			$\mathcal{O}^{12}_{\phi \tilde{B}}$&
			\textcolor{purple}{$(\phi_1^{\dagger}\phi_2)B_{\mu\nu}\tilde{B}^{\mu\nu}$}&
			$\mathcal{O}^{21}_{\phi B}$&
			\textcolor{purple}{$(\phi_2^{\dagger}\phi_1)B_{\mu\nu}B^{\mu\nu}$}\\
			
			$\mathcal{O}^{21}_{\phi \tilde{B}}$&
			\textcolor{purple}{$(\phi_2^{\dagger}\phi_1)B_{\mu\nu}\tilde{B}^{\mu\nu}$}&
			$\mathcal{O}^{12}_{\phi WB}$&
			\textcolor{purple}{$(\phi_1^{\dagger}\tau^I\phi_2)W^I_{\mu\nu}B^{\mu\nu}$}&
			$\mathcal{O}^{12}_{\phi \tilde{W}B}$&
			\textcolor{purple}{$(\phi_1^{\dagger}\tau^I\phi_2)\tilde{W}^I_{\mu\nu}B^{\mu\nu}$}\\
			
			$\mathcal{O}^{21}_{\phi WB}$&
			\textcolor{purple}{$(\phi_2^{\dagger}\tau^I\phi_1)W^I_{\mu\nu}B^{\mu\nu}$}&
			$\mathcal{O}^{21}_{\phi \tilde{W}B}$&
			\textcolor{purple}{$(\phi_2^{\dagger}\tau^I\phi_1)\tilde{W}^I_{\mu\nu}B^{\mu\nu}$}&
			&
			\\
			
			\hline
	\end{tabular}}
	\caption{2HDM: $\phi^2 X^2$ \textcolor{red}{[32]} class of operators.} \label{tab:2HDM-phi2X2}
\end{table}

\begin{table}[h!]
	\centering
	\renewcommand{\arraystretch}{1.5}
	{\small\begin{tabular}{|c|c|}
			\hline
			
			$\mathcal{O}_G$&
			$f^{ABC}G^{A\nu}_{\mu}G^{B\rho}_{\nu}G^{C\mu}_{\rho}$\\
			
			$\mathcal{O}_{\tilde{G}}$&
			$f^{ABC}\tilde{G}^{A\nu}_{\mu}G^{B\rho}_{\nu}G^{C\mu}_{\rho}$\\
			
			$\mathcal{O}_W$&
			$\epsilon^{IJK}W^{I\nu}_{\mu}W^{J\rho}_{\nu}W^{K\mu}_{\rho}$\\
			
			$\mathcal{O}_{\tilde{W}}$&
			$\epsilon^{IJK}\tilde{W}^{I\nu}_{\mu}W^{J\rho}_{\nu}W^{K\mu}_{\rho}$\\    
			
			\hline
	\end{tabular}}
	\caption{2HDM: $X^3$ \textcolor{red}{[4]}  class of operators.} 
	\label{tab:2HDM-X3}
\end{table}

\begin{table}[h!]
	\centering
	\renewcommand{\arraystretch}{1.5}
	{\small\begin{tabular}{|c|c|c|c|}
			\hline
			
			$\mathcal{O}_{dG}^1$&
			\textcolor{purple}{$(\bar{Q}\sigma^{\mu\nu}T^Ad)\phi_1G^A_{\mu\nu}$}&
			$\mathcal{O}_{dG}^2$&
			$(\bar{Q}\sigma^{\mu\nu}T^Ad)\phi_2G^A_{\mu\nu}$\\
			
			$\mathcal{O}_{uG}^{\tilde{1}}$&        
			\textcolor{purple}{$(\bar{Q}\sigma^{\mu\nu}T^Au)\tilde{\phi_1}G^A_{\mu\nu}$}&
			$\mathcal{O}_{uG}^{\tilde{2}}$&
			$(\bar{Q}\sigma^{\mu\nu}T^Au)\tilde{\phi_2}G^A_{\mu\nu}$
			\\
			
			$\mathcal{O}_{dW}^1$&        
			\textcolor{purple}{$(\bar{Q}\sigma^{\mu\nu}d)\tau^I\phi_1W^I_{\mu\nu}$}&
			$\mathcal{O}_{dW}^2$&
			$(\bar{Q}\sigma^{\mu\nu}d)\tau^I\phi_2W^I_{\mu\nu}$\\
			
			$\mathcal{O}_{uW}^{\tilde{1}}$&
			\textcolor{purple}{$(\bar{Q}\sigma^{\mu\nu}u)\tau^I\tilde{\phi_1}W^I_{\mu\nu}$}&
			$\mathcal{O}_{uW}^{\tilde{2}}$&
			$(\bar{Q}\sigma^{\mu\nu}u)\tau^I\tilde{\phi_2}W^I_{\mu\nu}$
			\\
			
			$\mathcal{O}_{dB}^1$&
			\textcolor{purple}{$(\bar{Q}\sigma^{\mu\nu}d)\phi_1B_{\mu\nu}$}&
			$\mathcal{O}_{dB}^2$&
			$(\bar{Q}\sigma^{\mu\nu}d)\phi_2B_{\mu\nu}$\\
			
			$\mathcal{O}_{uB}^{\tilde{1}}$&
			\textcolor{purple}{$(\bar{Q}\sigma^{\mu\nu}u)\tilde{\phi_1}B_{\mu\nu}$}&
			$\mathcal{O}_{uB}^{\tilde{2}}$&
			$(\bar{Q}\sigma^{\mu\nu}u)\tilde{\phi_2}B_{\mu\nu}$
			\\
			
			$\mathcal{O}_{eW}^1$&        
			\textcolor{purple}{$(\bar{L}\sigma^{\mu\nu}e)\tau^I\phi_1W^I_{\mu\nu}$}&
			$\mathcal{O}_{eW}^2$&
			$(\bar{L}\sigma^{\mu\nu}e)\tau^I\phi_2W^I_{\mu\nu}$\\
			
			$\mathcal{O}_{eB}^1$&
			\textcolor{purple}{$(\bar{L}\sigma^{\mu\nu}e)\phi_1B_{\mu\nu}$}&
			$\mathcal{O}_{eB}^2$&
			$(\bar{L}\sigma^{\mu\nu}e)\phi_2B_{\mu\nu}$
			\\
			\hline
	\end{tabular}}
	\caption{2HDM: $\psi^2\phi X$ \textcolor{red}{[16$\times$2 = 32]}  class of operators. Each of these operators also has a distinct Hermitian Conjugate.}
	\label{tab:2HDM-psi2phiX}
\end{table}

\begin{table}[h!]
	\centering
	\renewcommand{\arraystretch}{1.5}
	{\small\begin{tabular}{|c|c|c|c|c|c|}
			\hline
			$\mathcal{O}_{Le}^{1(21)}$&
			$(\bar{L}e\phi_1)(\phi_2^{\dagger}\phi_1)$&
			$\mathcal{O}_{Le}^{2(22)}$&
			$(\bar{L}e\phi_2)(\phi_2^{\dagger}\phi_2)$&
			$\mathcal{O}_{Le}^{2(11)}$&        
			$(\bar{L}e\phi_2)(\phi_1^{\dagger}\phi_1)$\\
			
			$\mathcal{O}_{Le}^{1(12)}$&
			$(\bar{L}e\phi_1)(\phi_1^{\dagger}\phi_2)$&
			$\mathcal{O}_{Qd}^{1(21)}$&        
			$(\bar{Q}d\phi_1)(\phi_2^{\dagger}\phi_1)$&
			$\mathcal{O}_{Qd}^{2(22)}$&
			$(\bar{Q}d\phi_2)(\phi_2^{\dagger}\phi_2)$\\
			
			$\mathcal{O}_{Qd}^{2(11)}$&
			$(\bar{Q}d\phi_2)(\phi_1^{\dagger}\phi_1)$&
			$\mathcal{O}_{Qd}^{1(12)}$&
			$(\bar{Q}d\phi_1)(\phi_1^{\dagger}\phi_2)$&
			$\mathcal{O}_{Qu}^{\tilde{2}(22)}$&        
			$(\bar{Q}u\tilde{\phi_2})(\phi_2^{\dagger}\phi_2)$\\
			
			$\mathcal{O}_{Qu}^{\tilde{1}(12)}$&
			$(\bar{Q}u\tilde{\phi_1})(\phi_1^{\dagger}\phi_2)$&
			$\mathcal{O}_{Qu}^{\tilde{2}(11)}$&        
			$(\bar{Q}u\tilde{\phi_2})(\phi_1^{\dagger}\phi_1)$&
			$\mathcal{O}_{Qu}^{\tilde{1}(21)}$&
			$(\bar{Q}u\tilde{\phi_1})(\phi_2^{\dagger}\phi_1)$\\
			
			$\mathcal{O}_{Le}^{1(11)}$&
			\textcolor{purple}{$(\bar{L}e\phi_1)(\phi_1^{\dagger}\phi_1)$}&
			$\mathcal{O}_{Le}^{2(12)}$&
			\textcolor{purple}{$(\bar{L}e\phi_2)(\phi_1^{\dagger}\phi_2)$}&
			$\mathcal{O}_{Le}^{1(22)}$&
			\textcolor{purple}{$(\bar{L}e\phi_1)(\phi_2^{\dagger}\phi_2)$}\\
			
			$\mathcal{O}_{Le}^{2(21)}$&
			\textcolor{purple}{$(\bar{L}e\phi_2)(\phi_2^{\dagger}\phi_1)$}&
			$\mathcal{O}_{Qd}^{1(11)}$&
			\textcolor{purple}{$(\bar{Q}d\phi_1)(\phi_1^{\dagger}\phi_1)$}&
			$\mathcal{O}_{Qd}^{2(12)}$&
			\textcolor{purple}{$(\bar{Q}d\phi_2)(\phi_1^{\dagger}\phi_2)$}\\
			
			$\mathcal{O}_{Qd}^{1(22)}$&
			\textcolor{purple}{$(\bar{Q}d\phi_1)(\phi_2^{\dagger}\phi_2)$}&
			$\mathcal{O}_{Qd}^{2(21)}$&
			\textcolor{purple}{$(\bar{Q}d\phi_2)(\phi_2^{\dagger}\phi_1)$}&
			$\mathcal{O}_{Qu}^{\tilde{1}(11)}$&
			\textcolor{purple}{$(\bar{Q}u\tilde{\phi_1})(\phi_1^{\dagger}\phi_1)$}\\
			
			$\mathcal{O}_{Qu}^{\tilde{2}(21)}$&
			\textcolor{purple}{$(\bar{Q}u\tilde{\phi_2})(\phi_2^{\dagger}\phi_1)$}&        
			$\mathcal{O}_{Qu}^{\tilde{1}(22)}$&
			\textcolor{purple}{$(\bar{Q}u\tilde{\phi_1})(\phi_2^{\dagger}\phi_2)$}&
			$\mathcal{O}_{Qu}^{\tilde{2}(12)}$&
			\textcolor{purple}{$(\bar{Q}u\tilde{\phi_2})(\phi_1^{\dagger}\phi_2)$}\\
			
			\hline
	\end{tabular}}
	\caption{2HDM: $\psi^2\phi^3$ \textcolor{red}{[24$\times$2 = 48]} class of operators. Each of these operators also has a distinct Hermitian Conjugate.}
	\label{tab:2HDM-psi2phi3}
\end{table}    

\begin{table}[h!]
	\centering
	\renewcommand{\arraystretch}{1.5}
	{\small\begin{tabular}{|c|c|c|c|c|c|}
			\hline
			
			$\mathcal{O}_{L\phi D}^{11[1]}$&
			$i(\phi_1^{\dagger}\overleftrightarrow{D_\mu}\phi_1)(\bar{L}\gamma^\mu L) $&
			$\mathcal{O}_{L\phi D}^{22[1]}$&
			$i(\phi_2^{\dagger}\overleftrightarrow{D_\mu}\phi_2)(\bar{L}\gamma^\mu L) $&
			$\mathcal{O}_{L\phi D}^{12[1]}$&
			\textcolor{purple}{$i(\phi_1^{\dagger}\overleftrightarrow{D_\mu}\phi_2)(\bar{L}\gamma^\mu L) + h.c. $}\\
			
			$\mathcal{O}_{Q\phi D}^{11[1]}$&        
			$i(\phi_1^{\dagger}\overleftrightarrow{D_\mu}\phi_1)(\bar{Q}\gamma^\mu Q) $ &
			$\mathcal{O}_{Q\phi D}^{22[1]}$&
			$i(\phi_2^{\dagger}\overleftrightarrow{D_\mu}\phi_2)(\bar{Q}\gamma^\mu Q) $ &
			$\mathcal{O}_{Q\phi D}^{12[1]}$&
			\textcolor{purple}{$i(\phi_1^{\dagger}\overleftrightarrow{D_\mu}\phi_2)(\bar{Q}\gamma^\mu Q) + h.c. $}\\
			
			$\mathcal{O}_{e\phi D}^{11}$&        
			$i(\phi_1^{\dagger}\overleftrightarrow{D_\mu}\phi_1)(\bar{e}\gamma^\mu e)$ &
			$\mathcal{O}_{e\phi D}^{22}$& 
			$i(\phi_2^{\dagger}\overleftrightarrow{D_\mu}\phi_2)(\bar{e}\gamma^\mu e)$&
			$\mathcal{O}_{e\phi D}^{12}$& 
			\textcolor{purple}{$i(\phi_1^{\dagger}\overleftrightarrow{D_\mu}\phi_2)(\bar{e}\gamma^\mu e) + h.c.$}\\
			
			$\mathcal{O}_{d\phi D}^{11}$&     
			$i(\phi_1^{\dagger}\overleftrightarrow{D_\mu}\phi_1)(\bar{d}\gamma^\mu d)$ &
			$\mathcal{O}_{d\phi D}^{22}$&
			$i(\phi_2^{\dagger}\overleftrightarrow{D_\mu}\phi_2)(\bar{d}\gamma^\mu d)$ & 
			$\mathcal{O}_{d\phi D}^{12}$&
			\textcolor{purple}{$i(\phi_1^{\dagger}\overleftrightarrow{D_\mu}\phi_2)(\bar{d}\gamma^\mu d) + h.c.$} \\
			
			$\mathcal{O}_{u\phi D}^{11}$&
			$i(\phi_1^{\dagger}\overleftrightarrow{D_\mu}\phi_1)(\bar{u}\gamma^\mu u) $ &
			$\mathcal{O}_{u\phi D}^{22}$&
			$i(\phi_2^{\dagger}\overleftrightarrow{D_\mu}\phi_2)(\bar{u}\gamma^\mu u) $&
			$\mathcal{O}_{u\phi D}^{12}$&
			\textcolor{purple}{$i(\phi_1^{\dagger}\overleftrightarrow{D_\mu}\phi_2)(\bar{u}\gamma^\mu u) + h.c. $}\\
			
			$\mathcal{O}_{L\phi D}^{11[3]}$&
			$i(\phi_1^{\dagger}\overleftrightarrow{D_\mu^I}\phi_1)(\bar{L}\tau^I\gamma^\mu L) $ &
			$\mathcal{O}_{L\phi D}^{22[3]}$&
			$i(\phi_2^{\dagger}\overleftrightarrow{D_\mu^I}\phi_2)(\bar{L}\tau^I\gamma^\mu L) $&
			$\mathcal{O}_{L\phi D}^{12[3]}$&
			\textcolor{purple}{$i(\phi_1^{\dagger}\overleftrightarrow{D_\mu^I}\phi_2)(\bar{L}\tau^I\gamma^\mu L) + h.c.$}\\
			
			$\mathcal{O}_{Q\phi D}^{11[3]}$&
			$i(\phi_1^{\dagger}\overleftrightarrow{D_\mu^I}\phi_1)(\bar{Q}\tau^I\gamma^\mu Q) $&
			$\mathcal{O}_{Q\phi D}^{22[3]}$&
			$i(\phi_2^{\dagger}\overleftrightarrow{D_\mu^I}\phi_2)(\bar{Q}\tau^I\gamma^\mu Q) $&
			$\mathcal{O}_{Q\phi D}^{12[3]}$&
			\textcolor{purple}{$i(\phi_1^{\dagger}\overleftrightarrow{D_\mu^I}\phi_2)(\bar{Q}\tau^I\gamma^\mu Q) + h.c.$}\\
			
			$\mathcal{O}_{ud\phi D}^{\tilde{1}1}$&        
			$i(\tilde{\phi_1}^{\dagger}\overleftrightarrow{D_\mu}\phi_1)(\bar{u}\gamma^\mu d) + h.c. $&
			$\mathcal{O}_{ud\phi D}^{\tilde{2}2}$&
			$i(\tilde{\phi_2}^{\dagger}\overleftrightarrow{D_\mu}\phi_2)(\bar{u}\gamma^\mu d) + h.c. $&
			$\mathcal{O}_{ud\phi D}^{\tilde{2}1}$&
			\textcolor{purple}{$i(\tilde{\phi_2}^{\dagger}\overleftrightarrow{D_\mu}\phi_1)(\bar{u}\gamma^\mu d) + h.c. $}\\

			\hline
	\end{tabular}}
	\caption{2HDM: $\psi^2\phi^2 D$ \textcolor{red}{[14 + 10$\times$2 = 20]} class of operators.}
	\label{tab:2HDM-psi2phi2D}
	
\end{table}

\begin{table}[h!]
	\centering
	\renewcommand{\arraystretch}{1.5}
	{\small\begin{tabular}{|c|c|c|c|}
			\hline
			$\mathcal{O}^{11(11)}_{\square}$&
			$(\phi_1^{\dagger}\phi_1)\square(\phi_1^{\dagger}\phi_1)$&
			$\mathcal{O}_{\square }^{22(22)}$&
			$(\phi_2^{\dagger}\phi_2)\square(\phi_2^{\dagger}\phi_2)$\\
			
			$\mathcal{O}_{\square }^{11(22)}$&
			$(\phi_1^{\dagger}\phi_1)\square(\phi_2^{\dagger}\phi_2)$&
			$\mathcal{O}_{\square }^{22(11)}$&
			$(\phi_2^{\dagger}\phi_2)\square(\phi_1^{\dagger}\phi_1)$\\
			
			$\mathcal{O}^{21(21)}_{\square}$&
			$(\phi_2^{\dagger}\phi_1)\square(\phi_2^{\dagger}\phi_1) + h.c.$&
			$\mathcal{O}^{12(1)(2)}_{\phi D}$&
			$(\phi_1^{\dagger}\phi_2)\big[(D^{\mu}\phi_1)^{\dagger}(D_{\mu}\phi_2)\big] +h.c.$\\
			
			$\mathcal{O}^{(1)11(1)}_{\phi D}$&
			$\big[(D^{\mu}\phi_1)^{\dagger}\phi_1\big]\big[\phi_1^{\dagger}(D_{\mu}\phi_1)\big]$&
			$\mathcal{O}^{(2)22(2)}_{\phi D}$&
			$\big[(D^{\mu}\phi_2)^{\dagger}\phi_2\big]\big[\phi_2^{\dagger}(D_{\mu}\phi_2)\big]$\\
			
			$\mathcal{O}^{(1)22(1)}_{\phi D}$&
			$\big[(D^{\mu}\phi_1)^{\dagger}\phi_2\big]\big[\phi_2^{\dagger}(D_{\mu}\phi_1)\big]$&
			$\mathcal{O}^{(2)11(2)}_{\phi D}$&
			$\big[(D^{\mu}\phi_2)^{\dagger}\phi_1\big]\big[\phi_1^{\dagger}(D_{\mu}\phi_2)\big]$\\
			
			$\mathcal{O}^{(2)22(1)}_{\phi D}$&
			\textcolor{purple}{$\big[(D^{\mu}\phi_2)^{\dagger}\phi_2\big]\big[\phi_2^{\dagger}(D_{\mu}\phi_1)\big] + h.c.$}&
			$\mathcal{O}^{(1)21(1)}_{\phi D}$&
			\textcolor{purple}{$\big[(D^{\mu}\phi_1)^{\dagger}\phi_2\big]\big[\phi_1^{\dagger}(D_{\mu}\phi_1)\big] + h.c.$}
			\\
			
			$\mathcal{O}^{21(2)(2)}_{\phi D}$&
			\textcolor{purple}{$(\phi_2^{\dagger}\phi_1)\big[(D^{\mu}\phi_2)^{\dagger}((D_{\mu}\phi_2))\big] +h.c.$}&
			$\mathcal{O}^{12(1)(1)}_{\phi D}$&
			\textcolor{purple}{$(\phi_1^{\dagger}\phi_2)\big[(D^{\mu}\phi_1)^{\dagger}((D_{\mu}\phi_1))\big] + h.c.$}
			\\
			
			\hline
	\end{tabular}}
	\caption{2HDM: $\phi^4 D^2$ \textcolor{red}{[8 + 6$\times$2 = 20]} class of operators. Note the presence of  $O_{\square}^{21(21)}$ operator  instead of operators ($ Q_{\varphi D}^{12(12)}$ and $Q_{\varphi D}^{(1)21(2)} $) as given in \cite{Crivellin:2016ihg}.}
	\label{tab:2HDM-phi4D2}
\end{table}    

\begin{table}[h!]
	\centering
	\renewcommand{\arraystretch}{1.5}
	{\small\begin{tabular}{|c|c|c|c|}
			\hline
			$\mathcal{O}_{dd}$&
			$(\bar{d}\gamma_{\mu}d)(\bar{d}\gamma^{\mu}d)$ &
			$\mathcal{O}_{uu}$&
			$(\bar{u}\gamma_{\mu}u)(\bar{u}\gamma^{\mu}u)$ \\
			
			$\mathcal{O}_{Le}$&                
			$(\bar{L}\gamma_{\mu}L)(\bar{e}\gamma^{\mu}e)$ &
			$\mathcal{O}_{Qe}$&
			$(\bar{Q}\gamma_{\mu}Q)(\bar{e}\gamma^{\mu}e)$\\
			
			$\mathcal{O}_{ee}$&
			$(\bar{e}\gamma_{\mu}e)(\bar{e}\gamma^{\mu}e)$ &
			$\mathcal{O}_{LL}$&
			$(\bar{L}\gamma_{\mu}L)(\bar{L}\gamma^{\mu}L)$\\
			
			$\mathcal{O}_{eu}$&        
			$(\bar{e}\gamma_{\mu}e)(\bar{u}\gamma^{\mu}u)$ &
			$\mathcal{O}_{ed}$&
			$(\bar{e}\gamma_{\mu}e)(\bar{d}\gamma^{\mu}d)$ \\
			
			$\mathcal{O}_{Lu}$&
			$(\bar{L}\gamma_{\mu}L)(\bar{u}\gamma^{\mu}u)$&
			$\mathcal{O}_{Ld}$&
			$(\bar{L}\gamma_{\mu}L)(\bar{d}\gamma^{\mu}d)$ \\
			
			$\mathcal{O}_{LQ}^{[1]}$&
			$(\bar{L}\gamma_{\mu}L)(\bar{Q}\gamma^{\mu}Q)$&
			$\mathcal{O}_{LQ}^{[3]}$&
			$(\bar{L}\gamma_{\mu}\tau^I L)(\bar{Q}\gamma^{\mu}\tau^I Q)$\\
			
			$\mathcal{O}_{QQ}^{[1]}$&
			$(\bar{Q}\gamma_{\mu}Q)(\bar{Q}\gamma^{\mu}Q)$ &
			$\mathcal{O}_{QQ}^{[3]}$&
			$(\bar{Q}\gamma_{\mu}\tau^I Q)(\bar{Q}\gamma^{\mu}\tau^I Q)$\\
			
			$\mathcal{O}_{ud}^{[1]}$&
			$(\bar{u}\gamma_{\mu}u)(\bar{d}\gamma^{\mu}d)$ &
			$\mathcal{O}_{ud}^{[8]}$&
			$(\bar{u}\gamma_{\mu}T^A u)(\bar{u}\gamma^{\mu}T^A u)$ \\
			
			$\mathcal{O}_{Qu}^{[1]}$&
			$(\bar{Q}\gamma_{\mu}Q)(\bar{u}\gamma^{\mu}u)$ &
			$\mathcal{O}_{Qu}^{[8]}$&
			$(\bar{Q}\gamma_{\mu}T^A Q)(\bar{u}\gamma^{\mu}T^A u)$ \\
			
			$\mathcal{O}_{Qd}^{[1]}$&
			$(\bar{Q}\gamma_{\mu}Q)(\bar{d}\gamma^{\mu}d)$&
			$\mathcal{O}_{Qd}^{[8]}$&
			$(\bar{Q}\gamma_{\mu}T^A Q)(\bar{d}\gamma^{\mu}T^A d)$ \\
			
			\hline
			
			$\mathcal{O}_{QuQd}^{[1]}$&
			\multicolumn{3}{c|}{$\epsilon_{jk}(\bar{Q}^j u)(\bar{Q}^k d) + h.c.$}\\
			
			$\mathcal{O}_{QuQd}^{[8]}$&
			\multicolumn{3}{c|}{$\epsilon_{jk}(\bar{Q}^j T^A u)(\bar{Q}^k T^A d) + h.c.$}\\
			
			$\mathcal{O}_{LeQu}^{[1]}$&
			\multicolumn{3}{c|}{$\epsilon_{jk}(\bar{L}^j e)(\bar{Q}^k u) + h.c.$}\\
			
			$\mathcal{O}_{LeQu}^{[3]}$&
			\multicolumn{3}{c|}{$\epsilon_{jk}(\bar{L}^j \sigma_{\mu\nu} e)(\bar{Q}^k \sigma^{\mu\nu} u) + h.c.$}\\
			
			$\mathcal{O}_{LedQ}$&
			\multicolumn{3}{c|}{$(\bar{L^j}e)(\bar{d}Q^j) + h.c.$}
			\\    
			\hline
	\end{tabular}}
	\caption{2HDM: Baryon Number Conserving $\psi^4\;$\textcolor{red}{[20 + 5$\times$2 = 30]} class of operators. Here $j,k$ are $SU(2)$ indices.}
	\label{tab:2HDM-psi4}
\end{table}    
\begin{table}[ht]
	\centering
	\renewcommand{\arraystretch}{1.5}
	{\small\begin{tabular}{|c|c|}
			\hline
			
			$\mathcal{O}_{QQQ}$&
			$\epsilon^{\alpha\beta\gamma}\epsilon_{jn}\epsilon_{km}\big[(Q^{\alpha j})^T C Q^{\beta k}\big] \big[(Q^{\gamma m})^T C L^n\big]$\\
			
			$\mathcal{O}_{QQu}$&
			$\epsilon^{\alpha\beta\gamma}\epsilon_{jk}\big[(Q^{\alpha j})^T C Q^{\beta k}\big] \big[(u^{\gamma})^T C e\big]$
			\\
			
			$\mathcal{O}_{duu}$&
			$\epsilon^{\alpha\beta\gamma}\big[(d^{\alpha})^T C u^{\beta}\big] \big[(u^{\gamma})^T C e\big]$ \\
			
			$\mathcal{O}_{duQ}$&
			$\epsilon^{\alpha\beta\gamma}\epsilon_{jk}\big[(d^{\alpha})^T C u^{\beta}\big] \big[(Q^{\gamma j})^T C L^k\big]$
			\\
			
			\hline
	\end{tabular}}
	\caption{2HDM: Baryon Number Violating $\psi^4$ \textcolor{red}{[4$\times$2 = 8]} class of operators. Each operator has a distinct Hermitian Conjugate. Here, $C=i\gamma^2\gamma^0$ in Dirac representation. Here $\alpha,\beta, \gamma$ are $SU(3)$ indices and $j,k$ are $SU(2)$ indices.}
	\label{tab:2HDM-psi4-BNV}
\end{table}

\vskip 0.2cm
\underline{\large{Majorana  mass of neutrinos from $\phi^2L^2$ operators:}}\\

The neutrino mass term [$\overline{(\nu_{L})^{c}} \nu_{_L}$] is generated from the dimension-5 operators, see Table~\ref{table:2HDM-dim5}, once the Higgs fields acquire VEVs. The coefficient of this term which is essentially the Majorana mass of the neutrino is given as:
\begin{equation}\label{eq:2HDM-neutrino-mass-dim5}
	\Delta m_{\nu} = \frac{v_{1}^{2}}{2\Lambda}\mathcal{C}^{11}_{\phi L}+\frac{v_{2}^{2}}{2\Lambda}\mathcal{C}^{22}_{\phi L}\textcolor{purple}{\;+\;\frac{v_{1} v_{2}}{2\Lambda}\mathcal{C}^{12}_{\phi L}}.
\end{equation}
\clearpage

\underline{\large{$\phi^6$ operators: Modification in the scalar potential}} \\

The inclusion of $\phi^6$ operators modifies the scalar potential as $V(\phi_1, \phi_2)$ + ${\mathcal{L}_{\phi^6}}$ where 
\begin{align}\label{eq:2HDM-scalar-pot-dim6}
{\mathcal{L}_{\phi^6}}&=\sum^{20}_{j=1}\mathcal{C}^{j}_{6} \mathcal{O}^{j}_{\phi}.
\end{align}

The additional minimization criteria along with Eq.~\ref{eq:2HDM-scalar-pot-min-tree}, are noted as:
\begin{eqnarray}\label{eq:2HDM-scalar-min-condition-dim6}
\frac{\partial\mathcal{L}_{\phi^6}}{\partial v_{1}}&= & 3 \mathcal{C}_{6}^{111111} v_{1}^4+2 \mathcal{C}_{6}^{111122} v_{1}^2 v_{2}^2+4 \mathcal{C}_{6}^{121211} v_{1}^2 v_{2}^2+2 \mathcal{C}_{6}^{121222} v_{2}^4+\mathcal{C}_{6}^{112222} v_{2}^4  +2 \mathcal{C}_{6}^{122111} v_{1}^2 v_{2}^2    \\ 
&+& \mathcal{C}_{6}^{122122} v_{2}^4{ \color{purple} +  3 \mathcal{C}_{6}^{121221} v_{1} v_{2}^3   +\frac{v_{2}^5}{v_{1}}\mathcal{C}_{6}^{122222} +5 \mathcal{C}_{6}^{111112} v_{1}^3 v_{2} +3 \mathcal{C}_{6}^{112212} v_{1} v_{2}^3 +3 \mathcal{C}_{6}^{121212} v_{1} v_{2}^3} \nonumber\\
&=& 0, \nonumber \\
\frac{\partial\mathcal{L}_{\phi^6}}{\partial v_{2}}&=& \mathcal{C}_{6}^{111122} v_{1}^4
+2 \mathcal{C}_{6}^{121211} v_{1}^4
+4 \mathcal{C}_{6}^{121222} v_{2}^2 v_{1}^2
+2 \mathcal{C}_{6}^{112222} v_{2}^2 v_{1}^2  
+\mathcal{C}_{6}^{122111} v_{1}^4
+ 3 \mathcal{C}_{6}^{222222} v_{2}^4  \\ 
& + & 2 \mathcal{C}_{6}^{122122} v_{2}^2 v_{1}^2 { \color{purple} + 3 \mathcal{C}_{6}^{121221} v_{2} v_{1}^3 
 + \frac{v_{1}^5}{v_{2}}\mathcal{C}_{6}^{111112}
	+5 \mathcal{C}_{6}^{122222} v_{2}^3 v_{1} 
	+3\mathcal{C}_{6}^{112212} v_{2} v_{1}^3
	+3 \mathcal{C}_{6}^{121212} v_{2} v_{1}^3 } \nonumber\\
&=& 0. \nonumber
\label{eq:mindim62}
\end{eqnarray}
In passing, we would like to mention that our minimization criteria are in well agreement with the Ref.~\cite{Crivellin:2016ihg} in the absence of the $Z_2$ violating operators.

\vskip 0.4cm
\underline{\large{$\phi^{4}D^{2}$ and $\phi^6$ operators: Scalar field redefinitions}}\\

The scalar-kinetic lagrangian, see Eq.~\ref{eq:2HDM-ren-lag}, gets modified in presence of additional $\phi^{4}D^{2}$ operators, Table~\ref{tab:2HDM-phi4D2}, as :
\small{\begin{eqnarray}\label{eq:2HDM-scalar-kinetic-full}
	\mathcal{L}_{kin}^{(4)+(6)}&=& \frac{1}{2} \begin{pmatrix}
	\partial^{\mu}h_{1} \ \ \partial^{\mu}h_{2} 
	\end{pmatrix}
	\begin{pmatrix} 
	1+\frac{A_{r}^{11}}{\Lambda^{2}} \hspace{0.5cm}  \frac{A_{r}^{12}}{\Lambda^{2}}\\
	\frac{A_{r}^{12}}{\Lambda^{2}} \hspace{0.5cm}
	1+\frac{A_{r}^{11}}{\Lambda^{2}}\\
	\end{pmatrix}
	\begin{pmatrix}
	\partial_{\mu}h_{1} \\ \partial_{\mu}h_{2}
	\end{pmatrix} \nonumber \\ 
	\vspace{0.5cm} \nonumber \\
	&+& \frac{1}{2} \begin{pmatrix}
	\partial^{\mu}a_{1} \ \  \partial^{\mu} a_{2} 
	\end{pmatrix}
	\begin{pmatrix} 
	1+\frac{A_{i}^{11}}{\Lambda^{2}} \hspace{0.5cm} \frac{A_{i}^{12}}{\Lambda^{2}}\\
	\frac{A_{i}^{12}}{\Lambda^{2}} \hspace{0.5cm} 
	1+\frac{A_{i}^{11}}{\Lambda^{2}}\\
	\end{pmatrix}
	\begin{pmatrix}
	\partial_{\mu}a_{1}  \\ \partial_{\mu} a_{2} 
	\end{pmatrix}  \\ 
	\vspace{0.5cm} \nonumber  \\
	&+& \begin{pmatrix}
	\partial^{\mu}\phi_{1}^{-} \ \ \partial^{\mu}\phi_{2}^{-} 
	\end{pmatrix}
	\begin{pmatrix} 
	1+\frac{A_{+}^{11}}{\Lambda^{2}} \hspace{0.5cm} \frac{A_{+}^{12}}{\Lambda^{2}} \\
	\frac{A_{+}^{12}}{\Lambda^{2}} \hspace{0.5cm} 1+\frac{A_{+}^{22}}{\Lambda^{2}} \\
	\end{pmatrix} 
	\begin{pmatrix}
	\partial_{\mu}\phi_{1}^{+} \\ \partial_{\mu}\phi_{2}^{+} 
	\end{pmatrix},  \nonumber 
	\end{eqnarray}}
\clearpage
where (note the additional contributions due to the $Z_2$ violating operators in comparison with \cite{Crivellin:2016ihg}),
{\small\begin{eqnarray*}
	A_{r}^{11}& =&\frac{1}{2}\mathcal{C}_{\phi D}^{(1)11(1)} v_1^2+\frac{1}{2}\mathcal{C}_{\phi D}^{(1)22(1)} v_2^2-\mathcal{C}_{\square }^{21(21)} v_2^2-2 \mathcal{C}_{\square }^{11(11)} v_1^2{\color{purple}+\mathcal{C}_{\phi D}^{(1)21(1)} v_2 v_1+\mathcal{C}_{\phi D}^{12(1)(1)} v_2 v_1}, \\
	A_{r}^{12}& =&\frac{1}{2} \mathcal{C}_{\phi D}^{12(1)(2)} v_1 v_2-\mathcal{C}_{\square }^{11(22)} v_1 v_2-\mathcal{C}_{\square }^{22(11)} v_1 v_2-\mathcal{C}_{\square }^{21(21)} v_1 v_2{\color{purple}+\frac{1}{2}\mathcal{C}_{\phi D}^{(2)22(1)} v_2^2},  \\
	A_{r}^{22}& =&\frac{1}{2}\mathcal{C}_{\phi D}^{(2)11(2)} v_1^2+\frac{1}{2}\mathcal{C}_{\phi D}^{(2)22(2)} v_2^2-\mathcal{C}_{\square }^{21(21)} v_1^2-2 \mathcal{C}_{\square }^{22(22)} v_2^2{\color{purple}+\mathcal{C}_{\phi D}^{21(2)(2)} v_2 v_1},  \\
	A_{i}^{11}& =&\frac{1}{2}\mathcal{C}_{\phi D}^{(1)11(1)} v_1^2+\frac{1}{2}\mathcal{C}_{\phi D}^{(1)22(1)} v_2^2+\mathcal{C}_{\square }^{21(21)} v_2^2 {\color{purple}+\mathcal{C}_{\phi D}^{(1)21(1)} v_2 v_1+\mathcal{C}_{\phi D}^{12(1)(1)} v_2 v_1}, \\
	A_{i}^{12}&=&\frac{1}{2} \mathcal{C}_{\phi D}^{12(1)(2)} v_1 v_2-\mathcal{C}_{\square }^{21(21)} v_1 v_2{\color{purple}+\frac{1}{2}\mathcal{C}_{\phi D}^{(2)22(1)} v_2^2} , \\
	A_{i}^{22}&=&\frac{1}{2}\mathcal{C}_{\phi D}^{(2)11(2)} v_1^2+\frac{1}{2}\mathcal{C}_{\phi D}^{(2)22(2)} v_2^2+\mathcal{C}_{\square }^{21(21)} v_1^2{\color{purple}+\mathcal{C}_{\phi D}^{21(2)(2)} v_2 v_1} ,  \\
	A_{+}^{11}&=&{\color{purple} \mathcal{C}_{\phi D}^{12(1)(1)} v_1 v_2} ,  \\
	A_{+}^{22}&=&{\color{purple} \mathcal{C}_{\phi D}^{21(2)(2)} v_1 v_2} .  \\
	A_{+}^{12}&=&\frac{1}{2} \mathcal{C}_{\phi D}^{12(1)(2)} v_1 v_2 .  
	\end{eqnarray*}}

In order to reduce the scalar kinetic terms to their canonical forms, scalar fields are redefined as \cite{Crivellin:2016ihg}:
{\small\begin{eqnarray}\label{eq:2HDM-scalar-field-redef}
	h_{1} & \rightarrow &\left( 1-\frac{A_{r}^{11}}{2\Lambda^{2}}\right)h_{1}-\left( \frac{A_{r}^{12}}{2\Lambda^{2}} \right)h_{2} , \nonumber \\
	h_{2} &\rightarrow &\left( 1-\frac{A_{r}^{22}}{2\Lambda^{2}}\right)h_{2}-\left( \frac{A_{r}^{12}}{2\Lambda^{2}} \right)h_{1} , \nonumber \\
	a_{1} &\rightarrow &\left( 1-\frac{A_{i}^{11}}{2\Lambda^{2}}\right)a_{1}-\left( \frac{A_{i}^{12}}{2\Lambda^{2}} \right)a_{2} , \\
	a_{2} & \rightarrow &\left( 1-\frac{A_{i}^{22}}{2\Lambda^{2}}\right)a_{2}-\left( \frac{A_{i}^{12}}{2\Lambda^{2}} \right)a_{1} , \nonumber \\
	\phi_{1}^{+} &\rightarrow& \left( 1-\frac{A_{+}^{11}}{2\Lambda^{2}}\right)\phi_{1}^{+}-\left( \frac{A_{+}^{12}}{2\Lambda^{2}} \right) \phi_{2}^{+}, \nonumber \\
	\phi_{2}^{+} &\rightarrow &\left( 1-\frac{A_{+}^{22}}{2\Lambda^{2}}\right)\phi_{2}^{+}-\left( \frac{A_{+}^{12}}{2\Lambda^{2}} \right) \phi_{1}^{+} . \nonumber
	\end{eqnarray}}

These redefinitions induce modifications in the scalar mass matrices \cite{Crivellin:2016ihg}
{\small\begin{eqnarray}\label{eq:2HDM-scalar-mass-lag-full}
	\mathcal{L}_{M_{H}}^{(4)+(6)}& =& \frac{1}{2} \begin{bmatrix}
	h_{1} \ \ h_{2} 
	\end{bmatrix} \mathcal{M}_{r}^2 \begin{bmatrix}
	h_{1} \\ h_{2}
	\end{bmatrix} + \frac{1}{2} \begin{bmatrix}
	a_{1} \ \ a_{2} 
	\end{bmatrix}\mathcal{M}_{i}^2 \begin{bmatrix}
	a_{1} \\ a_{2}
	\end{bmatrix} + \begin{bmatrix}
	\phi_{1}^{-} \ \ \phi_{2}^{-} 
	\end{bmatrix}\mathcal{M}_{+}^2 \begin{bmatrix}
	\phi_{1}^{+} \\ \phi_{2}^{+} 
	\end{bmatrix}.
	\end{eqnarray}}

We have to remember that there will be further contributions to the scalar masses from the $\phi^6$ operators, e.g., when four out of six scalar fields acquire VEVs. Encapsulating all these contributions, the final scalar mass matrix can be expressed as \cite{Crivellin:2016ihg}:
{\small\begin{eqnarray}\label{eq:2HDM-scalar-mass-matrix-full}
	\mathcal{M}_{j}^2& =& \begin{pmatrix}
	(m^{2}_{j}+\Delta m^{2}_{j,\phi^{6}} + \Delta m^{2}_{j,\phi D})_{11} & (m^{2}_{j}+\Delta m^{2}_{j,\phi^{6}} + \Delta m^{2}_{j,\phi D})_{12} \\
	(m^{2}_{j}+\Delta m^{2}_{j,\phi^{6}} + \Delta m^{2}_{j,\phi D})_{12} & (m^{2}_{j}+\Delta m^{2}_{j,\phi^{6}} + \Delta m^{2}_{j,\phi D})_{22} &
	\end{pmatrix}, \ \ \  \ \ \ \  \{j \equiv r,i,+\},
	\end{eqnarray}}
where, $\Delta m^{2}_{j,\phi D}$ and $\Delta m^{2}_{j,\phi^6}$ are the corrections to the scalar spectrum due to $\phi^4 D^2$ and $\phi^6$ operators respectively. First, we have discussed the modification in the scalar mass spectrum due to the field redefinitions ($\Delta m^{2}_{\phi D}$), and then we have computed the contributions from $\phi^6$ operators ($\Delta m^{2}_{\phi^6}$).

\vskip 0.2cm
\underline{\large{CP-even scalar mass matrix}}
{\small\begin{eqnarray}
	(\Delta m^{2}_{r,\phi D})_{11}&=& \frac{m_{12}^2 \left(v_1 A_{r }^{12}-v_2 A_{r }^{11}\right)}{\Lambda ^2 v_1}-\frac{\left(\lambda _3+\lambda _4+\lambda _5\right) v_1 v_2 A_{r }^{12}}{\Lambda ^2}-\frac{2 \lambda _1 v_1^2 A_{r }^{11}}{\Lambda ^2} \nonumber \\
	&-&\frac{3 \lambda _6 v_1 \left(v_1 A_{r }^{12}+v_2 A_{r }^{11}\right)}{2 \Lambda ^2}+\frac{\lambda _7 v_2^2 \left(v_2 A_{r }^{11}-3 v_1 A_{r }^{12}\right)}{2 \Lambda ^2 v_1} , \\ \vspace{1cm} 
	(\Delta m^{2}_{r,\phi D})_{12}&=&\frac{m_{12}^2 \left(v_1 v_2 \left(A_{r }^{22}+A_{r }^{11}\right)-v^2 A_{r }^{12}\right)}{2 \Lambda ^2 v_1 v_2}-\frac{\lambda _1 v_1^2 A_{r }^{12}}{\Lambda ^2}-\frac{\lambda _2 v_2^2 A_{r }^{12}}{\Lambda ^2}\nonumber \\
	&-&\frac{\left(\lambda _3+\lambda _4+\lambda _5\right) v_1 v_2 \left(A_{r }^{22}+A_{r }^{11}\right)}{2 \Lambda ^2}\nonumber \\
	&+&\frac{\lambda _6 v_1 \left(\left(v_1^2-3 v_2^2\right) A_{r }^{12}-3 v_1 v_2 \left(A_{r }^{22}+A_{r }^{11}\right)\right)}{4 \Lambda ^2 v_2}\nonumber \\
	&+&\frac{\lambda _7 v_2 \left(\left(v_2^2-3 v_1^2\right) A_{r }^{12}-3 v_1 v_2 \left(A_{r }^{22}+A_{r }^{11}\right)\right)}{4 \Lambda ^2 v_1} , \\
	(\Delta m^{2}_{r,\phi D})_{22}&=&\frac{m_{12}^2 \left(A_{r }^{12}-\frac{v_1 A_{r }^{22}}{v_2}\right)}{\Lambda ^2}-\frac{2 \lambda _2 v_2^2 A_{r }^{22}}{\Lambda ^2}-\frac{\left(\lambda _3+\lambda _4+\lambda _5\right) v_1 v_2 A_{r }^{12}}{\Lambda ^2}\nonumber \\
	&+&\frac{\lambda _6 v_1^2 \left(v_1 A_{r }^{22}-3 v_2 A_{r }^{12}\right)}{2 \Lambda ^2 v_2}-\frac{3 \lambda _7 v_2 \left(v_1 A_{r }^{22}+v_2 A_{h }^{12}\right)}{2 \Lambda ^2},
	\end{eqnarray}
	
\vskip 0.2cm
\underline{\large{CP-odd scalar mass matrix}}
{\small\begin{eqnarray}
	(\Delta m^{2}_{i,\phi D})_{11}&=& \frac{m_{12}^2 \left(A_{i }^{12}-\frac{v_2 A_{i }^{11}}{v_1}\right)}{\Lambda ^2}+\frac{\lambda _5 v_2 \left(v_2 A_{i }^{11}-v_1 A_{i }^{12}\right)}{\Lambda ^2}+\nonumber \\
	&+&\frac{\lambda _6 v_1 \left(v_2 A_{i }^{11}-v_1 A_{i }^{12}\right)}{2 \Lambda ^2}+\frac{\lambda _7 v_2^2 \left(v_2 A_{i }^{11}-v_1 A_{i }^{12}\right)}{2 \Lambda ^2 v_1} , \\
	(\Delta m^{2}_{i,\phi D})_{12}&=& \frac{m_{12}^2 \left(v_1 v_2 \left(A_{i }^{22}+A_{i }^{11}\right)-v^2 A_{i }^{12}\right)}{2 \Lambda ^2 v_1 v_2}+\frac{\lambda _5 \left(v^2 A_{i }^{12}\right)}{2 \Lambda ^2}\nonumber \\
	&-&\frac{\lambda _5 \left(v_1 v_2 \left(A_{i }^{22}+A_{i }^{11}\right)\right)}{2 \Lambda ^2}+\frac{\lambda _6 v_1 \left(v^2 A_{i }^{12}-v_1 v_2 \left(A_{i }^{22}+A_{i }^{11}\right)\right)}{4 \Lambda ^2 v_2}\nonumber \\
	&+&\frac{\lambda _7 v_2 \left(v^2 A_{i }^{12}-v_1 v_2 \left(A_{i }^{22}+A_{i }^{11}\right)\right)}{4 \Lambda ^2 v_1} , \\
	(\Delta m^{2}_{i,\phi D})_{22}&=&\frac{m_{12}^2 \left(v_2 A_{i }^{12}-v_1 A_{i }^{22}\right)}{\Lambda ^2 v_2}+\frac{\lambda _5 v_1 \left(v_1 A_{i }^{22}-v_2 A_{i }^{12}\right)}{\Lambda ^2}\nonumber \\
	&+&\frac{\lambda _6 v_1^2 \left(v_1 A_{i }^{22}-v_2 A_{i }^{12}\right)}{2 \Lambda ^2 v_2}+\frac{\lambda _7 v_2 \left(v_1 A_{i }^{22}-v_2 A_{i }^{12}\right)}{2 \Lambda ^2},
	\end{eqnarray}}
\pagebreak

\vskip 0.1cm
\underline{\large{Singly charged scalar mass matrix}}\\
{\small\begin{eqnarray*}
	(\Delta m^{2}_{+,\phi D})_{11}&=&-\frac{m_{12}^2 v_2 A_{+}^{11}}{\Lambda ^2 v_1}+\frac{\left(\lambda _4+\lambda _5\right) v_2^2 A_{+}^{11}}{2 \Lambda ^2}+\frac{\lambda _6 v_1 v_2 A_{+}^{11}}{2 \Lambda ^2}+\frac{\lambda _7 v_2^3 A_{+}^{11}}{2 \Lambda ^2 v_1}, \\
	(\Delta m^{2}_{+,\phi D})_{12}&=&\frac{m_{12}^2 \left(A_{+}^{22}+A_{+}^{11}\right)}{2 \Lambda ^2}-\frac{\left(\lambda _4+\lambda _5\right) v_1 v_2 \left(A_{+}^{22}+A_{+}^{11}\right)}{4 \Lambda ^2}\nonumber \\
	&-&\frac{\lambda _6 v_1^2 \left(A_{+}^{22}+A_{+}^{11}\right)}{4 \Lambda ^2}-\frac{\lambda _7 v_2^2 \left(A_{+}^{22}+A_{+}^{11}\right)}{4 \Lambda ^2} , \\
	(\Delta m^{2}_{+,\phi D})_{22}&=&-\frac{m_{12}^2 v_1 A_{+}^{22}}{\Lambda ^2 v_2}+\frac{\left(\lambda _4+\lambda _5\right) v_1^2 A_{+}^{22}}{2 \Lambda ^2}+\frac{\lambda _6 v_1^3 A_{+}^{22}}{2 \Lambda ^2 v_2}+\frac{\lambda _7 v_1 v_2 A_{+}^{22}}{2 \Lambda ^2}.
	\end{eqnarray*}}


\begin{center}
	\underline{\large{\bf {Contributions from $\phi^{6}$ operators}}}
\end{center}

Here, we would like to emphasize that the additional contributions due to the $Z_2$ violating $\phi^6$-operators  in comparison with \cite{Crivellin:2016ihg} are given in color throughout the text.

\vskip 0.2cm
\underline{\large{CP-even scalar mass matrix}}

\begin{eqnarray*}
(\Delta m^{2}_{r,\phi^{6}})_{11} &=& \frac{1}{\Lambda^{2} } \Big[  3 \mathcal{C}_{6}^{111111} v_1^4+\mathcal{C}_{6}^{111122} v_1^2 v_2^2+2 \mathcal{C}_{6}^{121211} v_1^2 v_2^2  +\mathcal{C}_{6}^{122111} v_1^2 v_2^2 \\ 
& &{ \color{purple}   + \frac{3}{4} \mathcal{C}_{6}^{121221} v_1 v_2^3-\frac{v_2^5}{4 v_1}\mathcal{C}_{6}^{122222} +\frac{15}{4} \mathcal{C}_{6}^{111112} v_1^3 v_2 + \frac{3}{4} \mathcal{C}_{6}^{112212} v_1 v_2^3 +\frac{3}{4} \mathcal{C}_{6}^{121212} v_1 v_2^3} \Big],\nonumber    \\
(\Delta m^{2}_{r,\phi^{6}})_{12} &=& \frac{1}{\Lambda^{2} } \left[\mathcal{C}_{6}^{111122} v_2 v_1^3
+2 \mathcal{C}_{6}^{121211} v_2 v_1^3
+2 \mathcal{C}_{6}^{121222} v_2^3 v_1+\mathcal{C}_{6}^{112222} v_2^3 v_1+\mathcal{C}_{6}^{122111} v_2 v_1^3 +\mathcal{C}_{6}^{122122} v_2^3 v_1 \right. \nonumber  \\
& &\left. {\color{purple}+\frac{9}{4} \mathcal{C}_{6}^{121221} v_2^2 v_1^2 +\frac{5 v_2^4}{4}\mathcal{C}_{6}^{122222} +\frac{9}{4} \mathcal{C}_{6}^{112212} v_2^2 v_1^2+\frac{9}{4} \mathcal{C}_{6}^{121212} v_2^2 v_1^2+\frac{5 v_1^4}{4}\mathcal{C}_{6}^{111112}}  \right],  \\
(\Delta m^{2}_{r,\phi^{6}})_{22}& =& \frac{1}{\Lambda^{2} } \left[2 \mathcal{C}_{6}^{121222} v_2^2 v_1^2+\mathcal{C}_{6}^{112222} v_2^2 v_1^2 +\mathcal{C}_{6}^{122122} v_2^2 v_1^2 +3 \mathcal{C}_{6}^{222222} v_2^4   \right. \\
& & \left. {\color{purple}+\frac{3}{4} \mathcal{C}_{6}^{121221} v_2 v_1^3 +\frac{15}{4} \mathcal{C}_{6}^{122222} v_2^3 v_1+\frac{3}{4} \mathcal{C}_{6}^{112212} v_2 v_1^3+\frac{3}{4} \mathcal{C}_{6}^{121212} v_2 v_1^3 -\frac{v_1^5}{4 v_2}\mathcal{C}_{6}^{111112}} \right],\nonumber
\end{eqnarray*}
\vskip 0.1cm
\underline{\large{CP-odd scalar mass matrix}}

\begin{eqnarray}
(\Delta m^{2}_{i,\phi^{6}})_{11}& =& \frac{1}{\Lambda^{2} } \left[ -\mathcal{C}_{6}^{121211} v_1^2 v_2^2-\mathcal{C}_{6}^{121222} v_2^4  \right.  \\
& &\left. { \color{purple} -\frac{1}{4} \mathcal{C}_{6}^{121221} v_1 v_2^3 -\frac{v_2^5}{4 v_1}\mathcal{C}_{6}^{122222} -\frac{1}{4} \mathcal{C}_{6}^{111112} v_1^3 v_2 -\frac{1}{4} \mathcal{C}_{6}^{112212} v_{1} v_{2}^{3} -\frac{9}{4} \mathcal{C}_{6}^{121212} v_1 v_2^3}   \right] \nonumber  \\
(\Delta m^{2}_{i,\phi^{6}})_{12} &=& \frac{1}{\Lambda^{2} } \left[ \mathcal{C}_{6}^{121211} v_2 v_1^3 +\mathcal{C}_{6}^{121222} v_2^3 v_1 \right.  \\
& &\left. {\color{purple}+ \frac{v_1^4}{4}\mathcal{C}_{6}^{111112} +\frac{1}{4} \mathcal{C}_{6}^{112212} v_2^2 v_1^2+\frac{9}{4} \mathcal{C}_{6}^{121212} v_2^2 v_1^2  +\frac{1}{4} \mathcal{C}_{6}^{121221} v_2^2 v_1^2+\frac{v_2^4}{4}\mathcal{C}_{6}^{122222}}  \right] \nonumber \\
(\Delta m^{2}_{i,\phi^{6}})_{22}& =& \frac{1}{\Lambda^{2} } \left[ -\mathcal{C}_{6}^{121211} v_1^4 -\mathcal{C}_{6}^{121222} v_2^2 v_1^2 \right.  \\
& & \left. {\color{purple} -\frac{v_1^5}{4 v_2}\mathcal{C}_{6}^{111112} -\frac{1}{4} \mathcal{C}_{6}^{112212} v_2 v_1^3-\frac{9}{4} \mathcal{C}_{6}^{121212} v_2 v_1^3 -\frac{1}{4} \mathcal{C}_{6}^{121221} v_2 v_1^3-\frac{1}{4} \mathcal{C}_{6}^{122222} v_2^3 v_1} \right].\nonumber 
\end{eqnarray}
\vskip 0.1cm
\underline{\large{Charged scalar mass matrix}}
\begin{eqnarray}
(\Delta m^{2}_{+,\phi^{6}})_{11}& =& \frac{1}{\Lambda^{2} } \left[-\frac{1}{2} \mathcal{C}_{6}^{121211} v_1^2 v_2^2-\frac{v_2^4}{2}\mathcal{C}_{6}^{121222} -\frac{1}{4} \mathcal{C}_{6}^{122111} v_1^2 v_2^2 -\frac{v_2^4}{4}\mathcal{C}_{6}^{122122} \right.  \\
& & \left. { \color{purple} -\frac{3}{4} \mathcal{C}_{6}^{121221} v_1 v_2^3 -\frac{v_2^5}{4 v_1}\mathcal{C}_{6}^{122222}-\frac{1}{4} \mathcal{C}_{6}^{111112} v_1^3 v_2 -\frac{1}{4} \mathcal{C}_{6}^{112212} v_{1} v_{2}^{3} -\frac{3}{4} \mathcal{C}_{6}^{121212} v_1 v_2^3}  \right] ,\nonumber \\
(\Delta m^{2}_{+,\phi^{6}})_{12}& =& \frac{1}{\Lambda^{2} } \left[ \frac{1}{4} \mathcal{C}_{6}^{122122} v_2^3 v_1+\frac{1}{2} \mathcal{C}_{6}^{121211} v_2 v_1^3+\frac{1}{2} \mathcal{C}_{6}^{121222} v_2^3 v_1 +\frac{1}{4} \mathcal{C}_{6}^{122111} v_2 v_1^3 \right.  \\
&& \left.  { \color{purple} +\frac{3}{4} \mathcal{C}_{6}^{121212} v_2^2 v_1^2 +\frac{3}{4} \mathcal{C}_{6}^{121221} v_2^2 v_1^2+\frac{v_2^4}{4}\mathcal{C}_{6}^{122222} + \frac{v_1^4}{4}\mathcal{C}_{6}^{111112} +\frac{1}{4} \mathcal{C}_{6}^{112212} v_2^2 v_1^2} \right] \nonumber ,\\
(\Delta m^{2}_{+,\phi^{6}})_{22}& =& \frac{1}{\Lambda^{2} } \left[ -\frac{v_1^4}{2}\mathcal{C}_{6}^{121211} -\frac{1}{2} \mathcal{C}_{6}^{121222} v_2^2 v_1^2 -\frac{v_1^4}{4}\mathcal{C}_{6}^{122111}-\frac{1}{4} \mathcal{C}_{6}^{122122} v_2^2 v_1^2  \right. \\
& & \left. { \color{purple}-\frac{3}{4} \mathcal{C}_{6}^{121212} v_2 v_1^3  -\frac{3}{4} \mathcal{C}_{6}^{121221} v_2 v_1^3-\frac{1}{4} \mathcal{C}_{6}^{122222} v_2^3 v_1 -\frac{v_1^5}{4 v_2}\mathcal{C}_{6}^{111112} -\frac{1}{4} \mathcal{C}_{6}^{112212} v_2 v_1^3} \right].\nonumber 
\end{eqnarray}
\noindent
These scalar mass matrices $\mathcal{M}_{j}^2$, Eq.~\ref{eq:2HDM-scalar-mass-matrix-full}, can be diagonalized by the new rotation angles $\bar{\beta_j}$ which are related to the older ones $\beta_j$ (for $\{ j\equiv r,i,+\}$) as:
{\small\begin{align}\label{eq:2HDM-scalar-rot-angle-modified}
	\tan2\bar{\beta}_{j}=\tan2\beta_{j} \left(1-\frac{(\Delta m^{2}_{j,\phi^{6}} + \Delta m^{2}_{j,\phi D})_{12}}{(m^{2}_{j})_{12}}-\frac{(\Delta m^{2}_{j,\phi^{6}} + \Delta m^{2}_{j,\phi D})_{22}-(\Delta m^{2}_{j,\phi^{6}} + \Delta m^{2}_{j,\phi D})_{11}}{(m^{2}_{j})_{22}-(m^{2}_{j})_{11}}\right).
	\end{align}}
We would like to mention that through out our calculation we have dropped terms which are beyond $\mathcal{O}(\frac{1}{\Lambda^2})$.
After diagonalizing the scalar mass matrices, we can express the physical scalar spectrum as \cite{Crivellin:2016ihg}: \\

\underline{\large{CP-even scalar ($H_1,H_2$)}} 
{\small\begin{eqnarray}\label{eq:2HDM-scalar-physical-mass-full}
	\mathcal{M}^2_{H_{1}}&=& (m^{2}_{r}+\Delta m^{2}_{r,\phi^{6}} + \Delta m^{2}_{r,\phi D})_{11}  \cos ^2\left(\bar{\beta}_{r}\right)-(m^{2}_{r}+\Delta m^{2}_{r,\phi^{6}} + \Delta m^{2}_{r,\phi D})_{12} \sin \left(2 \bar{\beta}_{r}\right)\nonumber \\
	&+&(m^{2}_{r}+\Delta m^{2}_{r,\phi^{6}} + \Delta m^{2}_{r,\phi D})_{22} \sin ^2\left(\bar{\beta}_{r}\right),\\
	\mathcal{M}^2_{H_{2}}&=& (m^{2}_{r}+\Delta m^{2}_{r,\phi^{6}} + \Delta m^{2}_{r,\phi D})_{11} \sin ^2\left(\bar{\beta}_{r}\right)+(m^{2}_{r}+\Delta m^{2}_{r,\phi^{6}} + \Delta m^{2}_{r,\phi D})_{12} \sin \left(2 \bar{\beta_{r}}\right)\nonumber \\
	&+&(m^{2}_{r}+\Delta m^{2}_{r,\phi^{6}} + \Delta m^{2}_{r,\phi D})_{22} \cos ^2\left(\bar{\beta}_{r}\right),
	\end{eqnarray}}

\underline{\large{CP-odd scalar ($A$)}} 
{\small\begin{eqnarray}
	\mathcal{M}^2_{A}&=& (m^{2}_{i}+\Delta m^{2}_{i,\phi^{6}} + \Delta m^{2}_{i,\phi D})_{11} \cos ^2\left(\bar{\beta}_{i}\right)-(m^{2}_{i}+\Delta m^{2}_{i,\phi^{6}} + \Delta m^{2}_{i,\phi D})_{12} \sin \left(2 \bar{\beta}_{i}\right)\nonumber \\
	&+&(m^{2}_{i}+\Delta m^{2}_{i,\phi^{6}} + \Delta m^{2}_{i,\phi D})_{22} \sin ^2\left(\bar{\beta}_{i}\right),
	\end{eqnarray}}

\underline{\large{Charged scalar ($H^{\pm}$)}} 
{\small\begin{eqnarray}
	\mathcal{M}^2_{H^{\pm}}&=& (m^{2}_{+}+\Delta m^{2}_{+,\phi^{6}} + \Delta m^{2}_{+,\phi D})_{11} \cos ^2\left(\bar{\beta}_{+}\right)-(m^{2}_{+}+\Delta m^{2}_{+,\phi^{6}} + \Delta m^{2}_{+,\phi D})_{12} \sin \left(2 \bar{\beta}_{+}\right)\nonumber \\
	&+&(m^{2}_{+}+\Delta m^{2}_{+,\phi^{6}} + \Delta m^{2}_{+,\phi D})_{22} \sin ^2\left(\bar{\beta}_{+}\right).
	\end{eqnarray}}
\clearpage

\vskip 0.2cm
\underline{\large{$\phi^{2}X^{2}$ and $\phi^{4}D^{2}$ operators: Gauge field redefinition and spectrum}}\\

The gauge-kinetic lagrangian, see Eq.~\ref{eq:2HDM-ren-lag}, gets modified in the presence of $\phi^{2}X^{2}$ operators, Table~\ref{tab:2HDM-phi2X2}, as \cite{Crivellin:2016ihg}:
\begin{equation}\label{eq:2HDM-GB-kinetic-full}
\mathcal{L}^{(4)+(6)}_{gauge,kin}= -\left(1 - \frac{2\Theta_{WW}}{\Lambda^2}\right)(\partial_{\mu}W^{+}_{\nu})(\partial^{\mu}W^{-\nu}) -
\frac{1}{2}\begin{pmatrix}
\partial_{\mu}W_{\nu}^{3} \ \ \partial_{\mu}B_{\nu}
\end{pmatrix}
\begin{pmatrix} 
1-\frac{2\Theta_{WW}}{\Lambda^{2}} \hspace{0.5cm} \frac{\Theta_{W_{3}B}}{\Lambda^{2}}\\
\frac{\Theta_{W_{3}B}}{\Lambda^{2}} \hspace{0.5cm} 
1-\frac{2\Theta_{BB}}{\Lambda^{2}}\\
\end{pmatrix}
\begin{pmatrix}
\partial^{\mu}W^{3\nu}  \\ \partial^{\mu}B^{\nu}
\end{pmatrix}, \nonumber
\end{equation}
where,
\begin{eqnarray}\label{eqn:gaugeredef_2HDM}
\Theta_{WW} &=& v_1^2\mathcal{C}^{11}_{\phi W} + v_2^2\mathcal{C}^{22}_{\phi W} + \textcolor{purple}{v_1v_2(\mathcal{C}^{12}_{\phi W}+\mathcal{C}^{21}_{\phi W})},\nonumber\\
\Theta_{BB} &=& v_1^2\mathcal{C}^{11}_{\phi B} + v_2^2\mathcal{C}^{22}_{\phi B} + \textcolor{purple}{v_1v_2(\mathcal{C}^{12}_{\phi B}+\mathcal{C}^{21}_{\phi B})},\\
\Theta_{W3B} &=& v_1^2\mathcal{C}^{11}_{\phi WB} + v_2^2\mathcal{C}^{22}_{\phi WB} + \textcolor{purple}{v_1v_2(\mathcal{C}^{12}_{\phi WB}+\mathcal{C}^{21}_{\phi WB})}.\nonumber
\end{eqnarray}

Similar to the scalar sector, we need to redefine the gauge fields such that the above gauge kinetic lagrangian can be brought into the canonical form. The redefined gauge fields can be written as  \cite{Crivellin:2016ihg}:
\begin{eqnarray}\label{eq:2HDM-GB-redef}
W_{\mu}^{\pm} &\rightarrow &W_{\mu}^{\pm}\Big( 1 + \frac{\Theta_{WW}}{\Lambda^2}\Big),\\
W_{\mu}^{3} &\rightarrow& W_{\mu}^{3}\Big( 1 + \frac{\Theta_{WW}}{\Lambda^2}\Big) - \frac{\Theta_{W_{3}B}}{2\Lambda^2}B_{\mu},\\
B_{\mu} &\rightarrow& B_{\mu}\Big( 1 + \frac{\Theta_{BB}}{\Lambda^2}\Big) - \frac{\Theta_{W_{3}B}}{2\Lambda^2}W_{\mu}^{3}.
\end{eqnarray}

The redefinitions of the gauge fields lead to  corrections in the gauge boson mass matrix as :
\vskip 0.2cm
\underline{\large{Charged gauge boson mass}}
{\small\begin{eqnarray}
	m^2_W + (\Delta m^2_W)_{\phi^2X^2} = \frac{1}{4}g^2v^2\left(1 + \frac{2\Theta_{WW}}{\Lambda^2}\right).
	\end{eqnarray}}

\underline{\large{Neutral gauge boson mass}}
\begin{eqnarray}
(m^2_0 + (\Delta m^2_0)_{\phi^2X^2})_{11}& =&\frac{1}{4}v^2\left[g^2\left(1+\frac{2\Theta_{WW}}{\Lambda^2}\right) + gg'\frac{\Theta_{W_{3}B}}{\Lambda^2}\right],\nonumber \\ \nonumber \\
(m^2_0 + (\Delta m^2_0)_{\phi^2X^2})_{12} &=&\frac{1}{4}v^2\left[-gg'\left(1+\frac{\Theta_{BB}+\Theta_{WW}}{\Lambda^2}\right)-(g^2+g'^2)\frac{\Theta_{W_{3}B}}{2\Lambda^2}\right]\nonumber\\ \nonumber \\
&=& (m^2_0 + (\Delta m^2_0)_{\phi^2X^2})_{21},\\ \nonumber \\
(m^2_0 + (\Delta m^2_0)_{\phi^2X^2})_{22}&=&\frac{1}{4}v^2\left[g'^2\left(1+\frac{2\Theta_{BB}}{\Lambda^2}\right) + gg'\frac{\Theta_{W_{3}B}}{\Lambda^2}\right].\nonumber
\end{eqnarray}

There will be additional contributions to these mass matrices from $\phi^4D^2$ operators, Table~\ref{tab:2HDM-phi4D2}. These shifts can be given as  (note the additional contributions due to the $Z_2$ violating operators in comparison with \cite{Crivellin:2016ihg}):
\begin{eqnarray}
(\Delta m_W^2)_{\phi^4D^2} &=& \frac{1}{4\Lambda^2}g^2v_1^2v_2^2\mathcal{C}_{\phi D}^{12(1)(2)}+\textcolor{purple}{\frac{1}{4\Lambda^2}g^2v_1v_2\left(v_1^2\mathcal{C}_{\phi D}^{12(1)(1)}+v_2^2\mathcal{C}_{\phi D}^{21(2)(2)}\right)}, \\ \\
(\Delta m_0^2)_{\phi^4D^2}& =& \frac{\Theta_6^{WB}}{8\Lambda^2}
\begin{pmatrix}
\hspace{0.5cm}g^2 \ \ \hspace{0.2cm} -gg'\\
-gg' \ \ \hspace{0.5cm} g'^2 
\end{pmatrix},
\end{eqnarray}
where
{\small\begin{eqnarray}
	\Theta_6^{WB} &=&  v^2_1\big(v^2_1 \mathcal{C}^{(1)11(1)}_{\varphi D} + v^2_2 \mathcal{C}^{(1)22(1)}_{\varphi D}\big) + v^2_2\big(v^2_1 \mathcal{C}^{(2)11(2)}_{\varphi D} + v^2_2 \mathcal{C}^{(2)22(2)}_{\varphi D}\big)+ 2 v_1^2 v_2^2 \mathcal{C}^{12(1)(2)}_{\varphi D}\nonumber\\
	&& \textcolor{purple}{+2v_1v_2\Big( v^2_1 (\mathcal{C}^{(1)21(1)}_{\varphi D} + \mathcal{C}^{12(1)(1)}_{\varphi D}) + v^2_2 (\mathcal{C}^{(2)22(1)}_{\varphi D} + \mathcal{C}^{21(2)(2)}_{\varphi D})\Big)}.
	\end{eqnarray}}
The neutral gauge boson mass matrix can be diagonalized using the following relation:
\begin{equation}
\begin{pmatrix}
W^3_{\mu} \\ B_{\mu} 
\end{pmatrix}=
\begin{pmatrix}
\cos\tilde{\theta}_{\text{w}} \ \ \sin\tilde{\theta}_{\text{w}} \\
-\sin\tilde{\theta}_{\text{w}} \ \ \cos\tilde{\theta}_{\text{w}}
\end{pmatrix}\begin{pmatrix}
Z_{\mu} \\ A_{\mu} 
\end{pmatrix},
\end{equation}
where the modified rotation angle $\tilde{\theta}_{\text{w}}$ is related to ${\theta}_{\text{w}}$ by the following relation:
{\small\begin{eqnarray}\label{eq:2HDM-GB-rot-angle-modified}
	\tan 2\tilde{\theta}_{\text{w}} &=& \tan 2\theta_{\text{w}}\Bigg[1+\frac{1}{\Lambda^2}\Big(\sec 2\theta_{\text{w}}(\Theta_{BB}-\Theta_{WW})+\csc 2\theta_{\text{w}} \Theta_{W3B} \Big) 
	\Bigg]\\    
	& =& \frac{2gg'}{g^2-g'^2}\Bigg[1+\frac{1}{\Lambda^2}\Big(\frac{g^2+g'^2}{g^2-g'^2}(\Theta_{BB}-\Theta_{WW})
	+\frac{g^2+g'^2}{2gg'} \Theta_{W3B} \Big) \Bigg]. 
	\end{eqnarray}}
Here, $\theta_{\text{w}}$ is the Weak mixing angle and $\tan\theta_{\text{w}} = g'/g$.

After diagonalizing the gauge boson mass matrices, we have noted the physical gauge boson spectrum as \cite{Crivellin:2016ihg}:
\begin{eqnarray}\label{eq:2HDM-GB-physical-mass-full}
\mathcal{M}^2_{W} & = & \frac{1}{4}g^2\Big[v^2+\frac{2v^2\Theta_{WW}}{\Lambda^2}+\frac{v_1^2v_2^2}{\Lambda^2}\mathcal{C}^{12(1)(2)}_{\phi D}+\textcolor{purple}{\frac{v_1v_2}{\Lambda^2}(v_1^2\mathcal{C}^{12(1)(1)}_{\phi D}+v_2^2\mathcal{C}^{21(2)(2)}_{\phi D})}\Big], \nonumber \\
\mathcal{M}^2_{Z} & = & \frac{1}{4}(g^2+g'^2)\Big[v^2 + \frac{2v^2}{\Lambda^2}(\cos^2\theta_{\text{w}}\Theta_{WW}+\sin^2\theta_{\text{w}}\Theta_{BB}+\cos\theta_{\text{w}}\sin\theta_{\text{w}}\Theta_{W3B})+\frac{\Theta^{WB}_6}{2\Lambda^2}\Big], \nonumber\\
\mathcal{M}^2_{A} & = & 0.
\end{eqnarray}
Here, we would like to draw the reader's attention towards the fact that even after incorporating all these corrections, photon is massless, i.e., $\mathcal{M}_{A}^2$ is zero, which is expected.
\clearpage

\underline{\large{$ \psi^2 \phi^3$ operators: Modification in spectrum}}\\

The $\psi^2 \phi^3$ operators lead to the additional contributions to the fermion masses as \cite{Crivellin:2016ihg}:
{\small\begin{eqnarray}\label{eq:2HDM-fermion-mass-shift}
	\Delta m_e & =& \frac{1}{2\sqrt{2}\Lambda^2}\big[\textcolor{purple}{\mathcal{C}_{Le}^{1(11)} v_1^3}+v_1^2v_2(\mathcal{C}_{Le}^{1(12)}+\mathcal{C}_{Le}^{1(21)}+\mathcal{C}_{Le}^{2(11)})\textcolor{purple}{+v_1v_2^2(\mathcal{C}_{Le}^{1(22)}+\mathcal{C}_{Le}^{2(12)}+\mathcal{C}_{Le}^{2(21)})}+\mathcal{C}_{Le}^{2(22)}v_2^3\big],\nonumber \\
	\Delta m_u& =& \frac{1}{2\sqrt{2}\Lambda^2}\big[\textcolor{purple}{\mathcal{C}_{Qu}^{\tilde{1}(11)} v_1^3}+v_1^2v_2(\mathcal{C}_{Qu}^{\tilde{1}(12)}+\mathcal{C}_{Qu}^{\tilde{1}(21)}+\mathcal{C}_{Qu}^{\tilde{2}(11)})\textcolor{purple}{+v_1v_2^2(\mathcal{C}_{Qu}^{\tilde{1}(22)}+\mathcal{C}_{Qu}^{\tilde{2}(12)}+\mathcal{C}_{Qu}^{\tilde{2}(21)})}+\mathcal{C}_{Qu}^{\tilde{2}(22)}v_2^3\big],\nonumber \\
	\Delta m_d& =& \frac{1}{2\sqrt{2}\Lambda^2}\big[\textcolor{purple}{\mathcal{C}_{Qd}^{1(11)} v_1^3}+v_1^2v_2(\mathcal{C}_{Qd}^{1(12)}+\mathcal{C}_{Qd}^{1(21)}+\mathcal{C}_{Qd}^{2(11)})\textcolor{purple}{+v_1v_2^2(\mathcal{C}_{Qd}^{1(22)}+\mathcal{C}_{Qd}^{2(12)}+\mathcal{C}_{Qd}^{2(21)})}+\mathcal{C}_{Qd}^{2(22)}v_2^3\big]. \nonumber
	\end{eqnarray}}
In the presence of these dimension-6 operators, the total fermion spectrum is given as:
\begin{eqnarray}\label{eq:2HDM-fermion-mass-full}
\mathcal{M}_{f}&=& \frac{1}{\sqrt{2}}\left(v_1y_1^f + v_2y_2^f\right)-\Delta m_{f},\hspace{0.5cm} {\text{with}} \;\; f\equiv \{e,u,d\}.
\end{eqnarray}

%% file: MLRSM.tex
\section{Minimal Left-Right Symmetric Model (MLRSM)}
\label{sec:MLRSM}
The Minimal Left-Right Symmetric Model (MLRSM) is a widely discussed beyond SM scenario where parity (P) and Charge conjugation (CP) symmetries are given the same status as the gauge ones. In this framework, P and CP symmetries are  broken simultaneously with the right-handed gauge symmetry. Unlike the 2HDM case, here, the gauge sector is extended in the form of $SU(3)_C\otimes SU(2)_L\otimes SU(2)_R \otimes U(1)_{B-L}$ \cite{Mohapatra:1974gc,Senjanovic:1975rk, Mohapatra:1979ia,Mohapatra:1980yp}. There are 16 fermions per generation, unlike the SM as the right-handed neutrino is a natural inclusion in this model. The particle content and their representations, and the charges under this gauge group are given in Table~\ref{tab:MLRSM-fields}.
\begin{table}[h!]
	\centering
	\renewcommand{\arraystretch}{1.5}
	\begin{tabular}{|c|c|c|c|c|c|c|}
		\hline
		\textbf{Field} & \textbf{$SU(3)_C$} & \textbf{$SU(2)_{L}$} & \textbf{$SU(2)_{R}$} &\textbf{$U(1)_{B-L}$}&Spin & Lorentz Group\\
		\hline
		$\Phi$    &1&2&2&0&0 & Scalar\\
		$\Delta_L$&1&3&1&2&0 & Scalar \\
		$\Delta_R$&1&1&3&2&0 & Scalar \\
		\hline
		$Q_L$     &3&2&1&1/3&1/2 & Spinor \\
		$Q_R$     &3&1&2&1/3&1/2 & Spinor\\
		$L_L$     &1&2&1&-1&1/2 & Spinor\\
		$L_R$     &1&1&2&-1&1/2 & Spinor\\
		\hline
		$G^A_{\mu}$ &8&1&1&0&1 & Vector\\
		$W^I_{\mu, L}$ &1&3&1&0&1 & Vector\\
		$W^I_{\mu, R}$ &1&1&3&0&1 & Vector\\
		$B_{\mu}$ &1&1&1&0 &1 & Vector\\
		\hline
	\end{tabular}
	\caption{MLRSM: Quantum number of the fields.}
	\label{tab:MLRSM-fields}
\end{table}     

\subsubsection*{The Renormalizable Lagrangian}
Based upon the quantum numbers and transformation properties of the particles, see Table~\ref{tab:MLRSM-fields}, the renormalizable lagrangian can be written as \cite{Mohapatra:1980yp,Gunion:1989in,Deshpande:1990ip}:
\small{\begin{align}\label{eq:MLRSM-ren-lag}
		\mathcal{L}^{(4)}&=-\frac{1}{4}Tr[G^{\mu\nu}G_{\mu\nu}]-\frac{1}{4}Tr[W^{\mu\nu}_{L}W_{\mu\nu,L}]-\frac{1}{4}Tr[W^{\mu\nu}_{R}W_{\mu\nu,R}]-\frac{1}{4}(B^{\mu \nu}B_{\mu \nu}) \nonumber \\
		&+Tr\left[\left(D_\mu \Delta_{L}\right)^\dagger\left(D^\mu\Delta_{L}\right)\right]+Tr\left[\left(D_\mu \Delta_{R}\right)^\dagger\left(D^\mu\Delta_{R}\right)\right]+Tr\left[\left(D_\mu \Phi\right)^\dagger\left(D^\mu \Phi\right)\right]\nonumber \\
		&    +\bar{L}_{L} i \slashed{D} L_L + \bar{L}_{R} i \slashed{D} L_R +\bar{Q}_{L} i \slashed{D}Q_L+ \bar{Q}_{R} i \slashed{D} Q_R+V(\Phi,\Delta_{L},\Delta_{R})+\mathcal{L}_{Y} + h.c.,
\end{align}}
where the components of field strength tensors corresponding to $SU(3)_C, SU(2)_L, SU(2)_R, U(1)_{B-L}$ gauge groups are given as \cite{Mohapatra:1980yp,Gunion:1989in,Deshpande:1990ip}:
\begin{eqnarray}\label{eq:MLRSM-field-tensor}
	G_{\mu\nu}^{A}&=&\partial_{\mu} G_{\nu}^{A}-\partial_{\nu} G_{\mu}^{A}+g_3f^{ABC}G_{\mu}^{B}G_{\nu}^{C},\nonumber\\
	W_{L,\mu\nu}^{I}&=&\partial_{\mu} W_{L,\nu}^{I}-\partial_{\nu} W_{L,\mu}^{I}+g_L\epsilon^{IJK}W_{L,\mu}^{J}W_{L,\nu}^{K},\nonumber\\
	W_{R,\mu\nu}^{I}&=&\partial_{\mu} W_{R,\nu}^{I}-\partial_{\nu} W_{R,\mu}^{I}+g_R\epsilon^{IJK}W_{R,\mu}^{J}W_{R,\nu}^{K},\nonumber\\
	B_{\mu\nu}&=&\partial_{\mu} B_{\nu}-\partial_{\nu} B_{\mu},\nonumber
\end{eqnarray}
with $\{ A,B,C\} \in (1,..,8)$,  $\{ I,J,K\} \in (1,2,3)$. 
In this scenario defining covariant derivative is not very straight forward unlike the 2HDM case. Thus we have explicitly mentioned them for fermions ($ L,Q$) and scalars ($\Phi, \Delta$) as \cite{Mohapatra:1980yp,Gunion:1989in,Deshpande:1990ip}:
\begin{eqnarray}\label{eq:MLRSM-cov-der}
	D^{\mu} L_{L,R} & = & \big(\partial^{\mu} -ig_{L,R} \frac{\tau^I}{2}W^{I,\mu}_{L,R}-i\tilde{g}\frac{Y}{2}B^{\mu}\big) L_{L,R}, \\
	D^{\mu} Q_{L,R} & = & \big(\partial^{\mu} - ig_3 \frac{T^A}{2}G^A_\mu  -ig_{L,R} \frac{\tau^I}{2}W^{I,\mu}_{L,R}-i\tilde{g}\frac{Y}{2}B^{\mu}\big) Q_{L,R}, \\
	D^{\mu} \Phi&=&\partial^{\mu}\Phi-ig_{L}W^{I,\mu}_{L} \frac{\tau^I}{2}\Phi+ig_{R}\Phi\frac{\tau^I}{2}W^{I,\mu}_{R}, \nonumber\\ 
	D^{\mu}\Delta_{L,R}&=&\partial^{\mu}\Delta_{L,R}-ig_{L,R}\left[\frac{\tau^I}{2}W^{I,\mu}_{L,R},\Delta_{L,R}\right]-i\tilde{g}B^{\mu} \Delta_{L,R},
\end{eqnarray}
where $g_3$, $g_L$, $g_R$, and $\tilde{g}$ are the $SU(3)_C, SU(2)_L, SU(2)_R,$ and $U(1)_{B-L}$  gauge couplings respectively. Here, we would like to note that the indices $``L,R"$ appearing in the subscript of the fields, e.g., $Q_{L,R}$, are to signify whether that field is transforming under $SU(2)_L$ or $SU(2)_R$ gauge groups respectively.  We would like to further add that as we are working within MLRSM, we have considered $g_L=g_R=g$. In the rest of our analysis we will follow these conventions.

\vskip0.2cm
The scalar potential for MLRSM is given as \cite{Senjanovic:1978ee,Grifols:1978wk,Olness:1985bg,Frank:1991sy,Chang:1992bg,Maalampi:1993tj,Gluza:1994ad,Bhattacharyya:1995nt,Boyarkina:2000bn,Barenboim:2001vu,Gogoladze:2003bb,Azuelos:2004mwa,Kiers:2005gh,Jung:2008pz,Guadagnoli:2010sd,Blanke:2011ry,Mohapatra:2013cia,Aydemir:2014ama,Maiezza:2015lza,Maiezza:2015qbd,Dev:2016dja,Chakrabortty:2010rq,Chakrabortty:2010zk,Maiezza:2016ybz,Deppisch:2017xhv,Dev:2018foq}:
\small{\begin{eqnarray}\label{eq:MLRSM-scalar-pot-tree}
	V(\l,\r,\1)&=&-\mu _1^2 \left({Tr}\left[\Phi ^{\dagger } \Phi \right]\right)-\mu _2^2 \left({Tr}\left[\tilde{\Phi }\Phi^{\dagger}\right]+{Tr}\left[\tilde{\Phi }^{\dagger } \Phi \right]\right)-\mu _3^2 \left({Tr}\left[\Delta _L \left(\Delta _L\right){}^{\dagger }\right]+{Tr}\left[\Delta _R \left(\Delta _R\right){}^{\dagger }\right]\right) \nonumber \\
	&+&\lambda _1 {Tr}\left[\Phi  \Phi ^{\dagger }\right]^2+\lambda _2 {Tr}\left[\tilde{\Phi } \Phi ^{\dagger }\right]^2+\lambda _3 {Tr}\left[\tilde{\Phi } \Phi ^{\dagger }\right] {Tr}\left[\tilde{\Phi }^{\dagger } \Phi \right]+\lambda _2 {Tr}\left[\tilde{\Phi }^{\dagger } \Phi \right]^2\nonumber \\
	&+&\lambda _4 {Tr}\left[\Phi  \Phi ^{\dagger }\right] \left({Tr}\left[\tilde{\Phi } \Phi ^{\dagger }\right]+{Tr}\left[\tilde{\Phi }^{\dagger } \Phi \right]\right)\nonumber \\
	&+&\rho _1 \left({Tr}\left[\Delta _L \left(\Delta _L\right){}^{\dagger }\right]{}^2+{Tr}\left[\Delta _R \left(\Delta _R\right){}^{\dagger }\right]{}^2\right)\nonumber \\
	&+&\rho _2 \left({Tr}\left[\Delta _R \Delta _R\right] {Tr}\left[\left(\Delta _R\right){}^{\dagger } \left(\Delta _R\right){}^{\dagger }\right]+ {Tr}\left[\Delta _L \Delta _L\right] {Tr}\left[\left(\Delta _L\right){}^{\dagger } \left(\Delta _L\right){}^{\dagger }\right]\right)\nonumber \\
	&+&\rho _3 {Tr}\left[\Delta _L \left(\Delta _L\right){}^{\dagger }\right] {Tr}\left[\Delta _R \left(\Delta _R\right){}^{\dagger }\right]\nonumber \\
	&+&\rho _4 \left({Tr}\left[\Delta _R \Delta _R\right] {Tr}\left[\left(\Delta _L\right){}^{\dagger } \left(\Delta _L\right){}^{\dagger }\right]+ {Tr}\left[\Delta _L \Delta _L\right] {Tr}\left[\left(\Delta _R\right){}^{\dagger } \left(\Delta _R\right){}^{\dagger }\right]\right) \nonumber \\
	&+&\alpha _1 \left({Tr}\left[\Phi  \Phi ^{\dagger }\right] {Tr}\left[\Delta _L \left(\Delta _L\right){}^{\dagger }\right]+{Tr}\left[\Phi  \Phi ^{\dagger }\right] {Tr}\left[\Delta _R \left(\Delta _R\right){}^{\dagger}\right]\right)\nonumber \\	
	&+&\alpha _2 \left({Tr}\left[\Phi  \tilde{\Phi }^{\dagger }\right] {Tr}\left[\Delta _R \left(\Delta _R\right){}^{\dagger }\right]+{Tr}\left[\Phi ^{\dagger } \tilde{\Phi }\right] {Tr}\left[\Delta _L \left(\Delta _L\right){}^{\dagger }\right]\right.\nonumber \\
	&+& \left.{Tr}\left[\Phi ^{\dagger } \tilde{\Phi }\right] {Tr}\left[\Delta _R \left(\Delta _R\right){}^{\dagger }\right]+ {Tr}\left[\tilde{\Phi }^{\dagger } \Phi \right] {Tr}\left[\Delta _L \left(\Delta _L\right){}^{\dagger }\right]\right)\nonumber \\
	&+&\alpha _3 \left({Tr}\left[\Phi  \Phi ^{\dagger } \Delta _L \left(\Delta _L\right){}^{\dagger }\right]+{Tr}\left[\Phi ^{\dagger } \Phi  \Delta _R \left(\Delta _R\right){}^{\dagger }\right]\right)\nonumber \\
	&+&\beta _1 \left({Tr}\left[\Phi  \Delta _R \Phi ^{\dagger } \left(\Delta _L\right){}^{\dagger }\right]+{Tr}\left[\Phi ^{\dagger } \Delta _L \Phi  \left(\Delta _R\right){}^{\dagger }\right]\right)\nonumber \\
	&+&\beta _2 \left({Tr}\left[\tilde{\Phi } \Delta _R \Phi ^{\dagger } \left(\Delta _L\right){}^{\dagger }\right]+{Tr}\left[\tilde{\Phi }^{\dagger } \Delta _L \Phi  \left(\Delta _R\right){}^{\dagger }\right]\right)\nonumber \\
	&+&\beta _3 \left({Tr}\left[\Phi ^{\dagger } \Delta _L \tilde{\Phi } \left(\Delta _R\right){}^{\dagger }\right]+{Tr}\left[\Phi  \Delta _R \tilde{\Phi }^{\dagger } \left(\Delta _L\right){}^{\dagger }\right]\right),
	\end{eqnarray}}
where $\3 = \tau_{2}\Phi^{*}\tau_{2}$. We have assumed the potential parameters $\mu_{i}$, $\lambda_{i}$, $\rho_{i}$, $\beta_{i}$, and $\alpha_{2}$  to be real to avoid any kind of explicit CP-violation in this scenario.
\vskip0.2cm
The Yukawa lagrangian for this model can be written as \cite{Deshpande:1990ip}:
\begin{eqnarray}\label{eq:MLRSM-Yukawa-lag}
\mathcal{L}_{Y} & = & -\left\lbrace y_{_D}\ \bar{L}_{L} \Phi L_{R} + \tilde{y}_{_D}\ \bar{L}_{L} \tilde{\Phi} L_{R}+{h.c.}\right\rbrace- \left\lbrace y_q \bar{Q}_L \Phi Q_R + \tilde{y}_q \bar{Q}_L \tilde{\Phi} Q_R +{h.c.}\right\rbrace\nonumber \\ 
& & -y_{_M}\Big\lbrace L_{L}^{T} C i\tau_{2}\Delta_{L} L_{L}+\ L_{R}^{T} C i\tau_{2}\Delta_{R} L_{R} + {h.c.}\Big\rbrace, 
\end{eqnarray}
where the fermion doublets are expressed as:
\begin{eqnarray}\label{eq:MLRSM-fermion}
L_{L,R} &=& \begin{pmatrix}
\nu_{_{L,R}} \\ e_{_{L,R}}
\end{pmatrix},
\ \ \ \ Q_{L,R} = \begin{pmatrix}
u_{_{L,R}} \\ d_{_{L,R}}
\end{pmatrix}.
\end{eqnarray}
Here, $y_D, \tilde{y}_D, y_q, \tilde{y}_q$, $y_M$ are the Yukawa couplings. The first two terms in Eq. \ref{eq:MLRSM-Yukawa-lag} are Chiral (Lepton Number Conserving) in nature, whereas the last term is a source of Lepton Number Violation (LNV).

\vskip0.5cm
In this model the spontaneous gauge symmetry breaking occurs at two levels: first $SU(2)_R\otimes U(1)_{B-L}$ is broken to $U(1)_Y$, and then $SU(2)_L\otimes U(1)_Y$ is broken to $U(1)_{em}$. 
The scalar fields after acquiring VEVs can be given as \cite{Mohapatra:1980yp,Gunion:1989in,Deshpande:1990ip,Duka:1999uc}:
 {\small\begin{align}\label{eq:MLRSM-scalar-ssb}
 	\Phi  = \begin{pmatrix}
 	(\kappa_{1}+h_{1}+ i a_{1})/ \sqrt{2} &\phi_{1}^{+} \\
 	\phi_{2}^{-} & (\kappa_{2}+h_{2}+ i a_{2})/ \sqrt{2} 
 	\end{pmatrix}, \hspace{0.2cm}
 	&   \Delta_{L,R} = \begin{pmatrix}
 	\delta^{+}/ \sqrt{2}  & \delta^{++} \\ (v+\delta^{0r}+ i\delta^{0i})/ \sqrt{2} \;\; & \;\; -\delta^{+}/ \sqrt{2} 
 	\end{pmatrix}_{L,R}.
 	\end{align}}

Following similar steps as in 2HDM, scalar potential minimization conditions are read as:
{\small\begin{eqnarray}\label{eq:MLRSM-min-condition-tree}
	\frac{\partial V}{\partial \kappa_{1}}=\frac{\partial V}{\partial \kappa_{2}}=\frac{\partial V}{\partial v_{R}}=\frac{\partial V}{\partial v_{L}}=0 . 
	\end{eqnarray}}
Simultaneously solving these equations, we obtain following expressions for $ \mu_{1}^2 $, $ \mu_{2}^2 $, $ \mu_{3}^2$, and $\beta_{2}$:
{\small\begin{eqnarray}\label{eq:MLRSM-min-relation-tree}
	\mu _1^2 &=& \frac{1}{2 \kappa_-^2}\Big(2 \kappa_-^2 \left(\kappa _1^2 \lambda _1+2 \kappa _2 \kappa _1 \lambda _4+\kappa _2^2 \lambda _1\right)+v_L^2 \left(-\alpha _3 \kappa _2^2+\alpha _1 \kappa_-^2+\left(4 \rho _1-2 \rho _3\right) v_R^2\right)\Big. \nonumber \\
	&&\Big.+v_R^2 \left(\alpha _1 \kappa_-^2-\alpha _3 \kappa _2^2\right) -2 \kappa _2 \left(\beta _1 \kappa _1+2 \beta _3 \kappa _2\right) v_L v_R \Big), \\ 
	\mu _2^2& =& \frac{1}{4 \kappa_1\kappa_-^2} \Big(v_L^2 \left(\kappa _1 \left(\alpha _3 \kappa _1 \kappa _2+2 \alpha _2 \kappa_-^2\right)+2 \kappa _2 \left(\rho _3-2 \rho _1\right) v_R^2\right)+\kappa_+^2 \left(\beta _1 \kappa _1+2 \beta _3 \kappa _2\right) v_L v_R \Big. \nonumber \\
	&& \Big. +\kappa _1 \left(2 \kappa_-^2 \left(\kappa _1^2 \lambda _4+2 \kappa _2 \kappa _1 \left(2 \lambda _2+\lambda _3\right)+\kappa _2^2 \lambda _4\right)+v_R^2 \left(\alpha _3 \kappa _1 \kappa _2+2 \alpha _2 \kappa_-^2\right)\right) \Big),\\
	\mu _3^2& =& \frac{1}{2} \left(\alpha _1 \kappa _1^2+\alpha _1 \kappa _2^2+\alpha _3 \kappa _2^2+4 \alpha _2 \kappa _1 \kappa _2+2 \rho _1 v_L^2+2 \rho _1 v_R^2\right), \\
	\beta _2 &=& \frac{1}{\kappa _1^2}\left(2 \rho _1-\rho _3\right) v_L v_R-\kappa _2 \left(\beta _1 \kappa _1+\beta _3 \kappa _2\right).
	\end{eqnarray}}
To keep our results in the same footing with \cite{Deshpande:1990ip,Duka:1999uc}, we have performed a sub-rotation in the scalar sector as:
\small{\begin{eqnarray}\label{eq:MLRSM-scalar-field-semi-redef}
	\phi_{1}^{0r}&=&\frac{1}{\kappa_+}(\kappa_{1}h_{1} + \kappa_{2}h_{2}),\hspace{0.2cm} 
	\phi_{2}^{0r}=\frac{1}{\kappa_+}(-\kappa_{2}h_{1} + \kappa_{1}h_{2}), \\
	\phi_{1}^{0i}&=&\frac{1}{\kappa_+}(\kappa_{1}a_{1}-\kappa_{2}a_{2}), \hspace{0.2cm}
	\phi_{2}^{0i}=\frac{1}{\kappa_+}(-\kappa_{2}a_{1}-\kappa_{1}a_{2}), \nonumber \\
	\phi_{1}'^{+}&=&\frac{1}{\kappa_{+}}(\kappa_{1}\phi_{1}^{+}+\kappa_{2}\phi_{2}^{+}), \hspace{0.2cm}
	\phi_{2}'^{+}=\frac{1}{\kappa_{+}}(\kappa_{1}\phi_{2}^{+}-\kappa_{2}\phi_{1}^{+}).\nonumber 
	\end{eqnarray}}
This allows us to define the following basis to proceed for further computation:
\begin{eqnarray}
\{\phi_{1}^{0r},\phi_{2}^{0r},\delta_{R}^{0r},\delta_{L}^{0r}\} \;\; &\Rightarrow& \text{CP-even scalar}, \\
\{\phi_{1}^{0i},\phi_{2}^{0i},\delta_{R}^{0i},\delta_{L}^{0i}\} \;\; &\Rightarrow& \text{CP-odd scalar}, \\
\{\phi_{1}'^{+},\phi_{2}'^{+},\delta_{R}^{+},\delta_{L}^{+}\}   \;\; &\Rightarrow& \text{Singly charged scalar}, \\
\{\delta_{R}^{++},\delta_{L}^{++}\}  \;\; &\Rightarrow& \text{Doubly charged scalar}. 
\end{eqnarray}
After SSB, the part of the lagrangian containing the scalar mass terms is given as \cite{Mohapatra:1980yp,Gunion:1989in,Deshpande:1990ip,Duka:1999uc}:
{\small\begin{align}\label{eq:MLRSM-scalar-mass-matrix-tree}
	\mathcal{L}_{M_{H}}^{(4)} =& \frac{1}{2} \begin{bmatrix}
	\phi^{0r}_{1} \\ \phi^{0r}_{2}  \\ \delta^{0r}_{R} \\ \delta^{0r}_{L}
	\end{bmatrix}^{T} m_{r}^{2} \begin{bmatrix}
	\phi^{0r}_{1} \\ \phi^{0r}_{2}  \\ \delta^{0r}_{R} \\ \delta^{0r}_{L}
	\end{bmatrix} + \frac{1}{2} \begin{bmatrix}
	\phi^{0i}_{1} \\ \phi^{0i}_{2}  \\ \delta^{0i}_{R} \\ \delta^{0i}_{L}
	\end{bmatrix}^{T}m_{i}^{2} \begin{bmatrix}
	\phi^{0i}_{1} \\ \phi^{0i}_{2}  \\ \delta^{0i}_{R} \\ \delta^{0i}_{L}
	\end{bmatrix} + \begin{bmatrix}
	\phi^{'+}_{1} \\ \phi^{'+}_{2}  \\ \delta^{+}_{R} \\ \delta^{+}_{L} 
	\end{bmatrix}^{\dagger}m_{+}^{2} \begin{bmatrix}
	\phi^{'+}_{1} \\ \phi^{'+}_{2}  \\ \delta^{+}_{R} \\ \delta^{+}_{L} 
	\end{bmatrix}
	+ \begin{bmatrix}
	\delta^{++}_{R} \\ \delta^{++}_{L} 
	\end{bmatrix}^{\dagger}m_{++}^{2} \begin{bmatrix}
	\delta^{++}_{R} \\ \delta^{++}_{L} 
	\end{bmatrix}. \nonumber
	\end{align}}
Here, $\,m_{r}^{2},m_{i}^{2}$, and $m_{+}^{2}$ are $(4\times4)$ matrices while $m_{++}^{2}$ is $(2\times2)$ matrix.
\vskip0.5cm
\underline{\large{CP-even scalar mass matrix ($m_{r}^2$)}} 
{\small\begin{eqnarray}\label{eq:MLRSM-scalar-mass-CP-even-tree}
	(m_{r}^2)_{11} &= &\frac{2}{\kappa_+^2}\left(\kappa _1^4 \lambda _1+4 \kappa _2 \kappa _1^3 \lambda _4+2 \kappa _2^2 \kappa _1^2 \left(\lambda _1+4 \lambda _2+2 \lambda _3\right)+4 \kappa _2^3 \kappa _1 \lambda _4+\kappa _2^4 \lambda _1\right),\nonumber\\
	(m_{r}^2)_{12} &=&\frac{2 \kappa_-^2}{\kappa_+^2} \left(\kappa _1^2 \lambda _4+2 \kappa _2 \kappa _1 \left(2 \lambda _2+\lambda _3\right)+\kappa _2^2 \lambda _4\right),\nonumber\\
	(m_{r}^2)_{13} &=&\frac{v_R}{\kappa_+} \left(\kappa _2 \left(4 \alpha _2 \kappa _1+\alpha _3 \kappa _2\right)+\alpha _1 \kappa_+^2+\left(2 \rho _1-\rho _3\right) v_L^2\right),\nonumber\\
	(m_{r}^2)_{14}&=&\frac{v_L}{\kappa_+} \left(\kappa _2 \left(4 \alpha _2 \kappa _1+\alpha _3 \kappa _2\right)+\alpha _1 \kappa _+^2+\left(2 \rho _1-\rho _3\right) v_R^2\right),\nonumber\\
	(m_{r}^2)_{22} &=&\frac{\kappa_+^4 v_L^2}{2 \kappa _1^2 \kappa_+^2\kappa_-^2} \left(\alpha _3 \kappa _1^2+\left(2 \rho _3-4 \rho _1\right) v_R^2\right)+2 \kappa_+^4 v_L v_R \left(\beta _1 \kappa _1 \kappa _2+\beta _3 \kappa_+^2\right)\nonumber \\&+&\kappa _1^2 \left(4 \left(2 \lambda _2+\lambda _3\right) \kappa_-^6+\alpha _3 \kappa_+^4 v_R^2\right),\nonumber\\
	(m_{r}^2)_{23}& =&\frac{1}{2 \kappa _1 \kappa_+}\Big(2 v_R \left(\kappa _1 \left(\alpha _3 \kappa _1 \kappa _2+2 \alpha _2 \kappa_-^2\right)+\kappa _2 \left(\rho _3-2 \rho _1\right) v_L^2\right)+\kappa_+^2 \left(\beta _1 \kappa _1+2 \beta _3 \kappa _2\right) v_L\Big),\nonumber\\
	(m_{r}^2)_{24}& =&\frac{1}{2 \kappa _1 \kappa_+}\Big(2 v_L \left(\kappa _1 \left(\alpha _3 \kappa _1 \kappa _2+2 \alpha _2 \kappa_-^2\right)+\kappa _2 \left(\rho _3-2 \rho _1\right) v_R^2\right)+\kappa_+^2 \left(\beta _1 \kappa _1+2 \beta _3 \kappa _2\right) v_R\Big),\nonumber\\
	(m_{r}^2)_{33}& =& \frac{1}{2} \left(\rho _3-2 \rho _1\right) v_L^2+2 \rho _1 v_R^2,\nonumber\\
	(m_{r}^2)_{34}& =&\left(2 \rho _1+\rho _3\right) v_L v_R,\nonumber\\
	(m_{r}^2)_{44}& =&2 \rho _1 v_L^2+\frac{1}{2} \left(\rho _3-2 \rho _1\right) v_R^2.
	\end{eqnarray}}

\underline{\large{CP-odd scalar mass matrix ($m_{i}^2$)}} 
{\small\begin{eqnarray}\label{eq:MLRSM-scalar-mass-CP-odd-tree}
	(m_{i}^2)_{11}&= &\frac{2 v_L^2 v_R^2}{\kappa_+^2} \left(\rho _3-2 \rho _1\right),\nonumber\\
	(m_{i}^2)_{12}&=&\frac{v_L v_R }{\kappa _1}\left(-\beta _1 \kappa _1-2 \beta _3 \kappa _2+\frac{2 \kappa _2 \left(2 \rho _1-\rho _3\right) v_L v_R}{\kappa_+^2}\right),\nonumber\\
	(m_{i}^2)_{13}&=&\frac{1}{\kappa_+}\left(2 \rho _1-\rho _3\right) v_L^2 v_R,\nonumber\\
	(m_{i}^2)_{14}&=&\frac{ v_L v_R^2}{\kappa_+}\left(\rho _3-2 \rho _1\right),\nonumber\\
	(m_{i}^2)_{22}&=&\frac{1}{2 \kappa _1^2\kappa_+^2\kappa_-^2}\Big(v_L^2 \left(\alpha _3 \kappa _1^2 \kappa_+^4-2 \left(\kappa _1^4+4 \kappa _2^2 \kappa _1^2-\kappa _2^4\right) \left(2 \rho _1-\rho _3\right) v_R^2\right)-2 \kappa_+^2 \left(\beta _1 \kappa _1 \kappa _2 \left(\kappa _2^2-3 \kappa _1^2\right)\right. \nonumber \\
	&&\left.+\beta _3 \left(\kappa _1^4-6 \kappa _2^2 \kappa _1^2+\kappa _2^4\right)\right) v_L v_R+\kappa _1^2 \kappa_+^4 \left(\alpha _3 v_R^2-4 \kappa_-^2 \left(2 \lambda _2-\lambda _3\right)\right)\Big),\nonumber\\
	(m_{i}^2)_{23}&=&\frac{v_L}{2 \kappa _1 \kappa_+} \left(\kappa_+^2 \left(\beta _1 \kappa _1+2 \beta _3 \kappa _2\right)+2 \kappa _2 \left(\rho _3-2 \rho _1\right) v_L v_R\right),\nonumber\\
	(m_{i}^2)_{24}&=&-\frac{v_R}{2 \kappa _1 \kappa_+} \left(\kappa_+^2 \left(\beta _1 \kappa _1+2 \beta _3 \kappa _2\right)+2 \kappa _2 \left(\rho _3-2 \rho _1\right) v_L v_R\right),\nonumber\\
	(m_{i}^2)_{33} &=& \frac{1}{2} \left(\rho _3-2 \rho _1\right) v_L^2,\nonumber\\
	(m_{i}^2)_{34} &=&\frac{1}{2} \left(2 \rho _1-\rho _3\right) v_L v_R,\nonumber\\
	(m_{i}^2)_{44} &=&\frac{1}{2} \left(\rho _3-2 \rho _1\right) v_R^2.
	\end{eqnarray}}

\underline{\large{Singly charged scalar mass matrix ($m_{+}^2$)}} 
{\small\begin{eqnarray}\label{eq:MLRSM-scalar-mass-scs-tree}
	(m_{+}^2)_{11}&= &\frac{1}{\kappa _+^2 \kappa _{-}^2}\Big(v_R^2 \left(\alpha _3 \kappa_+^4-2 \left(\kappa _1^2+3 \kappa _2^2\right) \left(2 \rho _1-\rho _3\right) v_L^2\right)+4 \kappa _2 \kappa_+^2 \left(\beta _1 \kappa _1+2 \beta _3 \kappa _2\right) v_L v_R\nonumber \\
	&+&4 \alpha _3 \kappa _1^2 \kappa _2^2 v_L^2\Big),\nonumber\\
	(m_{+}^2)_{12}& =&\frac{v_L}{\kappa _1 \kappa_+^2} \left(\kappa_+^2 \left(\beta _1 \kappa _1+2 \beta _3 \kappa _2\right) v_R+2 \kappa _2 \left(\rho _3-2 \rho _1\right) v_L v_R^2+2 \alpha _3 \kappa _1^2 \kappa _2 v_L\right),\nonumber\\
	(m_{+}^2)_{13}& =&\frac{1}{\sqrt{2} \kappa_+}v_R \left(\alpha _3 \kappa_+^2+\left(2 \rho _3-4 \rho _1\right) v_L^2\right)+2 \kappa _2 \left(\beta _1 \kappa _1+2 \beta _3 \kappa _2\right) v_L,\nonumber\\
	(m_{+}^2)_{14}& =&\frac{1}{\sqrt{2} \kappa _1 \kappa_+}\Big(\kappa_+^2 \left(\beta _1 \kappa _1+2 \beta _3 \kappa _2\right) v_R+2 \kappa _2 \left(\rho _3-2 \rho _1\right) v_L v_R^2+2 \alpha _3 \kappa _1^2 \kappa _2 v_L\Big),\nonumber\\
	(m_{+}^2)_{22}&=&\frac{v_L^2}{\kappa_+^2} \left(\alpha _3 \kappa_-^2+\left(2 \rho _3-4 \rho _1\right) v_R^2\right),\nonumber\\
	(m_{+}^2)_{23}&=&\frac{v_L}{\sqrt{2} \kappa _1 \kappa_+} \left(\kappa_-^2 \left(\beta _1 \kappa _1+2 \beta _3 \kappa _2\right)+2 \kappa _2 \left(2 \rho _1-\rho _3\right) v_L v_R\right),\nonumber\\
	(m_{+}^2)_{24}&=&\frac{v_L}{\sqrt{2} \kappa_+} \left(\alpha _3 \kappa_-^2+\left(2 \rho _3-4 \rho _1\right) v_R^2\right),\nonumber\\
	(m_{+}^2)_{33}&=&\frac{1}{2} \left(\alpha _3 \kappa_-^2+\left(2 \rho _3-4 \rho _1\right) v_L^2\right),\nonumber\\
	(m_{+}^2)_{34}&=&\frac{1}{2 \kappa _1}\left(\kappa_-^2 \left(\beta _1 \kappa _1+2 \beta _3 \kappa _2\right)+2 \kappa _2 \left(2 \rho _1-\rho _3\right) v_L v_R\right),\nonumber\\
	(m_{+}^2)_{44}&=&\frac{1}{2} \left(\alpha _3 \kappa_-^2+\left(2 \rho _3-4 \rho _1\right) v_R^2\right).
	\end{eqnarray}}

\underline{\large{Doubly charged scalar mass matrix ($m_{++}^2$)}} 
{\small\begin{eqnarray}\label{eq:MLRSM-scalar-mass-dcs-tree}
(m_{++}^2)_{11} &=& \alpha _3 \kappa_-^2+\left(\rho _3-2 \rho _1\right) v_L^2+4 \rho _2 v_R^2,\\
(m_{++}^2)_{12} &=&\frac{1}{\kappa _1^2}\left(\kappa_-^2 \left(\beta _1 \kappa _1 \kappa _2+\beta _3 \kappa_+^2\right)+\left(4 \kappa _1^2 \rho _4+\kappa _2^2 \left(2 \rho _1-\rho _3\right)\right) v_L v_R\right),\nonumber\\
(m_{++}^2)_{22} &=& \alpha _3 \kappa_-^2+4 \rho _2 v_L^2+\left(\rho _3-2 \rho _1\right) v_R^2.
\end{eqnarray}}

Once the cascade of  spontaneous symmetry is completed, two charged and  two neutral gauge bosons become massive, while one neutral gauge boson is massless identified as photon.  As color symmetry remains intact, we will focus on the uncolored sector ($SU(2)_L\otimes SU(2)_R\otimes U(1)_{B-L}$) only for this purpose. The unphysical  states  $\{ W^\mu_{1L},  W^\mu_{2L},  W^\mu_{3L} \} \in  W^\mu_{L}$, $\{ W^\mu_{1R},  W^\mu_{2R},  W^\mu_{3R} \} \in  W^\mu_{R}$, and $B^\mu$ are the gauge bosons corresponding to the  $SU(2)_L, SU(2)_R, U(1)_{B-L}$  groups respectively. 
After the SSB, the lagrangian containing gauge boson mass matrices are written as \cite{Deshpande:1990ip,Duka:1999uc}:
{\small\begin{equation}\label{eq:MLRSM-GB-lag-tree}
	\mathcal{L}^{gauge}_{M} = \begin{pmatrix}
	W_{L \mu }^{-} \ \ W_{R \mu }^{-}
	\end{pmatrix}\tilde{m}^2_W \begin{pmatrix}
	W^{+\mu}_{L} \\ W^{+\mu}_{R}
	\end{pmatrix} + \frac{1}{2}\begin{pmatrix}
	W_{3L \mu} \ \ W_{3R \mu} \ \ B_{\mu}
	\end{pmatrix}\tilde{m}^2_{0} \begin{pmatrix}
	W^{\mu}_{3L} \\ W^{\mu}_{3R} \\ B^{\mu}
	\end{pmatrix},
	\end{equation}} 
where the charged states are defined as 
{\small\begin{eqnarray}
	W^{\pm \mu}_{L,R} &=& \frac{1}{\sqrt{2}}\left(W^{\mu}_{1} \mp i  W^{\mu}_{2} \right)_{L,R}.
	\end{eqnarray}}
The charged and neutral gauge boson mass matrices have the forms :
{\small\begin{equation}\label{eq:MLRSM-cGB-mass-matrix-tree}
	\tilde{m}^2_W = \frac{g^2}{4}\begin{pmatrix}
	\kappa_{+}^2 + 2v^2_L \ \ -2\kappa_1 \kappa_2\\
	-2\kappa_1 \kappa_2   \hspace{0.4cm} \ \ \kappa_{+}^2 + 2v^2_R
	\end{pmatrix},
	\end{equation}}
and
{\small\begin{equation}\label{eq:MLRSM-nGB-mass-matrix-tree}
	\tilde{m}^2_0 = \frac{1}{2}\begin{pmatrix}
	\frac{1}{2}g^2(\kappa_{+}^2 + 4v^2_L) \hspace{0.8cm}\ \ -\frac{1}{2}g^2 \kappa_{+}^2 \hspace{1.5cm}\ \ -2g \tilde{g}v^2_L\\
	-\frac{1}{2}g^2 \kappa_{+}^2 \hspace{1.6cm}\ \ \frac{1}{2}g^2(\kappa_{+}^2 + 4v^2_L) \hspace{0.8cm}\ \ -2g \tilde{g}v^2_R\\
	\ \ -2g \tilde{g}v^2_L \hspace{1.6cm}\ \ -2g \tilde{g}v^2_R \hspace{1.5cm}\ \ 2\tilde{g}^2(v^2_L + v^2_R)
	\end{pmatrix},
	\end{equation}}
respectively.
Here, we have defined $\kappa^2_{\pm} = \kappa^2_1 \pm \kappa^2_2$, to express our result with the same convention  of \cite{Deshpande:1990ip,Duka:1999uc}.

Now to find the physical spectrum, we need to diagonalize the mass matrices. We have introduced the following rotation matrices to connect the physical (charged: $W^{\pm \mu}_1, W^{\pm \mu}_2$, neutral: $Z^\mu_1,Z^\mu_2,A^\mu$), and unphysical basis of gauge bosons \cite{Deshpande:1990ip,Duka:1999uc}:
{\small\begin{equation}\label{eq:MLRSM-GB-rot}
	\begin{pmatrix}
	W^{\pm \mu}_L \\ W^{\pm \mu}_R
	\end{pmatrix} = {\mathcal{R}_a}
	\begin{pmatrix}
	W^{\pm \mu}_1 \\ W^{\pm \mu}_2
	\end{pmatrix},
	\hspace{1.5cm}
	\begin{pmatrix}
	W^{\mu}_{3L} \\ W^{\mu}_{3R} \\ B^{\mu}
	\end{pmatrix} = \mathcal{R}_b
	\begin{pmatrix}
	Z^{\mu}_{1} \\ Z^{\mu}_{2} \\ A^{\mu}
	\end{pmatrix}.
	\end{equation}}
Here, the rotation matrices are given as:

\begin{equation}\label{eq:MLRSM-cGB-rot-matrix}
\mathcal{R}_a =\begin{pmatrix}
\cos\xi \ \ \sin\xi \\
-\sin\xi \ \ \cos\xi
\end{pmatrix},\\
\end{equation}
{\small\begin{equation}\label{eq:MLRSM-nGB-rot-matrix}
	\mathcal{R}_b =
	\begin{pmatrix}
	\hspace{0.8cm} \cos\theta_{\text{w}} \cos\theta_2 \hspace{3.0cm}\ \ \cos\theta_{\text{w}} \sin\theta_2 \hspace{2.7cm}\ \ \sin\theta_{\text{w}} \\
	-\sin\theta_{\text{w}}\sin\theta_1\cos\theta_2-\cos\theta_1\sin\theta_2 \hspace{0.2cm}\ \
	-\sin\theta_{\text{w}}\sin\theta_1\sin\theta_2+\cos\theta_1\cos\theta_2 \hspace{0.2cm}\ \
	\cos\theta_{\text{w}}\sin\theta_1 \\
	-\sin\theta_{\text{w}}\cos\theta_1\cos\theta_2+\sin\theta_1\sin\theta_2 \hspace{0.2cm}\ \
	-\sin\theta_{\text{w}}\cos\theta_1\sin\theta_2-\sin\theta_1\cos\theta_2 \hspace{0.2cm}\ \
	\cos\theta_{\text{w}}\cos\theta_1
	\end{pmatrix}.\nonumber
	\end{equation}}
The rotation  ``angles" are related to each other as \cite{Deshpande:1990ip,Duka:1999uc}:
{\small\begin{equation}\label{eq:MLRSM-nGB-rot-params-tree}
	\cos\theta_1 = \frac{\sqrt{\cos2\theta_{\text{w}}}}{\cos\theta_{\text{w}}}, \hspace{1.2cm} \sin\theta_1 = \tan\theta_{\text{w}}, \hspace{1.2cm}
	g = \frac{e}{\sin\theta_{\text{w}}}, \hspace{1.2cm} \tilde{g} = \frac{e}{\sqrt{\cos2\theta_{\text{w}}}},
	\end{equation}}
and 
{\small\begin{equation}\label{eq:MLRSM-cGB-rot-params-tree}
	\tan 2\xi = \frac{-2k_1 k_2}{v^2_R - v^2_L} ,\;\;\;\;\;	\tan 2\theta_2 = \frac{a}{b},
	\end{equation}}
where
{\small\begin{eqnarray}
	a &=& \Big( \frac{1}{4}g^2\kappa_{+}^2 -\tilde{g}^2v^2_L \Big)\sqrt{\cos2\theta_{\text{w}}},\\
	b &=& \Big( \frac{1}{4}g^2\kappa_{+}^2\sin^2\theta_{\text{w}} + \frac{1}{2}g^2 v_L^2 - \frac{1}{2}\tilde{g}^2v_L^2\sin^2\theta_{\text{w}} - \frac{1}{2}(g^2 +\tilde{g}^2)v_R^2\cos^2\theta_{\text{w}} \Big).
\end{eqnarray}}	
We would like to note, here, that $\theta_{\text{w}}$ is the Weak mixing angle.
	
	The physical gauge boson spectrum can be written as:
{\small\begin{eqnarray}\label{eq:MLRSM-nGB-rot-params-tree}
	m^2_{W_{1,2}} &= &\frac{g^2}{4}\Big[\kappa^2_{+} + v^2_L + v^2_R \mp \sqrt{(v^2_R - v^2_L)^2 + 4\kappa_1^2 \kappa_2^2}\Big],\\
	m^2_{Z_{1,2}}& =& \frac{1}{4}g^2\kappa_{+}^2 + \frac{1}{2}(g^2 +\tilde{g}^2)(v_L^2 + v_R^2) \mp \frac{1}{\cos^2\theta_{\text{w}}}\sqrt{a^2 + b^2}.
	\end{eqnarray}}

After the SSB, the charged fermion masses are generated  through the Yukawa couplings, Eq.~\ref{eq:MLRSM-Yukawa-lag} in the following form \cite{Deshpande:1990ip,Duka:1999uc}:
\small{\begin{eqnarray}\label{eq:MLRSM-fermion-mass-tree}
	m_{e} &=&\frac{1}{\sqrt{2}}\left(\kappa _1 \tilde{y}_D+\kappa _2 y_D\right) \Rightarrow \text{charged lepton mass},\\
	 m_{u} &=& \frac{1}{\sqrt{2}}\left(\kappa _2 \tilde{y}_q+\kappa _1 y_q\right)\Rightarrow \text{up-type quark mass},\\
	m_{d} &=& \frac{1}{\sqrt{2}}\left(\kappa _1 \tilde{y}_q+\kappa _2 y_q\right)\Rightarrow \text{down-type quark mass}.	
	\end{eqnarray}}

In this model, left- and right-handed neutrinos also acquire masses. We have defined a Majorana basis for light and heavy neutrinos as \cite{Deshpande:1990ip,Duka:1999uc} :
\small{\begin{align}\label{eq:MLRSM-neutrino-mass-basis-tree}
	\nu = \frac{\nu_{L} + (\nu_{L})^{c}}{\sqrt{2}}, \ \ \ \ N = \frac{\nu_{R} + (\nu_{R})^{c}}{\sqrt{2}},
	\end{align}}
where the charge conjugated state is defined as $\psi^c = C \bar{\psi}^T$ using the charge conjugation operator $C=i\gamma_2 \gamma_0$. Once the bi-doublet ($\Phi$) gets VEV, see Eq.~\ref{eq:MLRSM-scalar-ssb}, the Dirac-mass terms for the neutrinos are generated as: $\frac{1}{\sqrt{2}}\left(\kappa _2 \tilde{y}_D+\kappa _1 y_D\right)$. The VEVs of $\Delta_{L},\Delta_{R}$ lead to the Majorana mass terms for $\nu$ and $N$ respectively. Including all these, we can define the neutrino mass matrix in the basis $\{ \nu\; N\}$ as \cite{Deshpande:1990ip,Duka:1999uc}:
\begin{eqnarray}\label{eq:MLRSM-neutrino-mass-matrix-tree}
m_{\nu}=\begin{pmatrix}
\sqrt{2} y_M v_L & \frac{1}{\sqrt{2}}\left(\kappa _2 \tilde{y}_D+\kappa _1 y_D\right) \\
\frac{1}{\sqrt{2}}\left(\kappa _2 \tilde{y}_D+\kappa _1 y_D\right) & \sqrt{2} y_M v_R \\
\end{pmatrix}.
\end{eqnarray}
\clearpage



\section{Effective operators for MLRSM, their categorization and their impact on spectrum}
\label{sec:MLRSM-HDO}

\begin{figure}[h!]
	    \centering
	{
		\includegraphics[trim={2.0cm 0 2.0cm 0},scale=0.48]{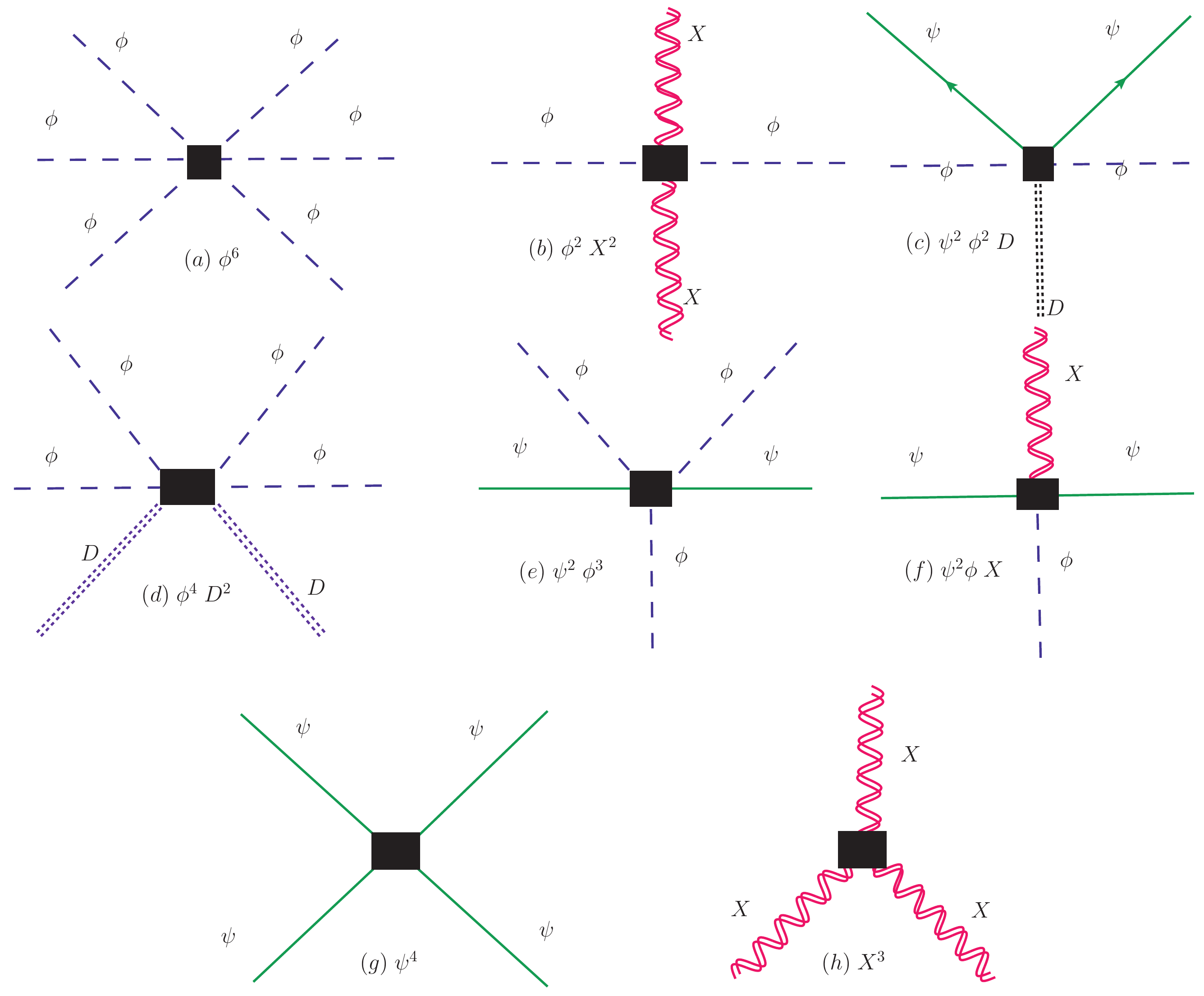}
	}
	\caption{Effective operators in ``Warsaw"-like basis representing the following class of operators:
		(a) $\phi^6\;$\textcolor{red}{[42$\times$2 + 28 = 112]}, (b) $\phi^2X^2\;$\textcolor{red}{[62]}, (c) $\psi^2\phi^2D\;$\textcolor{red}{[20 + 8$\times$2 = 36]}, (d) $\phi^4D^2\;$\textcolor{red}{[20 + 11$\times$2 = 42]}, (e) $ \psi^2\phi^3\;$\textcolor{red}{[22$\times$2 = 44]} (Majorana type) \textcolor{red}{+ [36 $\times$2 = 72]} (Dirac type), (f)  $\psi^2\phi X\;$\textcolor{red}{[18$\times$2 = 36]}, (g) $\psi^4\;$\textcolor{red}{[15 + 6$\times$2 = 27]} (Baryon Number Conserving) + \textcolor{red}{ [4$\times$2 = 8]} (Baryon Number Violating), (h) $ X^3\;$\textcolor{red}{[6]}. }\label{fig:operator-class-MLRSM}
\end{figure}

In this section, we have listed all the dimension-6 operators that are computed using the Hilbert Series method. These operators are computed using the information given in   Tables~\ref{tab:MLRSM-fields}, and following the thumb rules mentioned in the flowchart of Fig.~\ref{fig:HS}.   We have classified all the dimension-6 operators in the following categories: $\{ \phi^6, \phi^2X^2,  \psi^2\phi^2D, \phi^4D^2, \psi^2\phi^3, \psi^2\phi X, \psi^4, X^3 \}$. For MLRSM, the Hilbert Series yields \textcolor{red}{445} independent operators of mass dimension-6. This includes the distinct Hermitian Conjugates (h.c.) as well. Similar to the 2HDM case, the Wilson Coefficients are also taken to be real for this particular scenario. These operators are tabulated in Tables~\ref{tab:MLRSM-phi6-dhc}-\ref{tab:MLRSM-phi4D2}, and also schematically depicted in Fig.~\ref{fig:operator-class-MLRSM} where the number of operators is also mentioned for each class as ($\mathcal{O}\;[..]$). While writing the explicit structures of these operators along with the field variables we have also employed few tensors, e.g. $\delta_{\mu \nu}$ (Kronecker delta),  $\epsilon^{\mu \nu \rho}$ (Levi-Civita) etc. to construct the invariants. 
We have provided a suitable table, similar to the 2HDM case,  where we have outlined the nomenclature of the operators. This would be helpful for the readers to follow the operators ($\mathcal{O}$), and the respective Wilson Coefficients ($\mathcal{C}$) which are also defined following the same rule, see Tables~\ref{tab:MLRSM-operator-nomenclature-I}.
\vspace{0.5cm}

\begin{table}[h!]
	\centering
	\renewcommand{\arraystretch}{2.0}
	{\scriptsize\begin{tabular}{|c|c|c|c|c|}
			\hline
			\multirow{2}{*}{\textbf{Operator}}&
			\multirow{2}{*}{\textbf{Symbol}}&
			\multirow{2}{*}{\textbf{Superscript}}&
			\multirow{2}{*}{\textbf{Subscript}}&
			\multirow{2}{*}{\textbf{Remarks}}\\
			
			\textbf{Class}&&&&\\
			
			\hline
			\multirow{6}{*}{$\phi^6$}&
			\multirow{3}{*}{$\mathcal{O}^{ijklmn}_{6}$}&
			\multirow{3}{*}{$\{i,j,k,l,m,n\}$}&
			\multirow{6}{*}{6$\rightarrow$ mass dimension of $\phi^6$}
			&
			\\
			
			&&
			\multirow{2}{*}{$\downarrow$}&&
			1 $\rightarrow \phi$; 2 $\rightarrow \phi^{\dagger}$; \\
			
			&&
			$\{1,2,3,4,L,l,R,r\}$&&
			3$\rightarrow \tilde{\phi}$; 4$\rightarrow \tilde{\phi}^{\dagger};$\\
			
			&
			\multirow{3}{*}{$\mathcal{O}^{\{ij\}\{klmn\}}_{6}$}&
			\multirow{2}{*}{$\{ \}\implies$ ``Tr" }&&
			$l\rightarrow \Delta_L$; $L\rightarrow \Delta_L^{\dagger};$\\
			
			&&
			\multirow{2}{*}{no $\{ \}\implies$ ``$Tr$" over }&&
			$r\rightarrow \Delta_R$; $R\rightarrow \Delta_R^{\dagger};$\\
			
			&&
			the full operator&&
			\\
			\cline{1-5}
			
			$X^3$&
			$\mathcal{O}^{X^3/\tilde{X}^3}_{L,R}$&
			$X \rightarrow \{W, G\}$&
			$L,R$ for $W_L, W_R$&$X\equiv WB$, $\tilde{X}\rightarrow \tilde{W}B$\\
			\cline{1-5}
			
			\multirow{5}{*}{$\phi^2X^2$}&
			$\mathcal{O}^{ij}_{H X/H \tilde{X}}$&
			\multirow{2}{*}{$(i,j) \equiv (1,2),(2,3),$}&
			$H\equiv \phi, \Delta_L, \Delta_R$&
			\multirow{5}{*}{``$Tr$" is over the full operator} \\
			
			&&
			$(4,1),(L,l),(R,r)$&
			$X \rightarrow \{B, G, W_L, W_R\}$&\\
			
			&
			$\mathcal{O}^{ij}_{H X_1 X_2}$&&
			$\{X_1, X_2\} \rightarrow \{B, W_L, W_R\}$&\\
			\cline{2-4}
			&
			\multirow{2}{*}{$\mathcal{O}^{iX_1jX_2}_{\Delta W}$}&
			\multicolumn{2}{c|}{$(i,j) \equiv (L,l),(L,r),(R,l),(R,r)$}&\\
			
			&&
			\multicolumn{2}{c|}{$\{X_1, X_2\} \rightarrow \{W_L, W_R\}$}&\\
			\cline{1-5}
			
			\multirow{3}{*}{$\psi^2\phi X$}&
			\multirow{3}{*}{$\mathcal{O}^{i X}_{f_L f_R}$}&
			\multirow{2}{*}{$i = 1 \implies \phi$}&
			\multirow{3}{*}{$f\rightarrow q ({quark}),$}&
			\multirow{3}{*}{**within the operator a} \\
			
			&&
			$i = 3\implies \tilde{\phi}$&
			\multirow{2}{*}{$f\rightarrow l ({lepton})$}& \multirow{2}{*}{quark is denoted by Q} \\ 
			
			&&
			$X \rightarrow \{B, G, W_L, W_R\}$&&and a lepton by L \\
			
			\cline{1-5}
			
			\multirow{4}{*}{$\psi^2\phi^3$}&
			\multirow{2}{*}{$\mathcal{O}^{i(jk)}_{f}$/$\mathcal{O}^{ijk}_{f}$}&
			\multirow{3}{*}{$\{i,j,k\}$}&
			\multirow{2}{*}{$f\rightarrow \{l_L, l_R\}$}&
			\multirow{2}{*}{The field outside ``( )" are coupled }
			\\ 
			
			&&
			\multirow{2}{*}{$\downarrow$}&
			& directly to the fermions,\\ 
			\cline{2-2} \cline{4-4}
			&
			\multirow{2}{*}{$\mathcal{O}^{i(jk)}_{\tilde{f}_L\tilde{f}_R}$/$\mathcal{O}^{ijk}_{\tilde{f}_L\tilde{f}_R}$}&
			$\{1,2,3,4,L,l,R,r\}$&
			\multirow{2}{*}{$\tilde{f}\rightarrow \{q,l\}$}&
			\multirow{2}{*}{``$Tr$" is taken over the fields}\\
			
			&&&& 
			inside ``( )".\\
			
			\cline{1-5}
			
			\hline
			\multirow{3}{*}{$\psi^2\phi^2D$}&
			$\mathcal{O}^{f}_{iDj}$&
			\multirow{2}{*}{$f\rightarrow \{l_L, l_R, q_L, q_R\}$}&
			\multirow{3}{*}{$\{i,j\}$}&
			\multirow{2}{*}{[1]$ \implies SU(2)$ Singlet}\\
			
			&
			$\mathcal{O}^{f[1]/f[3]}_{iDj}$&
			&
			\multirow{2}{*}{$\downarrow $}&
			[3] $ \implies SU(2)$ Triplet (involves $\tau^I$)\\
			\cline{2-3}
			
			&
			$\mathcal{O}^{f_Lf_R}_{iDj}$&$f\rightarrow \{l,q\} $&
			$\{1,2,3,4,L,l,R,r\}$&
			``$Tr$" is to be taken over the ``$iDj$" part\\
			\cline{1-5}
			
			\multirow{6}{*}{$\phi^4D^2$}&
			$\mathcal{O}^{ij(kl)}_{\square}$&
			\multirow{3}{*}{$\{i,j,k,l\}$}&
			&
			\multirow{4}{*}{(i) $\implies$ $D$ acting on $\phi_i$}\\
			
			&
			$\mathcal{O}^{(i)jk(l)}_{\phi D}$&
			\multirow{2}{*}{$\downarrow $}&
			\multirow{4}{*}{$\square$ for terms like }&
			\\
			
			&
			$\mathcal{O}^{ij(k)(l)}_{\phi D}$&
			$\{1,2,L, l, R, r\}$&
			&
			\\
			
			&
			$\mathcal{O}^{i(j)k(l)}_{\phi D}$&
			$\{ \} \implies$ ``Tr"&
			$(\phi_{i}^{\dagger}\phi_{j})\square(\phi_{k}^{\dagger}\phi_{l})$, else $\phi D$&
			$\phi_i \rightarrow \{\phi, \phi^{\dagger}, \Delta_{L}, \Delta_{L}^{\dagger} \Delta_{R}, \Delta_{R}^{\dagger} \}$\\
			
			&
			$\mathcal{O}^{(ij)(kl)}_{\phi D}$&
			\multirow{2}{*}{no $\{ \}\implies$ ``Tr" over }&
			&\\
			
			&
			$\mathcal{O}^{\{ij\}\{(k)(l)\}}_{\phi D}$&
			the full operator&&
			\\
			\cline{1-5}
			
			\multirow{3}{*}{$\psi^4$}&
			\multirow{3}{*}{$\mathcal{O}^{[i]}_{f_1 f_2}$/$\mathcal{O}^{[i]}_{f_1 f_2 f_3 f_4}$}&
			\multirow{3}{*}{$i\rightarrow \{1,3,8\}$}&
			\multirow{3}{*}{$\{f_1,..f_4\}\rightarrow \{q_L, q_R, l_L, l_R\}$}&
			[1] $ \implies SU(2)$, $SU(3)$ Singlet\\
			
			&
			&&
			&
			[3] $ \implies SU(2)$ Triplet (involves $\tau^I$)\\
			
			&
			&&
			&
			[8] $\implies SU(3)$ Octet (involves $T^A$)\\
			
			\hline
	\end{tabular}}
	\caption{MLRSM: Summary of the nomenclature of operators. Note: ** - applies to all operator classes. ``Tr" - trace must be taken over internal symmetry indices.}
	\label{tab:MLRSM-operator-nomenclature-I}    
\end{table}

\newpage

\begin{table}[h!]
	\centering
	\renewcommand{\arraystretch}{1.5}
	{\small\begin{tabular}{|c|c|c|c|}
			\hline
			$\mathcal{O}_{6}^{414141}$&$Tr\big[ \tilde{\Phi}^{\dagger}\Phi \ \tilde{\Phi}^{\dagger}\Phi \ \tilde{\Phi}^{\dagger}\Phi \big]$&
			$\mathcal{O}_{6}^{412121}$&$Tr\big[\tilde{\Phi}^{\dagger}\Phi \ \Phi^{\dagger}\Phi \ \Phi^{\dagger}\Phi\big]$\\
			
			$\mathcal{O}_{6}^{414123}$&$Tr[\4\1\4\1\2\3]$&
			$\mathcal{O}_{6}^{414121}$&$Tr[\4\1\4\1\2\1]$\\
			
			$\mathcal{O}_{6}^{4141Rr}$&$Tr[\4\1\4\1\R\r]$&
			$\mathcal{O}_{6}^{41R2l1}$&$Tr[\4\1\R\2\l\1]$\\
			
			$\mathcal{O}_{6}^{4141Ll}$&$Tr[\4\1\4\1\L\l]$&
			$\mathcal{O}_{6}^{41R2r1}$&$Tr[\4\1\R\2\r\1]$\\
			
			$\mathcal{O}_{6}^{2L41r3}$&$Tr[\2\L\4\1\r\3]$&
			$\mathcal{O}_{6}^{21L4r1}$&$Tr[\2\1\L\4\r\1]$\\
			
			$\mathcal{O}_{6}^{LR4rr1}$&$Tr[\L\R\4\r\r\1]$&
			$\mathcal{O}_{6}^{41L2l1}$&$Tr[\4\1\L\2\l\1]$\\
			
			$\mathcal{O}_{6}^{LR4lr1}$&$Tr[\L\R\4\l\r\1]$&
			$\mathcal{O}_{6}^{41R41l}$&$Tr[\4\1\R\4\1\l]$\\
			
			$\mathcal{O}_{6}^{R4Rlr1}$&$Tr[\R\4\R\l\r\1]$&
			$\mathcal{O}_{6}^{41L41r}$&$Tr[\4\1\L\4\1\r]$\\
			
			$\mathcal{O}_{6}^{L4Lrl1}$&$Tr[\L\4\L\r\l\1]$&
			$\mathcal{O}_{6}^{21L2r1}$&$Tr[\2\1\L\2\r\1]$\\
			
			$\mathcal{O}_{6}^{LR4ll1}$&$Tr[\L\R\4\l\l\1]$&
			$\mathcal{O}_{6}^{41L2r3}$&$Tr[\4\1\L\2\r\3]$\\
			
			$\mathcal{O}_{6}^{LR2LL1}$&$Tr[\L\R\2\L\L\1]$&
			$\mathcal{O}_{6}^{RrL4r1}$&$Tr[\R\r\L\4\r\1]$\\                
			
			$\mathcal{O}_{6}^{LR2rr1}$&$Tr[\L\R\2\r\r\1]$&
			$\mathcal{O}_{6}^{RrR4l1}$&$Tr[\R\r\R\4\l\1]$\\            
			$\mathcal{O}_{6}^{21R4l1}$&$Tr[\2\1\R\4\l\1]$&
			$\mathcal{O}_{6}^{LlL4r1}$&$Tr[\L\l\L\4\r\1]$\\        
			$\mathcal{O}_{6}^{41l2r1}$&$Tr[\4\1\l\2\r\1]$&
			$\mathcal{O}_{6}^{LlR4l1}$&$Tr[\L\l\R\4\l\1]$\\    
			
			$\mathcal{O}_{6}^{41RrRr}$&$Tr[\4\1\R\r\R\r]$&
			$\mathcal{O}_{6}^{LlR2l1}$&$Tr[\L\l\R\2\l\1]$\\
			
			$\mathcal{O}_{6}^{41LlLl}$&$Tr[\4\1\L\l\L\l]$&
			$\mathcal{O}_{6}^{RrL2r1}$&$Tr[\R\r\L\2\r\1]$\\
			
			$\mathcal{O}_{6}^{2141Rr}$&$Tr[\2\1\4\1\R\r]$&
			$\mathcal{O}_{6}^{2141Ll}$&$Tr[\2\1\4\1\L\l]$\\
			
			$\mathcal{O}_{6}^{41RRrr}$&$Tr[\4\1\R\R\r\r]$&
			$\mathcal{O}_{6}^{41LlRr}$&$Tr[\4\1\L\l\R\r]$\\
			
			$\mathcal{O}_{6}^{41RRll}$&$Tr[\4\1\R\R\l\l]$&
			$\mathcal{O}_{6}^{41LLll}$&$Tr[\4\1\L\L\l\l]$\\
			
			$\mathcal{O}_{6}^{21LLrr}$&$Tr[\2\1\L\L\r\r]$&
			$\mathcal{O}_{6}^{21LLrr}$&$Tr[\2\1\L\L\r\r]$\\
			
			$\mathcal{O}_{6}^{llRRRr}$&$Tr[\l\l\R\R\R\r]$&
			$\mathcal{O}_{6}^{LlLLrr}$&$Tr[\L\l\L\L\r\r]$\\        
			\hline
	\end{tabular}}
	\caption{MLRSM: $\phi^6\;$\textcolor{red}{[42$\times$2 = 84]} class of operators. Each of these operators also has a distinct Hermitian Conjugate.}
	\label{tab:MLRSM-phi6-dhc}
\end{table}

\begin{table}[h!]
	\centering
	\renewcommand{\arraystretch}{1.5}
	{\small\begin{tabular}{|c|c|c|c|c|c|}
			\hline
			$\mathcal{O}_{l_Ll_R}^{1B}$&
			$(\bar{L}_L\sigma_{\mu\nu}\Phi L_R)B^{\mu\nu}$&
			$\mathcal{O}_{l_Ll_R}^{3B}$&
			$(\bar{L}_L\sigma_{\mu\nu}\tilde{\Phi} L_R)B^{\mu\nu}$&
			$\mathcal{O}_{l_Ll_R}^{1W_L}$&
			$(\bar{L}_L\sigma_{\mu\nu}\tau^I\Phi L_R)W^{I\mu\nu}_L$
			\\
			
			$\mathcal{O}_{l_Ll_R}^{3W_L}$&
			$(\bar{L}_L\sigma_{\mu\nu}\tau^I\tilde{\Phi} L_R)W^{I\mu\nu}_L$&
			$\mathcal{O}_{l_Ll_R}^{1W_R}$&
			$(\bar{L}_L\sigma_{\mu\nu}\tau^I\Phi L_R)W^{I\mu\nu}_R$&
			$\mathcal{O}_{l_Ll_R}^{3W_R}$&
			$(\bar{L}_L\sigma_{\mu\nu}\tau^I\tilde{\Phi} L_R)W^{I\mu\nu}_R$
			\\
			
			$\mathcal{O}_{q_Lq_R}^{1B}$&
			$(\bar{Q}_L\sigma_{\mu\nu}\Phi Q_R)B^{\mu\nu}$&
			$\mathcal{O}_{q_Lq_R}^{3B}$&
			$(\bar{Q}_L\sigma_{\mu\nu}\tilde{\Phi} Q_R)B^{\mu\nu}$&
			$\mathcal{O}_{q_Lq_R}^{1W_L}$&
			$(\bar{Q}_L\sigma_{\mu\nu}\tau^I\Phi Q_R)W^{I\mu\nu}_L$
			\\
			
			$\mathcal{O}_{q_Lq_R}^{3W_L}$&
			$(\bar{Q}_L\sigma_{\mu\nu}\tau^I\tilde{\Phi} Q_R)W^{I\mu\nu}_L$&
			$\mathcal{O}_{q_Lq_R}^{1W_R}$&
			$(\bar{Q}_L\sigma_{\mu\nu}\tau^I\Phi Q_R)W^{I\mu\nu}_R$&
			$\mathcal{O}_{q_Lq_R}^{3W_R}$&
			$(\bar{Q}_L\sigma_{\mu\nu}\tau^I\tilde{\Phi} Q_R)W^{I\mu\nu}_R$
			\\
			
			$\mathcal{O}_{q_Lq_R}^{1G}$&
			$(\bar{Q}_L\sigma_{\mu\nu}T^A\Phi Q_R)G^{A\mu\nu}$&
			$\mathcal{O}_{q_Lq_R}^{3G}$&
			$(\bar{Q}_L\sigma_{\mu\nu}T^A\tilde{\Phi} Q_R)G^{A\mu\nu}$&
			&\\
			\hline
			
			$\mathcal{O}_{l_L l_L}^{\Delta_{L}W_L}$&
			$(L^T_LC\sigma_{\mu\nu}\tau^I\Sigma_{L}L_L)W^{I\mu\nu}_L$&
			$\mathcal{O}_{l_L l_L}^{\Delta_{R}W_R}$&
			$(L^T_LC\sigma_{\mu\nu}\tau^I\Sigma_{R}L_L)W^{I\mu\nu}_R$&
			$\mathcal{O}_{l_R l_R}^{\Delta_{L}W_L}$&
			$(L^T_RC\sigma_{\mu\nu}\tau^I\Sigma_{L}L_R)W^{I\mu\nu}_L$\\
			
			$\mathcal{O}_{l_R l_R}^{\Delta_{R}W_R}$&
			$(L^T_RC\sigma_{\mu\nu}\tau^I\Sigma_{R}L_R)W^{I\mu\nu}_R$&
			&&&\\
			\hline
	\end{tabular}}
	\caption{MLRSM: $\psi^2\phi X\;$\textcolor{red}{[18$\times$2 =  36]} class of operators. Each of these operators also has a distinct Hermitian Conjugate. Note $I \in \{1,2,3\}$ and $A \in \{1,2,...,8\}$ are the $SU(2)$ and $SU(3)$ indices respectively. Here, $\Sigma_{L,R}= i \tau_2 \Delta_{L,R}$.}
	\label{tab:MLRSM-psi2phiX}
\end{table}

\begin{table}[h!]
	\centering
	\renewcommand{\arraystretch}{1.5}
	{\small\begin{tabular}{|c|c|c|c|c|c|}
			\hline
			$\mathcal{O}^{G^3}$&
			$f^{ABC}G^{A\nu}_{\mu}G^{B\rho}_{\nu}G^{C\mu}_{\rho}$&
			$\mathcal{O}^{W^3}_L$&
			$\epsilon^{IJK}W^{I\nu}_{L\mu}W^{J\rho}_{L\nu}W^{K\mu}_{L\rho}$&
			$\mathcal{O}^{W^3}_R$&
			$\epsilon^{IJK}W^{I\nu}_{R\mu}W^{J\rho}_{R\nu}W^{K\mu}_{R\rho}$\\
			
			$\mathcal{O}^{\tilde{G}^3}$&
			$f^{ABC}\tilde{G}^{A\nu}_{\mu}G^{B\rho}_{\nu}G^{C\mu}_{\rho}$&
			$\mathcal{O}^{\tilde{W}^3}_L$&
			$\epsilon^{IJK}\tilde{W}^{I\nu}_{L\mu}W^{J\rho}_{L\nu}W^{K\mu}_{L\rho}$&
			$\mathcal{O}^{\tilde{W}^3}_R$&$\epsilon^{IJK}\tilde{W}^{I\nu}_{R\mu}W^{J\rho}_{R\nu}W^{K\mu}_{R\rho}$\\
			
			\hline
	\end{tabular}}
	\caption{MLRSM: $X^3\;$\textcolor{red}{[6]} class of operators.}
	\label{tab:MLRSM-X3}
\end{table}

\begin{table}[h!]
	\centering
	\renewcommand{\arraystretch}{1.5}
	{\small\begin{tabular}{|c|c|c|c|}
			\hline
			$\mathcal{O}_{6}^{212121}$&$Tr[\2\1\2\1\2\1]$&
			$\mathcal{O}_{6}^{\{21\}\{2121\}}$ &$Tr[\2\1]Tr[\2\1\2\1]$ \\
			
			$\mathcal{O}_{6}^{LlLLll}$&$Tr[\L\l\L\L\l\l]$&
			$\mathcal{O}_{6}^{LlLlLl}$&$Tr[\L\l\L\l\L\l]$\\
			
			$\mathcal{O}_{6}^{RrRRrr}$&$Tr[\R\r\R\R\r\r]$&
			$\mathcal{O}_{6}^{RrRrRr}$&$Tr[\R\r\R\r\R\r]$\\
			
			$\mathcal{O}_{6}^{RrLLll}$&$Tr[\R\r\L\L\l\l]$&
			$\mathcal{O}_{6}^{LlLlRr}$&$Tr[\L\l\L\l\R\r]$\\
			
			$\mathcal{O}_{6}^{LlRRrr}$&$Tr[\L\l\R\R\r\r]$&
			$\mathcal{O}_{6}^{RrRrLl}$&$Tr[\R\r\R\r\L\l]$\\
			
			$\mathcal{O}_{6}^{21LlLl}$&$Tr[\2\1\L\l\L\l]$&
			$\mathcal{O}_{6}^{\{21\}\{LlLl\}}$ &$Tr[\2\1]Tr[\L\l\L\l]$ \\
			
			$\mathcal{O}_{6}^{LlL21l}$&$Tr[\L\l\L\2\1\l]$&
			$\mathcal{O}_{6}^{21RrRr}$&$Tr[\2\1\R\r\R\r]$\\
			
			$\mathcal{O}_{6}^{RrR21r}$&$Tr[\R\r\R\2\1\r]$&
			$\mathcal{O}_{6}^{\{21\}\{RrRr\}}$ &$Tr[\2\1]Tr[\R\r\R\r]$ \\
			
			$\mathcal{O}_{6}^{2121Ll}$&$Tr[\2\1\2\1\L\l]$&
			$\mathcal{O}_{6}^{2211Ll}$&$Tr[\2\2\1\1\L\l]$\\
			
			$\mathcal{O}_{6}^{23Ll41}$&$Tr[\2\3\L\l\4\1]$&
			$\mathcal{O}_{6}^{2L21l1}$&$Tr[\2\L\2\1\l\1]$\\
			
			$\mathcal{O}_{6}^{2121Rr}$&$Tr[\2\1\2\1\R\r]$&
			$\mathcal{O}_{6}^{2211Rr}$&$Tr[\2\2\1\1\R\r]$\\
			
			$\mathcal{O}_{6}^{23Rr41}$&$Tr[\2\3\R\r\4\1]$&
			$\mathcal{O}_{6}^{2R21r1}$&$Tr[\2\R\2\1\r\1]$\\
			
			$\mathcal{O}_{6}^{LlR21r}$&$Tr[\L\l\R\2\1\r]$&$\mathcal{O}_{6}^{RrL21l}$&$Tr[\R\r\L\2\1\l]$\\
			
			$\mathcal{O}_{6}^{LR21rl}$& $Tr[\L\R\2\1\r\l]$& $\mathcal{O}_{6}^{21LlRr}$&	$Tr[\2\1\L\l\R\r]$\\
			\hline
	\end{tabular}}
	\caption{MLRSM: $\phi^6\;$\textcolor{red}{[28]} class of operators. These operators are Self-Hermitian.}
	\label{tab:MLRSM-phi6-shc}
\end{table}

\begin{table}[h!]
	\centering
	\renewcommand{\arraystretch}{1.5}
	{\small\begin{tabular}{|c|c|c|c|}
			\hline
			$\mathcal{O}_{l_L l_L}$&
			$(\bar{L}_L\gamma_{\mu}L_L)(\bar{L}_L\gamma^{\mu}L_L)$ &
			$\mathcal{O}_{l_R l_R}$&
			$(\bar{L}_R\gamma_{\mu}L_R)(\bar{L}_R\gamma^{\mu}L_R)$\\
			
			$\mathcal{O}_{l_R q_L}$&
			$(\bar{L}_R\gamma_{\mu}L_R)(\bar{Q}_L\gamma^{\mu}Q_L)$&
			$\mathcal{O}_{l_L q_R}$&
			$(\bar{L}_L\gamma_{\mu}L_L)(\bar{Q}_R\gamma^{\mu}Q_R)$\\
			
			$\mathcal{O}_{l_L l_R}$&
			$(\bar{L}_L\gamma_{\mu}L_L)(\bar{L}_R\gamma^{\mu}L_R)$&&\\
			\hline
			$\mathcal{O}_{q_L q_L}^{[1]}$&
			$(\bar{Q}_L\gamma_{\mu}Q_L)(\bar{Q}_L\gamma^{\mu}Q_L)$ &
			$\mathcal{O}_{q_L q_L}^{[3]}$&
			$(\bar{Q}_L\gamma_{\mu}\tau^I Q_L)(\bar{Q}_L\gamma^{\mu}\tau^I Q_L)$\\
			
			$\mathcal{O}_{l_L q_L}^{[1]}$&
			$(\bar{L}_L\gamma_{\mu}L_L)(\bar{Q}_L\gamma^{\mu}Q_L)$&
			$\mathcal{O}_{l_L q_L}^{[3]}$&
			$(\bar{L}_L\gamma_{\mu}\tau^I L_L)(\bar{Q}_L\gamma^{\mu}\tau^I Q_L)$\\
			
			$\mathcal{O}_{q_R q_R}^{[1]}$&
			$(\bar{Q}_R\gamma_{\mu}Q_R)(\bar{Q}_R\gamma^{\mu}Q_R)$ &
			$\mathcal{O}_{q_R q_R}^{[3]}$&
			$(\bar{Q}_R\gamma_{\mu}\tau^I Q_R)(\bar{Q}_R\gamma^{\mu}\tau^I Q_R)$\\
			
			$\mathcal{O}_{l_R q_R}^{[1]}$&
			$(\bar{L}_R\gamma_{\mu}L_R)(\bar{Q}_R\gamma^{\mu}Q_R)$&
			$\mathcal{O}_{l_R q_R}^{[3]}$&
			$(\bar{L}_R\gamma_{\mu}\tau^I L_R)(\bar{Q}_R\gamma^{\mu}\tau^I Q_R)$\\
			
			$\mathcal{O}_{q_L q_R}^{[1]}$&
			$(\bar{Q}_L\gamma_{\mu}Q_L)(\bar{Q}_R\gamma^{\mu}Q_R)$ &
			$\mathcal{O}_{q_L q_R}^{[8]}$&
			$(\bar{Q}_L\gamma_{\mu}T^A Q_L)(\bar{Q}_R\gamma^{\mu}T^A Q_R)$\\
			\hline
			
			$\mathcal{O}_{l_L l_R q_L q_R}^{(1)}$&
			\multicolumn{3}{c|}{$\epsilon_{jk}\epsilon_{lm}(\bar{L}_L^jL_R^l)(\bar{Q}_L^kQ_R^m) + h.c.$}\\
			
			$\mathcal{O}_{l_L l_R q_L q_R}^{(3)}$&
			\multicolumn{3}{c|}{$\epsilon_{jk}\epsilon_{lm}(\bar{L}_L^j\sigma_{\mu\nu}L_R^l)(\bar{Q}_L^k\sigma^{\mu\nu}Q_R^m) + h.c.$}\\
			
			$\mathcal{O}_{l_L l_R q_R q_L}$&
			\multicolumn{3}{c|}{$\epsilon_{jk}\epsilon_{lm}(\bar{L}_L^jL_R^l)(\bar{Q}_R^mQ_L^k) + h.c.$} \\
			
			$\mathcal{O}_{l_L l_R l_L l_R}$&
			\multicolumn{3}{c|}{$\epsilon_{jk}\epsilon_{lm}(\bar{L}_L^j L_R^l)(\bar{L}_L^k L_R^m) + h.c.$}\\
			
			$\mathcal{O}_{q_L q_R q_L q_R}^{(1)}$&
			\multicolumn{3}{c|}{$\epsilon_{jk}\epsilon_{lm}(\bar{Q}_L^j Q_R^l)(\bar{Q}_L^k Q_R^m) + h.c.$}\\
			
			$\mathcal{O}_{q_L q_R q_L q_R}^{(8)}$&
			\multicolumn{3}{c|}{$\epsilon_{jk}\epsilon_{lm}(\bar{Q}_L^jT^A Q_R^l)(\bar{Q}_L^kT^A Q_R^m) + h.c.$}\\
			
			\hline
	\end{tabular}}
	\caption{MLRSM: Baryon Number Conserving $\psi^4\;$\textcolor{red}{[15 + 6$\times$2 = 27]} class of operators.}
	\label{tab:MLRSM-psi4}
\end{table}

\begin{table}[h!]
	\centering
	\renewcommand{\arraystretch}{1.5}
	{\small\begin{tabular}{|c|c|c|c|}
			\hline
			
			$\mathcal{O}_{q_L q_L q_L l_L}$&
			$\epsilon_{\alpha\beta\gamma}\epsilon_{jn}\epsilon_{km}[(Q^{\alpha j}_L)^T C Q^{\beta k}_L][(Q^{\gamma m}_L)^T C L^{n}_L]$&
			$\mathcal{O}_{q_R q_R q_R l_R}$&
			$\epsilon_{\alpha\beta\gamma}\epsilon_{jn}\epsilon_{km}[(Q^{\alpha j}_R)^T C Q^{\beta k}_R][(Q^{\gamma m}_R)^T C L^{n}_R]$\\
			
			$\mathcal{O}_{q_L q_L q_R l_R}$&
			$\epsilon_{\alpha\beta\gamma}\epsilon_{jk}\epsilon_{lm}[(Q^{\alpha j}_L)^T C Q^{\beta k}_L][(Q^{\gamma l}_R)^T C L^{m}_R]$&
			$\mathcal{O}_{q_R q_R q_L l_L}$&
			$\epsilon_{\alpha\beta\gamma}\epsilon_{jk}\epsilon_{lm}[(Q^{\alpha j}_R)^T C Q^{\beta k}_R][(Q^{\gamma l}_L)^T C L^{m}_L]$\\
			
			\hline
	\end{tabular}}
	\caption{MLRSM: Baryon Number Violating $\psi^4\;$\textcolor{red}{[4$\times$2 = 8]} class of operators. Each operator has a distinct Hermitian Conjugate.}
	\label{tab:MLRSM-psi4-b-viol}
\end{table}

\begin{table}[h!]
	\centering
	\renewcommand{\arraystretch}{1.5}
	{\small\begin{tabular}{|c|c|c|c|c|c|}
			
			\hline
			$\mathcal{O}_{l_L}^{rLr}$&$L_L^TC  \Sigma_{R} \Sigma_{L}^{\dagger} \Sigma_{R}L_L$&$\mathcal{O}_{l_L}^{r41}$&$L_L^TC \Sigma_{R}\tilde{\Phi}^{\dagger}\Phi L_L$&$\mathcal{O}_{l_L }^{l(23)}$&$[L_L^TC\Sigma_L L_L]Tr\big[\Phi^{\dagger}\tilde{\Phi}\big]$\\
			
			$\mathcal{O}_{l_L}^{lLl}$&$L_L^TC  \Sigma_{L} \Sigma_{L}^{\dagger} \Sigma_{L} L_L$&$\mathcal{O}_{l_L}^{r21}$&$L_L^TC \Sigma_{R}\Phi^{\dagger}\Phi L_L$&$\mathcal{O}_{l_L }^{l(41)}$&$[L_L^TC\Sigma_L L_L]Tr\big[\tilde{\Phi}^{\dagger}\Phi\big]$\\
			
			$\mathcal{O}_{l_L}^{l(Rr)}$&$\big[L_L^TC \Sigma_{L} L_L] Tr\big[\Sigma_{R}^{\dagger}\Sigma_{R}]$&$\mathcal{O}_{l_L}^{l21}$&$L_L^TC \Sigma_{L}\Phi^{\dagger}\Phi L_L$&$\mathcal{O}_{l_L }^{l(21)}$&
			$[L_L^TC\Sigma_L L_L]Tr\big[\Phi^{\dagger}\Phi\big]$\\
			
			$\mathcal{O}_{l_L}^{l(Ll)}$&$[L_L^TC \Sigma_{L} L_L]Tr[\Sigma_{L}^{\dagger}\Sigma_{L}]$&$\mathcal{O}_{l_L}^{r23}$&$L_L^TC \Sigma_{R}\Phi^{\dagger}\tilde{\Phi} L_L$&$\mathcal{O}_{l_R }^{r(41)}$&
			$[L_R^TC\Sigma_R L_R]Tr\big[\tilde{\Phi}^{\dagger}\Phi\big]$\\
			
			$\mathcal{O}_{l_R}^{r(Ll)}$&$[L_R^TC \Sigma_{R} L_R]Tr[\Sigma_{L}^{\dagger}\Sigma_{L}]$&$\mathcal{O}_{l_R}^{l23}$&$L_R^TC \Sigma_{L}\Phi^{\dagger}\tilde{\Phi} L_R$&$\mathcal{O}_{l_R }^{r(23)}$&
			$[L_R^TC\Sigma_R L_R]Tr\big[\Phi^{\dagger}\tilde{\Phi}\big]$\\
			
			$\mathcal{O}_{l_R}^{r(Rr)}$&$[L_R^TC \Sigma_{R} L_R]Tr[\Sigma_{R}^{\dagger}\Sigma_{R}]$&$\mathcal{O}_{l_R}^{r21}$&$L_R^TC \Sigma_{R}\Phi^{\dagger}\Phi L_R$&$\mathcal{O}_{l_R }^{r(21)}$&
			$[L_R^TC\Sigma_R L_R]Tr\big[\Phi^{\dagger}\Phi\big]$\\
			
			$\mathcal{O}_{l_R}^{lRl}$&$L_R^TC  \Sigma_{L} \Sigma_{R}^{\dagger} \Sigma_{L} L_R$&$\mathcal{O}_{l_R}^{l41}$&$L_R^TC \Sigma_{L}\tilde{\Phi}^{\dagger}\Phi L_R$&&\\
			
			$\mathcal{O}_{l_R}^{rRr}$&$L_R^TC  \Sigma_{R} \Sigma_{R}^{\dagger} \Sigma_{R} L_R$&$\mathcal{O}_{l_R}^{l21}$&$L_R^TC \Sigma_{L}\Phi^{\dagger}\Phi L_R$&&\\
			
			\hline
	\end{tabular}}
	\caption{MLRSM: $\psi^2\phi^3\;$\textcolor{red}{[22$\times$2 = 44]} (Majorana type) class of operators. Each of these operators also has a distinct Hermitian Conjugate. Note $\Sigma_{L,R}= i \tau_2 \Delta_{L,R}$.}
	\label{tab:MLRSM-psi2phi3-M}
\end{table}

\begin{table}[h!]
	\centering
	\renewcommand{\arraystretch}{1.5}
	{\small\begin{tabular}{|c|c|c|c|c|c|}
			\hline
			$\mathcal{O}^{1(41)}_{l_L l_R}$&
			$(\bar{L}_L\Phi L_R)Tr[\tilde{\Phi}^{\dagger}\Phi]$&
			$\mathcal{O}^{3(23)}_{l_L l_R}$&
			$(\bar{L}_L\tilde{\Phi} L_R)Tr[\Phi^{\dagger}\tilde{\Phi}]$&
			$\mathcal{O}^{1(21)}_{l_L l_R}$&
			$(\bar{L}_L\Phi L_R)Tr[\Phi^{\dagger}\Phi]$\\
			
			$\mathcal{O}^{3(41)}_{l_L l_R}$&
			$(\bar{L}_L\tilde{\Phi} L_R)Tr[\tilde{\Phi}^{\dagger}\Phi]$&
			$\mathcal{O}^{3(21)}_{l_L l_R}$&
			$(\bar{L}_L\tilde{\Phi} L_R)Tr[\Phi^{\dagger}\Phi]$&
			$\mathcal{O}^{1(23)}_{l_L l_R}$&
			$(\bar{L}_L\Phi L_R)Tr[\Phi^{\dagger}\tilde{\Phi}]$\\
			
			$\mathcal{O}^{1(41)}_{q_L q_R}$&
			$(\bar{Q}_L\Phi Q_R)Tr[\tilde{\Phi}^{\dagger}\Phi]$&
			$\mathcal{O}^{3(23)}_{q_L q_R}$&
			$(\bar{Q}_L\tilde{\Phi} Q_R)Tr[\Phi^{\dagger}\tilde{\Phi}]$&
			$\mathcal{O}^{1(21)}_{q_L q_R}$&
			$(\bar{Q}_L\Phi Q_R)Tr[\Phi^{\dagger}\Phi]$\\
			
			$\mathcal{O}^{3(41)}_{q_L q_R}$&
			$(\bar{Q}_L\tilde{\Phi} Q_R)Tr[\tilde{\Phi}^{\dagger}\Phi]$&
			$\mathcal{O}^{3(21)}_{q_L q_R}$&
			$(\bar{Q}_L\tilde{\Phi} Q_R)Tr[\Phi^{\dagger}\Phi]$&
			$\mathcal{O}^{1(23)}_{l_L l_R}$&
			$(\bar{Q}_L\Phi Q_R)Tr[\Phi^{\dagger}\tilde{\Phi}]$\\
			
			$\mathcal{O}^{L1l}_{l_L l_R}$&
			$\bar{L}_L \Delta_{L}^{\dagger}\Phi\Delta_{L} L_R$&
			$\mathcal{O}^{l1L}_{l_L l_R}$&
			$\bar{L}_L \Delta_{L}\Phi\Delta_{L}^{\dagger} L_R$&
			$\mathcal{O}^{L1r}_{l_L l_R}$&
			$\bar{L}_L \Delta_{L}^{\dagger}\Phi\Delta_{R} L_R$\\
			
			$\mathcal{O}^{R1r}_{l_L l_R}$&
			$\bar{L}_L \Delta_{R}^{\dagger}\Phi\Delta_{R} L_R$&
			$\mathcal{O}^{r1R}_{l_L l_R}$&
			$\bar{L}_L \Delta_{R}\Phi\Delta_{R}^{\dagger} L_R$&
			$\mathcal{O}^{l1R}_{l_L l_R}$&
			$\bar{L}_L \Delta_{L}\Phi\Delta_{R}^{\dagger} L_R$\\
			
			$\mathcal{O}^{L3l}_{l_L l_R}$&
			$\bar{L}_L \Delta_{L}^{\dagger}\tilde{\Phi}\Delta_{L} L_R$&
			$\mathcal{O}^{l3L}_{l_L l_R}$&
			$\bar{L}_L \Delta_{L}\tilde{\Phi}\Delta_{L}^{\dagger} L_R$&
			$\mathcal{O}^{L3r}_{l_L l_R}$&
			$\bar{L}_L \Delta_{L}^{\dagger}\tilde{\Phi}\Delta_{R} L_R$\\
			
			$\mathcal{O}^{R3r}_{l_L l_R}$&
			$\bar{L}_L \Delta_{R}^{\dagger}\tilde{\Phi}\Delta_{R} L_R$&
			$\mathcal{O}^{r3R}_{l_L l_R}$&
			$\bar{L}_L \Delta_{R}\tilde{\Phi}\Delta_{R}^{\dagger} L_R$&
			$\mathcal{O}^{l3R}_{l_L l_R}$&
			$\bar{L}_L \Delta_{L}\tilde{\Phi}\Delta_{R}^{\dagger} L_R$\\
			
			$\mathcal{O}^{L1l}_{q_L q_R}$&
			$\bar{Q}_L \Delta_{L}^{\dagger}\Phi\Delta_{L} Q_R$&
			$\mathcal{O}^{l1L}_{q_L q_R}$&
			$\bar{Q}_L \Delta_{L}\Phi\Delta_{L}^{\dagger} Q_R$&
			$\mathcal{O}^{L1r}_{l_L l_R}$&
			$\bar{Q}_L \Delta_{L}^{\dagger}\Phi\Delta_{R} Q_R$\\
			
			$\mathcal{O}^{R1r}_{q_L q_R}$&
			$\bar{Q}_L \Delta_{R}^{\dagger}\Phi\Delta_{R} Q_R$&
			$\mathcal{O}^{r1R}_{q_L q_R}$&
			$\bar{Q}_L \Delta_{R}\Phi\Delta_{R}^{\dagger} Q_R$&
			$\mathcal{O}^{l1R}_{q_L q_R}$&
			$\bar{Q}_L \Delta_{L}\Phi\Delta_{R}^{\dagger} Q_R$\\
			
			$\mathcal{O}^{L3l}_{q_L q_R}$&
			$\bar{Q}_L \Delta_{L}^{\dagger}\tilde{\Phi}\Delta_{L} Q_R$&
			$\mathcal{O}^{l3L}_{q_L q_R}$&
			$\bar{Q}_L \Delta_{L}\tilde{\Phi}\Delta_{L}^{\dagger} Q_R$&
			$\mathcal{O}^{L3r}_{q_L q_R}$&
			$\bar{Q}_L \Delta_{L}^{\dagger}\tilde{\Phi}\Delta_{R} Q_R$\\
			
			$\mathcal{O}^{R3r}_{q_L q_R}$&
			$\bar{Q}_L \Delta_{R}^{\dagger}\tilde{\Phi}\Delta_{R} Q_R$&
			$\mathcal{O}^{r3R}_{q_L q_R}$&
			$\bar{Q}_L \Delta_{R}\tilde{\Phi}\Delta_{R}^{\dagger} Q_R$&
			$\mathcal{O}^{l3R}_{q_L q_R}$&
			$\bar{Q}_L \Delta_{L}\tilde{\Phi}\Delta_{R}^{\dagger} Q_R$\\
			\hline
	\end{tabular}}
	\caption{MLRSM: $\psi^2\phi^3\;$\textcolor{red}{[36$\times$2 = 72]} (Dirac type) class of operators. Each of these operators also has a distinct Hermitian Conjugate.}
	\label{tab:MLRSM-psi2phi3-D}
\end{table}

\begin{table}[h!]
	\centering
	\renewcommand{\arraystretch}{1.5}
	{\small\begin{tabular}{|c|c|c|c|}
			\hline
			$\mathcal{O}_{2D1}^{l_R[1]}$&
			$Tr[\Phi^{\dagger}i\overleftrightarrow{D_\mu}\Phi](\bar{L}_R\gamma^\mu L_R)$ &
			$\mathcal{O}_{2D1}^{l_R[3]}$&
			$Tr[\Phi^{\dagger}i\overleftrightarrow{D_\mu^I}\Phi](\bar{L}_R\gamma^\mu \tau^I L_R)$\\
			
			$\mathcal{O}_{2D1}^{l_L[1]}$&
			$Tr[\Phi^{\dagger}i\overleftrightarrow{D_\mu}\Phi](\bar{L}_L\gamma^\mu L_L)$ &
			$\mathcal{O}_{2D1}^{l_L[3]}$&
			$Tr[\Phi^{\dagger}i\overleftrightarrow{D_\mu^I}\Phi](\bar{L}_L\gamma^\mu \tau^I L_L)$ \\
			
			$\mathcal{O}_{2D1}^{q_R[1]}$&
			$Tr[\Phi^{\dagger}i\overleftrightarrow{D_\mu}\Phi](\bar{Q}_R\gamma^\mu Q_R)$ &
			$\mathcal{O}_{2D1}^{q_R[3]}$&
			$Tr[\Phi^{\dagger}i\overleftrightarrow{D_\mu^I}\Phi](\bar{Q}_R\gamma^\mu \tau^I Q_R)$\\
			
			$\mathcal{O}_{2D1}^{q_L[1]}$&
			$Tr[\Phi^{\dagger}i\overleftrightarrow{D_\mu}\Phi](\bar{Q}_L\gamma^\mu Q_L)$ &
			$\mathcal{O}_{2D1}^{q_L[3]}$&
			$Tr[\Phi^{\dagger}i\overleftrightarrow{D_\mu^I}\Phi](\bar{Q}_L\gamma^\mu \tau^I Q_L)$ \\
			
			$\mathcal{O}_{LDl}^{l_L[1]}$&
			$Tr[\Delta_L^{\dagger}i\overleftrightarrow{D_\mu}\Delta_L](\bar{L}_L\gamma^\mu L_L)$ &
			$\mathcal{O}_{LDl}^{l_L[3]}$&
			$Tr[\Delta_L^{\dagger}i\overleftrightarrow{D_\mu^I}\Delta_L](\bar{L}_L\gamma^\mu \tau^I L_L)$ \\
			
			$\mathcal{O}_{RDr}^{l_L}$&
			$Tr[\Delta_R^{\dagger}i\overleftrightarrow{D_\mu}\Delta_R](\bar{L}_L\gamma^\mu L_L)$ &
			$\mathcal{O}_{LDl}^{l_R}$&
			$Tr[\Delta_L^{\dagger}i\overleftrightarrow{D_\mu}\Delta_L](\bar{L}_R\gamma^\mu L_R)$ \\
			
			$\mathcal{O}_{RDr}^{l_R[1]}$&
			$Tr[\Delta_R^{\dagger}i\overleftrightarrow{D_\mu}\Delta_R](\bar{L}_R\gamma^\mu L_R)$ &
			$\mathcal{O}_{RDr}^{l_R[3]}$&
			$Tr[\Delta_R^{\dagger}i\overleftrightarrow{D_\mu^I}\Delta_R](\bar{L}_R\gamma^\mu \tau^I L_R)$ \\
			
			$\mathcal{O}_{LDl}^{q_L[1]}$&
			$Tr[\Delta_L^{\dagger}i\overleftrightarrow{D_\mu}\Delta_L](\bar{Q}_L\gamma^\mu Q_L)$ &
			$\mathcal{O}_{LDl}^{q_L[3]}$&
			$Tr[\Delta_L^{\dagger}i\overleftrightarrow{D_\mu^I}\Delta_L](\bar{Q}_L\gamma^\mu \tau^I Q_L)$ \\ 
			
			$\mathcal{O}_{RDr}^{q_L}$&
			$Tr[\Delta_R^{\dagger}i\overleftrightarrow{D_\mu}\Delta_R](\bar{Q}_L\gamma^\mu Q_L)$ &
			$\mathcal{O}_{LDl}^{q_R}$&
			$Tr[\Delta_L^{\dagger}i\overleftrightarrow{D_\mu}\Delta_L](\bar{Q}_R\gamma^\mu Q_R)$ \\
			
			$\mathcal{O}_{Rr}^{q_R[1]}$&
			$Tr[\Delta_R^{\dagger}i\overleftrightarrow{D_\mu}\Delta_R](\bar{Q}_R\gamma^\mu Q_R)$ &
			$\mathcal{O}_{Rr}^{q_R[3]}$&
			$Tr[\Delta_R^{\dagger}i\overleftrightarrow{D_\mu^I}\Delta_R](\bar{Q}_R\gamma^\mu \tau^I Q_R)$ \\
			
			$\mathcal{O}_{4D1}^{l_L}$&
			$Tr[\tilde{\Phi}^{\dagger}i\overleftrightarrow{D_\mu}\Phi](\bar{L}_L\gamma^\mu L_L) + h.c.$&
			$\mathcal{O}_{4D1}^{l_R}$&
			$Tr[\tilde{\Phi}^{\dagger}i\overleftrightarrow{D_\mu}\Phi](\bar{L}_R\gamma^\mu L_R) + h.c.$\\
			
			$\mathcal{O}_{4D1}^{q_L}$&
			$Tr[\tilde{\Phi}^{\dagger}i\overleftrightarrow{D_\mu}\Phi](\bar{Q}_L\gamma^\mu Q_L) + h.c.$&
			$\mathcal{O}_{4D1}^{q_R}$&
			$Tr[\tilde{\Phi}^{\dagger}i\overleftrightarrow{D_\mu}\Phi](\bar{Q}_R\gamma^\mu Q_R) + h.c.$\\
			\hline
			$\mathcal{O}^{l_Ll_R}_{lD1}$&
			$L_L^TCi\tau_2\Delta_L(\gamma_{\mu}D^{\mu}\Phi) L_R + h.c.$&
			$\mathcal{O}^{l_Ll_R}_{rD1}$&
			$L_R^TCi\tau_2\Delta_R(\gamma_{\mu}D^{\mu}\Phi) L_L + h.c.$\\
			
			$\mathcal{O}^{l_Ll_R}_{lD3}$&
			$L_L^TCi\tau_2\Delta_L(\gamma_{\mu}D^{\mu}\tilde{\Phi}) L_R + h.c.$&
			$\mathcal{O}^{l_Ll_R}_{rD3}$&
			$L_R^TCi\tau_2\Delta_R(\gamma_{\mu}D^{\mu}\tilde{\Phi}) L_L + h.c.$\\
			
			\hline
	\end{tabular}}
	\caption{MLRSM: $\psi^2\phi^2D \;$\textcolor{red}{[20 + 8$\times$2 = 36]} class of operators.}
	\label{tab:MLRSM-psi2phi2D}
\end{table}

\begin{table}[h!]
	\centering
	\renewcommand{\arraystretch}{1.5}
	{\small\begin{tabular}{|c|c|c|c|}
			\hline
			$\mathcal{O}_{\phi D}^{2(1)2(1)}$&
			$Tr[(\Phi^{\dagger}i\overleftrightarrow{D_\mu}\Phi)(\Phi^{\dagger}i\overleftrightarrow{D^\mu}\Phi)]$&
			$\mathcal{O}_{\phi D}^{(2)(1)21}$&
			$Tr[(D_{\mu}\Phi)^{\dagger}(D^{\mu}\Phi)\Phi^{\dagger}\Phi]$\\
			
			$\mathcal{O}_{\phi D}^{(11)(22)}$&
			$Tr[D_{\mu}(\Phi\Phi)(D^{\mu}(\Phi\Phi))^{\dagger}]$&
			$\mathcal{O}_{\phi D}^{\{(2)(1)\}\{21\}}$&
			$Tr[(D_{\mu}\Phi)^{\dagger}(D^{\mu}\Phi)]Tr[\Phi^{\dagger}\Phi]$\\
			
			$\mathcal{O}_{\phi D}^{2(1)1(1)}$&
			$Tr[\Phi^{\dagger}(D_{\mu}\Phi)\Phi( D^{\mu}\Phi)\big] + h.c.$&
			$\mathcal{O}_{\phi D}^{1(1)1(1)}$&
			$Tr[\Phi(D_{\mu}\Phi)\Phi(D^{\mu}\Phi)] + h.c.$\\
			\hline
			
			$\mathcal{O}_{\phi D}^{Ll(2)(1)}$&
			$Tr[(\Delta_L^{\dagger}\Delta_L)(D^{\mu}\Phi)^{\dagger}(D_{\mu}\Phi)]$&
			$\mathcal{O}_{\phi D}^{Rr(2)(1)}$&
			$Tr[(\Delta_R^{\dagger}\Delta_R)(D^{\mu}\Phi)^{\dagger}(D_{\mu}\Phi)]$\\
			
			$\mathcal{O}_{\phi D}^{21(L)(l)}$&
			$Tr[(\Phi^{\dagger}\Phi)(D_{\mu}\Delta_L)^{\dagger}(D^{\mu}\Delta_L)]$&
			$\mathcal{O}_{\phi D}^{21(R)(r)}$&
			$Tr[(\Phi^{\dagger}\Phi)(D_{\mu}\Delta_R)^{\dagger}(D^{\mu}\Delta_R)]$\\
			
			$\mathcal{O}_{\phi D}^{\{Ll\}\{(2)(1)\}}$&
			$Tr[\Delta_L^{\dagger}\Delta_L]Tr[(D_{\mu}\Phi)^{\dagger}(D^{\mu}\Phi)]$&
			$\mathcal{O}_{\phi D}^{\{Rr\}\{(2)(1)\}}$&
			$Tr[\Delta_R^{\dagger}\Delta_R]Tr[(D_{\mu}\Phi)^{\dagger}(D^{\mu}\Phi)]$\\
			
			$\mathcal{O}_{\phi D}^{\{21\}\{(L)(l)\}}$&
			$Tr[\Phi^{\dagger}\Phi]Tr[(D_{\mu}\Delta_L)^{\dagger}(D^{\mu}\Delta_L)]$&
			$\mathcal{O}_{\phi D}^{\{21\}\{(R)(r)\}}$&
			$Tr[\Phi^{\dagger}\Phi]Tr[(D_{\mu}\Delta_R)^{\dagger}(D^{\mu}\Delta_R)]$
			\\
			\hline
			
			$\mathcal{O}_{\square}^{Rr(Rr)}$&
			$Tr[(\Delta_R^{\dagger}\Delta_R)\square(\Delta_R^{\dagger}\Delta_R)]$&
			$\mathcal{O}_{\square}^{Ll(Ll)}$&
			$Tr[(\Delta_L^{\dagger}\Delta_L)\square(\Delta_L^{\dagger}\Delta_L)]$\\
			
			$\mathcal{O}_{\phi D}^{R(r)R(r)}$&
			$Tr[(\Delta_R^{\dagger}i\overleftrightarrow{D_\mu}\Delta_R)(\Delta_R^{\dagger}i\overleftrightarrow{D^\mu}\Delta_R)]$&
			$\mathcal{O}_{\phi D}^{L(l)L(l)}$&
			$Tr[(\Delta_L^{\dagger}i\overleftrightarrow{D_\mu}\Delta_L)(\Delta_L^{\dagger}i\overleftrightarrow{D^\mu}\Delta_L)]$\\
			
			$\mathcal{O}_{\phi D}^{\{Rr\}\{(R)(r)\}}$&
			$Tr[\Delta_{R}^{\dagger}\Delta_{R}]Tr[(D_{\mu}\Delta_{R})^{\dagger}(D^{\mu}\Delta_{R})]$&
			$\mathcal{O}_{\phi D}^{\{Ll\}\{(L)(l)\}}$&
			$Tr[\Delta_{L}^{\dagger}\Delta_{L}]Tr[(D_{\mu}\Delta_{L})^{\dagger}(D^{\mu}\Delta_{L})]$\\
			\hline
			
			$\mathcal{O}_{\phi D}^{R(r)1(1)}$&
			$Tr[\Delta_R^{\dagger}(D_\mu\Delta_R)\Phi(D^{\mu}\Phi)] + h.c.$&
			$\mathcal{O}_{\phi D}^{L(l)1(1)}$&
			$Tr[\Delta_L^{\dagger}(D_\mu\Delta_L)\Phi(D^{\mu}\Phi)] + h.c.$\\

			$\mathcal{O}_{\phi D}^{(R)11(r)}$&
			$Tr\{\big[(D_{\mu}\Delta_R)^{\dagger}\Phi\big]\big[\Phi(D^{\mu}\Delta_R)\big]\} + h.c.$&
			$\mathcal{O}_{\phi D}^{(L)11(l)}$&
			$Tr\{\big[(D_{\mu}\Delta_L)^{\dagger}\Phi\big]\big[\Phi(D^{\mu}\Delta_L)\big]\} + h.c.$\\
			
			$\mathcal{O}_{\phi D}^{(R)11(l)}$&
			$Tr\{\big[(D_{\mu}\Delta_R)^{\dagger}\Phi\big]\big[\Phi(D^{\mu}\Delta_L)\big]\} + h.c.$&
			$\mathcal{O}_{\phi D}^{(L)11(r)}$&
			$Tr\{\big[(D_{\mu}\Delta_L)^{\dagger}\Phi\big]\big[\Phi(D^{\mu}\Delta_R)\big]\} + h.c.$\\
			
			$\mathcal{O}_{\phi D}^{(2)rL(1)}$&
			$Tr\{\big[(D_{\mu}\Phi)^{\dagger}\Delta_R\big]\big[\Delta_L^{\dagger}(D^{\mu}\Phi)\big]\} + h.c.$&
			$\mathcal{O}_{\phi D}^{(L)12(r)}$&
			$Tr\{\big[(D_{\mu}\Delta_L)^{\dagger}\Phi\big]\big[\Phi^{\dagger}(D^{\mu}\Delta_R)\big]\} + h.c.$\\
			\hline
			
			$\mathcal{O}_{\phi D}^{(R)lL(r)}$&
			$Tr\{\big[(D_{\mu}\Delta_R)^{\dagger}\Delta_L\big]\big[\Delta_L^{\dagger}(D^{\mu}\Delta_R)\big]\}$&
			$\mathcal{O}_{\phi D}^{(L)rR(l)}$&
			$Tr\{\big[(D_{\mu}\Delta_L)^{\dagger}\Delta_R\big]\big[\Delta_R^{\dagger}(D^{\mu}\Delta_L)\big]\}$\\
			
			$\mathcal{O}_{\phi D}^{(L)rL(r)}$&
			$Tr\{\big[(D_{\mu}\Delta_L)^{\dagger}\Delta_R\big]\big[\Delta_L^{\dagger}(D^{\mu}\Delta_R)\big]\} + h.c.$&
			&\\
			\hline			
	\end{tabular}}
	\caption{MLRSM: $\phi^4D^2\;$\textcolor{red}{[20 + 11$\times$2 = 42]} class of operators.}
	\label{tab:MLRSM-phi4D2}
\end{table}

\begin{table}[h!]
	\centering
	\renewcommand{\arraystretch}{1.5}
	{\small\begin{tabular}{|c|c|c|c|}
			\hline
			$\mathcal{O}_{\phi B}^{41}$&
			$Tr\big[\tilde{\Phi}^{\dagger}\Phi\big]B_{\mu\nu}B^{\mu\nu}$&
			$\mathcal{O}_{\phi \tilde{B}}^{41}$&
			$Tr\big[\tilde{\Phi}^{\dagger}\Phi\big]B_{\mu\nu}\tilde{B}^{\mu\nu}$\\
			
			$\mathcal{O}_{\phi B}^{23}$&
			$Tr\big[\Phi^{\dagger}\tilde{\Phi}\big]B_{\mu\nu}B^{\mu\nu}$&
			$\mathcal{O}_{\phi \tilde{B}}^{23}$&
			$Tr\big[\Phi^{\dagger}\tilde{\Phi}\big]B_{\mu\nu}\tilde{B}^{\mu\nu}$\\
			
			$\mathcal{O}_{\phi B}^{21}$&
			$Tr\big[\Phi^{\dagger}\Phi\big]B_{\mu\nu}B^{\mu\nu}$&
			$\mathcal{O}_{\phi \tilde{B}}^{21}$&
			$Tr\big[\Phi^{\dagger}\Phi\big]B_{\mu\nu}\tilde{B}^{\mu\nu}$\\
			
			$\mathcal{O}_{\phi G}^{21}$&
			$Tr\big[\Phi^{\dagger}\Phi\big]G^a_{\mu\nu}G^{a\mu\nu}$&
			$\mathcal{O}_{\phi \tilde{G}}^{21}$&		
			$Tr\big[\Phi^{\dagger}\Phi\big]G^a_{\mu\nu}\tilde{G}^{a\mu\nu}$\\
			
			$\mathcal{O}_{\phi G}^{23}$&		
			$Tr\big[\Phi^{\dagger}\tilde{\Phi}\big]G^a_{\mu\nu}G^{a\mu\nu}$&
			$\mathcal{O}_{\phi\tilde{G}}^{23}$&				
			$Tr\big[\Phi^{\dagger}\tilde{\Phi}\big]G^a_{\mu\nu}\tilde{G}^{a\mu\nu}$\\
			
			$\mathcal{O}_{\phi G}^{41}$&
			$Tr\big[\tilde{\Phi}^{\dagger}\Phi\big]G^a_{\mu\nu}G^{a\mu\nu}$&
			$\mathcal{O}_{\phi \tilde{G}}^{41}$&		
			$Tr\big[\tilde{\Phi}^{\dagger}\Phi\big]G^a_{\mu\nu}\tilde{G}^{a\mu\nu}$\\
			
			$\mathcal{O}_{\phi W_L}^{23}$&		
			$Tr\big[\Phi^{\dagger}\tilde{\Phi}W_{L\mu\nu}W_L^{\mu\nu}\big]$&
			$\mathcal{O}_{\phi\tilde{W}_L}^{23}$&			
			$Tr\big[\Phi^{\dagger}\tilde{\Phi}W_{L\mu\nu}\tilde{W}_L^{\mu\nu}\big]$\\
			
			$\mathcal{O}_{\phi W_L}^{41}$&
			$Tr\big[\tilde{\Phi}^{\dagger}\Phi W_{L\mu\nu}W_L^{\mu\nu}\big]$&
			$\mathcal{O}_{\phi \tilde{W}_L}^{41}$&		
			$Tr\big[\tilde{\Phi}^{\dagger}\Phi W_{L\mu\nu}\tilde{W}_L^{\mu\nu}\big]$\\
			
			$\mathcal{O}_{\phi W_L}^{21}$&
			$Tr\big[\Phi^{\dagger}\Phi W_{L\mu\nu}W_L^{\mu\nu}\big]$&
			$\mathcal{O}_{\phi \tilde{W}_L}^{21}$&		
			$Tr\big[\Phi^{\dagger}\Phi W_{L\mu\nu}\tilde{W}_L^{\mu\nu}\big]$\\
			
			$\mathcal{O}_{\phi W_R}^{23}$&		
			$Tr\big[\Phi^{\dagger}\tilde{\Phi}W_{R\mu\nu}W_R^{\mu\nu}\big]$&
			$\mathcal{O}_{\phi\tilde{W}_R}^{23}$&			
			$Tr\big[\Phi^{\dagger}\tilde{\Phi}W_{R\mu\nu}\tilde{W}_R^{\mu\nu}\big]$\\
			
			$\mathcal{O}_{\phi W_R}^{41}$&
			$Tr\big[\tilde{\Phi}^{\dagger}\Phi W_{R\mu\nu}W_R^{\mu\nu}\big]$&
			$\mathcal{O}_{\phi \tilde{W}_R}^{41}$&		
			$Tr\big[\tilde{\Phi}^{\dagger}\Phi W_{R\mu\nu}\tilde{W}_R^{\mu\nu}\big]$\\
			
			$\mathcal{O}_{\phi W_R}^{21}$&
			$Tr\big[\Phi^{\dagger}\Phi W_{R\mu\nu}W_R^{\mu\nu}\big]$&
			$\mathcal{O}_{\phi \tilde{W}_R}^{21}$&		
			$Tr\big[\Phi^{\dagger}\Phi W_{R\mu\nu}\tilde{W}_R^{\mu\nu}\big]$\\
			
			$\mathcal{O}_{\Delta B}^{Ll}$&
			$Tr\big[\Delta_L^{\dagger}\Delta_L\big]B_{\mu\nu}B^{\mu\nu}$&
			$\mathcal{O}_{\Delta \tilde{B}}^{Ll}$&		
			$Tr\big[\Delta_L^{\dagger}\Delta_L\big]B_{\mu\nu}\tilde{B}^{\mu\nu}$\\
			
			$\mathcal{O}_{\Delta B}^{Rr}$&
			$Tr\big[\Delta_R^{\dagger}\Delta_R\big]B_{\mu\nu}B^{\mu\nu}$&
			$\mathcal{O}_{\Delta \tilde{B}}^{Rr}$&		
			$Tr\big[\Delta_R^{\dagger}\Delta_R\big]B_{\mu\nu}\tilde{B}^{\mu\nu}$\\
			
			$\mathcal{O}_{\Delta G}^{Ll}$&
			$Tr\big[\Delta_L^{\dagger}\Delta_L\big]G^a_{\mu\nu}G^{a\mu\nu}$&
			$\mathcal{O}_{\Delta \tilde{G}}^{Ll}$&		
			$Tr\big[\Delta_L^{\dagger}\Delta_L\big]G^a_{\mu\nu}\tilde{G}^{a\mu\nu}$\\
			
			$\mathcal{O}_{\Delta G}^{Rr}$&
			$Tr\big[\Delta_R^{\dagger}\Delta_R\big]G^a_{\mu\nu}G^{a\mu\nu}$&
			$\mathcal{O}_{\Delta \tilde{G}}^{Rr}$&		
			$Tr\big[\Delta_R^{\dagger}\Delta_R\big]G^a_{\mu\nu}\tilde{G}^{a\mu\nu}$\\

			$\mathcal{O}_{\phi W_L B}^{21}$&
			$Tr\big[\Phi^{\dagger}W_L^{\mu\nu}\Phi\big]B_{\mu\nu}$&
			$\mathcal{O}_{\phi \tilde{W}_L B}^{21}$&		
			$Tr\big[\Phi^{\dagger}\tilde{W}_L^{\mu\nu}\Phi\big]B_{\mu\nu}$\\
			
			$\mathcal{O}_{\phi W_R B}^{21}$&
			$Tr\big[\Phi^{\dagger}W_R^{\mu\nu}\Phi\big]B_{\mu\nu}$&
			$\mathcal{O}_{\phi \tilde{W}_R B}^{21}$&		
			$Tr\big[\Phi^{\dagger}\tilde{W}_R^{\mu\nu}\Phi\big]B_{\mu\nu}$\\
			
			$\mathcal{O}_{\phi W_L W_R}^{21}$&
			$Tr\big[\Phi^{\dagger} W_{L\mu\nu} \Phi W_R^{\mu\nu}\big]$&
			$\mathcal{O}_{\phi W_L \tilde{W}_R}^{21}$&		
			$Tr\big[\Phi^{\dagger} W_{L\mu\nu} \Phi \tilde{W}_R^{\mu\nu}\big]$\\
			
			$\mathcal{O}_{\phi W_L W_R}^{23}$&
			$Tr\big[\Phi^{\dagger} W_{L\mu\nu} \tilde{\Phi} W_R^{\mu\nu}\big]$&
			$\mathcal{O}_{\phi W_L \tilde{W}_R}^{23}$&		
			$Tr\big[\Phi^{\dagger} W_{L\mu\nu} \tilde{\Phi} \tilde{W}_R^{\mu\nu}\big]$\\
			
			$\mathcal{O}_{\phi W_L W_R}^{41}$&
			$Tr\big[\tilde{\Phi}^{\dagger} W_{L\mu\nu} \Phi W_R^{\mu\nu}\big]$&
			$\mathcal{O}_{\phi W_L \tilde{W}_R}^{41}$&		
			$Tr\big[\tilde{\Phi}^{\dagger} W_{L\mu\nu} \Phi \tilde{W}_R^{\mu\nu}\big]$\\
			
			$\mathcal{O}_{\Delta W_L B}^{Ll}$&
			$Tr\big[\Delta_L^{\dagger}W_L^{\mu\nu}\Delta_L\big]B_{\mu\nu}$&
			$\mathcal{O}_{\Delta \tilde{W}_L B}^{Ll}$&		
			$Tr\big[\Delta_L^{\dagger}\tilde{W}_L^{\mu\nu}\Delta_L\big]B_{\mu\nu}$\\
			
			$\mathcal{O}_{\Delta W_R B}^{Rr}$&
			$Tr\big[\Delta_R^{\dagger}W_R^{\mu\nu}\Delta_R\big]B_{\mu\nu}$&
			$\mathcal{O}_{\Delta \tilde{W}_R B}^{Rr}$&		
			$Tr\big[\Delta_R^{\dagger}\tilde{W}_R^{\mu\nu}\Delta_R\big]B_{\mu\nu}$\\
			
			$\mathcal{O}_{\Delta W}^{LlW_LW_L}$&
			$Tr\big[\Delta_L^{\dagger}\Delta_L W_{L\mu\nu}W_L^{\mu\nu}\big]$&
			$\mathcal{O}_{\Delta \tilde{W}}^{LlW_LW_L}$&		
			$Tr\big[\Delta_L^{\dagger}\Delta_L W_{L\mu\nu}\tilde{W}_L^{\mu\nu}\big]$\\
			
			$\mathcal{O}_{\Delta W}^{RrW_LW_L}$&
			$Tr\big[\Delta_R^{\dagger}\Delta_R W_{L\mu\nu}W_L^{\mu\nu}\big]$&
			$\mathcal{O}_{\Delta \tilde{W}}^{RrW_LW_L}$&		
			$Tr\big[\Delta_R^{\dagger}\Delta_R W_{L\mu\nu}\tilde{W}_L^{\mu\nu}\big]$\\
			
			$\mathcal{O}_{\Delta W}^{LlW_RW_R}$&
			$Tr\big[\Delta_L^{\dagger}\Delta_L W_{R\mu\nu}W_R^{\mu\nu}\big]$&
			$\mathcal{O}_{\Delta \tilde{W}}^{LlW_RW_R}$&		
			$Tr\big[\Delta_L^{\dagger}\Delta_L W_{R\mu\nu}\tilde{W}_R^{\mu\nu}\big]$\\
			
			$\mathcal{O}_{\Delta W}^{RrW_RW_R}$&
			$Tr\big[\Delta_R^{\dagger}\Delta_R W_{R\mu\nu}W_R^{\mu\nu}\big]$&
			$\mathcal{O}_{\Delta \tilde{W}}^{RrW_RW_R}$&		
			$Tr\big[\Delta_R^{\dagger}\Delta_R W_{R\mu\nu}\tilde{W}_R^{\mu\nu}\big]$\\
			
			$\mathcal{O}_{\Delta W}^{LW_LlW_L}$&
			$Tr\big[\Delta_LW_{L\mu\nu}\Delta_L^{\dagger}W_L^{\mu\nu}\big]$&
			$\mathcal{O}_{\Delta \tilde{W}}^{LW_LlW_L}$&		
			$Tr\big[\Delta_LW_{L\mu\nu}\Delta_L^{\dagger}\tilde{W}_L^{\mu\nu}\big]$\\
			
			$\mathcal{O}_{\Delta W}^{LW_RrW_L}$&
			$Tr\big[\Delta_R W_{R\mu\nu} \Delta_L^{\dagger} W_L^{\mu\nu}\big]$&
			$\mathcal{O}_{\Delta \tilde{W}}^{LW_RrW_L}$&		
			$Tr\big[\Delta_R W_{R\mu\nu} \Delta_L^{\dagger} \tilde{W}_L^{\mu\nu}\big]$\\
			
			$\mathcal{O}_{\Delta W}^{RW_LlW_R}$&
			$Tr\big[\Delta_L W_{L\mu\nu} \Delta_R^{\dagger} W_R^{\mu\nu}\big]$&
			$\mathcal{O}_{\Delta \tilde{W}}^{RW_LlW_R}$&		
			$Tr\big[\Delta_L W_{L\mu\nu} \Delta_R^{\dagger} \tilde{W}_R^{\mu\nu}\big]$\\
			
			$\mathcal{O}_{\Delta W}^{RW_RrW_R}$&
			$Tr\big[\Delta_RW_{R\mu\nu}\Delta_R^{\dagger}W_R^{\mu\nu}\big]$&
			$\mathcal{O}_{\Delta \tilde{W}}^{RW_RrW_R}$&		
			$Tr\big[\Delta_RW_{R\mu\nu}\Delta_R^{\dagger}\tilde{W}_R^{\mu\nu}\big]$\\
			
			\hline
	\end{tabular}}
	\caption{MLRSM: $\phi^2X^2\;$\textcolor{red}{[62]} class of operators. Note $W^{\mu\nu}_{L,R} = W^{I\mu\nu}_{L,R}\cdot\tau^I, I = 1,2,3.$}
	\label{tab:MLRSM-phi2X2}
\end{table}
\clearpage


\newpage

\begin{center}
	\underline{\bf { Scalar Potential in the presence of $\phi^6$ operators:}}
\end{center}

In the presence of  $\phi^6$ operators, the  scalar potential is modified as $V(\phi_1, \phi_2)$ + ${\mathcal{L}_{\phi^6}}$ where 
\begin{align}
{\mathcal{L}_{\phi^6}}&=\sum^{112}_{j=1}\mathcal{C}^{j}_{6} \mathcal{O}^{j}_{6}.
\end{align}

Apart from the  earlier minimisation conditions Eq.~\ref{eq:MLRSM-min-condition-tree}, we have further noted the following minimisation criteria:
\begin{eqnarray}\label{eq:MLRSM-min-condition-dim6-I}
	\frac{\partial \mathcal{L}_{\phi^6}}{\partial v_R}&=&\frac{1}{4 \Lambda ^2}\big[\kappa _1 v_L \big(3 v_R^2 \big(\kappa _2 \mathcal{C}_6^{{RrL2r1}}+\kappa _1 (\mathcal{C}_6^{{RrL4r1}}+\mathcal{C}_6^{{RrR4l1}})\big)+\kappa _2 \kappa _1^2 \mathcal{C}_6^{21{L2r1}}+\kappa _2^3 \mathcal{C}_6^{41{L2r3}}+ \mathcal{C}_6^{{RrRrRr}} v_R^3 \nonumber\\
	&+&\kappa _1^3 (\mathcal{C}_6^{21{L4r1}}+\mathcal{C}_6^{21{R4l1}})+\kappa _2^2 \kappa _1 (\mathcal{C}_6^{2{L41r3}}+\mathcal{C}_6^{41{L41r}}+\mathcal{C}_6^{41{R2l1}}+\mathcal{C}_6^{41{R41l}})\big)+v_L^2 v_R \big(2 \mathcal{C}_6^{{RrRrLl}} v_R^2\nonumber\\
	&+&\kappa _1^2 \mathcal{C}_6^{21{LlRr}}+2 \kappa _1 \kappa _2 \mathcal{C}_6^{41{LlRr}}+\kappa _2^2 (\mathcal{C}_6^{{LlR21r}}+\mathcal{C}_6^{{RrL21l}})\big)+\mathcal{C}_6^{{LlLlRr}} v_L^4 v_R+\kappa _1 v_L^3 \big(\kappa _1 (\mathcal{C}_6^{{LlL4r1}}+\mathcal{C}_6^{{LlR4l1}})\nonumber\\
	&+&\kappa _2 \mathcal{C}_6^{{LlR2l1}}\big)+v_R \big(v_R^2 \big(\kappa _1^2 (\mathcal{C}_6^{\{21\}\{{RrRr}\}}+2 \mathcal{C}_6^{21{RrRr}})+\kappa _2^2 (\mathcal{C}_6^{\{21\}\{{RrRr}\}}+2 \mathcal{C}_6^{{RrR21r}})+4 \kappa _2 \kappa _1 \mathcal{C}_6^{41{RrRr}}\big)\nonumber\\
	&+&\kappa _1^2 \big(\kappa _1^2 (\mathcal{C}_6^{2121{Rr}}+\mathcal{C}_6^{2211{Rr}})+2 \kappa _2 \kappa _1 \mathcal{C}_6^{2141{Rr}}+\kappa _2^2 (\mathcal{C}_6^{23{Rr41}}+\mathcal{C}_6^{2{R21r1}}+2 (\mathcal{C}_6^{4141{Rr}}+\mathcal{C}_6^{41{R2r1}}))\big)\big)\big] \nonumber \\
	&=&0,\\
	\frac{\partial \mathcal{L}_{\phi^6}}{\partial v_L}&=&\frac{1}{4 \Lambda ^2}\big[v_L \big(v_R^2 \big(\kappa _1^2 \mathcal{C}_6^{21{LlRr}}+2 \kappa _2 \kappa _1 \mathcal{C}_6^{41{LlRr}}+\kappa _2^2 (\mathcal{C}_6^{{LlR21r}}+\mathcal{C}_6^{{RrL21l}})\big)+\mathcal{C}_6^{{RrRrLl}} v_R^4\nonumber\\
	&+&\kappa _1^2 \big(\kappa _1^2 (\mathcal{C}_6^{2121{Ll}}+\mathcal{C}_6^{2211{Ll}})+2 \kappa _2 \kappa _1 \mathcal{C}_6^{2141{Ll}}+\kappa _2^2 (\mathcal{C}_6^{23{Ll41}}+\mathcal{C}_6^{2{L21l1}}+2 (\mathcal{C}_6^{4141{Ll}}+\mathcal{C}_6^{41{L2l1}}))\big)\big)\nonumber\\
	&+&2 v_L^3 \big(\mathcal{C}_6^{{LlLlRr}} v_R^2+\kappa _1^2 (\mathcal{C}_6^{\{21\}\{{LlLl}\}}+\mathcal{C}_6^{21{LlLl}})+\kappa _2^2 (\mathcal{C}_6^{\{21\}\{{LlLl}\}}+\mathcal{C}_6^{{LlL21l}})+2 \kappa _1 \kappa _2 \mathcal{C}_6^{41{LlLl}}\big)\nonumber\\
	&+&3 \kappa _1 v_L^2 v_R \big(\kappa _1 (\mathcal{C}_6^{{LlL4r1}}+\mathcal{C}_6^{{LlR4l1}})+\kappa _2 \mathcal{C}_6^{{LlR2l1}}\big)+3 \mathcal{C}_6^{{LlLlLl}} v_L^5+\kappa _1 v_R \big(v_R^2 \big(\kappa _2 \mathcal{C}_6^{{RrL2r1}}+\kappa _1 (\mathcal{C}_6^{{RrL4r1}}\nonumber\\
	&+&\mathcal{C}_6^{{RrR4l1}})\big)+\kappa _2 \kappa _1^2 \mathcal{C}_6^{21{L2r1}}+\kappa _1^3 (\mathcal{C}_6^{21{L4r1}}+\mathcal{C}_6^{21{R4l1}})+\kappa _2^2 \kappa _1 (\mathcal{C}_6^{2{L41r3}}+\mathcal{C}_6^{41{L41r}}+\mathcal{C}_6^{41{R2l1}}+\mathcal{C}_6^{41{R41l}})\nonumber \\
	&+&\kappa _2^3 \mathcal{C}_6^{41{L2r3}}\big)\big]\nonumber \\&=& 0,  \\
	\frac{\partial \mathcal{L}_{\phi^6}}{\partial \kappa_1}&=&\frac{1}{4 \Lambda ^2}\big[v_L^2 \big(v_R^2 \big(\kappa _1 \mathcal{C}_6^{21{LlRr}}+\kappa _2 \mathcal{C}_6^{41{LlRr}}\big)+\kappa _1 \big(2 \kappa _1^2 (\mathcal{C}_6^{2121{Ll}}+\mathcal{C}_6^{2211{Ll}})+3 \kappa _2 \kappa _1 \mathcal{C}_6^{2141{Ll}}+\kappa _2^2 (\mathcal{C}_6^{23{Ll41}}\nonumber\\
	&+&\mathcal{C}_6^{2{L21l1}}+2 (\mathcal{C}_6^{4141{Ll}}+\mathcal{C}_6^{41{L2l1}}))\big)\big)+v_L v_R \big(v_R^2 \big(\kappa _2 \mathcal{C}_6^{{RrL2r1}}+2 \kappa _1 (\mathcal{C}_6^{{RrL4r1}}+\mathcal{C}_6^{{RrR4l1}})\big)+3 \kappa _2 \kappa _1^2 \mathcal{C}_6^{21{L2r1}}\nonumber\\
	&+&4 \kappa _1^3 (\mathcal{C}_6^{21{L4r1}}+\mathcal{C}_6^{21{R4l1}})+2 \kappa _2^2 \kappa _1 (\mathcal{C}_6^{2{L41r3}}+\mathcal{C}_6^{41{L41r}}+\mathcal{C}_6^{41{R2l1}}+\mathcal{C}_6^{41{R41l}})+\kappa _2^3 \mathcal{C}_6^{41{L2r3}}\big)\nonumber\\
	&+&v_L^3 v_R \big(2 \kappa _1 (\mathcal{C}_6^{{LlL4r1}}+\mathcal{C}_6^{{LlR4l1}})+\kappa _2 \mathcal{C}_6^{{LlR2l1}}\big)+v_L^4 \big(\kappa _1 (\mathcal{C}_6^{\{21\}\{{LlLl}\}}+\mathcal{C}_6^{21{LlLl}})+\kappa _2 \mathcal{C}_6^{41{LlLl}}\big)\nonumber\\
	&+&\kappa _1 v_R^2 \big(2 \kappa _1^2 (\mathcal{C}_6^{2121{Rr}}+\mathcal{C}_6^{2211{Rr}})+3 \kappa _2 \kappa _1 \mathcal{C}_6^{2141{Rr}}+\kappa _2^2 (\mathcal{C}_6^{23{Rr41}}+\mathcal{C}_6^{2{R21r1}}+2 (\mathcal{C}_6^{4141{Rr}}+\mathcal{C}_6^{41{R2r1}}))\big)\nonumber\\
	&+&v_R^4 \big(\kappa _1 (\mathcal{C}_6^{\{21\}\{{RrRr}\}}+\mathcal{C}_6^{21{RrRr}})+\kappa _2 \mathcal{C}_6^{41{RrRr}}\big)+3 \kappa _1^5 \mathcal{C}_6^{\{21\}\{2121\}}+2 \kappa _2^2 \kappa _1^3 \mathcal{C}_6^{\{21\}\{2121\}}+\kappa _2^4 \kappa _1 \mathcal{C}_6^{\{21\}\{2121\}}\nonumber\\
	&+&3 \kappa _1^5 \mathcal{C}_6^{212121}+5 \kappa _2 \kappa _1^4 \mathcal{C}_6^{412121}+\kappa _2^5 \mathcal{C}_6^{412121}+4 \kappa _2^2 \kappa _1^3 \mathcal{C}_6^{414121}+2 \kappa _2^4 \kappa _1 \mathcal{C}_6^{414121}+6 \kappa _2^3 \kappa _1^2 \mathcal{C}_6^{414123}+6 \kappa _2^3 \kappa _1^2 \mathcal{C}_6^{414141}\big] \nonumber \\
	&=& 0,\\
		\frac{\partial \mathcal{L}_{\phi^6}}{\partial \kappa_2}&=&\frac{1}{4 \Lambda ^2}\big[v_L^2 \big(v_R^2 \big(\kappa _1 \mathcal{C}_6^{41{LlRr}}+\kappa _2 (\mathcal{C}_6^{{LlR21r}}+\mathcal{C}_6^{{RrL21l}})\big)+\kappa _1^2 \big(\kappa _1 \mathcal{C}_6^{2141{Ll}}+\kappa _2 (\mathcal{C}_6^{23{Ll41}}+\mathcal{C}_6^{2{L21l1}} \nonumber\\
	&+&2 (\mathcal{C}_6^{4141{Ll}}+\mathcal{C}_6^{41{L2l1}}))\big)\big)+\kappa _1 v_L v_R \big(\mathcal{C}_6^{{RrL2r1}} v_R^2+\kappa _1^2 \mathcal{C}_6^{21{L2r1}}+2 \kappa _1 \kappa _2 (\mathcal{C}_6^{2{L41r3}}+\mathcal{C}_6^{41{L41r}}+\mathcal{C}_6^{41{R2l1}}\nonumber\\
	&+&\mathcal{C}_6^{41{R41l}})+3 \kappa _2^2 \mathcal{C}_6^{41{L2r3}}\big)+\kappa _1 \mathcal{C}_6^{{LlR2l1}} v_L^3 v_R+v_L^4 \big(\kappa _2 (\mathcal{C}_6^{\{21\}\{{LlLl}\}}+\mathcal{C}_6^{{LlL21l}})+\kappa _1 \mathcal{C}_6^{41{LlLl}}\big)\nonumber\\
	&+&\kappa _1^2 v_R^2 \big(\kappa _1 \mathcal{C}_6^{2141{Rr}}+\kappa _2 (\mathcal{C}_6^{23{Rr41}}+\mathcal{C}_6^{2{R21r1}}+2 (\mathcal{C}_6^{4141{Rr}}+\mathcal{C}_6^{41{R2r1}}))\big)+v_R^4 \big(\kappa _2 (\mathcal{C}_6^{\{21\}\{{RrRr}\}}+\mathcal{C}_6^{{RrR21r}})\nonumber\\
	&+&\kappa _1 \mathcal{C}_6^{41{RrRr}}\big)+\kappa _2 \kappa _1^4 \mathcal{C}_6^{\{21\}\{2121\}}+2 \kappa _2^3 \kappa _1^2 \mathcal{C}_6^{\{21\}\{2121\}}+3 \kappa _2^5 \mathcal{C}_6^{\{21\}\{2121\}}+3 \kappa _2^5 \mathcal{C}_6^{212121}+\kappa _1^5 \mathcal{C}_6^{412121}\nonumber\\
	&+&5 \kappa _2^4 \kappa _1 \mathcal{C}_6^{412121}+2 \kappa _2 \kappa _1^4 \mathcal{C}_6^{414121}+4 \kappa _2^3 \kappa _1^2 \mathcal{C}_6^{414121}+6 \kappa _2^2 \kappa _1^3 \mathcal{C}_6^{414123}+6 \kappa _2^2 \kappa _1^3 \mathcal{C}_6^{414141}\big] \nonumber \\
	&=& 0.
	\end{eqnarray}
\clearpage
\pagebreak

 \underline{\large{$\phi^{4}D^{2}$ and $\phi^6$ operators: Redefinitions of the Scalar fields and modifications in spectrum }}: \\

The scalar kinetic lagrangian,  Eq.~\ref{eq:MLRSM-ren-lag}, gets modified  in presence of the $\phi^{4}D^{2}$ operators, Table~\ref{tab:MLRSM-phi4D2}, as:
\small{\begin{eqnarray}\label{eq:MLRSM-scalar-kinetic-lag-full}
	\mathcal{L}^{(4)+(6)}&=& \frac{1}{2}
	\begin{pmatrix}
	\partial^{\mu}\phi_{1}^{0r} \\ \partial^{\mu}\phi_{2}^{0r} \\ \partial^{\mu}\delta_{R}^{0r} \\ \partial^{\mu}\delta_{L}^{0r}
	\end{pmatrix}^{T} \begin{pmatrix}
	1+\frac{A_{r}^{11}}{\Lambda^{2}} & \frac{A_{r}^{12}}{\Lambda^{2}} & \frac{A_{r}^{13}}{\Lambda^{2}} & \frac{A_{r}^{14}}{\Lambda^{2}} \\
	\frac{A_{r}^{12}}{\Lambda^{2}} & 1+\frac{A_{r}^{22}}{\Lambda^{2}} & \frac{A_{r}^{23}}{\Lambda^{2}} & \frac{A_{r}^{24}}{\Lambda^{2}} \\
	\frac{A_{r}^{13}}{\Lambda^{2}} & \frac{A_{r}^{23}}{\Lambda^{2}} & 1+\frac{A_{r}^{33}}{\Lambda^{2}} & \frac{A_{r}^{34}}{\Lambda^{2}} \\
	\frac{A_{r}^{14}}{\Lambda^{2}} & \frac{A_{r}^{24}}{\Lambda^{2}} & \frac{A_{r}^{34}}{\Lambda^{2}} & 1+\frac{A_{r}^{44}}{\Lambda^{2}} \\
	\end{pmatrix}
	\begin{pmatrix}
	\partial_{\mu}\phi_{1}^{0r} \\ \partial_{\mu}\phi_{2}^{0r} \\ \partial_{\mu}\delta_{R}^{0r} \\ \partial_{\mu}\delta_{L}^{0r}
	\end{pmatrix} \nonumber \\
	\nonumber \\
	& + &\frac{1}{2}
	\begin{pmatrix}
	\partial^{\mu}\phi_{1}^{0i} \\ \partial^{\mu}\phi_{2}^{0i} \\ \partial^{\mu}\delta_{R}^{0i} \\ \partial^{\mu}\delta_{L}^{0i}
	\end{pmatrix}^{T}
	\begin{pmatrix}
	1+\frac{A_{i}^{11}}{\Lambda^{2}} & \frac{A_{i}^{12}}{\Lambda^{2}} & \frac{A_{i}^{13}}{\Lambda^{2}} & \frac{A_{i}^{14}}{\Lambda^{2}} \\
	\frac{A_{i}^{12}}{\Lambda^{2}} & 1+\frac{A_{i}^{22}}{\Lambda^{2}} & \frac{A_{i}^{23}}{\Lambda^{2}} & \frac{A_{i}^{24}}{\Lambda^{2}} \\
	\frac{A_{i}^{13}}{\Lambda^{2}} & \frac{A_{i}^{23}}{\Lambda^{2}} & 1+\frac{A_{i}^{33}}{\Lambda^{2}} & \frac{A_{i}^{34}}{\Lambda^{2}} \\
	\frac{A_{i}^{14}}{\Lambda^{2}} & \frac{A_{i}^{24}}{\Lambda^{2}} & \frac{A_{i}^{34}}{\Lambda^{2}} & 1+\frac{A_{i}^{44}}{\Lambda^{2}} \\
	\end{pmatrix}
	\begin{pmatrix}
	\partial_{\mu}\phi_{1}^{0i} \\ \partial_{\mu}\phi_{2}^{0i} \\ \partial_{\mu}\delta_{R}^{0i} \\ \partial_{\mu}\delta_{L}^{0i}
	\end{pmatrix}\nonumber \\
	\\
	&+&\begin{pmatrix}
	\partial^{\mu}\phi_{1}^{'+} \\ \partial^{\mu}\phi_{2}^{'+} \\ \partial^{\mu}\delta_{R}^{+} \\ \partial^{\mu}\delta_{L}^{+}
	\end{pmatrix}^{\dagger}
	\begin{pmatrix}
	1+\frac{A_{+}^{11}}{\Lambda^{2}} & \frac{A_{+}^{12}}{\Lambda^{2}} & \frac{A_{+}^{13}}{\Lambda^{2}} & \frac{A_{+}^{14}}{\Lambda^{2}} \\
	\frac{A_{+}^{12}}{\Lambda^{2}} & 1+\frac{A_{+}^{22}}{\Lambda^{2}} & \frac{A_{+}^{23}}{\Lambda^{2}} & \frac{A_{+}^{24}}{\Lambda^{2}} \\
	\frac{A_{+}^{13}}{\Lambda^{2}} & \frac{A_{+}^{23}}{\Lambda^{2}} & 1+\frac{A_{+}^{33}}{\Lambda^{2}} & \frac{A_{+}^{34}}{\Lambda^{2}} \\
	\frac{A_{+}^{14}}{\Lambda^{2}} & \frac{A_{+}^{24}}{\Lambda^{2}} & \frac{A_{+}^{34}}{\Lambda^{2}} & 1+\frac{A_{+}^{44}}{\Lambda^{2}} \\
	\end{pmatrix}
	\begin{pmatrix}
	\partial_{\mu}\phi_{1}^{'+} \\ \partial_{\mu}\phi_{2}^{'+} \\ \partial_{\mu}\delta_{R}^{+} \\ \partial_{\mu}\delta_{L}^{+}
	\end{pmatrix}\nonumber \\
	\nonumber\\
	&+&\begin{pmatrix}
	\partial^{\mu}\delta_{R}^{++} \\ \partial^{\mu}\delta_{L}^{++}
	\end{pmatrix}^{\dagger}
	\begin{pmatrix}
	1+\frac{A_{++}^{11}}{\Lambda^{2}} & \frac{A_{++}^{12}}{\Lambda^{2}}  \\
	\frac{A_{++}^{12}}{\Lambda^{2}} & 1+\frac{A_{++}^{22}}{\Lambda^{2}} \\
	\end{pmatrix}
	\begin{pmatrix}
	\partial_{\mu}\delta_{R}^{++} \\ \partial_{\mu}\delta_{L}^{++}
	\end{pmatrix}.\nonumber 
	\end{eqnarray}}
The components of the scalar mass matrices are given as:
{\begin{eqnarray*}
		A_{r}^{11}& =&\frac{1}{2 \kappa _+^2}\Big[2 \kappa _2^2 \mathcal{C}_{\phi D}^{(2){rL}(1)} v_L v_R+v_L^2 \left(\kappa _+^2 \mathcal{C}_{\phi D}^{\{{Ll}\}\{(2)(1)\}}+\kappa _1^2 \mathcal{C}_{\phi D}^{{Ll}(2)(1)}\right)+v_R^2 \left(\kappa _+^2 \mathcal{C}_{\phi D}^{\{{Rr}\}\{(2)(1)\}}+\kappa _1^2 \mathcal{C}_{\phi D}^{{Rr}(2)(1)}\right)\nonumber \\
		&+&2 \kappa _1^4 \mathcal{C}_{\phi D}^{1(1)1(1)}+2 \kappa _2^4 \mathcal{C}_{\phi D}^{1(1)1(1)}+4 \kappa _1^4 \mathcal{C}_{\phi D}^{(11)(22)}+4 \kappa _2^4 \mathcal{C}_{\phi D}^{(11)(22)}+2 \kappa _1^4 \mathcal{C}_{\phi D}^{2(1)1(1)}+2 \kappa _2^4 \mathcal{C}_{\phi D}^{2(1)1(1)}\nonumber \\
		&+&\kappa _1^4 \mathcal{C}_{\phi D}^{\{(2)(1)\}\{21\}}+2 \kappa _2^2 \kappa _1^2 \mathcal{C}_{\phi D}^{\{(2)(1)\}\{21\}}+\kappa _2^4 \mathcal{C}_{\phi D}^{\{(2)(1)\}\{21\}}+\kappa _1^4 \mathcal{C}_{\phi D}^{(2)(1)21}+\kappa _2^4 \mathcal{C}_{\phi D}^{(2)(1)21}\Big],\\
		A_{r}^{12}& =&-\frac{\kappa _1 \kappa _2 }{2 \kappa _+^2}\Big[-2 \mathcal{C}_{\phi D}^{(2){rL}(1)} v_L v_R+\mathcal{C}_{\phi D}^{{Ll}(2)(1)} v_L^2+\mathcal{C}_{\phi D}^{{Rr}(2)(1)} v_R^2+\kappa _-^2 (2 \mathcal{C}_{\phi D}^{1(1)1(1)}+4 \mathcal{C}_{\phi D}^{(11)(22)}+2 \mathcal{C}_{\phi D}^{2(1)1(1)}\nonumber\\
		&+&\mathcal{C}_{\phi D}^{(2)(1)21})\Big],\\
		A_{r}^{13}& =&\frac{\kappa _1^2 \mathcal{C}_{\phi D}^{R(r)1(1)} v_R}{2 \kappa _+},\\
		A_{r}^{14}& =&\frac{\kappa _1^2 \mathcal{C}_{\phi D}^{L(l)1(1)} v_L}{2 \kappa _+},\\
				A_{r}^{22}& =&\frac{1}{2 \kappa _+^2}\Big[2 \kappa _1^2 \mathcal{C}_{\phi D}^{(2){rL}(1)} v_L v_R+v_L^2 \left(\kappa _+^2 \mathcal{C}_{\phi D}^{\{{Ll}\}\{(2)(1)\}}+\kappa _2^2 \mathcal{C}_{\phi D}^{{Ll}(2)(1)}\right)+v_R^2 \left(\kappa _+^2 \mathcal{C}_{\phi D}^{\{{Rr}\}\{(2)(1)\}}+\kappa _2^2 \mathcal{C}_{\phi D}^{{Rr}(2)(1)}\right)\nonumber \\
		&+&4 \kappa _2^2 \kappa _1^2 \mathcal{C}_{\phi D}^{1(1)1(1)}+8 \kappa _2^2 \kappa _1^2 \mathcal{C}_{\phi D}^{(11)(22)}+4 \kappa _2^2 \kappa _1^2 \mathcal{C}_{\phi D}^{2(1)1(1)}+\kappa _1^4 \mathcal{C}_{\phi D}^{\{(2)(1)\}\{21\}}+2 \kappa _2^2 \kappa _1^2 \mathcal{C}_{\phi D}^{\{(2)(1)\}\{21\}}\nonumber \\
		&+&\kappa _2^4 \mathcal{C}_{\phi D}^{\{(2)(1)\}\{21\}}+2 \kappa _2^2 \kappa _1^2 \mathcal{C}_{\phi D}^{(2)(1)21}\Big],\\
				A_{r}^{23}& =&-\frac{\kappa _1 \kappa _2 \mathcal{C}_{\phi D}^{R(r)1(1)} v_R}{2 \kappa _+},
		\end{eqnarray*}
		
		{\small{\begin{eqnarray*}
		A_{r}^{24}& =&-\frac{\kappa _1 \kappa _2 \mathcal{C}_{\phi D}^{L(l)1(1)} v_L}{2 \kappa _+},\\
		A_{r}^{33}& =&\frac{1}{2} \Big[\mathcal{C}_{\phi D}^{(R){lL}(r)} v_L^2+ v_R^2 (\mathcal{C}_{\phi D}^{\{{Rr}\}\{{(R)(r)}\}}-4\mathcal{C}_{\square }^{{Rr(Rr)}})+\kappa _1^2 \mathcal{C}_{\phi D}^{\{21\}\{(R)(r)\}}+\kappa _2^2 \mathcal{C}_{\phi D}^{\{21\}\{(R)(r)\}}+\kappa _1^2 \mathcal{C}_{\phi D}^{21(R)(r)}\nonumber \\
		&+&2 \kappa _2^2 \mathcal{C}_{\phi D}^{(R)11(r)}\Big],\\
		A_{r}^{34}& =&\frac{1}{2} \Big[\mathcal{C}_{\phi D}^{(L){rL}(r)} v_L v_R+\kappa _2^2 (\mathcal{C}_{\phi D}^{(L)11(r)}+\mathcal{C}_{\phi D}^{(L)12(r)}+\mathcal{C}_{\phi D}^{(R)11(l)})\Big],\\
		A_{r}^{44}& =&\frac{1}{2} \Big[v_L^2 (\mathcal{C}_{\phi D}^{\{{Ll}\}\{{(L)(l)}\}}-4\mathcal{C}_{\square }^{{Ll(Ll)}})+\mathcal{C}_{\phi D}^{(L){rR}(l)} v_R^2+\kappa _1^2 \mathcal{C}_{\phi D}^{\{21\}\{(L)(l)\}}+\kappa _2^2 \mathcal{C}_{\phi D}^{\{21\}\{(L)(l)\}}+\kappa _1^2 \mathcal{C}_{\phi D}^{21(L)(l)}\nonumber ]\\
		&+&2 \kappa _2^2 \mathcal{C}_{\phi D}^{(L)11(l)}\Big],\\
		\\ \\
			A_{i}^{11}& =&\frac{1}{2 \kappa _+^2}\Big[2 \kappa _2^2 \mathcal{C}_{\phi D}^{(2){rL}(1)} v_L v_R+v_L^2 \left(\kappa _+^2 \mathcal{C}_{\phi D}^{\{{Ll}\}\{(2)(1)\}}+\kappa _1^2 \mathcal{C}_{\phi D}^{{Ll}(2)(1)}\right)+v_R^2 \left(\kappa _+^2 \mathcal{C}_{\phi D}^{\{{Rr}\}\{(2)(1)\}}+\kappa _1^2 \mathcal{C}_{\phi D}^{{Rr}(2)(1)}\right)\nonumber \\
		&+&\kappa _1^2 \mathcal{C}_{\phi D}^{{Rr}(2)(1)}-2 \kappa _1^4 \mathcal{C}_{\phi D}^{1(1)1(1)}-2 \kappa _2^4 \mathcal{C}_{\phi D}^{1(1)1(1)}+4 \kappa _1^4 \mathcal{C}_{\phi D}^{(11)(22)}+4 \kappa _2^4 \mathcal{C}_{\phi D}^{(11)(22)}-2 \kappa _1^4 \mathcal{C}_{\phi D}^{2(1)1(1)}-2 \kappa _2^4 \mathcal{C}_{\phi D}^{2(1)1(1)}\nonumber \\
		&+&\kappa _1^4 \mathcal{C}_{\phi D}^{\{(2)(1)\}\{21\}}+2 \kappa _2^2 \kappa _1^2 \mathcal{C}_{\phi D}^{\{(2)(1)\}\{21\}}+\kappa _2^4 \mathcal{C}_{\phi D}^{\{(2)(1)\}\{21\}}+\kappa _1^4 \mathcal{C}_{\phi D}^{(2)(1)21}+\kappa _2^4 \mathcal{C}_{\phi D}^{(2)(1)21}+4 \kappa _1^4 \mathcal{C}_{\phi D}^{2(1)2(1)}\nonumber \\
		&+&4 \kappa _2^4 \mathcal{C}_{\phi D}^{2(1)2(1)}\Big],\\
		A_{i}^{12}& =&\frac{\kappa _1 \kappa _2}{2 \kappa _+^2} \Big[2 \mathcal{C}_{\phi D}^{(2){rL}(1)} v_L v_R-\mathcal{C}_{\phi D}^{{Ll}(2)(1)} v_L^2-\mathcal{C}_{\phi D}^{{Rr}(2)(1)} v_R^2\nonumber \\
		&+&\kappa _-^2 (2 \mathcal{C}_{\phi D}^{1(1)1(1)}-4 \mathcal{C}_{\phi D}^{(11)(22)}+2 \mathcal{C}_{\phi D}^{2(1)1(1)}-\mathcal{C}_{\phi D}^{(2)(1)21}-4 \mathcal{C}_{\phi D}^{2(1)2(1)})\Big],\\
		A_{i}^{13}& =&-\frac{\kappa _1^2 \mathcal{C}_{\phi D}^{R(r)1(1)} v_R}{2 \kappa _+},\\
		A_{i}^{14}& =&-\frac{\kappa _1^2 \mathcal{C}_{\phi D}^{L(l)1(1)} v_L}{2 \kappa _+},\\
		A_{i}^{22}& =&\frac{1}{2 \kappa _+^2}\Big[2 \kappa _1^2 \mathcal{C}_{\phi D}^{(2){rL}(1)} v_L v_R+v_L^2 \left(\kappa _+^2 \mathcal{C}_{\phi D}^{\{{Ll}\}\{(2)(1)\}}+\kappa _2^2 \mathcal{C}_{\phi D}^{{Ll}(2)(1)}\right)+v_R^2 \left(\kappa _+^2 \mathcal{C}_{\phi D}^{\{{Rr}\}\{(2)(1)\}}+\kappa _2^2 \mathcal{C}_{\phi D}^{{Rr}(2)(1)}\right)\nonumber \\
		&-&4 \kappa _2^2 \kappa _1^2 \mathcal{C}_{\phi D}^{1(1)1(1)}+8 \kappa _2^2 \kappa _1^2 \mathcal{C}_{\phi D}^{(11)(22)}-4 \kappa _2^2 \kappa _1^2 \mathcal{C}_{\phi D}^{2(1)1(1)}+\kappa _1^4 \mathcal{C}_{\phi D}^{\{(2)(1)\}\{21\}}+2 \kappa _2^2 \kappa _1^2 \mathcal{C}_{\phi D}^{\{(2)(1)\}\{21\}}\nonumber \\
		&+&\kappa _2^4 \mathcal{C}_{\phi D}^{\{(2)(1)\}\{21\}}+2 \kappa _2^2 \kappa _1^2 \mathcal{C}_{\phi D}^{(2)(1)21}+8 \kappa _2^2 \kappa _1^2 \mathcal{C}_{\phi D}^{2(1)2(1)}\Big],\\
		A_{i}^{23}&=& \frac{\kappa _1 \kappa _2 \mathcal{C}_{\phi D}^{R(r)1(1)} v_R}{2 \kappa _+},\\
		A_{i}^{24}& =&\frac{\kappa _1 \kappa _2 \mathcal{C}_{\phi D}^{L(l)1(1)} v_L}{2 \kappa _+},\\
		A_{i}^{33}& =&\frac{1}{2} \Big[\mathcal{C}_{\phi D}^{(R){lL}(r)} v_L^2+ v_R^2 (\mathcal{C}_{\phi D}^{\{{Rr}\}\{{(R)(r)}\}}+4\mathcal{C}_{\phi D}^{R(r)R(r)})+\kappa _1^2 \mathcal{C}_{\phi D}^{\{21\}\{(R)(r)\}}+\kappa _2^2 \mathcal{C}_{\phi D}^{\{21\}\{(R)(r)\}}+\kappa _1^2 \mathcal{C}_{\phi D}^{21(R)(r)}\nonumber \\&+&2 \kappa _2^2 \mathcal{C}_{\phi D}^{(R)11(r)}\Big],\\
						A_{i}^{34}& =&\frac{1}{2} \Big[\mathcal{C}_{\phi D}^{(L){rL}(r)} v_L v_R+\kappa _2^2 (\mathcal{C}_{\phi D}^{(L)11(r)}+\mathcal{C}_{\phi D}^{(L)12(r)}+\mathcal{C}_{\phi D}^{(R)11(l)})\Big],\\
		       A_{i}^{44}& =&\frac{1}{2} \Big[ v_L^2 (\mathcal{C}_{\phi D}^{\{{Ll}\}\{{(L)(l)}\}}+4 \mathcal{C}_{\phi D}^{L(l)L(l)})+\mathcal{C}_{\phi D}^{(L){rR}(l)} v_R^2+\kappa _1^2 \mathcal{C}_{\phi D}^{\{21\}\{(L)(l)\}}+\kappa _2^2 \mathcal{C}_{\phi D}^{\{21\}\{(L)(l)\}}+\kappa _1^2 \mathcal{C}_{\phi D}^{21(L)(l)}\nonumber \\&+&2 \kappa _2^2 \mathcal{C}_{\phi D}^{(L)11(l)}\Big],
			\end{eqnarray*}}
		
		{\small{\begin{eqnarray*}
				A_{+}^{11}& =&\frac{1}{2 \kappa _+^2}\Big[2 \kappa _2^2 \mathcal{C}_{\phi D}^{(2){rL}(1)} v_L v_R+v_L^2 \left(\kappa _+^2 \mathcal{C}_{\phi D}^{\{{Ll}\}\{(2)(1)\}}+\kappa _2^2 \mathcal{C}_{\phi D}^{{Ll}(2)(1)}\right)+v_R^2 \left(\kappa _+^2 \mathcal{C}_{\phi D}^{\{{Rr}\}\{(2)(1)\}}+\kappa _2^2 \mathcal{C}_{\phi D}^{{Rr}(2)(1)}\right)\nonumber \\
				&+&4 \kappa _2^2 \kappa _1^2 \mathcal{C}_{\phi D}^{1(1)1(1)}+\kappa _1^4 \mathcal{C}_{\phi D}^{(11)(22)}+2 \kappa _2 \kappa _1^3 \mathcal{C}_{\phi D}^{(11)(22)}+2 \kappa _2^2 \kappa _1^2 \mathcal{C}_{\phi D}^{(11)(22)}+2 \kappa _2^3 \kappa _1 \mathcal{C}_{\phi D}^{(11)(22)}+\kappa _2^4 \mathcal{C}_{\phi D}^{(11)(22)}\nonumber \\
				&+&4 \kappa _2^2 \kappa _1^2 \mathcal{C}_{\phi D}^{2(1)1(1)}+\kappa _1^4 \mathcal{C}_{\phi D}^{\{(2)(1)\}\{21\}}+2 \kappa _2^2 \kappa _1^2 \mathcal{C}_{\phi D}^{\{(2)(1)\}\{21\}}+\kappa _2^4 \mathcal{C}_{\phi D}^{\{(2)(1)\}\{21\}}+2 \kappa _2^2 \kappa _1^2 \mathcal{C}_{\phi D}^{(2)(1)21}\nonumber \\
				&+&2 \kappa _1^4 \mathcal{C}_{\phi D}^{2(1)2(1)}-4 \kappa _2^2 \kappa _1^2 \mathcal{C}_{\phi D}^{2(1)2(1)}+2 \kappa _2^4 \mathcal{C}_{\phi D}^{2(1)2(1)}\Big],\\
				A_{+}^{12}& =&\frac{\kappa _1 \kappa _2 }{2 \kappa _+^2}\Big[2 \mathcal{C}_{\phi D}^{(2){rL}(1)} v_L v_R+\mathcal{C}_{\phi D}^{{Ll}(2)(1)} v_L^2+\mathcal{C}_{\phi D}^{{Rr}(2)(1)} v_R^2+\kappa _-^2 (2 \mathcal{C}_{\phi D}^{1(1)1(1)}+2 \mathcal{C}_{\phi D}^{2(1)1(1)}+\mathcal{C}_{\phi D}^{(2)(1)21}\nonumber \\&-&4 \mathcal{C}_{\phi D}^{2(1)2(1)})\Big],\\
				A_{+}^{13}& =&-\frac{\kappa _2^2 \mathcal{C}_{\phi D}^{R(r)1(1)} v_R}{2 \sqrt{2} \kappa _+},\\
				A_{+}^{14}& =&-\frac{\kappa _2^2 \mathcal{C}_{\phi D}^{L(l)1(1)} v_L}{2 \sqrt{2} \kappa _+},\\
				A_{+}^{22}& =&\frac{1}{2 \kappa _+^2}\Big[2 \kappa _1^2 \mathcal{C}_{\phi D}^{(2){rL}(1)} v_L v_R+v_L^2 \left(\kappa _+^2 \mathcal{C}_{\phi D}^{\{{Ll}\}\{(2)(1)\}}+\kappa _1^2 \mathcal{C}_{\phi D}^{{Ll}(2)(1)}\right)+v_R^2 \left(\kappa _+^2 \mathcal{C}_{\phi D}^{\{{Rr}\}\{(2)(1)\}}+\kappa _1^2 \mathcal{C}_{\phi D}^{{Rr}(2)(1)}\right)\nonumber \\
				&-&4 \kappa _2^2 \kappa _1^2 \mathcal{C}_{\phi D}^{1(1)1(1)}+\kappa _1^4 \mathcal{C}_{\phi D}^{(11)(22)}+2 \kappa _2 \kappa _1^3 \mathcal{C}_{\phi D}^{(11)(22)}+2 \kappa _2^2 \kappa _1^2 \mathcal{C}_{\phi D}^{(11)(22)}+2 \kappa _2^3 \kappa _1 \mathcal{C}_{\phi D}^{(11)(22)}+\kappa _2^4 \mathcal{C}_{\phi D}^{(11)(22)}\nonumber \\
				&-&4 \kappa _2^2 \kappa _1^2 \mathcal{C}_{\phi D}^{2(1)1(1)}+\kappa _1^4 \mathcal{C}_{\phi D}^{\{(2)(1)\}\{21\}}+2 \kappa _2^2 \kappa _1^2 \mathcal{C}_{\phi D}^{\{(2)(1)\}\{21\}}+\kappa _2^4 \mathcal{C}_{\phi D}^{\{(2)(1)\}\{21\}}+\kappa _1^4 \mathcal{C}_{\phi D}^{(2)(1)21}+\kappa _2^4 \mathcal{C}_{\phi D}^{(2)(1)21}\nonumber \\
				&+&8 \kappa _2^2 \kappa _1^2 \mathcal{C}_{\phi D}^{2(1)2(1)}\Big],\\
				A_{+}^{23}& =&-\frac{\kappa _1 \kappa _2 \mathcal{C}_{\phi D}^{R(r)1(1)} v_R}{2 \sqrt{2} \kappa _+},\\
				A_{+}^{24}& =&-\frac{\kappa _1 \kappa _2 \mathcal{C}_{\phi D}^{L(l)1(1)} v_L}{2 \sqrt{2} \kappa _+},\\
				A_{+}^{33}& =&\frac{1}{4} \Big[\mathcal{C}_{\phi D}^{(R){lL}(r)} v_L^2+ v_R^2 (2\mathcal{C}_{\phi D}^{\{{Rr}\}\{{(R)(r)}\}}+2\mathcal{C}_{\phi D}^{R(r)R(r)}-2\mathcal{C}_{\square }^{Rr(Rr)})+\kappa _+^2 (2 \mathcal{C}_{\phi D}^{\{21\}\{(R)(r)\}}+\mathcal{C}_{\phi D}^{21(R)(r)}\nonumber \\&+&2 \mathcal{C}_{\phi D}^{(R)11(r)})\Big],\\
				A_{+}^{34}& =&\frac{1}{4} \Big[\mathcal{C}_{\phi D}^{(L){rL}(r)} v_L v_R+\kappa _+^2 (\mathcal{C}_{\phi D}^{(L)11(r)}+\mathcal{C}_{\phi D}^{(L)12(r)}+\mathcal{C}_{\phi D}^{(R)11(l)})\Big],\\
				A_{+}^{44}& =&\frac{1}{4} \Big[ v_L^2 (2\mathcal{C}_{\phi D}^{\{{Ll}\}\{{(L)(l)}\}}+2\mathcal{C}_{\phi D}^{L(l)L(l)}-2\mathcal{C}_{\square }^{Ll(Ll)})+\mathcal{C}_{\phi D}^{(L){rR}(l)} v_R^2+\kappa _+^2 (2 \mathcal{C}_{\phi D}^{\{21\}\{(L)(l)\}}+\mathcal{C}_{\phi D}^{21(L)(l)}\nonumber \\&+&2 \mathcal{C}_{\phi D}^{(L)11(l)})\Big],
				\end{eqnarray*}}
		
			{\small\begin{eqnarray*}
				A_{++}^{11}& =&\frac{1}{2} \Big[\kappa _+^2 \mathcal{C}_{\phi D}^{\{21\}\{(R)(r)\}}+\kappa _2^2 \mathcal{C}_{\phi D}^{21(R)(r)}+2 \kappa _1^2 \mathcal{C}_{\phi D}^{(R)11(r)}+v_{R}^2 \mathcal{C}_{\phi D}^{\{{Rr}\}\{{(R)(r)}\}}\Big],\\
				A_{++}^{12}&=& \frac{1}{2} \kappa _1^2 (\mathcal{C}_{\phi D}^{(L)11(r)}+\mathcal{C}_{\phi D}^{(L)12(r)\}}+\mathcal{C}_{\phi D}^{(R)11(l)}),\\
				A_{++}^{22}& =&\frac{1}{2} \Big[\kappa _+^2 \mathcal{C}_{\phi D}^{\{21\}\{(L)(l)\}}+\kappa _2^2 \mathcal{C}_{\phi D}^{21(L)(l)}+2 \kappa _1^2 \mathcal{C}_{\phi D}^{(L)11(l)}+v_{L}^2 \mathcal{C}_{\phi D}^{\{{Ll}\}\{{(L)(l)}\}}\Big].
				\end{eqnarray*}}

			In order to reduce the scalar kinetic terms in their canonical forms,  we have redefined the scalar fields as:
			{\small\begin{eqnarray}\label{eq:MLRSM-scalar-field-CP-even-redef-full}
				\phi_{1}^{0r}&\rightarrow& \left(1-\frac{A_{r}^{11}}{2\Lambda^{2}}\right)\phi_{1}^{0r}-\left(\frac{A_{r}^{12}}{2\Lambda^{2}}\right)\phi_{2}^{0r}-\left(\frac{A_{r}^{13}}{2\Lambda^{2}}\right)\delta_{R}^{0r}-\left(\frac{A_{r}^{14}}{2\Lambda^{2}}\right)\delta_{L}^{0r}, \nonumber \\
				\phi_{2}^{0r}&\rightarrow& \left(1-\frac{A_{r}^{22}}{2\Lambda^{2}}\right)\phi_{2}^{0r}-\left(\frac{A_{r}^{12}}{2\Lambda^{2}}\right)\phi_{1}^{0r}-\left(\frac{A_{r}^{23}}{2\Lambda^{2}}\right)\delta_{R}^{0r}-\left(\frac{A_{r}^{24}}{2\Lambda^{2}}\right)\delta_{L}^{0r},\\
				\delta_{R}^{0r}&\rightarrow& \left(1-\frac{A_{r}^{33}}{2\Lambda^{2}}\right)\delta_{R}^{0r}-\left(\frac{A_{r}^{13}}{2\Lambda^{2}}\right)\phi_{1}^{0r}-\left(\frac{A_{r}^{23}}{2\Lambda^{2}}\right)\phi_{2}^{0r}-\left(\frac{A_{r}^{34}}{2\Lambda^{2}}\right)\delta_{L}^{0r}, \nonumber \\
				\delta_{L}^{0r}&\rightarrow& \left(1-\frac{A_{r}^{44}}{2\Lambda^{2}}\right)\delta_{L}^{0r}-\left(\frac{A_{r}^{14}}{2\Lambda^{2}}\right)\phi_{1}^{0r}-\left(\frac{A_{r}^{24}}{2\Lambda^{2}}\right)\phi_{2}^{0r}-\left(\frac{A_{r}^{34}}{2\Lambda^{2}}\right)\delta_{R}^{0r}, \nonumber \\
				\nonumber \\ 	\nonumber \\
				\phi_{1}^{0i}&\rightarrow& \left(1-\frac{A_{i}^{11}}{2\Lambda^{2}}\right)\phi_{1}^{0i}-\left(\frac{A_{i}^{12}}{2\Lambda^{2}}\right)\phi_{2}^{0i}-\left(\frac{A_{i}^{13}}{2\Lambda^{2}}\right)\delta_{R}^{0i}-\left(\frac{A_{i}^{14}}{2\Lambda^{2}}\right)\delta_{L}^{0i}, \nonumber \\
				\phi_{2}^{0i}&\rightarrow& \left(1-\frac{A_{i}^{22}}{2\Lambda^{2}}\right)\phi_{2}^{0i}-\left(\frac{A_{i}^{12}}{2\Lambda^{2}}\right)\phi_{1}^{0i}-\left(\frac{A_{i}^{23}}{2\Lambda^{2}}\right)\delta_{R}^{0i}-\left(\frac{A_{i}^{24}}{2\Lambda^{2}}\right)\delta_{L}^{0i},\\
				\delta_{R}^{0i}&\rightarrow& \left(1-\frac{A_{i}^{33}}{2\Lambda^{2}}\right)\delta_{R}^{0i}-\left(\frac{A_{i}^{13}}{2\Lambda^{2}}\right)\phi_{1}^{0i}-\left(\frac{A_{i}^{23}}{2\Lambda^{2}}\right)\phi_{2}^{0i}-\left(\frac{A_{i}^{34}}{2\Lambda^{2}}\right)\delta_{L}^{0i}, \nonumber \\
				\delta_{L}^{0i}&\rightarrow& \left(1-\frac{A_{i}^{44}}{2\Lambda^{2}}\right)\delta_{L}^{0i}-\left(\frac{A_{i}^{14}}{2\Lambda^{2}}\right)\phi_{1}^{0i}-\left(\frac{A_{i}^{24}}{2\Lambda^{2}}\right)\phi_{2}^{0i}-\left(\frac{A_{i}^{34}}{2\Lambda^{2}}\right)\delta_{R}^{0i}, \nonumber \\
			\nonumber \\ 	\nonumber \\
				\phi_{1}'^{+}&\rightarrow& \left(1-\frac{A_{+}^{11}}{2\Lambda^{2}}\right)\phi_{1}'^{+}-\left(\frac{A_{+}^{12}}{2\Lambda^{2}}\right)\phi_{2}'^{+}-\left(\frac{A_{+}^{13}}{2\Lambda^{2}}\right)\delta_{R}^{+}-\left(\frac{A_{+}^{14}}{2\Lambda^{2}}\right)\delta_{L}^{+}, \nonumber \\
				\phi_{2}'^{+}&\rightarrow& \left(1-\frac{A_{+}^{22}}{2\Lambda^{2}}\right)\phi_{2}'^{+}-\left(\frac{A_{+}^{12}}{2\Lambda^{2}}\right)\phi_{1}'^{+}-\left(\frac{A_{i}^{23}}{2\Lambda^{2}}\right)\delta_{R}^{+}-\left(\frac{A_{i}^{24}}{2\Lambda^{2}}\right)\delta_{L}^{+}, \\
				\delta_{R}^{+}&\rightarrow& \left(1-\frac{A_{+}^{33}}{2\Lambda^{2}}\right)\delta_{R}^{+}-\left(\frac{A_{+}^{13}}{2\Lambda^{2}}\right)\phi_{1}'^{+}-\left(\frac{A_{+}^{23}}{2\Lambda^{2}}\right)\phi_{2}'^{+}-\left(\frac{A_{+}^{34}}{2\Lambda^{2}}\right)\delta_{L}^{+}, \nonumber \\
				\delta_{L}^{+}&\rightarrow& \left(1-\frac{A_{+}^{44}}{2\Lambda^{2}}\right)\delta_{L}^{+}-\left(\frac{A_{+}^{14}}{2\Lambda^{2}}\right)\phi_{1}'^{+}-\left(\frac{A_{+}^{24}}{2\Lambda^{2}}\right)\phi_{2}'^{+}-\left(\frac{A_{+}^{34}}{2\Lambda^{2}}\right)\delta_{R}^{+},\nonumber \\
			    \nonumber \\ \nonumber \\
				\delta_{R}^{++}&\rightarrow& \left(1-\frac{A_{++}^{11}}{2\Lambda^{2}}\right)\delta_{R}^{++}-\left(\frac{A_{++}^{12}}{2\Lambda^{2}}\right)\delta_{L}^{++}, \nonumber \\
				\delta_{L}^{++}&\rightarrow& \left(1-\frac{A_{++}^{22}}{2\Lambda^{2}}\right)\delta_{L}^{++}-\left(\frac{A_{++}^{12}}{2\Lambda^{2}}\right)\delta_{R}^{++}.
				\end{eqnarray}}


These redefined scalar fields modify the scalar mass matrices. There will be further contributions to the scalar masses from the $\phi^6$ operators. Encapsulating all these contributions, the total scalar mass matrices can be expressed as:
{\small\begin{align}\label{eq:MLRSM-scalar-mass-matrix-full}
	\mathcal{L}_{M_{H}}^{(4)+(6)} =& \frac{1}{2} \begin{bmatrix}
	\phi^{0r}_{1} \\ \phi^{0r}_{2}  \\ \delta^{0r}_{R} \\ \delta^{0r}_{L}
	\end{bmatrix}^{T} \mathcal{M}_{r}^{2} \begin{bmatrix}
	\phi^{0r}_{1} \\ \phi^{0r}_{2}  \\ \delta^{0r}_{R} \\ \delta^{0r}_{L}
	\end{bmatrix} + \frac{1}{2} \begin{bmatrix}
	\phi^{0i}_{1} \\ \phi^{0i}_{2}  \\ \delta^{0i}_{R} \\ \delta^{0i}_{L}
	\end{bmatrix}^{T}\mathcal{M}_{i}^{2} \begin{bmatrix}
	\phi^{0i}_{1} \\ \phi^{0i}_{2}  \\ \delta^{0i}_{R} \\ \delta^{0i}_{L}
	\end{bmatrix} + \begin{bmatrix}
	\phi^{'+}_{1} \\ \phi^{'+}_{2}  \\ \delta^{+}_{R} \\ \delta^{+}_{L} 
	\end{bmatrix}^{\dagger}\mathcal{M}_{+}^{2} \begin{bmatrix}
	\phi^{'+}_{1} \\ \phi^{'+}_{2}  \\ \delta^{+}_{R} \\ \delta^{+}_{L} 
	\end{bmatrix}
	+ \begin{bmatrix}
	\delta^{++}_{R} \\ \delta^{++}_{L} 
	\end{bmatrix}^{\dagger}\mathcal{M}_{++}^{2} \begin{bmatrix}
	\delta^{++}_{R} \\ \delta^{++}_{L} 
	\end{bmatrix}. \nonumber
	\end{align}}
Here, $\,\mathcal{M}_{r}^{2},\mathcal{M}_{i}^{2}$ and $\mathcal{M}_{+}^{2}$ are $(4\times4)$ matrices while $\mathcal{M}_{++}^{2}$ is $(2\times2)$ matrix. Elements of these mass matrices can be written as :
{\small\begin{align}
	(\mathcal{M}_{j})_{mn}^{2}=\left(m^2_{j}+ (\Delta m_{j,\phi D}^2)+(\Delta m_{j,\phi^6}^2)\right)_{mn} ; \ \ \ \ \ \ {j\equiv\{r,i,+,++\}}.
	\end{align}}
Here, we have provided the individual matrix elements :
\\ \\
\underline{\large{CP-even scalar mass matrix}} 
{\small\begin{eqnarray}
	(\Delta m_{r,\phi D}^2)_{11}&=&\frac{1}{\kappa _+^3 \Lambda ^2}\Big[-\kappa _1^2 \big(\kappa _2^2 \big(2 \alpha _1 A_r^{14} v_L+\alpha _3 A_r^{14} v_L+2 \alpha _1 A_r^{13} v_R+\alpha _3 A_r^{13} v_R+4 \kappa _+ \lambda _1  A_r^{11}+16 \kappa _+ \lambda _2  A_r^{11}\big)\nonumber \\
	&+&8 \kappa _+ \lambda _3  A_r^{11}+\big(2 \rho _1-\rho _3\big) v_L v_R \big(A_r^{13} v_L+A_r^{14} v_R\big)\big)-\kappa _2^2 \big(\kappa _2^2 \big(\big(\alpha _1+\alpha _3\big) \big(A_r^{14} v_L+A_r^{13} v_R\big)\nonumber \\
	&-&2 \kappa _+ \lambda _4 A_r^{12}+2 \kappa _+ \lambda _1  A_r^{11}\big)+\big(2 \rho _1-\rho _3\big) v_L v_R \big(A_r^{13} v_L+A_r^{14} v_R\big)\big)+\kappa _1^4 \big(-\big(\alpha _1 \big(A_r^{14} v_L+A_r^{13} v_R\big)\nonumber \\
	&+&2 \kappa _+ \lambda _4 A_r^{12}+2 \kappa _+ \lambda _1  A_r^{11}\big)\big)-4 \kappa _2 \kappa _1^3 \big(\alpha _2 A_r^{14} v_L+2 \kappa _+ \lambda _2 A_r^{12}+\kappa _+ \lambda _3 A_r^{12}+\alpha _2 A_r^{13} v_R+2 \kappa _+ \lambda _4  A_r^{11}\big)\nonumber \\ 
	&-&4 \kappa _2^3 \kappa _1 \big(\alpha _2 A_r^{14} v_L-2 \kappa _+ \lambda _2 A_r^{12}-\kappa _+ \lambda _3 A_r^{12}+\alpha _2 A_r^{13} v_R+2 \kappa _+ \lambda _4  A_r^{11}\big)\Big],\\ \nonumber \nonumber \\
	(\Delta m_{r,\phi D}^2)_{12}&=&\frac{1}{4 \Lambda ^2 \kappa _-^2 \kappa _+^3 \kappa _1^2}\Big[-2 \kappa _+ v_L v_R \kappa _2^4 \big(\beta _3 \kappa _2^2+v_L v_R \big(\rho _3-2 \rho _1\big)\big) A_r^{12}-2 \kappa _1^7 \kappa _2 \big(4 v_L \alpha _2 A_r^{24}+4 v_R \alpha _2 A_r^{23}\nonumber \\
	&+&v_L \alpha _3 A_r^{14}+v_R \beta _3 A_r^{14}+v_R \alpha _3 A_r^{13}+v_L \beta _3 A_r^{13}+8 \kappa _+ \lambda _4 A_r^{12}+8 \kappa _+ \big(A_r^{22}+ A_r^{11}\big) \lambda _2\nonumber \\
	&+&4 \kappa _+ \big(A_r^{22}+ A_r^{11}\big) \lambda _3\big)-\kappa _1^8 \big(2 v_L \alpha _1 A_r^{24}+2 v_R \alpha _1 A_r^{23}+4 v_L \alpha _2 A_r^{14}+v_R \beta _1 A_r^{14}+4 v_R \alpha _2 A_r^{13}\nonumber \\
	&+&v_L \beta _1 A_r^{13}+4 \kappa _+ \lambda _1 A_r^{12}+8 \kappa _+ \lambda _2 A_r^{12}+4 \kappa _+ \lambda _3 A_r^{12}+4 \kappa _+ \big(A_r^{22}+ A_r^{11}\big) \lambda _4\big)+\kappa _1^3 \big(2 \kappa _2^5 \big(4 v_L \alpha _2 A_r^{24}\nonumber \\
	&+&4 v_R \alpha _2 A_r^{23}+v_L \alpha _3 A_r^{14}+v_R \beta _3 A_r^{14}+v_R \alpha _3 A_r^{13}+v_L \beta _3 A_r^{13}+8 \kappa _+ \lambda _4 A_r^{12}-8 \kappa _+ \big(A_r^{22}+ A_r^{11}\big) \lambda _2\nonumber \\
	&-&4 \kappa _+ \big(A_r^{22}+ A_r^{11}\big) \lambda _3\big)-4 \kappa _+ A_r^{12} v_L v_R \beta _1 \kappa _2^3\big)+\kappa _1^6 \big(-\kappa _+ v_R^2 \alpha _3 A_r^{12}+\kappa _2^2 \big(-2 v_L \alpha _1 A_r^{24}-2 v_L \alpha _3 A_r^{24}\nonumber \\
	&-&2 v_R \alpha _1 A_r^{23}-2 v_R \alpha _3 A_r^{23}+4 v_L \alpha _2 A_r^{14}-v_R \beta _1 A_r^{14}+4 v_R \alpha _2 A_r^{13}-v_L \beta _1 A_r^{13}-4 \kappa _+ \lambda _1 A_r^{12}-8 \kappa _+ \lambda _2 A_r^{12}\nonumber \\
	&-&4 \kappa _+ \lambda _3 A_r^{12}+4 \kappa _+ \big(A_r^{22}+ A_r^{11}\big) \lambda _4\big)-v_L^2 \big(2 v_R \big(2 \rho _1-\rho _3\big) A_r^{23}+\kappa _+ \alpha _3 A_r^{12}\big)-2 v_L v_R \big(v_R \big(2 \rho _1-\rho _3\big) A_r^{24}\nonumber \\
	&+&\kappa _+ \beta _3 A_r^{12}\big)\big)+2 \kappa _1^5 \big(\big(-\big(v_R A_r^{14}+v_L A_r^{13}\big) \beta _3+16 \kappa _+ \big(A_r^{22}+ A_r^{11}\big) \lambda _2+8 \kappa _+ \big(A_r^{22}+ A_r^{11}\big) \lambda _3\big) \kappa _2^3\nonumber \\
	&+&v_L v_R \kappa _2 \big(\big(v_R A_r^{14}+v_L A_r^{13}\big) \big(2 \rho _1-\rho _3\big)-\kappa _+ A_r^{12} \beta _1\big) \big)+\kappa _1^4 \big(-2 \kappa _+ \big(\alpha _3 v_L^2+3 v_R \beta _3 v_L+v_R^2 \alpha _3\big) \kappa _2^2 A_r^{12}\nonumber \\
	&+&2 \kappa _+ v_L^2 v_R^2 \big(2 \rho _1-\rho _3\big) A_r^{12}+\kappa _2^4 \big(2 v_L \alpha _1 A_r^{24}+2 v_R \alpha _1 A_r^{23}+4 v_L \alpha _2 A_r^{14}+v_R \beta _1 A_r^{14}+4 v_R \alpha _2 A_r^{13}\nonumber \\
	&+&v_L \beta _1 A_r^{13}+4 \kappa _+ \lambda _1 A_r^{12}+8 \kappa _+ \lambda _2 A_r^{12}+4 \kappa _+ \lambda _3 A_r^{12}+4 \kappa _+ \big(A_r^{22}+ A_r^{11}\big) \lambda _4\big)\big)+2 \kappa _1 \kappa _2^5 \big(v_R \beta _3 \kappa _2^2 A_r^{14}\nonumber \\
	&+&v_L^2 v_R \big(\rho _3-2 \rho _1\big) A_r^{13}+v_L \big(v_R^2 \big(\rho _3-2 \rho _1\big) A_r^{14}+\beta _3 \kappa _2^2 A_r^{13}-\kappa _+ v_R \beta _1 A_r^{12}\big)\big)+\kappa _1^2 \big(4 \kappa _+ v_L^2 v_R^2 \kappa _2^2 \big(2 \rho _1-\rho _3\big) A_r^{12}\nonumber \\
	&+&\kappa _2^6 \big(2 v_L \alpha _1 A_r^{24}+2 v_L \alpha _3 A_r^{24}+2 v_R \alpha _1 A_r^{23}+2 v_R \alpha _3 A_r^{23}-4 v_L \alpha _2 A_r^{14}+v_R \beta _1 A_r^{14}-4 v_R \alpha _2 A_r^{13}+v_L \beta _1 A_r^{13}\nonumber \\
	&+&4 \kappa _+ \lambda _1 A_r^{12}+8 \kappa _+ \lambda _2 A_r^{12}+4 \kappa _+ \lambda _3 A_r^{12}-4 \kappa _+ \big(A_r^{22}+ A_r^{11}\big) \lambda _4\big)-\kappa _2^4 \big(\kappa _+ v_R^2 \alpha _3 A_r^{12}\nonumber \\
	&+&v_L^2 \big(2 v_R \big(\rho _3-2 \rho _1\big) A_r^{23}+\kappa _+ \alpha _3 A_r^{12}\big)+2 v_L v_R \big(v_R \big(\rho _3-2 \rho _1\big) A_r^{24}+3 \kappa _+ \beta _3 A_r^{12}\big)\big)\big)\Big],
	\end{eqnarray}
	\begin{eqnarray}	%
	(\Delta m_{r,\phi D}^2)_{13}&=&\frac{1}{4 \kappa _+^3 \kappa _1 \Lambda ^2}\Big[-\kappa _+^2 v_R \big(4 \kappa _1 \rho _1 A_r^{33} v_L^2-2 \kappa _1 \rho _3 A_r^{33} v_L^2+2 \kappa _+ \kappa _1 \rho _1 A_r^{14} v_L+\kappa _+ \kappa _1 \rho _3 A_r^{14} v_L-4 \kappa _2 \rho _1 A_r^{12} v_L^2 \nonumber \\
	&+&2 \kappa _2 \rho _3 A_r^{12} v_L^2+2 \alpha _3 \kappa _1 \kappa _2^2 A_r^{33}+2 \alpha _3 \kappa _1^2 \kappa _2 A_r^{12}+2 \alpha _1 \kappa _+^2 \kappa _1 \big(A_r^{33}+ A_r^{11}\big)\nonumber \\
	&+&4 \alpha _2 \kappa _1 \big(\kappa _1^2 A_r^{12}-\kappa _2^2 A_r^{12}+2 \kappa _1 \kappa _2 \big(A_r^{33}+ A_r^{11}\big)\big)+4 \kappa _1 \rho _1  A_r^{11} v_L^2-2 \kappa _1 \rho _3  A_r^{11} v_L^2+2 \alpha _3 \kappa _1 \kappa _2^2  A_r^{11}\big)\nonumber \\
	&-&2 \kappa _+^2 \kappa _1 v_R^2 \big(2 \rho _1 \big(A_r^{34} v_L+\kappa _+ A_r^{13}\big)-\rho _3 A_r^{34} v_L\big)-\kappa _+^2 v_L \big(2 \kappa _1 \big(\alpha _1 \kappa _+^2+\kappa _2 \big(4 \alpha _2 \kappa _1+\alpha _3 \kappa _2\big)\big) A_r^{34}+\beta _1 \kappa _+^2 \kappa _1 A_r^{12}\nonumber \\
	&+&2 \beta _3 \kappa _+^2 \kappa _2 A_r^{12}\big)+\kappa _+^3 \kappa _1 \big(2 \rho _1-\rho _3\big) A_r^{13} v_L^2-4 \kappa _+ \kappa _1 \big(2 \kappa _1^2 \kappa _2^2 \big(\lambda _1+4 \lambda _2+2 \lambda _3\big) A_r^{13}\nonumber \\
	&+&2 \kappa _1 \kappa _2^3 \big(2 \lambda _4 A_r^{13}-\big(2 \lambda _2+\lambda _3\big) A_r^{23}\big)+2 \kappa _1^3 \kappa _2 \big(\big(2 \lambda _2+\lambda _3\big) A_r^{23}+2 \lambda _4 A_r^{13}\big)+\kappa _2^4 \big(\lambda _1 A_r^{13}-\lambda _4 A_r^{23}\big)\nonumber \\
	&+&\kappa _1^4 \big(\lambda _4 A_r^{23}+\lambda _1 A_r^{13}\big)\big)\Big],\\
	\nonumber \\
	(\Delta m_{r,\phi D}^2)_{14}&=&\frac{1}{4 \kappa _+^3 \kappa _1 \Lambda ^2}\Big[-\kappa _+^2 v_L \big(2 \alpha _3 \kappa _1 \kappa _2^2 A_r^{44}+2 \alpha _3 \kappa _1^2 \kappa _2 A_r^{12}+4 \kappa _1 \rho _1 A_r^{44} v_R^2-2 \kappa _1 \rho _3 A_r^{44} v_R^2+2 \kappa _+ \kappa _1 \rho _1 A_r^{13} v_R \\
	&+&\kappa _+ \kappa _1 \rho _3 A_r^{13} v_R-4 \kappa _2 \rho _1 A_r^{12} v_R^2+2 \kappa _2 \rho _3 A_r^{12} v_R^2+2 \alpha _1 \kappa _+^2 \kappa _1 \big(A_r^{44}+ A_r^{11}\big)+4 \alpha _2 \kappa _1 \big(\kappa _1^2 A_r^{12}-\kappa _2^2 A_r^{12}\nonumber \\
	&+&2 \kappa _1 \kappa _2 \big(A_r^{44}+ A_r^{11}\big)\big)+4 \kappa _1 \rho _1  A_r^{11} v_R^2-2 \kappa _1 \rho _3  A_r^{11} v_R^2+2 \alpha _3 \kappa _1 \kappa _2^2  A_r^{11}\big)-2 \kappa _+^2 \kappa _1 v_L^2 \big(2 \rho _1 \big(\kappa _+ A_r^{14}+A_r^{34} v_R\big)\nonumber \\
	&-&\rho _3 A_r^{34} v_R\big)-4 \kappa _+ \kappa _1 \big(2 \kappa _1^2 \kappa _2^2 \big(\lambda _1+4 \lambda _2+2 \lambda _3\big) A_r^{14}+2 \kappa _1 \kappa _2^3 \big(2 \lambda _4 A_r^{14}-\big(2 \lambda _2+\lambda _3\big) A_r^{24}\big)\nonumber \\
	&+&2 \kappa _1^3 \kappa _2 \big(\big(2 \lambda _2+\lambda _3\big) A_r^{24}+2 \lambda _4 A_r^{14}\big)+\kappa _2^4 \big(\lambda _1 A_r^{14}-\lambda _4 A_r^{24}\big)+\kappa _1^4 \big(\lambda _4 A_r^{24}+\lambda _1 A_r^{14}\big)\big)\nonumber \\
	&-&\kappa _+^2 v_R \big(2 \kappa _1 \big(\alpha _1 \kappa _+^2+\kappa _2 \big(4 \alpha _2 \kappa _1+\alpha _3 \kappa _2\big)\big) A_r^{34}+\beta _1 \kappa _+^2 \kappa _1 A_r^{12}+2 \beta _3 \kappa _+^2 \kappa _2 A_r^{12}\big)+\kappa _+^3 \kappa _1 \big(2 \rho _1-\rho _3\big) A_r^{14} v_R^2\Big],\nonumber
	\\ \nonumber \\
	(\Delta m_{r,\phi D}^2)_{22}&=&\frac{1}{2 \Lambda ^2 \kappa _-^2 \kappa _+^3 \kappa _1^2}\Big[-2 \kappa _+ v_L v_R \kappa _2^4 \big(\beta _3 \kappa _2^2+v_L v_R \big(\rho _3-2 \rho _1\big)\big) A_r^{22}-2 \kappa _1^7 \kappa _2 \big(v_L \alpha _3 A_r^{24}+v_R \beta _3 A_r^{24}+v_R \alpha _3 A_r^{23}\nonumber \\
	&+&v_L \beta _3 A_r^{23}+8 \kappa _+ \lambda _2 A_r^{12}+4 \kappa _+ \lambda _3 A_r^{12}\big)+\kappa _1^3 \big(2 \kappa _2^5 \big(v_L \alpha _3 A_r^{24}+v_R \beta _3 A_r^{24}+v_R \alpha _3 A_r^{23}+v_L \beta _3 A_r^{23}-8 \kappa _+ \lambda _2 A_r^{12}\nonumber \\
	&-&4 \kappa _+ \lambda _3 A_r^{12}\big)-4 \kappa _+ A_r^{22} v_L v_R \beta _1 \kappa _2^3\big)-\kappa _1^8 \big(4 v_L \alpha _2 A_r^{24}+v_R \beta _1 A_r^{24}+4 v_R \alpha _2 A_r^{23}+v_L \beta _1 A_r^{23}+8 \kappa _+ \lambda _2 A_r^{22}\nonumber \\
	&+&4 \kappa _+ \lambda _3 A_r^{22}+4 \kappa _+ \lambda _4 A_r^{12}\big)+\kappa _1^6 \big(\kappa _2^2 \big(4 v_L \alpha _2 A_r^{24}-v_R \beta _1 A_r^{24}+4 v_R \alpha _2 A_r^{23}-v_L \beta _1 A_r^{23}+24 \kappa _+ \lambda _2 A_r^{22}\nonumber \\
	&+&12 \kappa _+ \lambda _3 A_r^{22}+4 \kappa _+ \lambda _4 A_r^{12}\big)-\kappa _+ A_r^{22} \big(\alpha _3 v_L^2+2 v_R \beta _3 v_L+v_R^2 \alpha _3\big)\big)+2 \kappa _1^5 \big(\big(16 \kappa _+ \lambda _2 A_r^{12}+8 \kappa _+ \lambda _3 A_r^{12}\nonumber \\
	&-&\big(v_R A_r^{24}+v_L A_r^{23}\big) \beta _3\big) \kappa _2^3+v_L v_R \big(\big(v_R A_r^{24}+v_L A_r^{23}\big) \big(2 \rho _1-\rho _3\big)-\kappa _+ A_r^{22} \beta _1\big) \kappa _2\big)\nonumber \\
	&+&\kappa _1^4 \big(-2 \kappa _+ \big(\alpha _3 v_L^2+3 v_R \beta _3 v_L+v_R^2 \alpha _3\big) \kappa _2^2 A_r^{22}+2 \kappa _+ v_L^2 v_R^2 \big(2 \rho _1-\rho _3\big) A_r^{22}+\kappa _2^4 \big(4 v_L \alpha _2 A_r^{24}+v_R \beta _1 A_r^{24}\nonumber \\
	&+&4 v_R \alpha _2 A_r^{23}+v_L \beta _1 A_r^{23}-24 \kappa _+ \lambda _2 A_r^{22}-12 \kappa _+ \lambda _3 A_r^{22}+4 \kappa _+ \lambda _4 A_r^{12}\big)\big)\nonumber \\
	&-&\kappa _1^2 \big(\kappa _+ \big(\alpha _3 v_L^2+6 v_R \beta _3 v_L+v_R^2 \alpha _3\big) \kappa _2^4 A_r^{22}+4 \kappa _+ v_L^2 v_R^2 \kappa _2^2 \big(\rho _3-2 \rho _1\big) A_r^{22}+\kappa _2^6 \big(4 v_L \alpha _2 A_r^{24}-v_R \beta _1 A_r^{24}\nonumber \\
	&+&4 v_R \alpha _2 A_r^{23}-v_L \beta _1 A_r^{23}-8 \kappa _+ \lambda _2 A_r^{22}-4 \kappa _+ \lambda _3 A_r^{22}+4 \kappa _+ \lambda _4 A_r^{12}\big)\big)\nonumber \\
	&+&2 \kappa _1 \kappa _2^5 \big(v_R \beta _3 \kappa _2^2 A_r^{24}+v_L^2 v_R \big(\rho _3-2 \rho _1\big) A_r^{23}+v_L \big(v_R^2 \big(\rho _3-2 \rho _1\big) A_r^{24}+\beta _3 \kappa _2^2 A_r^{23}-\kappa _+ v_R \beta _1 A_r^{22}\big)\big)\Big], 
	\end{eqnarray}}
{\small\begin{eqnarray}
	(\Delta m_{r,\phi D}^2)_{23}&=&\frac{1}{4 \Lambda ^2 \kappa _-^2 \kappa _+^3 \kappa _1^2}\Big[-v_R^2 \big(-2 \kappa _-^2 v_L \kappa _1 \kappa _2 \big(2 \rho _1-\rho _3\big) A_r^{34}+\kappa _+^3 \alpha _3 \kappa _1^2 A_r^{23}+4 \kappa _-^2 \kappa _+ \kappa _1^2 \rho _1 A_r^{23}\nonumber \\
	&-&2 \kappa _+^3 v_L^2 \big(2 \rho _1-\rho _3\big) A_r^{23}\big) \kappa _+^2-v_R \big(\beta _1 \kappa _1^6 A_r^{34}-2 \beta _3 \kappa _1 \kappa _2^5 A_r^{34}-\beta _1 \kappa _1^2 \kappa _2^4 A_r^{34}+2 \beta _3 \kappa _1^5 \kappa _2 A_r^{34}-2 \alpha _3 \kappa _1^3 \kappa _2^3 A_r^{33}\nonumber \\
	&+&2 \alpha _3 \kappa _1^5 \kappa _2 A_r^{33}+4 v_L^2 \kappa _1 \kappa _2^3 \rho _1 A_r^{33}-4 v_L^2 \kappa _1^3 \kappa _2 \rho _1 A_r^{33}-2 v_L^2 \kappa _1 \kappa _2^3 \rho _3 A_r^{33}+2 v_L^2 \kappa _1^3 \kappa _2 \rho _3 A_r^{33}+2 \kappa _+ v_L \kappa _1^4 \rho _1 A_r^{24}\nonumber \\
	&-&2 \kappa _+ v_L \kappa _1^2 \kappa _2^2 \rho _1 A_r^{24}+\kappa _+ v_L \kappa _1^4 \rho _3 A_r^{24}-\kappa _+ v_L \kappa _1^2 \kappa _2^2 \rho _3 A_r^{24}+2 \kappa _+ v_L \beta _3 \kappa _1^4 A_r^{23}+2 \kappa _+ v_L \beta _3 \kappa _2^4 A_r^{23}\nonumber \\
	&+&2 \kappa _+ v_L \beta _1 \kappa _1 \kappa _2^3 A_r^{23}+4 \kappa _+ v_L \beta _3 \kappa _1^2 \kappa _2^2 A_r^{23}+2 \kappa _+ v_L \beta _1 \kappa _1^3 \kappa _2 A_r^{23}-2 \alpha _3 \kappa _1^3 \kappa _2^3 A_r^{22}+2 \alpha _3 \kappa _1^5 \kappa _2 A_r^{22}\nonumber \\
	&+&4 v_L^2 \kappa _1 \kappa _2^3 \rho _1 A_r^{22}-4 v_L^2 \kappa _1^3 \kappa _2 \rho _1 A_r^{22}-2 v_L^2 \kappa _1 \kappa _2^3 \rho _3 A_r^{22}+2 v_L^2 \kappa _1^3 \kappa _2 \rho _3 A_r^{22}-2 \alpha _3 \kappa _1^2 \kappa _2^4 A_r^{12}+2 \alpha _3 \kappa _1^4 \kappa _2^2 A_r^{12}\nonumber \\
	&+&2 \alpha _1 \kappa _1^2 \big(\kappa _1^4-\kappa _2^4\big) A_r^{12}+4 v_L^2 \kappa _1^4 \rho _1 A_r^{12}-4 v_L^2 \kappa _1^2 \kappa _2^2 \rho _1 A_r^{12}-2 v_L^2 \kappa _1^4 \rho _3 A_r^{12}+2 v_L^2 \kappa _1^2 \kappa _2^2 \rho _3 A_r^{12}\nonumber \\
	&+&4 \kappa _-^2 \alpha _2 \kappa _1^2 \big(2 \kappa _1 \kappa _2 A_r^{12}+\big(A_r^{33}+A_r^{22}\big) \kappa _1^2-\big(A_r^{33}+A_r^{22}\big) \kappa _2^2\big)\big) \kappa _+^2-\kappa _1 \big(\big(4 \kappa _+ \big(2 \lambda _2+\lambda _3\big) A_r^{23}+4 \kappa _+ \lambda _4 A_r^{13}\nonumber \\
	&+&v_L \big(4 \alpha _2 A_r^{34}+\big(A_r^{33}+A_r^{22}\big) \beta _1\big)\big) \kappa _1^7+2 \kappa _2 \big(8 \kappa _+ \lambda _2 A_r^{13}+4 \kappa _+ \lambda _3 A_r^{13}+v_L \big(\alpha _3 A_r^{34}+\big(A_r^{33}+A_r^{22}\big) \beta _3\big)\big) \kappa _1^6\nonumber \\
	&+&\kappa _1^5\big(\kappa _+ v_L^2 \big(\alpha _3-2 \rho _1+\rho _3\big) A_r^{23}+\kappa _2^2 \big(-12 \kappa _+ \big(2 \lambda _2+\lambda _3\big) A_r^{23}-4 \kappa _+ \lambda _4 A_r^{13}+v_L \big(\big(A_r^{33}+A_r^{22}\big) \beta _1-4 A_r^{34} \alpha _2\big)\big)\big) \nonumber \\
	&+&2 \kappa _2^3 \big(-16 \kappa _+ \lambda _2 A_r^{13}-8 \kappa _+ \lambda _3 A_r^{13}+\big(A_r^{33}+A_r^{22}\big) v_L \beta _3\big) \kappa _1^4-\big(\kappa _2^4 \big(-12 \kappa _+ \big(2 \lambda _2+\lambda _3\big) A_r^{23}+4 \kappa _+ \lambda _4 A_r^{13}\nonumber \\
	&+&v_L \big(4 \alpha _2 A_r^{34}+\big(A_r^{33}+A_r^{22}\big) \beta _1\big)\big)-2 \kappa _+ A_r^{23} v_L^2 \alpha _3 \kappa _2^2\big) \kappa _1^3+2 \kappa _2^5 \big(8 \kappa _+ \lambda _2 A_r^{13}+4 \kappa _+ \lambda _3 A_r^{13}\nonumber \\
	&-&v_L \big(\alpha _3 A_r^{34}+\big(A_r^{33}+A_r^{22}\big) \beta _3\big)\big) \kappa _1^2+\kappa _2^4 \big(\kappa _+ v_L^2 \big(\alpha _3+2 \rho _1-\rho _3\big) A_r^{23}+\kappa _2^2 \big(-4 \kappa _+ \big(2 \lambda _2+\lambda _3\big) A_r^{23}\nonumber \\
	&+&4 \kappa _+ \lambda _4 A_r^{13}-v_L \big(\big(A_r^{33}+A_r^{22}\big) \beta _1-4 A_r^{34} \alpha _2\big)\big)\big) \kappa _1-2 \big(A_r^{33}+A_r^{22}\big) v_L \beta _3 \kappa _2^7\big)\Big],\\
	\nonumber \\
	(\Delta m_{r,\phi D}^2)_{24}&=&\frac{1}{4 \Lambda ^2 \kappa _-^2 \kappa _+^3 \kappa _1^2}\Big[-v_L^2 \big(-2 \kappa _-^2 v_R \kappa _1 \kappa _2 \big(2 \rho _1-\rho _3\big) A_r^{34}+\kappa _+^3 \alpha _3 \kappa _1^2 A_r^{24}+4 \kappa _-^2 \kappa _+ \kappa _1^2 \rho _1 A_r^{24}\nonumber \\
	&-&2 \kappa _+^3 v_R^2 \big(2 \rho _1-\rho _3\big) A_r^{24}\big) \kappa _+^2-v_L \big(-2 \alpha _3 \kappa _1^3 \kappa _2^3 A_r^{44}+2 \alpha _3 \kappa _1^5 \kappa _2 A_r^{44}+4 v_R^2 \kappa _1 \kappa _2^3 \rho _1 A_r^{44}-4 v_R^2 \kappa _1^3 \kappa _2 \rho _1 A_r^{44}\nonumber \\
	&-&2 v_R^2 \kappa _1 \kappa _2^3 \rho _3 A_r^{44}+2 v_R^2 \kappa _1^3 \kappa _2 \rho _3 A_r^{44}+\beta _1 \kappa _1^6 A_r^{34}-2 \beta _3 \kappa _1 \kappa _2^5 A_r^{34}-\beta _1 \kappa _1^2 \kappa _2^4 A_r^{34}+2 \beta _3 \kappa _1^5 \kappa _2 A_r^{34}\nonumber \\
	&+&2 \kappa _+ v_R \beta _3 \kappa _1^4 A_r^{24}+2 \kappa _+ v_R \beta _3 \kappa _2^4 A_r^{24}+2 \kappa _+ v_R \beta _1 \kappa _1 \kappa _2^3 A_r^{24}+4 \kappa _+ v_R \beta _3 \kappa _1^2 \kappa _2^2 A_r^{24}+2 \kappa _+ v_R \beta _1 \kappa _1^3 \kappa _2 A_r^{24}\nonumber \\
	&+&2 \kappa _+ v_R \kappa _1^4 \rho _1 A_r^{23}-2 \kappa _+ v_R \kappa _1^2 \kappa _2^2 \rho _1 A_r^{23}+\kappa _+ v_R \kappa _1^4 \rho _3 A_r^{23}-\kappa _+ v_R \kappa _1^2 \kappa _2^2 \rho _3 A_r^{23}-2 \alpha _3 \kappa _1^3 \kappa _2^3 A_r^{22}+2 \alpha _3 \kappa _1^5 \kappa _2 A_r^{22}\nonumber \\
	&+&4 v_R^2 \kappa _1 \kappa _2^3 \rho _1 A_r^{22}-4 v_R^2 \kappa _1^3 \kappa _2 \rho _1 A_r^{22}-2 v_R^2 \kappa _1 \kappa _2^3 \rho _3 A_r^{22}+2 v_R^2 \kappa _1^3 \kappa _2 \rho _3 A_r^{22}-2 \alpha _3 \kappa _1^2 \kappa _2^4 A_r^{12}+2 \alpha _3 \kappa _1^4 \kappa _2^2 A_r^{12}\nonumber \\
	&+&2 \alpha _1 \kappa _1^2 \big(\kappa _1^4-\kappa _2^4\big) A_r^{12}+4 v_R^2 \kappa _1^4 \rho _1 A_r^{12}-4 v_R^2 \kappa _1^2 \kappa _2^2 \rho _1 A_r^{12}-2 v_R^2 \kappa _1^4 \rho _3 A_r^{12}+2 v_R^2 \kappa _1^2 \kappa _2^2 \rho _3 A_r^{12}\nonumber \\
	&+&4 \kappa _-^2 \alpha _2 \kappa _1^2 \big(2 \kappa _1 \kappa _2 A_r^{12}+\big(A_r^{44}+A_r^{22}\big) \kappa _1^2-\big(A_r^{44}+A_r^{22}\big) \kappa _2^2\big)\big) \kappa _+^2-\kappa _1 \big(\big(4 \kappa _+ \big(2 \lambda _2+\lambda _3\big) A_r^{24}\nonumber \\
	&+&4 \kappa _+ \lambda _4 A_r^{14}+v_R \big(4 \alpha _2 A_r^{34}+\big(A_r^{44}+A_r^{22}\big) \beta _1\big)\big) \kappa _1^7+2 \kappa _2 \big(8 \kappa _+ \lambda _2 A_r^{14}+4 \kappa _+ \lambda _3 A_r^{14}\nonumber \\
	&+&v_R \big(\alpha _3 A_r^{34}+\big(A_r^{44}+A_r^{22}\big) \beta _3\big)\big) \kappa _1^6+\big(\kappa _+ v_R^2 \big(\alpha _3-2 \rho _1+\rho _3\big) A_r^{24}+\kappa _2^2 \big(-12 \kappa _+ \big(2 \lambda _2+\lambda _3\big) A_r^{24}\nonumber \\
	&-&4 \kappa _+ \lambda _4 A_r^{14}+v_R \big(\big(A_r^{44}+A_r^{22}\big) \beta _1-4 A_r^{34} \alpha _2\big)\big)\big) \kappa _1^5+2 \kappa _2^3 \kappa _1^4 \big(-16 \kappa _+ \lambda _2 A_r^{14}-8 \kappa _+ \lambda _3 A_r^{14}\nonumber \\
	&+&\big(A_r^{44}+A_r^{22}\big) v_R \beta _3\big)-\big(\kappa _2^4 \big(-12 \kappa _+ \big(2 \lambda _2+\lambda _3\big) A_r^{24}+4 \kappa _+ \lambda _4 A_r^{14}+v_R \big(4 \alpha _2 A_r^{34}+\big(A_r^{44}+A_r^{22}\big) \beta _1\big)\big)\nonumber \\
	&-&2 \kappa _+ A_r^{24} v_R^2 \alpha _3 \kappa _2^2\big) \kappa _1^3+2 \kappa _2^5 \big(8 \kappa _+ \lambda _2 A_r^{14}+4 \kappa _+ \lambda _3 A_r^{14}-v_R \big(\alpha _3 A_r^{34}+\big(A_r^{44}+A_r^{22}\big) \beta _3\big)\big) \kappa _1^2\nonumber \\
	&+&\kappa _2^4 \big(\kappa _+ v_R^2 \big(\alpha _3+2 \rho _1-\rho _3\big) A_r^{24}+\kappa _2^2 \big(-4 \kappa _+ \big(2 \lambda _2+\lambda _3\big) A_r^{24}+4 \kappa _+ \lambda _4 A_r^{14}\nonumber \\
	&-&v_R \big(\big(A_r^{44}+A_r^{22}\big) \beta _1-4 A_r^{34} \alpha _2\big)\big)\big) \kappa _1-2 \big(A_r^{44}+A_r^{22}\big) v_R \beta _3 \kappa _2^7\big)\Big],\\
		\nonumber \\
	(\Delta m_{r,\phi D}^2)_{33}&=&\frac{1}{2 \kappa _+ \kappa _1 \Lambda ^2}\Big[-v_R \big(2 \kappa _+ \kappa _1 \rho _1 A_r^{34} v_L+\kappa _+ \kappa _1 \rho _3 A_r^{34} v_L-4 \kappa _2 \rho _1 A_r^{23} v_L^2+2 \kappa _2 \rho _3 A_r^{23} v_L^2+4 \kappa _1 \rho _1 A_r^{13} v_L^2\nonumber \\
	&-&2 \kappa _1 \rho _3 A_r^{13} v_L^2+2 \alpha _3 \kappa _1^2 \kappa _2 A_r^{23}+2 \alpha _3 \kappa _1 \kappa _2^2 A_r^{13}+2 \alpha _1 \kappa _+^2 \kappa _1 A_r^{13}+4 \alpha _2 \kappa _1 \big(\kappa _1^2 A_r^{23}-\kappa _2^2 A_r^{23}+2 \kappa _1 \kappa _2 A_r^{13}\big)\big)\nonumber \\
	&+&v_L \big(\kappa _+ \kappa _1 \big(2 \rho _1-\rho _3\big) A_r^{33} v_L-\beta _1 \kappa _+^2 \kappa _1 A_r^{23}-2 \beta _3 \kappa _+^2 \kappa _2 A_r^{23}\big)-4 \kappa _+ \kappa _1 \rho _1 A_r^{33} v_R^2\Big],
	\end{eqnarray}}

{\small\begin{eqnarray}
	(\Delta m_{r,\phi D}^2)_{34}&=&\frac{1}{4 \kappa _+ \kappa _1 \Lambda ^2}\Big[-v_L \big(2 \alpha _3 \kappa _1^2 \kappa _2 A_r^{23}+2 \alpha _3 \kappa _1 \kappa _2^2 A_r^{13}+2 \alpha _1 \kappa _+^2 \kappa _1 A_r^{13}+4 \alpha _2 \kappa _1 \big(\kappa _1^2 A_r^{23}-\kappa _2^2 A_r^{23}+2 \kappa _1 \kappa _2 A_r^{13}\big)\nonumber \\
	&+&\beta _1 \kappa _1^3 A_r^{24}+2 \beta _3 \kappa _2^3 A_r^{24}+\beta _1 \kappa _1 \kappa _2^2 A_r^{24}+2 \beta _3 \kappa _1^2 \kappa _2 A_r^{24}+2 \kappa _+ \kappa _1 \rho _1 A_r^{44} v_R+\kappa _+ \kappa _1 \rho _3 A_r^{44} v_R+2 \kappa _+ \kappa _1 \rho _1 A_r^{33} v_R\nonumber \\
	&+&\kappa _+ \kappa _1 \rho _3 A_r^{33} v_R-4 \kappa _2 \rho _1 A_r^{23} v_R^2+2 \kappa _2 \rho _3 A_r^{23} v_R^2+4 \kappa _1 \rho _1 A_r^{13} v_R^2-2 \kappa _1 \rho _3 A_r^{13} v_R^2\big)\nonumber \\
	&+&v_L^2 \big(-\big(\kappa _+ \kappa _1 \big(2 \rho _1+\rho _3\big) A_r^{34}+2 \big(2 \rho _1-\rho _3\big) v_R \big(\kappa _1 A_r^{14}-\kappa _2 A_r^{24}\big)\big)\big)\nonumber \\
	&-&v_R \big(2 \alpha _3 \kappa _1^2 \kappa _2 A_r^{24}+2 \alpha _3 \kappa _1 \kappa _2^2 A_r^{14}+2 \alpha _1 \kappa _+^2 \kappa _1 A_r^{14}+4 \alpha _2 \kappa _1 \big(\kappa _1^2 A_r^{24}-\kappa _2^2 A_r^{24}+2 \kappa _1 \kappa _2 A_r^{14}\big)+\beta _1 \kappa _1^3 A_r^{23}\nonumber \\
	&+&2 \beta _3 \kappa _2^3 A_r^{23}+\beta _1 \kappa _1 \kappa _2^2 A_r^{23}+2 \beta _3 \kappa _1^2 \kappa _2 A_r^{23}+2 \kappa _+ \kappa _1 \rho _1 A_r^{34} v_R+\kappa _+ \kappa _1 \rho _3 A_r^{34} v_R\big)\Big],\\
	\nonumber \\
	(\Delta m_{r,\phi D}^2)_{44}&=&\frac{1}{2 \kappa _+ \kappa _1 \Lambda ^2}\Big[-v_L \big(2 \alpha _3 \kappa _1^2 \kappa _2 A_r^{24}+2 \alpha _3 \kappa _1 \kappa _2^2 A_r^{14}+2 \alpha _1 \kappa _+^2 \kappa _1 A_r^{14}+4 \alpha _2 \kappa _1 \big(\kappa _1^2 A_r^{24}-\kappa _2^2 A_r^{24}+2 \kappa _1 \kappa _2 A_r^{14}\big)\nonumber \\
	&+&2 \kappa _+ \kappa _1 \rho _1 A_r^{34} v_R+\kappa _+ \kappa _1 \rho _3 A_r^{34} v_R-4 \kappa _2 \rho _1 A_r^{24} v_R^2+2 \kappa _2 \rho _3 A_r^{24} v_R^2+4 \kappa _1 \rho _1 A_r^{14} v_R^2-2 \kappa _1 \rho _3 A_r^{14} v_R^2\big)\nonumber \\
	&-&4 \kappa _+ \kappa _1 \rho _1 A_r^{44} v_L^2+v_R \big(-\beta _1 \kappa _+^2 \kappa _1 A_r^{24}-2 \beta _3 \kappa _+^2 \kappa _2 A_r^{24}+\kappa _+ \kappa _1 \big(2 \rho _1-\rho _3\big) A_r^{44} v_R\big)\Big].
	\end{eqnarray}}

\underline{\large{CP-odd scalar mass matrix}}

{\small\begin{eqnarray}
	(\Delta m_{i,\phi D}^2)_{11}&=&\frac{v_L v_R}{\kappa _+^3 \kappa _1 \Lambda ^2} \Big[\big(2 \rho _1-\rho _3\big) v_L \big(2 \kappa _+ v_R \big(\kappa _1 A_i^{11}-\kappa _2 A_i^{12}\big)-\kappa _+^2 \kappa _1 A_i^{13}\big)+\kappa _+^2 \big(\beta _1 \kappa _+ \kappa _1 A_i^{12}+2 \beta _3 \kappa _+ \kappa _2 A_i^{12}\nonumber \\
	&+&\kappa _1 \big(2 \rho _1-\rho _3\big) A_i^{14} v_R\big)\Big],\\
	\nonumber \\
	(\Delta m_{i,\phi D}^2)_{12}&=&\frac{1}{4 \kappa _-^2 \kappa _+^3 \kappa _1^2 \Lambda ^2}\Big[-v_L^2 \big(\alpha _3 \kappa _+^5 \kappa _1^2 A_i^{12}+2 \kappa _1 \kappa _+^2 \kappa _-^2 \big(2 \rho _1-\rho _3\big) v_R \big(\kappa _1 A_i^{23}-\kappa _2 A_i^{13}\big)\nonumber \\
	&-&2 \kappa _+ \big(2 \rho _1-\rho _3\big) v_R^2 \big(3 \kappa _1^4 A_i^{12}-\kappa _2^4 A_i^{12}+2 \kappa _1^2 \kappa _2^2 A_i^{12}+2 \kappa _1 \kappa _2^3 \big(A_i^{22}+A_i^{11}\big)-2 \kappa _1^3 \kappa _2 \big(A_i^{22}+A_i^{11}\big)\big)\big)\nonumber \\
	&+&\kappa _+^2 v_L \big(-\kappa _1 \big(\kappa _1^4-\kappa _2^4\big) A_i^{13} \big(\beta _1 \kappa _1+2 \beta _3 \kappa _2\big)+2 \kappa _+ v_R \big(\beta _1 \kappa _1 \big(\kappa _2^3 A_i^{12}-3 \kappa _1^2 \kappa _2 A_i^{12}+\kappa _1^3 \big(A_i^{22}+A_i^{11}\big)\nonumber \\
	&-&\kappa _1 \kappa _2^2 \big(A_i^{22}+A_i^{11}\big)\big)+\beta _3 \big(\kappa _1^4 A_i^{12}+\kappa _2^4 A_i^{12}-6 \kappa _1^2 \kappa _2^2 A_i^{12}-2 \kappa _1 \kappa _2^3 \big(A_i^{22}+A_i^{11}\big)+2 \kappa _1^3 \kappa _2 \big(A_i^{22}+A_i^{11}\big)\big)\big)\nonumber \\
	&+&2 \kappa _-^2 \kappa _1 \big(2 \rho _1-\rho _3\big) v_R^2 \big(\kappa _1 A_i^{24}-\kappa _2 A_i^{14}\big)\big)+\kappa _1 \big(4 \kappa _-^2 \kappa _+^5 \kappa _1 \big(2 \lambda _2-\lambda _3\big) A_i^{12}-\alpha _3 \kappa _+^5 \kappa _1 A_i^{12} v_R^2\nonumber \\
	&+&\kappa _-^2 \kappa _+^4 A_i^{14} \big(\beta _1 \kappa _1+2 \beta _3 \kappa _2\big) v_R\big)\Big],\\
	\nonumber \\
	(\Delta m_{i,\phi D}^2)_{13}&=&\frac{v_L}{4 \kappa _+^3 \kappa _1 \Lambda ^2} \Big[\big(2 \rho _1-\rho _3\big) v_L \big(\kappa _+^3 \kappa _1 A_i^{13}-2 \kappa _+^2 v_R \big(\kappa _1 \big(A_i^{33}+A_i^{11}\big)-\kappa _2 A_i^{12}\big)+4 \kappa _+ v_R^2 \big(\kappa _1 A_i^{13}-\kappa _2 A_i^{23}\big)\big)\nonumber \\
	&-&\kappa _+^2 \big(\beta _1 \kappa _1 \big(\kappa _1^2 A_i^{12}+\kappa _2^2 A_i^{12}-2 \kappa _+ A_i^{23} v_R\big)+2 \beta _3 \kappa _2 \big(\kappa _1^2 A_i^{12}+\kappa _2^2 A_i^{12}-2 \kappa _+ A_i^{23} v_R\big)\nonumber \\
	&+&\kappa _1 \big(2 \rho _1-\rho _3\big) v_R \big(\kappa _+ A_i^{14}-2 A_i^{34} v_R\big)\big)\Big],\\
	\nonumber \\
	(\Delta m_{i,\phi D}^2)_{14}&=&\frac{v_R}{4 \kappa _+^3 \kappa _1 \Lambda ^2} \Big[\kappa _+^2 v_L \big(\kappa _+ \big(\kappa _1 \big(2 \beta _1 A_i^{24}-2 \rho _1 A_i^{13}+\rho _3 A_i^{13}\big)+4 \beta _3 \kappa _2 A_i^{24}\big)+2 \big(2 \rho _1-\rho _3\big) v_R \big(\kappa _1 \big(A_i^{44}+A_i^{11}\big)\nonumber \\
	&-&\kappa _2 A_i^{12}\big)\big)+2 \big(2 \rho _1-\rho _3\big) v_L^2 \big(2 \kappa _+ v_R \big(\kappa _1 A_i^{14}-\kappa _2 A_i^{24}\big)-\kappa _+^2 \kappa _1 A_i^{34}\big)\nonumber \\
	&+&\kappa _+^2 \big(\beta _1 \kappa _+^2 \kappa _1 A_i^{12}+2 \beta _3 \kappa _+^2 \kappa _2 A_i^{12}+\kappa _+ \kappa _1 \big(2 \rho _1-\rho _3\big) A_i^{14} v_R\big)\Big],
	\end{eqnarray}}

\small{\begin{eqnarray}
	(\Delta m_{i,\phi D}^2)_{22}&=&\frac{1}{2 \kappa _-^2 \kappa _+^3 \kappa _1^2 \Lambda ^2}\Big[-v_L^2 \big(\alpha _3 \kappa _+^5 \kappa _1^2 A_i^{22}-2 \kappa _1 \kappa _2 \big(\kappa _1^4-\kappa _2^4\big) \big(2 \rho _1-\rho _3\big) A_i^{23} v_R\nonumber \\
	&-&2 \kappa _+ \big(2 \rho _1-\rho _3\big) v_R^2 \big(\kappa _1^4 A_i^{22}-\kappa _2^4 A_i^{22}+4 \kappa _1^2 \kappa _2^2 A_i^{22}+2 \kappa _1 \kappa _2^3 A_i^{12}-2 \kappa _1^3 \kappa _2 A_i^{12}\big)\big)\nonumber \\
	&+&\kappa _+^2 v_L \big(-\kappa _1 \big(\kappa _1^4-\kappa _2^4\big) A_i^{23} \big(\beta _1 \kappa _1+2 \beta _3 \kappa _2\big)+2 \kappa _+ v_R \big(\beta _1 \kappa _1 \big(\kappa _2^3 A_i^{22}-3 \kappa _1^2 \kappa _2 A_i^{22}+\kappa _1^3 A_i^{12}-\kappa _1 \kappa _2^2 A_i^{12}\big)\nonumber \\
	&+&\beta _3 \big(\kappa _1^4 A_i^{22}+\kappa _2^4 A_i^{22}-6 \kappa _1^2 \kappa _2^2 A_i^{22}-2 \kappa _1 \kappa _2^3 A_i^{12}+2 \kappa _1^3 \kappa _2 A_i^{12}\big)\big)-2 \kappa _-^2 \kappa _1 \kappa _2 \big(2 \rho _1-\rho _3\big) A_i^{24} v_R^2\big)\nonumber \\
	&+&\kappa _1 \big(4 \kappa _-^2 \kappa _+^5 \kappa _1 \big(2 \lambda _2-\lambda _3\big) A_i^{22}-\alpha _3 \kappa _+^5 \kappa _1 A_i^{22} v_R^2+\kappa _-^2 \kappa _+^4 A_i^{24} \big(\beta _1 \kappa _1+2 \beta _3 \kappa _2\big) v_R\big)\Big],\\
	\nonumber \\
	(\Delta m_{i,\phi D}^2)_{23}&=&\frac{v_L^2}{4 \kappa _-^2 \kappa _+^3 \kappa _1^2 \Lambda ^2} \Big[\big(-\kappa _+^3 \kappa _1^2 A_i^{23} \big(\alpha _3 \kappa _+^2-\kappa _-^2 \big(2 \rho _1-\rho _3\big)\big)-2 \kappa _1 \kappa _+^2 \kappa _-^2 \big(2 \rho _1-\rho _3\big) v_R \big(\kappa _1 A_i^{12}\nonumber \\
	&-&\kappa _2 \big(A_i^{33}+A_i^{22}\big)\big)+2 \kappa _+ \big(2 \rho _1-\rho _3\big) v_R^2 \big(\kappa _1^4 A_i^{23}-\kappa _2^4 A_i^{23}+4 \kappa _1^2 \kappa _2^2 A_i^{23}+2 \kappa _1 \kappa _2^3 A_i^{13}-2 \kappa _1^3 \kappa _2 A_i^{13}\big)\big)\nonumber \\
	&+&\kappa _+^2 v_L \big(-\kappa _1 \kappa _+^2 \kappa _-^2\big(A_i^{33}+A_i^{22}\big) \big(\beta _1 \kappa _1+2 \beta _3 \kappa _2\big)+\kappa _+ v_R \big(2 \beta _1 \kappa _1 \big(\kappa _2^3 A_i^{23}-3 \kappa _1^2 \kappa _2 A_i^{23}+\kappa _1^3 A_i^{13}\nonumber \\
	&-&\kappa _1 \kappa _2^2 A_i^{13}\big)+2 \beta _3 \big(\kappa _1^4 A_i^{23}+\kappa _2^4 A_i^{23}-6 \kappa _1^2 \kappa _2^2 A_i^{23}-2 \kappa _1 \kappa _2^3 A_i^{13}+2 \kappa _1^3 \kappa _2 A_i^{13}\big)-\kappa _-^2 \kappa _1^2 \big(2 \rho _1-\rho _3\big) A_i^{24}\big)\nonumber \\
	&-&2 \kappa _-^2 \kappa _1 \kappa _2 \big(2 \rho _1-\rho _3\big) A_i^{34} v_R^2\big)+\kappa _1 \big(4 \kappa _-^2 \kappa _+^5 \kappa _1 \big(2 \lambda _2-\lambda _3\big) A_i^{23}-\alpha _3 \kappa _+^5 \kappa _1 A_i^{23} v_R^2\nonumber \\&+&\kappa _-^2 \kappa _+^4 A_i^{34} \big(\beta _1 \kappa _1+2 \beta _3 \kappa _2\big) v_R\big)\big],\\
	\nonumber \\
	(\Delta m_{i,\phi D}^2)_{24}&=&\frac{1}{4 \kappa _-^2 \kappa _+^3 \kappa _1^2 \Lambda ^2}\Big[-v_L^2 \big(\alpha _3 \kappa _+^5 \kappa _1^2 A_i^{24}-2 \kappa _1 \kappa _2 \kappa _+^2 \kappa _-^2  A_i^{34} v_R\big(2 \rho _1-\rho _3\big)\nonumber \\
	&-&2 \kappa _+ \big(2 \rho _1-\rho _3\big) v_R^2 \big(\kappa _1^4 A_i^{24}-\kappa _2^4 A_i^{24}+4 \kappa _1^2 \kappa _2^2 A_i^{24}+2 \kappa _1 \kappa _2^3 A_i^{14}-2 \kappa _1^3 \kappa _2 A_i^{14}\big)\big)\nonumber \\
	&+&\kappa _+^2 v_L \big(-\kappa _1 \kappa _+^2 \kappa _-^2A_i^{34} \big(\beta _1 \kappa _1+2 \beta _3 \kappa _2\big)+\kappa _+ v_R \big(2 \beta _1 \kappa _1 \big(\kappa _2^3 A_i^{24}-3 \kappa _1^2 \kappa _2 A_i^{24}+\kappa _1^3 A_i^{14}-\kappa _1 \kappa _2^2 A_i^{14}\big)\nonumber \\
	&+&2 \beta _3 \big(\kappa _1^4 A_i^{24}+\kappa _2^4 A_i^{24}-6 \kappa _1^2 \kappa _2^2 A_i^{24}-2 \kappa _1 \kappa _2^3 A_i^{14}+2 \kappa _1^3 \kappa _2 A_i^{14}\big)-\kappa _-^2 \kappa _1^2 \big(2 \rho _1-\rho _3\big) A_i^{23}\big)\nonumber \\
	&+&2 \kappa _-^2 \kappa _1 \big(2 \rho _1-\rho _3\big) v_R^2 \big(\kappa _1 A_i^{12}-\kappa _2 \big(A_i^{44}+A_i^{22}\big)\big)\big)-\kappa _1 \kappa _+^2 \big(-4 \kappa _+ \kappa _1 \kappa _+^2 \kappa _-^2s \big(2 \lambda _2-\lambda _3\big) A_i^{24}\nonumber \\
	&+&\kappa _+ \kappa _1 A_i^{24} v_R^2 \big(\alpha _3 \kappa _+^2-\kappa _-^2 \big(2 \rho _1-\rho _3\big)\big)-\kappa _+^2 \kappa _-^2 \big(A_i^{44}+A_i^{22}\big) \big(\beta _1 \kappa _1+2 \beta _3 \kappa _2\big) v_R\big)\Big],\\
	\nonumber \\
	(\Delta m_{i,\phi D}^2)_{33}&=&-\frac{v_L}{2 \kappa _+ \kappa _1 \Lambda ^2} \Big[2 \beta _3 \kappa _2^3 A_i^{23}+\beta _1 \kappa _+^2 \kappa _1 A_i^{23}+2 \beta _3 \kappa _1^2 \kappa _2 A_i^{23}+\big(2 \rho _1-\rho _3\big) v_L \big(2 v_R \big(\kappa _1 A_i^{13}-\kappa _2 A_i^{23}\big)\nonumber \\
	&-&\kappa _+ \kappa _1 A_i^{33}\big)+2 \kappa _+ \kappa _1 \rho _1 A_i^{34} v_R-\kappa _+ \kappa _1 \rho _3 A_i^{34} v_R\Big],\\
	\nonumber \\
	(\Delta m_{i,\phi D}^2)_{34}&=&\frac{1}{4 \kappa _+ \kappa _1 \Lambda ^2}\Big[v_L \big(-\kappa _+^2 A_i^{24} \big(\beta _1 \kappa _1+2 \beta _3 \kappa _2\big)-\kappa _+ \kappa _1 \big(2 \rho _1-\rho _3\big) \big(A_i^{44}+A_i^{33}\big) v_R\nonumber \\
	&+&2 \big(2 \rho _1-\rho _3\big) v_R^2 \big(\kappa _1 A_i^{13}-\kappa _2 A_i^{23}\big)\big)+\big(2 \rho _1-\rho _3\big) v_L^2 \big(-\big(2 v_R \big(\kappa _1 A_i^{14}-\kappa _2 A_i^{24}\big)-\kappa _+ \kappa _1 A_i^{34}\big)\big)\nonumber \\
	&+&v_R \big(\beta _1 \kappa _+^2 \kappa _1 A_i^{23}+2 \beta _3 \kappa _+^2 \kappa _2 A_i^{23}+\kappa _+ \kappa _1 \big(2 \rho _1-\rho _3\big) A_i^{34} v_R\big)\big],\\
	\nonumber \\
	(\Delta m_{i,\phi D}^2)_{44}&=&\frac{v_R}{2 \kappa _+ \kappa _1 \Lambda ^2} \Big[2 \beta _3 \kappa _2^3 A_i^{24}+\beta _1 \kappa _+^2 \kappa _1 A_i^{24}+2 \beta _3 \kappa _1^2 \kappa _2 A_i^{24}+\big(2 \rho _1-\rho _3\big) v_L \big(2 v_R \big(\kappa _1 A_i^{14}-\kappa _2 A_i^{24}\big)\nonumber \\
	&-&\kappa _+ \kappa _1 A_i^{34}\big)+2 \kappa _+ \kappa _1 \rho _1 A_i^{44} v_R-\kappa _+ \kappa _1 \rho _3 A_i^{44} v_R\Big].
	\end{eqnarray}}

\pagebreak

\underline{\large{Singly Charged scalar mass matrix}}

{\small\begin{eqnarray}
	(\Delta m_{+,\phi D}^2)_{11}&=&\frac{1}{4 \kappa _+^3 \kappa _1 \kappa _-^2 \Lambda ^2}\Big[-\kappa _+^2 v_R \big(\sqrt{2} \kappa _+^2 \kappa _-^2 \big(\alpha _3 A_+^{13} \kappa _1+A_+^{14} \big(\beta _1 \kappa _1+2 \beta _3 \kappa _2\big)\big)-2 \sqrt{2} A_+^{13} \kappa _-^2 \kappa _1 \big(2 \rho _1-\rho _3\big) v_L^2 \\
	&+&2 \kappa _+ \big(A_+^{12} \kappa _1^2-A_+^{12} \kappa _2^2+4 A_+^{11} \kappa _1 \kappa _2\big) \big(\beta _1 \kappa _1+2 \beta _3 \kappa _2\big) v_L\big)-2 v_R^2 \big(\alpha _3 A_+^{11} \kappa _+^5 \kappa _1\nonumber \\
	&-&\big(2 \rho _1-\rho _3\big) v_L \big(\sqrt{2} A_+^{14} \kappa _2 \kappa _+^2 \kappa _-^2+2 \kappa _+ \big(-A_+^{12} \kappa _2^3+A_+^{12} \kappa _1^2 \kappa _2+A_+^{11} \kappa _1^3+3 A_+^{11} \kappa _1 \kappa _2^2\big) v_L\big)\big) \nonumber \\
	&-&2 \kappa _1 \kappa _2 v_L \big(\sqrt{2} \big(\kappa _1^4-\kappa _2^4\big) \big(\alpha _3 A_+^{14} \kappa _1+A_+^{13} \beta _1 \kappa _1+2 A_+^{13} \beta _3 \kappa _2\big)+2 \alpha _3 \kappa _+ \kappa _1 \big(A_+^{12} \kappa _1^2-A_+^{12} \kappa _2^2+2 A_+^{11} \kappa _1 \kappa _2\big) v_L\big)\Big],\nonumber\\
	\nonumber \\
	(\Delta m_{+,\phi D}^2)_{12}&=&\frac{1}{8 \kappa _+^3\kappa _1 \kappa _-^2 \Lambda ^2}\Big[-\kappa _+^2 v_L \big(\sqrt{2} \kappa _-^2 \big(\alpha _3 \kappa _1 \big(2 A_+^{24} \kappa _1 \kappa _2+A_+^{14} \kappa _1^2-A_+^{14} \kappa _2^2\big)+\beta _1 \kappa _1 \big(2 A_+^{23} \kappa _1 \kappa _2+A_+^{13} \kappa _1^2-A_+^{13} \kappa _2^2\big)\nonumber \\
	&+&2 \beta _3 \kappa _2 \big(2 A_+^{23} \kappa _1 \kappa _2+A_+^{13} \kappa _1^2-A_+^{13} \kappa _2^2\big)\big)+2 \kappa _+ v_R\big(\beta _1 \kappa _1+2 \beta _3 \kappa _2\big)  \big(4 A_+^{12} \kappa _1 \kappa _2+\big(A_+^{22}+A_+^{11}\big) \kappa _-^2\big) \nonumber \\
	&-&2 \sqrt{2} \kappa _-^2 \big(2 \rho _1-\rho _3\big) \big(A_+^{24} \kappa _2+A_+^{14} \kappa _1\big) v_R^2\big)-2 v_L^2 \big(\alpha _3 \kappa _+ \kappa _1 \big(A_+^{12} \kappa _1^4+A_+^{12} \kappa _2^4+2 A_+^{12} \kappa _1^2 \kappa _2^2\nonumber \\
	&-&2 \big(A_+^{22}+A_+^{11}\big) \kappa _1 \kappa _2^3+2 \big(A_+^{22}+A_+^{11}\big) \kappa _1^3 \kappa _2\big)-\big(2 \rho _1-\rho _3\big) v_R \big(\sqrt{2}\kappa _+^2 \kappa _-^2 \big(A_+^{23} \kappa _1-A_+^{13} \kappa _2\big)\nonumber \\
	&+&2 \kappa _+ \big(2 A_+^{12} \kappa _1^3+2 A_+^{12} \kappa _1 \kappa _2^2-\big(A_+^{22}+A_+^{11}\big) \kappa _2^3+\big(A_+^{22}+A_+^{11}\big) \kappa _1^2 \kappa _2\big) v_R\big)\big)\nonumber \\
	&-&v_R \big(\sqrt{2} \kappa _-^2 \kappa _+^4 \big(\alpha _3 A_+^{23} \kappa _1+A_+^{24} \big(\beta _1 \kappa _1+2 \beta _3 \kappa _2\big)\big)+2 \alpha _3 A_+^{12} \kappa _+^5 \kappa _1 v_R\big)\Big],\\
	\nonumber \\
	(\Delta m_{+,\phi D}^2)_{13}&=&\frac{1}{8 \kappa _+^3\kappa _1 \kappa _-^2 \Lambda ^2}\Big[-\kappa _+^3 \kappa _-^4 \big(\alpha _3 A_+^{13} \kappa _1 + A_+^{14} \big(\beta _1 \kappa _1+2 \beta _3 \kappa _2\big)\big)-\kappa _+^2 v_R \big(\sqrt{2} \alpha _3 \kappa _1 \kappa _+^2 \kappa _-^2 \big(A_+^{33}+A_+^{11}\big)  \\
	&+&\sqrt{2} A_+^{34} \kappa _+^2 \kappa _-^2\big(\beta _1 \kappa _1+2 \beta _3 \kappa _2\big)+2 \kappa _+ v_L \big(\kappa _2 \big(2 \beta _3 \big(A_+^{23} \kappa _1^2-A_+^{23} \kappa _2^2+4 A_+^{13} \kappa _1 \kappa _2\big)+A_+^{14} \kappa _-^2 \big(2 \rho _1-\rho _3\big)\big)\nonumber \\
	&+&\beta _1 \kappa _1 \big(A_+^{23} \kappa _1^2-A_+^{23} \kappa _2^2+4 A_+^{13} \kappa _1 \kappa _2\big)\big)-2 \sqrt{2} \kappa _-^2 \big(2 \rho _1-\rho _3\big) \big(\big(A_+^{33}+A_+^{11}\big) \kappa _1-A_+^{12} \kappa _2\big) v_L^2\big)\nonumber \\
	&-&2 v_R^2 \big(\alpha _3 A_+^{13} \kappa _+^5 \kappa _1-\big(2 \rho _1-\rho _3\big) v_L \big(\sqrt{2} A_+^{34} \kappa _2 \kappa _+^2 \kappa _-^2+2 \kappa _+ \big(-A_+^{23} \kappa _2^3+A_+^{23} \kappa _1^2 \kappa _2+A_+^{13} \kappa _1^3+3 A_+^{13} \kappa _1 \kappa _2^2\big) v_L\big)\big)\nonumber \\
	&-&\sqrt{2} \kappa _+^2 \kappa _-^2 v_L \big(2 \kappa _2 \big(\alpha _3 A_+^{34} \kappa _1^2+\beta _3 \big(A_+^{12} \kappa _1^2-A_+^{12} \kappa _2^2+2 \big(A_+^{33}+A_+^{11}\big) \kappa _1 \kappa _2\big)\big)+\beta _1 \kappa _1 \big(A_+^{12} \kappa _1^2-A_+^{12} \kappa _2^2 \nonumber \\
	&+&2 \big(A_+^{33}+A_+^{11}\big) \kappa _1 \kappa _2\big)\big)+2 \kappa _+ \kappa _1 v_L^2 \big(A_+^{13} \big(\kappa _1^4-\kappa _2^4\big) \big(2 \rho _1-\rho _3\big)-2 \alpha _3 \kappa _1 \kappa _2 \big(A_+^{23} \kappa _1^2-A_+^{23} \kappa _2^2+2 A_+^{13} \kappa _1 \kappa _2\big)\big)\Big],\nonumber\\
	\nonumber \\
	(\Delta m_{+,\phi D}^2)_{14}&=&\frac{1}{8 \kappa _+^3\kappa _1 \kappa _-^2 \Lambda ^2}\Big[-\kappa _+^3 \kappa _-^4 \big(\alpha _3 A_+^{14} \kappa _1+A_+^{13} \beta _1 \kappa _1+2 A_+^{13} \beta _3 \kappa _2\big) \\
	&-&\kappa _+^2 v_R \big(\kappa _-^2 \big(\sqrt{2} \alpha _3 A_+^{34} \kappa _+^2 \kappa _1+2 \sqrt{2} A_+^{34} \kappa _1 \big(\rho _3-2 \rho _1\big) v_L^2+2 A_+^{13} \kappa _+ \kappa _2 \big(2 \rho _1-\rho _3\big) v_L\big)\nonumber \\
	&+&\beta _1 \kappa _1 \big(\sqrt{2} \big(A_+^{44}+A_+^{11}\big) \kappa _1^4-\sqrt{2} \big(A_+^{44}+A_+^{11}\big) \kappa _2^4+2 A_+^{24} \kappa _+ \kappa _1^2 v_L-2 A_+^{24} \kappa _+ \kappa _2^2 v_L+8 A_+^{14} \kappa _+ \kappa _1 \kappa _2 v_L\big)\nonumber\\
	&+&2 \beta _3 \kappa _2 \big(\sqrt{2} \big(A_+^{44}+A_+^{11}\big) \kappa _1^4-\sqrt{2} \big(A_+^{44}+A_+^{11}\big) \kappa _2^4+2 A_+^{24} \kappa _+ \kappa _1^2 v_L-2 A_+^{24} \kappa _+ \kappa _2^2 v_L+8 A_+^{14} \kappa _+ \kappa _1 \kappa _2 v_L\big)\big)\nonumber \\
	&+&2 v_R^2 \big(-A_+^{14} \kappa _+^3 \kappa _1 \big(\alpha _3 \kappa _+^2-\kappa _-^2 \big(2 \rho _1-\rho _3\big)\big)+2 \kappa _+ \big(2 \rho _1-\rho _3\big) \big(-A_+^{24} \kappa _2^3+A_+^{24} \kappa _1^2 \kappa _2+A_+^{14} \kappa _1^3+3 A_+^{14} \kappa _1 \kappa _2^2\big) v_L^2 \nonumber \\
	&+&\sqrt{2} \kappa _+^2 \kappa _-^2 \big(2 \rho _1-\rho _3\big) \big(A_+^{12} \kappa _1+\big(A_+^{44}+A_+^{11}\big) \kappa _2\big) v_L\big)\nonumber \\
	&-&\sqrt{2} \kappa _1 \kappa _+^2 \kappa _-^2 v_L \big(\alpha _3 \big(A_+^{12} \kappa _1^2-A_+^{12} \kappa _2^2+2 \big(A_+^{44}+A_+^{11}\big) \kappa _1 \kappa _2\big)+2 A_+^{34} \kappa _2 \big(\beta _1 \kappa _1+2 \beta _3 \kappa _2\big)\big)\nonumber \\
	&-&4 \alpha _3 \kappa _+ \kappa _1^2 \kappa _2 \big(A_+^{24} \kappa _1^2-A_+^{24} \kappa _2^2+2 A_+^{14} \kappa _1 \kappa _2\big) v_L^2\Big],\nonumber\\
	\nonumber \\
	(\Delta m_{+,\phi D}^2)_{22}&=&-\frac{v_L }{4 \kappa _+^3 \kappa _1 \Lambda ^2}\Big[\sqrt{2} \big(\kappa _1^4-\kappa _2^4\big) \big(\alpha _3 A_+^{24} \kappa _1+A_+^{23} \beta _1 \kappa _1+2 A_+^{23} \beta _3 \kappa _2\big)+2 \kappa _+^2 v_R \big(A_+^{12} \beta _1 \kappa _+ \kappa _1 \\
	&+&\kappa _2 \big(2 A_+^{12} \beta _3 \kappa _++\sqrt{2} A_+^{23} \big(2 \rho _1-\rho _3\big) v_L\big)\big)-2 \big(2 \rho _1-\rho _3\big) v_R^2 \big(\sqrt{2} A_+^{24} \kappa _+^2 \kappa _1+2 \kappa _+ \big(A_+^{22} \kappa _1+A_+^{12} \kappa _2\big) v_L\big)\nonumber \\
	&+&2 \alpha _3 \kappa _1 \kappa _+ \big(A_+^{22} \kappa _1^2-A_+^{22} \kappa _2^2+2 A_+^{12} \kappa _1 \kappa _2\big) v_L\Big],\nonumber\\
    \end{eqnarray}}
{\small\begin{eqnarray}
	(\Delta m_{+,\phi D}^2)_{23}&=&\frac{1}{8 \kappa _+^3 \kappa _1 \Lambda ^2}\Big[-\kappa _+^3 \kappa _-^2  \big(\alpha _3 A_+^{23} \kappa _1+A_+^{24} \big(\beta _1 \kappa _1+2 \beta _3 \kappa _2\big)\big)-\kappa _+^2 v_R \big(\sqrt{2} \alpha _3 A_+^{12} \kappa _+^2 \kappa _1 \\
	&+&2 v_L \big(\kappa _+ \big(\kappa _2 \big(2 A_+^{13} \beta _3+A_+^{24} \big(2 \rho _1-\rho _3\big)\big)+A_+^{13} \beta _1 \kappa _1\big)-\sqrt{2} \big(2 \rho _1-\rho _3\big) \big(A_+^{12} \kappa _1-\big(A_+^{33}+A_+^{22}\big) \kappa _2\big) v_L\big)\big)\nonumber \\
	&+&2 \big(2 \rho _1-\rho _3\big) v_L v_R^2 \big(\sqrt{2} A_+^{34} \kappa _+^2 \kappa _1+2 \kappa _+ \big(A_+^{23} \kappa _1+A_+^{13} \kappa _2\big) v_L\big)-\sqrt{2} \kappa _+^2 v_L \big(\alpha _3 A_+^{34} \kappa _-^2 \kappa _1\nonumber \\
	&+&\beta _1 \kappa _1 \big(2 A_+^{12} \kappa _1 \kappa _2+\big(A_+^{33}+A_+^{22}\big) \kappa _1^2-\big(A_+^{33}+A_+^{22}\big) \kappa _2^2\big)\nonumber \\
	&+&2 \beta _3 \kappa _2 \big(2 A_+^{12} \kappa _1 \kappa _2+\big(A_+^{33}+A_+^{22}\big) \kappa _1^2-\big(A_+^{33}+A_+^{22}\big) \kappa _2^2\big)\big)-2 \kappa _1 \kappa _+ v_L^2 \big(\alpha _3 \big(A_+^{23} \kappa _1^2-A_+^{23} \kappa _2^2+2 A_+^{13} \kappa _1 \kappa _2\big)\nonumber \\
    &-&A_+^{23} \kappa _+^2 \big(2 \rho _1-\rho _3\big)\big)\Big]\nonumber,\\
    	\nonumber \\
	(\Delta m_{+,\phi D}^2)_{24}&=&\frac{1}{8 \kappa _+^3 \kappa _1 \Lambda ^2}\Big[-\kappa _+^3\kappa _-^2  \big(\alpha _3 A_+^{24} \kappa _1+A_+^{23} \beta _1 \kappa _1+2 A_+^{23} \beta _3 \kappa _2\big) \\
	&-&\kappa _+^2 v_R \big(2 \kappa _2 \big(\beta _3 \big(\sqrt{2} A_+^{12} \kappa _1^2+\sqrt{2} A_+^{12} \kappa _2^2+2 A_+^{14} \kappa _+ v_L\big)+\big(2 \rho _1-\rho _3\big) v_L \big(A_+^{23} \kappa _++\sqrt{2} A_+^{34} v_L\big)\big)\nonumber \\
	&+&\beta _1 \kappa _1 \big(\sqrt{2} A_+^{12} \kappa _1^2+\sqrt{2} A_+^{12} \kappa _2^2+2 A_+^{14} \kappa _+ v_L\big)\big)+ \big(4 \rho _1-2\rho _3\big) v_R^2 \big(A_+^{24} \kappa _+^3 \kappa _1 \nonumber \\
	&+&\sqrt{2} \kappa _+^2 \big(A_+^{12} \kappa _2+\big(A_+^{44}+A_+^{22}\big) \kappa _1\big) v_L+2 \kappa _+ \big(A_+^{24} \kappa _1+A_+^{14} \kappa _2\big) v_L^2\big)\nonumber \\
	&-&\sqrt{2} \kappa _+^2 v_L \big(\alpha _3 \kappa _1 \big(2 A_+^{12} \kappa _1 \kappa _2+\big(A_+^{44}+A_+^{22}\big) \kappa _1^2-\big(A_+^{44}+A_+^{22}\big) \kappa _2^2\big)+A_+^{34} \kappa _-^2 \big(\beta _1 \kappa _1+2 \beta _3 \kappa _2\big)\big)\nonumber \\
	&-&2 \alpha _3 \kappa _1 \kappa _+ \big(A_+^{24} \kappa _1^2-A_+^{24} \kappa _2^2+2 A_+^{14} \kappa _1 \kappa _2\big) v_L^2\Big],\nonumber\\
	\nonumber \\
	(\Delta m_{+,\phi D}^2)_{33}&=&\frac{1}{4 \kappa _+ \kappa _1 \Lambda ^2}\Big[-\kappa _-^2 \kappa _+ \big(\alpha _3 A_+^{33} \kappa _1+A_+^{34} \big(\beta _1 \kappa _1+2 \beta _3 \kappa _2\big)\big) \\
	&-&v_R \big(\sqrt{2} \alpha _3 A_+^{13} \kappa _+^2 \kappa _1-2 \big(2 \rho _1-\rho _3\big) v_L \big(\sqrt{2} \big(A_+^{13} \kappa _1-A_+^{23} \kappa _2\big) v_L-A_+^{34} \kappa _+ \kappa _2\big)\big)\nonumber \\
	&-&\sqrt{2} \big(A_+^{23} \kappa _1^2-A_+^{23} \kappa _2^2+2 A_+^{13} \kappa _1 \kappa _2\big) \big(\beta _1 \kappa _1+2 \beta _3 \kappa _2\big) v_L+2 A_+^{33} \kappa _+ \kappa _1 \big(2 \rho _1-\rho _3\big) v_L^2\Big],\nonumber\\
	\nonumber \\
	(\Delta m_{+,\phi D}^2)_{34}&=&\frac{1}{8 \kappa _+ \kappa _1 \Lambda ^2}\Big[-\kappa _-^2 \kappa _+ \big(2 \alpha _3 A_+^{34} \kappa _1+\big(A_+^{44}+A_+^{33}\big) \beta _1 \kappa _1+2 \big(A_+^{44}+A_+^{33}\big) \beta _3 \kappa _2\big)
	-v_R \big(\sqrt{2} \alpha _3 A_+^{14} \kappa _1^3 \nonumber \\
	&+&\sqrt{2} \alpha _3 A_+^{14} \kappa _1 \kappa _2^2+\sqrt{2} A_+^{13} \beta _1 \kappa _+^2 \kappa _1+2 \sqrt{2} A_+^{13} \beta _3 \kappa _+^2 \kappa _2+4 A_+^{44} \kappa _+ \kappa _2 \rho _1 v_L-2 A_+^{44} \kappa _+ \kappa _2 \rho _3 v_L\nonumber \\
	&+&4 A_+^{33} \kappa _+ \kappa _2 \rho _1 v_L-2 A_+^{33} \kappa _+ \kappa _2 \rho _3 v_L+4 \sqrt{2} A_+^{24} \kappa _2 \rho _1 v_L^2-2 \sqrt{2} A_+^{24} \kappa _2 \rho _3 v_L^2-4 \sqrt{2} A_+^{14} \kappa _1 \rho _1 v_L^2\nonumber \\
	&+&2 \sqrt{2} A_+^{14} \kappa _1 \rho _3 v_L^2\big)+2 \big(2 \rho _1-\rho _3\big) v_R^2 \big(A_+^{34} \kappa _+ \kappa _1+\sqrt{2} \big(A_+^{23} \kappa _1+A_+^{13} \kappa _2\big) v_L\big)\nonumber \\
	&-&\sqrt{2} v_L \big(\alpha _3 \kappa _1 \big(A_+^{23} \kappa _1^2-A_+^{23} \kappa _2^2+2 A_+^{13} \kappa _1 \kappa _2\big)+\big(A_+^{24} \kappa _1^2-A_+^{24} \kappa _2^2+2 A_+^{14} \kappa _1 \kappa _2\big) \big(\beta _1 \kappa _1+2 \beta _3 \kappa _2\big)\big)\nonumber \\
	&+&2 A_+^{34} \kappa _+ \kappa _1 \big(2 \rho _1-\rho _3\big) v_L^2\Big],\\
	\nonumber \\
	(\Delta m_{+,\phi D}^2)_{44}&=&\frac{1}{4 \kappa _+ \kappa _1 \Lambda ^2}\Big[-\kappa _+ \kappa _-^2 \big(\alpha _3 A_+^{44} \kappa _1+A_+^{34} \beta _1 \kappa _1+2 A_+^{34} \beta _3 \kappa _2\big)-v_R \big(\sqrt{2} A_+^{14} \beta _1 \kappa _+^2 \kappa _1\nonumber \\
	&+&2 \kappa _2 \big(\sqrt{2} A_+^{14} \beta _3 \kappa _+^2+A_+^{34} \kappa _+ \big(2 \rho _1-\rho _3\big) v_L\big)\big)+2 \big(2 \rho _1-\rho _3\big) v_R^2 \big(A_+^{44} \kappa _+ \kappa _1+\sqrt{2} \big(A_+^{24} \kappa _1+A_+^{14} \kappa _2\big) v_L\big)\nonumber \\
	&+&\sqrt{2} \alpha _3 \kappa _1 \big(A_+^{24} \big(-\kappa _1^2\big)+A_+^{24} \kappa _2^2-2 A_+^{14} \kappa _1 \kappa _2\big) v_L\Big].
	\end{eqnarray}}
\pagebreak

\underline{\large{Doubly charged scalar mass matrix}}
{\small\begin{eqnarray}
	(\Delta m_{++,\phi D}^2)_{11}&=&\frac{1}{2 \kappa _1^2 \Lambda ^2}\Big[A_{++}^{12} \big(\kappa _2^2 \big(\rho _3-2 \rho _1\big)-4 \kappa _1^2 \rho _4\big) v_L v_R+\kappa _1^2 \big(2 \rho _1-\rho _3\big) A_{++}^{11} v_L^2+\alpha _3 \kappa _1^2 \big(\kappa _2^2-\kappa _1^2\big) A_{++}^{11}\nonumber \\
	&-&\beta _3 \kappa _1^4 A_{++}^{12}+\beta _3 \kappa _2^4 A_{++}^{12}+\beta _1 \kappa _1 \kappa _2^3 A_{++}^{12}-\beta _1 \kappa _1^3 \kappa _2 A_{++}^{12}-4 \kappa _1^2 \rho _2 A_{++}^{11} v_R^2\Big],\\
	(\Delta m_{++,\phi D}^2)_{12}&=&-\frac{1}{4 \kappa _1^2 \Lambda ^2}\Big[2 \kappa _2^2 \rho _1 A_{++}^{22} v_L v_R-\kappa _2^2 \rho _3 A_{++}^{22} v_L v_R+4 \kappa _1^2 \rho _4 A_{++}^{22} v_L v_R+2 \kappa _2^2 \rho _1 A_{++}^{11} v_L v_R-\kappa _2^2 \rho _3 A_{++}^{11} v_L v_R\nonumber \\
	&+&4 \kappa _1^2 \rho _4 A_{++}^{11} v_L v_R-2 \kappa _1^2 \rho _1 A_{++}^{12} v_L^2+4 \kappa _1^2 \rho _2 A_{++}^{12} v_L^2+\kappa _1^2 \rho _3 A_{++}^{12} v_L^2+2 \alpha _3 \kappa _1^4 A_{++}^{12}-2 \alpha _3 \kappa _1^2 \kappa _2^2 A_{++}^{12}\nonumber \\
	&+&\beta _1 \kappa _-^2 \kappa _1 \kappa _2 \big(A_{++}^{22}+A_{++}^{11}\big)+\beta _3 \big(\kappa _1^4-\kappa _2^4\big) \big(A_{++}^{22}+A_{++}^{11}\big)-2 \kappa _1^2 \rho _1 A_{++}^{12} v_R^2+4 \kappa _1^2 \rho _2 A_{++}^{12} v_R^2\nonumber \\
	&+&\kappa _1^2 \rho _3 A_{++}^{12} v_R^2\Big],\\
	(\Delta m_{++,\phi D}^2)_{22}&=&\frac{1}{2 \kappa _1^2 \Lambda ^2}\Big[-2 \kappa _2^2 \rho _1 A_{++}^{12} v_L v_R+\kappa _2^2 \rho _3 A_{++}^{12} v_L v_R-4 \kappa _1^2 \rho _4 A_{++}^{12} v_L v_R-4 \kappa _1^2 \rho _2 A_{++}^{22} v_L^2-\alpha _3 \kappa _1^4 A_{++}^{22}\nonumber \\
	&+&\alpha _3 \kappa _1^2 \kappa _2^2 A_{++}^{22}+\beta _1 \kappa _1 \kappa _2 \big(\kappa _2^2-\kappa _1^2\big) A_{++}^{12}+\beta _3 \big(\kappa _2^4-\kappa _1^4\big) A_{++}^{12}+2 \kappa _1^2 \rho _1 A_{++}^{22} v_R^2-\kappa _1^2 \rho _3 A_{++}^{22} v_R^2\Big].
	\end{eqnarray}}


\begin{center}
	\underline{\large{\bf {Contributions from $\phi^{6}$ operators }}}
\end{center}


\underline{\large{CP-even scalar mass matrix}}

{\small\begin{eqnarray}
	(\Delta m^2_{r,\phi^6})_{11}&=&\frac{1}{4 \kappa _+^2 \Lambda ^2}\big[v_L^2 \big(v_R^2 \big(\kappa _1^2 \mathcal{C}_6^{21{LlRr}}+2 \kappa _2 \kappa _1 \mathcal{C}_6^{41{LlRr}}+\kappa _2^2 (\mathcal{C}_6^{{LlR21r}}+\mathcal{C}_6^{{RrL21l}})\big)+6 \kappa _1^2 \big(\kappa _1^2 (\mathcal{C}_6^{2121{Ll}}\nonumber\\
	&+&\mathcal{C}_6^{2211{Ll}})+2 \kappa _2 \kappa _1 \mathcal{C}_6^{2141{Ll}}+\kappa _2^2 (\mathcal{C}_6^{23{Ll41}}+\mathcal{C}_6^{2{L21l1}}+2 (\mathcal{C}_6^{4141{Ll}}+\mathcal{C}_6^{41{L2l1}}))\big)\big)+2 \kappa _1 v_L v_R \big(v_R^2 \big(\kappa _2 \mathcal{C}_6^{{RrL2r1}}\nonumber\\
	&+&\kappa _1 (\mathcal{C}_6^{{RrL4r1}}+\mathcal{C}_6^{{RrR4l1}})\big)+6 \big(\kappa _2 \kappa _1^2 \mathcal{C}_6^{21{L2r1}}+\kappa _1^3 (\mathcal{C}_6^{21{L4r1}}+\mathcal{C}_6^{21{R4l1}})+\kappa _2^2 \kappa _1 (\mathcal{C}_6^{2{L41r3}}+\mathcal{C}_6^{41{L41r}}\nonumber\\
	&+&\mathcal{C}_6^{41{R2l1}}+\mathcal{C}_6^{41{R41l}})+\kappa _2^3 \mathcal{C}_6^{41{L2r3}}\big)\big)+2 \kappa _1 v_L^3 v_R \big(\kappa _1 (\mathcal{C}_6^{{LlL4r1}}+\mathcal{C}_6^{{LlR4l1}})+\kappa _2 \mathcal{C}_6^{{LlR2l1}}\big)+v_L^4 \big(\kappa _1^2 (\mathcal{C}_6^{\{21\}\{{LlLl}\}}\nonumber\\
	&+&\mathcal{C}_6^{21{LlLl}})+\kappa _2^2 (\mathcal{C}_6^{\{21\}\{{LlLl}\}}+\mathcal{C}_6^{{LlL21l}})+2 \kappa _2 \kappa _1 \mathcal{C}_6^{41{LlLl}}\big)+6 \kappa _1^2 v_R^2 \big(\kappa _1^2 (\mathcal{C}_6^{2121{Rr}}+\mathcal{C}_6^{2211{Rr}})\nonumber\\
	&+&2 \kappa _2 \kappa _1 \mathcal{C}_6^{2141{Rr}}+\kappa _2^2 (\mathcal{C}_6^{23{Rr41}}+\mathcal{C}_6^{2{R21r1}}+2 (\mathcal{C}_6^{4141{Rr}}+\mathcal{C}_6^{41{R2r1}}))\big)+v_R^4 \big(\kappa _1^2 (\mathcal{C}_6^{\{21\}\{{RrRr}\}}+\mathcal{C}_6^{21{RrRr}})\nonumber\\
	&+&\kappa _2^2 (\mathcal{C}_6^{\{21\}\{{RrRr}\}}+\mathcal{C}_6^{{RrR21r}})+2 \kappa _2 \kappa _1 \mathcal{C}_6^{41{RrRr}}\big)+15 \big(\kappa _1^6 (\mathcal{C}_6^{\{21\}\{2121\}}+\mathcal{C}_6^{212121})+\kappa _2^6 (\mathcal{C}_6^{\{21\}\{2121\}}+\mathcal{C}_6^{212121})\nonumber\\
	&+&\kappa _2^2 \kappa _1^4 (\mathcal{C}_6^{\{21\}\{2121\}}+2 \mathcal{C}_6^{414121})+\kappa _2^4 \kappa _1^2 (\mathcal{C}_6^{\{21\}\{2121\}}+2 \mathcal{C}_6^{414121})+2 \kappa _2 \kappa _1^5 \mathcal{C}_6^{412121}+2 \kappa _2^5 \kappa _1 \mathcal{C}_6^{412121}\nonumber\\
	&+&4 \kappa _2^3 \kappa _1^3 (\mathcal{C}_6^{414123}+\mathcal{C}_6^{414141})\big)\big], \\
    \nonumber \\
	(\Delta m^2_{r,\phi^6})_{13}&=&\frac{1}{2\kappa _+ \Lambda ^2}\big[\kappa _1 v_L \big(3 v_R^2 \big(\kappa _2 \mathcal{C}_6^{{RrL2r1}}+\kappa _1 (\mathcal{C}_6^{{RrL4r1}}+\mathcal{C}_6^{{RrR4l1}})\big)+2 \big(\kappa _2 \kappa _1^2 \mathcal{C}_6^{21{L2r1}}\nonumber\\
	&+&\kappa _1^3 (\mathcal{C}_6^{21{L4r1}}+\mathcal{C}_6^{21{R4l1}})+\kappa _2^2 \kappa _1 (\mathcal{C}_6^{2{L41r3}}+\mathcal{C}_6^{41{L41r}}+\mathcal{C}_6^{41{R2l1}}+\mathcal{C}_6^{41{R41l}})+\kappa _2^3 \mathcal{C}_6^{41{L2r3}}\big)\big)+v_L^2 v_R \big(\kappa _1^2 \mathcal{C}_6^{21{LlRr}}\nonumber\\
	&+&2 \kappa _2 \kappa _1 \mathcal{C}_6^{41{LlRr}}+\kappa _2^2 (\mathcal{C}_6^{{LlR21r}}+\mathcal{C}_6^{{RrL21l}})\big)+\kappa _1 v_L^3 \big(\kappa _1 (\mathcal{C}_6^{{LlL4r1}}+\mathcal{C}_6^{{LlR4l1}})+\kappa _2 \mathcal{C}_6^{{LlR2l1}}\big)\nonumber\\
	&+&v_R \big(v_R^2 \big(\kappa _1^2 (\mathcal{C}_6^{\{21\}\{{RrRr}\}}+2 \mathcal{C}_6^{21{RrRr}})+\kappa _2^2 (\mathcal{C}_6^{\{21\}\{{RrRr}\}}+2 \mathcal{C}_6^{{RrR21r}})+4 \kappa _2 \kappa _1 \mathcal{C}_6^{41{RrRr}}\big)+2 \kappa _1^2 \big(\kappa _1^2 (\mathcal{C}_6^{2121{Rr}}\nonumber\\
	&+&\mathcal{C}_6^{2211{Rr}})+2 \kappa _2 \kappa _1 \mathcal{C}_6^{2141{Rr}}+\kappa _2^2 (\mathcal{C}_6^{23{Rr41}}+\mathcal{C}_6^{2{R21r1}}+2 (\mathcal{C}_6^{4141{Rr}}+\mathcal{C}_6^{41{R2r1}}))\big)\big)\big],
	 \end{eqnarray}}
    {\small\begin{eqnarray}
    (\Delta m^2_{r,\phi^6})_{12}&=&\frac{1}{4 \kappa _+^2 \Lambda ^2}\big[v_L^2 \big(v_R^2 \big(\kappa _2 \kappa _1 (-\mathcal{C}_6^{21{LlRr}}+\mathcal{C}_6^{{LlR21r}}+\mathcal{C}_6^{{RrL21l}})+\kappa _1^2 \mathcal{C}_6^{41{LlRr}}-\kappa _2^2 \mathcal{C}_6^{41{LlRr}}\big)+3 \kappa _1 \big(\kappa _2 \kappa _1^2 (-2 \mathcal{C}_6^{2121{Ll}}\nonumber\\
    	&-&2 \mathcal{C}_6^{2211{Ll}}+\mathcal{C}_6^{23{Ll41}}+\mathcal{C}_6^{2{L21l1}}+2 \mathcal{C}_6^{4141{Ll}}+2 \mathcal{C}_6^{41{L2l1}})+\kappa _1^3 \mathcal{C}_6^{2141{Ll}}-3 \kappa _2^2 \kappa _1 \mathcal{C}_6^{2141{Ll}}-\kappa _2^3 (\mathcal{C}_6^{23{Ll41}}+\mathcal{C}_6^{2{L21l1}}\nonumber\\
    	&+&2 (\mathcal{C}_6^{4141{Ll}}+\mathcal{C}_6^{41{L2l1}}))\big)\big)+v_L v_R \big(v_R^2 \big(\kappa _1^2 \mathcal{C}_6^{{RrL2r1}}-\kappa _2^2 \mathcal{C}_6^{{RrL2r1}}-2 \kappa _2 \kappa _1 (\mathcal{C}_6^{{RrL4r1}}+\mathcal{C}_6^{{RrR4l1}})\big)\nonumber\\
    	&+&3 \big(\kappa _1^4 \mathcal{C}_6^{21{L2r1}}+3 \kappa _2^2 \kappa _1^2 (\mathcal{C}_6^{41{L2r3}}-\mathcal{C}_6^{21{L2r1}})+2 \kappa _2 \kappa _1^3 (-2 \mathcal{C}_6^{21{L4r1}}-2 \mathcal{C}_6^{21{R4l1}}+\mathcal{C}_6^{2{L41r3}}\nonumber\\
    	&+&\mathcal{C}_6^{41{L41r}}+\mathcal{C}_6^{41{R2l1}}+\mathcal{C}_6^{41{R41l}})-2 \kappa _2^3 \kappa _1 (\mathcal{C}_6^{2{L41r3}}+\mathcal{C}_6^{41{L41r}}+\mathcal{C}_6^{41{R2l1}}+\mathcal{C}_6^{41{R41l}})-\kappa _2^4 \mathcal{C}_6^{41{L2r3}}\big)\big)\nonumber\\
    	&+&v_L^3 v_R \big(-2 \kappa _2 \kappa _1 (\mathcal{C}_6^{{LlL4r1}}+\mathcal{C}_6^{{LlR4l1}})+\kappa _1^2 \mathcal{C}_6^{{LlR2l1}}-\kappa _2^2 \mathcal{C}_6^{{LlR2l1}}\big)+v_L^4 \big(\kappa _2 \kappa _1 (\mathcal{C}_6^{{LlL21l}}-\mathcal{C}_6^{21{LlLl}})\nonumber\\
    	&+&\kappa _1^2 \mathcal{C}_6^{41{LlLl}}-\kappa _2^2 \mathcal{C}_6^{41{LlLl}}\big)+3 \kappa _1 v_R^2 \big(\kappa _2 \kappa _1^2 (-2 \mathcal{C}_6^{2121{Rr}}-2 \mathcal{C}_6^{2211{Rr}}+\mathcal{C}_6^{23{Rr41}}+\mathcal{C}_6^{2{R21r1}}+2 \mathcal{C}_6^{4141{Rr}}+2 \mathcal{C}_6^{41{R2r1}})\nonumber\\
    	&+&\kappa _1^3 \mathcal{C}_6^{2141{Rr}}-3 \kappa _2^2 \kappa _1 \mathcal{C}_6^{2141{Rr}}-\kappa _2^3 (\mathcal{C}_6^{23{Rr41}}+\mathcal{C}_6^{2{R21r1}}+2 (\mathcal{C}_6^{4141{Rr}}+\mathcal{C}_6^{41{R2r1}}))\big)+v_R^4 \big(\kappa _2 \kappa _1 (\mathcal{C}_6^{{RrR21r}}\nonumber\\
    	&-&\mathcal{C}_6^{21{RrRr}})+\kappa _1^2 \mathcal{C}_6^{41{RrRr}}-\kappa _2^2 \mathcal{C}_6^{41{RrRr}}\big)+5 \kappa _-^2 \big(\kappa _2 \kappa _1^3 (-2 \mathcal{C}_6^{\{21\}\{2121\}}-3 \mathcal{C}_6^{212121}+2 \mathcal{C}_6^{414121})\nonumber\\
    	&+&\kappa _2^3 \kappa _1 (-2 \mathcal{C}_6^{\{21\}\{2121\}}-3 \mathcal{C}_6^{212121}+2 \mathcal{C}_6^{414121})+\kappa _1^4 \mathcal{C}_6^{412121}+\kappa _2^4 \mathcal{C}_6^{412121}+2 \kappa _2^2 \kappa _1^2 (3 (\mathcal{C}_6^{414123}+\mathcal{C}_6^{414141})\nonumber\\
    	&-&2 \mathcal{C}_6^{412121})\big)\big],\\ \nonumber\\
    (\Delta m^2_{r,\phi^6})_{14}&=&\frac{1}{2 \kappa_+ \Lambda ^2}\big[v_L \big(v_R^2 \big(\kappa _1^2 \mathcal{C}_6^{21{LlRr}}+2 \kappa _2 \kappa _1 \mathcal{C}_6^{41{LlRr}}+\kappa _2^2 (\mathcal{C}_6^{{LlR21r}}+\mathcal{C}_6^{{RrL21l}})\big)+2 \kappa _1^2 \big(\kappa _1^2 (\mathcal{C}_6^{2121{Ll}}\nonumber\\
    &+&\mathcal{C}_6^{2211{Ll}})+2 \kappa _2 \kappa _1 \mathcal{C}_6^{2141{Ll}}+\kappa _2^2 (\mathcal{C}_6^{23{Ll41}}+\mathcal{C}_6^{2{L21l1}}+2 (\mathcal{C}_6^{4141{Ll}}+\mathcal{C}_6^{41{L2l1}}))\big)\big)+3 \kappa _1 v_L^2 v_R \big(\kappa _1 (\mathcal{C}_6^{{LlL4r1}}\nonumber\\
    &+&\mathcal{C}_6^{{LlR4l1}})+\kappa _2 \mathcal{C}_6^{{LlR2l1}}\big)+2 v_L^3 \big(\kappa _1^2 (\mathcal{C}_6^{\{21\}\{{LlLl}\}}+\mathcal{C}_6^{21{LlLl}})+\kappa _2^2 (\mathcal{C}_6^{\{21\}\{{LlLl}\}}+\mathcal{C}_6^{{LlL21l}})+2 \kappa _2 \kappa _1 \mathcal{C}_6^{41{LlLl}}\big)\nonumber\\
    &+&\kappa _1 v_R \big(v_R^2 \big(\kappa _2 \mathcal{C}_6^{{RrL2r1}}+\kappa _1 (\mathcal{C}_6^{{RrL4r1}}+\mathcal{C}_6^{{RrR4l1}})\big)+2 \big(\kappa _2 \kappa _1^2 \mathcal{C}_6^{21{L2r1}}+\kappa _1^3 (\mathcal{C}_6^{21{L4r1}}+\mathcal{C}_6^{21{R4l1}})\nonumber\\
    &+&\kappa _2^2 \kappa _1 (\mathcal{C}_6^{2{L41r3}}+\mathcal{C}_6^{41{L41r}}+\mathcal{C}_6^{41{R2l1}}+\mathcal{C}_6^{41{R41l}})+\kappa _2^3 \mathcal{C}_6^{41{L2r3}}\big)\big)\big], \nonumber \\
    \nonumber \\
    (\Delta m^2_{r,\phi^6})_{22}&=&\frac{1}{4 \kappa _+^2 \Lambda ^2}\big[v_L^2 \big(v_R^2 \big(\kappa _2^2 \mathcal{C}_6^{21{LlRr}}-2 \kappa _2 \kappa _1 \mathcal{C}_6^{41{LlRr}}+\kappa _1^2 (\mathcal{C}_6^{{LlR21r}}+\mathcal{C}_6^{{RrL21l}})\big)+2 \kappa _2^2 \kappa _1^2 (3 \mathcal{C}_6^{2121{Ll}}+3 \mathcal{C}_6^{2211{Ll}}\nonumber\\
    &-&2 (\mathcal{C}_6^{23{Ll41}}+\mathcal{C}_6^{2{L21l1}}+2 (\mathcal{C}_6^{4141{Ll}}+\mathcal{C}_6^{41{L2l1}})))-6 \kappa _2 \kappa _1^3 \mathcal{C}_6^{2141{Ll}}+6 \kappa _2^3 \kappa _1 \mathcal{C}_6^{2141{Ll}}+\kappa _1^4 (\mathcal{C}_6^{23{Ll41}}+\mathcal{C}_6^{2{L21l1}}\nonumber\\
    &+&2 (\mathcal{C}_6^{4141{Ll}}+\mathcal{C}_6^{41{L2l1}}))+\kappa _2^4 (\mathcal{C}_6^{23{Ll41}}+\mathcal{C}_6^{2{L21l1}}+2 (\mathcal{C}_6^{4141{Ll}}+\mathcal{C}_6^{41{L2l1}}))\big)+2 v_L v_R \big(-\kappa _1 \big(\kappa _2 \mathcal{C}_6^{{RrL2r1}} v_R^2\nonumber\\
    &+&3 \kappa _2^3 (\mathcal{C}_6^{41{L2r3}}-\mathcal{C}_6^{21{L2r1}})\big)+\kappa _2^2 \big(v_R^2 (\mathcal{C}_6^{{RrL4r1}}+\mathcal{C}_6^{{RrR4l1}})+\kappa _2^2 (\mathcal{C}_6^{2{L41r3}}+\mathcal{C}_6^{41{L41r}}+\mathcal{C}_6^{41{R2l1}}\nonumber\\
    &+&\mathcal{C}_6^{41{R41l}})\big)+3 \kappa _2 \kappa _1^3 (\mathcal{C}_6^{41{L2r3}}-\mathcal{C}_6^{21{L2r1}})+2 \kappa _2^2 \kappa _1^2 (3 \mathcal{C}_6^{21{L4r1}}+3 \mathcal{C}_6^{21{R4l1}}-2 (\mathcal{C}_6^{2{L41r3}}+\mathcal{C}_6^{41{L41r}}+\mathcal{C}_6^{41{R2l1}}\nonumber\\
    &+&\mathcal{C}_6^{41{R41l}}))+\kappa _1^4 (\mathcal{C}_6^{2{L41r3}}+\mathcal{C}_6^{41{L41r}}+\mathcal{C}_6^{41{R2l1}}+\mathcal{C}_6^{41{R41l}})\big)+2 \kappa _2 v_L^3 v_R \big(\kappa _2 (\mathcal{C}_6^{{LlL4r1}}+\mathcal{C}_6^{{LlR4l1}})-\kappa _1 \mathcal{C}_6^{{LlR2l1}}\big)\nonumber\\
    &+&v_L^4 \big(\kappa _2^2 (\mathcal{C}_6^{\{21\}\{{LlLl}\}}+\mathcal{C}_6^{21{LlLl}})+\kappa _1^2 (\mathcal{C}_6^{\{21\}\{{LlLl}\}}+\mathcal{C}_6^{{LlL21l}})-2 \kappa _2 \kappa _1 \mathcal{C}_6^{41{LlLl}}\big)+v_R^2 \big(2 \kappa _2^2 \kappa _1^2 (3 \mathcal{C}_6^{2121{Rr}}\nonumber\\
    &+&3 \mathcal{C}_6^{2211{Rr}}-2 (\mathcal{C}_6^{23{Rr41}}+\mathcal{C}_6^{2{R21r1}}+2 (\mathcal{C}_6^{4141{Rr}}+\mathcal{C}_6^{41{R2r1}})))-6 \kappa _2 \kappa _1^3 \mathcal{C}_6^{2141{Rr}}+6 \kappa _2^3 \kappa _1 \mathcal{C}_6^{2141{Rr}}+\kappa _1^4 (\mathcal{C}_6^{23{Rr41}}\nonumber\\
    &+&\mathcal{C}_6^{2{R21r1}}+2 (\mathcal{C}_6^{4141{Rr}}+\mathcal{C}_6^{41{R2r1}}))+\kappa _2^4 (\mathcal{C}_6^{23{Rr41}}+\mathcal{C}_6^{2{R21r1}}+2 (\mathcal{C}_6^{4141{Rr}}+\mathcal{C}_6^{41{R2r1}}))\big)+v_R^4 \big(\kappa _2^2 (\mathcal{C}_6^{\{21\}\{{RrRr}\}}\nonumber\\
    &+&\mathcal{C}_6^{21{RrRr}})+\kappa _1^2 (\mathcal{C}_6^{\{21\}\{{RrRr}\}}+\mathcal{C}_6^{{RrR21r}})-2 \kappa _2 \kappa _1 \mathcal{C}_6^{41{RrRr}}\big)+\kappa _1^6 \mathcal{C}_6^{\{21\}\{2121\}}+13 \kappa _2^2 \kappa _1^4 \mathcal{C}_6^{\{21\}\{2121\}}\nonumber\\
    &+&13 \kappa _2^4 \kappa _1^2 \mathcal{C}_6^{\{21\}\{2121\}}+\kappa _2^6 \mathcal{C}_6^{\{21\}\{2121\}}+15 \kappa _2^2 \kappa _1^4 \mathcal{C}_6^{212121}+15 \kappa _2^4 \kappa _1^2 \mathcal{C}_6^{212121}-10 \kappa _2 \kappa _1^5 \mathcal{C}_6^{412121}+40 \kappa _2^3 \kappa _1^3 \mathcal{C}_6^{412121}\nonumber\\
    &-&10 \kappa _2^5 \kappa _1 \mathcal{C}_6^{412121}+2 \kappa _1^6 \mathcal{C}_6^{414121}-4 \kappa _2^2 \kappa _1^4 \mathcal{C}_6^{414121}-4 \kappa _2^4 \kappa _1^2 \mathcal{C}_6^{414121}+2 \kappa _2^6 \mathcal{C}_6^{414121}+12 \kappa _2 \kappa _1^5 \mathcal{C}_6^{414123}\nonumber\\
    &-&36 \kappa _2^3 \kappa _1^3 \mathcal{C}_6^{414123}+12 \kappa _2^5 \kappa _1 \mathcal{C}_6^{414123}+12 \kappa _2 \kappa _1^5 \mathcal{C}_6^{414141}-36 \kappa _2^3 \kappa _1^3 \mathcal{C}_6^{414141}+12 \kappa _2^5 \kappa _1 \mathcal{C}_6^{414141}\big],
  \end{eqnarray}}
{\small\begin{eqnarray}
	(\Delta m^2_{r,\phi^6})_{23}&=&\frac{1}{4 \kappa_+ \Lambda ^2}\big[v_L \big(3 v_R^2 \big(\kappa _1^2 \mathcal{C}_6^{{RrL2r1}}-\kappa _2^2 \mathcal{C}_6^{{RrL2r1}}-2 \kappa _2 \kappa _1 (\mathcal{C}_6^{{RrL4r1}}+\mathcal{C}_6^{{RrR4l1}})\big)+\kappa _1^4 \mathcal{C}_6^{21{L2r1}}\nonumber\\
	&+&3 \kappa _2^2 \kappa _1^2 (\mathcal{C}_6^{41{L2r3}}-\mathcal{C}_6^{21{L2r1}})+2 \kappa _2 \kappa _1^3 (-2 \mathcal{C}_6^{21{L4r1}}-2 \mathcal{C}_6^{21{R4l1}}+\mathcal{C}_6^{2{L41r3}}+\mathcal{C}_6^{41{L41r}}+\mathcal{C}_6^{41{R2l1}}+\mathcal{C}_6^{41{R41l}})\nonumber\\
	&-&2 \kappa _2^3 \kappa _1 (\mathcal{C}_6^{2{L41r3}}+\mathcal{C}_6^{41{L41r}}+\mathcal{C}_6^{41{R2l1}}+\mathcal{C}_6^{41{R41l}})-\kappa _2^4 \mathcal{C}_6^{41{L2r3}}\big)+2 v_L^2 v_R \big(\kappa _2 \kappa _1 (-\mathcal{C}_6^{21{LlRr}}+\mathcal{C}_6^{{LlR21r}}\nonumber\\
	&+&\mathcal{C}_6^{{RrL21l}})+\kappa _1^2 \mathcal{C}_6^{41{LlRr}}-\kappa _2^2 \mathcal{C}_6^{41{LlRr}}\big)+v_L^3 \big(-2 \kappa _2 \kappa _1 (\mathcal{C}_6^{{LlL4r1}}+\mathcal{C}_6^{{LlR4l1}})+\kappa _1^2 \mathcal{C}_6^{{LlR2l1}}-\kappa _2^2 \mathcal{C}_6^{{LlR2l1}}\big)\nonumber\\
	&-&2 v_R \big(\kappa _1 \big(\kappa _2 \kappa _1^2 (2 \mathcal{C}_6^{2121{Rr}}+2 \mathcal{C}_6^{2211{Rr}}-\mathcal{C}_6^{23{Rr41}}-\mathcal{C}_6^{2{R21r1}}-2 \mathcal{C}_6^{4141{Rr}}-2 \mathcal{C}_6^{41{R2r1}})+\kappa _1^3 (-\mathcal{C}_6^{2141{Rr}})\nonumber\\
	&+&3 \kappa _2^2 \kappa _1 \mathcal{C}_6^{2141{Rr}}+\kappa _2^3 (\mathcal{C}_6^{23{Rr41}}+\mathcal{C}_6^{2{R21r1}}+2 (\mathcal{C}_6^{4141{Rr}}+\mathcal{C}_6^{41{R2r1}}))\big)-2 v_R^2 \big(\kappa _2 \kappa _1 (\mathcal{C}_6^{{RrR21r}}\nonumber\\
	&-&\mathcal{C}_6^{21{RrRr}})+\kappa _1^2 \mathcal{C}_6^{41{RrRr}}-\kappa _2^2 \mathcal{C}_6^{41{RrRr}}\big)\big)\big],\\ \nonumber\\
	(\Delta m^2_{r,\phi^6})_{24}&=&\frac{1}{4 \kappa_+ \Lambda ^2}\big[2 v_L \big(v_R^2 \big(\kappa _2 \kappa _1 (-\mathcal{C}_6^{21{LlRr}}+\mathcal{C}_6^{{LlR21r}}+\mathcal{C}_6^{{RrL21l}})+\kappa _1^2 \mathcal{C}_6^{41{LlRr}}-\kappa _2^2 \mathcal{C}_6^{41{LlRr}}\big)+\kappa _1 \big(\kappa _2 \kappa _1^2 (-2 \mathcal{C}_6^{2121{Ll}}\nonumber\\
	&-&2 \mathcal{C}_6^{2211{Ll}}+\mathcal{C}_6^{23{Ll41}}+\mathcal{C}_6^{2{L21l1}}+2 \mathcal{C}_6^{4141{Ll}}+2 \mathcal{C}_6^{41{L2l1}})+\kappa _1^3 \mathcal{C}_6^{2141{Ll}}-3 \kappa _2^2 \kappa _1 \mathcal{C}_6^{2141{Ll}}-\kappa _2^3 (\mathcal{C}_6^{23{Ll41}}\nonumber\\
	&+&\mathcal{C}_6^{2{L21l1}}+2 (\mathcal{C}_6^{4141{Ll}}+\mathcal{C}_6^{41{L2l1}}))\big)\big)+3 v_L^2 v_R \big(-2 \kappa _2 \kappa _1 (\mathcal{C}_6^{{LlL4r1}}+\mathcal{C}_6^{{LlR4l1}})+\kappa _1^2 \mathcal{C}_6^{{LlR2l1}}-\kappa _2^2 \mathcal{C}_6^{{LlR2l1}}\big)\nonumber\\
	&+&4 v_L^3 \big(\kappa _2 \kappa _1 (\mathcal{C}_6^{{LlL21l}}-\mathcal{C}_6^{21{LlLl}})+\kappa _1^2 \mathcal{C}_6^{41{LlLl}}-\kappa _2^2 \mathcal{C}_6^{41{LlLl}}\big)+v_R \big(v_R^2 \big(\kappa _1^2 \mathcal{C}_6^{{RrL2r1}}-\kappa _2^2 \mathcal{C}_6^{{RrL2r1}}\nonumber\\
	&-&2 \kappa _2 \kappa _1 (\mathcal{C}_6^{{RrL4r1}}+\mathcal{C}_6^{{RrR4l1}})\big)+\kappa _1^4 \mathcal{C}_6^{21{L2r1}}+3 \kappa _2^2 \kappa _1^2 (\mathcal{C}_6^{41{L2r3}}-\mathcal{C}_6^{21{L2r1}})+2 \kappa _2 \kappa _1^3 (-2 \mathcal{C}_6^{21{L4r1}}\nonumber\\
	&-&2 \mathcal{C}_6^{21{R4l1}}+\mathcal{C}_6^{2{L41r3}}+\mathcal{C}_6^{41{L41r}}+\mathcal{C}_6^{41{R2l1}}+\mathcal{C}_6^{41{R41l}})-2 \kappa _2^3 \kappa _1 (\mathcal{C}_6^{2{L41r3}}+\mathcal{C}_6^{41{L41r}}+\mathcal{C}_6^{41{R2l1}}+\mathcal{C}_6^{41{R41l}})\nonumber\\
	&-&\kappa _2^4 \mathcal{C}_6^{41{L2r3}}\big)\big],\\ \nonumber\\
	(\Delta m^2_{r,\phi^6})_{33}&=&\frac{1}{4 \Lambda ^2}\big[v_L^2 \big(6 \mathcal{C}_6^{{RrRrLl}} v_R^2+\kappa _1^2 \mathcal{C}_6^{21{LlRr}}+2 \kappa _1 \kappa _2 \mathcal{C}_6^{41{LlRr}}+\kappa _2^2 (\mathcal{C}_6^{{LlR21r}}+\mathcal{C}_6^{{RrL21l}})\big)+6 \kappa _1 v_L v_R \big(\kappa _2 \mathcal{C}_6^{{RrL2r1}}\nonumber\\
	&+&\kappa _1 (\mathcal{C}_6^{{RrL4r1}}+\mathcal{C}_6^{{RrR4l1}})\big)+\mathcal{C}_6^{{LlLlRr}} v_L^4+v_R^2 \big(\kappa _1^2 (\mathcal{C}_6^{\{21\}\{{RrRr}\}}+6 \mathcal{C}_6^{21{RrRr}})+\kappa _2^2 (\mathcal{C}_6^{\{21\}\{{RrRr}\}}+6 \mathcal{C}_6^{{RrR21r}})\nonumber\\
	&+&12 \kappa _2 \kappa _1 \mathcal{C}_6^{41{RrRr}}\big)+15 \mathcal{C}_6^{{RrRrRr}} v_R^4+\kappa _1^2 \big(\kappa _1^2 (\mathcal{C}_6^{2121{Rr}}+\mathcal{C}_6^{2211{Rr}})+2 \kappa _2 \kappa _1 \mathcal{C}_6^{2141{Rr}}+\kappa _2^2 (\mathcal{C}_6^{23{Rr41}}\nonumber\\
	&+&\mathcal{C}_6^{2{R21r1}}+2 (\mathcal{C}_6^{4141{Rr}}+\mathcal{C}_6^{41{R2r1}}))\big)\big],\\ \nonumber\\
	(\Delta m^2_{r,\phi^6})_{34}&=&\frac{1}{4 \Lambda ^2}\big[2 v_L v_R \big(2 \mathcal{C}_6^{{RrRrLl}} v_R^2+\kappa _1^2 \mathcal{C}_6^{21{LlRr}}+2 \kappa _1 \kappa _2 \mathcal{C}_6^{41{LlRr}}+\kappa _2^2 (\mathcal{C}_6^{{LlR21r}}+\mathcal{C}_6^{{RrL21l}})\big)+4 \mathcal{C}_6^{{LlLlRr}} v_L^3 v_R\nonumber\\
	&+&3 \kappa _1 v_L^2 \big(\kappa _1 (\mathcal{C}_6^{{LlL4r1}}+\mathcal{C}_6^{{LlR4l1}})+\kappa _2 \mathcal{C}_6^{{LlR2l1}}\big)+\kappa _1 \big(3 v_R^2 \big(\kappa _2 \mathcal{C}_6^{{RrL2r1}}+\kappa _1 (\mathcal{C}_6^{{RrL4r1}}+\mathcal{C}_6^{{RrR4l1}})\big)\\
	&+&\kappa _2 \kappa _1^2 \mathcal{C}_6^{21{L2r1}}+\kappa _1^3 (\mathcal{C}_6^{21{L4r1}}+\mathcal{C}_6^{21{R4l1}})+\kappa _2^2 \kappa _1 (\mathcal{C}_6^{2{L41r3}}+\mathcal{C}_6^{41{L41r}}+\mathcal{C}_6^{41{R2l1}}+\mathcal{C}_6^{41{R41l}})+\kappa _2^3 \mathcal{C}_6^{41{L2r3}}\big)\big],\nonumber
	\end{eqnarray}}
{\small\begin{eqnarray}
	(\Delta m^2_{r,\phi^6})_{44}&=&\frac{1}{4 \Lambda ^2}\big[6 v_L^2 \big(\mathcal{C}_6^{{LlLlRr}} v_R^2+\kappa _1^2 (\mathcal{C}_6^{\{21\}\{{LlLl}\}}+\mathcal{C}_6^{21{LlLl}})+\kappa _2^2 (\mathcal{C}_6^{\{21\}\{{LlLl}\}}+\mathcal{C}_6^{{LlL21l}})+2 \kappa _1 \kappa _2 \mathcal{C}_6^{41{LlLl}}\big)\nonumber\\
	&+&6 \kappa _1 v_L v_R \big(\kappa _1 (\mathcal{C}_6^{{LlL4r1}}+\mathcal{C}_6^{{LlR4l1}})+\kappa _2 \mathcal{C}_6^{{LlR2l1}}\big)+15 \mathcal{C}_6^{{LlLlLl}} v_L^4+v_R^2 \big(\kappa _1^2 \mathcal{C}_6^{21{LlRr}}+2 \kappa _2 \kappa _1 \mathcal{C}_6^{41{LlRr}}\nonumber\\
	&+&\kappa _2^2 (\mathcal{C}_6^{{LlR21r}}+\mathcal{C}_6^{{RrL21l}})\big)+\mathcal{C}_6^{{RrRrLl}} v_R^4+\kappa _1^2 \big(\kappa _1^2 (\mathcal{C}_6^{2121{Ll}}+\mathcal{C}_6^{2211{Ll}})+2 \kappa _2 \kappa _1 \mathcal{C}_6^{2141{Ll}}+\kappa _2^2 (\mathcal{C}_6^{23{Ll41}}\nonumber\\
	&+&\mathcal{C}_6^{2{L21l1}}+2 (\mathcal{C}_6^{4141{Ll}}+\mathcal{C}_6^{41{L2l1}}))\big)\big].
	\end{eqnarray}}

\pagebreak

\underline{\large{CP-odd scalar mass matrix}}

{\small\begin{eqnarray}
	(\Delta m^2_{i,\phi^6})_{11}&=&\frac{1}{4 \kappa _+^2 \Lambda ^2}\big[v_L^2 \big(v_R^2 \big(\kappa _1^2 \mathcal{C}_6^{21{LlRr}}+2 \kappa _2 \kappa _1 \mathcal{C}_6^{41{LlRr}}+\kappa _2^2 (\mathcal{C}_6^{{LlR21r}}+\mathcal{C}_6^{{RrL21l}})\big)+2 \kappa _1^2 \big(\kappa _1^2 (\mathcal{C}_6^{2121{Ll}}\nonumber\\
	&+&\mathcal{C}_6^{2211{Ll}})+2 \kappa _2 \kappa _1 \mathcal{C}_6^{2141{Ll}}+\kappa _2^2 (\mathcal{C}_6^{23{Ll41}}+\mathcal{C}_6^{2{L21l1}}+2 \mathcal{C}_6^{4141{Ll}})\big)\big)-2 \kappa _1 v_L v_R \big(v_R^2 \big(\kappa _2 \mathcal{C}_6^{{RrL2r1}}\nonumber\\
	&+&\kappa _1 (\mathcal{C}_6^{{RrL4r1}}+\mathcal{C}_6^{{RrR4l1}})\big)-2 \kappa _1 \kappa _2^2 (\mathcal{C}_6^{2{L41r3}}+\mathcal{C}_6^{41{L41r}}+\mathcal{C}_6^{41{R41l}})\big)-2 \kappa _1 v_L^3 v_R \big(\kappa _1 (\mathcal{C}_6^{{LlL4r1}}+\mathcal{C}_6^{{LlR4l1}})\nonumber\\
	&+&\kappa _2 \mathcal{C}_6^{{LlR2l1}}\big)+v_L^4 \big(\kappa _1^2 (\mathcal{C}_6^{\{21\}\{{LlLl}\}}+\mathcal{C}_6^{21{LlLl}})+\kappa _2^2 (\mathcal{C}_6^{\{21\}\{{LlLl}\}}+\mathcal{C}_6^{{LlL21l}})+2 \kappa _2 \kappa _1 \mathcal{C}_6^{41{LlLl}}\big)\nonumber\\
	&+&2 \kappa _1^2 v_R^2 \big(\kappa _1^2 (\mathcal{C}_6^{2121{Rr}}+\mathcal{C}_6^{2211{Rr}})+2 \kappa _2 \kappa _1 \mathcal{C}_6^{2141{Rr}}+\kappa _2^2 (\mathcal{C}_6^{23{Rr41}}+\mathcal{C}_6^{2{R21r1}}+2 \mathcal{C}_6^{4141{Rr}})\big)\nonumber\\
	&+&v_R^4 \big(\kappa _1^2 (\mathcal{C}_6^{\{21\}\{{RrRr}\}}+\mathcal{C}_6^{21{RrRr}})+\kappa _2^2 (\mathcal{C}_6^{\{21\}\{{RrRr}\}}+\mathcal{C}_6^{{RrR21r}})+2 \kappa _2 \kappa _1 \mathcal{C}_6^{41{RrRr}}\big)+3 \big(2 \kappa _2 \kappa _1^5 \mathcal{C}_6^{412121}\nonumber\\
	&+&\kappa _1^6 (\mathcal{C}_6^{\{21\}\{2121\}}+\mathcal{C}_6^{212121})+\kappa _2^6 (\mathcal{C}_6^{\{21\}\{2121\}}+\mathcal{C}_6^{212121})+\kappa _2^2 \kappa _1^4 (\mathcal{C}_6^{\{21\}\{2121\}}+2 \mathcal{C}_6^{414121})\nonumber\\
	&+&\kappa _2^4 \kappa _1^2 (\mathcal{C}_6^{\{21\}\{2121\}}+2 \mathcal{C}_6^{414121})+2 \kappa _2^5 \kappa _1 \mathcal{C}_6^{412121}+4 \kappa _2^3 \kappa _1^3 (\mathcal{C}_6^{414123}+\mathcal{C}_6^{414141})\big)\big],\\ \nonumber\\
	(\Delta m^2_{i,\phi^6})_{12}&=&\frac{1}{4 \kappa _+^2 \Lambda^2}\big[v_L^2 \big(v_R^2 \big(\kappa _2 \kappa _1 (-\mathcal{C}_6^{21{LlRr}}+\mathcal{C}_6^{{LlR21r}}+\mathcal{C}_6^{{RrL21l}})+\kappa _1^2 \mathcal{C}_6^{41{LlRr}}-\kappa _2^2 \mathcal{C}_6^{41{LlRr}}\big)\nonumber \\
	&+&\kappa _1 \big(\kappa _2 \kappa _1^2 (-2 \mathcal{C}_6^{2121{Ll}}-2 \mathcal{C}_6^{2211{Ll}}+\mathcal{C}_6^{23{Ll41}}+\mathcal{C}_6^{2{L21l1}}+2 \mathcal{C}_6^{4141{Ll}}+2 \mathcal{C}_6^{41{L2l1}})+\kappa _1^3 \mathcal{C}_6^{2141{Ll}}\nonumber \\
	&-&3 \kappa _2^2 \kappa _1 \mathcal{C}_6^{2141{Ll}}-\kappa _2^3 (\mathcal{C}_6^{23{Ll41}}+\mathcal{C}_6^{2{L21l1}}+2 \mathcal{C}_6^{4141{Ll}}-2 \mathcal{C}_6^{41{L2l1}})\big)\big)+v_L v_R \big(v_R^2 \big(-\kappa _1^2 \mathcal{C}_6^{{RrL2r1}}\nonumber \\
	&+&\kappa _2^2 \mathcal{C}_6^{{RrL2r1}}+2 \kappa _2 \kappa _1 (\mathcal{C}_6^{{RrL4r1}}+\mathcal{C}_6^{{RrR4l1}})\big)-\kappa _1^4 \mathcal{C}_6^{21{L2r1}}+\kappa _2^2 \kappa _1^2 (\mathcal{C}_6^{41{L2r3}}-\mathcal{C}_6^{21{L2r1}})+\kappa _2^4 \mathcal{C}_6^{41{L2r3}}\nonumber \\
	&+&2 \kappa _2 \kappa _1^3 (\mathcal{C}_6^{2{L41r3}}+\mathcal{C}_6^{41{L41r}}+\mathcal{C}_6^{41{R2l1}}+\mathcal{C}_6^{41{R41l}})-2 \kappa _2^3 \kappa _1 (\mathcal{C}_6^{2{L41r3}}+\mathcal{C}_6^{41{L41r}}-\mathcal{C}_6^{41{R2l1}}+\mathcal{C}_6^{41{R41l}})\big)\nonumber \\
	&+&v_L^3 v_R \big(2 \kappa _2 \kappa _1 (\mathcal{C}_6^{{LlL4r1}}+\mathcal{C}_6^{{LlR4l1}})-\kappa _1^2 \mathcal{C}_6^{{LlR2l1}}+\kappa _2^2 \mathcal{C}_6^{{LlR2l1}}\big)+v_L^4 \big(\kappa _2 \kappa _1 (\mathcal{C}_6^{{LlL21l}}-\mathcal{C}_6^{21{LlLl}})\nonumber \\
	&+&\kappa _1^2 \mathcal{C}_6^{41{LlLl}}-\kappa _2^2 \mathcal{C}_6^{41{LlLl}}\big)+\kappa _1 v_R^2 \big(\kappa _2 \kappa _1^2 (-2 \mathcal{C}_6^{2121{Rr}}-2 \mathcal{C}_6^{2211{Rr}}+\mathcal{C}_6^{23{Rr41}}+\mathcal{C}_6^{2{R21r1}}+2 \mathcal{C}_6^{4141{Rr}}\nonumber \\
	&+&2 \mathcal{C}_6^{41{R2r1}})+\kappa _1^3 \mathcal{C}_6^{2141{Rr}}-3 \kappa _2^2 \kappa _1 \mathcal{C}_6^{2141{Rr}}-\kappa _2^3 (\mathcal{C}_6^{23{Rr41}}+\mathcal{C}_6^{2{R21r1}}+2 \mathcal{C}_6^{4141{Rr}}-2 \mathcal{C}_6^{41{R2r1}})\big)\nonumber \\
	&+&v_R^4 \big(\kappa _2 \kappa _1 (\mathcal{C}_6^{{RrR21r}}-\mathcal{C}_6^{21{RrRr}})+\kappa _1^2 \mathcal{C}_6^{41{RrRr}}-\kappa _2^2 \mathcal{C}_6^{41{RrRr}}\big)+\kappa _-^2 \big(\kappa _2 \kappa _1^3 (-2 \mathcal{C}_6^{\{21\}\{2121\}}-3 \mathcal{C}_6^{212121}\nonumber \\
	&+&2 \mathcal{C}_6^{414121})+\kappa _2^3 \kappa _1 (-2 \mathcal{C}_6^{\{21\}\{2121\}}-3 \mathcal{C}_6^{212121}+2 \mathcal{C}_6^{414121})+\kappa _1^4 \mathcal{C}_6^{412121}+\kappa _2^4 \mathcal{C}_6^{412121}\nonumber \\&+&2 \kappa _2^2 \kappa _1^2 (3 (\mathcal{C}_6^{414123}+\mathcal{C}_6^{414141})-2 \mathcal{C}_6^{412121})\big)\big],\\ \nonumber\\
	(\Delta m^2_{i,\phi^6})_{13}&=&-\frac{\kappa _1 v_L}{2 \kappa_+ \Lambda^2} \big[v_L^2 \big(\kappa _1 (\mathcal{C}_6^{{LlL4r1}}-\mathcal{C}_6^{{LlR4l1}})-\kappa _2 \mathcal{C}_6^{{LlR2l1}}\big)+v_R^2 \big(\kappa _2 \mathcal{C}_6^{{RrL2r1}}+\kappa _1 (\mathcal{C}_6^{{RrL4r1}}-\mathcal{C}_6^{{RrR4l1}})\big)\nonumber \\
	&+&\kappa _2 \kappa _1^2 \mathcal{C}_6^{21{L2r1}}+\kappa _1^3 \mathcal{C}_6^{21{L4r1}}-\kappa _1^3 \mathcal{C}_6^{21{R4l1}}+\kappa _2^3 \mathcal{C}_6^{41{L2r3}}-\kappa _2^2 \kappa _1 \mathcal{C}_6^{41{R2l1}}\big],\\ \nonumber\\
	(\Delta m^2_{i,\phi^6})_{14}&=&\frac{\kappa _1 v_R }{2 \kappa_+ \Lambda^2}\big[v_L^2 \big(\kappa _1 (\mathcal{C}_6^{{LlL4r1}}-\mathcal{C}_6^{{LlR4l1}})-\kappa _2 \mathcal{C}_6^{{LlR2l1}}\big)+v_R^2 \big(\kappa _2 \mathcal{C}_6^{{RrL2r1}}+\kappa _1 (\mathcal{C}_6^{{RrL4r1}}-\mathcal{C}_6^{{RrR4l1}})\big)\nonumber \\
	&+&\kappa _2 \kappa _1^2 \mathcal{C}_6^{21{L2r1}}+\kappa _1^3 \mathcal{C}_6^{21{L4r1}}-\kappa _1^3 \mathcal{C}_6^{21{R4l1}}+\kappa _2^3 \mathcal{C}_6^{41{L2r3}}-\kappa _2^2 \kappa _1 \mathcal{C}_6^{41{R2l1}}\big],
	\end{eqnarray}}
{\small\begin{eqnarray}
	(\Delta m^2_{i,\phi^6})_{22}&=&\frac{1}{4 \kappa _+^2 \Lambda ^2}\big[v_L^2 \big(v_R^2 \big(\kappa _2^2 \mathcal{C}_6^{21{LlRr}}-2 \kappa _2 \kappa _1 \mathcal{C}_6^{41{LlRr}}+\kappa _1^2 (\mathcal{C}_6^{{LlR21r}}+\mathcal{C}_6^{{RrL21l}})\big)+2 \kappa _2^2 \kappa _1^2 (\mathcal{C}_6^{2121{Ll}}\nonumber\\
	&+&\mathcal{C}_6^{2211{Ll}}-4 \mathcal{C}_6^{4141{Ll}})-2 \kappa _2 \kappa _1^3 \mathcal{C}_6^{2141{Ll}}+2 \kappa _2^3 \kappa _1 \mathcal{C}_6^{2141{Ll}}+\kappa _1^4 (\mathcal{C}_6^{23{Ll41}}+\mathcal{C}_6^{2{L21l1}}-2 \mathcal{C}_6^{4141{Ll}}+2 \mathcal{C}_6^{41{L2l1}})\nonumber\\
	&+&\kappa _2^4 (\mathcal{C}_6^{23{Ll41}}+\mathcal{C}_6^{2{L21l1}}-2 (\mathcal{C}_6^{4141{Ll}}+\mathcal{C}_6^{41{L2l1}}))\big)+2 v_L v_R \big(\kappa _1 \big(\kappa _2 \mathcal{C}_6^{{RrL2r1}} v_R^2+\kappa _2^3 (\mathcal{C}_6^{21{L2r1}}+\mathcal{C}_6^{41{L2r3}})\big)\nonumber\\
	&+&\kappa _2^2 \big(\kappa _2^2 (\mathcal{C}_6^{2{L41r3}}-\mathcal{C}_6^{41{L41r}}-\mathcal{C}_6^{41{R2l1}}-\mathcal{C}_6^{41{R41l}})-v_R^2 (\mathcal{C}_6^{{RrL4r1}}+\mathcal{C}_6^{{RrR4l1}})\big)+\kappa _2 \kappa _1^3 (\mathcal{C}_6^{21{L2r1}}\nonumber\\
	&+&\mathcal{C}_6^{41{L2r3}})+\kappa _1^4 (\mathcal{C}_6^{2{L41r3}}-\mathcal{C}_6^{41{L41r}}+\mathcal{C}_6^{41{R2l1}}-\mathcal{C}_6^{41{R41l}})-4 \kappa _2^2 \kappa _1^2 (\mathcal{C}_6^{41{L41r}}+\mathcal{C}_6^{41{R41l}})\big)\nonumber\\
	&+&\mathcal{C}_6^{{LlR4l1}})-\kappa _1 \mathcal{C}_6^{{LlR2l1}}\big)+v_L^4 \big(\kappa _2^2 (\mathcal{C}_6^{\{21\}\{{LlLl}\}}+\mathcal{C}_6^{21{LlLl}})+\kappa _1^2 (\mathcal{C}_6^{\{21\}\{{LlLl}\}}+\mathcal{C}_6^{{LlL21l}})-2 \kappa _2 \kappa _1 \mathcal{C}_6^{41{LlLl}}\big)\nonumber\\
	&+&v_R^2 \big(2 \kappa _2^2 \kappa _1^2 (\mathcal{C}_6^{2121{Rr}}+\mathcal{C}_6^{2211{Rr}}-4 \mathcal{C}_6^{4141{Rr}})-2 \kappa _2 \kappa _1^3 \mathcal{C}_6^{2141{Rr}}+2 \kappa _2^3 \kappa _1 \mathcal{C}_6^{2141{Rr}}+\kappa _1^4 (\mathcal{C}_6^{23{Rr41}}+\mathcal{C}_6^{2{R21r1}}\nonumber\\
	&-&2 \mathcal{C}_6^{4141{Rr}}+2 \mathcal{C}_6^{41{R2r1}})+\kappa _2^4 (\mathcal{C}_6^{23{Rr41}}+\mathcal{C}_6^{2{R21r1}}-2 (\mathcal{C}_6^{4141{Rr}}+\mathcal{C}_6^{41{R2r1}}))\big)+v_R^4 \big(\kappa _2^2 (\mathcal{C}_6^{\{21\}\{{RrRr}\}}+\mathcal{C}_6^{21{RrRr}})\nonumber\\
	&+&\kappa _1^2 (\mathcal{C}_6^{\{21\}\{{RrRr}\}}+\mathcal{C}_6^{{RrR21r}})-2 \kappa _2 \kappa _1 \mathcal{C}_6^{41{RrRr}}\big)+\kappa _1^6 \mathcal{C}_6^{\{21\}\{2121\}}+5 \kappa _2^2 \kappa _1^4 \mathcal{C}_6^{\{21\}\{2121\}}+5 \kappa _2^4 \kappa _1^2 \mathcal{C}_6^{\{21\}\{2121\}}\nonumber\\
	&+&\kappa _2^6 \mathcal{C}_6^{\{21\}\{2121\}}+3 \kappa _2^2 \kappa _1^4 \mathcal{C}_6^{212121}+3 \kappa _2^4 \kappa _1^2 \mathcal{C}_6^{212121}-2 \kappa _2 \kappa _1^5 \mathcal{C}_6^{412121}+8 \kappa _2^3 \kappa _1^3 \mathcal{C}_6^{412121}-2 \kappa _2^5 \kappa _1 \mathcal{C}_6^{412121}\nonumber\\
	&-&2 \kappa _1^6 \mathcal{C}_6^{414121}-8 \kappa _2^2 \kappa _1^4 \mathcal{C}_6^{414121}-8 \kappa _2^4 \kappa _1^2 \mathcal{C}_6^{414121}-2 \kappa _2^6 \mathcal{C}_6^{414121}+4 \kappa _2 \kappa _1^5 \mathcal{C}_6^{414123}-4 \kappa _2^3 \kappa _1^3 \mathcal{C}_6^{414123}\nonumber\\
	&+&4 \kappa _2^5 \kappa _1 \mathcal{C}_6^{414123}-12 \kappa _2 \kappa _1^5 \mathcal{C}_6^{414141}-36 \kappa _2^3 \kappa _1^3 \mathcal{C}_6^{414141}-12 \kappa _2^5 \kappa _1 \mathcal{C}_6^{414141}\big],\\
	\nonumber \\
	(\Delta m^2_{i,\phi^6})_{23}&=&\frac{v_L}{4 \kappa_+ \Lambda^2} \big[v_L^2 \big(2 \kappa _2 \kappa _1 (\mathcal{C}_6^{{LlL4r1}}-\mathcal{C}_6^{{LlR4l1}})+\kappa _1^2 \mathcal{C}_6^{{LlR2l1}}-\kappa _2^2 \mathcal{C}_6^{{LlR2l1}}\big)+v_R^2 \big(-\kappa _1^2 \mathcal{C}_6^{{RrL2r1}}+\kappa _2^2 \mathcal{C}_6^{{RrL2r1}}\nonumber \\
	&+&2 \kappa _2 \kappa _1 (\mathcal{C}_6^{{RrL4r1}}-\mathcal{C}_6^{{RrR4l1}})\big)-\kappa _1^4 \mathcal{C}_6^{21{L2r1}}+\kappa _2^2 \kappa _1^2 \mathcal{C}_6^{21{L2r1}}+2 \kappa _2 \kappa _1^3 \mathcal{C}_6^{21{L4r1}}-2 \kappa _2 \kappa _1^3 \mathcal{C}_6^{21{R4l1}}-\kappa _2^2 \kappa _1^2 \mathcal{C}_6^{41{L2r3}}\nonumber \\
	&+&\kappa _2^4 \mathcal{C}_6^{41{L2r3}}+2 \kappa _2 \kappa _1^3 \mathcal{C}_6^{41{L41r}}+2 \kappa _2^3 \kappa _1 \mathcal{C}_6^{41{L41r}}-2 \kappa _2^3 \kappa _1 \mathcal{C}_6^{41{R2l1}}-2 \kappa _2 \kappa _1^3 \mathcal{C}_6^{41{R41l}}-2 \kappa _2^3 \kappa _1 \mathcal{C}_6^{41{R41l}}\big],\\
	\nonumber \\
	(\Delta m^2_{i,\phi^6})_{24}&=&\frac{v_R}{4 \kappa_+ \Lambda^2} \big[v_L^2 \big(2 \kappa _2 \kappa _1 (\mathcal{C}_6^{{LlR4l1}}-\mathcal{C}_6^{{LlL4r1}})-\kappa _1^2 \mathcal{C}_6^{{LlR2l1}}+\kappa _2^2 \mathcal{C}_6^{{LlR2l1}}\big)+v_R^2 \big(\kappa _1^2 \mathcal{C}_6^{{RrL2r1}}-\kappa _2^2 \mathcal{C}_6^{{RrL2r1}}\nonumber \\
	&+&2 \kappa _2 \kappa _1 (\mathcal{C}_6^{{RrR4l1}}-\mathcal{C}_6^{{RrL4r1}})\big)+\kappa _1^4 \mathcal{C}_6^{21{L2r1}}-\kappa _2^2 \kappa _1^2 \mathcal{C}_6^{21{L2r1}}-2 \kappa _2 \kappa _1^3 \mathcal{C}_6^{21{L4r1}}+2 \kappa _2 \kappa _1^3 \mathcal{C}_6^{21{R4l1}}+\kappa _2^2 \kappa _1^2 \mathcal{C}_6^{41{L2r3}}\nonumber \\
	&-&\kappa _2^4 \mathcal{C}_6^{41{L2r3}}-2 \kappa _2 \kappa _1^3 \mathcal{C}_6^{41{L41r}}-2 \kappa _2^3 \kappa _1 \mathcal{C}_6^{41{L41r}}+2 \kappa _2^3 \kappa _1 \mathcal{C}_6^{41{R2l1}}+2 \kappa _2 \kappa _1^3 \mathcal{C}_6^{41{R41l}}+2 \kappa _2^3 \kappa _1 \mathcal{C}_6^{41{R41l}}\big],\\
	\nonumber \\
	(\Delta m^2_{i,\phi^6})_{33}&=&\frac{1}{4 \Lambda ^2}\big[v_L^2 \big(2 \mathcal{C}_6^{{RrRrLl}} v_R^2+\kappa _1^2 \mathcal{C}_6^{21{LlRr}}+2 \kappa _1 \kappa _2 \mathcal{C}_6^{41{LlRr}}+\kappa _2^2 (\mathcal{C}_6^{{LlR21r}}+\mathcal{C}_6^{{RrL21l}})\big)+2 \kappa _1 v_L v_R \big(\kappa _2 \mathcal{C}_6^{{RrL2r1}}\nonumber\\
	&+&\kappa _1 (\mathcal{C}_6^{{RrL4r1}}+\mathcal{C}_6^{{RrR4l1}})\big)+\mathcal{C}_6^{{LlLlRr}} v_L^4+v_R^2 \big(\kappa _1^2 (-(\mathcal{C}_6^{\{21\}\{{RrRr}\}}-2 \mathcal{C}_6^{21{RrRr}}))-\kappa _2^2 (\mathcal{C}_6^{\{21\}\{{RrRr}\}}-2 \mathcal{C}_6^{{RrR21r}})\nonumber\\
	&+&4 \kappa _2 \kappa _1 \mathcal{C}_6^{41{RrRr}}\big)+3 \mathcal{C}_6^{{RrRrRr}} v_R^4+\kappa _1^2 \big(\kappa _1^2 (\mathcal{C}_6^{2121{Rr}}+\mathcal{C}_6^{2211{Rr}})+2 \kappa _2 \kappa _1 \mathcal{C}_6^{2141{Rr}}+\kappa _2^2 (\mathcal{C}_6^{23{Rr41}}\nonumber\\
	&+&\mathcal{C}_6^{2{R21r1}}+2 (\mathcal{C}_6^{4141{Rr}}+\mathcal{C}_6^{41{R2r1}}))\big)\big],\\
	\nonumber \\
	(\Delta m^2_{i,\phi^6})_{34}&=&\frac{ \kappa _1}{4\Lambda^2} \big[v_L^2 \big(\kappa _1 (\mathcal{C}_6^{{LlL4r1}}+\mathcal{C}_6^{{LlR4l1}})+\kappa _2 \mathcal{C}_6^{{LlR2l1}}\big)+v_R^2 \big(\kappa _2 \mathcal{C}_6^{{RrL2r1}}+\kappa _1 (\mathcal{C}_6^{{RrL4r1}}+\mathcal{C}_6^{{RrR4l1}})\big)\\
	&+&\kappa _2 \kappa _1^2 \mathcal{C}_6^{21{L2r1}}+\kappa _1^3 (\mathcal{C}_6^{21{L4r1}}+ \mathcal{C}_6^{21{R4l1}})+\kappa _2^3 \mathcal{C}_6^{41{L2r3}}+\kappa _2^2 \kappa _1 (\mathcal{C}_6^{2{L41r3}}+ \mathcal{C}_6^{41{L41r}}+ \mathcal{C}_6^{41{R2l1}}+\mathcal{C}_6^{41{R41l}})\big],\nonumber\\
	\nonumber \\
	(\Delta m^2_{i,\phi^6})_{44}&=&\frac{1}{4 \Lambda^2} \big[2 v_L^2 \big(\mathcal{C}_6^{{LlLlRr}} v_R^2+\kappa _1^2 (\mathcal{C}_6^{\{21\}\{{LlLl}\}}+\mathcal{C}_6^{21{LlLl}})+\kappa _2^2 (\mathcal{C}_6^{\{21\}\{{LlLl}\}}+\mathcal{C}_6^{{LlL21l}})+2 \kappa _1 \kappa _2 \mathcal{C}_6^{41{LlLl}}\big)\nonumber \\
	&+&2 \kappa _1 v_L v_R \big(\kappa _1 (\mathcal{C}_6^{{LlL4r1}}+\mathcal{C}_6^{{LlR4l1}})+\kappa _2 \mathcal{C}_6^{{LlR2l1}}\big)+3 \mathcal{C}_6^{{LlLlLl}} v_L^4+v_R^2 \big(\kappa _1^2 \mathcal{C}_6^{21{LlRr}}+2 \kappa _2 \kappa _1 \mathcal{C}_6^{41{LlRr}}\nonumber \\
	&+&\kappa _2^2 (\mathcal{C}_6^{{LlR21r}}+\mathcal{C}_6^{{RrL21l}})\big)+\mathcal{C}_6^{{RrRrLl}} v_R^4+\kappa _1^2 \big(\kappa _1^2 (\mathcal{C}_6^{2121{Ll}}+\mathcal{C}_6^{2211{Ll}})+2 \kappa _2 \kappa _1 \mathcal{C}_6^{2141{Ll}}\nonumber \\
	&+&\kappa _2^2 (\mathcal{C}_6^{23{Ll41}}+\mathcal{C}_6^{2{L21l1}}+2 (\mathcal{C}_6^{4141{Ll}}+\mathcal{C}_6^{41{L2l1}}))\big)\big].
	\end{eqnarray}}

\pagebreak
\underline{\large{Singly Charged scalar mass matrix}}

{\small\begin{eqnarray}
	(\Delta m^2_{+,\phi^6})_{11}&=&\frac{1}{4 \kappa _+^2 \Lambda ^2}\big[v_L^2 \big(v_R^2 \big(\kappa _2^2 \mathcal{C}_6^{21{LlRr}}-2 \kappa _2 \kappa _1 \mathcal{C}_6^{41{LlRr}}+\kappa _1^2 (\mathcal{C}_6^{{LlR21r}}+\mathcal{C}_6^{{RrL21l}})\big)+\kappa _2^4 (\mathcal{C}_6^{2121{Ll}}\nonumber\\
	&+&\mathcal{C}_6^{2211{Ll}})+\kappa _2^2 \kappa _1^2 (4 \mathcal{C}_6^{2121{Ll}}+\mathcal{C}_6^{2211{Ll}}-2 \mathcal{C}_6^{23{Ll41}}+\mathcal{C}_6^{2{L21l1}}-4 \mathcal{C}_6^{4141{Ll}}-2 \mathcal{C}_6^{41{L2l1}})+\kappa _1^4 (\mathcal{C}_6^{2121{Ll}}\nonumber\\
	&+&\mathcal{C}_6^{2{L21l1}})+2 \kappa _2 \kappa _1^3 (\mathcal{C}_6^{2211{Ll}}-\mathcal{C}_6^{2141{Ll}})+2 \kappa _2^3 \kappa _1 (\mathcal{C}_6^{2141{Ll}}+\mathcal{C}_6^{2211{Ll}})\big)+2 \kappa _1 v_L v_R \big(\kappa _2 \kappa _1^2 \mathcal{C}_6^{21{L2r1}}\nonumber\\
	&+&\kappa _2^3 (2 \mathcal{C}_6^{21{L2r1}}-\mathcal{C}_6^{41{L2r3}})+\kappa _1^3 (\mathcal{C}_6^{21{L4r1}}+\mathcal{C}_6^{21{R4l1}})+\kappa _2^2 \kappa _1 (2 \mathcal{C}_6^{21{L4r1}}+2 \mathcal{C}_6^{21{R4l1}}-2 \mathcal{C}_6^{2{L41r3}}+\mathcal{C}_6^{41{l2r1}}\nonumber\\
	&-&2 \mathcal{C}_6^{41{L41r}}-\mathcal{C}_6^{41{R2l1}}-2 \mathcal{C}_6^{41{R41l}})\big)+v_L^4 \big(\kappa _2^2 (\mathcal{C}_6^{\{21\}\{{LlLl}\}}+\mathcal{C}_6^{21{LlLl}})+\kappa _1^2 (\mathcal{C}_6^{\{21\}\{{LlLl}\}}+\mathcal{C}_6^{{LlL21l}})\nonumber\\
	&-&2 \kappa _2 \kappa _1 \mathcal{C}_6^{41{LlLl}}\big)+v_R^2 \big(\kappa _2^4 (\mathcal{C}_6^{2121{Rr}}+\mathcal{C}_6^{2211{Rr}})+\kappa _2^2 \kappa _1^2 (4 \mathcal{C}_6^{2121{Rr}}+\mathcal{C}_6^{2211{Rr}}-2 \mathcal{C}_6^{23{Rr41}}+\mathcal{C}_6^{2{R21r1}}\nonumber\\
	&-&4 \mathcal{C}_6^{4141{Rr}}-2 \mathcal{C}_6^{41{R2r1}})+\kappa _1^4 (\mathcal{C}_6^{2121{Rr}}+\mathcal{C}_6^{2{R21r1}})+2 \kappa _2 \kappa _1^3 (\mathcal{C}_6^{2211{Rr}}-\mathcal{C}_6^{2141{Rr}})+2 \kappa _2^3 \kappa _1 (\mathcal{C}_6^{2141{Rr}}\nonumber\\
	&+&\mathcal{C}_6^{2211{Rr}})\big)+v_R^4 \big(\kappa _2^2 (\mathcal{C}_6^{\{21\}\{{RrRr}\}}+\mathcal{C}_6^{21{RrRr}})+\kappa _1^2 (\mathcal{C}_6^{\{21\}\{{RrRr}\}}+\mathcal{C}_6^{{RrR21r}})-2 \kappa _2 \kappa _1 \mathcal{C}_6^{41{RrRr}}\big)\nonumber\\
	&+&3 \kappa _1^6 \mathcal{C}_6^{\{21\}\{2121\}}+11 \kappa _2^2 \kappa _1^4 \mathcal{C}_6^{\{21\}\{2121\}}+11 \kappa _2^4 \kappa _1^2 \mathcal{C}_6^{\{21\}\{2121\}}+3 \kappa _2^6 \mathcal{C}_6^{\{21\}\{2121\}}+3 \kappa _1^6 \mathcal{C}_6^{212121}\nonumber\\
	&+&12 \kappa _2^2 \kappa _1^4 \mathcal{C}_6^{212121}+12 \kappa _2^4 \kappa _1^2 \mathcal{C}_6^{212121}+3 \kappa _2^6 \mathcal{C}_6^{212121}+2 \kappa _2 \kappa _1^5 \mathcal{C}_6^{412121}+16 \kappa _2^3 \kappa _1^3 \mathcal{C}_6^{412121}+2 \kappa _2^5 \kappa _1 \mathcal{C}_6^{412121}\nonumber\\
	&-&2 \kappa _2^2 \kappa _1^4 \mathcal{C}_6^{414121}-2 \kappa _2^4 \kappa _1^2 \mathcal{C}_6^{414121}-12 \kappa _2^3 \kappa _1^3 \mathcal{C}_6^{414123}-12 \kappa _2^3 \kappa _1^3 \mathcal{C}_6^{414141}\big],\\
	(\Delta m^2_{+,\phi^6})_{12}&=&\frac{1}{4 \kappa _+^2 \Lambda ^2}\big[v_L^2 \big(\kappa _1 \big(\kappa _2 \kappa _1^2 (2 \mathcal{C}_6^{2121{Ll}}+\mathcal{C}_6^{2211{Ll}}-\mathcal{C}_6^{23{Ll41}}-\mathcal{C}_6^{2{L21l1}}-2 \mathcal{C}_6^{4141{Ll}}-\mathcal{C}_6^{41{L2l1}})+\kappa _1^3 \mathcal{C}_6^{2211{Ll}}\nonumber \\
	&-&\mathcal{C}_6^{2141{Ll}}+\kappa _2^2 \kappa _1 (3 \mathcal{C}_6^{2141{Ll}}+\mathcal{C}_6^{2211{Ll}})+\kappa _2^3 (\mathcal{C}_6^{2211{Ll}}+\mathcal{C}_6^{23{Ll41}}-\mathcal{C}_6^{2{L21l1}}+2 \mathcal{C}_6^{4141{Ll}}+\mathcal{C}_6^{41{L2l1}})\big)\nonumber \\
	&-&v_R^2 \big(\kappa _2 \kappa _1 (-\mathcal{C}_6^{21{LlRr}}+\mathcal{C}_6^{{LlR21r}}+\mathcal{C}_6^{{RrL21l}})+\kappa _1^2 \mathcal{C}_6^{41{LlRr}}-\kappa _2^2 \mathcal{C}_6^{41{LlRr}}\big)\big)\nonumber \\
	&+&\kappa _-^2 \kappa _2 v_L v_R \big(\kappa _2 (\mathcal{C}_6^{21{L2r1}}-\mathcal{C}_6^{41{L2r3}})+\kappa _1 (\mathcal{C}_6^{21{L4r1}}+\mathcal{C}_6^{21{R4l1}}-2 \mathcal{C}_6^{2{L41r3}}+\mathcal{C}_6^{41{l2r1}}\nonumber \\
	&-&2 \mathcal{C}_6^{41{L41r}}-\mathcal{C}_6^{41{R2l1}}-2 \mathcal{C}_6^{41{R41l}})\big)+v_L^4 \big(\kappa _2 \kappa _1 (\mathcal{C}_6^{21{LlLl}}-\mathcal{C}_6^{{LlL21l}})+\kappa _1^2 (-\mathcal{C}_6^{41{LlLl}})+\kappa _2^2 \mathcal{C}_6^{41{LlLl}}\big)\nonumber \\
	&+&\kappa _1 v_R^2 \big(\kappa _2 \kappa _1^2 (2 \mathcal{C}_6^{2121{Rr}}+\mathcal{C}_6^{2211{Rr}}-\mathcal{C}_6^{23{Rr41}}-\mathcal{C}_6^{2{R21r1}}-2 \mathcal{C}_6^{4141{Rr}}-\mathcal{C}_6^{41{R2r1}})+\kappa _1^3 (\mathcal{C}_6^{2211{Rr}}-\mathcal{C}_6^{2141{Rr}})\nonumber \\
	&+&\kappa _2^2 \kappa _1 (3 \mathcal{C}_6^{2141{Rr}}+\mathcal{C}_6^{2211{Rr}})+\kappa _2^3 (\mathcal{C}_6^{2211{Rr}}+\mathcal{C}_6^{23{Rr41}}-\mathcal{C}_6^{2{R21r1}}+2 \mathcal{C}_6^{4141{Rr}}+\mathcal{C}_6^{41{R2r1}})\big)\nonumber \\
	&+&v_R^4 \big(\kappa _2 \kappa _1 (\mathcal{C}_6^{21{RrRr}}-\mathcal{C}_6^{{RrR21r}})+\kappa _1^2 (-\mathcal{C}_6^{41{RrRr}})+\kappa _2^2  \mathcal{C}_6^{41{RrRr}}\big)\nonumber \\
	&-&\kappa _-^2 \big(\kappa _2 \kappa _1^3 (-2 \mathcal{C}_6^{\{21\}\{2121\}}-3 \mathcal{C}_6^{212121}+2 \mathcal{C}_6^{414121})+\kappa _2^3 \kappa _1 (-2 \mathcal{C}_6^{\{21\}\{2121\}}-3 \mathcal{C}_6^{212121}+2 \mathcal{C}_6^{414121})\nonumber \\
	&+&\kappa _1^4 \mathcal{C}_6^{412121}+\kappa _2^4 \mathcal{C}_6^{412121}+2 \kappa _2^2 \kappa _1^2 (3 (\mathcal{C}_6^{414123}+\mathcal{C}_6^{414141})-2 \mathcal{C}_6^{412121})\big)\big],\\
	(\Delta m^2_{+,\phi^6})_{13}&=&-\frac{1}{8 \sqrt{2} \kappa _+ \Lambda ^2}\big[2 v_L \big(v_R^2 \big(\kappa _2 \kappa _1 (\mathcal{C}_6^{{R4Rlr1}}+\mathcal{C}_6^{{RrL2r1}}+2 \mathcal{C}_6^{{RrL4r1}})+\kappa _2^2 \mathcal{C}_6^{{RrL2r1}}+\kappa _1^2 (-\mathcal{C}_6^{{RrL2r1}}+\mathcal{C}_6^{{RrL4r1}}\nonumber\\
	&+&2 \mathcal{C}_6^{{RrR4l1}})\big)+\kappa _2^4 \mathcal{C}_6^{21{L2r1}}+\kappa _1^4 (-\mathcal{C}_6^{21{L2r1}}+2 \mathcal{C}_6^{21{R4l1}})+2 \kappa _2^2 \kappa _1^2 (\mathcal{C}_6^{21{L2r1}}+\mathcal{C}_6^{21{R4l1}}-\mathcal{C}_6^{41{L2r3}})\nonumber \\
	&+&\kappa _2 \kappa _1^3 (3 \mathcal{C}_6^{21{L4r1}}-\mathcal{C}_6^{41{l2r1}})+\kappa _2^3 \kappa _1 (\mathcal{C}_6^{21{L4r1}}+\mathcal{C}_6^{41{l2r1}})\big)+\kappa _+^2 v_L^2 v_R (\mathcal{C}_6^{21{LlRr}}-2 \mathcal{C}_6^{{LlR21r}})\nonumber \\
	&+&2 \kappa _1 v_L^3 \big(\kappa _1 \mathcal{C}_6^{{LlR4l1}}-\kappa _2 (\mathcal{C}_6^{{L4Lrl1}}-2 \mathcal{C}_6^{{LlL4r1}}+\mathcal{C}_6^{{LlR2l1}})\big)+2 v_R \big(\kappa _+^2 v_R^2 (\mathcal{C}_6^{21{RrRr}}-\mathcal{C}_6^{{RrR21r}})\nonumber \\
	&+&\kappa _2^4 (\mathcal{C}_6^{2121{Rr}}+\mathcal{C}_6^{2211{Rr}})+\kappa _1^4 (\mathcal{C}_6^{2121{Rr}}+\mathcal{C}_6^{2211{Rr}}-\mathcal{C}_6^{2{R21r1}})+\kappa _2^2 \kappa _1^2 (2 \mathcal{C}_6^{2121{Rr}}-\mathcal{C}_6^{2{R21r1}})\nonumber \\
	&+&\kappa _2^3 \kappa _1 (2 \mathcal{C}_6^{2141{Rr}}+\mathcal{C}_6^{2211{Rr}}+2 \mathcal{C}_6^{2{R21r1}}+\mathcal{C}_6^{41{R2r1}})+\kappa _2 \kappa _1^3 (2 \mathcal{C}_6^{2141{Rr}}+\mathcal{C}_6^{2211{Rr}}-\mathcal{C}_6^{41{R2r1}})\big)\big],\\
	(\Delta m^2_{+,\phi^6})_{14}&=&-\frac{1}{8 \sqrt{2} \kappa _+ \Lambda ^2}\big[v_L \big(\kappa _+^2 v_R^2 (\mathcal{C}_6^{21{LlRr}}-2 \mathcal{C}_6^{{RrL21l}})+2 \big(\kappa _2^4 (\mathcal{C}_6^{2121{Ll}}+\mathcal{C}_6^{2211{Ll}})+\kappa _1^4 (\mathcal{C}_6^{2121{Ll}}+\mathcal{C}_6^{2211{Ll}}\nonumber \\
	&-&\mathcal{C}_6^{2{L21l1}})+\kappa _2^2 \kappa _1^2 (2 \mathcal{C}_6^{2121{Ll}}-\mathcal{C}_6^{2{L21l1}})+\kappa _2^3 \kappa _1 (2 \mathcal{C}_6^{2141{Ll}}+\mathcal{C}_6^{2211{Ll}}+2 \mathcal{C}_6^{2{L21l1}}+\mathcal{C}_6^{41{L2l1}})\nonumber \\
	&+&\kappa _2 \kappa _1^3 (2 \mathcal{C}_6^{2141{Ll}}+\mathcal{C}_6^{2211{Ll}}-\mathcal{C}_6^{41{L2l1}})\big)\big)+2 v_L^2 v_R \big(\kappa _2 \kappa _1 (\mathcal{C}_6^{{L4Lrl1}}+\mathcal{C}_6^{{LlR2l1}}+2 \mathcal{C}_6^{{LlR4l1}})\nonumber \\
	&+&\kappa _1^2 (2 \mathcal{C}_6^{{LlL4r1}}-\mathcal{C}_6^{{LlR2l1}}+\mathcal{C}_6^{{LlR4l1}})+\kappa _2^2 \mathcal{C}_6^{{LlR2l1}}\big)+2 \kappa _+^2 v_L^3 (\mathcal{C}_6^{21{LlLl}}-\mathcal{C}_6^{{LlL21l}})\nonumber \\
	&+&2 \kappa _1 v_R \big(v_R^2 \big(\kappa _1 \mathcal{C}_6^{{RrL4r1}}-\kappa _2 (\mathcal{C}_6^{{R4Rlr1}}+\mathcal{C}_6^{{RrL2r1}}-2 \mathcal{C}_6^{{RrR4l1}})\big)+\kappa _2^3 (2 \mathcal{C}_6^{21{L2r1}}+\mathcal{C}_6^{21{R4l1}}\nonumber \\
	&-&2 \mathcal{C}_6^{41{L2r3}}+\mathcal{C}_6^{41{R2l1}})+2 \kappa _1^3 \mathcal{C}_6^{21{L4r1}}+2 \kappa _2^2 \kappa _1 \mathcal{C}_6^{21{L4r1}}+\kappa _2 \kappa _1^2 (3 \mathcal{C}_6^{21{R4l1}}-\mathcal{C}_6^{41{R2l1}})\big)\big],
	\end{eqnarray}}
{\small\begin{eqnarray}
	(\Delta m^2_{+,\phi^6})_{22}&=&\frac{1}{4 \kappa _+^2 \Lambda ^2}\big[v_L^2 \big(v_R^2 \big(\kappa _1^2 \mathcal{C}_6^{21{LlRr}}+2 \kappa _2 \kappa _1 \mathcal{C}_6^{41{LlRr}}+\kappa _2^2 (\mathcal{C}_6^{{LlR21r}}+\mathcal{C}_6^{{RrL21l}})\big)+\kappa _1^4 (2 \mathcal{C}_6^{2121{Ll}}\\
	&+&\mathcal{C}_6^{2211{Ll}})+4 \kappa _2 \kappa _1^3 \mathcal{C}_6^{2141{Ll}}+\kappa _2^2 \kappa _1^2 (\mathcal{C}_6^{2211{Ll}}+2 \mathcal{C}_6^{23{Ll41}}+\mathcal{C}_6^{2{L21l1}}+4 \mathcal{C}_6^{4141{Ll}}+2 \mathcal{C}_6^{41{L2l1}})+\kappa _2^4 \mathcal{C}_6^{2{L21l1}}\big)\nonumber\\
	&+&2 \kappa _1 v_L v_R \big(\kappa _2 \kappa _1^2 \mathcal{C}_6^{21{L2r1}}+\kappa _1^3 (\mathcal{C}_6^{21{L4r1}}+\mathcal{C}_6^{21{R4l1}})+\kappa _2^2 \kappa _1 (2 \mathcal{C}_6^{2{L41r3}}-\mathcal{C}_6^{41{l2r1}}+2 \mathcal{C}_6^{41{L41r}}+\mathcal{C}_6^{41{R2l1}}\nonumber\\
	&+&2 \mathcal{C}_6^{41{R41l}})+\kappa _2^3 \mathcal{C}_6^{41{L2r3}}\big)+v_L^4 \big(\kappa _1^2 (\mathcal{C}_6^{\{21\}\{{LlLl}\}}+\mathcal{C}_6^{21{LlLl}})+\kappa _2^2 (\mathcal{C}_6^{\{21\}\{{LlLl}\}}+\mathcal{C}_6^{{LlL21l}})+2 \kappa _2 \kappa _1 \mathcal{C}_6^{41{LlLl}}\big)\nonumber\\
	&+&v_R^2 \big(\kappa _1^4 (2 \mathcal{C}_6^{2121{Rr}}+\mathcal{C}_6^{2211{Rr}})+4 \kappa _2 \kappa _1^3 \mathcal{C}_6^{2141{Rr}}+\kappa _2^2 \kappa _1^2 (\mathcal{C}_6^{2211{Rr}}+2 \mathcal{C}_6^{23{Rr41}}+\mathcal{C}_6^{2{R21r1}}+4 \mathcal{C}_6^{4141{Rr}}\nonumber\\
	&+&2 \mathcal{C}_6^{41{R2r1}})+\kappa _2^4 \mathcal{C}_6^{2{R21r1}}\big)+v_R^4 \big(\kappa _1^2 (\mathcal{C}_6^{\{21\}\{{RrRr}\}}+\mathcal{C}_6^{21{RrRr}})+\kappa _2^2 (\mathcal{C}_6^{\{21\}\{{RrRr}\}}+\mathcal{C}_6^{{RrR21r}})+2 \kappa _2 \kappa _1 \mathcal{C}_6^{41{RrRr}}\big)\nonumber\\
	&+&3 \big(\kappa _1^6 (\mathcal{C}_6^{\{21\}\{2121\}}+\mathcal{C}_6^{212121})+\kappa _2^6 (\mathcal{C}_6^{\{21\}\{2121\}}+\mathcal{C}_6^{212121})+\kappa _2^2 \kappa _1^4 (\mathcal{C}_6^{\{21\}\{2121\}}+2 \mathcal{C}_6^{414121})\nonumber\\
	&+&\kappa _2^4 \kappa _1^2 (\mathcal{C}_6^{\{21\}\{2121\}}+2 \mathcal{C}_6^{414121})+2 \kappa _2 \kappa _1^5 \mathcal{C}_6^{412121}+2 \kappa _2^5 \kappa _1 \mathcal{C}_6^{412121}+4 \kappa _2^3 \kappa _1^3 (\mathcal{C}_6^{414123}+\mathcal{C}_6^{414141})\big)\big],\nonumber\\
	(\Delta m^2_{+,\phi^6})_{23}&=&\frac{1}{4 \sqrt{2} \kappa _+ \Lambda ^2}\big[v_L \big(v_R^2 \big(\kappa _1^2 (-(\mathcal{C}_6^{{R4Rlr1}}+2 \mathcal{C}_6^{{RrL4r1}}))+\kappa _2^2 \mathcal{C}_6^{{RrL2r1}}+\kappa _2 \kappa _1 (-2 \mathcal{C}_6^{{RrL2r1}}+\mathcal{C}_6^{{RrL4r1}}+2 \mathcal{C}_6^{{RrR4l1}})\big)\nonumber \\
	&+&\kappa _1 \big(\kappa _2 \kappa _1^2 (-2 \mathcal{C}_6^{21{L2r1}}+\mathcal{C}_6^{21{R4l1}}+\mathcal{C}_6^{41{R2l1}})-2 \kappa _1^3 \mathcal{C}_6^{21{L4r1}}+\kappa _2^3 (\mathcal{C}_6^{21{R4l1}}-2 \mathcal{C}_6^{41{L2r3}}+\mathcal{C}_6^{41{R2l1}})\nonumber \\
	&-&2 \kappa _2^2 \kappa _1 \mathcal{C}_6^{41{l2r1}}\big)\big)+\kappa _1 v_L^3 \big(\kappa _1 (\mathcal{C}_6^{{L4Lrl1}}-2 \mathcal{C}_6^{{LlL4r1}}+\mathcal{C}_6^{{LlR2l1}})+\kappa _2 \mathcal{C}_6^{{LlR4l1}}\big)\nonumber \\
	&+&\big(\kappa _1-\kappa _2\big) \kappa _2 v_R \big(\kappa _2 \kappa _1 (\mathcal{C}_6^{2211{Rr}}-\mathcal{C}_6^{2{R21r1}}-\mathcal{C}_6^{41{R2r1}})+\kappa _1^2 (\mathcal{C}_6^{2211{Rr}}+\mathcal{C}_6^{41{R2r1}})-\kappa _2^2 \mathcal{C}_6^{2{R21r1}}\big)\big],\\
	(\Delta m^2_{+,\phi^6})_{24}&=&\frac{1}{4 \sqrt{2} \kappa _+ \Lambda ^2}\big[v_L^2 v_R \big(\kappa _1^2 (-(\mathcal{C}_6^{{L4Lrl1}}+2 \mathcal{C}_6^{{LlR4l1}}))+\kappa _2 \kappa _1 (2 \mathcal{C}_6^{{LlL4r1}}-2 \mathcal{C}_6^{{LlR2l1}}+\mathcal{C}_6^{{LlR4l1}})+\kappa _2^2 \mathcal{C}_6^{{LlR2l1}}\big)\nonumber \\
	&+&\big(\kappa _1-\kappa _2\big) \kappa _2 v_L \big(\kappa _2 \kappa _1 (\mathcal{C}_6^{2211{Ll}}-\mathcal{C}_6^{2{L21l1}}-\mathcal{C}_6^{41{L2l1}})+\kappa _1^2 (\mathcal{C}_6^{2211{Ll}}+\mathcal{C}_6^{41{L2l1}})-\kappa _2^2 \mathcal{C}_6^{2{L21l1}}\big)\nonumber \\
	&+&v_R \big(\kappa _1 v_R^2 \big(\kappa _1 (\mathcal{C}_6^{{R4Rlr1}}+\mathcal{C}_6^{{RrL2r1}}-2 \mathcal{C}_6^{{RrR4l1}})+\kappa _2 \mathcal{C}_6^{{RrL4r1}}\big)+\kappa _2^4 \mathcal{C}_6^{21{L2r1}}+\kappa _1^4 (\mathcal{C}_6^{21{L2r1}}-2 \mathcal{C}_6^{21{R4l1}})\nonumber \\
	&+&\kappa _2 \kappa _1^3 (\mathcal{C}_6^{21{L4r1}}+\mathcal{C}_6^{41{l2r1}})+\kappa _2^3 \kappa _1 (\mathcal{C}_6^{21{L4r1}}+\mathcal{C}_6^{41{l2r1}})+2 \kappa _2^2 \kappa _1^2 (\mathcal{C}_6^{41{L2r3}}-\mathcal{C}_6^{41{R2l1}})\big)\big],\\
	(\Delta m^2_{+,\phi^6})_{33}&=&\frac{1}{8 \Lambda ^2}\big[v_L^2 \big(3 \mathcal{C}_6^{{RrRrLl}} v_R^2+\kappa _1^2 (\mathcal{C}_6^{21{LlRr}}+\mathcal{C}_6^{{LlR21r}}-2 \mathcal{C}_6^{{LR4lr1}})+2 \kappa _1 \kappa _2 \mathcal{C}_6^{41{LlRr}}+\kappa _2^2 (\mathcal{C}_6^{{LR21rl}}+\mathcal{C}_6^{{RrL21l}})\big)\nonumber \\
	&+&2 v_L v_R \big(\kappa _2 \kappa _1 (\mathcal{C}_6^{{R4Rlr1}}+\mathcal{C}_6^{{RrL2r1}}+\mathcal{C}_6^{{RrL4r1}})+\kappa _1^2 (-\mathcal{C}_6^{{R4Rlr1}}+\mathcal{C}_6^{{RrL4r1}}+2 \mathcal{C}_6^{{RrR4l1}})+\kappa _2^2 \mathcal{C}_6^{{RrL2r1}}\big)\nonumber \\
	&+&\mathcal{C}_6^{{LlLlRr}} v_L^4+v_R^2 \big(\kappa _1^2 (3 \mathcal{C}_6^{21{RrRr}}+\mathcal{C}_6^{{RrR21r}})+\kappa _2^2 (\mathcal{C}_6^{21{RrRr}}+3 \mathcal{C}_6^{{RrR21r}})+8 \kappa _2 \kappa _1 \mathcal{C}_6^{41{RrRr}}\big)\nonumber \\
	&+&6 \mathcal{C}_6^{{RrRrRr}} v_R^4+\kappa _1^4 \mathcal{C}_6^{2121{Rr}}+\kappa _2^4 \mathcal{C}_6^{2121{Rr}}+2 \kappa _1 \kappa _2^3 \mathcal{C}_6^{2141{Rr}}+2 \kappa _1^3 \kappa _2 \mathcal{C}_6^{2141{Rr}}+\kappa _1^4 \mathcal{C}_6^{2211{Rr}}+\kappa _2^4 \mathcal{C}_6^{2211{Rr}}\nonumber \\
	&+&2 \kappa _1^2 \kappa _2^2 \mathcal{C}_6^{23{Rr41}}+\kappa _1^4 \mathcal{C}_6^{2{R21r1}}+\kappa _2^4 \mathcal{C}_6^{2{R21r1}}+4 \kappa _1^2 \kappa _2^2 \mathcal{C}_6^{4141{Rr}}+2 \kappa _1 \kappa _2^3 \mathcal{C}_6^{41{R2r1}}+2 \kappa _1^3 \kappa _2 \mathcal{C}_6^{41{R2r1}}\big],\\
	(\Delta m^2_{+,\phi^6})_{34}&=&\frac{1}{16 \Lambda ^2}\big[v_L v_R \big(2 \mathcal{C}_6^{{RrRrLl}} v_R^2+\kappa _2^2 (\mathcal{C}_6^{21{LlRr}}+2 (\mathcal{C}_6^{{LlR21r}}-\mathcal{C}_6^{{LR21rl}}+\mathcal{C}_6^{{RrL21l}}))+\kappa _1^2 (\mathcal{C}_6^{21{LlRr}}+4 \mathcal{C}_6^{{LR4lr1}})\nonumber \\
	&+&4 \kappa _1 \kappa _2 \mathcal{C}_6^{41{LlRr}}\big)+2 \mathcal{C}_6^{{LlLlRr}} v_L^3 v_R+2 \kappa _1 v_L^2 \big(\kappa _2 (-\mathcal{C}_6^{{L4Lrl1}}+2 \mathcal{C}_6^{{LlL4r1}}+\mathcal{C}_6^{{LlR2l1}}+\mathcal{C}_6^{{LlR4l1}})\nonumber \\
	&+&\kappa _1 (\mathcal{C}_6^{{L4Lrl1}}+\mathcal{C}_6^{{LlR2l1}}+\mathcal{C}_6^{{LlR4l1}})\big)+2 \big(\kappa _1 v_R^2 \big(\kappa _1 (\mathcal{C}_6^{{R4Rlr1}}+\mathcal{C}_6^{{RrL2r1}}+\mathcal{C}_6^{{RrL4r1}})\nonumber  \\
	&+&\kappa _2 (-\mathcal{C}_6^{{R4Rlr1}}+\mathcal{C}_6^{{RrL2r1}}+\mathcal{C}_6^{{RrL4r1}}+2 \mathcal{C}_6^{{RrR4l1}})\big)+\kappa _1^4 \mathcal{C}_6^{21{L2r1}}+\kappa _2^4 \mathcal{C}_6^{21{L2r1}}\nonumber \\
	&+&\kappa _2 \kappa _1^3 (\mathcal{C}_6^{21{L4r1}}+\mathcal{C}_6^{21{R4l1}}+\mathcal{C}_6^{41{l2r1}}+\mathcal{C}_6^{41{R2l1}})+\kappa _2^3 \kappa _1 (\mathcal{C}_6^{21{L4r1}}+\mathcal{C}_6^{21{R4l1}}+\mathcal{C}_6^{41{l2r1}}+\mathcal{C}_6^{41{R2l1}})\nonumber  \\
	&+&2 \kappa _2^2 \kappa _1^2 (\mathcal{C}_6^{2{L41r3}}+\mathcal{C}_6^{41{L2r3}}+\mathcal{C}_6^{41{L41r}}+\mathcal{C}_6^{41{R41l}})\big)\big],\\
	(\Delta m^2_{+,\phi^6})_{44}&=&\frac{1}{8 \Lambda ^2}\big[v_L^2 \big(3 \mathcal{C}_6^{{LlLlRr}} v_R^2+\kappa _1^2 (4 \mathcal{C}_6^{\{21\}\{{LlLl}\}}+3 \mathcal{C}_6^{21{LlLl}}+\mathcal{C}_6^{{LlL21l}})+\kappa _2^2 (4 \mathcal{C}_6^{\{21\}\{{LlLl}\}}+\mathcal{C}_6^{21{LlLl}}+3 \mathcal{C}_6^{{LlL21l}})\nonumber \\
	&+&8 \kappa _1 \kappa _2 \mathcal{C}_6^{41{LlLl}}\big)+2 v_L v_R \big(\kappa _1^2 (-\mathcal{C}_6^{{L4Lrl1}}+2 \mathcal{C}_6^{{LlL4r1}}+\mathcal{C}_6^{{LlR4l1}})+\kappa _2 \kappa _1 (\mathcal{C}_6^{{L4Lrl1}}+\mathcal{C}_6^{{LlR2l1}}+\mathcal{C}_6^{{LlR4l1}})\nonumber \\
	&+&\kappa _2^2 \mathcal{C}_6^{{LlR2l1}}\big)+6 \mathcal{C}_6^{{LlLlLl}} v_L^4+v_R^2 \big(\kappa _1^2 (\mathcal{C}_6^{21{LlRr}}-2 \mathcal{C}_6^{{LR4lr1}}+\mathcal{C}_6^{{RrL21l}})+2 \kappa _2 \kappa _1 \mathcal{C}_6^{41{LlRr}}+\kappa _2^2 (\mathcal{C}_6^{{LlR21r}}\nonumber \\
	&+&\mathcal{C}_6^{{LR21rl}})\big)+\mathcal{C}_6^{{RrRrLl}} v_R^4+\kappa _1^4 \mathcal{C}_6^{2121{Ll}}+\kappa _2^4 \mathcal{C}_6^{2121{Ll}}+2 \kappa _1 \kappa _2^3 \mathcal{C}_6^{2141{Ll}}+2 \kappa _1^3 \kappa _2 \mathcal{C}_6^{2141{Ll}}+\kappa _1^4 \mathcal{C}_6^{2211{Ll}}\nonumber  \\
	&+&\kappa _2^4 \mathcal{C}_6^{2211{Ll}}+2 \kappa _1^2 \kappa _2^2 \mathcal{C}_6^{23{Ll41}}+\kappa _1^4 \mathcal{C}_6^{2{L21l1}}+\kappa _2^4 \mathcal{C}_6^{2{L21l1}}+4 \kappa _1^2 \kappa _2^2 \mathcal{C}_6^{4141{Ll}}+2 \kappa _1 \kappa _2\kappa _+^2 \mathcal{C}_6^{41{L2l1}}\big].
	\end{eqnarray}}

\pagebreak

\underline{\large{Doubly Charged scalar mass matrix}}

{\small\begin{eqnarray}
	(\Delta m^2_{++,\phi^6})_{11}&=&\frac{1}{4 \Lambda ^2}\big[v_L^2 \big(\mathcal{C}_6^{{RrLLll}} v_R^2+4 \kappa _1 \kappa _2 \mathcal{C}_6^{41{LLll}}\big)+2 \kappa _2 v_L v_R \big(\kappa _2 (\mathcal{C}_6^{{L4Lrl1}}+\mathcal{C}_6^{{LR2LL1}})+\kappa _1 \mathcal{C}_6^{{LR4ll1}}\big)\nonumber \\
	&+&\mathcal{C}_6^{{LlLLll}} v_L^4+\kappa _2^2 \big(\mathcal{C}_6^{{LR21rl}} v_R^2+\kappa _2^2 (\mathcal{C}_6^{2121{Ll}}+\mathcal{C}_6^{2211{Ll}})+2 \kappa _1 \kappa _2 \mathcal{C}_6^{2141{Ll}}\nonumber \\
	&+&\kappa _1^2 (\mathcal{C}_6^{23{Ll41}}+\mathcal{C}_6^{2{L21l1}}+2 (\mathcal{C}_6^{4141{Ll}}+\mathcal{C}_6^{41{L2l1}}))\big)\big],\\
	\nonumber \\
	(\Delta m^2_{++,\phi^6})_{12}&=&\frac{1}{4 \Lambda ^2}\big[v_L v_R \big(\mathcal{C}_6^{{llRRRr}} v_R^2+2 \big(\kappa _1^2 \mathcal{C}_6^{21{LLrr}}+\kappa _2^2 \mathcal{C}_6^{21{LLrr}}+\kappa _2 \kappa _1 (\mathcal{C}_6^{41{RRll}}+\mathcal{C}_6^{{LR4lr1}})\big)\big)+\mathcal{C}_6^{{LlLLrr}} v_L^3 v_R \nonumber \\
	&+&\kappa _1 v_L^2 \big(\kappa _1 (\mathcal{C}_6^{{L4Lrl1}}+\mathcal{C}_6^{{LR2LL1}})+\kappa _2 \mathcal{C}_6^{{LR4ll1}}\big)+v_R^2 \big(\kappa _2^2 \mathcal{C}_6^{{LR2rr1}}+\kappa _2 \kappa _1 \mathcal{C}_6^{{LR4rr1}}+\kappa _1^2 \mathcal{C}_6^{{R4Rlr1}}\big)\nonumber \\
	&+&\kappa _2 \big(\kappa _2^2 \kappa _1 \mathcal{C}_6^{21{L2r1}}+\kappa _2^3 (\mathcal{C}_6^{21{L4r1}}+\mathcal{C}_6^{21{R4l1}})+\kappa _2 \kappa _1^2 (\mathcal{C}_6^{2{L41r3}}+\mathcal{C}_6^{41{L41r}}+\mathcal{C}_6^{41{R2l1}}+\mathcal{C}_6^{41{R41l}})\nonumber \\&+&\kappa _1^3 \mathcal{C}_6^{41{L2r3}}\big)\big],\\
	\nonumber \\
	(\Delta m^2_{++,\phi^6})_{22}&=&\frac{1}{4 \Lambda ^2}\big[v_L^2 \big(\mathcal{C}_6^{{LlRRrr}} v_R^2+\kappa _1^2 \mathcal{C}_6^{{LR21rl}}\big)+2 v_L v_R \big(\kappa _1^2 \mathcal{C}_6^{{LR2rr1}}+\kappa _2 \kappa _1 \mathcal{C}_6^{{LR4rr1}}+\kappa _2^2 \mathcal{C}_6^{{R4Rlr1}}\big)\nonumber \\
	&+&4 \kappa _1 \kappa _2 \mathcal{C}_6^{41{RRrr}} v_R^2+\mathcal{C}_6^{{RrRRrr}} v_R^4+\kappa _2^2 \big(\kappa _2^2 (\mathcal{C}_6^{2121{Rr}}+\mathcal{C}_6^{2211{Rr}})+2 \kappa _2 \kappa _1 \mathcal{C}_6^{2141{Rr}}\nonumber \\
	&+&\kappa _1^2 (\mathcal{C}_6^{23{Rr41}}+\mathcal{C}_6^{2{R21r1}}+2 (\mathcal{C}_6^{4141{Rr}}+\mathcal{C}_6^{41{R2r1}}))\big)\big].
	\end{eqnarray}}


\underline{\large{$\phi^{2}X^2$ and $\phi^{4}D^{2}$ operators: Gauge field redefinitions and spectrum }} \\

The gauge-kinetic terms, Eq.~\ref{eq:MLRSM-ren-lag}, get modified in presence of $\phi^2 X^2$ operators as:
\begin{eqnarray}\label{eq:MLRSM-GB-kinetic-full}
\mathcal{L}^{(4)+(6)}_{gauge,kin}&=& -\begin{pmatrix}
\partial_{\mu}W^{-}_{L\nu}& \partial_{\mu}W^{-}_{R \nu}
\end{pmatrix}
\begin{pmatrix} 
1-\frac{2\Theta_{W_{LL}}}{\Lambda^{2}} \hspace{0.5cm} -\frac{2\Theta_{W_{LR}}}{\Lambda^{2}}\\
-\frac{2\Theta_{W_{RL}}}{\Lambda^{2}} \hspace{0.5cm} 
1-\frac{2\Theta_{W_{RR}}}{\Lambda^{2}}\\
\end{pmatrix}
\begin{pmatrix}
\partial^{\mu}W^{+\nu}_{L}\\ \partial^{\mu}W^{+\nu}_{R }
\end{pmatrix} \nonumber \\
&-&
\frac{1}{2}\begin{pmatrix}
\partial_{\mu}W_{3L \nu} \\ \partial_{\mu}W_{3R \nu} \\ \partial_{\mu}B_{\nu}
\end{pmatrix}^T
\begin{pmatrix} 
	1-\frac{2\Theta_{3L3L}}{\Lambda^{2}}  & -\frac{2\Theta_{3L3R}}{\Lambda^{2}} & -\frac{2\Theta_{3LB}}{\Lambda^{2}} \\
-\frac{2\Theta_{3L3R}}{\Lambda^{2}} & 1-\frac{2\Theta_{3R3R}}{\Lambda^{2}} & -\frac{2\Theta_{3RB}}{\Lambda^{2}} \\
-\frac{2\Theta_{3LB}}{\Lambda^{2}} & -\frac{2\Theta_{3RB}}{\Lambda^{2}} & 1-\frac{2\Theta_{BB}}{\Lambda^{2}} \\
\end{pmatrix}
\begin{pmatrix}
\partial^{\mu}W_{3L}^{\nu} \\ \partial^{\mu}W_{3R}^{\nu} \\ \partial^{\mu}B^{\nu}
\end{pmatrix}. \nonumber
\end{eqnarray}
Here, the parameters are given as:
\small\begin{eqnarray}
\Theta_{W_{LL}} &=& \kappa_+^2 \mathcal{C}^{21}_{\phi W_L} + 2\kappa_1 \kappa_2 (\mathcal{C}^{23}_{\phi W_L} + \mathcal{C}^{41}_{\phi W_L}) + v_L^2 \mathcal{C}^{Ll W_L W_L}_{\Delta W} + v_R^2 \mathcal{C}^{Rr W_L W_L}_{\Delta W},\nonumber\\
\Theta_{W_{LR}} &=&\kappa_2^2 \mathcal{C}^{23}_{\phi W_L W_R} + \kappa_1\kappa_2 \mathcal{C}^{21}_{\phi W_L W_R} + \kappa_1^2 \mathcal{C}^{41}_{\phi W_L W_R},\nonumber\\
\Theta_{W_{RL}} &=& \kappa_2^2 \mathcal{C}^{41}_{\phi W_L W_R} + \kappa_1\kappa_2 \mathcal{C}^{21}_{\phi W_L W_R} + \kappa_1^2 \mathcal{C}^{23}_{\phi W_L W_R},\nonumber\\
\Theta_{W_{RR}} &=& \kappa_+^2 \mathcal{C}^{21}_{\phi W_R} + 2\kappa_1 \kappa_2 (\mathcal{C}^{23}_{\phi W_R} + \mathcal{C}^{41}_{\phi W_R}) + v_L^2 \mathcal{C}^{Ll W_R W_R}_{\Delta W} + v_R^2 \mathcal{C}^{Rr W_R W_R}_{\Delta W},\nonumber\\
\Theta_{3L3L} &=& \kappa_+^2 \mathcal{C}^{21}_{\phi W_L} + 2\kappa_1 \kappa_2 (\mathcal{C}^{23}_{\phi W_L} + \mathcal{C}^{41}_{\phi W_L}) + v_L^2(\mathcal{C}_{\Delta W}^{LlW_L W_L} - \mathcal{C}_{\Delta W}^{LW_LlW_L}) + v_R^2 \mathcal{C}_{\Delta W}^{RrW_L W_L},\nonumber\\
\Theta_{3L3R} &=& \frac{1}{2}\kappa_+^2 \mathcal{C}^{21}_{\phi W_L W_R} + \kappa_1 \kappa_2 (\mathcal{C}^{23}_{\phi W_L W_R} + \mathcal{C}^{41}_{\phi W_L W_R}) - \frac{1}{2} v_Lv_R (\mathcal{C}_{\Delta W}^{LW_RrW_L}+\mathcal{C}_{\Delta W}^{RW_LlW_R}),\nonumber\\
\Theta_{3R3R} &=& \kappa_+^2 \mathcal{C}^{21}_{\phi W_R} + 2\kappa_1 \kappa_2 (\mathcal{C}^{23}_{\phi W_R} + \mathcal{C}^{41}_{\phi W_R}) + v_R^2(\mathcal{C}_{\Delta W}^{RrW_R W_R} - \mathcal{C}_{\Delta W}^{RW_RrW_R}) + v_L^2 \mathcal{C}_{\Delta W}^{LlW_R W_R},\nonumber\\
\Theta_{3LB} &= &\frac{1}{2}\kappa_-^2 \mathcal{C}_{\phi W_L B}^{21} - \frac{1}{2}v_L^2\mathcal{C}_{\Delta W_L B}^{Ll},\nonumber\\
\Theta_{3RB} &= &\frac{1}{2}\kappa_-^2 \mathcal{C}_{\phi W_R B}^{21} - \frac{1}{2}v_R^2\mathcal{C}_{\Delta W_R B}^{Rr},\nonumber\\
\Theta_{BB} &= &\kappa_+^2 \mathcal{C}^{21}_{\phi B} + 2\kappa_1 \kappa_2 (\mathcal{C}^{23}_{\phi B} + \mathcal{C}^{41}_{\phi B}) + v_L^2\mathcal{C}_{\Delta B}^{Ll} + v_R^2 \mathcal{C}_{\Delta B}^{Rr}.
\end{eqnarray}}
\vskip 0.5cm
 Now similar to the scalar sector, we need to redefine the gauge fields such that the modified gauge kinetic lagrangian can be brought into the canonical form. The redefined gauge fields are written as:
{\small\begin{eqnarray}\label{eq:MLRSM-GB-redef-full}
    W_L^{\pm\mu} &\rightarrow& \Big(1 + \frac{\Theta_{W_{LL}}}{\Lambda^2}\Big) W_L^{\pm\mu} + \frac{\Theta_{W_{LR}}}{\Lambda^2}\hspace{0.1cm} W_R^{\pm\mu},\nonumber\\
    W_R^{\pm\mu} &\rightarrow& \Big(1 + \frac{\Theta_{W_{RR}}}{\Lambda^2}\Big) W_R^{\pm\mu} + \frac{\Theta_{W_{RL}}}{\Lambda^2}\hspace{0.1cm} W_L^{\pm\mu}\nonumber,\\
	W_{3L}^{\mu} &\rightarrow &\Big(1 + \frac{\Theta_{3L3L}}{\Lambda^2}\Big)W_{3L}^{\mu} + \frac{\Theta_{3L3R}}{\Lambda^2}\hspace{0.1cm}W_{3R}^{\mu} + \frac{\Theta_{3LB}}{\Lambda^2}\hspace{0.1cm}B^{\mu},\nonumber\\
	W_{3R}^{\mu} &\rightarrow &\Big(1 + \frac{\Theta_{3R3R}}{\Lambda^2}\Big)W_{3R}^{\mu} + \frac{\Theta_{3L3R}}{\Lambda^2}\hspace{0.1cm}W_{3L}^{\mu} + \frac{\Theta_{3RB}}{\Lambda^2}\hspace{0.1cm}B^{\mu},\nonumber\\
	B^{\mu} &\rightarrow &\Big(1 + \frac{\Theta_{BB}}{\Lambda^2}\Big)B^{\mu}\hspace{0.2cm} + \hspace{0.2cm} \frac{\Theta_{3LB}}{\Lambda^2}\hspace{0.1cm}W_{3L}^{\mu} + \frac{\Theta_{3RB}}{\Lambda^2}\hspace{0.1cm}W_{3R}^{\mu}.
	\end{eqnarray}}

After incorporating the redefined gauge fields in the renormalizable gauge kinetic terms, the gauge boson masses are shifted. The shifts in the charged and neutral gauge boson mass matrices can be recast in the following forms respectively:
\small{\begin{eqnarray}
		(\Delta m_{W^{\pm}}^2)_{11}^{\phi^2X^2} &=& \frac{g^2}{\Lambda^2}\left(\frac{1}{2}(\kappa_{+}^2 + 2v_L^2)\Theta_{W_{LL}} - \kappa_{1}\kappa_{2}\Theta_{W_{RL}}\right), \nonumber\\
		(\Delta m_{W^{\pm}}^2)_{12}^{\phi^2X^2} & = &\frac{g^2}{\Lambda^2}\left(\frac{1}{4}\kappa_{+}^2 (\Theta_{W_{LR}}+\Theta_{W_{RL}})+\frac{1}{2}(v_L^2\Theta_{W_{LR}} + v_R^2\Theta_{W_{RL}}) - \frac{1}{2}\kappa_1\kappa_2(\Theta_{W_{LL}} + \Theta_{W_{RR}})\right),\nonumber \\
		(\Delta m_{W^{\pm}}^2)_{22}^{\phi^2X^2} &=& \frac{g^2}{\Lambda^2}\left(\frac{1}{2}(\kappa_{+}^2 + 2v_R^2)\Theta_{W_{RR}} - \kappa_{1}\kappa_{2}\Theta_{W_{LR}}\right), 
\end{eqnarray}}
and
\small{\begin{eqnarray}
		(\Delta m_{W_3 B}^2)_{11}^{\phi^2X^2} & = & \frac{g}{\Lambda ^2} \left(v_L^2 \left(-2 \tilde{g} \Theta_{3LB}+2g\Theta_{3L3L}\right)+\frac{g}{2} \kappa_+^2\left(\Theta_{3L3L}-\Theta_{3L3R}\right)\right),\nonumber\\ \nonumber\\ 
		(\Delta m_{W_3 B}^2)_{12}^{\phi^2X^2} & = & \frac{g}{\Lambda ^2} \left(v_L^2 \left(-\tilde{g}\Theta_{3RB}+g\Theta_{3L3R}\right)+ v_R^2\left(-\tilde{g}\Theta_{3LB}+g\Theta_{3L3R}\right)\right.\nonumber \\
		&-&\left.\frac{g}{4}\kappa_+^2(\Theta_{3L3L}+\Theta_{3R3R}-2\Theta_{3L3R})\right), \nonumber\\ \nonumber\\ 
		(\Delta m_{W_3 B}^2)_{13}^{\phi^2X^2} & = & \frac{1}{\Lambda^2}\Big(\frac{g^2}{4}\kappa_+^2(\Theta_{3LB}-\Theta_{3RB})+(\tilde{g}^2v_R^2+(g^2+\tilde{g}^2)v_L^2)\Theta_{3LB}\nonumber\\ &-&g\tilde{g}(v_L^2(\Theta_{BB}+\Theta_{3L3L})+v_R^2\Theta_{3L3R})\Big),\nonumber\\ \nonumber\\ 
		(\Delta m_{W_3 B}^2)_{22}^{\phi^2X^2} & = & \frac{g}{\Lambda ^2} \left(v_R^2 \left(-2 \tilde{g} \Theta_{3RB}+2g\Theta_{3R3R}\right)+\frac{g}{2} \kappa_+^2\left(\Theta_{3R3R}-\Theta_{3L3R}\right)\right),\nonumber\\ \nonumber\\ 
		(\Delta m_{W_3 B}^2)_{23}^{\phi^2X^2} & = &\frac{1}{\Lambda^2}\Big(\frac{g^2}{4}\kappa_+^2(\Theta_{3RB}-\Theta_{3LB})+(\tilde{g}^2v_L^2+(g^2+\tilde{g}^2)v_R^2)\Theta_{3RB}\nonumber\\ &-&g\tilde{g}(v_R^2(\Theta_{BB}+\Theta_{3R3R})+v_L^2\Theta_{3L3R})\Big),\nonumber\\ \nonumber\\ 
		(\Delta m_{W_3 B}^2)_{33}^{\phi^2X^2} & = &\frac{1}{\Lambda ^2} \left(2\tilde{g}^2\left(v_L^2+v_R^2\right)\Theta_{BB} - 2g\tilde{g}\left(v_L^2\Theta_{3LB}+v_R^2\Theta_{3RB}\right)\right).
\end{eqnarray}}
\clearpage

There will be an additional contribution to the gauge boson masses due to the  $\phi^4 D^2$ operators, and their effects are captured as:
{\small\begin{eqnarray}
		(\Delta m_{W^{\pm}}^{2})^{\phi^4D^2}_{11}&=&\frac{g^2}{8 \Lambda^2}\Big[\mathcal{C}_{\phi D}^{(L)rR(l)} v_L^2v_R^2+2v_L^2 \kappa_+^2 \mathcal{C}_{\phi D}^{\{21\}\{(L)(l)\}}+v_L^2\kappa_{+}^{2} \mathcal{C}_{\phi D}^{21(L)(l)}+2v_L^2 \kappa_{+}^{2} \mathcal{C}_{\phi D}^{(L)11(l)} \nonumber \\
		&+&2v_L^2 \kappa_1 \kappa_2 \mathcal{C}_{\phi D}^{L(l)1(1)}+v_L^2\kappa_{+}^{2} \mathcal{C}_{\phi D}^{\{Ll\}\{(2)(1)\}}+v_L^2\kappa_1^2 \mathcal{C}_{\phi D}^{Ll(2)(1)}+2\kappa_1^2 \mathcal{C}_{\phi D}^{(2)rL(1)} v_L v_R\nonumber \\
		&+&2 v_L^4 (\mathcal{C}_{\phi D}^{\{Ll\}\{(L)(l)\}}+\mathcal{C}_{\phi D}^{L(l)L(l)}+\mathcal{C}_{\square}^{Ll(Ll)})+ v_R^2\kappa_{+}^{2} \mathcal{C}_{\phi D}^{\{Rr\}\{(2)(1)\}}+v_R^2\kappa_1^2 \mathcal{C}_{\phi D}^{Rr(2)(1)}\nonumber \\
		&-& 4 \kappa_1^2 \kappa_2^2 \mathcal{C}_{\phi D}^{1(1)1(1)}+\kappa_+^2(\kappa_1+\kappa_2)^2 \mathcal{C}_{\phi D}^{(11)(22)}+ \kappa_+^4\mathcal{C}_{\phi D}^{\{(2)(1)\}\{21\}}+(\kappa_1^4+\kappa_2^4) \mathcal{C}_{\phi D}^{(2)(1)21} \nonumber\\
		&-& 4 \kappa_1^2 \kappa_2^2 \mathcal{C}_{\phi D}^{2(1)1(1)}+ 8 \kappa_1^2 \kappa_2^2 \mathcal{C}_{\phi D}^{2(1)2(1)}\Big],\\
		(\Delta m_{W^{\pm}}^{2})^{\phi^4D^2}_{12}&=& \frac{g^2}{8 \Lambda ^2}\Big[v_L v_R \left(-2 \kappa_2 \kappa_1 \mathcal{C}_{\phi D}^{(2)rL(1)}+\kappa_{+}^{2} \mathcal{C}_{\phi D}^{(L)11(r)}+\kappa_{+}^{2} \mathcal{C}_{\phi D}^{(L)12(r)}+\kappa_+^2 \mathcal{C}_{\phi D}^{(R)11(l)}\right) \nonumber \\
		&-&v_L^2 \left(\kappa_2 \left(\kappa_2 \mathcal{C}_{\phi D}^{L(l)1(1)}+2 \kappa_1 \mathcal{C}_{\phi D}^{\{Ll\}\{(2)(1)\}}+\kappa_1 \mathcal{C}_{\phi D}^{Ll(2)(1)}\right)-\mathcal{C}_{\phi D}^{L(r)L(r)} v_R^2\right) \\
		&+&\kappa_1 \kappa_2 \left(v_R^2 (\mathcal{C}_{\phi D}^{R(r)1(1)}-2 \mathcal{C}_{\phi D}^{\{Rr\}\{(2)(1)\}}-\mathcal{C}_{\phi D}^{Rr(2)(1)})-4 \kappa_1 \kappa_2 \mathcal{C}_{\phi D}^{(11)(22)}\right)\nonumber\\
		&+&\kappa_1 \kappa_2 \kappa_+^2\left(2  \mathcal{C}_{\phi D}^{2(1)1(1)}-2 \mathcal{C}_{\phi D}^{\{(2)(1)\}\{21\}}-\mathcal{C}_{\phi D}^{(2)(1)21}-4 \mathcal{C}_{\phi D}^{2(1)2(1)}+2\mathcal{C}_{\phi D}^{1(1)1(1)}-2\mathcal{C}_{\phi D}^{(11)(22)}\right)\Big], \nonumber \\ \nonumber\\
		(\Delta m_{W^{\pm}}^{2})^{\phi^4D^2}_{22}&=&\frac{g^2}{8 \Lambda ^2} \Big[2\kappa_2^2 \mathcal{C}_{\phi D}^{(2)rL(1)} v_L v_R+v_L^2 \left(\mathcal{C}_{\phi D}^{(R)lL(r)} v_R^2+\kappa_{+}^{2} \mathcal{C}_{\phi D}^{\{Ll\}\{(2)(1)\}}+\kappa_2^2 \mathcal{C}_{\phi D}^{Ll(2)(1)}\right)\nonumber \\
		&+&v_R^2 \kappa_{+}^{2}\left(2  \mathcal{C}_{\phi D}^{\{21\}\{(R)(r)\}}+ \mathcal{C}_{\phi D}^{\{Rr\}\{(2)(1)\}}+ \mathcal{C}_{\phi D}^{21(R)(r)}+2\mathcal{C}_{\phi D}^{(R)11(r)}\right)+v_R^2 \left(-2 \kappa_2^2 \mathcal{C}_{\phi D}^{R(r)1(1)}+\kappa_2^2 \mathcal{C}_{\phi D}^{Rr(2)(1)}\right) \nonumber\\
		&+&2 v_R^4 \left(\mathcal{C}_{\phi D}^{{\{Rr\}\{(R)(r)\}}}+\mathcal{C}_{\phi D}^{R(r)R(r)}+\mathcal{C}_{\square}^{Rr(Rr)}\right)+2 \kappa_1^2 \kappa_2^2 \left(-2 \mathcal{C}_{\phi D}^{1(1)1(1)}-2 \mathcal{C}_{\phi D}^{2(1)1(1)}+\mathcal{C}_{\phi D}^{(2)(1)21}\right)\nonumber\\
		&+&\left(\kappa_1+\kappa_2\right)^2 \kappa_{+}^{2} \mathcal{C}_{\phi D}^{(11)(22)}+\kappa_{+}^{4} \mathcal{C}_{\phi D}^{\{(2)(1)\}\{21\}}+2 \kappa_{+}^{4} \mathcal{C}_{\phi D}^{2(1)2(1)}\Big],
\end{eqnarray}}
{\small\begin{eqnarray}
		(\Delta m_{W_3B}^{2})^{\phi^4D^2}_{11}&=& \frac{g^2}{8 \Lambda ^2}\Big[v_L^2 \left(4 \mathcal{C}_{\phi D}^{(L)rR(l)} v_R^2+4 \kappa_+^2 \mathcal{C}_{\phi D}^{\{21\}\{(L)(l)\}}+4 \kappa_1^2 \mathcal{C}_{\phi D}^{21(L)(l)}+8 \kappa_2^2 \mathcal{C}_{\phi D}^{(L)11(l)}+4 \kappa_1^2 \mathcal{C}_{\phi D}^{L(l)1(1)}\right)\nonumber \\
		&+&v_L^2 \left(\kappa_+^2 \mathcal{C}_{\phi D}^{\{Ll\}\{(2)(1)\}}+\kappa_1^2 \mathcal{C}_{\phi D}^{Ll(2)(1)}\right)+2 \kappa_2^2 \mathcal{C}_{\phi D}^{(2)rL(1)} v_L v_R+4 v_L^4 \left(\mathcal{C}_{\phi D}^{\{Ll\}\{(L)(l)\}}+4 \mathcal{C}_{\phi D}^{L(l)L(l)}\right) \nonumber \\
		&+&v_R^2 \left(\kappa_+^2 \mathcal{C}_{\phi D}^{\{Rr\}\{(2)(1)\}}+\kappa_1^2 \mathcal{C}_{\phi D}^{Rr(2)(1)}\right)+\kappa_+^4 \mathcal{C}_{\phi D}^{\{(2)(1)\}\{21\}}\nonumber \\
		&+&\left(\kappa_1^4+\kappa_2^4\right) \left(\mathcal{C}_{\phi D}^{(2)(1)21}-2\mathcal{C}_{\phi D}^{1(1)1(1)}-2\mathcal{C}_{\phi D}^{2(1)1(1)}+ 4\mathcal{C}_{\phi D}^{2(1)2(1)}+ 4\mathcal{C}_{\phi D}^{(11)(22)}\right)\Big], \\
			(\Delta m_{W_3B}^{2})^{\phi^4D^2}_{12}&= &\frac{g^2}{8 \Lambda ^2}\Big[2 \kappa_2^2 v_L v_R \left(2 (\mathcal{C}_{\phi D}^{(L)11(r)}+\mathcal{C}_{\phi D}^{(L)12(r)}+\mathcal{C}_{\phi D}^{(R)11(l)})-\mathcal{C}_{\phi D}^{(2)rL(1)}\right)-\kappa_+^4 \mathcal{C}_{\phi D}^{\{(2)(1)\}\{21\}}\nonumber \\
		&-&v_L^2 \left(\kappa_+^2 \mathcal{C}_{\phi D}^{\{Ll\}\{(2)(1)\}}+\kappa_1^2 \mathcal{C}_{\phi D}^{Ll(2)(1)}\right)+v_R^2 \left(2 \kappa_1^2 \mathcal{C}_{\phi D}^{R(r)1(1)}-\kappa_{+}^{2}\mathcal{C}_{\phi D}^{\{Rr\}\{(2)(1)\}}-\kappa_1^2 \mathcal{C}_{\phi D}^{Rr(2)(1)}\right) \nonumber \\
		&+&\left(\kappa_1^4+\kappa_2^4 \right) \left(-\mathcal{C}_{\phi D}^{(2)(1)21}+2\mathcal{C}_{\phi D}^{1(1)1(1)}+2\mathcal{C}_{\phi D}^{2(1)1(1)}-4\mathcal{C}_{\phi D}^{(11)(22)}-4\mathcal{C}_{\phi D}^{2(1)2(1)}\right)\nonumber \\
		&-&v_L^2 \left(-4 \mathcal{C}_{\phi D}^{L(r)L(r)} v_R^2+2 \kappa_1^2 \mathcal{C}_{\phi D}^{L(l)1(1)}\right)\Big],
		\\
		(\Delta m_{W_3B}^{2})^{\phi^4D^2}_{13}&=& -\frac{g\tilde{g}}{4 \Lambda ^2}\Big[v_L^2 \big(2 v_R^2 (\mathcal{C}_{\phi D}^{L(r)L(r)}+\mathcal{C}_{\phi D}^{(L)rR(l)})+2 \kappa_{+}^{2} \mathcal{C}_{\phi D}^{\{21\}\{(L)(l)\}}+2 \kappa_1^2 \mathcal{C}_{\phi D}^{21(L)(l)}+4 \kappa_2^2 \mathcal{C}_{\phi D}^{(L)11(l)} \nonumber\\
		&+&\kappa_1^2 \mathcal{C}_{\phi D}^{L(l)1(1)}\big)+2 \kappa_2^2 v_L v_R \left(\mathcal{C}_{\phi D}^{(L)11(r)}+\mathcal{C}_{\phi D}^{(L)12(r)}+\mathcal{C}_{\phi D}^{(R)11(l)}\right)+ 2v_L^4 \left(\mathcal{C}_{\phi D}^{\{Ll\}\{(L)(l)\}}+4 \mathcal{C}_{\phi D}^{L(l)L(l)}\right)\nonumber\\
		&+&\kappa_1^2 \mathcal{C}_{\phi D}^{R(r)1(1)} v_R^2 \Big], 
\end{eqnarray}}

{\small\begin{eqnarray}
		(\Delta m_{W_3B}^{2})^{\phi^4D^2}_{22}&=& \frac{g^2}{8 \Lambda ^2}\Big[2 \kappa_2^2 \mathcal{C}_{\phi D}^{(2)rL(1)} v_L v_R+v_L^2 \left(4 \mathcal{C}_{\phi D}^{(R)lL(r)} v_R^2+\kappa_{+}^{2} \mathcal{C}_{\phi D}^{\{Ll\}\{(2)(1)\}}+\kappa_1^2 \mathcal{C}_{\phi D}^{Ll(2)(1)}\right) \nonumber \\
		&+&\kappa_+^4 \mathcal{C}_{\phi D}^{\{(2)(1)\}\{21\}}+\left(\kappa_1^4+\kappa_2^4\right) \left(\mathcal{C}_{\phi D}^{(2)(1)21}-2\mathcal{C}_{\phi D}^{1(1)1(1)}- 2\mathcal{C}_{\phi D}^{2(1)1(1)}+4 \mathcal{C}_{\phi D}^{2(1)2(1)}+4 \mathcal{C}_{\phi D}^{(11)(22)}\right)\nonumber\\
		&+&v_R^2 \left(4 \kappa_{+}^{2} \mathcal{C}_{\phi D}^{\{21\}\{(R)(r)\}}+4 \kappa_1^2 \mathcal{C}_{\phi D}^{21(R)(r)}+8 \kappa_2^2 \mathcal{C}_{\phi D}^{(R)11(r)}-4 \kappa_1^2 \mathcal{C}_{\phi D}^{R(r)1(1)}\right) \nonumber \\
		&+&v_R^2 \left(\kappa_+^2 \mathcal{C}_{\phi D}^{\{Rr\}\{(2)(1)\}}+\kappa_1^2 \mathcal{C}_{\phi D}^{Rr(2)(1)}\right)+4 v_R^4 \left(\mathcal{C}_{\phi D}^{\{Rr\}\{(R)(r)\}}+4 \mathcal{C}_{\phi D}^{R(r)R(r)}\right)\Big], \\ \nonumber \\		
		(\Delta m_{W_3B}^{2})^{\phi^4D^2}_{23}&=& -\frac{g \tilde{g}}{4 \Lambda^2}\Big[2 \kappa_2^2 v_L v_R \left(\mathcal{C}_{\phi D}^{(L)11(r)}+\mathcal{C}_{\phi D}^{(L)12(r)}+\mathcal{C}_{\phi D}^{(R)11(l)}\right)+v_R^2 \left(4 \kappa_2^2 \mathcal{C}_{\phi D}^{(R)11(r)}-\kappa_1^2 \mathcal{C}_{\phi D}^{R(r)1(1)}\right) \nonumber \\
		&+&v_R^2 \left(2 v_R^2 (\mathcal{C}_{\phi D}^{\{Rr\}\{(R)(r)\}}+4 \mathcal{C}_{\phi D}^{R(r)R(r)})+2 \kappa_{+}^{2} \mathcal{C}_{\phi D}^{\{21\}\{(R)(r)\}}+2 \kappa_1^2 \mathcal{C}_{\phi D}^{21(R)(r)}\right)  \\
		&+&v_L^2 \left(2 v_R^2 (\mathcal{C}_{\phi D}^{L(r)L(r)}+\mathcal{C}_{\phi D}^{(R)lL(r)})-\kappa_1^2 \mathcal{C}_{\phi D}^{L(l)1(1)}\right)\Big],\nonumber \\ \nonumber \\
		(\Delta m_{W_3B}^{2})^{\phi^4D^2}_{33}&=& \frac{\tilde{g}^2}{2\Lambda^2}\Big[2 \kappa_2^2 v_L v_R \left(\mathcal{C}_{\phi D}^{(L)11(r)}+\mathcal{C}_{\phi D}^{(L)12(r)}+\mathcal{C}_{\phi D}^{(R)11(l)}\right)+ v_L^4 \left(\mathcal{C}_{\phi D}^{\{Ll\}\{(L)(l)\}}+4\mathcal{C}_{\phi D}^{L(l)L(l)}\right) \\ 
		&+&v_L^2 \left(v_R^2 (2 \mathcal{C}_{\phi D}^{L(r)L(r)}+\mathcal{C}_{\phi D}^{(L)rR(l)}+\mathcal{C}_{\phi D}^{(R)lL(r)})+\kappa_{+}^{2} \mathcal{C}_{\phi D}^{\{21\}\{(L)(l)\}}+\kappa_1^2 \mathcal{C}_{\phi D}^{21(L)(l)}+2 \kappa_2^2 \mathcal{C}_{\phi D}^{(L)11(l)}\right) \nonumber \\
		&+&v_R^2 \left(v_R^2 (\mathcal{C}_{\phi D}^{\{Rr\}\{(R)(r)\}}+4\mathcal{C}_{\phi D}^{R(r)R(r)})+\kappa_{+}^{2} \mathcal{C}_{\phi D}^{\{21\}\{(R)(r)\}}+\kappa_1^2 \mathcal{C}_{\phi D}^{21(R)(r)}+2 \kappa_2^2 \mathcal{C}_{\phi D}^{(R)11(r)}\right)\Big].\nonumber
\end{eqnarray}}

Once we incorporate the contributions from the dimension-6 operators, then the full gauge boson mass matrices can be written as:

{\small\begin{equation}\label{eq:MLRSM-cGB-mass-matrix-full}
		\mathcal{M}_{W^{\pm}}^2 = 
		\begin{pmatrix}
			(\tilde{m}_W^2)_{11} + (\Delta m_{W^{\pm}}^2)_{11} \ \ \hspace{1.cm} (\tilde{m}_W^2)_{12} +(\Delta m_{W^{\pm}}^2)_{12} \\ \\
			(\tilde{m}_W^2)_{21} +(\Delta m_{W^{\pm}}^2)_{21} \ \ \hspace{1.cm} (\tilde{m}_W^2)_{22} +(\Delta m_{W^{\pm}}^2)_{22}
		\end{pmatrix},
\end{equation}}
and 
{\small\begin{equation}\label{eq:MLRSM-nGB-mass-matrix-full}
		\mathcal{M}_{W_3 B}^2 =
		\begin{pmatrix}
			(\tilde{m}_0^2)_{11}+ (\Delta m_{W_3 B}^{2})_{11} \ \ \hspace{.6cm} (\tilde{m}_0^2)_{12}+ (\Delta m_{W_3 B}^{2})_{12} \ \ \hspace{.6cm} (\tilde{m}_0^2)_{13}+ (\Delta m_{W_3 B}^{2})_{13}\\ \\
			(\tilde{m}_0^2)_{21}+ (\Delta m_{W_3 B}^{2})_{21} \ \ \hspace{.6cm} (\tilde{m}_0^2)_{22}+ (\Delta m_{W_3 B}^{2})_{22} \ \ \hspace{.6cm} (\tilde{m}_0^2)_{23}+ (\Delta m_{W_3 B}^{2})_{23}\\ \\
			(\tilde{m}_0^2)_{31}+ (\Delta m_{W_3 B}^{2})_{31} \ \ \hspace{.6cm} (\tilde{m}_0^2)_{32}+ (\Delta m_{W_3 B}^{2})_{32} \ \ \hspace{.6cm} (\tilde{m}_0^2)_{33}+ (\Delta m_{W_3 B}^{2})_{33} 
		\end{pmatrix},
\end{equation}}
where  $(\tilde{m}_W^2)_{ij}$, and $(\tilde{m}_0^2)_{ij}$ are the elements of tree level charged and neutral gauge boson mass matrices respectively, see Eq.~\ref{eq:MLRSM-cGB-mass-matrix-tree} and Eq.~\ref{eq:MLRSM-nGB-mass-matrix-tree}.
Here, the shifted elements of the full mass matrices are:
\vskip 0.2cm
\underline{\large{Charged gauge boson mass matrix}}
\small{\begin{eqnarray}
		(\Delta m_{W^{\pm}}^2)_{11} &=&(\Delta m^2_{W^{\pm}})^{\phi^2 X^2}_{11}+(\Delta m^2_{W^{\pm}})^{\phi^4 D^2}_{11},\nonumber\\ 
		(\Delta m_{W^{\pm}}^2)_{22} &=&(\Delta m^2_{W^{\pm}})^{\phi^2 X^2}_{22}+(\Delta m^2_{W^{\pm}})^{\phi^4 D^2}_{22}, \nonumber \\ 
		(\Delta m_{W^{\pm}}^2)_{12} &=& (\Delta m_{W^{\pm}}^2)_{21} = (\Delta m^2_{W^{\pm}})^{\phi^2 X^2}_{12}+(\Delta m^2_{W^{\pm}})^{\phi^4 D^2}_{12}.
		\end{eqnarray}
		\vskip 0.2cm
		\underline{\large{Neutral gauge boson mass matrix}}
\small{\begin{eqnarray}
		(\Delta m_{W_3 B}^2)_{11} &=&(\Delta m^2_{W_3 B})^{\phi^2 X^2}_{11}+(\Delta m^2_{W_3 B})^{\phi^4 D^2}_{11},\nonumber\\
		(\Delta m_{W_3 B}^2)_{22} &=&(\Delta m^2_{W_3 B})^{\phi^2 X^2}_{22}+(\Delta m^2_{W_3 B})^{\phi^4 D^2}_{22},\nonumber\\
		(\Delta m_{W_3 B}^2)_{33} &=&(\Delta m^2_{W_3 B})^{\phi^2 X^2}_{33}+(\Delta m^2_{W_3 B})^{\phi^4 D^2}_{33},\nonumber\\
		(\Delta m_{W_3 B}^2)_{12} &=& (\Delta m_{W_3 B}^2)_{21} = (\Delta m^2_{W_3 B})^{\phi^2 X^2}_{12}+(\Delta m^2_{W_3 B})^{\phi^4 D^2}_{12},\nonumber\\
		(\Delta m_{W_3 B}^2)_{13} &=& (\Delta m_{W_3 B}^2)_{31} = (\Delta m^2_{W_3 B})^{\phi^2 X^2}_{13}+(\Delta m^2_{W_3 B})^{\phi^4 D^2}_{13},\nonumber\\
		(\Delta m_{W_3 B}^2)_{23} &=& (\Delta m_{W_3 B}^2)_{32} = (\Delta m^2_{W_3 B})^{\phi^2 X^2}_{23}+(\Delta m^2_{W_3 B})^{\phi^4 D^2}_{23}.
\end{eqnarray}}
The following matrix diagonalizes the modified charged gauge boson mass matrix:
\begin{equation}\label{eq:MLRSM-cGB-rot-matrix-full}
	\tilde{\mathcal{R}}_a =    \begin{pmatrix}
		\cos \tilde{\xi} \ \ \sin \tilde{\xi} \\
		-\sin \tilde{\xi}  \ \ \cos \tilde{\xi}
	\end{pmatrix},
\end{equation}
where this new ``angle" $\tilde{\xi}$ is related to the older one ${\xi}$ as:
{\small\begin{equation}\label{eq:MLRSM-cGB-rot-angle-full}
		\tan 2\tilde{\xi} = \tan 2\xi \left[ 1 +\left(\frac{(\Delta\mathcal{M}_{W^{\pm}}^2)_{12}}{(\tilde{M}_W^2)_{12}}-\frac{(\Delta\mathcal{M}_{W^{\pm}}^2)_{22}-(\Delta\mathcal{M}_{W^{\pm}}^2)_{11}}{(\tilde{M}_W^2)_{22}-(\tilde{M}_W^2)_{11}}\right) + O\left(\frac{1}{\Lambda^4}\right) \right].
\end{equation}}}
After performing the diagonalization, the charged gauge boson masses are given as:
\small{\begin{eqnarray}\label{eq:MLRSM-cGB-physical-mass-full}
	\mathcal{M}^2_{W_1} &=& m^2_{W_1} + ((\Delta m_{W^{\pm}}^2)_{11}\cos^2 \tilde{\xi} - (\Delta m_{W^{\pm}}^2)_{12}\sin 2 \tilde{\xi} + (\Delta m_{W^{\pm}}^2)_{22}\sin^2 \tilde{\xi}),\\
	\mathcal{M}^2_{W_2} &=& m^2_{W_2} + ((\Delta m_{W^{\pm}}^2)_{22}\cos^2 \tilde{\xi} + (\Delta m_{W^{\pm}}^2)_{12}\sin 2 \tilde{\xi} + (\Delta m_{W^{\pm}}^2)_{11}\sin^2 \tilde{\xi}).
	\label{eqn:MWLR}
\end{eqnarray}

The modified neutral gauge boson mass matrix is diagonalized by a similar rotation matrix $\tilde{\mathcal{R}}_{b}$ which will be of the similar form as $\mathcal{R}_{b}$ see Eq.~\ref{eq:MLRSM-cGB-rot-matrix}. The only difference is that the earlier rotation angle $\theta_2\in \mathcal{R}_{b}$ is replaced by $\tilde{\theta}_2\in \tilde{\mathcal{R}}_{b}$ where as the other two angles remain unaltered. 
The neutral gauge boson masses are given as:
\small{\begin{eqnarray}\label{eq:MLRSM-nGB-physical-mass-full-1}
		\mathcal{M}^2_{Z_1} &=& m^2_{Z_1} + \sin^2 \tilde{\theta}_2\Big[(\Delta m_{W_3 B}^{2})_{22}\cos^2\theta_1 + (\Delta m_{W_3 B}^{2})_{33}\sin^2\theta_1-(\Delta m_{W_3 B}^{2})_{23}\sin 2\theta_1\Big]\nonumber\\
		&+&\sin 2 \tilde{\theta}_2\Big[\sin\theta_{\text{w}}\big((\Delta m_{W_3 B}^{2})_{23}\cos 2\theta_1+\frac{1}{2}((\Delta m_{W_3 B}^{2})_{22}-(\Delta m_{W_3 B}^{2})_{33})\sin 2\theta_1\big)\nonumber\\
		&+&(\Delta m_{W_3 B}^{2})_{13}\sin\theta_{\text{w}}-(\Delta m_{W_3 B}^{2})_{12}\sqrt{\cos 2\theta_{\text{w}}} \hspace{0.1cm}\Big]\nonumber\\
		&+&\cos^2 \tilde{\theta}_2 \Big[(\Delta m_{W_3B}^{2})_{11}\cos^2\theta_{\text{w}}-2\sin\theta_{\text{w}}\big((\Delta m_{W_3 B}^{2})_{12}\sin\theta_{\text{w}}+(\Delta m_{W_3 B}^{2})_{13}\sqrt{\cos 2\theta_{\text{w}}}\big)\nonumber\\
		&+&\sin^2\theta_{\text{w}}\big((\Delta m_{W_3 B}^{2})_{33}\cos^2\theta_1 + (\Delta m_{W_3 B}^{2})_{22}\sin^2\theta_1+(\Delta m_{W_3 B}^{2})_{23}\sin 2\theta_1\big)\Big],\\
		\mathcal{M}^2_{Z_2} &=& m^2_{Z_2} +  \cos^2 \tilde{\theta}_2 \Big[(\Delta m_{W_3 B}^{2})_{22}\cos^2\theta_1 + (\Delta m_{W_3 B}^{2})_{33}\sin^2\theta_1-(\Delta m_{W_3 B}^{2})_{23}\sin 2\theta_1\Big]\nonumber\\
		&-&\sin 2 \tilde{\theta}_2 \Big[\sin\theta_{\text{w}}\big((\Delta m_{W_3 B}^{2})_{23}\cos 2\theta_1+\frac{1}{2}((\Delta m_{W_3 B}^{2})_{22}-(\Delta m_{W_3 B}^{2})_{33})\sin 2\theta_1\big)\nonumber\\
		&+&(\Delta m_{W_3 B}^{2})_{13}\sin\theta_{\text{w}}-(\Delta m_{W_3 B}^{2})_{12}\sqrt{\cos 2\theta_{\text{w}}} \hspace{0.1cm}\Big]\nonumber\\
		&+&\sin^2 \tilde{\theta}_2 \Big[(\Delta m_{W_3B}^{2})_{11}\cos^2\theta_{\text{w}}-2\sin\theta_{\text{w}}\big((\Delta m_{W_3 B}^{2})_{12}\sin\theta_{\text{w}}+(\Delta m_{W_3 B}^{2})_{13}\sqrt{\cos 2\theta_{\text{w}}}\big)\nonumber\\
		&+&\sin^2\theta_{\text{w}}\big((\Delta m_{W_3 B}^{2})_{33}\cos^2\theta_1 + (\Delta m_{W_3 B}^{2})_{22}\sin^2\theta_1+(\Delta m_{W_3 B}^{2})_{23}\sin 2\theta_1\big)\Big],\\ 
		\mathcal{M}^2_{A}  & =& 0.
	\end{eqnarray}
Note that even after inclusion of all the corrections due to the effective operators the mass of photon  turns out to be zero ($\mathcal{M}_{A}^2=0$) which is expected.\\
The modified rotation angle $\tilde{\theta}_2$ can be written in terms of ${\theta}_2$ as
	{\small\begin{equation}\label{eq:MLRSM-nGB-rot-angle-full}
			\tan2\tilde{\theta}_2 = \tan2\theta_2\Big[ 1 +\left(\frac{p}{q}-\frac{r}{s}\right)+O\left(\frac{1}{\Lambda^4}\right) \Big],
	\end{equation}}
	where,
	{\small\begin{eqnarray*}
			p &=& 2\cos 2\theta_{\text{w}}((\Delta m_{W_3 B}^{2})_{13}\cos\theta_1+(\Delta m_{W_3 B}^{2})_{12}\sin\theta_1)-\frac{1}{2}\sin 2\theta_{\text{w}}((\Delta m_{W_3 B}^{2})_{22}\\
			&+&(\Delta m_{W_3 B}^{2})_{33}-2(\Delta m_{W_3 B}^{2})_{11}+((\Delta m_{W_3 B}^{2})_{33}-(\Delta m_{W_3 B}^{2})_{22})\cos 2\theta_1+2(\Delta m_{W_3 B}^{2})_{23}\sin 2\theta_1),\\
			q &= &2\cos 2\theta_{\text{w}}((\tilde{m}_0^2)_{13}\cos\theta_1+(\tilde{m}_0^2)_{12}\sin\theta_1)\\
			&-&\frac{1}{2}\sin 2\theta_{\text{w}}((\tilde{m}_0^2)_{22}+(\tilde{m}_0^2)_{33}-2(\tilde{m}_0^2)_{11}+((\tilde{m}_0^2)_{33}-(\tilde{m}_0^2)_{22})\cos 2\theta_1+2(\tilde{m}_0^2)_{23}\sin 2\theta_1),\\
			r &=&\cos\theta_{\text{w}} (2(\Delta m_{W_3 B}^{2})_{23}\cos 2\theta_1+((\Delta m_{W_3 B}^{2})_{22}-(\Delta m_{W_3 B}^{2})_{33})\sin 2\theta_1\\ 
			& -&2\sin\theta_{\text{w}}((\Delta m_{W_3 B}^{2})_{13}\sin\theta_1-(\Delta m_{W_3 B}^{2})_{12}\cos\theta_1)),\\
			s &=&\cos\theta_{\text{w}}\left(2(\tilde{m}_0^2)_{23}\cos 2\theta_1+((\tilde{m}_0^2)_{22}-(\tilde{m}_0^2)_{33})\sin 2\theta_1-2\sin\theta_{\text{w}}((\tilde{m}_0^2)_{13}\sin\theta_1-(\tilde{m}_0^2)_{12}\cos\theta_1)\right).
	\end{eqnarray*}}
	Here, we would like to specify that the eigenvalues of the gauge boson masses remain unchanged upto $\mathcal{O}(1/\Lambda^2)$ even if we use the old rotation angles in the rotation matrices. But then the photon mass will have contributions   $\mathcal{O}(1/\Lambda^4)$ which needs to be ignored.
	\vskip0.5cm
	
	\underline{\large{$\psi^2 \phi^3 $ operators: Modification in  spectrum in the fermion sector }} \\
	
	The $ \psi^2 \phi^3$ operators generate the additional contributions to the charged fermion masses as:
	\small{\begin{eqnarray}\label{eq:MLRSM-fermion-mass-shift}
			\Delta m_{e} &=& \frac{1}{2\sqrt{2}\Lambda^2}\Big[\kappa_{2}\Big(2\kappa_{1}\kappa_{2}(\mathcal{C}^{1(41)}_{l_L l_R} +\mathcal{C}^{1(23)}_{l_L l_R}) + (\kappa_{1}^2 + \kappa_{2}^2)\mathcal{C}^{1(21)}_{l_L l_R}\Big) \big]\nonumber\\
			& & + \frac{1}{2\sqrt{2}\Lambda^2}\Big[ \kappa_{1}\Big(2\kappa_{1}\kappa_{2}(\mathcal{C}^{3(23)}_{l_L l_R} + \mathcal{C}^{3(41)}_{l_L l_R}) + (\kappa_{1}^2 + \kappa_{2}^2)\mathcal{C}^{3(21)}_{l_L l_R} \Big)\big]\\
			&    &+\frac{1}{2\sqrt{2}\Lambda^2}\Big[\kappa_{2}\big(v_L^2\mathcal{C}^{l3L}_{l_L l_R} + v_Lv_R\mathcal{C}^{{l3R}}_{l_L l_R} + v_R^2\mathcal{C}^{r3R}_{l_L l_R}\big) + \kappa_{1}\big(v_L^2\mathcal{C}^{l1L}_{l_L l_R} + v_Lv_R\mathcal{C}^{l1R}_{l_L l_R} + v_R^2\mathcal{C}^{r1R}_{l_L l_R}\big)\Big],\nonumber\\ \nonumber \\
			\Delta m_{u} &=& \frac{1}{2\sqrt{2}\Lambda^2}\Big[\kappa_{1}\Big(2\kappa_{1}\kappa_{2}(\mathcal{C}^{1(41)}_{q_L q_R} +\mathcal{C}^{1(23)}_{q_L q_R}) + (\kappa_{1}^2 + \kappa_{2}^2)\mathcal{C}^{1(21)}_{q_L q_R}\Big) \Big]\\
			& &  + \frac{1}{2\sqrt{2}\Lambda^2}\Big[\kappa_{2}\Big(2\kappa_{1}\kappa_{2}(\mathcal{C}^{3(23)}_{q_L q_R} + \mathcal{C}^{3(41)}_{q_L q_R}) + (\kappa_{1}^2 + \kappa_{2}^2)\mathcal{C}^{3(21)}_{q_L q_R} \Big)\Big]\nonumber\\
			&    &+\frac{1}{2\sqrt{2}\Lambda^2}\Big[\kappa_{1}\big(v_L^2\mathcal{C}^{L3l}_{q_L q_R} + v_Lv_R\mathcal{C}^{L3r}_{q_L q_R} + v_R^2\mathcal{C}^{R3r}_{q_L q_R}\big) + \kappa_{2}\big(v_L^2\mathcal{C}^{L1l}_{q_L q_R} + v_Lv_R\mathcal{C}^{L1r}_{q_L q_R} + v_R^2\mathcal{C}^{R1r}_{q_L q_R}\big)\Big],\nonumber \\  \nonumber \\
				\Delta m_{d} &=& \frac{1}{2\sqrt{2}\Lambda^2}\Big[\kappa_{2}\Big(2\kappa_{1}\kappa_{2}(\mathcal{C}^{1(41)}_{q_L q_R} +\mathcal{C}^{1(23)}_{q_L q_R}) + (\kappa_{1}^2 + \kappa_{2}^2)\mathcal{C}^{1(21)}_{q_L q_R}\Big)  \Big]\\
			& & + \frac{1}{2\sqrt{2}\Lambda^2}\Big[\kappa_{1}\Big(2\kappa_{1}\kappa_{2}(\mathcal{C}^{3(23)}_{q_L q_R} + \mathcal{C}^{3(41)}_{q_L q_R}) + (\kappa_{1}^2 + \kappa_{2}^2)\mathcal{C}^{3(21)}_{q_L q_R} \Big)\Big]\nonumber\\
			&    &+\frac{1}{2\sqrt{2}\Lambda^2}\Big[\kappa_{2}\big(v_L^2\mathcal{C}^{l3L}_{q_L q_R} + v_Lv_R\mathcal{C}^{{l3R}}_{q_L q_R} + v_R^2\mathcal{C}^{r3R}_{q_L q_R}\big) + \kappa_{1}\big(v_L^2\mathcal{C}^{l1L}_{q_L q_R} + v_Lv_R\mathcal{C}^{l1R}_{q_L q_R} + v_R^2\mathcal{C}^{r1R}_{q_L q_R}\big)\Big].\nonumber
		\end{eqnarray}
\clearpage
\noindent
		After incorporating these corrections, the resulting charged fermion mass matrices are written as:
		\small{\begin{eqnarray}\label{eq:MLRSM-fermion-mass-full}
				\mathcal{M}_{e}&=&\frac{1}{\sqrt{2}}\left(\kappa _1 \tilde{y}_D+\kappa _2 y_D\right)-\Delta m_{e},\\
				\mathcal{M}_{u}&=&\frac{1}{\sqrt{2}}\left(\kappa _2 \tilde{y}_q+\kappa _1 y_q\right)-\Delta m_{u},\\
				\mathcal{M}_{d}&=&\frac{1}{\sqrt{2}}\left(\kappa _1 \tilde{y}_q+\kappa _2 y_q\right)-\Delta m_{d}.
		\end{eqnarray}}
		
		In the process, the Dirac-mass term for neutrinos also achieve correction ($\Delta m_{\nu}^{D} $):
		\small{\begin{eqnarray}\label{eq:MLRSM-neutrino-Dirac-mass-shift}
				\Delta m_{\nu}^{D} &=& \frac{1}{2\sqrt{2}\Lambda^2}\Big[\kappa_{1}\Big(2\kappa_{1}\kappa_{2}(\mathcal{C}^{1(41)}_{l_L l_R} +\mathcal{C}^{1(23)}_{l_L l_R}) + (\kappa_{1}^2 + \kappa_{2}^2)\mathcal{C}^{1(21)}_{l_L l_R}\Big) \Big]\nonumber\\
				&+& \frac{1}{2\sqrt{2}\Lambda^2}\Big[\kappa_{2}\Big(2\kappa_{1}\kappa_{2}(\mathcal{C}^{3(23)}_{l_L l_R} + \mathcal{C}^{3(41)}_{l_L l_R}) + (\kappa_{1}^2 + \kappa_{2}^2)\mathcal{C}^{3(21)}_{l_L l_R} \Big)\Big] \\
				&+&\frac{1}{2\sqrt{2}\Lambda^2}\Big[\kappa_{1}\big(v_L^2\mathcal{C}^{L3l}_{l_L l_R} + v_Lv_R\mathcal{C}^{L3r}_{l_L l_R} + v_R^2\mathcal{C}^{R3r}_{l_L l_R}\big) + \kappa_{2}\big(v_L^2\mathcal{C}^{L1l}_{l_L l_R} + v_Lv_R\mathcal{C}^{L1r}_{l_L l_R} + v_R^2\mathcal{C}^{R1r}_{l_L l_R}\big)\Big],\nonumber
		\end{eqnarray}}
		and the resulting Dirac-mass term for the neutrinos is read as:
		\small{\begin{eqnarray}\label{eq:MLRSM-neutrino-Dirac-mass-full}
				\mathcal{M}_{\nu}^{D}&=&\frac{1}{\sqrt{2}}\left(\kappa _2 \tilde{y}_D+\kappa _1 y_D\right)-\Delta m_{\nu}^{D}.
		\end{eqnarray}}
		The lepton number violating $ \psi^2 \phi^3$ operators will provide additional contributions to the Majorana masses for the light ($\nu$) and heavy ($N$) neutrinos, Eq.~\ref{eq:MLRSM-neutrino-mass-basis-tree}, respectively:
		{\small\begin{eqnarray}\label{eq:MLRSM-Majorana-neutrino-mass-shift}
				\Delta m_{\nu}^{M} & = &\frac{1}{ \sqrt{2} \Lambda^2}\big[v_L \big(v_R^2 (\mathcal{C}_{l_L}^{rLr}+\mathcal{C}_{l_L}^{l(Rr)})+\kappa _1^2 (\mathcal{C}_{l_L}^{2l1}+\mathcal{C}_{l_L}^{l(21)})+\kappa _2^2 \mathcal{C}_{l_L}^{l(21)}+2 \kappa _1 \kappa _2 (\mathcal{C}_{l_L}^{l(23)}+\mathcal{C}_{l_L}^{l(41)})\big) \nonumber\\
				&    & +v_L^3 (\mathcal{C}_{l_L}^{lLl} +\mathcal{C}_{l_L}^{l(Ll)})+\kappa _1 v_R \big(\kappa _1 \mathcal{C}_{l_L}^{r21}+\kappa _2 (\mathcal{C}_{l_L}^{r23}+\mathcal{C}_{l_L}^{r41})\big)\big],\\
				\Delta m_{_N}^{M} &= &\frac{1}{ \sqrt{2} \Lambda^2}\big[v_L^2 v_R (\mathcal{C}_{l_R}^{lRl}+\mathcal{C}_{l_R}^{r(Ll)})+\kappa _1 v_L \big(\kappa _1 \mathcal{C}_{l_R}^{l21}+\kappa _2 (\mathcal{C}_{l_R}^{l23}+\mathcal{C}_{l_R}^{l41})\big)+v_R \big(v_R^2 (\mathcal{C}_{l_R}^{rRr}+\mathcal{C}_{l_R}^{r(Rr)})\nonumber\\
				&    &+\kappa _1^2 (\mathcal{C}_{l_R}^{r21}+\mathcal{C}_{l_R}^{r(21)})+\kappa _2^2 \mathcal{C}_{l_R}^{r(21)}+2 \kappa _1 \kappa _2 (\mathcal{C}_{l_R}^{r(23)}+\mathcal{C}_{l_R}^{r(41)})\big)\big].
		\end{eqnarray}}
		After incorporating the corrections due to the effective operators, the full neutrino mass matrix, Eq.~\ref{eq:MLRSM-neutrino-mass-matrix-tree}, can be written as:
		\begin{eqnarray}\label{eq:MLRSM-neutrino-mass-matrix-full}
			m_{\nu}=\begin{pmatrix}
				\sqrt{2} y_M v_L-\Delta m_{\nu}^{M} & \mathcal{M}_{\nu}^{D} \\
				\mathcal{M}_{\nu}^{D} & \sqrt{2} y_M v_R-\Delta m_{_N}^{M}  \\
			\end{pmatrix}.
		\end{eqnarray}
\clearpage

%% file: Weak_mixing_angle.tex
\subsection{Weak mixing angle}

The Weak mixing angle is measured using the low energy data of nucleon-neutrino scattering. 
This scattering is driven by the charged and neutral currents. One can define a ratio ($\widehat{R}$) using this scattering information as \cite{Buchmuller:1985jz}:
{\small\begin{align}
		\widehat{R}=\frac{\sigma^{\nu NC}-\sigma^{\bar{\nu} NC}}{\sigma^{\nu CC}-\sigma^{\bar{\nu} CC}},
\end{align}}
which can be recast in terms of the Weak mixing angle  \cite{Buchmuller:1985jz}:
{\small\begin{align}\label{eqn:R}
		\widehat{R}=\frac{1}{2}-\sin^2\bar{\theta}_\text{w}.
\end{align}}
Here, $\sigma^{\nu(\bar{\nu}) NC}$ and $\sigma^{\nu(\bar{\nu}) CC}$ represent the cross section of neutrino (anti-neutrino)-nucleon scattering through neutral (NC) and charged (CC) currents respectively.

\begin{figure}[h!]
	\centering
	{
		\includegraphics[trim={1cm 0 1.2cm 0},scale=0.48]{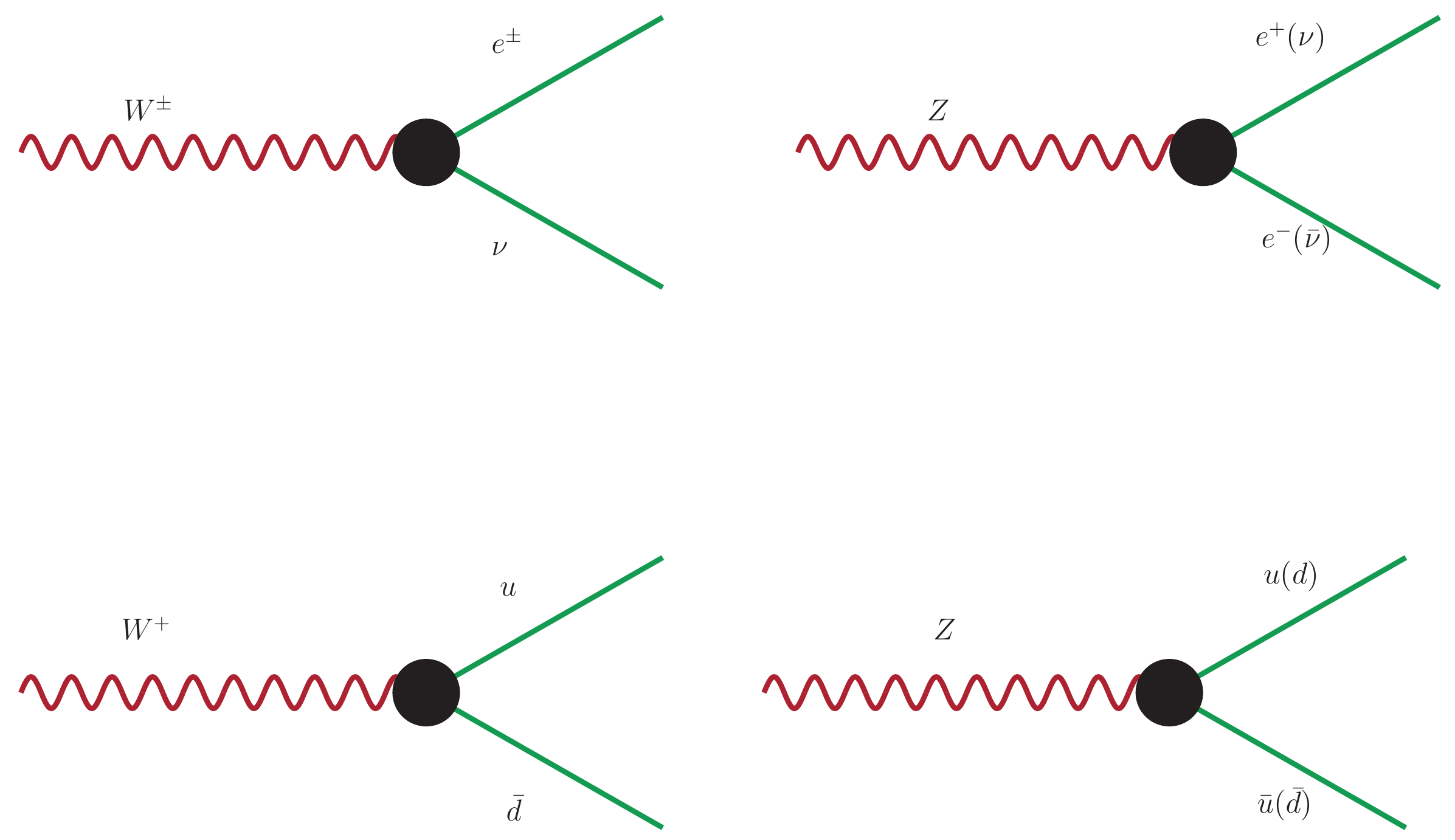}
	}
	\caption{The effective interactions among gauge bosons with the fermions. The [{\Large $\bullet$}]  depicts the modified Feynman vertex (renormalizable + corrections due to dimension-6 operators) factors which are tabulated in the following sections for both 2HDM and MLRSM effective scenarios.}\label{fig:GB-lepton-quark}
\end{figure}

\section*{2HDM}

For 2HDM scenario, the scattering cross-sections ($\sigma^{\nu(\bar{\nu}) NC}$, and $\sigma^{\nu(\bar{\nu}) CC}$) are derived using the following part of the lagrangian \cite{Buchmuller:1985jz}:
{\small\begin{eqnarray}
		\mathcal{L}^{CC}_{\nu Q}&=& \frac{g^2}{2\mathcal{M}^2_W}  \overline{e_{L}}\gamma ^{\mu } \nu_{L}\Big(\epsilon_{\nu*(ud)_{L}} \overline{u_{L}}\gamma ^{\mu } d_{L}+(\epsilon_{\nu*(ud)_{R}} \overline{u_{R}}\gamma ^{\mu } d_{R}\Big)+h.c.,\\
		\mathcal{L}^{NC}_{\nu Q}&=& \frac{g^2}{ \cos^2\theta_{\text{w}} \mathcal{M}^2_{Z}} \overline{\nu_{L}}\gamma ^{\mu } \nu_{L}\Big(\zeta_{\nu*u_{L}} \overline{u_{L}}\gamma ^{\mu } u_{L}+\zeta_{\nu*u_{R}} \overline{u_{R}}\gamma ^{\mu } u_{R}\nonumber \\
		&&+\zeta_{\nu*d_{L}}\overline{d_{L}}\gamma ^{\mu } d_{L}+\zeta_{\nu*d_{R}} \overline{d_{R}}\gamma ^{\mu } d_{R}\Big),
\end{eqnarray}}
where $    \epsilon_{\nu*(ud)_{L}}, \epsilon_{\nu*(ud)_{R}}, \zeta_{\nu*u_{L}},     \zeta_{\nu*d_{L}}, \zeta_{\nu*u_{R}}, \zeta_{\nu*d_{R}} $ are given as:
{\small\begin{align*}
		&    \epsilon_{\nu*(ud)_{L}}=  \epsilon_{(e \nu )_L} * \epsilon_{(ud)_L}-\frac{4  \mathcal{M}_W^2}{g^2 \Lambda ^2}\mathcal{C}_{LQ}^{[3]},& &
		\epsilon_{\nu*(ud)_{R}}= \epsilon_{(e \nu )_L} * \epsilon_{(ud)_R},\\
		&    \zeta_{\nu*u_{L}}=\zeta_{\nu_L} \zeta_{u_L}- \frac{\cos^2 \theta_{\text{w}} \mathcal{M}_Z^2 }{g^2 \Lambda ^2}\Big[\mathcal{C}_{LQ}^{[1]}+\mathcal{C}_{LQ}^{[3]}\Big],& &
		\zeta_{\nu*d_{L}}= \zeta_{d_L} * \zeta_{\nu_L}-\frac{\cos^2 \theta_{\text{w}} \mathcal{M}_Z^2 }{g^2 \Lambda ^2}\Big[\mathcal{C}_{LQ}^{[1]}-\mathcal{C}_{LQ}^{[3]}\Big],\\
		&    \zeta_{\nu*u_{R}}= \zeta_{\nu_L} * \zeta_{u_R}-\frac{\cos^2 \theta_{\text{w}} \mathcal{M}_Z^2}{g^2 \Lambda ^2}\mathcal{C}_{Lu},& &
		\zeta_{\nu*d_{R}}=\zeta_{\nu_l}* \zeta_{d_R}- \frac{\cos^2 \theta_{\text{w}}\mathcal{M}_Z^2}{g^2 \Lambda ^2}\mathcal{C}_{Ld},
\end{align*}}
and their associated parameters $\epsilon$ and $\zeta$ which encapsulate the features of the relevant modified interactions (schematically depicted in Fig.~\ref{fig:GB-lepton-quark}.) are given in Table~\ref{tab:2HDM-GB-F-F-I}.
\\ \\
The parameters $a^{i}_{L/R}$ for $ i = u,d,\nu,e$ in Table~\ref{tab:2HDM-GB-F-F-I}, are given as:
\small{\begin{align*}
		a_L^{\nu }&= \frac{1}{2}+\frac{1}{2\Lambda ^2}\left( \Theta _{\text{BB}} \sin ^2\theta_{\text{w}}+ \Theta _{\text{WW}} \cos ^2\theta_{\text{w}}-2\Theta_{W_{3}B} \sin \theta_{\text{w}} \cos \theta_{\text{w}}\right),\\
		a_L^e &= -\frac{1}{2}+\sin ^2\theta_{\text{w}}+\frac{1}{2\Lambda ^2}\left( \Theta _{\text{BB}} \sin ^2\theta_{\text{w}}- \Theta_{\text{WW}} \cos ^2\theta_{\text{w}}\right),\\
		a_R^e &= \sin ^2\theta_{\text{w}}+\frac{1}{\Lambda ^2}\left(\Theta _{\text{BB}} \sin ^2\theta_{\text{w}}-\Theta_{W_{3}B} \sin \theta_{\text{w}} \cos \theta_{\text{w}}\right),\\
		a_L^u&= \frac{1}{2}-\frac{2}{3} \sin ^2\theta_{\text{w}}+\frac{1}{\Lambda^2}\left(\frac{1}{2} \Theta _{\text{WW}} \cos ^2\theta_{\text{w}}-\frac{1}{3} \Theta_{W_{3}B} \sin \theta_{\text{w}} \cos \theta_{\text{w}}-\frac{1}{6} \Theta _{\text{BB}} \sin ^2\theta_{\text{w}}\right),\\
		a_R^u&=-\frac{2}{3} \sin ^2\theta_{\text{w}}+ \frac{2}{3\Lambda ^2}\left( \Theta_{W_{3}B} \sin \theta_{\text{w}} \cos \theta_{\text{w}}- \Theta _{\text{BB}} \sin ^2\theta_{\text{w}}\right),\\
		a_L^d&= -\frac{1}{2}+\frac{1}{3} \sin ^2\theta_{\text{w}}+\frac{1}{\Lambda ^2}\left(-\frac{1}{2} \Theta _{\text{WW}} \cos ^2\theta_{\text{w}}-\frac{1}{6} \Theta _{\text{BB}} \sin ^2\theta_{\text{w}}+\frac{2}{3} \Theta_{W_{3}B} \sin \theta_{\text{w}} \cos \theta_{\text{w}}\right),\\
		a_R^d&= \frac{1}{3} \sin ^2\theta_{\text{w}}+\frac{1}{3\Lambda ^2}\left( \Theta _{\text{BB}} \sin ^2\theta_{\text{w}}- \Theta_{W_{3}B} \sin \theta_{\text{w}} \cos \theta_{\text{w}}\right).
\end{align*}}

\begin{table}[h!]
	\centering
	\renewcommand{\arraystretch}{2.0}
	{\small\begin{tabular}{|c | c|}
			\hline
			$\overline{\psi}\gamma ^{\mu } \psi  X_{\mu} $& {\bf2HDM: Charged current vertex factor}\\
			\hline
			$\overline{\nu_L}\gamma ^{\mu } e_L W^{+}_{\mu }$   &  $\frac{g}{\sqrt{2}} \epsilon_{(e\nu)_L}=\frac{g}{\sqrt{2}}\Big[1+\frac{\Theta_{\text{WW}}}{\Lambda ^2}+\frac{\sqrt{2}}{g}A_{e_L \nu _L W}\Big]$ \\
			$\overline{u_L}\gamma ^{\mu } d_LW^{+}_{\mu }$  & $\frac{g}{\sqrt{2}} \epsilon_{(ud)_L}=\frac{g}{\sqrt{2}}\Big[1+\frac{\Theta_{\text{WW}}}{\Lambda ^2}+\frac{\sqrt{2}}{g}A_{u_L  d_L W}\Big] $\\
			$\overline{u_R}\gamma ^{\mu } d_R W^{+}_{\mu } $ & $ \frac{g}{\sqrt{2}} \epsilon_{(ud)_R}= A_{u_R d_R W}$ \\
			\hline
			 &{\bf 2HDM: Neutral current vertex factor}\\
			\hline
			$\overline{\nu_{_L}}\gamma ^{\mu } \nu _L Z_{\mu }$& $ \frac{g}{\cos \theta_{\text{w}}} \zeta _{\nu_L}= \frac{g}{\cos \theta_{\text{w}}} \Big[a^{\nu}_{L}+\frac{\cos \theta_{\text{w}}}{g} A_{\nu_L \nu _L Z}\Big]$ \\
			$\overline{e_L}\gamma ^{\mu } e_LZ_{\mu }$  & $\frac{g}{\cos \theta_{\text{w}}} \zeta _{e_L}=\frac{g}{\cos \theta_{\text{w}}}\Big[ a^{e}_{L}+\frac{\cos \theta_{\text{w}}}{g} A_{e_L e_L Z}\Big]$\\
			$\overline{e_R}\gamma ^{\mu } e_R  Z_{\mu } $ & $ \frac{g }{\cos \theta_{\text{w}}}\zeta _{e_R}=\frac{g}{\cos \theta_{\text{w}}} \Big[a^{e}_{R}+\frac{\cos \theta_{\text{w}}}{g} A_{e_R e_R Z} \Big]$ \\
			$\overline{u_L}\gamma ^{\mu } u_L  Z_{\mu } $& $ \frac{g}{\cos \theta_{\text{w}}} \zeta _{u_L}=\frac{g}{\cos \theta_{\text{w}}} \Big[a^{u}_{L}+\frac{\cos \theta_{\text{w}}}{g} A_{u_L u_L Z} \Big]$ \\
			$\overline{u_R}\gamma ^{\mu } u_R  Z_{\mu }$& $ \frac{g}{\cos \theta_{\text{w}}} \zeta _{u_R} =\frac{g}{\cos \theta_{\text{w}}}\Big[ a^{u}_{R}+\frac{\cos \theta_{\text{w}}}{g} A_{u_R u_R Z}\Big]$\\
			$\overline{d_L}\gamma ^{\mu } d_L Z_{\mu } $& $ \frac{g}{\cos \theta_{\text{w}}}\zeta _{d_L}=\frac{g}{\cos \theta_{\text{w}}}\Big[ a^{d}_{L}+\frac{\cos \theta_{\text{w}}}{g}A_{d_L d_L Z} \Big]$ \\
			$\overline{d_R}\gamma ^{\mu } d_R  Z_{\mu }$& $ \frac{g}{\cos \theta_{\text{w}}} \zeta _{d_R} =\frac{g}{\cos \theta_{\text{w}}}\Big[ a^{d}_{R}+\frac{\cos \theta_{\text{w}}}{g} A_{d_R d_R Z} \Big]$\\
			\hline
	\end{tabular}}
	\caption{2HDM: Coupling of gauge bosons and fermions. The detailed structures of the $A$'s are given in Table~\ref{tab:2HDM-GB-F-F-II}. Hermitian conjugate vertices are not separately mentioned.}
	\label{tab:2HDM-GB-F-F-I}
\end{table}

Implementing all these corrections due to the dimension-6 operators we can estimate the new Weak mixing angle as $\sin^2\bar{\theta}_\text{w}	=\frac{1}{2}-\widehat{R}$, where 
{\small\begin{align}\label{eqn:R6}
	\widehat{R}=\frac{4\mathcal{M}_W^4}{\cos^4 \theta_{\text{w}} \mathcal{M}_Z^4}\Bigg(\frac{\zeta_{\nu*u_{L}}^2+\zeta_{\nu*d_{L}}^2-\zeta_{\nu*u_{R}}^2-\zeta_{\nu*d_{R}}^2}{\epsilon_{\nu*(ud)_{L}}^2-\epsilon_{\nu*(ud)_{R}}^2}\Bigg).
	\end{align}}

\begin{table}[h!]
	\centering
	\renewcommand{\arraystretch}{2.50}
	\begin{tabular}{|c|cccc|}
			\hline
			\hspace{0.2cm}
$ \overline{\nu _L} \gamma ^{\mu } e_L W^{+}_{\mu }$ 		&  &$ A_{\nu_L e_L W}$ & = &$ \frac{g \mathcal{C}_{L\phi D}^{11[3]} v_1^2}{\sqrt{2} \Lambda ^2}\textcolor{purple}{+\frac{\sqrt{2} g \mathcal{C}_{L\phi D}^{12[3]} v_2 v_1}{\Lambda ^2}}+\frac{g \mathcal{C}_{L\phi D}^{22[3]} v_2^2}{\sqrt{2} \Lambda ^2}$ \\
			 	\hspace{0.2cm}
			$\overline{u_L}\gamma ^{\mu } d_L W^{+}_{\mu } $ &  & $A _{u_L d_L W}$ & = &$ \frac{g \mathcal{C}_{Q\phi D}^{11[3]} v_1^2}{\sqrt{2} \Lambda ^2}\textcolor{purple}{+\frac{\sqrt{2} g \mathcal{C}_{Q\phi D}^{12[3]} v_2 v_1}{\Lambda ^2}}+\frac{g \mathcal{C}_{Q\phi D}^{22[3]} v_2^2}{\sqrt{2} \Lambda ^2}$ \\
			 	\hspace{0.2cm}
			$\overline{u_R} \gamma ^{\mu } d_R W^{+}_{\mu } $ &  &$ A_{u_R d_R W} $ & = &$ \frac{g \mathcal{C}_{ud\phi D}^{\tilde{1}1} v_1^2}{\sqrt{2} \Lambda ^2}+\frac{g \mathcal{C}_{ud\phi D}^{\tilde{2}2} v_2^2}{\sqrt{2} \Lambda ^2}\textcolor{purple}{+\frac{g \mathcal{C}_{ud\phi D}^{\tilde{2}1} v_2 v_1}{\sqrt{2} \Lambda ^2}} $\\
			
			$\overline{\nu _L}\gamma ^{\mu } \nu _L Z_{\mu } $ &  & $A_{\nu_L \nu_L Z} $& = & $-\frac{g \mathcal{C}_{L\phi D}^{11[1]} v_1^2}{\sqrt{2} \Lambda ^2}+\frac{g \mathcal{C}_{L\phi D}^{11[3]} v_1^2}{\sqrt{2} \Lambda ^2}\textcolor{purple}{-\frac{\sqrt{2} g \mathcal{C}_{L\phi D}^{12[1]} v_2 v_1}{\Lambda ^2}}$\\
			& & & &$\textcolor{purple}{+\frac{\sqrt{2} g \mathcal{C}_{L\phi D}^{12[3]} v_2 v_1}{\Lambda ^2}}-\frac{g \mathcal{C}_{L\phi D}^{22[1]} v_2^2}{\sqrt{2} \Lambda ^2}+\frac{g \mathcal{C}_{L\phi D}^{22[3]} v_2^2}{\sqrt{2} \Lambda ^2}$ \\
			
			$\overline{e_L}\gamma ^{\mu } e_L Z_{\mu }$  &  &$ A_{e_L e_L Z} $& = & $-\frac{g \mathcal{C}_{L\phi D}^{11[1]} v_1^2}{\sqrt{2} \Lambda ^2}-\frac{g \mathcal{C}_{L\phi D}^{11[3]} v_1^2}{\sqrt{2} \Lambda ^2}\textcolor{purple}{-\frac{\sqrt{2} g \mathcal{C}_{L\phi D}^{12[1]} v_2 v_1}{\Lambda ^2}}$\\
			& & & &$\textcolor{purple}{-\frac{\sqrt{2} g \mathcal{C}_{L\phi D}^{12[3]} v_2 v_1}{\Lambda ^2}}-\frac{g \mathcal{C}_{L\phi D}^{22[1]} v_2^2}{\sqrt{2} \Lambda ^2}-\frac{g \mathcal{C}_{L\phi D}^{22[3]} v_2^2}{\sqrt{2} \Lambda ^2} $\\
			
			$\overline{e_R} \gamma ^{\mu } e_R Z_{\mu }  $&  &$ A_{e_R e_R Z}$ & = & $-\frac{g \mathcal{C}_{e\phi D}^{11} v_1^2}{\sqrt{2} \Lambda ^2}\textcolor{purple}{-\frac{\sqrt{2} g \mathcal{C}_{e\phi D}^{12} v_2 v_1}{\Lambda ^2}}-\frac{g \mathcal{C}_{e\phi D}^{22} v_2^2}{\sqrt{2} \Lambda ^2}$ \\
			
			$\overline{u_L}\gamma ^{\mu } u_L Z_{\mu } $ &  & $A _{u_L u_L Z} $& = &$ -\frac{g \mathcal{C}_{Q\phi D}^{11[1]} v_1^2}{\sqrt{2} \Lambda ^2}+\frac{g \mathcal{C}_{Q\phi D}^{11[3]} v_1^2}{\sqrt{2} \Lambda ^2}\textcolor{purple}{-\frac{\sqrt{2} g \mathcal{C}_{Q\phi D}^{12[1]} v_2 v_1}{\Lambda ^2}}$\\
			& & & &$\textcolor{purple}{+\frac{\sqrt{2} g \mathcal{C}_{Q\phi D}^{12[3]} v_2 v_1}{\Lambda ^2}}-\frac{g \mathcal{C}_{Q\phi D}^{22[1]} v_2^2}{\sqrt{2} \Lambda ^2}+\frac{g \mathcal{C}_{Q\phi D}^{22[3]} v_2^2}{\sqrt{2} \Lambda ^2}$ \\
			
			$\overline{d_L}\gamma ^{\mu } d_L Z_{\mu } $ &  & $A_{d_L d_L Z} $& = &$ -\frac{g \mathcal{C}_{Q\phi D}^{11[1]} v_1^2}{\sqrt{2} \Lambda ^2}-\frac{g \mathcal{C}_{Q\phi D}^{11[3]} v_1^2}{\sqrt{2} \Lambda ^2}\textcolor{purple}{-\frac{\sqrt{2} g \mathcal{C}_{Q\phi D}^{12[1]} v_2 v_1}{\Lambda ^2}}$\\
			& & & &$\textcolor{purple}{-\frac{\sqrt{2} g \mathcal{C}_{Q\phi D}^{12[3]} v_2 v_1}{\Lambda ^2}}-\frac{g \mathcal{C}_{Q\phi D}^{22[1]} v_2^2}{\sqrt{2} \Lambda ^2}-\frac{g \mathcal{C}_{Q\phi D}^{22[3]} v_2^2}{\sqrt{2} \Lambda ^2} $\\
			
			$\overline{u_R} \gamma ^{\mu } u_R Z_{\mu }$ &  &$ A_{u_R u_R Z} $& = & $-\frac{g \mathcal{C}_{u\phi D}^{11} v_1^2}{\sqrt{2} \Lambda ^2}\textcolor{purple}{-\frac{\sqrt{2} g \mathcal{C}_{u\phi D}^{12} v_2 v_1}{\Lambda ^2}}-\frac{g \mathcal{C}_{u\phi D}^{22} v_2^2}{\sqrt{2} \Lambda ^2} $\\
			
			$\overline{d_R} \gamma ^{\mu } d_R Z_{\mu }$  &  &$ A_{d_R d_R Z}$ & = & $-\frac{g \mathcal{C}_{d\phi D}^{11} v_1^2}{\sqrt{2} \Lambda ^2}\textcolor{purple}{-\frac{\sqrt{2} g \mathcal{C}_{d\phi D}^{12} v_2 v_1}{\Lambda ^2}}-\frac{g \mathcal{C}_{d\phi D}^{22} v_2^2}{\sqrt{2} \Lambda ^2}$ \\
			\hline
	\end{tabular}
	\caption{2HDM: Corrections to gauge bosons and fermion couplings from $ \psi^2 \phi^2 D$ operators.}
	\label{tab:2HDM-GB-F-F-II}
\end{table}
\clearpage



\section*{MLRSM}

Similar to the 2HDM case, we have constructed the lagrangian which contains the neutral and charged current interactions :
{\small
	\begin{eqnarray}
	\mathcal{L}^{CC}_{\nu Q}&=& \frac{g^2}{2\mathcal{M}^2_{W_1}} \overline{e_{L}}\gamma^{\mu} \nu_{L}\Big(\epsilon^1_{( e \nu)_L*(ud)_{L}} \overline{u_{L}}\gamma^{\mu} d_{L}+\epsilon^1_{(e \nu )_L*(ud)_{R}} \overline{u_{R}}\gamma^{\mu} d_{R}\Big) + h.c. \nonumber\\
	&&+\frac{g^2}{2\mathcal{M}^2_{W_2}} \overline{e_{L}}\gamma^{\mu} \nu_{L}\Big(\epsilon^2_{(e \nu )_L*(ud)_{L}} \overline{u_{L}}\gamma^{\mu} d_{L}+\epsilon^2_{( e \nu )_L*(ud)_{R}} \overline{u_{R}}\gamma^{\mu} d_{R}\Big)+ h.c.,\nonumber\\
	\mathcal{L}^{NC}_{\nu Q}&=& \frac{g^2}{4 \cos^2\theta_{\text{w}} \mathcal{M}^2_{Z_1}} \overline{\nu_{L}}\gamma ^{\mu } \nu_{L}\Big(\zeta^1_{\nu_L*u_{L}} \overline{u_{L}}\gamma ^{\mu } u_{L}+\zeta^1_{\nu_L*u_{R}} \overline{u_{R}}\gamma ^{\mu } u_{R}+\zeta^1_{\nu_L*d_{L}}\overline{d_{L}}\gamma ^{\mu } d_{L}+\zeta^1_{\nu_L*d_{R}} \overline{d_{R}}\gamma ^{\mu } d_{R}\Big)\nonumber\\
	&&+\frac{g^2}{4 \cos^2\theta_{\text{w}} \mathcal{M}^2_{Z_2}} \overline{\nu_{L}}\gamma ^{\mu } \nu_{L}\Big(\zeta^2_{\nu_L*u_{L}} \overline{u_{L}}\gamma ^{\mu } u_{L}+\zeta^2_{\nu_L*u_{R}} \overline{u_{R}}\gamma ^{\mu } u_{R}+\zeta^2_{\nu_L*d_{L}}\overline{d_{L}}\gamma ^{\mu } d_{L}+\zeta^2_{\nu_L*d_{R}} \overline{d_{R}}\gamma ^{\mu } d_{R}\nonumber\Big).
	\end{eqnarray}}

The parameters [$\epsilon^1_{(e\nu)_{L}*(ud)_{L}}, \epsilon^1_{(e\nu)_{L}*(ud)_{R}}, \epsilon^2_{(e\nu)_{L}*(ud)_{L}}, \epsilon^2_{(e\nu)_{L}*(ud)_{R}}$] contained in the charged current are expressed as:
\begin{eqnarray}
	\epsilon^1_{(e\nu)_{L}*(ud)_{L}}&= \epsilon^1_{(e\nu)_L}\epsilon^1_{(ud)_L}-\frac{4\mathcal{M}^2_{W_1}}{g^2 \Lambda^2}\mathcal{C}^{[3]}_{l_L q_L},\hspace{1cm}\epsilon^1_{(e\nu)_{L}*(ud)_{R}}=  \epsilon^1_{(e\nu)_L}\epsilon^1_{(ud)_R},\\
	\epsilon^2_{(e\nu)_{L}*(ud)_{L}}&=  \epsilon^2_{(e\nu)_L}\epsilon^2_{(ud)_L}-\frac{4\mathcal{M}^2_{W_2}}{g^2 \Lambda^2}\mathcal{C}^{[3]}_{l_L q_L},\hspace{1cm}\epsilon^2_{(e\nu)_{L}*(ud)_{R}}=  \epsilon^1_{(e\nu)_L}\epsilon^1_{(ud)_R}.
	\end{eqnarray}

\begin{table}[h!]
	\centering
	\renewcommand{\arraystretch}{2.2}
	\begin{tabular}{|c|c|}
		\hline
		$\overline{\psi}\gamma ^{\mu } \psi  X_{\mu} $&{\bf MLRSM: Charged current vertex factors}\\
		\hline
		$\overline{\nu_L}\gamma ^{\mu } e_L  W^{+}_{1\mu}$& $ \frac{g}{\sqrt{2}}\epsilon^1_{(e\nu)_L} =\frac{g}{\sqrt{2}}\left[\cos\xi\left(1+\frac{\Theta_{W_{LL}}}{\Lambda^2}\right)-\sin\xi\frac{\Theta_{W_{LR}}}{\Lambda^2} + \frac{\sqrt{2}}{g}A_{\nu_L e_L W_2}\right]$\\
		$\overline{\nu_L}\gamma ^{\mu } e_L  W^{+}_{2\mu}$& $ \frac{g}{\sqrt{2}}\epsilon^2_{(e\nu)_L} =\frac{g}{\sqrt{2}}\left[\sin\xi\left(1+\frac{\Theta_{W_{LL}}}{\Lambda^2}\right)+\cos\xi\frac{\Theta_{W_{LR}}}{\Lambda^2} + \frac{\sqrt{2}}{g}A_{\nu_L e_L W_2}\right]$\\
		$\overline{\nu_R}\gamma ^{\mu } e_R  W^{+}_{1\mu}$& $ \frac{g}{\sqrt{2}}\epsilon^1_{(e\nu)_R} =\frac{g}{\sqrt{2}}\left[-\sin\xi\left(1+\frac{\Theta_{W_{RR}}}{\Lambda^2}\right)+\cos\xi\frac{\Theta_{W_{RL}}}{\Lambda^2} + \frac{\sqrt{2}}{g}A_{\nu_R e_R W_1}\right]$\\
		$\overline{\nu_R}\gamma ^{\mu } e_R  W^{+}_{2\mu}$& $ \frac{g}{\sqrt{2}}\epsilon^2_{(e\nu)_R} =\frac{g}{\sqrt{2}}\left[\cos\xi\left(1+\frac{\Theta_{W_{RR}}}{\Lambda^2}\right)+\sin\xi\frac{\Theta_{W_{RL}}}{\Lambda^2} + \frac{\sqrt{2}}{g}A_{\nu_R e_R W_2}\right]$\\
		$\overline{u_L}\gamma ^{\mu } d_L  W^+_{1\mu}$& $ \frac{g}{\sqrt{2}}\epsilon^1_{(ud)_L} =\frac{g}{\sqrt{2}}\left[\cos\xi\left(1+\frac{\Theta_{W_{LL}}}{\Lambda^2}\right)-\sin\xi\frac{\Theta_{W_{LR}}}{\Lambda^2} + \frac{\sqrt{2}}{g}A_{u_L d_L W_1}\right]$\\
		$\overline{u_L}\gamma ^{\mu } d_L  W^+_{2\mu}$& $ \frac{g}{\sqrt{2}}\epsilon^2_{(ud)_L} =\frac{g}{\sqrt{2}}\left[\sin\xi\left(1+\frac{\Theta_{W_{LL}}}{\Lambda^2}\right)+\cos\xi\frac{\Theta_{W_{LR}}}{\Lambda^2} + \frac{\sqrt{2}}{g}A_{u_L d_L W_2}\right]$\\
		$\overline{u_R}\gamma ^{\mu } d_R  W^+_{1\mu}$& $ \frac{g}{\sqrt{2}}\epsilon^1_{(ud)_R} =\frac{g}{\sqrt{2}}\left[-\sin\xi\left(1+\frac{\Theta_{W_{RR}}}{\Lambda^2}\right)+\cos\xi\frac{\Theta_{W_{RL}}}{\Lambda^2} + \frac{\sqrt{2}}{g}A_{u_R d_R W_1}\right]$\\
		$\overline{u_R}\gamma ^{\mu } d_R  W^+_{2\mu}$& $ \frac{g}{\sqrt{2}}\epsilon^2_{(ud)_R} =\frac{g}{\sqrt{2}}\left[\cos\xi\left(1+\frac{\Theta_{W_{RR}}}{\Lambda^2}\right)+\sin\xi\frac{\Theta_{W_{RL}}}{\Lambda^2} + \frac{\sqrt{2}}{g}A_{u_R d_R W_2}\right]$\\
		\hline	\end{tabular}
	\caption{MLRSM: Coupling of charged gauge boson and fermions. Here, $\tilde{e} = \frac{e}{2\cos\theta_{\text{w}}\sin\theta_{\text{w}}}$. Hermitian conjugate of these vertices are not separately mentioned. The detailed structures of $A$'s are depicted in Table~\ref{tab:MLRSM-cGB-F-F-II}.}
	\label{tab:MLRSM-cGB-F-F-I}
	
\end{table}
\clearpage

\begin{table}[h!]
	\centering
	\renewcommand{\arraystretch}{2.5}
	\begin{tabular}{|c|ccc|}
			\hline
			$\overline{\nu _L}\gamma ^{\mu } e_L  W_{1\mu }^+$ & $A_{\nu_L e_L W_1}$ & = & $\frac{g}{2\sqrt{2} \Lambda ^2} \left(\mathcal{R}^{a}_{11} \left(\mathcal{C}_{LDl}^{l_L[3]} v_L^2+\left(\kappa _1^2+\kappa _2^2\right) \mathcal{C}_{2D1}^{l_L[3]}\right)-2 \mathcal{R}^{a}_{21} \kappa _1 \kappa _2 \mathcal{C}_{2D1}^{l_L[3]}\right)$ \\
			$\overline{\nu _L}\gamma ^{\mu } e_L  W_{2\mu }^+ $ & $A_{\nu_L e_L W_2}$ & = & $\frac{g}{2\sqrt{2} \Lambda ^2} \left(\mathcal{R}^{a}_{12} \left(\mathcal{C}_{LDl}^{l_L[3]} v_L^2+\left(\kappa _1^2+\kappa _2^2\right) \mathcal{C}_{2D1}^{l_L[3]}\right)-2 \mathcal{R}^{a}_{22} \kappa _1 \kappa _2 \mathcal{C}_{2D1}^{l_L[3]}\right)$\\
			$\overline{\nu _R}\gamma ^{\mu } e_R  W_{1\mu }^+$ & $A_{\nu_R e_R W_1}$ & = & $\frac{g}{2\sqrt{2} \Lambda ^2} \left(\mathcal{R}^{a}_{11} \left(\kappa _1^2+\kappa _2^2\right) \mathcal{C}_{2D1}^{l_R[3]}+\mathcal{R}^{a}_{21} \left(\mathcal{C}_{RDr}^{l_R[3]} v_R^2-2 \kappa _1 \kappa _2 \mathcal{C}_{2D1}^{l_R[3]}\right)\right)$ \\
			$\overline{\nu _R}\gamma ^{\mu } e_R  W_{2\mu }^+ $ & $A_{\nu_R e_R W_2}$ & = & $\frac{g}{2\sqrt{2} \Lambda ^2} \left(\mathcal{R}^{a}_{12} \left(\kappa _1^2+\kappa _2^2\right) \mathcal{C}_{2D1}^{l_R[3]}+\mathcal{R}^{a}_{22} \left(\mathcal{C}_{RDr}^{l_R[3]} v_R^2-2 \kappa _1 \kappa _2 \mathcal{C}_{2D1}^{l_R[3]}\right)\right)$\\
			$\overline{u_L}\gamma ^{\mu } d_L  W_{1\mu }^+$ & $A_{u_L d_L W_1}$ & = & $\frac{g}{2\sqrt{2} \Lambda ^2} \left(\mathcal{R}^{a}_{11} \left(\mathcal{C}_{LDl}^{q_L[3]} v_L^2+\left(\kappa _1^2+\kappa _2^2\right) \mathcal{C}_{2D1}^{q_L[3]}\right)-2 \mathcal{R}^{a}_{21} \kappa _1 \kappa _2 \mathcal{C}_{2D1}^{q_L[3]}\right)$\\
			$\overline{u_L}\gamma ^{\mu } d_L  W_{2\mu }^+$ & $A_{u_L d_L W_2}$ & = & $\frac{g}{2\sqrt{2} \Lambda ^2} \left(\mathcal{R}^{a}_{12} \left(\mathcal{C}_{LDl}^{q_L[3]} v_L^2+\left(\kappa _1^2+\kappa _2^2\right) \mathcal{C}_{2D1}^{q_L[3]}\right)-2 \mathcal{R}^{a}_{22} \kappa _1 \kappa _2 \mathcal{C}_{2D1}^{q_L[3]}\right)$\\
			$\overline{u_R}\gamma ^{\mu } d_R  W_{1\mu }^+ $& $A_{u_R d_R W_1}$ & = & $\frac{g}{2\sqrt{2} \Lambda ^2} \left(\mathcal{R}^{a}_{11} \left(\kappa _1^2+\kappa _2^2\right) \mathcal{C}_{2D1}^{q_R[3]}+\mathcal{R}^{a}_{21} \left(\mathcal{C}_{RDr}^{q_R[3]} v_R^2-2 \kappa _1 \kappa _2 \mathcal{C}_{2D1}^{q_R[3]}\right)\right)$\\
			$\overline{u_R}\gamma ^{\mu } d_R  W_{2\mu }^+ $ & $A_{u_R d_R W_2}$& = &$ \frac{g}{2\sqrt{2} \Lambda ^2} \left(\mathcal{R}^{a}_{12} \left(\kappa _1^2+\kappa _2^2\right) \mathcal{C}_{2D1}^{q_R[3]}+\mathcal{R}^{a}_{22} \left(\mathcal{C}_{RDr}^{q_R[3]} v_R^2-2 \kappa _1 \kappa _2 \mathcal{C}_{2D1}^{q_R[3]}\right)\right)$\\
			\hline
	\end{tabular}
	\caption{MLRSM: Corrections to the coupling of charged gauge bosons to fermions due to $\psi^2\phi^2 D$ operators.}
	\label{tab:MLRSM-cGB-F-F-II}
	
\end{table}


The parameters [$\zeta^1_{\nu_L*u_{L}}, \zeta^1_{\nu_L*u_{R}}, \zeta^2_{\nu_L*u_{R}}, \zeta^1_{\nu_L*d_{R}}, \zeta^2_{\nu_L*d_{R}}$] contained in the neutral current are expressed as:
{\small\begin{eqnarray}
	\zeta^1_{\nu_L*u_{L}}&=&\zeta^1_{\nu_L} \zeta^1_{u_L} - \frac{4\mathcal{M}^2_{Z_1}\cos^2 \theta_{\text{w}}}{g^2 \Lambda^2}\left(\mathcal{C}^{[1]}_{l_L q_L}+\mathcal{C}^{[3]}_{l_L q_L}\right),
	\hspace{0.5cm}
	\zeta^1_{\nu_L*u_{R}}=\zeta^1_{\nu_L} \zeta^1_{u_R} - \frac{4\mathcal{M}^2_{Z_1}\cos^2 \theta_{\text{w}}}{g^2 \Lambda^2}\mathcal{C}_{l_L q_R},\nonumber\\
	\zeta^2_{\nu_L*u_{L}}&=&\zeta^2_{\nu_L} \zeta^2_{u_L} - \frac{4\mathcal{M}^2_{Z_2}\cos^2 \theta_{\text{w}}}{g^2 \Lambda^2}\left(\mathcal{C}^{[1]}_{l_L q_L}+\mathcal{C}^{[3]}_{l_L q_L}\right),
	\hspace{0.5cm}
	\zeta^2_{\nu_L*u_{R}}=\zeta^2_{\nu_L} \zeta^2_{u_R} - \frac{4\mathcal{M}^2_{Z_2}\cos^2 \theta_{\text{w}}}{g^2 \Lambda^2}\mathcal{C}_{l_L q_R},\nonumber\\
	\zeta^1_{\nu_L*d_{L}}&=&\zeta^1_{\nu_L} \zeta^1_{d_L} - \frac{4\mathcal{M}^2_{Z_1}\cos^2 \theta_{\text{w}}}{g^2\Lambda^2}\left(\mathcal{C}^{[1]}_{l_L q_L}-\mathcal{C}^{[3]}_{l_L q_L}\right),
	\hspace{0.5cm}
	\zeta^1_{\nu_L*d_{R}}=\zeta^1_{\nu_L} \zeta^1_{d_R} - \frac{4\mathcal{M}^2_{Z_1}\cos^2 \theta_{\text{w}}}{g^2 \Lambda^2}\mathcal{C}_{l_L q_R},\nonumber\\
	\zeta^2_{\nu_L*d_{L}}&=&\zeta^2_{\nu_L} \zeta^2_{d_L} - \frac{4\mathcal{M}^2_{Z_2}\cos^2 \theta_{\text{w}}}{g^2\Lambda^2}\left(\mathcal{C}^{[1]}_{l_L q_L}-\mathcal{C}^{[3]}_{l_L q_L}\right),
	\hspace{0.5cm}
	\zeta^2_{\nu_L*d_{R}}=\zeta^2_{\nu_L} \zeta^2_{d_R} - \frac{4\mathcal{M}^2_{Z_2}\cos^2 \theta_{\text{w}}}{g^2 \Lambda^2}\mathcal{C}_{l_L q_R}.\nonumber
	\end{eqnarray}}

The $``a"$ and $``b"$ parameters in  Table~\ref{tab:MLRSM-nGB-F-F-I} are given as: 
\small{\begin{eqnarray}
	a^{i}_L &=& \left(A^{iL}_1\cos^2\theta_{\text{w}}-A^{iL}_2\sin^2\theta_{\text{w}}\right)+\frac{1}{\Lambda^2}\Big[A^{iL}_1\left(\cos^2\theta_{\text{w}}\Theta_{3L3L}-\sin^2\theta_{\text{w}}\Theta_{3L3R}-\sin\theta_{\text{w}}\sqrt{\cos 2\theta_{\text{w}}}\Theta_{3LB}\right)\nonumber \\
	&&+A^{iL}_2\left(\frac{\sin\theta_{\text{w}}\cos^2\theta_{\text{w}}}{\sqrt{\cos 2 \theta_{\text{w}}}}\Theta_{3LB} - \frac{\sin^3\theta_{\text{w}}}{ \sqrt{\cos 2\theta_{\text{w}}}}\Theta_{3RB}-\sin^2\theta_{\text{w}}\Theta_{BB}\right)\Big],\\
	b^{i}_L &=& A^{iL}_2\frac{\sin^2\theta_{\text{w}}}{\sqrt{\cos 2\theta_{\text{w}}}}+\frac{1}{\Lambda^2}\Big[A^{iL}_1\left(-\sqrt{\cos 2\theta_{\text{w}}}\Theta_{3L3R}+\sin\theta_{\text{w}}\Theta_{3LB}\right)+A^{iL}_2\big(-\sin\theta_{\text{w}}\Theta_{3RB}\\
	&&+\frac{\sin^2\theta_{\text{w}}}{\sqrt{\cos 2\theta_{\text{w}}}}\Theta_{BB}\big)\Big],\nonumber
	\end{eqnarray}}
    \small{\begin{eqnarray}
	a^{i}_R &=& -\left(A^{iR}_1+A^{iR}_2\right)\sin^2\theta_{\text{w}} +\frac{1}{\Lambda^2}\Big[A^{iR}_1\left(\cos^2\theta_{\text{w}}\Theta_{3L3R}-\sin^2\theta_{\text{w}}\Theta_{3R3R}-\sin\theta_{\text{w}}\sqrt{\cos 2\theta_{\text{w}}}\Theta_{3RB}\right)\nonumber\\
	&&+A^{iR}_2\left(\frac{\sin\theta_{\text{w}}\cos^2\theta_{\text{w}}}{\sqrt{\cos 2 \theta_{\text{w}}}}\Theta_{3LB} - \frac{\sin^3\theta_{\text{w}}}{ \sqrt{\cos 2\theta_{\text{w}}}}\Theta_{3RB}-\sin^2\theta_{\text{w}}\Theta_{BB}\right)\Big],\\
	b^{i}_R &=& \left(-A^{iR}_1\sqrt{\cos 2\theta_{\text{w}}}+A^{iR}_2\frac{\sin^2\theta_{\text{w}}}{\sqrt{\cos 2\theta_{\text{w}}}}\right)\\
	&&+\frac{1}{\Lambda^2}\Big[A^{iR}_1\left(-\sqrt{\cos 2\theta_{\text{w}}}\Theta_{3R3R}+\sin\theta_{\text{w}}\Theta_{3RB}\right)+A^{iR}_2\left(-\sin\theta_{\text{w}}\Theta_{3RB}+\frac{\sin^2\theta_{\text{w}}}{\sqrt{\cos 2\theta_{\text{w}}}}\Theta_{BB}\right)\Big],\nonumber
	\end{eqnarray}}
where  
{\small
	\begin{equation}
	A^{iL,R}_1 = 2 T_{3L,R}^i, \hspace{0.7cm} A^{iL,R}_2 = 2 (Q^i-T_{3L,R}^i); \hspace{0.7cm}
	i = u,d,\nu,e.
	\end{equation}
	The $T_{3L}, T_{3R}, Q$ are the quantum numbers corresponding to $SU(2)_L$, $SU(2)_R$, and $U(1)_{em}$ gauge groups respectively. These are provided in Table~\ref{tab:MLRSM-quantum-F}. 
\begin{table}[h!]
	\centering
	\renewcommand{\arraystretch}{2.2}
	{\small\begin{tabular}{|c|c|c|c|c|c|c|c|c|}\hline
		Quantum no.	& $\nu_{L}$ &$\nu_{R}$& $e_{L}$ &$e_{R}$&$u_{L}$& $u_{R}$ & $d_{L}$&$d_{R}$ \\
			\hline
			$T_{3L}$ & $\frac{1}{2}$ &0& $-\frac{1}{2}$&0 & $\frac{1}{2}$&0& $-\frac{1}{2}$&0 \\
			\hline
			$T_{3R}$ & 0&$\frac{1}{2}$ &0& $-\frac{1}{2}$&0 & $\frac{1}{2}$&0& $-\frac{1}{2}$ \\
			\hline
			$Q$ &0 &0 &-1&-1&$\frac{2}{3}$ &$\frac{2}{3}$& $-\frac{1}{3}$&$-\frac{1}{3}$ \\
			\hline
	\end{tabular}}
	\caption{MLRSM: $T_{3L}$, $T_{3R}$, $Q$ quantum number for fermions.}
	\label{tab:MLRSM-quantum-F}
\end{table}

	Now incorporating all the above mentioned corrections, we can express $\widehat{R}$ as:	
{\small\begin{align}\label{eqn:R6LR}
	\widehat{R}=\frac{4\mathcal{M}_{W_1}^4}{\cos^4\theta_{\text{w}}\mathcal{M}_{Z_1}^4}\Big[\frac{(\zeta^1_{\nu_L*u_{L}})^2+(\zeta^1_{\nu_L*d_{L}})^2-(\zeta^1_{\nu_L*u_{R}})^2-(\zeta^1_{\nu_L*d_{R}})^2}{(\epsilon^1_{(\nu e)_L*(ud)_{L}})^2-(\epsilon^1_{(\nu e)_L*(ud)_{R}})^2}\Big].
	\end{align}}
The modified  Weak mixing angle can be derived using
$\sin^2\bar{\theta}_\text{w}	=\frac{1}{2}-\widehat{R}$.


\begin{table}[h!]
	\centering
	\renewcommand{\arraystretch}{2.2}
	\begin{tabular}{|c|c|}
		\hline
		$\overline{\psi}\gamma ^{\mu } \psi  X_{\mu} $&{\bf MLRSM: Neutral current vertex factors}\\
		\hline		
		$\overline{\nu_L}\gamma ^{\mu } \nu_L  Z_{1\mu}$& $ \tilde{e}\zeta^1_{\nu_L} =\tilde{e}\left[a^{\nu}_L\cos\theta_2+b^{\nu}_L\sin\theta_2+\frac{1}{\tilde{e}}A_{\nu_L Z_1}\right] $\\
		$\overline{\nu_L}\gamma ^{\mu} \nu_L  Z_{2\mu}$& $ \tilde{e}\zeta^2_{\nu_L} =\tilde{e}\left[a^{\nu}_L\sin\theta_2-b^{\nu}_L\cos\theta_2+\frac{1}{\tilde{e}}A_{\nu_L Z_2}\right] $\\
		$\overline{e_L}\gamma ^{\mu } e_L  Z_{1\mu}$& $ \tilde{e}\zeta^1_{e_L} =\tilde{e}\left[a^{e}_L\cos\theta_2+b^{e}_L\sin\theta_2+\frac{1}{\tilde{e}}A_{e_L Z_1}\right] $\\
		$\overline{e_L}\gamma ^{\mu} e_L  Z_{2\mu}$& $ \tilde{e}\zeta^2_{e_L} =\tilde{e}\left[a^{e}_L\sin\theta_2-b^{e}_L\cos\theta_2+\frac{1}{\tilde{e}}A_{e_L Z_2}\right] $\\
		$\overline{e_R}\gamma ^{\mu} e_R  Z_{1\mu}$& $ \tilde{e}\zeta^1_{e_R} =\tilde{e}\left[a^{e}_R\cos\theta_2+b^{e}_R\sin\theta_2+\frac{1}{\tilde{e}}A_{e_R Z_1}\right] $\\
		$\overline{e_R}\gamma ^{\mu} e_R  Z_{2\mu}$& $ \tilde{e}\zeta^2_{e_R} =\tilde{e} \left[a^{e}_R\sin\theta_2-b^{e}_R\cos\theta_2+\frac{1}{\tilde{e}}A_{e_R Z_2}\right]$\\
		$\overline{u_L}\gamma ^{\mu } u_L  Z_{1\mu}$& $ \tilde{e}\zeta^1_{u_L} =\tilde{e}\left[a^{u}_L\cos\theta_2+b^{u}_L\sin\theta_2+\frac{1}{\tilde{e}}A_{u_L Z_1}\right] $\\
		$\overline{u_L}\gamma ^{\mu} u_L  Z_{2\mu}$& $ \tilde{e}\zeta^2_{u_L} =\tilde{e} \left[a^{u}_L\sin\theta_2-b^{u}_L\cos\theta_2+\frac{1}{\tilde{e}}A_{u_L Z_2}\right]$\\
		$\overline{u_R}\gamma ^{\mu} u_R  Z_{1\mu}$& $ \tilde{e}\zeta^1_{d_R} =\tilde{e} \left[a^{u}_R\cos\theta_2+b^{u}_R\sin\theta_2+\frac{1}{\tilde{e}}A_{u_R Z_1}\right]$\\
		$\overline{u_R}\gamma ^{\mu} u_R  Z_{2\mu}$& $ \tilde{e}\zeta^2_{u_R} =\tilde{e} \left[a^{u}_R\sin\theta_2-b^{u}_R\cos\theta_2+\frac{1}{\tilde{e}}A_{u_R Z_2}\right]$\\
		$\overline{d_L}\gamma ^{\mu } d_L  Z_{1\mu}$& $ \tilde{e}\zeta^1_{d_L} =\tilde{e} \left[a^{d}_L\cos\theta_2+b^{d}_L\sin\theta_2+\frac{1}{\tilde{e}}A_{d_L Z_1}\right]$\\
		$\overline{d_L}\gamma ^{\mu} d_L  Z_{2\mu}$& $ \tilde{e}\zeta^2_{d_L} =\tilde{e}\left[a^{d}_L\sin\theta_2-b^{d}_L\cos\theta_2+\frac{1}{\tilde{e}}A_{d_L Z_2}\right] $\\
		$\overline{d_R}\gamma ^{\mu} d_R  Z_{1\mu}$& $ \tilde{e}\zeta^1_{d_R} =\tilde{e}\left[a^{d}_R\cos\theta_2+b^{d}_R\sin\theta_2+\frac{1}{\tilde{e}}A_{d_R Z_1}\right] $\\
		$\overline{d_R}\gamma ^{\mu} d_R  Z_{2\mu}$& $ \tilde{e}\zeta^2_{d_R} =\tilde{e} \left[a^{d}_R\sin\theta_2-b^{d}_R\cos\theta_2 +\frac{1}{\tilde{e}}A_{d_R Z_2}\right]$\\
		\hline
	\end{tabular}
	\caption{ MLRSM: Coupling of neutral gauge bosons  and fermions: Here, $\tilde{e} = \frac{e}{2\cos\theta_{\text{w}}\sin\theta_{\text{w}}}$. The detailed structures of $A$'s are mentioned in Tables~\ref{tab:MLRSM-nGB-F-F-II}, \ref{tab:MLRSM-nGB-F-F-III}.}
	\label{tab:MLRSM-nGB-F-F-I}
\end{table}

\begin{table}[h!]
	\centering
	\renewcommand{\arraystretch}{2.4}
	{\small\begin{tabular}{|c|cl|}\hline
			$\overline{\nu _L}\gamma ^{\mu } \nu _L  Z_{1\mu }$ & $A_{\nu_L Z_1}$&= $\frac{\mathcal{R}^{b}_{31} \mathcal{C}_{LDl}^{l_L[1]} \tilde{g} v_L^2}{\Lambda ^2}-\frac{\mathcal{R}^{b}_{31} \mathcal{C}_{LDl}^{l_L[3]} \tilde{g} v_L^2}{\Lambda ^2}+\frac{\mathcal{R}^{b}_{31} \mathcal{C}_{RDr}^{l_L} \tilde{g} v_R^2}{\Lambda ^2}-\frac{\mathcal{R}^{b}_{11} g \mathcal{C}_{LDl}^{l_L[1]} v_L^2}{\Lambda ^2}+\frac{\mathcal{R}^{b}_{11} g \mathcal{C}_{LDl}^{l_L[3]} v_L^2}{\Lambda ^2}$\\
			& & $-\frac{\mathcal{R}^{b}_{21} g \mathcal{C}_{RDr}^{l_L} v_R^2}{\Lambda ^2}+\frac{\mathcal{R}^{b}_{11} g \kappa _1^2 \mathcal{C}_{2D1}^{l_L[1]}}{2 \Lambda ^2}+\frac{\mathcal{R}^{b}_{21} g \kappa _2^2 \mathcal{C}_{2D1}^{l_L[1]}}{2 \Lambda ^2}-\frac{\mathcal{R}^{b}_{21} g \kappa _1^2 \mathcal{C}_{2D1}^{l_L[1]}}{2 \Lambda ^2}-\frac{\mathcal{R}^{b}_{11} g \kappa _2^2 \mathcal{C}_{2D1}^{l_L[1]}}{2 \Lambda ^2}$\\
			& & $+\frac{\mathcal{R}^{b}_{11} g \kappa _1^2 \mathcal{C}_{2D1}^{l_L[3]}}{2 \Lambda ^2}+\frac{\mathcal{R}^{b}_{11} g \kappa _2^2 \mathcal{C}_{2D1}^{l_L[3]}}{2 \Lambda ^2}-\frac{\mathcal{R}^{b}_{21} g \kappa _1^2 \mathcal{C}_{2D1}^{l_L[3]}}{2 \Lambda ^2}-\frac{\mathcal{R}^{b}_{21} g \kappa _2^2 \mathcal{C}_{2D1}^{l_L[3]}}{2 \Lambda ^2}$ \\
			\hline
			$\overline{\nu _L}\gamma ^{\mu } \nu _L  Z_{2\mu }$ & $A_{\nu_L Z_2}$& = $\frac{\mathcal{R}^{b}_{32} \mathcal{C}_{LDl}^{l_L[1]} \tilde{g} v_L^2}{\Lambda ^2}-\frac{\mathcal{R}^{b}_{32} \mathcal{C}_{LDl}^{l_L[3]} \tilde{g} v_L^2}{\Lambda ^2}+\frac{\mathcal{R}^{b}_{32} \mathcal{C}_{RDr}^{l_L} \tilde{g} v_R^2}{\Lambda ^2}-\frac{\mathcal{R}^{b}_{12} g \mathcal{C}_{LDl}^{l_L[1]} v_L^2}{\Lambda ^2}+\frac{\mathcal{R}^{b}_{12} g \mathcal{C}_{LDl}^{l_L[3]} v_L^2}{\Lambda ^2}$
			\\&&$-\frac{\mathcal{R}^{b}_{22} g \mathcal{C}_{RDr}^{l_L} v_R^2}{\Lambda ^2}+\frac{\mathcal{R}^{b}_{12} g \kappa _1^2 \mathcal{C}_{2D1}^{l_L[1]}}{2 \Lambda ^2}+\frac{\mathcal{R}^{b}_{22} g \kappa _2^2 \mathcal{C}_{2D1}^{l_L[1]}}{2 \Lambda ^2}-\frac{\mathcal{R}^{b}_{22} g \kappa _1^2 \mathcal{C}_{2D1}^{l_L[1]}}{2 \Lambda ^2}-\frac{\mathcal{R}^{b}_{12} g \kappa _2^2 \mathcal{C}_{2D1}^{l_L[1]}}{2 \Lambda ^2}$\\&&$+\frac{\mathcal{R}^{b}_{12} g \kappa _1^2 \mathcal{C}_{2D1}^{l_L[3]}}{2 \Lambda ^2}+\frac{\mathcal{R}^{b}_{12} g \kappa _2^2 \mathcal{C}_{2D1}^{l_L[3]}}{2 \Lambda ^2}-\frac{\mathcal{R}^{b}_{22} g \kappa _1^2 \mathcal{C}_{2D1}^{l_L[3]}}{2 \Lambda ^2}-\frac{\mathcal{R}^{b}_{22} g \kappa _2^2 \mathcal{C}_{2D1}^{l_L[3]}}{2 \Lambda ^2}$ \\
			\hline
			$\overline{e_L}\gamma ^{\mu } e_L  Z_{1\mu }$ & $A_{e_L Z_1}$& = $\frac{\mathcal{R}^{b}_{31} \mathcal{C}_{LDl}^{l_L[1]} \tilde{g} v_L^2}{\Lambda ^2}+\frac{\mathcal{R}^{b}_{31} \mathcal{C}_{LDl}^{l_L[3]} \tilde{g} v_L^2}{\Lambda ^2}+\frac{\mathcal{R}^{b}_{31} \mathcal{C}_{RDr}^{l_L} \tilde{g} v_R^2}{\Lambda ^2}-\frac{\mathcal{R}^{b}_{11} g \mathcal{C}_{LDl}^{l_L[1]} v_L^2}{\Lambda ^2}-\frac{\mathcal{R}^{b}_{11} g \mathcal{C}_{LDl}^{l_L[3]} v_L^2}{\Lambda ^2}$\\
			&&$-\frac{\mathcal{R}^{b}_{21} g \mathcal{C}_{RDr}^{l_L} v_R^2}{\Lambda ^2}+\frac{\mathcal{R}^{b}_{11} g \kappa _1^2 \mathcal{C}_{2D1}^{l_L[1]}}{2 \Lambda ^2}+\frac{\mathcal{R}^{b}_{21} g \kappa _2^2 \mathcal{C}_{2D1}^{l_L[1]}}{2 \Lambda ^2}-\frac{\mathcal{R}^{b}_{21} g \kappa _1^2 \mathcal{C}_{2D1}^{l_L[1]}}{2 \Lambda ^2}-\frac{\mathcal{R}^{b}_{11} g \kappa _2^2 \mathcal{C}_{2D1}^{l_L[1]}}{2 \Lambda ^2}$\\&&$+\frac{\mathcal{R}^{b}_{21} g \kappa _1^2 \mathcal{C}_{2D1}^{l_L[3]}}{2 \Lambda ^2}+\frac{\mathcal{R}^{b}_{21} g \kappa _2^2 \mathcal{C}_{2D1}^{l_L[3]}}{2 \Lambda ^2}-\frac{\mathcal{R}^{b}_{11} g \kappa _1^2 \mathcal{C}_{2D1}^{l_L[3]}}{2 \Lambda ^2}-\frac{\mathcal{R}^{b}_{11} g \kappa _2^2 \mathcal{C}_{2D1}^{l_L[3]}}{2 \Lambda ^2}$ \\
			\hline
			$\overline{e_L}\gamma ^{\mu } e_L  Z_{2\mu }$ & $A_{e_L Z_2}$& = $\frac{\mathcal{R}^{b}_{32} \mathcal{C}_{LDl}^{l_L[1]} \tilde{g} v_L^2}{\Lambda ^2}+\frac{\mathcal{R}^{b}_{32} \mathcal{C}_{LDl}^{l_L[3]} \tilde{g} v_L^2}{\Lambda ^2}+\frac{\mathcal{R}^{b}_{32} \mathcal{C}_{RDr}^{l_L} \tilde{g} v_R^2}{\Lambda ^2}-\frac{\mathcal{R}^{b}_{12} g \mathcal{C}_{LDl}^{l_L[1]} v_L^2}{\Lambda ^2}-\frac{\mathcal{R}^{b}_{12} g \mathcal{C}_{LDl}^{l_L[3]} v_L^2}{\Lambda ^2}$\\&&$-\frac{\mathcal{R}^{b}_{22} g \mathcal{C}_{RDr}^{l_L} v_R^2}{\Lambda ^2}+\frac{\mathcal{R}^{b}_{12} g \kappa _1^2 \mathcal{C}_{2D1}^{l_L[1]}}{2 \Lambda ^2}+\frac{\mathcal{R}^{b}_{22} g \kappa _2^2 \mathcal{C}_{2D1}^{l_L[1]}}{2 \Lambda ^2}-\frac{\mathcal{R}^{b}_{22} g \kappa _1^2 \mathcal{C}_{2D1}^{l_L[1]}}{2 \Lambda ^2}-\frac{\mathcal{R}^{b}_{12} g \kappa _2^2 \mathcal{C}_{2D1}^{l_L[1]}}{2 \Lambda ^2}$\\&&$+\frac{\mathcal{R}^{b}_{22} g \kappa _1^2 \mathcal{C}_{2D1}^{l_L[3]}}{2 \Lambda ^2}+\frac{\mathcal{R}^{b}_{22} g \kappa _2^2 \mathcal{C}_{2D1}^{l_L[3]}}{2 \Lambda ^2}-\frac{\mathcal{R}^{b}_{12} g \kappa _1^2 \mathcal{C}_{2D1}^{l_L[3]}}{2 \Lambda ^2}-\frac{\mathcal{R}^{b}_{12} g \kappa _2^2 \mathcal{C}_{2D1}^{l_L[3]}}{2 \Lambda ^2}$ \\
			\hline
			$\overline{e_R}\gamma ^{\mu } e_R  Z_{1\mu }$ & $A_{e_R Z_1}$ & =  $\frac{\mathcal{R}^{b}_{31} \mathcal{C}_{LDl}^{l_R} \tilde{g} v_L^2}{\Lambda ^2}+\frac{\mathcal{R}^{b}_{31} \mathcal{C}_{RDr}^{l_R[1]} \tilde{g} v_R^2}{\Lambda ^2}+\frac{\mathcal{R}^{b}_{31} \mathcal{C}_{RDr}^{l_R[3]} \tilde{g} v_R^2}{\Lambda ^2}-\frac{\mathcal{R}^{b}_{11} g \mathcal{C}_{LDl}^{l_R} v_L^2}{\Lambda ^2}-\frac{\mathcal{R}^{b}_{21} g \mathcal{C}_{RDr}^{l_R[1]} v_R^2}{\Lambda ^2}$\\&&$-\frac{\mathcal{R}^{b}_{21} g \mathcal{C}_{RDr}^{l_R[3]} v_R^2}{\Lambda ^2}+\frac{\mathcal{R}^{b}_{11} g \kappa _1^2 \mathcal{C}_{2D1}^{l_R[1]}}{2 \Lambda ^2}+\frac{\mathcal{R}^{b}_{21} g \kappa _2^2 \mathcal{C}_{2D1}^{l_R[1]}}{2 \Lambda ^2}-\frac{\mathcal{R}^{b}_{21} g \kappa _1^2 \mathcal{C}_{2D1}^{l_R[1]}}{2 \Lambda ^2}-\frac{\mathcal{R}^{b}_{11} g \kappa _2^2 \mathcal{C}_{2D1}^{l_R[1]}}{2 \Lambda ^2}$\\&&$+\frac{\mathcal{R}^{b}_{21} g \kappa _1^2 \mathcal{C}_{2D1}^{l_R[3]}}{2 \Lambda ^2}+\frac{\mathcal{R}^{b}_{21} g \kappa _2^2 \mathcal{C}_{2D1}^{l_R[3]}}{2 \Lambda ^2}-\frac{\mathcal{R}^{b}_{11} g \kappa _1^2 \mathcal{C}_{2D1}^{l_R[3]}}{2 \Lambda ^2}-\frac{\mathcal{R}^{b}_{11} g \kappa _2^2 \mathcal{C}_{2D1}^{l_R[3]}}{2 \Lambda ^2}$ \\
			\hline
			$\overline{e_R}\gamma ^{\mu } e_R  Z_{2\mu }$ & $A_{e_R Z_2}$ & = $\frac{\mathcal{R}^{b}_{32} \mathcal{C}_{LDl}^{l_R} \tilde{g} v_L^2}{\Lambda ^2}+\frac{\mathcal{R}^{b}_{32} \mathcal{C}_{RDr}^{l_R[1]} \tilde{g} v_R^2}{\Lambda ^2}+\frac{\mathcal{R}^{b}_{32} \mathcal{C}_{RDr}^{l_R[3]} \tilde{g} v_R^2}{\Lambda ^2}-\frac{\mathcal{R}^{b}_{12} g \mathcal{C}_{LDl}^{l_R} v_L^2}{\Lambda ^2}-\frac{\mathcal{R}^{b}_{22} g \mathcal{C}_{RDr}^{l_R[1]} v_R^2}{\Lambda ^2}$\\&&$-\frac{\mathcal{R}^{b}_{22} g \mathcal{C}_{RDr}^{l_R[3]} v_R^2}{\Lambda ^2}+\frac{\mathcal{R}^{b}_{12} g \kappa _1^2 \mathcal{C}_{2D1}^{l_R[1]}}{2 \Lambda ^2}+\frac{\mathcal{R}^{b}_{22} g \kappa _2^2 \mathcal{C}_{2D1}^{l_R[1]}}{2 \Lambda ^2}-\frac{\mathcal{R}^{b}_{22} g \kappa _1^2 \mathcal{C}_{2D1}^{l_R[1]}}{2 \Lambda ^2}-\frac{\mathcal{R}^{b}_{12} g \kappa _2^2 \mathcal{C}_{2D1}^{l_R[1]}}{2 \Lambda ^2}$\\&&$+\frac{\mathcal{R}^{b}_{22} g \kappa _1^2 \mathcal{C}_{2D1}^{l_R[3]}}{2 \Lambda ^2}+\frac{\mathcal{R}^{b}_{22} g \kappa _2^2 \mathcal{C}_{2D1}^{l_R[3]}}{2 \Lambda ^2}-\frac{\mathcal{R}^{b}_{12} g \kappa _1^2 \mathcal{C}_{2D1}^{l_R[3]}}{2 \Lambda ^2}-\frac{\mathcal{R}^{b}_{12} g \kappa _2^2 \mathcal{C}_{2D1}^{l_R[3]}}{2 \Lambda ^2}$ \\
			\hline
	\end{tabular}}
	\caption{MLRSM: Corrections to the coupling of neutral gauge bosons and fermions due to $\psi^2\phi^2 D$ operators.}
	\label{tab:MLRSM-nGB-F-F-II}
\end{table}

\begin{table}[h!]
	\centering
	\renewcommand{\arraystretch}{2.}
	{\small\begin{tabular}{|c|cl|}
			\hline
			$\overline{u_L}\gamma ^{\mu } u_L  Z_{1\mu }$ & $A_{u_L Z_1}$ & =  $\frac{\mathcal{R}^{b}_{31} \mathcal{C}_{LDl}^{q_L[1]} \tilde{g} v_L^2}{\Lambda ^2}-\frac{\mathcal{R}^{b}_{31} \mathcal{C}_{LDl}^{q_L[3]} \tilde{g} v_L^2}{\Lambda ^2}+\frac{\mathcal{R}^{b}_{31} \mathcal{C}_{RDr}^{q_L} \tilde{g} v_R^2}{\Lambda ^2}-\frac{\mathcal{R}^{b}_{11} g \mathcal{C}_{LDl}^{q_L[1]} v_L^2}{\Lambda ^2}+\frac{\mathcal{R}^{b}_{11} g \mathcal{C}_{LDl}^{q_L[3]} v_L^2}{\Lambda ^2}$\\&&$-\frac{\mathcal{R}^{b}_{21} g \mathcal{C}_{RDr}^{q_L} v_R^2}{\Lambda ^2}+\frac{\mathcal{R}^{b}_{11} g \kappa _1^2 \mathcal{C}_{2D1}^{q_L[1]}}{2 \Lambda ^2}+\frac{\mathcal{R}^{b}_{21} g \kappa _2^2 \mathcal{C}_{2D1}^{q_L[1]}}{2 \Lambda ^2}-\frac{\mathcal{R}^{b}_{21} g \kappa _1^2 \mathcal{C}_{2D1}^{q_L[1]}}{2 \Lambda ^2}-\frac{\mathcal{R}^{b}_{11} g \kappa _2^2 \mathcal{C}_{2D1}^{q_L[1]}}{2 \Lambda ^2}$\\&&$+\frac{\mathcal{R}^{b}_{11} g \kappa _1^2 \mathcal{C}_{2D1}^{q_L[3]}}{2 \Lambda ^2}+\frac{\mathcal{R}^{b}_{11} g \kappa _2^2 \mathcal{C}_{2D1}^{q_L[3]}}{2 \Lambda ^2}-\frac{\mathcal{R}^{b}_{21} g \kappa _1^2 \mathcal{C}_{2D1}^{q_L[3]}}{2 \Lambda ^2}-\frac{\mathcal{R}^{b}_{21} g \kappa _2^2 \mathcal{C}_{2D1}^{q_L[3]}}{2 \Lambda ^2}$ \\
			\hline
			$\overline{u_L}\gamma ^{\mu } u_L  Z_{2\mu }$ & $A_{u_L Z_2}$ & = $\frac{\mathcal{R}^{b}_{32} \mathcal{C}_{LDl}^{q_L[1]} \tilde{g} v_L^2}{\Lambda ^2}-\frac{\mathcal{R}^{b}_{32} \mathcal{C}_{LDl}^{q_L[3]} \tilde{g} v_L^2}{\Lambda ^2}+\frac{\mathcal{R}^{b}_{32} \mathcal{C}_{RDr}^{q_L} \tilde{g} v_R^2}{\Lambda ^2}-\frac{\mathcal{R}^{b}_{12} g \mathcal{C}_{LDl}^{q_L[1]} v_L^2}{\Lambda ^2}+\frac{\mathcal{R}^{b}_{12} g \mathcal{C}_{LDl}^{q_L[3]} v_L^2}{\Lambda ^2}$\\&&$-\frac{\mathcal{R}^{b}_{22} g \mathcal{C}_{RDr}^{q_L} v_R^2}{\Lambda ^2}+\frac{\mathcal{R}^{b}_{12} g \kappa _1^2 \mathcal{C}_{2D1}^{q_L[1]}}{2 \Lambda ^2}+\frac{\mathcal{R}^{b}_{22} g \kappa _2^2 \mathcal{C}_{2D1}^{q_L[1]}}{2 \Lambda ^2}-\frac{\mathcal{R}^{b}_{22} g \kappa _1^2 \mathcal{C}_{2D1}^{q_L[1]}}{2 \Lambda ^2}-\frac{\mathcal{R}^{b}_{12} g \kappa _2^2 \mathcal{C}_{2D1}^{q_L[1]}}{2 \Lambda ^2}$\\&&$+\frac{\mathcal{R}^{b}_{12} g \kappa _1^2 \mathcal{C}_{2D1}^{q_L[3]}}{2 \Lambda ^2}+\frac{\mathcal{R}^{b}_{12} g \kappa _2^2 \mathcal{C}_{2D1}^{q_L[3]}}{2 \Lambda ^2}-\frac{\mathcal{R}^{b}_{22} g \kappa _1^2 \mathcal{C}_{2D1}^{q_L[3]}}{2 \Lambda ^2}-\frac{\mathcal{R}^{b}_{22} g \kappa _2^2 \mathcal{C}_{2D1}^{q_L[3]}}{2 \Lambda ^2}$ \\
			\hline
			$\overline{u_R}\gamma ^{\mu } u_R  Z_{1\mu }$ & $A_{u_R Z_1}$ & =  $\frac{\mathcal{R}^{b}_{31} \mathcal{C}_{LDl}^{q_R} \tilde{g} v_L^2}{\Lambda ^2}+\frac{\mathcal{R}^{b}_{31} \mathcal{C}_{RDr}^{q_R[1]} \tilde{g} v_R^2}{\Lambda ^2}-\frac{\mathcal{R}^{b}_{31} \mathcal{C}_{RDr}^{q_R[3]} \tilde{g} v_R^2}{\Lambda ^2}-\frac{\mathcal{R}^{b}_{11} g \mathcal{C}_{LDl}^{q_R} v_L^2}{\Lambda ^2}-\frac{\mathcal{R}^{b}_{21} g \mathcal{C}_{RDr}^{q_R[1]} v_R^2}{\Lambda ^2}$\\&&$+\frac{\mathcal{R}^{b}_{21} g \mathcal{C}_{RDr}^{q_R[3]} v_R^2}{\Lambda ^2}+\frac{\mathcal{R}^{b}_{11} g \kappa _1^2 \mathcal{C}_{2D1}^{q_R[1]}}{2 \Lambda ^2}+\frac{\mathcal{R}^{b}_{21} g \kappa _2^2 \mathcal{C}_{2D1}^{q_R[1]}}{2 \Lambda ^2}-\frac{\mathcal{R}^{b}_{21} g \kappa _1^2 \mathcal{C}_{2D1}^{q_R[1]}}{2 \Lambda ^2}-\frac{\mathcal{R}^{b}_{11} g \kappa _2^2 \mathcal{C}_{2D1}^{q_R[1]}}{2 \Lambda ^2}$\\&&$+\frac{\mathcal{R}^{b}_{11} g \kappa _1^2 \mathcal{C}_{2D1}^{q_R[3]}}{2 \Lambda ^2}+\frac{\mathcal{R}^{b}_{11} g \kappa _2^2 \mathcal{C}_{2D1}^{q_R[3]}}{2 \Lambda ^2}-\frac{\mathcal{R}^{b}_{21} g \kappa _1^2 \mathcal{C}_{2D1}^{q_R[3]}}{2 \Lambda ^2}-\frac{\mathcal{R}^{b}_{21} g \kappa _2^2 \mathcal{C}_{2D1}^{q_R[3]}}{2 \Lambda ^2}$ \\
			\hline
			$\overline{u_R}\gamma ^{\mu } u_R  Z_{2\mu }$ & $A_{u_R Z_2}$ & = $\frac{\mathcal{R}^{b}_{32} \mathcal{C}_{LDl}^{q_R} \tilde{g} v_L^2}{\Lambda ^2}+\frac{\mathcal{R}^{b}_{32} \mathcal{C}_{RDr}^{q_R[1]} \tilde{g} v_R^2}{\Lambda ^2}-\frac{\mathcal{R}^{b}_{32} \mathcal{C}_{RDr}^{q_R[3]} \tilde{g} v_R^2}{\Lambda ^2}-\frac{\mathcal{R}^{b}_{12} g \mathcal{C}_{LDl}^{q_R} v_L^2}{\Lambda ^2}-\frac{\mathcal{R}^{b}_{22} g \mathcal{C}_{RDr}^{q_R[1]} v_R^2}{\Lambda ^2}$\\&&$+\frac{\mathcal{R}^{b}_{22} g \mathcal{C}_{RDr}^{q_R[3]} v_R^2}{\Lambda ^2}+\frac{\mathcal{R}^{b}_{12} g \kappa _1^2 \mathcal{C}_{2D1}^{q_R[1]}}{2 \Lambda ^2}+\frac{\mathcal{R}^{b}_{22} g \kappa _2^2 \mathcal{C}_{2D1}^{q_R[1]}}{2 \Lambda ^2}-\frac{\mathcal{R}^{b}_{22} g \kappa _1^2 \mathcal{C}_{2D1}^{q_R[1]}}{2 \Lambda ^2}-\frac{\mathcal{R}^{b}_{12} g \kappa _2^2 \mathcal{C}_{2D1}^{q_R[1]}}{2 \Lambda ^2}$\\&&$+\frac{\mathcal{R}^{b}_{12} g \kappa _1^2 \mathcal{C}_{2D1}^{q_R[3]}}{2 \Lambda ^2}+\frac{\mathcal{R}^{b}_{12} g \kappa _2^2 \mathcal{C}_{2D1}^{q_R[3]}}{2 \Lambda ^2}-\frac{\mathcal{R}^{b}_{22} g \kappa _1^2 \mathcal{C}_{2D1}^{q_R[3]}}{2 \Lambda ^2}-\frac{\mathcal{R}^{b}_{22} g \kappa _2^2 \mathcal{C}_{2D1}^{q_R[3]}}{2 \Lambda ^2}$ \\
			\hline
			$\overline{d_L}\gamma ^{\mu } d_L  Z_{1\mu }$ & $A_{d_L Z_1}$ & = $\frac{\mathcal{R}^{b}_{31} \mathcal{C}_{LDl}^{q_L[1]} \tilde{g} v_L^2}{\Lambda ^2}+\frac{\mathcal{R}^{b}_{31} \mathcal{C}_{LDl}^{q_L[3]} \tilde{g} v_L^2}{\Lambda ^2}+\frac{\mathcal{R}^{b}_{31} \mathcal{C}_{RDr}^{q_L} \tilde{g} v_R^2}{\Lambda ^2}-\frac{\mathcal{R}^{b}_{11} g \mathcal{C}_{LDl}^{q_L[1]} v_L^2}{\Lambda ^2}-\frac{\mathcal{R}^{b}_{11} g \mathcal{C}_{LDl}^{q_L[3]} v_L^2}{\Lambda ^2}$\\&&$-\frac{\mathcal{R}^{b}_{21} g \mathcal{C}_{RDr}^{q_L} v_R^2}{\Lambda ^2}+\frac{\mathcal{R}^{b}_{11} g \kappa _1^2 \mathcal{C}_{2D1}^{q_L[1]}}{2 \Lambda ^2}+\frac{\mathcal{R}^{b}_{21} g \kappa _2^2 \mathcal{C}_{2D1}^{q_L[1]}}{2 \Lambda ^2}-\frac{\mathcal{R}^{b}_{21} g \kappa _1^2 \mathcal{C}_{2D1}^{q_L[1]}}{2 \Lambda ^2}-\frac{\mathcal{R}^{b}_{11} g \kappa _2^2 \mathcal{C}_{2D1}^{q_L[1]}}{2 \Lambda ^2}$\\&&$+\frac{\mathcal{R}^{b}_{21} g \kappa _1^2 \mathcal{C}_{2D1}^{q_L[3]}}{2 \Lambda ^2}+\frac{\mathcal{R}^{b}_{21} g \kappa _2^2 \mathcal{C}_{2D1}^{q_L[3]}}{2 \Lambda ^2}-\frac{\mathcal{R}^{b}_{11} g \kappa _1^2 \mathcal{C}_{2D1}^{q_L[3]}}{2 \Lambda ^2}-\frac{\mathcal{R}^{b}_{11} g \kappa _2^2 \mathcal{C}_{2D1}^{q_L[3]}}{2 \Lambda ^2}$ \\
			\hline
			$\overline{d_L}\gamma ^{\mu } d_L  Z_{2\mu }$ & $A_{d_L Z_2}$ & = $\frac{\mathcal{R}^{b}_{32} \mathcal{C}_{LDl}^{q_L[1]} \tilde{g} v_L^2}{\Lambda ^2}+\frac{\mathcal{R}^{b}_{32} \mathcal{C}_{LDl}^{q_L[3]} \tilde{g} v_L^2}{\Lambda ^2}+\frac{\mathcal{R}^{b}_{32} \mathcal{C}_{RDr}^{q_L} \tilde{g} v_R^2}{\Lambda ^2}-\frac{\mathcal{R}^{b}_{12} g \mathcal{C}_{LDl}^{q_L[1]} v_L^2}{\Lambda ^2}-\frac{\mathcal{R}^{b}_{12} g \mathcal{C}_{LDl}^{q_L[3]} v_L^2}{\Lambda ^2}$\\&&$-\frac{\mathcal{R}^{b}_{22} g \mathcal{C}_{RDr}^{q_L} v_R^2}{\Lambda ^2}+\frac{\mathcal{R}^{b}_{12} g \kappa _1^2 \mathcal{C}_{2D1}^{q_L[1]}}{2 \Lambda ^2}+\frac{\mathcal{R}^{b}_{22} g \kappa _2^2 \mathcal{C}_{2D1}^{q_L[1]}}{2 \Lambda ^2}-\frac{\mathcal{R}^{b}_{22} g \kappa _1^2 \mathcal{C}_{2D1}^{q_L[1]}}{2 \Lambda ^2}-\frac{\mathcal{R}^{b}_{12} g \kappa _2^2 \mathcal{C}_{2D1}^{q_L[1]}}{2 \Lambda ^2}$\\&&$+\frac{\mathcal{R}^{b}_{22} g \kappa _1^2 \mathcal{C}_{2D1}^{q_L[3]}}{2 \Lambda ^2}+\frac{\mathcal{R}^{b}_{22} g \kappa _2^2 \mathcal{C}_{2D1}^{q_L[3]}}{2 \Lambda ^2}-\frac{\mathcal{R}^{b}_{12} g \kappa _1^2 \mathcal{C}_{2D1}^{q_L[3]}}{2 \Lambda ^2}-\frac{\mathcal{R}^{b}_{12} g \kappa _2^2 \mathcal{C}_{2D1}^{q_L[3]}}{2 \Lambda ^2}$ \\
			\hline
			$\overline{d_R}\gamma ^{\mu } d_R  Z_{1\mu }$ & $A_{d_R Z_1}$ & = $\frac{\mathcal{R}^{b}_{31} \mathcal{C}_{LDl}^{q_R} \tilde{g} v_L^2}{\Lambda ^2}+\frac{\mathcal{R}^{b}_{31} \mathcal{C}_{RDr}^{q_R[1]} \tilde{g} v_R^2}{\Lambda ^2}+\frac{\mathcal{R}^{b}_{31} \mathcal{C}_{RDr}^{q_R[3]} \tilde{g} v_R^2}{\Lambda ^2}-\frac{\mathcal{R}^{b}_{11} g \mathcal{C}_{LDl}^{q_R} v_L^2}{\Lambda ^2}-\frac{\mathcal{R}^{b}_{21} g \mathcal{C}_{RDr}^{q_R[1]} v_R^2}{\Lambda ^2}$\\&&$-\frac{\mathcal{R}^{b}_{21} g \mathcal{C}_{RDr}^{q_R[3]} v_R^2}{\Lambda ^2}+\frac{\mathcal{R}^{b}_{11} g \kappa _1^2 \mathcal{C}_{2D1}^{q_R[1]}}{2 \Lambda ^2}+\frac{\mathcal{R}^{b}_{21} g \kappa _2^2 \mathcal{C}_{2D1}^{q_R[1]}}{2 \Lambda ^2}-\frac{\mathcal{R}^{b}_{21} g \kappa _1^2 \mathcal{C}_{2D1}^{q_R[1]}}{2 \Lambda ^2}-\frac{\mathcal{R}^{b}_{11} g \kappa _2^2 \mathcal{C}_{2D1}^{q_R[1]}}{2 \Lambda ^2}$\\&&$+\frac{\mathcal{R}^{b}_{21} g \kappa _1^2 \mathcal{C}_{2D1}^{q_R[3]}}{2 \Lambda ^2}+\frac{\mathcal{R}^{b}_{21} g \kappa _2^2 \mathcal{C}_{2D1}^{q_R[3]}}{2 \Lambda ^2}-\frac{\mathcal{R}^{b}_{11} g \kappa _1^2 \mathcal{C}_{2D1}^{q_R[3]}}{2 \Lambda ^2}-\frac{\mathcal{R}^{b}_{11} g \kappa _2^2 \mathcal{C}_{2D1}^{q_R[3]}}{2 \Lambda ^2}$ \\
			\hline
			$\overline{d_R}\gamma ^{\mu } d_R  Z_{2\mu }$ & $A_{d_R Z_2}$ & = $\frac{\mathcal{R}^{b}_{32} \mathcal{C}_{LDl}^{q_R} \tilde{g} v_L^2}{\Lambda ^2}+\frac{\mathcal{R}^{b}_{32} \mathcal{C}_{RDr}^{q_R[1]} \tilde{g} v_R^2}{\Lambda ^2}+\frac{\mathcal{R}^{b}_{32} \mathcal{C}_{RDr}^{q_R[3]} \tilde{g} v_R^2}{\Lambda ^2}-\frac{\mathcal{R}^{b}_{12} g \mathcal{C}_{LDl}^{q_R} v_L^2}{\Lambda ^2}-\frac{\mathcal{R}^{b}_{22} g \mathcal{C}_{RDr}^{q_R[1]} v_R^2}{\Lambda ^2}$\\&&$-\frac{\mathcal{R}^{b}_{22} g \mathcal{C}_{RDr}^{q_R[3]} v_R^2}{\Lambda ^2}+\frac{\mathcal{R}^{b}_{12} g \kappa _1^2 \mathcal{C}_{2D1}^{q_R[1]}}{2 \Lambda ^2}+\frac{\mathcal{R}^{b}_{22} g \kappa _2^2 \mathcal{C}_{2D1}^{q_R[1]}}{2 \Lambda ^2}-\frac{\mathcal{R}^{b}_{22} g \kappa _1^2 \mathcal{C}_{2D1}^{q_R[1]}}{2 \Lambda ^2}-\frac{\mathcal{R}^{b}_{12} g \kappa _2^2 \mathcal{C}_{2D1}^{q_R[1]}}{2 \Lambda ^2}$\\&&$+\frac{\mathcal{R}^{b}_{22} g \kappa _1^2 \mathcal{C}_{2D1}^{q_R[3]}}{2 \Lambda ^2}+\frac{\mathcal{R}^{b}_{22} g \kappa _2^2 \mathcal{C}_{2D1}^{q_R[3]}}{2 \Lambda ^2}-\frac{\mathcal{R}^{b}_{12} g \kappa _1^2 \mathcal{C}_{2D1}^{q_R[3]}}{2 \Lambda ^2}-\frac{\mathcal{R}^{b}_{12} g \kappa _2^2 \mathcal{C}_{2D1}^{q_R[3]}}{2 \Lambda ^2}$ \\
			\hline
	\end{tabular}}
	\caption{Table \ref{tab:MLRSM-nGB-F-F-II} continued.}
	\label{tab:MLRSM-nGB-F-F-III}
\end{table}
\clearpage





\subsection{Fermi Constant $(\mathcal{G}_{F})$ and $\rho$ parameter}

The Fermi constant $(\mathcal{G}_{F})_{_{SM}}$ is defined using the information from muon decay
$\mu^{-} \rightarrow e^{-} + \bar{\nu_{e}} + \nu_{\mu}$ as \cite{Buchmuller:1985jz}:
{\small\begin{align}
	(\mathcal{G}_{F})_{_{SM}}= \frac{g^2}{4\sqrt{2}M_{W}^2}.
	\end{align}} 

In the context of the Standard Model, the $\rho$ parameter is defined as \cite{Buchmuller:1985jz}:
{\small\begin{align}
	\rho=\frac{M^2_{W}}{M^2_{Z} \cos^2 \theta_\text{w}}.
	\end{align}}

\begin{figure}[h!]
	\centering
	{
	\includegraphics[trim={1.5cm 0 1.5cm 0},scale=0.5]{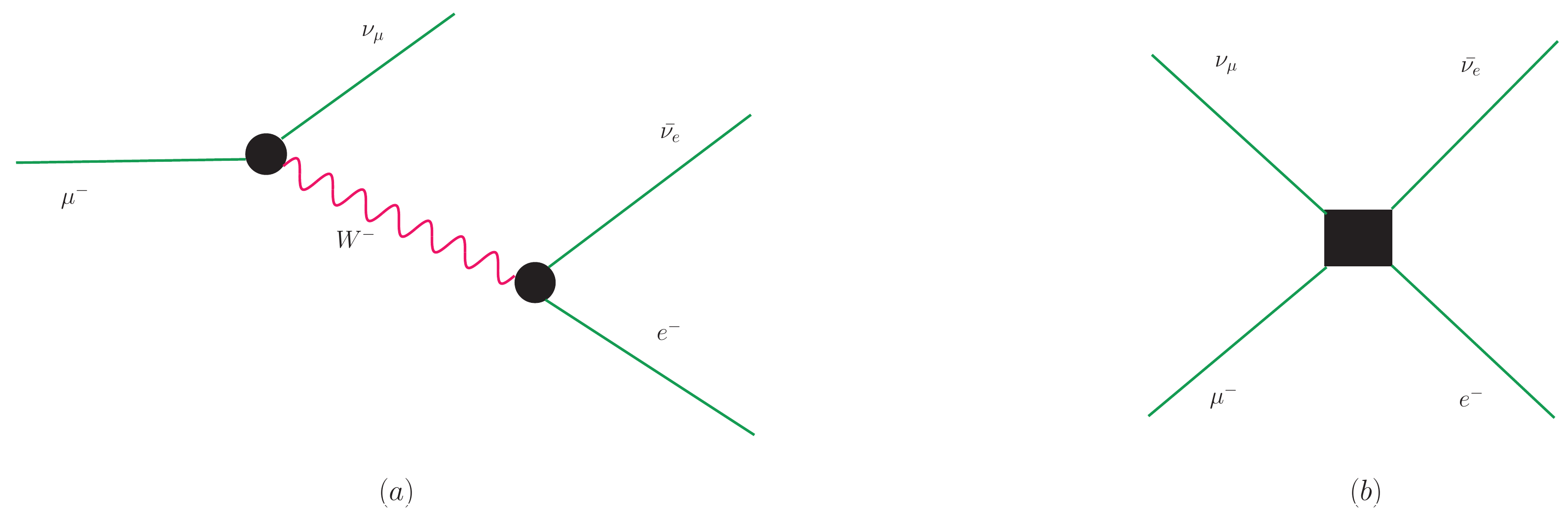}
	}
	\caption{Three-body decay of muon: $\mu^{-} \rightarrow e^{-} + \bar{\nu_{e}} + \nu_{\mu}$. This process gives an estimation for Fermi constant $(\mathcal{G}_{F})$. The (a) [{\Large $\bullet$}]  depicts the modified Feynman vertex (renormalizable + corrections due to dimension-6 operators), and (b) [$\blacksquare$] represents the effective four-Fermi vertex.}\label{fig:GF}
\end{figure}

In the presence of the effective operators, there can be two fold contributions to the Fermi constant:  (i) through the modified  $\mu-W-\nu_\mu, e-W-\bar{\nu_e}$ vertices and gauge boson mass $\mathcal{M}_{W}$, see Fig.~\ref{fig:GF}(a), and (ii) direct contributions from the dimension-6 four-fermion operators, Fig.~\ref{fig:GF}(b). We have encapsulated these contributions and computed the modified $G_F$ and $\rho$ parameters for both 2HDM and MLRSM effective theories.

	\subsection*{2HDM}
The modified Fermi constant for 2HDM-EFT can be written as:	
{\small\begin{equation}
	(\mathcal{G}_{F})_{_{2HDM}}=(\mathcal{G}_{F})_{_{SM}}\Big[1-\frac{\Delta{M}^2_{W}}{v^2}-\frac{1}{\Lambda^2}\left(\epsilon^e_{(e\nu)_L}+\epsilon^\mu_{(e\nu)_L}-\frac{2 \mathcal{C}_{LL} M_{_{W}}^2}{g^2}\right)\Big].
	\end{equation}}
Here, $\mathcal{C}_{LL}$ contains information regarding the four-fermion interactions involving $\{ e,\mu,\nu_{e},\nu_{\mu} \}$, see Table~\ref{tab:2HDM-psi4}. The vertex corrections $\epsilon^e_{(e\nu)_L}, \epsilon^\mu_{(e\nu)_L}$ are depicted in  Tables~\ref{tab:2HDM-GB-F-F-I}, \ref{tab:2HDM-GB-F-F-II}.
\\	
The rho-parameter for this effective scenario can be defined as 
{\small\begin{align}
	\bar{\rho}=\frac{\mathcal{M}_{W}^2}{\mathcal{M}_{Z}^2 (1-\sin^2\bar{\theta}_\text{w})}.
	\end{align}}

\subsection*{MLRSM}
The modified Fermi constant for MLRSM-EFT can be written as:	
{\small\begin{equation}
	(\mathcal{G}_{F})_{MLRSM}=\frac{g^2}{4\sqrt{2}\mathcal{M}^2_{W_1}}\left[\epsilon^{1,e}_{(e\nu)_L}+\epsilon^{1,\mu}_{(e\nu)_L}+\frac{1}{\Lambda^2}\frac{8\sqrt{2}\mathcal{C}_{l_L l_L}\mathcal{M}^2_{W_1}}{g^2}\right].
	\end{equation}}

Here, $\mathcal{C}_{l_L l_L}$ contains information regarding the four-fermion interactions involving $\{ e,\mu,\nu_{e},\nu_{\mu} \}$, see Table~\ref{tab:MLRSM-psi4}. The vertex corrections $\epsilon^{1,e}_{(e\nu)_L},\epsilon^{1,\mu}_{(e\nu)_L}$ are depicted Tables~\ref{tab:MLRSM-cGB-F-F-I}, \ref{tab:MLRSM-cGB-F-F-II}.
\\
The modified  rho-parameter for this effective scenario can be expressed as 
{\small\begin{align}
	\bar{\rho}=\frac{\mathcal{M}_{W_1}^2}{\mathcal{M}_{Z_1}^2 (1-\sin^2\bar{\theta}_\text{w})}.
	\end{align}}

%% file: Oblique.tex
\subsection{Oblique Parameters}

 The two-point vector boson correlation functions  after including the radiative corrections, i.e., the vacuum polarization can be written as \cite{Peskin:1991sw,Kundu:1996ah,Cacciapaglia:2006pk}: 
\small{\begin{eqnarray}\label{eq:2point-correlation-func}
	i  \Pi_{V_iV_j}^{\mu\nu}(p^2)= i \left(g^{\mu\nu}-\frac{p^\mu p^\nu}{p^2}\right)\Pi_{V_iV_j}(p^2)+\left(i\frac{p^\mu p^\nu}{p^2}\; \text{terms}\right),
	\end{eqnarray}
where $V_i$ represents the vector boson in either unphysical  ${V_i}\in $ $\{W_{1},W_{2},W_{3},B\}$ or physical  $\{W,Z,\gamma \}$ basis and $p^\mu$ is the external momentum, see Fig.~\ref{fig:two-point-function}. Note that $\gamma$ represents the photon field $A_{\mu}$.

\begin{figure}[h!]
	\centering
	{
		\includegraphics[trim={2.5cm 0 2.5cm 0},scale=0.9]{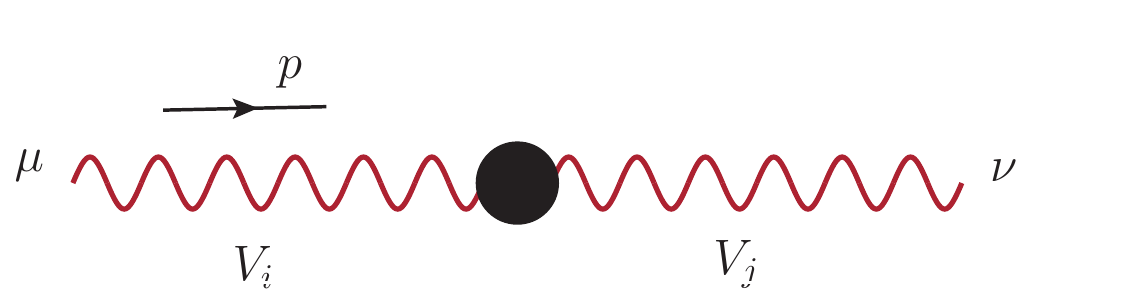}
	}
	\caption{The effective two-point vector boson correlation function. The  [{\Large $\bullet$}]  represents the corrections due to the dimension-6 operators.}\label{fig:two-point-function}
\end{figure}

The vacuum polarization amplitudes $\Pi_{V_iV_j}(p^2)$ can be expanded as a polynomial in external momentum ($p^2$) as \cite{Peskin:1991sw,Kundu:1996ah,Cacciapaglia:2006pk}:
\begin{align}\label{eqn:vacpol}
\Pi_{V_i V_j}(p^2)=  [\Pi_{0} + \Pi_{2}p^2+ \Pi_{4}p^4+\mathcal{O}(p^6)]_{V_iV_j}.
\end{align}

In \cite{Peskin:1991sw}, Peskin and  Takeuchi had defined the oblique parameters $S,T,U$ to grab the radiative corrections to the tree-level correlation functions. These variables also capture the new physics effects. 
\clearpage

These $S,T,U $ parameters can be defined in both unphysical and physical gauge boson basis as \cite{Peskin:1991sw,Cobanoglu:2010ie,Kundu:1996ah,Cacciapaglia:2006pk}:
\small{\begin{eqnarray}\label{eq:oblique-S} 
	S&=&-\frac{4 \cos\theta_{\text{w}}\sin\theta_{\text{w}}}{\alpha}\Pi'_{W_{3}B}(0) \\
	&=&-\frac{4 \cos^2\theta_{\text{w}}\sin^2\theta_{\text{w}}}{m^2_{Z}\alpha}\Big[\Pi_{ZZ}(m^2_{Z})-\Pi_{ZZ}(0)-\Pi_{\gamma \gamma}(m^2_{Z}) \nonumber \\
	&&-\frac{\cos^2\theta_{\text{w}}-\sin^2\theta_{\text{w}}}{\cos\theta_{\text{w}}\sin\theta_{\text{w}}}\Big(\Pi_{\gamma Z}(m^2_{Z})-\Pi_{\gamma Z}(0)\Big)\Big], \nonumber\\  \nonumber \\
	T &=& \frac{1}{\alpha m_W^2}\Big(\Pi_{W_{1}W_{1}}(0)-\Pi_{W_{3}W_{3}}(0)\Big) \label{eq:oblique-T}  \\
	&=& \frac{1}{\alpha}\Big[\frac{\Pi_{WW}(0)}{ m_W^2}-\frac{\Pi_{ZZ}(0)}{ m_Z^2}-\frac{2\sin\theta_{\text{w}}\Pi_{\gamma Z}(0)}{ \cos\theta_{\text{w}} m_Z^2}\Big],\nonumber \\  \nonumber \\
	U &=&  \frac{4 \sin^2\theta_{\text{w}}}{\alpha}\Big(\Pi'_{W_{1}W_{1}}(0)-\Pi'_{W_{3}W_{3}}(0)\Big) \label{eq:oblique-U} \\
	&=&-\frac{4\sin^2\theta_{\text{w}}}{\alpha}\Big[\Big(\frac{\Pi_{WW}(m^2_{W})-\Pi_{WW}(0)}{m_W^2}\Big)-\cos^2\theta_{\text{w}}\Big(\frac{\Pi_{ZZ}(m^2_{Z})-\Pi_{ZZ}(0)}{m_Z^2}\Big)\nonumber \\
	&&-2\cos\theta_{\text{w}}\sin\theta_{\text{w}}\Big(\frac{\Pi_{\gamma Z}(m^2_{Z})-\Pi_{\gamma Z}(0)}{m_Z^2}\Big)-\sin^2\theta_{\text{w}}\frac{\Pi_{\gamma \gamma }(m^2_{Z})}{m_Z^2}\Big],\nonumber
	\end{eqnarray}} 
where $\alpha=e^2/4\pi$ is the Fine structure constant.\\ \\
Here,  for unphysical basis
\begin{eqnarray*}
	\Pi'_{V_i V_j}(0) &= &\frac{d\Pi_{V_i V_j}}{dp^2}\at[\Big]{p^2 =0} \hspace{0.2cm }, \hspace{1.0cm }\Pi_{W_1 W_1}(p^2)=\Pi_{W_2 W_2}(p^2),
\end{eqnarray*}	
and for physical basis, we have
\small{\begin{eqnarray*}
	\Pi'_{V_{i}V_{j}}(0)&=&\Big(\frac{\Pi_{V_{i}V_{j}}(p^2)-\Pi_{V_{i}V_{J}}(0)}{p^2}\Big)\at[\Big]{p^2=0}\hspace{0.2cm } ,\hspace{1cm}\Pi_{\gamma \gamma}(0)=\Pi_{\gamma Z}(0)=0.
		\end{eqnarray*}}


\underline{\large{$\phi^{4} D^2$ and $\phi^2 X^2$ operators: Modifications in vacuum polarisation functions }}\\

In the earlier paragraph,  we have discussed the general structure of the oblique parameters. Based on that, the $S,T,U$ parameters are computed 
in the context of renormalizable 2HDM \cite{Funk:2011ad} and extended gauge scenarios \cite{Altarelli:1990wt,Holdom:1996bn}. These oblique parameters get additional contributions in the presence of the effective operators \cite{Kennedy:1988sn,Grinstein:1991cd,SanchezColon:1998xg}. Here, we will focus on the computation of $S,T,U$ parameters specifically for 2HDM- and MLRSM-EFT scenarios. 

We have noted how the $\phi^2 X^2$ operators redefine the gauge fields, and the $\phi^{4} D^2$ operators affect the gauge boson mass terms which are bilinear in gauge fields. In the process, both classes of operators modify the two-point gauge field correlation functions, see Fig.~\ref{fig:two-point-function}.

\section*{2HDM}

The part of the effective 2HDM lagrangian  that is bilinear in gauge fields is given in unphysical basis as :
\small{\begin{eqnarray}\label{eq:2HDM-bilinear-GB-lag}
		\mathcal{L}_{W_{1}W_{1}}&=& \frac{1}{2}W_{1\mu}\left(-\partial ^2 g^{\mu \nu }+\partial ^{\mu } \partial ^{\nu }\right)W_{1\nu}\left(1+\frac{2 \Theta_{WW}}{\Lambda ^2}\right)+(\frac{1}{2}W_{1\mu}W_{1}^{\mu})\Big[\frac{1}{4} g^2 (v_1^2+ v_2^2) \\
		&&+\frac{g^2}{4 \Lambda ^2}\big(\textcolor{purple}{\mathcal{C}_{\phi D}^{12(1)(1)} v_2 v_1^3}+ \mathcal{C}_{\phi D}^{12(1)(2)} v_2^2 v_1^2+ \textcolor{purple}{\mathcal{C}_{\phi D}^{21(2)(2)} v_2^3 v_1}+ 2 v_1^2 \Theta_{WW}+2 v_2^2 \Theta_{WW}\big)\Big],\nonumber\\  \nonumber \\
		\mathcal{L}_{W_{3}W_{3}}&=&  \frac{1}{2} W_{3\mu}\left(-\partial ^2 g^{\mu \nu }+\partial ^{\mu } \partial ^{\nu }\right)W_{3\nu}\left(1+\frac{2 \Theta_{WW}}{\Lambda ^2}\right)+(\frac{1}{2}W_{3\mu} W_{3}^{\mu})\Big[\frac{1}{4} g^2 (v_1^2+ v_2^2)+\frac{g g'\Theta_{W_{3}B}}{4 \Lambda ^2}\big( v_1^2 + v_2^2 \big)  \nonumber \\
		&&+\frac{g^2}{8 \Lambda ^2}\Big( \mathcal{C}_{\phi D}^{(1)11(1)} v_1^4+\textcolor{purple}{2 \mathcal{C}_{\phi D}^{(1)21(1)} v_2 v_1^3+2\mathcal{C}_{\phi D}^{12(1)(1)} v_2 v_1^3}+2 \mathcal{C}_{\phi D}^{12(1)(2)} v_2^2 v_1^2+\mathcal{C}_{\phi D}^{(1)22(1)} v_2^2 v_1^2\nonumber \\
		&&+ \mathcal{C}_{\phi D}^{(2)11(2)} v_2^2 v_1^2+ \textcolor{purple}{2 \mathcal{C}_{\phi D}^{21(2)(2)} v_2^3 v_1+2 \mathcal{C}_{\phi D}^{(2)22(1)} v_2^3 v_1}+ \mathcal{C}_{\phi D}^{2(22)2} v_2^4+4  \Theta_{WW}(v_1^2+ v_2^2) \Big)\Big],\\  \nonumber \\
		\mathcal{L}_{W_{3}B}&=& W_{3\mu}\left(-\partial ^2 g^{\mu \nu }+\partial ^{\mu } \partial ^{\nu }\right)B_{\nu}\left(-\frac{\Theta_{W_{3}B}}{\Lambda ^2}\right)\nonumber \\
		&&+W_{3\mu}B^{\mu}\Big[-\frac{1}{4} g g'( v_1^2+v_2^2 )-\frac{g g'  (\Theta_{BB}+\Theta_{WW})}{4 \Lambda ^2}(v_1^2+ v_2^2 )-\frac{\Theta_{W_{3}B}}{8 \Lambda ^2}(g^2+g'^2) (v_1^2+ v_2^2)\nonumber\\
		&&-\frac{g g'}{8 \Lambda ^2}\Big( \mathcal{C}_{\phi D}^{(1)11(1)} v_1^4\textcolor{purple}{+2 \mathcal{C}_{\phi D}^{(1)21(1)} v_2 v_1^3 +2\mathcal{C}_{\phi D}^{12(1)(1)} v_2 v_1^3} +2 \mathcal{C}_{\phi D}^{12(1)(2)} v_2^2 v_1^2 +\mathcal{C}_{\phi D}^{(1)22(1)} v_2^2 v_1^2  \nonumber \\
		&&+\mathcal{C}_{\phi D}^{(2)11(2)} v_2^2 v_1^2\textcolor{purple}{+2 \mathcal{C}_{\phi D}^{21(2)(2)} v_2^3 v_1 +2 \mathcal{C}_{\phi D}^{(2)22(1)} v_2^3 v_1} +\mathcal{C}_{\phi D}^{2(22)2} v_2^4\Big)\Big],\\  \nonumber \\
		\mathcal{L}_{BB}&=& \frac{1}{2} B_{\mu}\left(-\partial ^2 g^{\mu \nu }+\partial ^{\mu } \partial ^{\nu }\right) B_{\nu}\left(1+\frac{2 \Theta_{BB}}{\Lambda ^2}\right)\nonumber \\
		&&+(\frac{1}{2}B_{\mu} B^{\mu})\Big[\frac{1}{4} g'^2 (v_1^2+ v_2^2)+\frac{ \Theta_{BB} g'^2}{2 \Lambda ^2}(v_1^2+v_2^2) +\frac{g  g' \Theta_{W_{3}B}}{4 \Lambda ^2}(v_1^2+ v_2^2)\nonumber \\
		&&+\frac{g'^2}{8 \Lambda ^2}\Big(\mathcal{C}_{\phi D}^{(1)11(1)} v_1^4 \textcolor{purple}{+2\mathcal{C}_{\phi D}^{(1)21(1)} v_2 v_1^3 +2\mathcal{C}_{\phi D}^{12(1)(1)} v_2 v_1^3} +2\mathcal{C}_{\phi D}^{12(1)(2)} v_2^2 v_1^2 +\mathcal{C}_{\phi D}^{(1)22(1)} v_2^2 v_1^2 \nonumber \\
		&&+\mathcal{C}_{\phi D}^{(2)11(2)} v_2^2 v_1^2 \textcolor{purple}{+2\mathcal{C}_{\phi D}^{21(2)(2)} v_2^3 v_1 +2\mathcal{C}_{\phi D}^{(2)22(1)} v_2^3 v_1}+\mathcal{C}_{\phi D}^{2(22)2} v_2^4 \Big)\Big].
\end{eqnarray}}
We can relate the $\Pi_{VV}^{\mu \nu} (p^2)$ with the relevant part of  the Fourier-transformed ($FT$) lagrangian as: 
\begin{equation*}
	\Pi_{V_i V_j}^{\mu \nu} (p^2) V_{i\mu} V_{j\nu} = FT\;\{ \mathcal{L}_{V_i V_j} \},
\end{equation*} 
and these can be written in terms of  $\Pi_{V_{i}V_{j}}$:
\small{\begin{eqnarray}
		i \Pi_{W_{1}W_{1}}^{\mu\nu}(p^2)&=&i \left(g^{\mu\nu}-\frac{p^\mu p^\nu}{p^2}\right)\Pi_{W_1 W_1}(p^2),\\
		i \Pi_{W_{3}W_{3}}^{\mu\nu}(p^2)&=&i \left(g^{\mu\nu}-\frac{p^\mu p^\nu}{p^2}\right)\Pi_{W_3 W_3}(p^2),\\
		i \Pi_{W_{3}B}^{\mu\nu}(p^2)&=&i \left(g^{\mu\nu}-\frac{p^\mu p^\nu}{p^2}\right)\Pi_{W_3 B}(p^2),\\
		i \Pi_{BB}^{\mu\nu}(p^2)&=&i \left(g^{\mu\nu}-\frac{p^\mu p^\nu}{p^2}\right)\Pi_{BB}(p^2).
\end{eqnarray}}

Now the $\Pi_{V_iV_j}(p^ 2)$ can be expressed as: 
\begin{eqnarray}\label{eq:2HDM-Pi-function}
\Pi_{W_1 W_1}(p^2)&=&p^2 \left(1+\frac{2 \Theta_{WW}}{\Lambda ^2}\right)+\frac{g^2}{4} (v_1^2+ v_2^2)\nonumber \\
&&+\frac{g^2}{4 \Lambda ^2}\Big[\textcolor{purple}{\mathcal{C}_{\phi D}^{12(1)(1)} v_2 v_1^3}+ \mathcal{C}_{\phi D}^{12(1)(2)} v_2^2 v_1^2\textcolor{purple}{+ \mathcal{C}_{\phi D}^{21(2)(2)} v_2^3 v_1}+2 v_1^2 \Theta_{WW}+2 v_2^2 \Theta_{WW}\Big],\\  \nonumber \\
\Pi_{W_3 W_3}(p^2)&=&p^2 \left(1+\frac{2 \Theta_{WW}}{\Lambda ^2}\right)+\frac{g^2}{4} ( v_1^2+ v_2^2)+\frac{g g'\Theta_{W_{3}B}}{4 \Lambda ^2}\big(  v_1^2+ v_2^2\big)\nonumber \\
&&+\frac{g^2}{8 \Lambda ^2}\Big[ \mathcal{C}_{\phi D}^{(1)11(1)} v_1^4\textcolor{purple}{+2 \mathcal{C}_{\phi D}^{(1)21(1)} v_2 v_1^3+2\mathcal{C}_{\phi D}^{12(1)(1)} v_2 v_1^3}+2 \mathcal{C}_{\phi D}^{12(1)(2)} v_2^2 v_1^2+\mathcal{C}_{\phi D}^{(1)22(1)} v_2^2 v_1^2\nonumber \\
&&+ \mathcal{C}_{\phi D}^{(2)11(2)} v_2^2 v_1^2\textcolor{purple}{+ 2 \mathcal{C}_{\phi D}^{21(2)(2)} v_2^3 v_1+2 \mathcal{C}_{\phi D}^{(2)22(1)} v_2^3 v_1}+ \mathcal{C}_{\phi D}^{2(22)2} v_2^4+4  \Theta_{WW}(v_1^2+ v_2^2) \Big],\\  \nonumber \\
\Pi_{W_3 B}(p^2)&=&-\frac{p^2 \Theta_{W_{3}B}}{\Lambda ^2}-\frac{g g'}{4} (v_1^2 + v_2^2)-\frac{g g'  (\Theta_{BB}+\Theta_{WW})}{4 \Lambda ^2}(v_1^2+ v_2^2 )-\frac{\Theta_{W_{3}B}}{8 \Lambda ^2}(g^2+g'^2) (v_1^2+ v_2^2)  \nonumber \\
&&-\frac{g g'}{8 \Lambda ^2}\Big[ \mathcal{C}_{\phi D}^{(1)11(1)} v_1^4+\textcolor{purple}{2 \mathcal{C}_{\phi D}^{(1)21(1)} v_2 v_1^3 +2\mathcal{C}_{\phi D}^{12(1)(1)} v_2 v_1^3} +2 \mathcal{C}_{\phi D}^{12(1)(2)} v_2^2 v_1^2 +\mathcal{C}_{\phi D}^{(1)22(1)} v_2^2 v_1^2  \nonumber \\
&&+\mathcal{C}_{\phi D}^{(2)11(2)} v_2^2 v_1^2\textcolor{purple}{+2 \mathcal{C}_{\phi D}^{21(2)(2)} v_2^3 v_1 +2 \mathcal{C}_{\phi D}^{(2)22(1)} v_2^3 v_1} +\mathcal{C}_{\phi D}^{2(22)2} v_2^4\Big],\\  \nonumber \\
\Pi_{BB}(p^2)&=&p^2 \left(1+\frac{2 \Theta_{BB}}{\Lambda ^2}\right)+\frac{g'^2}{4} (v_1^2+ v_2^2) +\frac{ \Theta_{BB} g'^2}{2 \Lambda ^2}(v_1^2+v_2^2) +\frac{g  g' \Theta_{W_{3}B}}{4 \Lambda ^2}(v_1^2+ v_2^2)\nonumber \\
&&+\frac{g'^2}{8 \Lambda ^2}\Big[\mathcal{C}_{\phi D}^{(1)11(1)} v_1^4 \textcolor{purple}{+2\mathcal{C}_{\phi D}^{(1)21(1)} v_2 v_1^3 +2\mathcal{C}_{\phi D}^{12(1)(1)} v_2 v_1^3} +2\mathcal{C}_{\phi D}^{12(1)(2)} v_2^2 v_1^2 +\mathcal{C}_{\phi D}^{(1)22(1)} v_2^2 v_1^2 \nonumber \\
&&+\mathcal{C}_{\phi D}^{(2)11(2)} v_2^2 v_1^2 \textcolor{purple}{+2\mathcal{C}_{\phi D}^{21(2)(2)} v_2^3 v_1 +2\mathcal{C}_{\phi D}^{(2)22(1)} v_2^3 v_1}+\mathcal{C}_{\phi D}^{2(22)2} v_2^4 \Big].
\end{eqnarray}}

Then recasting these  $\Pi_{V_iV_j}(p^ 2)$ in Eqs.~\ref{eq:oblique-S}, \ref{eq:oblique-T}, \ref{eq:oblique-U}, we can construct the oblique parameters for 2HDM-EFT scenario as:
\small{\begin{eqnarray}\label{eq:2HDM-oblique}
	S&=& \frac{4 \sin \theta_{\text{w}} \cos \theta_{\text{w}}}{\alpha  \Lambda ^2} \Big[\mathcal{C}_{\phi WB}^{11} v_1^2\textcolor{purple}{+v_2 v_1 (\mathcal{C}_{\phi WB}^{12}+\mathcal{C}_{\phi WB}^{21})}+\mathcal{C}_{\phi WB}^{22} v_2^2\Big], \\  \nonumber \\
	T &=& -\frac{g}{8 \alpha  \Lambda ^2 M_{W}^2} \Big[g \mathcal{C}_{\phi D}^{(1)11(1)} v_1^4+2 \mathcal{C}_{\phi WB}^{11} \big(v_1^2+v_2^2\big) v_1^2 g'\nonumber \\ & & +v_2 \Big(2 \mathcal{C}_{\phi WB}^{22} v_2 v_1^2 g'+2 \mathcal{C}_{\phi WB}^{22} v_2^3 g'
	+g\mathcal{C}_{\phi D}^{(1)22(1)} v_2 v_1^2+g\mathcal{C}_{\phi D}^{(2)11(2)} v_2 v_1^2+g\mathcal{C}_{\phi D}^{2(22)2} v_2^3\nonumber \\
	&&+\textcolor{purple}{2 g\mathcal{C}_{\phi D}^{(1)21(1)} v_1^3}+
	\textcolor{purple}{2 \mathcal{C}_{\phi WB}^{12}  v_1 g'\big(v_1^2+v_2^2\big)+2 \mathcal{C}_{\phi WB}^{21} v_1^3 g'+2 \mathcal{C}_{\phi WB}^{21} v_2^2 v_1 g'}+\textcolor{purple}{2 g\mathcal{C}_{\phi D}^{(2)22(1)} v_2^2 v_1}\Big)\Big],\nonumber \\  \nonumber \\
	U &=& 0. \nonumber 
	\end{eqnarray}}
\clearpage


\section*{MLRSM}

Using the same principle followed in 2HDM-EFT, vacuum polarization amplitudes $\Pi_{V_iV_j}(p^ 2)$ in case of MLRSM-EFT can be given as:

\small{\begin{eqnarray}\label{eq:MLRSM-Pi-function-I}
	\Pi_{W_{L}W_{L}}(p^2)&=&p^2\Big(1+2\frac{ \Theta _{W_{{LL}}}}{\Lambda ^2}\Big)+\frac{1}{8} \left(2 g^2 \kappa _1^2+2 g^2 \kappa _2^2+4 g^2 v_L^2\right)\nonumber \\
	&&+\frac{1}{8\Lambda ^2} \Big[2 g^2 \kappa _1^2 \mathcal{C}_{\phi D}^{(2){rL}(1)} v_L v_R+g^2 \mathcal{C}_{\phi D}^{(L){rR}(l)} v_L^2 v_R^2+8 g^2 v_L^2 \Theta _{W_{{LL}}}+2 g^2 \kappa _1^2 \mathcal{C}_{\phi D}^{\{21\}\{(L)(l)\}} v_L^2\nonumber \\
	&&+2 g^2 \kappa _2^2 \mathcal{C}_{\phi D}^{\{21\}\{(L)(l)\}} v_L^2+g^2 \kappa _1^2 \mathcal{C}_{\phi D}^{21(L)(l)} v_L^2+g^2 \kappa _2^2 \mathcal{C}_{\phi D}^{21(L)(l)} v_L^2+2 g^2 \kappa _1^2 \mathcal{C}_{\phi D}^{(L)11(l)} v_L^2+2 g^2 \kappa _2^2 \mathcal{C}_{\phi D}^{(L)11(l)} v_L^2\nonumber \\
	&&+2 g^2 \kappa _1 \kappa _2 \mathcal{C}_{\phi D}^{L(l)1(1)} v_L^2+g^2 \kappa _1^2 \mathcal{C}_{\phi D}^{\{{Ll}\}\{(2)(1)\}} v_L^2+g^2 \kappa _2^2 \mathcal{C}_{\phi D}^{\{{Ll}\}\{(2)(1)\}} v_L^2+g^2 \kappa _1^2 \mathcal{C}_{\phi D}^{{Ll}(2)(1)} v_L^2\nonumber \\
	&&+2 g^2 v_L^4 (\mathcal{C}_{\phi D}^{\{{Ll}\}\{(L)(l)\}}+\mathcal{C}_{\phi D}^{L(l)L(l)}+\mathcal{C}_{\square }^{Ll(Ll)})+g^2 \kappa _1^2 \mathcal{C}_{\phi D}^{\{{Rr}\}\{(2)(1)\}} v_R^2+g^2 \kappa _2^2 \mathcal{C}_{\phi D}^{\{{Rr}\}\{(2)(1)\}} v_R^2\nonumber \\
	&&+g^2 \kappa _1^2 \mathcal{C}_{\phi D}^{{Rr}(2)(1)} v_R^2+4 g^2 \kappa _1^2 \Theta _{W_{{LL}}}+4 g^2 \kappa _2^2 \Theta _{W_{{LL}}}-8 g^2 \kappa _1 \kappa _2 \Theta _{W_{{RL}}}-4 g^2 \kappa _1^2 \kappa _2^2 \mathcal{C}_{\phi D}^{1(1)1(1)}+g^2 \kappa _1^4 \mathcal{C}_{\phi D}^{(11)(22)}\nonumber \\
	&&+g^2 \kappa _2^4 \mathcal{C}_{\phi D}^{(11)(22)}+2 g^2 \kappa _1 \kappa _2^3 \mathcal{C}_{\phi D}^{(11)(22)}+2 g^2 \kappa _1^2 \kappa _2^2 \mathcal{C}_{\phi D}^{(11)(22)}+2 g^2 \kappa _1^3 \kappa _2 \mathcal{C}_{\phi D}^{(11)(22)}-4 g^2 \kappa _1^2 \kappa _2^2 \mathcal{C}_{\phi D}^{2(1)1(1)}\nonumber \\
	&&+g^2 \kappa _1^4 \mathcal{C}_{\phi D}^{\{(2)(1)\}\{21\}}+g^2 \kappa _2^4 \mathcal{C}_{\phi D}^{\{(2)(1)\}\{21\}}+2 g^2 \kappa _1^2 \kappa _2^2 \mathcal{C}_{\phi D}^{\{(2)(1)\}\{21\}}+g^2 \kappa _1^4 \mathcal{C}_{\phi D}^{(2)(1)21}+g^2 \kappa _2^4 \mathcal{C}_{\phi D}^{(2)(1)21}\nonumber \\
	&&+8 g^2 \kappa _1^2 \kappa _2^2 \mathcal{C}_{\phi D}^{2(1)2(1)}\Big], \\  \nonumber \\
	\Pi_{W_{R}W_{R}}(p^2)&=&p^2\Big(1+2 \frac{ \Theta _{W_{{RR}}}}{\Lambda ^2}\Big)+\frac{1}{8} \left(2 g^2 \kappa _1^2+2 g^2 \kappa _2^2+4 g^2 v_R^2\right) \\
	&&+\frac{1}{8\Lambda ^2} \Big[2 g^2 \kappa _2^2 \mathcal{C}_{\phi D}^{(2){rL}(1)} v_L v_R+g^2 \mathcal{C}_{\phi D}^{(R){lL}(r)} v_L^2 v_R^2+g^2 \kappa _1^2 \mathcal{C}_{\phi D}^{\{{Ll}\}\{(2)(1)\}} v_L^2+g^2 \kappa _2^2 \mathcal{C}_{\phi D}^{\{{Ll}\}\{(2)(1)\}} v_L^2\nonumber \\
	&&+g^2 \kappa _2^2 \mathcal{C}_{\phi D}^{{Ll}(2)(1)} v_L^2+8 g^2 v_R^2 \Theta _{W_{{RR}}}+2 g^2 \kappa _1^2 \mathcal{C}_{\phi D}^{\{21\}\{(R)(r)\}} v_R^2+2 g^2 \kappa _2^2 \mathcal{C}_{\phi D}^{\{21\}\{(R)(r)\}} v_R^2\nonumber \\
	&&+g^2 \kappa _1^2 \mathcal{C}_{\phi D}^{21(R)(r)} v_R^2+g^2 \kappa _2^2 \mathcal{C}_{\phi D}^{21(R)(r)} v_R^2+2 g^2 \kappa _1^2 \mathcal{C}_{\phi D}^{(R)11(r)} v_R^2+2 g^2 \kappa _2^2 \mathcal{C}_{\phi D}^{(R)11(r)} v_R^2\nonumber \\
	&&-2 g^2 \kappa _2^2 \mathcal{C}_{\phi D}^{R(r)1(1)} v_R^2+g^2 \kappa _1^2 \mathcal{C}_{\phi D}^{\{{Rr}\}\{(2)(1)\}} v_R^2+g^2 \kappa _2^2 \mathcal{C}_{\phi D}^{\{{Rr}\}\{(2)(1)\}} v_R^2+g^2 \kappa _2^2 \mathcal{C}_{\phi D}^{{Rr}(2)(1)} v_R^2\nonumber \\
	&&+2 g^2 v_R^4 (\mathcal{C}_{\phi D}^{\{{Rr}\}\{(R)(r)\}}+\mathcal{C}_{\phi D}^{R(r)R(r)}+\mathcal{C}_{\square }^{Rr(Rr)})-8 g^2 \kappa _1 \kappa _2 \Theta _{W_{{LR}}}+4 g^2 \kappa _1^2 \Theta _{W_{{RR}}}+4 g^2 \kappa _2^2 \Theta _{W_{{RR}}}\nonumber \\
	&&-4 g^2 \kappa _1^2 \kappa _2^2 \mathcal{C}_{\phi D}^{1(1)1(1)}+g^2 \kappa _1^4 \mathcal{C}_{\phi D}^{(11)(22)}+g^2 \kappa _2^4 \mathcal{C}_{\phi D}^{(11)(22)}+2 g^2 \kappa _1 \kappa _2^3 \mathcal{C}_{\phi D}^{(11)(22)}+2 g^2 \kappa _1^2 \kappa _2^2 \mathcal{C}_{\phi D}^{(11)(22)}\nonumber \\
	&&+2 g^2 \kappa _1^3 \kappa _2 \mathcal{C}_{\phi D}^{(11)(22)}-4 g^2 \kappa _1^2 \kappa _2^2 \mathcal{C}_{\phi D}^{2(1)1(1)}+g^2 \kappa _1^4 \mathcal{C}_{\phi D}^{\{(2)(1)\}\{21\}}+g^2 \kappa _2^4 \mathcal{C}_{\phi D}^{\{(2)(1)\}\{21\}}\nonumber \\
	&&+2 g^2 \kappa _1^2 \kappa _2^2 \mathcal{C}_{\phi D}^{\{(2)(1)\}\{21\}}+2 g^2 \kappa _1^2 \kappa _2^2 \mathcal{C}_{\phi D}^{(2)(1)21}+2 g^2 \kappa _1^4 \mathcal{C}_{\phi D}^{2(1)2(1)}+2 g^2 \kappa _2^4 \mathcal{C}_{\phi D}^{2(1)2(1)}+4 g^2 \kappa _1^2 \kappa _2^2 \mathcal{C}_{\phi D}^{2(1)2(1)}\Big],\nonumber \\  \nonumber \\
		\Pi_{W_{L}W_{R}}(p^2)&=&-\frac{1}{8\Lambda ^2} p^2 \left(-8 \Theta _{W_{{LR}}}-8 \Theta _{W_{{RL}}}\right)-\frac{1}{2} g^2 \kappa _1 \kappa _2 \nonumber \\
	&&+\frac{1}{8\Lambda ^2} \Big[g^2 v_L^2 \left(\mathcal{C}_{\phi D}^{(L){rL}(r)} v_R^2+4 \Theta _{W_{{LR}}}-\kappa _2 \left(\kappa _2 \mathcal{C}_{\phi D}^{L(l)1(1)}+\kappa _1 (2 \mathcal{C}_{\phi D}^{\{{Ll}\}\{(2)(1)\}}+\mathcal{C}_{\phi D}^{{Ll}(2)(1)})\right)\right)\nonumber \\
	&&+g^2 v_L v_R \big(-2 \kappa _2 \kappa _1 \mathcal{C}_{\phi D}^{(2){rL}(1)}+\kappa _1^2 (\mathcal{C}_{\phi D}^{(L)11(r)}+\mathcal{C}_{\phi D}^{(L)12(r)}+\mathcal{C}_{\phi D}^{(R)11(l)})+\kappa _2^2 (\mathcal{C}_{\phi D}^{(L)11(r)}+\mathcal{C}_{\phi D}^{(L)12(r)}\nonumber \\
	&&+\mathcal{C}_{\phi D}^{(R)11(l)})\big)+4 g^2 v_R^2 \Theta _{W_{{RL}}}+g^2 \kappa _2 \kappa _1 \mathcal{C}_{\phi D}^{R(r)1(1)} v_R^2-2 g^2 \kappa _2 \kappa _1 \mathcal{C}_{\phi D}^{\{{Rr}\}\{(2)(1)\}} v_R^2-g^2 \kappa _2 \kappa _1 \mathcal{C}_{\phi D}^{{Rr}(2)(1)} v_R^2 \nonumber \\
	&&-4 g^2 \kappa _2 \kappa _1 \Theta _{W_{{LL}}}+2 g^2 \kappa _1^2 \Theta _{W_{{LR}}}+2 g^2 \kappa _2^2 \Theta _{W_{{LR}}}+2 g^2 \kappa _1^2 \Theta _{W_{{RL}}}+2 g^2 \kappa _2^2 \Theta _{W_{{RL}}}-4 g^2 \kappa _2 \kappa _1 \Theta _{W_{{RR}}}\nonumber \\
	&&+2 g^2 \kappa _2 \kappa _1^3 \mathcal{C}_{\phi D}^{1(1)1(1)}+2 g^2 \kappa _2^3 \kappa _1 \mathcal{C}_{\phi D}^{1(1)1(1)}-2 g^2 \kappa _2 \kappa _1^3 \mathcal{C}_{\phi D}^{(11)(22)}-4 g^2 \kappa _2^2 \kappa _1^2 \mathcal{C}_{\phi D}^{(11)(22)}-2 g^2 \kappa _2^3 \kappa _1 \mathcal{C}_{\phi D}^{(11)(22)}\nonumber \\
	&&+2 g^2 \kappa _2 \kappa _1^3 \mathcal{C}_{\phi D}^{2(1)1(1)}+2 g^2 \kappa _2^3 \kappa _1 \mathcal{C}_{\phi D}^{2(1)1(1)}-2 g^2 \kappa _2 \kappa _1^3 \mathcal{C}_{\phi D}^{\{(2)(1)\}\{21\}}-2 g^2 \kappa _2^3 \kappa _1 \mathcal{C}_{\phi D}^{\{(2)(1)\}\{21\}}\nonumber \\
	&&-g^2 \kappa _2 \kappa _1^3 \mathcal{C}_{\phi D}^{(2)(1)21}-g^2 \kappa _2^3 \kappa _1 \mathcal{C}_{\phi D}^{(2)(1)21}-4 g^2 \kappa _2 \kappa _1^3 \mathcal{C}_{\phi D}^{2(1)2(1)}-4 g^2 \kappa _2^3 \kappa _1 \mathcal{C}_{\phi D}^{2(1)2(1)}\Big],
		\end{eqnarray}}
\small{\begin{eqnarray}\label{eq:MLRSM-Pi-function-II}
	\Pi_{W_{3L}W_{3L}}(p^2)&=&p^2\left(1+2 \frac{ \Theta _{{3L3L}}}{\Lambda ^2}\right)+\frac{1}{8} \left(2 g^2 \kappa _1^2+2 g^2 \kappa _2^2+8 g^2 v_L^2\right)\nonumber \\
	&&+\frac{1}{8\Lambda ^2} \Big[4 g^2 \kappa _1^2 \Theta _{{3L3L}}+4 g^2 \kappa _2^2 \Theta _{{3L3L}}+16 g^2 \Theta _{{3L3L}} v_L^2-4 g^2 \kappa _1^2 \Theta _{{3L3R}}-4 g^2 \kappa _2^2 \Theta _{{3L3R}}-16 g \Theta _{{3LB}} \tilde{g} v_L^2\nonumber \\
	&&+2 g^2 \kappa _2^2 \mathcal{C}_{\phi D}^{(2){rL}(1)} v_L v_R+4 g^2 \mathcal{C}_{\phi D}^{(L){rR}(l)} v_L^2 v_R^2+4 g^2 \kappa _1^2 \mathcal{C}_{\phi D}^{\{21\}\{(L)(l)\}} v_L^2+4 g^2 \kappa _2^2 \mathcal{C}_{\phi D}^{\{21\}\{(L)(l)\}} v_L^2\nonumber \\
	&&+4 g^2 \kappa _1^2 \mathcal{C}_{\phi D}^{21(L)(l)} v_L^2+8 g^2 \kappa _2^2 \mathcal{C}_{\phi D}^{(L)11(l)} v_L^2+4 g^2 \kappa _1^2 \mathcal{C}_{\phi D}^{L(l)1(1)} v_L^2+g^2 \kappa _1^2 \mathcal{C}_{\phi D}^{\{{Ll}\}\{(2)(1)\}} v_L^2\nonumber \\
	&&+g^2 \kappa _2^2 \mathcal{C}_{\phi D}^{\{{Ll}\}\{(2)(1)\}} v_L^2+g^2 \kappa _1^2 \mathcal{C}_{\phi D}^{{Ll}(2)(1)} v_L^2+4 g^2 v_L^4 (\mathcal{C}_{\phi D}^{\{{Ll}\}\{(L)(l)\}}+4 \mathcal{C}_{\phi D}^{L(l)L(l)})+g^2 \kappa _1^2 \mathcal{C}_{\phi D}^{\{{Rr}\}\{(2)(1)\}} v_R^2\nonumber \\
	&&+g^2 \kappa _2^2 \mathcal{C}_{\phi D}^{\{{Rr}\}\{(2)(1)\}} v_R^2+g^2 \kappa _1^2 \mathcal{C}_{\phi D}^{{Rr}(2)(1)} v_R^2-2 g^2 \kappa _1^4 \mathcal{C}_{\phi D}^{1(1)1(1)}-2 g^2 \kappa _2^4 \mathcal{C}_{\phi D}^{1(1)1(1)}+4 g^2 \kappa _1^4 \mathcal{C}_{\phi D}^{(11)(22)}\nonumber \\
	&&+4 g^2 \kappa _2^4 \mathcal{C}_{\phi D}^{(11)(22)}-2 g^2 \kappa _1^4 \mathcal{C}_{\phi D}^{2(1)1(1)}-2 g^2 \kappa _2^4 \mathcal{C}_{\phi D}^{2(1)1(1)}+g^2 \kappa _1^4 \mathcal{C}_{\phi D}^{\{(2)(1)\}\{21\}}+g^2 \kappa _2^4 \mathcal{C}_{\phi D}^{\{(2)(1)\}\{21\}}\nonumber \\
	&&+2 g^2 \kappa _1^2 \kappa _2^2 \mathcal{C}_{\phi D}^{\{(2)(1)\}\{21\}}+g^2 \kappa _1^4 \mathcal{C}_{\phi D}^{(2)(1)21}+g^2 \kappa _2^4 \mathcal{C}_{\phi D}^{(2)(1)21}+4 g^2 \kappa _1^4 \mathcal{C}_{\phi D}^{2(1)2(1)}+4 g^2 \kappa _2^4 \mathcal{C}_{\phi D}^{2(1)2(1)}\Big],\\
	\Pi_{W_{3L}W_{3R}}(p^2)&=&4 p^2 \frac{\Theta _{{3L3R}}}{\Lambda ^2}+\frac{1}{4} \left(-2 g^2 \kappa _1^2-2 g^2 \kappa _2^2\right)\nonumber \\
	&&+\frac{1}{4\Lambda ^2} \Big[-2 g^2 \kappa _1^2 \Theta _{{3L3L}}-2 g^2 \kappa _2^2 \Theta _{{3L3L}}+g v_L^2 \big(8 g \Theta _{{3L3R}}-8 \Theta _{{3RB}} \tilde{g}+4 g \mathcal{C}_{\phi D}^{(L){rL}(r)} v_R^2-2 g \kappa _1^2 \mathcal{C}_{\phi D}^{L(l)1(1)}\nonumber \\
	&&-g \kappa _1^2 \mathcal{C}_{\phi D}^{\{{Ll}\}\{(2)(1)\}}-g \kappa _2^2 \mathcal{C}_{\phi D}^{\{{Ll}\}\{(2)(1)\}}-g \kappa _1^2 \mathcal{C}_{\phi D}^{{Ll}(2)(1)}\big)+4 g^2 \kappa _1^2 \Theta _{{3L3R}}+4 g^2 \kappa _2^2 \Theta _{{3L3R}}+8 g^2 \Theta _{{3L3R}} v_R^2\nonumber \\
	&&-8 g \Theta _{{3LB}} \tilde{g} v_R^2-2 g^2 \kappa _1^2 \Theta _{{3R3R}}-2 g^2 \kappa _2^2 \Theta _{{3R3R}}+2 g^2 \kappa _2^2 v_L v_R (2 (\mathcal{C}_{\phi D}^{(L)11(r)}+\mathcal{C}_{\phi D}^{(L)12(r)}+\mathcal{C}_{\phi D}^{(R)11(l)})\nonumber \\
	&&-\mathcal{C}_{\phi D}^{(2){rL}(1)})+g^2 \kappa _1^2 v_R^2 (2 \mathcal{C}_{\phi D}^{R(r)1(1)}-\mathcal{C}_{\phi D}^{\{{Rr}\}\{(2)(1)\}}-\mathcal{C}_{\phi D}^{{Rr}(2)(1)})-g^2 \kappa _2^2 \mathcal{C}_{\phi D}^{\{{Rr}\}\{(2)(1)\}} v_R^2\nonumber \\
	&&+g^2 \kappa _1^4 (2 \mathcal{C}_{\phi D}^{1(1)1(1)}-4 \mathcal{C}_{\phi D}^{(11)(22)}+2 \mathcal{C}_{\phi D}^{2(1)1(1)}-\mathcal{C}_{\phi D}^{\{(2)(1)\}\{21\}}-\mathcal{C}_{\phi D}^{(2)(1)21}-4 \mathcal{C}_{\phi D}^{2(1)2(1)})\nonumber \\
	&&+g^2 \kappa _2^4 (2 \mathcal{C}_{\phi D}^{1(1)1(1)}-4 \mathcal{C}_{\phi D}^{(11)(22)}+2 \mathcal{C}_{\phi D}^{2(1)1(1)}-\mathcal{C}_{\phi D}^{\{(2)(1)\}\{21\}}-\mathcal{C}_{\phi D}^{(2)(1)21}-4 \mathcal{C}_{\phi D}^{2(1)2(1)})\nonumber \\
	&&-2 g^2 \kappa _2^2 \kappa _1^2 \mathcal{C}_{\phi D}^{\{(2)(1)\}\{21\}}\Big],\\
	\Pi_{W_{3R}W_{3R}}(p^2)&=&p^2\left(1+2 \frac{\Theta _{{3R3R}}}{\Lambda ^2}\right)+\frac{1}{8} \left(2 g^2 \kappa _1^2+2 g^2 \kappa _2^2+8 g^2 v_R^2\right)\nonumber \\
	&&+\frac{1}{8\Lambda ^2} \Big[-4 g^2 \kappa _1^2 \Theta _{{3L3R}}-4 g^2 \kappa _2^2 \Theta _{{3L3R}}+4 g^2 \kappa _1^2 \Theta _{{3R3R}}+4 g^2 \kappa _2^2 \Theta _{{3R3R}}+16 g^2 \Theta _{{3R3R}} v_R^2\nonumber \\
	&&-16 g \Theta _{{3RB}} \tilde{g} v_R^2+2 g^2 \kappa _2^2 \mathcal{C}_{\phi D}^{(2){rL}(1)} v_L v_R+4 g^2 \mathcal{C}_{\phi D}^{(R){lL}(r)} v_L^2 v_R^2+g^2 \kappa _1^2 \mathcal{C}_{\phi D}^{\{{Ll}\}\{(2)(1)\}} v_L^2\nonumber \\
	&&+g^2 \kappa _2^2 \mathcal{C}_{\phi D}^{\{{Ll}\}\{(2)(1)\}} v_L^2+g^2 \kappa _1^2 \mathcal{C}_{\phi D}^{{Ll}(2)(1)} v_L^2+4 g^2 \kappa _1^2 \mathcal{C}_{\phi D}^{\{21\}\{(R)(r)\}} v_R^2+4 g^2 \kappa _2^2 \mathcal{C}_{\phi D}^{\{21\}\{(R)(r)\}} v_R^2\nonumber \\
	&&+4 g^2 \kappa _1^2 \mathcal{C}_{\phi D}^{21(R)(r)} v_R^2+8 g^2 \kappa _2^2 \mathcal{C}_{\phi D}^{(R)11(r)} v_R^2-4 g^2 \kappa _1^2 \mathcal{C}_{\phi D}^{R(r)1(1)} v_R^2+g^2 \kappa _1^2 \mathcal{C}_{\phi D}^{\{{Rr}\}\{(2)(1)\}} v_R^2\nonumber \\
	&&+g^2 \kappa _2^2 \mathcal{C}_{\phi D}^{\{{Rr}\}\{(2)(1)\}} v_R^2+g^2 \kappa _1^2 \mathcal{C}_{\phi D}^{{Rr}(2)(1)} v_R^2+4 g^2 v_R^4 (\mathcal{C}_{\phi D}^{\{{Rr}\}\{(R)(r)\}}+4 \mathcal{C}_{\phi D}^{R(r)R(r)})-2 g^2 \kappa _1^4 \mathcal{C}_{\phi D}^{1(1)1(1)}\nonumber \\
	&&-2 g^2 \kappa _2^4 \mathcal{C}_{\phi D}^{1(1)1(1)}+4 g^2 \kappa _1^4 \mathcal{C}_{\phi D}^{(11)(22)}+4 g^2 \kappa _2^4 \mathcal{C}_{\phi D}^{(11)(22)}-2 g^2 \kappa _1^4 \mathcal{C}_{\phi D}^{2(1)1(1)}-2 g^2 \kappa _2^4 \mathcal{C}_{\phi D}^{2(1)1(1)}\nonumber \\
	&&+g^2 \kappa _1^4 \mathcal{C}_{\phi D}^{\{(2)(1)\}\{21\}}+g^2 \kappa _2^4 \mathcal{C}_{\phi D}^{\{(2)(1)\}\{21\}}+2 g^2 \kappa _1^2 \kappa _2^2 \mathcal{C}_{\phi D}^{\{(2)(1)\}\{21\}}+g^2 \kappa _1^4 \mathcal{C}_{\phi D}^{(2)(1)21}+g^2 \kappa _2^4 \mathcal{C}_{\phi D}^{(2)(1)21}\nonumber \\
	&&+4 g^2 \kappa _1^4 \mathcal{C}_{\phi D}^{2(1)2(1)}+4 g^2 \kappa _2^4 \mathcal{C}_{\phi D}^{2(1)2(1)}\Big],\\
		\Pi_{W_{3L}B}(p^2)&=&p^2 \frac{\Theta _{{3LB}}}{\Lambda ^2}-g \tilde{g} v_L^2\nonumber \\
	&&+\frac{1}{4\Lambda ^2} \Big[-4 g \Theta _{{3L3L}} \tilde{g} v_L^2-g \tilde{g} v_R^2 \left(4 \Theta _{{3L3R}}+\kappa _1^2 \mathcal{C}_{\phi D}^{R(r)1(1)}\right)+g^2 \kappa _1^2 \Theta _{{3LB}}+g^2 \kappa _2^2 \Theta _{{3LB}}\nonumber \\
	&&+4 \Theta _{{3LB}} \left(\tilde{g}\right)^2 v_R^2+4 \Theta _{{3LB}} \left(g^2+\left(\tilde{g}\right)^2\right) v_L^2-g^2 \left(\kappa _1^2+\kappa _2^2\right) \Theta _{{3RB}}-4 g \Theta _{{BB}} \tilde{g} v_L^2\nonumber \\
	&&-2 g \kappa _2^2 \tilde{g} v_L v_R (\mathcal{C}_{\phi D}^{(L)11(r)}+\mathcal{C}_{\phi D}^{(L)12(r)}+\mathcal{C}_{\phi D}^{(R)11(l)})-2 g \tilde{g} v_L^2 v_R^2 (\mathcal{C}_{\phi D}^{(L){rL}(r)}+\mathcal{C}_{\phi D}^{(L){rR}(l)})\nonumber \\
	&&-2 g \kappa _1^2 \mathcal{C}_{\phi D}^{\{21\}\{(L)(l)\}} \tilde{g} v_L^2-2 g \kappa _2^2 \mathcal{C}_{\phi D}^{\{21\}\{(L)(l)\}} \tilde{g} v_L^2-2 g \kappa _1^2 \mathcal{C}_{\phi D}^{21(L)(l)} \tilde{g} v_L^2-4 g \kappa _2^2 \mathcal{C}_{\phi D}^{(L)11(l)} \tilde{g} v_L^2\nonumber \\
	&&-g \kappa _1^2 \mathcal{C}_{\phi D}^{L(l)1(1)} \tilde{g} v_L^2-2 g \tilde{g} v_L^4 (\mathcal{C}_{\phi D}^{\{{Ll}\}\{(L)(l)\}}+4 \mathcal{C}_{\phi D}^{L(l)L(l)})\Big],
		\end{eqnarray}}
\small{\begin{eqnarray}\label{eq:MLRSM-Pi-function-III}
	\Pi_{W_{3R}B}(p^2)&=&p^2 \frac{\Theta _{{3RB}}}{\Lambda ^2}-g \tilde{g} v_R^2\nonumber \\
	&&+\frac{1}{4\Lambda ^2} \Big[-g \tilde{g} v_L^2 \left(4 \Theta _{{3L3R}}-\kappa _1^2 \mathcal{C}_{\phi D}^{L(l)1(1)}\right)-g^2 \left(\kappa _1^2+\kappa _2^2\right) \Theta _{{3LB}}-4 g \Theta _{{3R3R}} \tilde{g} v_R^2+g^2 \kappa _1^2 \Theta _{{3RB}}\nonumber \\
	&&+g^2 \kappa _2^2 \Theta _{{3RB}}+4 \Theta _{{3RB}} \left(\tilde{g}\right)^2 v_L^2+4 \Theta _{{3RB}} \left(g^2+\left(\tilde{g}\right)^2\right) v_R^2-4 g \Theta _{{BB}} \tilde{g} v_R^2\nonumber \\
	&&-2 g \kappa _2^2 \tilde{g} v_L v_R (\mathcal{C}_{\phi D}^{(L)11(r)}+\mathcal{C}_{\phi D}^{(L)12(r)\}}+\mathcal{C}_{\phi D}^{(R)11(l)})-2 g \tilde{g} v_L^2 v_R^2 (\mathcal{C}_{\phi D}^{(L){rL}(r)}+\mathcal{C}_{\phi D}^{(R){lL}(r)})\nonumber \\
	&&-2 g \kappa _1^2 \mathcal{C}_{\phi D}^{\{21\}\{(R)(r)\}} \tilde{g} v_R^2-2 g \kappa _2^2 \mathcal{C}_{\phi D}^{\{21\}\{(R)(r)\}} \tilde{g} v_R^2-2 g \kappa _1^2 \mathcal{C}_{\phi D}^{21(R)(r)} \tilde{g} v_R^2-4 g \kappa _2^2 \mathcal{C}_{\phi D}^{(R)11(r)} \tilde{g} v_R^2\nonumber \\
	&&+g \kappa _1^2 \mathcal{C}_{\phi D}^{R(r)1(1)} \tilde{g} v_R^2-2 g \tilde{g} v_R^4 (\mathcal{C}_{\phi D}^{\{{Rr}\}\{(R)(r)\}}+4 \mathcal{C}_{\phi D}^{R(r)R(r)})\Big],\\  \nonumber \\
	\Pi_{BB}(p^2)&=&\frac{1}{2} \tilde{g} \left(2 \tilde{g} v_L^2+2 \tilde{g} v_R^2\right)+\frac{\tilde{g} }{2 \Lambda ^2}\Big[-4 g \Theta _{{3LB}} v_L^2-4 g \Theta _{{3RB}} v_R^2+4 \Theta _{{BB}} \tilde{g} v_L^2+4 \Theta _{{BB}} \tilde{g} v_R^2\nonumber \\
	&&+2 \kappa _2^2 \tilde{g} v_L v_R (\mathcal{C}_{\phi D}^{(L)11(r)}+\mathcal{C}_{\phi D}^{(L)12(r)}+\mathcal{C}_{\phi D}^{(R)11(l)})+\tilde{g} v_L^2 v_R^2 (2 \mathcal{C}_{\phi D}^{(L){rL}(r)}+\mathcal{C}_{\phi D}^{(L){rR}(l)}+\mathcal{C}_{\phi D}^{(R){lL}(r)})\nonumber \\
	&&+\kappa _1^2 \mathcal{C}_{\phi D}^{\{21\}\{(L)(l)\}} \tilde{g} v_L^2+\kappa _2^2 \mathcal{C}_{\phi D}^{\{21\}\{(L)(l)\}} \tilde{g} v_L^2+\kappa _1^2 \mathcal{C}_{\phi D}^{21(L)(l)} \tilde{g} v_L^2+2 \kappa _2^2 \mathcal{C}_{\phi D}^{(L)11(l)} \tilde{g} v_L^2+\tilde{g} v_L^4 (\mathcal{C}_{\phi D}^{\{{Ll}\}\{(L)(l)\}}\nonumber \\
	&&+4 \mathcal{C}_{\phi D}^{L(l)L(l)})+\kappa _1^2 \mathcal{C}_{\phi D}^{\{21\}\{(R)(r)\}} \tilde{g} v_R^2+\kappa _2^2 \mathcal{C}_{\phi D}^{\{21\}\{(R)(r)\}} \tilde{g} v_R^2+\kappa _1^2 \mathcal{C}_{\phi D}^{21(R)(r)} \tilde{g} v_R^2+2 \kappa _2^2 \mathcal{C}_{\phi D}^{(R)11(r)} \tilde{g} v_R^2\nonumber \\
	&&+\tilde{g} v_R^4 (\mathcal{C}_{\phi D}^{\{{Rr}\}\{(R)(r)\}}+4 \mathcal{C}_{\phi D}^{R(r)R(r)})\Big].
	\end{eqnarray}}

Using the rotation matrices $\tilde{\mathcal{R}}_{a,b}$, see Eq.~\ref{eq:MLRSM-cGB-rot-matrix}, we can define these $\Pi_{V_iV_j}(p^ 2)$ in the physical basis as follows:
\small{\begin{eqnarray}\label{eq:MLRSM-Pi-function-physical-basis}
	\Pi_{W_{1}W_{1}}&=& \Pi_{W_{L}W_{L}} (\tilde{\mathcal{R}}_{a}^{11})^{2}+2 \Pi_{W_{L}W_{R}} \tilde{\mathcal{R}}_{a}^{11} \tilde{\mathcal{R}}_{a}^{21}+\Pi_{W_{R}W_{R}} (\tilde{\mathcal{R}}_{a}^{21})^{2},\\
	\Pi_{W_{1}W_{2}}&=& \Pi_{W_{L}W_{L}} \tilde{\mathcal{R}}_{a}^{11} \tilde{\mathcal{R}}^{a}_{12}+\Pi_{W_{L}W_{R}} (\tilde{\mathcal{R}}^{a}_{11} \tilde{\mathcal{R}}^{a}_{22}+\tilde{\mathcal{R}}^{a}_{12} \tilde{\mathcal{R}}^{a}_{21})+\Pi_{W_{R}W_{R}} \tilde{\mathcal{R}}^{a}_{21} \tilde{\mathcal{R}}^{a}_{22},\nonumber\\
	\Pi_{W_{2}W_{2}}&=& \Pi_{W_{L}W_{L}} (\tilde{\mathcal{R}}^{a}_{12})^{2}+2 \Pi_{W_{L}W_{R}} \tilde{\mathcal{R}}^{a}_{11} \tilde{\mathcal{R}}^{a}_{22}+\Pi_{W_{R}W_{R}} (\tilde{\mathcal{R}}^{a}_{22})^{2},\nonumber\\  \nonumber \\
	\Pi_{Z_{1}Z_{1}}&=& \Pi_{BB} (\tilde{\mathcal{R}}^{b}_{31})^{2}+2 \Pi_{W_{3L}B} \tilde{\mathcal{R}}^{b}_{11} \tilde{\mathcal{R}}^{b}_{31}+\Pi_{W_{3L}W_{3L}} (\tilde{\mathcal{R}}^{b}_{11})^{2}+2 \Pi_{W_{3L}W_{3R}} \tilde{\mathcal{R}}^{b}_{11} \tilde{\mathcal{R}}^{b}_{21}+2 \Pi_{W_{3R}B} \tilde{\mathcal{R}}^{b}_{21} \tilde{\mathcal{R}}^{b}_{31}\nonumber \\
	&&+\Pi_{W_{3R}W_{3R}} (\tilde{\mathcal{R}}^{b}_{21})^{2}, \nonumber\\
	\Pi_{Z_{2}Z_{2}}&=& \Pi_{BB} (\tilde{\mathcal{R}}^{b}_{32})^{2}+2 \Pi_{W_{3L}B} \tilde{\mathcal{R}}^{b}_{12} \tilde{\mathcal{R}}^{b}_{32}+\Pi_{W_{3L}W_{3L}} (\tilde{\mathcal{R}}^{b}_{12})^{2}+2 \Pi_{W_{3L}W_{3R}} \tilde{\mathcal{R}}^{b}_{12} \tilde{\mathcal{R}}^{b}_{22}+2 \Pi_{W_{3R}B} \tilde{\mathcal{R}}^{b}_{22} \tilde{\mathcal{R}}^{b}_{32}\nonumber \\
	&&+\Pi_{W_{3R}W_{3R}} (\tilde{\mathcal{R}}^{b}_{22})^{2},\nonumber\\
	\Pi_{Z_{1}Z_{2}}&=& \Pi_{BB} \tilde{\mathcal{R}}^{b}_{31} \tilde{\mathcal{R}}^{b}_{32}+\Pi_{W_{3L}B} (\tilde{\mathcal{R}}^{b}_{11} \tilde{\mathcal{R}}^{b}_{32}+\tilde{\mathcal{R}}^{b}_{12} \tilde{\mathcal{R}}^{b}_{31})+\Pi_{W_{3L}W_{3L}} \tilde{\mathcal{R}}^{b}_{11} \tilde{\mathcal{R}}^{b}_{12}+\Pi{W_{3L}W_{3R}} (\tilde{\mathcal{R}}^{b}_{11} \tilde{\mathcal{R}}^{b}_{22}+\tilde{\mathcal{R}}^{b}_{12} \tilde{\mathcal{R}}^{b}_{21})\nonumber \\
	&&+\Pi_{W_{3R}B} (\tilde{\mathcal{R}}^{b}_{21} \tilde{\mathcal{R}}^{b}_{32}+\tilde{\mathcal{R}}^{b}_{22} \tilde{\mathcal{R}}^{b}_{31})+\Pi_{W_{3R}W_{3R}} \tilde{\mathcal{R}}^{b}_{21} \tilde{\mathcal{R}}^{b}_{22},\nonumber\\  \nonumber \\
	\Pi_{\gamma \gamma}&=& \Pi_{BB} (\tilde{\mathcal{R}}^{b}_{33})^{2}+2 \Pi_{W_{3L}B} \tilde{\mathcal{R}}^{b}_{13} \tilde{\mathcal{R}}^{b}_{33}+\Pi_{W_{3L}W_{3L}} (\tilde{\mathcal{R}}^{b}_{13})^{2}+2 \Pi_{W_{3L}W_{3R}} \tilde{\mathcal{R}}^{b}_{13} \tilde{\mathcal{R}}^{b}_{23}+2 \Pi_{W_{3R}B} \tilde{\mathcal{R}}^{b}_{23} \tilde{\mathcal{R}}^{b}_{33}\nonumber \\
	&&+\Pi_{W_{3R}W_{3R}} (\tilde{\mathcal{R}}^{b}_{23})^{2},\nonumber\\
	\Pi_{Z_{1}\gamma}&=& \Pi_{BB} \tilde{\mathcal{R}}^{b}_{31} \tilde{\mathcal{R}}^{b}_{33}+\Pi_{W_{3L}B} (\tilde{\mathcal{R}}^{b}_{11} \tilde{\mathcal{R}}^{b}_{33}+\tilde{\mathcal{R}}^{b}_{13} \tilde{\mathcal{R}}^{b}_{31})+\Pi_{W_{3L}W_{3L}} \tilde{\mathcal{R}}^{b}_{11} \tilde{\mathcal{R}}^{b}_{13}+\Pi_{W_{3L}W_{3R}} (\tilde{\mathcal{R}}^{b}_{11} \tilde{\mathcal{R}}^{b}_{23}+\tilde{\mathcal{R}}^{b}_{13} \tilde{\mathcal{R}}^{b}_{21})\nonumber \\
	&&+\Pi_{W_{3R}B} (\tilde{\mathcal{R}}^{b}_{21} \tilde{\mathcal{R}}^{b}_{33}+\tilde{\mathcal{R}}^{b}_{23} \tilde{\mathcal{R}}^{b}_{31})+\Pi_{W_{3R}W_{3R}} \tilde{\mathcal{R}}^{b}_{21} \tilde{\mathcal{R}}^{b}_{23},\nonumber\\
	\Pi_{Z_{2}\gamma}&=& \Pi_{BB} \tilde{\mathcal{R}}^{b}_{32} \tilde{\mathcal{R}}^{b}_{33}+\Pi_{W_{3L}B} (\tilde{\mathcal{R}}^{b}_{12} \tilde{\mathcal{R}}^{b}_{33}+\tilde{\mathcal{R}}^{b}_{13} \tilde{\mathcal{R}}^{b}_{32})+\Pi_{W_{3L}W_{3L}} \tilde{\mathcal{R}}^{b}_{12} \tilde{\mathcal{R}}^{b}_{13}+\Pi_{W_{3L}W_{3R}} (\tilde{\mathcal{R}}^{b}_{12} \tilde{\mathcal{R}}^{b}_{23}+\tilde{\mathcal{R}}^{b}_{13} \tilde{\mathcal{R}}^{b}_{22})\nonumber \\
	&&+\Pi_{W_{3R}B} (\tilde{\mathcal{R}}^{b}_{22} \tilde{\mathcal{R}}^{b}_{33}+\tilde{\mathcal{R}}^{b}_{23} \tilde{\mathcal{R}}^{b}_{32})+\Pi_{W_{3R}W_{3R}} \tilde{\mathcal{R}}^{b}_{22} \tilde{\mathcal{R}}^{b}_{23}.\nonumber
	\end{eqnarray}}
\clearpage
Now the oblique parameters $(S,T,U)$ for MLRSM-EFT scenario can be constructed using Eqs.~\ref{eq:oblique-S}, \ref{eq:oblique-T}, \ref{eq:oblique-U}. We need to keep in mind that in order to define the $S,T,U$ parameters, we should use only SM like gauge bosons $W_1^{\pm}, Z_1,\gamma$.

%% file: Magnetic-moment.tex
\subsection{Magnetic moments of charged fermions}

The charged fermions can couple to the gauge field strength tensors through effective $\psi^2 \phi X$ class of dimension-6 operators, see Fig.~\ref{fig:mag-moment}(a). Here, we are mostly interested in the coupling between the charged leptons and quarks with the electromagnetic field strength tensor ($\bar{f_L} \sigma_{\mu \nu} f_R A^{\mu \nu}$) which lead to their respective magnetic moments ($f$). There are contributions to the anomalous magnetic moments of the charged leptons ($\tau,\mu,e$) due to the modified interactions in the presence of charged scalars, see Figs.~\ref{fig:mag-moment}(b), (c).



\section*{2HDM}
The magnetic moment interactions between the charged leptons, and quarks with the elecromagnetic field tensor ($A_{\mu \nu})$ in presence of $\psi^2 \phi X$ operators (see Table~\ref{tab:2HDM-psi2phiX}) within 2HDM-EFT framework are given as:

\small{\begin{eqnarray}
		\bar{l}_L\sigma ^{\mu \nu }l_R~A_{\mu \nu }&:&\frac{1}{\sqrt{2} \Lambda ^2}\left[\left(\textcolor{purple}{\mathcal{C}_{eB}^1 v_1}+\mathcal{C}_{eB}^2 v_2\right) \cos \theta_{\text{w}}-\left(\textcolor{purple}{\mathcal{C}_{eW}^1 v_1}+\mathcal{C}_{eW}^2 v_2\right) \sin \theta_{\text{w}}\right],\nonumber \\
		\bar{u}_L\sigma ^{\mu \nu }u_RA_{\mu \nu } &:& \frac{1}{\sqrt{2} \Lambda ^2}\left[\left(\textcolor{purple}{\mathcal{C}_{uB}^{\tilde{1}} v_1} +\mathcal{C}_{uB}^{\tilde{2}} v_2\right) \cos \theta_{\text{w}}+\left(\textcolor{purple}{\mathcal{C}_{uW}^{\tilde{1}} v_1}+\mathcal{C}_{uW}^{\tilde{2}} v_2\right) \sin \theta_{\text{w}}\right],\nonumber \\
		\bar{d}_L\sigma ^{\mu \nu }d_RA_{\mu \nu }&:&\frac{1}{\sqrt{2} \Lambda ^2}\left[\left(\textcolor{purple}{\mathcal{C}_{dB}^1 v_1}+\mathcal{C}_{dB}^2 v_2\right) \cos \theta_{\text{w}}-\left(\textcolor{purple}{\mathcal{C}_{dW}^1 v_1}+\mathcal{C}_{dW}^2 v_2\right) \sin \theta_{\text{w}}\right].\nonumber
	\end{eqnarray}
	
	\begin{figure}
		\centering
		\qquad
		\subfloat[$\psi^2 \phi X $: effective magnetic moment vertex]{{\includegraphics[trim={.02cm 0.2cm .01cm 0},scale=0.53]{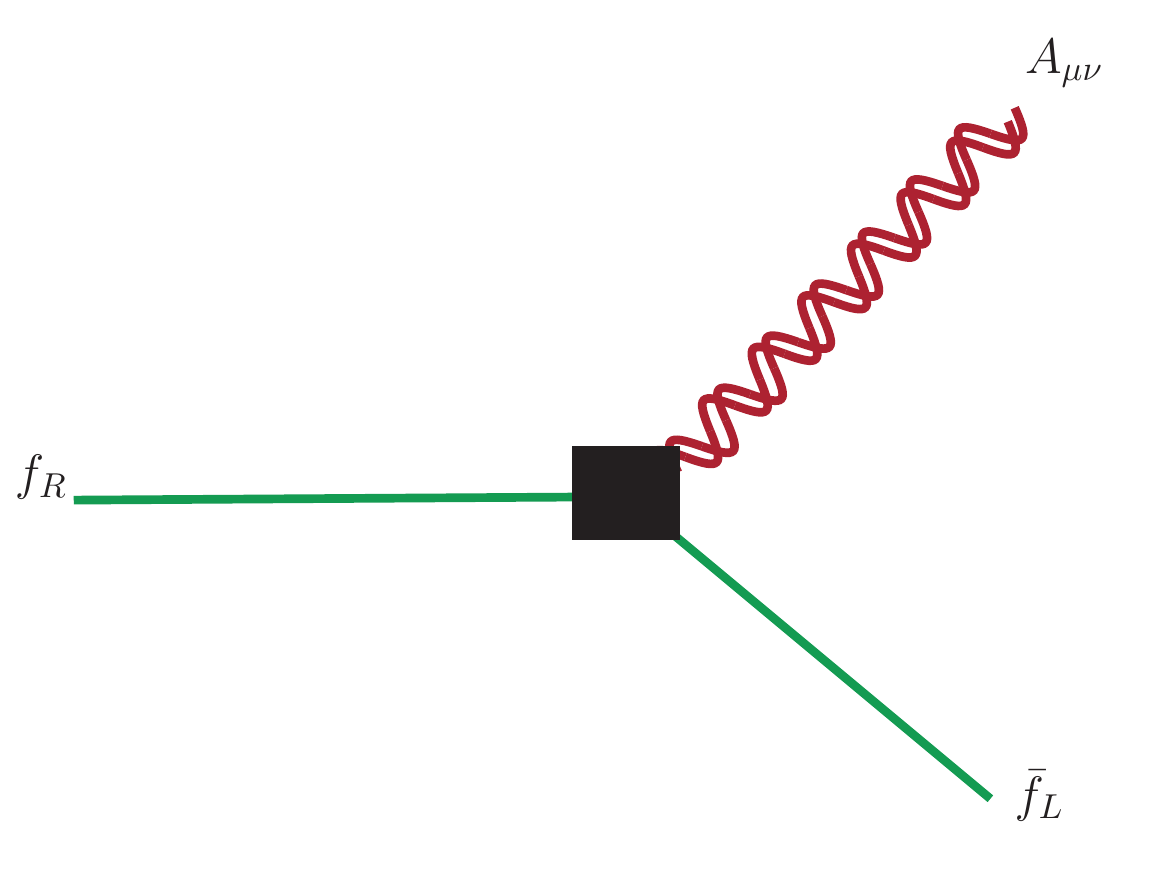} }}\\
		\qquad
		\subfloat[$l_i \to l_j \gamma$: singly and doubly charged scalars]{{\includegraphics[trim={1.0cm 5.0 1.5cm 0},scale=0.5]{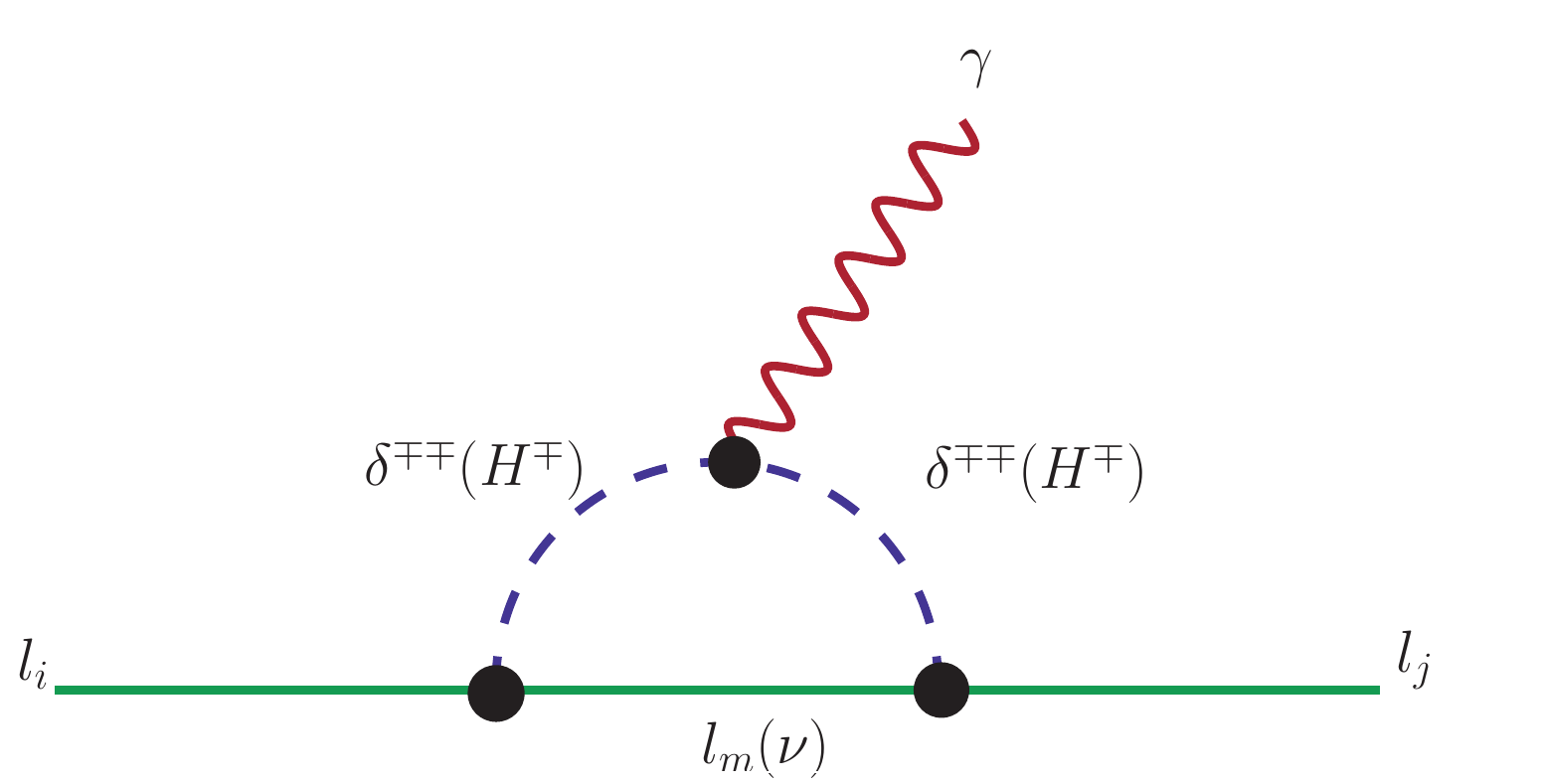} }}
		\qquad
		\subfloat[$l_i \to l_j \gamma$: doubly charged scalars]{{\includegraphics[trim={1.0cm 5.0 1.5cm 0},scale=0.5]{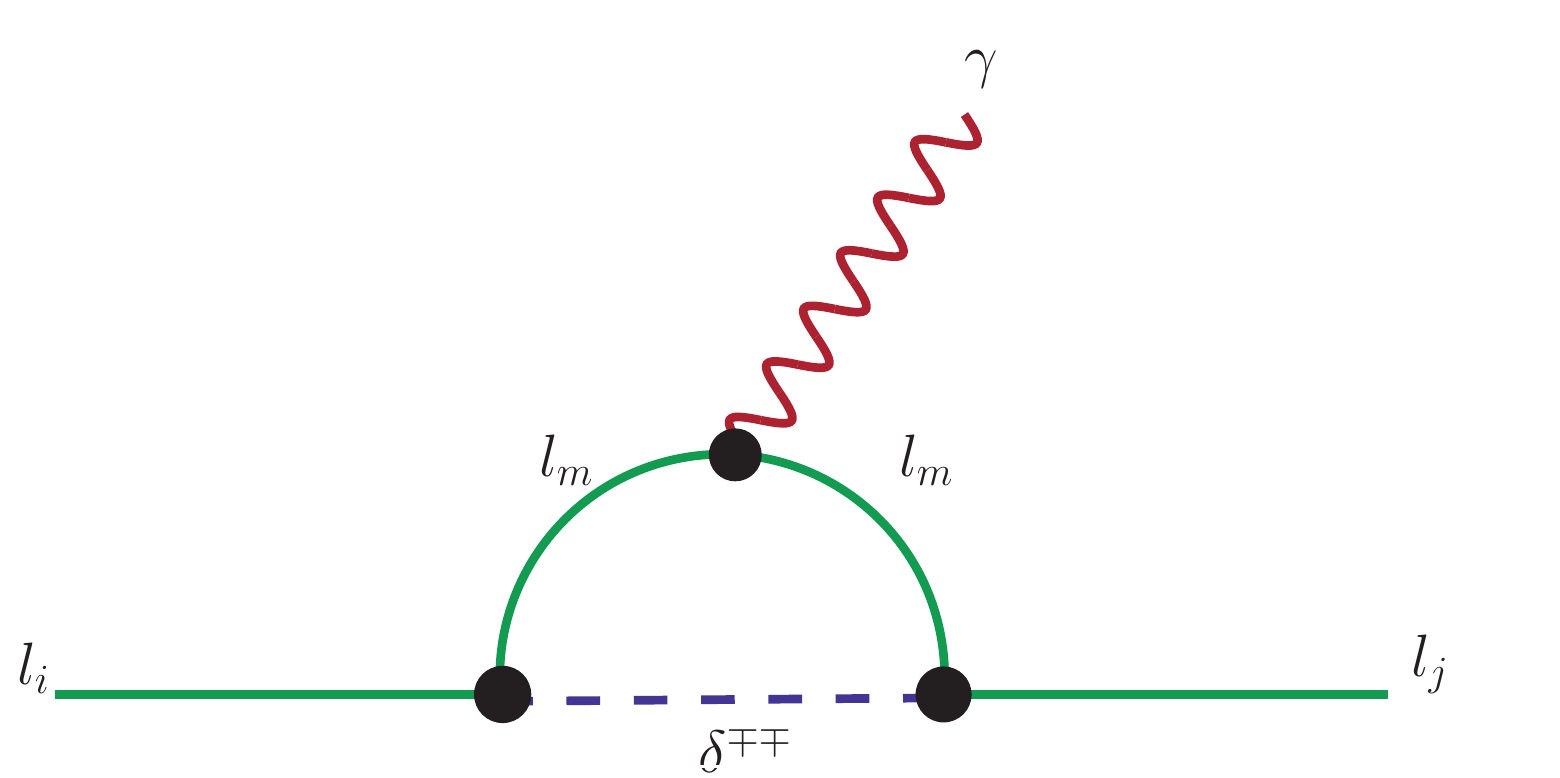} }}
		\caption{Feynman diagram representing the magnetic moment interactions of charged fermions: (a) $\psi^2 \phi X \Rightarrow$ class of effective operators [{\large $\blacksquare$}] leading to magnetic moment interactions $\bar{f_L} \sigma_{\mu \nu} f_R A^{\mu \nu}$ with $f$ representing charged fermions. (b) and (c) modified vertices [{\Large $\bullet$}] lead to radiative decay of charged leptons ($l_i \to l_j \gamma$). For $i=j=\tau,\mu,e$, both diagrams contribute to the magnetic moment of charged leptons.}
		\label{fig:mag-moment}
	\end{figure}
	
	The magnetic moments of the charged leptons get additional contributions due to their modified couplings with the singly charged scalars ($H^{\pm}$), see Table~\ref{tab:2HDM-scs-f-f} and Fig.~\ref{fig:mag-moment}(b). There will be similar contributions to the quark magnetic moments which are not explicitly mentioned here.
	
	\begin{table}[h!]
		{ \scriptsize
			\begin{center}
				\renewcommand{\arraystretch}{2.5}
				\begin{tabular}{|*{2}{l|}}
					\hline 
					{\bf Vertex} & {\bf2HDM: Effective vertex ($\bar{\nu}\,e^{-}\,H^+$) factor}\\
					\hline
					$\left[\bar{\nu}~e^{-}~H^+\right]_{lm}$  &  $\left[  y_2^e s_{_{\bar{\beta }_{+}}}- y_1^e c_{_{\bar{\beta }_{+}}}+\frac{1}{2 \Lambda ^2}\left(-A_{+}^{11} y_1^e s_{_{{\beta }_{+}}}-A_{+}^{22} y_2^e s_{_{{\beta }_{+}}}+A_{+}^{11} y_1^e c_{_{{\beta }_{+}}}+A_{+}^{12} y_2^e c_{_{{\beta }_{+}}}\right.\right.$\\
					&$\left.\left.+v_1^2 \left(\textcolor{purple}{\mathcal{C}_{{Le}}^{1(11)} c_{_{{\beta }_{+}}}}-\mathcal{C}_{{Le}}^{2(11)} s_{_{{\beta }_{+}}}\right)+v_1 v_2 \left((\mathcal{C}_{{Le}}^{1(12)}+\mathcal{C}_{{Le}}^{1(21)}) c_{_{{\beta }_{+}}}\textcolor{purple}{-(\mathcal{C}_{{Le}}^{2(12)}+\mathcal{C}_{{Le}}^{2(21)}) s_{_{{\beta }_{+}}}}\right)\right.\right.$\\
					&$\left.\left.+v_2^2 \left(\textcolor{purple}{\mathcal{C}_{{Le}}^{1(22)} c_{_{{\beta }_{+}}}}-\mathcal{C}_{{Le}}^{2(22)} s_{_{{\beta }_{+}}}\right)\right)\right]_{lm}$  \\
					\hline
				\end{tabular}%
			\end{center}
			\caption{ 2HDM: Coupling of Singly charged scalar with a charged lepton and a neutrino. Here, $\mathit{s}_{\theta_i}=\sin \theta_i$, $\mathit{c}_{\theta_i}=\cos \theta_i$ and $l,m$ are the fermion family indices.}
			\label{tab:2HDM-scs-f-f}
		}
	\end{table}

\subsubsection*{MLRSM}
The magnetic moment interactions between the charged leptons and quarks with the electromagnetic field strength tensor ($A_{\mu \nu})$ in the presence of $\psi^2 \phi X$ operators (see Table~\ref{tab:MLRSM-psi2phiX}) within MLRSM-EFT framework are given as:
 \small{\begin{eqnarray}
 	\bar{l}_L\sigma ^{\mu \nu }l_R A_{\mu \nu } &:& -\frac{1}{\sqrt{2} \Lambda ^2}\left[\left(\kappa _2 \mathcal{C}_{l_Ll_R}^{1W_L}+\kappa _1 \mathcal{C}_{l_Ll_R}^{3W_L}\right) \sin \theta_{\text{w}}+\left(\kappa _2 \mathcal{C}_{l_Ll_R}^{1W_R}+\kappa _1 \mathcal{C}_{l_Ll_R}^{3W_R}\right) \sin \theta_{1} \cos \theta_{\text{w}}\right],\nonumber\\
 	\bar{u}_L\sigma ^{\mu \nu }u_RA_{\mu \nu } &:& \frac{1}{\sqrt{2} \Lambda ^2}\left[\left(\kappa _1 \mathcal{C}_{q_Lq_R}^{1W_L}+\kappa _2 \mathcal{C}_{q_Lq_R}^{3W_L}\right) \sin \theta_{\text{w}}+\left(\kappa _1 \mathcal{C}_{q_Lq_R}^{1W_R}+\kappa _2 \mathcal{C}_{q_Lq_R}^{3W_R}\right) \sin \theta_{1} \cos \theta_{\text{w}}\right],\nonumber \\
 	\bar{d}_L\sigma ^{\mu \nu }d_R A_{\mu\nu }&:&-\frac{1}{\sqrt{2} \Lambda ^2}\left[\left(\kappa _2 \mathcal{C}_{q_Lq_R}^{1W_L}+\kappa _1 \mathcal{C}_{q_Lq_R}^{3W_L}\right) \sin \theta_{\text{w}}+ \left(\kappa _2 \mathcal{C}_{q_Lq_R}^{1W_R}+\kappa _1 \mathcal{C}_{q_Lq_R}^{3W_R}\right)\sin \theta_{1}  \cos \theta_{\text{w}}\right].\nonumber
\end{eqnarray}}

The modified interactions involving singly ($H^{\pm}$) and doubly ($\delta^{\pm\pm}$) charged scalars provide  additional  contributions to the magnetic moments of the charged leptons, see Tables~\ref{tab:MLRSM-scs-f-f},  \ref{tab:MLRSM-dcs-l-l} and Figs.~\ref{fig:mag-moment}(b), (c).
There will be similar contributions to the quark magnetic moments through their couplings with singly charged scalars. The vertices are not explicitly depicted here.

\begin{table}[h!]
	{ \scriptsize
		\begin{center}
			\renewcommand{\arraystretch}{2.}
			\begin{tabular}{|*{2}{l|}}
				\hline 
				{\bf Vertex} & {\bf MLRSM : Effective vertex ($\bar{\nu}\,e^{-}\,H^+$) factor}\\
				\hline
				&\bf{Majorana type interactions}	\\ \hline
				$\left[ \overline{(e_{L})^c}~\nu_{L}~ \delta_{L}^{+}\right]_{lm}$  &  $\left[\frac{y_M}{\sqrt{2}}+\frac{1}{2 \sqrt{2} \Lambda ^2}\left(-y_M {A}_+^{44}+v_L^2 (-\mathcal{C}_{l_L}^{l({Ll})}-\mathcal{C}_{l_L}^{{lLl}})-\mathcal{C}_{l_L}^{l({Rr})} v_R^2-\kappa _+^2 \mathcal{C}_{l_L}^{l(21)}-\kappa _1^2 \mathcal{C}_{l_L}^{{l21}}-2 \kappa _1 \kappa _2 \mathcal{C}_{l_L}^{l(23)}\right.\right.\nonumber$ \\
				&$\left.\left.-2 \kappa _1 \kappa _2 \mathcal{C}_{l_L}^{l(41)}\right)\right]_{lm}$  \\
				\hline
				$\left[\overline{(e_{R})^c}~\nu_{R} ~\delta_{R}^{+}\right]_{lm}$  &  $\left[\frac{y_M}{\sqrt{2}}+\frac{1}{2 \sqrt{2} \Lambda ^2}\left(-y_M {A}_+^{33}-\mathcal{C}_{l_R}^{r({Ll})} v_L^2+v_R^2 (-\mathcal{C}_{l_R}^{r({Rr})}-\mathcal{C}_{l_R}^{{rRr}})-\kappa _+^2 \mathcal{C}_{l_R}^{r(21)}-\kappa _1^2 \mathcal{C}_{l_R}^{{r21}}-2 \kappa _1 \kappa _2 \mathcal{C}_{l_R}^{r(23)}\right.\right.\nonumber$ \\
				&$\left.\left.-2 \kappa _1 \kappa _2 \mathcal{C}_{l_R}^{r(41)}\right)\right]_{lm}$  \\
				\hspace{1cm}	{\bf {\large$\vdots$}} & \hspace{6cm} {\bf{\large$\vdots$}} \\
				\hline \hline
				& \bf{Dirac type interactions}	\\ \hline
				\hspace{1cm}	{\bf {\large$\vdots$}} & \hspace{6cm} {\bf{\large$\vdots$}} \\
				$\left[\overline{\nu}_{L}~e_{R}~\phi^{'+}_{1}\right]_{lm}$  &  $\left[\frac{1}{ \kappa _+}\left( \kappa _1 \tilde{y}_D- \kappa _1 y_D\right)+\frac{1}{2 \kappa_+ \Lambda ^2}\left(\kappa _1 \left(y_D-\tilde{y}_D\right) {A}_+^{11}+\kappa _2\left(\tilde{y}_D-y_D\right)  {A}_+^{12}+\kappa _1^3 \mathcal{C}_{l_Ll_R}^{1(21)}\right.\right.\nonumber$ \\
				&$\left.\left.+\kappa _1 \kappa _2^2 \mathcal{C}_{l_Ll_R}^{1(21)}+2 \kappa _1^2 \kappa _2 \mathcal{C}_{l_Ll_R}^{1(23)}+2 \kappa _1^2 \kappa _2 \mathcal{C}_{l_Ll_R}^{1(41)}-\kappa _2^3 \mathcal{C}_{l_Ll_R}^{3(21)}-\kappa _1^2 \kappa _2 \mathcal{C}_{l_Ll_R}^{3(21)}-2 \kappa _1 \kappa _2^2 \mathcal{C}_{l_Ll_R}^{3(23)}-2 \kappa _1 \kappa _2^2 \mathcal{C}_{l_Ll_R}^{3(41)}\right)\right]_{lm}$  \\
				\hline
				$\left[\overline{\nu}_{R}~e_{L}~\phi^{'+}_{2}\right]_{lm}$  &  $\left[\frac{1}{\kappa _+}\left( \kappa _1 \tilde{y}_D-\kappa _1 y_D\right)+\frac{1}{2 \kappa _+ \Lambda ^2}\left(\kappa _2 \left(y_D-\tilde{y}_D\right) {A}_+^{12}+\kappa_1 \left(y_D-\tilde{y}_D\right) {A}_+^{22} +\kappa _1^3 \mathcal{C}_{l_Ll_R}^{1(21)}\right.\right.\nonumber$ \\
				&$\left.\left.+\kappa _1 \kappa _2^2 \mathcal{C}_{l_Ll_R}^{1(21)}+2 \kappa _1^2 \kappa _2 \mathcal{C}_{l_Ll_R}^{1(23)}+2 \kappa _1^2 \kappa _2 \mathcal{C}_{l_Ll_R}^{1(41)}+\kappa _2^3 \mathcal{C}_{l_Ll_R}^{3(21)}+\kappa _1^2 \kappa _2 \mathcal{C}_{l_Ll_R}^{3(21)}+2 \kappa _1 \kappa _2^2 \mathcal{C}_{l_Ll_R}^{3(23)}+2 \kappa _1 \kappa _2^2 \mathcal{C}_{l_Ll_R}^{3(41)}\right)\right]_{lm}$  \\
				\hline
			\end{tabular}%
		\end{center}
		\caption{ MLRSM: Couplings of Singly charged scalars with a charged lepton and a neutrino in unphysical basis. A few Majorana type  and Dirac type interactions are listed. Here, $L,R$ represent the fermions belonging to $SU(2)_L, SU(2)_R$ gauge groups respectively.}
		\label{tab:MLRSM-scs-f-f}
	}
\end{table}

\begin{table}[h!]
	{ \footnotesize
		\renewcommand{\arraystretch}{1.90}
		\begin{tabular}{|c|c|}
			\hline 
			{\bf Vertex} & {\bf MLRSM: Effective vertex $(e^{\mp}\,e^{\mp}\,\delta^{\pm \pm})$ factor}\\			
			\hline
			$ \left[e^{\mp}_{_L} e^{\mp}_{_L} \delta_{L}^{\pm \pm}\right]_{lm} $ & $\left[y_M-\frac{1}{2 \Lambda ^2}\left(y_M {A}_{++}^{22}+\mathcal{C}_{l_L}^{l({Ll})} v_L^2+\mathcal{C}_{l_L}^{l({Rr})} v_R^2+\kappa _1^2 \mathcal{C}_{l_L}^{l(21)}+\kappa _2^2 \mathcal{C}_{l_L}^{l(21)}+\kappa _2^2 \mathcal{C}_{l_L}^{l21}+2 \kappa _1 \kappa _2 \mathcal{C}_{l_L}^{l(23)}+2 \kappa _1 \kappa _2 \mathcal{C}_{l_L}^{l(41)}\right)\right]_{lm}$\\
			\hline
			$\left[e^{\mp}_{_R} e^{\mp}_{_R} \delta_{R}^{\pm \pm}\right]_{lm}$&$\left[y_M-\frac{1}{2 \Lambda ^2}\left(y_M {A}_{{++}}^{11}+\mathcal{C}_{l_R}^{r({Ll})} v_L^2+\mathcal{C}_{l_R}^{r({Rr})} v_R^2+\kappa _1^2 \mathcal{C}_{l_R}^{r(21)}+\kappa _2^2 \mathcal{C}_{l_R}^{r(21)}+\kappa _2^2 \mathcal{C}_{l_R}^{{r21}}+2 \kappa _1 \kappa _2 \mathcal{C}_{l_R}^{r(23)}+2 \kappa _1 \kappa _2 \mathcal{C}_{l_R}^{r(41)}\right)\right]_{lm}$\\
			\hline
			$\left[e^{\mp}_{_R} e^{\mp}_{_R} \delta_{L}^{\pm \pm}\right]_{lm} $&$\frac{1}{\Lambda^2}\left[-\frac{y_{_M}}{2}A_{++}^{12}-\frac{1}{2} \kappa _2^2 \mathcal{C}_{l_R}^{l21}-\frac{1}{2} \kappa _1 \kappa _2 \mathcal{C}_{l_R}^{l23}-\frac{1}{2} \kappa _1 \kappa _2 \mathcal{C}_{l_R}^{l41}\right]_{lm}$\\
			\hline
			$\left[e^{\mp}_{_L} e^{\mp}_{_L} \delta_{R}^{\pm \pm}\right]_{lm} $&$\frac{1}{\Lambda^2}\left[-\frac{y_{_M}}{2}A_{++}^{12}-\frac{1}{2} \kappa _2^2 \mathcal{C}_{l_L}^{r21}-\frac{1}{2} \kappa _1 \kappa _2 \mathcal{C}_{l_L}^{r23}-\frac{1}{2} \kappa _1 \kappa _2 \mathcal{C}_{l_L}^{r41}\right]_{lm}$\\
			\hline
		\end{tabular}%
		\caption{MLRSM: Coupling between the doubly charged scalar ($\delta^{\pm\pm}$) and a pair of same signed charged lepton ($e^\pm$). Here, the subscript $L,R$ in the charged lepton $(e)$ correspond to the $SU(2)_{L,R}$.}
		\label{tab:MLRSM-dcs-l-l}
	}
\end{table}
\clearpage

%% file: LFV-LNV.tex
\subsection{Phenomenology involving charged scalars}

\begin{figure}[h!]
	\centering
	{
		\includegraphics[trim={1.3cm 0 1.1cm 0},scale=0.44]{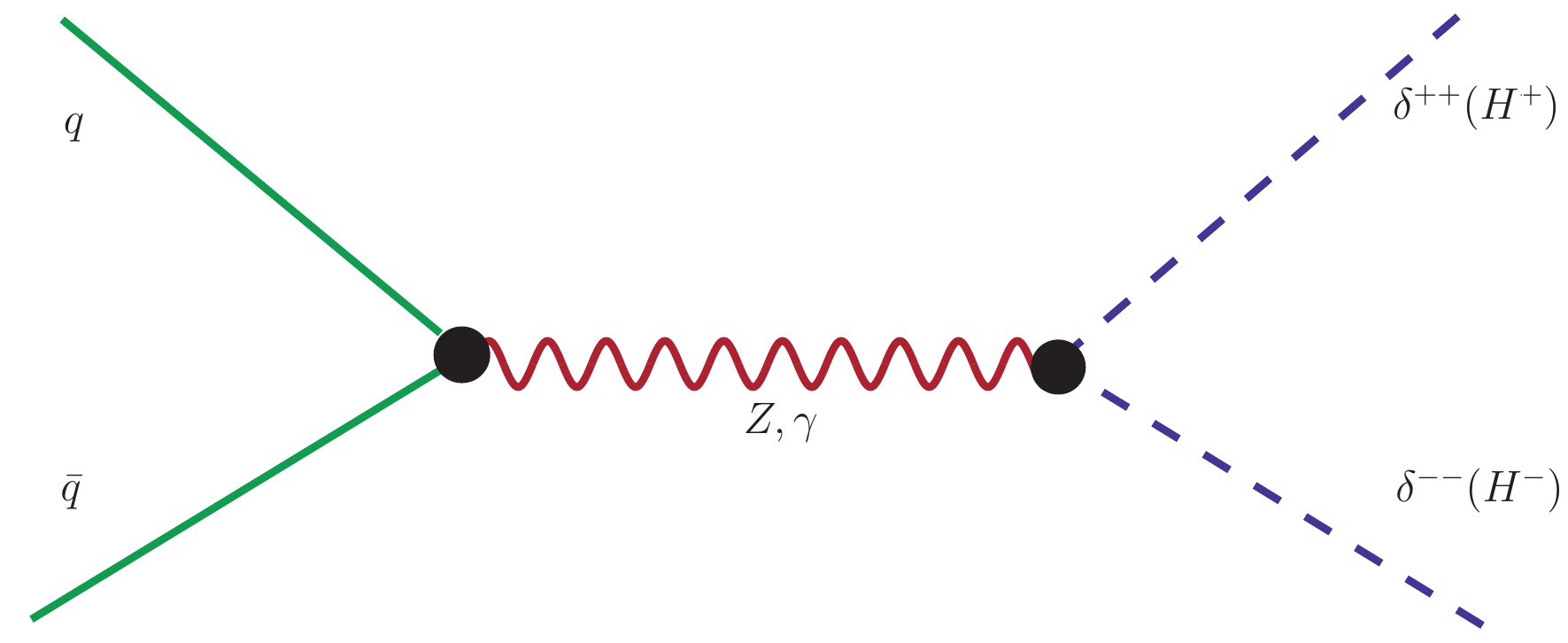}
	}
	\caption{Pair production of charged scalar (singly and doubly) through  neutral gauge boson propagator in $s$-channel. The  [{\Large $\bullet$}]  depicts the modified Feynman vertex (renormalizable + corrections due to dimension-6 operators).}\label{fig:s-channel}
\end{figure}

\begin{figure}[h!]
	\centering
	{
		\includegraphics[trim={2.5cm 0 2.3cm 0},scale=0.44]{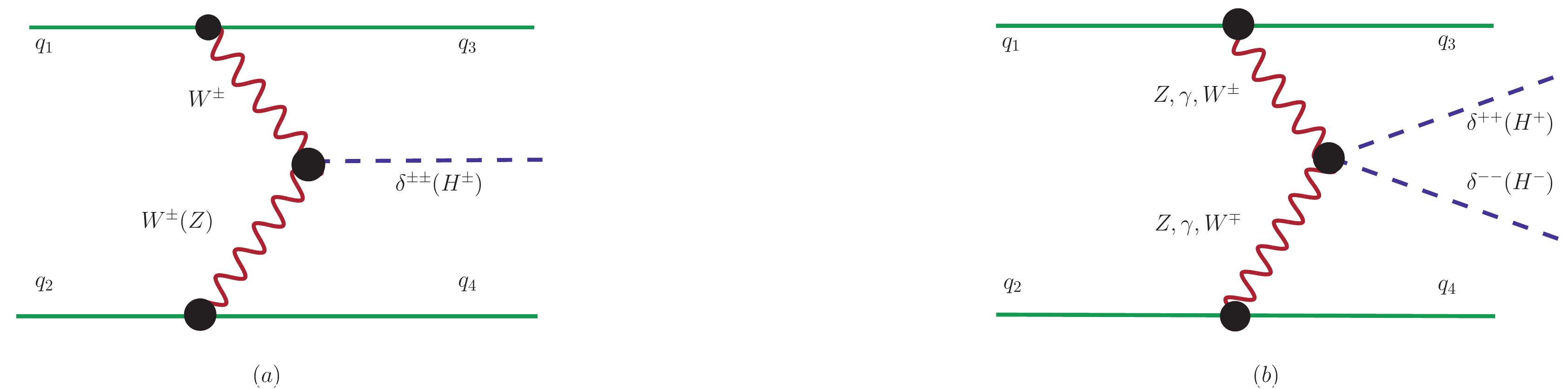}
	}
	\caption{(a) Production of single doubly (singly) charged scalar and (b) Pair production of doubly (singly) charged scalars through Vector Boson Fusion (VBF). The  [{\Large $\bullet$}] depicts the modified Feynman vertex (renormalizable + corrections due to dimension-6 operators).}\label{fig:VBF}
\end{figure}

The SM scalar sector contains only one Higgs doublet which takes part in the spontaneous electro-weak symmetry breaking, and the Higgs mechanism predicts only one real scalar. Thus any signature supporting new scalars will signify the presence of BSM physics. In this section, we have briefly outlined the interactions involving the charged scalars within the EFT framework. Both the models, 2HDM and MLRSM,  predict the existence of charged scalars. These charged scalars can be produced through $s$-channel and Vector Boson Fusion (VBF) processes, see Figs.~\ref{fig:s-channel}, and \ref{fig:VBF}.

\begin{center}
	\underline{\bf {2HDM: Feynman vertices involving singly charged scalars}}  
\end{center}




%
In the 2HDM scenario, due to the presence of an extra Higgs doublet, we have a physical singly charged scalar particle $(H^{\pm})$ which signifies a distinct departure from the Standard Model. Thus analysis related to this charged scalar $H^{\pm}$ could pave a path to explore this model in the light of present and future colliders.  In this context, their pair production $pp \to H^{\pm} H^{\pm}$, and leptonic decay modes, see Table~\ref{tab:2HDM-scs-f-f}, need to be analyzed. As the interaction vertices are modified in the presence of the dimension-6 operators, the predictions for cross sections and exclusion limits may alter significantly.  Thus it will be worthwhile to revisit the phenomenology involving the charged scalars in \cite{Gunion:1989we,Gunion:2002zf,Branco:2011iw,Carena:2013ooa,Haber:2013mia,Chen:2013jvg,Mrazek:2011iu,Dev:2014yca,Bhattacharyya:2015nca,Crivellin:2015hha,Coleppa:2013dya,Chen:2013rba} within the 2HDM-EFT framework.


\begin{center}
	\underline{\bf {MLRSM: Feynman vertices involving doubly charged scalars}}  
\end{center}


The MLRSM scenario has been explored extensively in the context of the LHC and FCC, considering only renormalizable interactions. In this paper, we have shown how those interactions get modified once we include the dimension-6 effective operators.  Here, instead of discussing all possible phenomenology, we have highlighted a  few channels which are considered to be the smoking gun features of this model. One of the most striking aspects of this scenario is the presence of left $(\Delta_{L})$ and right $(\Delta_{R})$ triplet Higgs multiplets that contain neutral, singly and doubly charged scalars.

\begin{figure}[h!]
	\centering
	{
		\includegraphics[trim={1.5cm 0 2.3cm 0},scale=0.44]{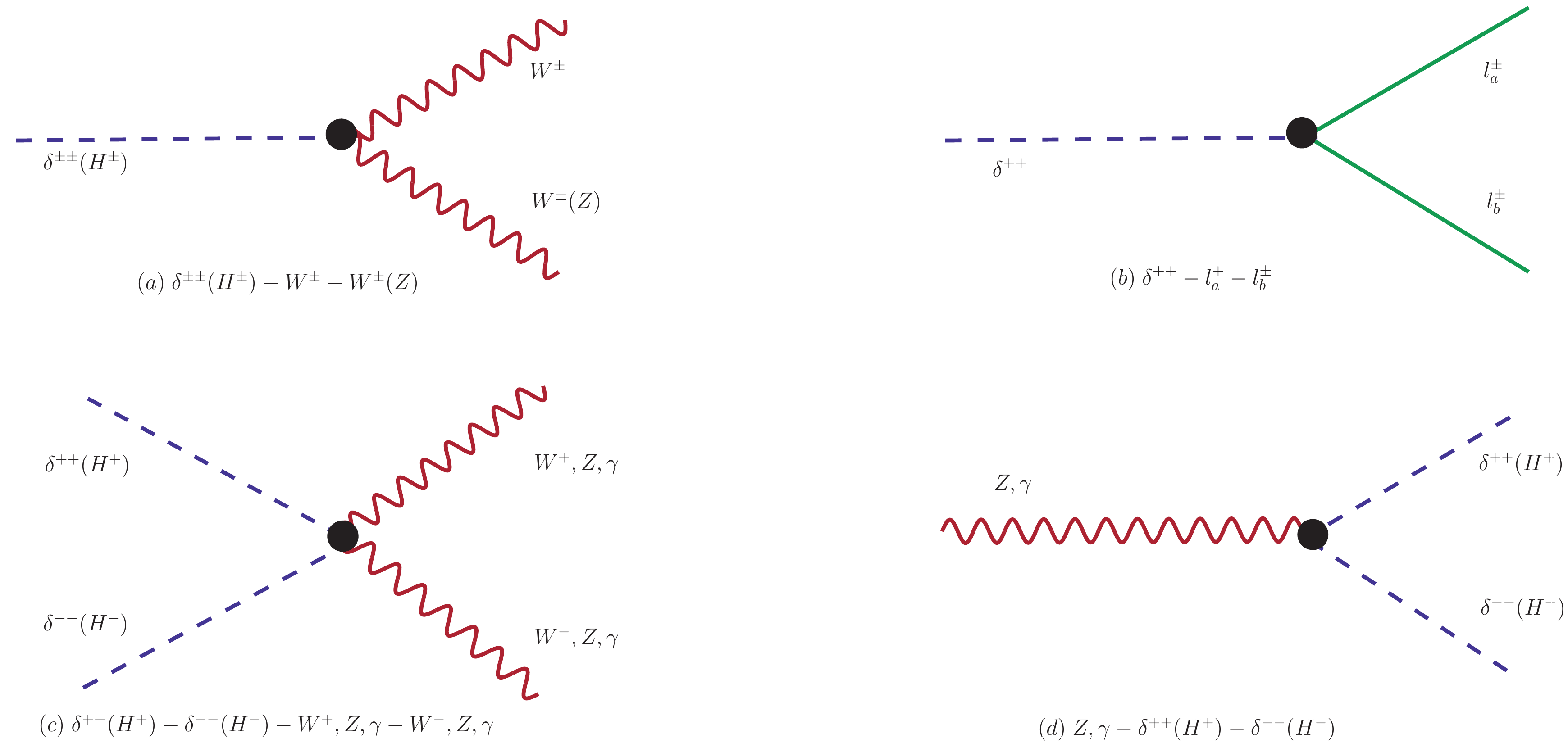}
	}
	\caption{Effective interaction vertices between charged scalars and gauge bosons in MLRSM model. Decay of doubly charged scalar is driven by: (a) $\delta^{\pm \pm}$-$W^{\mp}$-$W^{\mp}$ and (b) $\delta^{\pm \pm}$-$l_a^{\mp}$-$l_b^{\mp}$ vertices. The vertex (c) $\delta^{++} (H^+)$-$\delta^{--} (H^-)$-$W^{+}$-$W^{-}$ is relevant for VBF process. The vertex (d) $\delta^{++} (H^+)$-$ \delta^{--} (H^-)$-$Z(\gamma)$ participates in the $s$-channel pair production of doubly (singly) charged scalar. The [{\Large $\bullet$}]  depicts the modified Feynman vertex (renormalizable + corrections due to dimension-6 operators).}\label{fig:GB-DCS-vertex}
\end{figure}

The presence of doubly charged scalars makes this model phenomenologically attractive and distinct. Our focus will be on the doubly charged scalars: their production ($pp \to \delta_i^{++} \delta_j^{--}$) and their decay modes to a pair of (i) charged leptons $l^{\pm}_a l^{\pm}_b$ and (ii) charged gauge bosons $W^{\pm} W^{\pm}$.  One of the most significant final states in this model is hadronically quiet four leptons ($l_a^\pm l_b^\pm l_c^\mp l_d^\mp$) through neutral current $s$-channel processes \cite{Chakrabortty:2012pp,Bambhaniya:2013wza,Bambhaniya:2014cia}, see Fig.~\ref{fig:s-channel}. Apart from the $s$-channel process, through the Vector Boson Fusion (VBF) \cite{Cahn:1983ip, Rainwater:1998kj, Rainwater:1999sd}, the following final states can be achieved: (i) same-signed di-leptons  ($l_a^\pm l_b^\pm $) with 2 forward jets, see Fig.~\ref{fig:VBF}(a) and (ii) four leptons  with 2 forwards jets, see Fig.~\ref{fig:VBF}(b). In this paper, we have depicted the relevant Feynman vertices, see Fig.~\ref{fig:GB-DCS-vertex}, in Tables~\ref{tab:MLRSM-dcs-l-l}-\ref{tab:MLRSM-scs-scs-GB-GB}. In passing we would like to note that the phenomenology involving the singly charged scalars will be very similar to that for 2HDM case which is already discussed in the previous segment. The ``Golden channel" \cite{Maiezza:2010ic,Chakrabortty:2012pp} involving heavy right-handed ($N$) leading to same signed charged leptons and two jets also gets modified in presence of the effective interactions, see Tables~\ref{tab:MLRSM-cGB-F-F-I},~\ref{tab:MLRSM-cGB-F-F-II}. The phenomenology involving the charged scalars and right-handed neutrino within the renormlizable MLRSM framework \cite{Melfo:2011nx,Chakrabortty:2012pp,Bambhaniya:2013wza, Bambhaniya:2014cia,Deppisch:2014qpa,Deppisch:2015cua,Dobrescu:2015qna, Gluza:2015goa,Gluza:2016qqv,Dhuria:2015swa,Bambhaniya:2015wna, Dev:2016dja,Brehmer:2015cia,Dev:2015pga} needs to be revisited in the presence of the effective operators.

\begin{table}[h!]
	\centering
	{ \scriptsize
		\renewcommand{\arraystretch}{2.30}
		\begin{tabular}{|c|l|}
			\hline 
			{\bf Vertex} & {\bf 2HDM: Effective vertex $(W^{\pm}\,Z\,H^{\mp} )$ factor}\\			
			\hline
             $ W^{\pm\mu}~ Z_\mu~ H^{\mp} $ & $-\frac{g}{8 \sqrt{2}} \big(4 \tilde{g} \sin\tilde{\theta}_{\text{w}}\big)\big( v_1 \cos \bar{\beta }_{+}-v_2  \sin\bar{\beta }_{+}\big)-\frac{g}{8 \sqrt{2} \Lambda ^2}\big(4  \tilde{g} \sin \theta_{\text{w}} \Theta_{BB}  (-v_{2} \sin \beta_{+}+4 v_1 \cos \beta _{+})$\\
             &$+\cos \tilde{\theta}_{\text{w}} \big(v_1 \big(2 \tilde{g}  \cos \beta _{+}\Theta_{W_{3}B}+g v_2^2 \big((\mathcal{C}_{\phi D}^{(1)22(1)}+\mathcal{C}_{\phi D}^{(2)11(2)}) \cos \beta _{+}-3 \mathcal{C}_{\phi D}^{(2)22(1)} \sin \beta_{+}\big)\big)-2 v_2 \tilde{g}  \sin \beta_{+} \Theta_{W_{3}B}$\\
             &$+g v_1^3 \big(2 \mathcal{C}_{\phi D}^{(1)11(1)} \cos \beta _{+}-\mathcal{C}_{\phi D}^{(1)21(1)} \sin \beta_{+}\big)+g v_2 v_1^2 \big(3 \mathcal{C}_{\phi D}^{(1)21(1)} \cos \beta _{+}-(\mathcal{C}_{\phi D}^{(1)22(1)}+\mathcal{C}_{\phi D}^{(2)11(2)}) \sin \beta_{+}\big)$\\
             &$+g v_2^3 \big(\mathcal{C}_{\phi D}^{(2)22(1)} \cos \beta _{+}-2 \mathcal{C}_{\phi D}^{2(22)2} \sin \beta_{+}\big)\big)+v_1^3 \tilde{g} \sin \theta_{\text{w}}  \big(2 \mathcal{C}_{\phi D}^{(1)11(1)} \cos \beta _{+}-\mathcal{C}_{\phi D}^{(1)21(1)} \sin \beta_{+}\big)$\\
             &$+v_1^2 v_2 \tilde{g} \sin \theta_{\text{w}}  \big((3 \mathcal{C}_{\phi D}^{(1)21(1)}+4 \mathcal{C}_{\phi D}^{12(1)(1)}) \cos \beta _{+}-(2 \mathcal{C}_{\phi D}^{12(1)(2)}+\mathcal{C}_{\phi D}^{(1)22(1)}+\mathcal{C}_{\phi D}^{(2)11(2)}) \sin \beta_{+}\big)$\\
             &$+v_1 v_2^2 \tilde{g} \sin \theta_{\text{w}}  \big((2 \mathcal{C}_{\phi D}^{12(1)(2)}+\mathcal{C}_{\phi D}^{(1)22(1)}+\mathcal{C}_{\phi D}^{(2)11(2)}) \cos \beta _{+}-(4 \mathcal{C}_{\phi D}^{21(2)(2)}+3 \mathcal{C}_{\phi D}^{(2)22(1)}) \sin \beta_{+}\big)$\\
             &$+v_2^3 \tilde{g} \sin \theta_{\text{w}}  \big(\mathcal{C}_{\phi D}^{(2)22(1)} \cos \beta _{+}-2 \mathcal{C}_{\phi D}^{2(22)2} \sin \beta_{+}\big)+2 v_2 \tilde{g} A_{+}^{22} \sin \theta_{\text{w}}  \sin \beta_{+}+2 v_1 \tilde{g} A_{+}^{11} \sin \theta_{\text{w}}  \sin \beta_{+}$\\
             &$-2 v_2 \tilde{g} A_{+}^{12} \sin \theta_{\text{w}}  \cos \beta _{+}-2 v_1 \tilde{g} A_{+}^{11} \sin \theta_{\text{w}}  \cos \beta _{+}-4 \tilde{g} \sin \theta_{\text{w}}  \Theta_{WW} \big( v_2 \sin \beta_{+}- v_1  \cos \beta_{\phi ^+}\big)\big)$\\
        	\hline
        \end{tabular}%
     }
\caption{2HDM: Coupling of charged scalar ($H^{\mp}$) with  a charged gauge boson ($W^{\pm}$) and a neutral gauge boson ($Z$) .}
\label{tab:2HDM-cs-cGB-nGB}
\end{table}

\begin{table}[h!]
	\centering
	{ \scriptsize
		\renewcommand{\arraystretch}{2.30}
		\begin{tabular}{|c|c|}
			\hline 
			{\bf Vertex} & {\bf MLRSM: Effective vertex $(W^{\pm}\,W^{\pm}\,\delta^{\mp \mp} )$ factor}\\			
			\hline
			$ W^{\pm\mu}_{L}~ W^{\pm}_{L\mu}~ \delta_{L}^{\mp \mp} $ & $-\frac{g^2 v_L}{4 \sqrt{2} } \left[4+\frac{1}{\Lambda ^2}\left(-2 A_{\text{++}}^{22}+8 \Theta _{W_{\text{LL}}}+2 v_L^2 (\mathcal{C}_{\phi D}^{\{\text{Ll}\}\{(L)(l)\}}+2 \mathcal{C}_{\phi D}^{L(l)L(l)})+\mathcal{C}_{\phi D}^{(L)\text{rR}(l)} v_R^2+2 \kappa _+^2 \mathcal{C}_{\phi D}^{\{21\}\{(L)(l)\}}\right.\right.$\\
			&$\left.\left.+\kappa _+^2 \mathcal{C}_{\phi D}^{21(L)(l)}+2 \kappa _+^2 \mathcal{C}_{\phi D}^{(L)11(l)}+2 \kappa _1 \kappa _2 \mathcal{C}_{\phi D}^{L(l)1(1)}\right)\right]$\\
			\hline
			$ W^{\pm\mu}_{L}~ W^{\pm}_{R\mu}~ \delta_{L}^{\mp \mp} $&$-\frac{g^2}{4 \sqrt{2} \Lambda ^2}\left[\left(v_L \left(\mathcal{C}_{\phi D}^{(L)\text{rL}(r)} v_R^2+8 \Theta _{W_{\text{LR}}}-\kappa_+^2 \mathcal{C}_{\phi D}^{L(l)1(1)}\right)+\kappa_+^2 v_R (\mathcal{C}_{\phi D}^{(L)11(r)}+\mathcal{C}_{\phi D}^{(L)12(r)\}}+\mathcal{C}_{\phi D}^{(R)11(l)})\right)\right]$\\
			\hline
			$ W^{\pm\mu}_{R}~ W^{\pm}_{R\mu}~ \delta_{L}^{\mp \mp} $&$\frac{g^2 v_R A_{\text{++}}^{12}}{2 \sqrt{2} \Lambda ^2}$\\
			\hline \hline
			$ W^{\pm\mu}_{R}~ W^{\pm}_{R\mu}~ \delta_{R}^{\mp \mp} $&$-\frac{g^2 v_R}{4 \sqrt{2}}\left[4+\frac{1}{\Lambda ^2}\left(-2 A_{\text{++}}^{11}+8 \Theta _{W_{\text{RR}}}+\mathcal{C}_{\phi D}^{(R)\text{lL}(r)} v_L^2+2 \mathcal{C}_{\phi D}^{\{\text{Rr}\}\{(R)(r)\}} v_R^2+4 \mathcal{C}_{\phi D}^{R(r)R(r)} v_R^2+2 \kappa _+^2 \mathcal{C}_{\phi D}^{\{21\}\{(R)(r)\}}\right.\right.$\\
			&$\left.\left.+\kappa _+^2 \mathcal{C}_{\phi D}^{21(R)(r)}+2 \kappa _+^2 \mathcal{C}_{\phi D}^{(R)11(r)}-\kappa _+^2 \mathcal{C}_{\phi D}^{R(r)1(1)}\right)\right]$\\
			\hline
			$ W^{\pm\mu}_{R}~ W^{\pm}_{L\mu}~ \delta_{R}^{\mp \mp} $&$-\frac{g^2}{4 \sqrt{2} \Lambda ^2}\left[ \left(\mathcal{C}_{\phi D}^{(L)\text{rL}(r)} v_L^2 v_R+\kappa_+^2 v_L (\mathcal{C}_{\phi D}^{(L)11(r)}+\mathcal{C}_{\phi D}^{(L)12(r)}+\mathcal{C}_{\phi D}^{(R)11(l)})+2 v_R \left(4 \Theta _{W_{\text{RL}}}+\kappa _1 \kappa _2 \mathcal{C}_{\phi D}^{R(r)1(1)}\right)\right)\right]$\\
			\hline
			$W^{\pm\mu}_{L}~ W^{\pm}_{L\mu}~ \delta_{R}^{\mp \mp} $&$\frac{g^2 v_{L} A_{\text{++}}^{12}}{2 \sqrt{2} \Lambda ^2}$\\
			\hline
		\end{tabular}%
	}
\caption{MLRSM: Coupling of doubly charged scalar ($\delta^{\pm\pm}$) with a pair of same signed charged gauge bosons ($W^{\pm}$).}
\label{tab:MLRSM-dcs-cGB-cGB}
\end{table}

	\begin{table}[h!]
	\centering
	{ \scriptsize
		\renewcommand{\arraystretch}{2.0}
		\begin{tabular}{|c|l|}
			\hline 
			{\bf Vertex} & {\bf  MLRSM: Effective vertex $(\delta^{\mp \mp}\,\delta^{\pm \pm}\,Z_1)$ factor}\\
			\hline
			$\delta_{L}^{--} \delta_{L}^{++}Z_{1}^{\mu}$&$ \left(k_{--}^{\mu }+k_{++}^{\mu }\right) \left(- \tilde{g} c_{\theta _1} c_{\tilde{\theta}_2} s_{\theta _\text{w}}+ \tilde{g} s_{\theta _1} s_{\tilde{\theta} _2}+ g c_{\tilde{\theta}_2} c_{\theta _\text{w}}\right)+\frac{1}{4 \Lambda ^2} \left(k_{--}^{\mu }+k_{++}^{\mu }\right) \left(4 \tilde{g} c_{\theta _1} c_{\theta _2} {A}_{\text{++}}^{22} s_{\theta _\text{w}}-4 \tilde{g} s_{\theta _1} s_{\theta _2} A_{{++}}^{22} \right.$\\
			&$\left. -c_{\theta _2} s_{\theta _1} s_{\theta _\text{w}} \left(4 \tilde{g} \Theta _{{3RB}}+4 g \Theta _{{3L3R}}+g \kappa _2^2 (-\mathcal{C}_{\phi D}^{L(l)1(1)})\right)+4 \tilde{g} \Theta _{{3LB}} c_{\theta _2} c_{\theta _\text{w}}-4 \tilde{g} \Theta _{{3RB}} c_{\theta _1} s_{\theta _2}-4 \tilde{g} \Theta _{{BB}} c_{\theta _1} c_{\theta _2} s_{\theta _\text{w}} \right.$\\
			&$\left. +4 \tilde{g} \Theta _{{BB}} s_{\theta _1} s_{\theta _2}-2 \mathcal{C}_{\phi D}^{\{{Ll}\}\{(L)(l)\}} \tilde{g} c_{\theta _1} c_{\theta _2} v_L^2 s_{\theta _\text{w}}-2 \kappa _1^2 \tilde{g} c_{\theta _1} c_{\theta _2} (\mathcal{C}_{\phi D}^{\{21\}\{(L)(l)\}}+2 \mathcal{C}_{\phi D}^{(L)11(l)}) s_{\theta _\text{w}} \right.$\\
			&$\left. -2 \kappa _2^2 \mathcal{C}_{\phi D}^{21(L)(l)} \tilde{g} c_{\theta _1} c_{\theta _2} s_{\theta _\text{w}}+2 \mathcal{C}_{\phi D}^{\{{Ll}\}\{(L)(l)\}} \tilde{g} v_L^2 s_{\theta _1} s_{\theta _2}+2 \kappa _2^2 \mathcal{C}_{\phi D}^{\{21\}\{(L)(l)\}} \tilde{g} s_{\theta _1} s_{\theta _2}+2 \kappa _1^2 \tilde{g} s_{\theta _1} s_{\theta _2} (\mathcal{C}_{\phi D}^{\{21\}\{(L)(l)\}}+2 \mathcal{C}_{\phi D}^{(L)11(l)}) \right.$\\
			&$\left. +2 \kappa _2^2 \mathcal{C}_{\phi D}^{21(L)(l)} \tilde{g} s_{\theta _1} s_{\theta _2}-4 g c_{\theta _2} {A}_{\text{++}}^{22} c_{\theta _\text{w}}+4 g \Theta _{{3L3L}} c_{\theta _2} c_{\theta _\text{w}}+g s_{\theta _2} \left(c_{\theta _1} \left(\kappa _2^2 \mathcal{C}_{\phi D}^{L(l)1(1)}-4 \Theta _{{3L3R}}\right)+4 \Theta _{{3LB}} s_{\theta _1}\right) \right.$\\
			&$\left. -4 g \Theta _{{3LB}} c_{\theta _1} c_{\theta _2} s_{\theta _\text{w}}+2 g \mathcal{C}_{\phi D}^{\{{Ll}\}\{(L)(l)\}} c_{\theta _2} v_L^2 c_{\theta _\text{w}}+g \kappa _2^2 c_{\theta _2} c_{\theta _\text{w}} (2 \mathcal{C}_{\phi D}^{\{21\}\{(L)(l)\}}+2 \mathcal{C}_{\phi D}^{21(L)(l)}+\mathcal{C}_{\phi D}^{L(l)1(1)}) \right.$\\
			&$\left. +2 g \kappa _1^2 c_{\theta _2} (\mathcal{C}_{\phi D}^{\{21\}\{(L)(l)\}}  +2 \mathcal{C}_{\phi D}^{(L)11(l)}) c_{\theta _\text{w}}-2 \kappa _2^2 \mathcal{C}_{\phi D}^{\{21\}\{(L)(l)\}} \tilde{g} c_{\theta _1} c_{\theta _2} s_{\theta _\text{w}}\right)$\\			
			\hline
			$\delta_{R}^{--} \delta_{R}^{++}Z_{1}^{\mu}$&$- \left(p_{--}^{\mu }+p_{++}^{\mu }\right) \left( \tilde{g} c_{\theta _1} c_{\tilde{\theta} _2} s_{\theta _\text{w}}- \tilde{g} s_{\theta _1} s_{\tilde{\theta} _2}+ g c_{\theta _1} s_{\tilde{\theta} _2}+ g c_{\tilde{\theta} _2} s_{\theta _1} s_{\theta _\text{w}}\right)-\frac{1}{4 \Lambda ^2}\left(p_{--}^{\mu }+p_{++}^{\mu }\right) \left(-4 \tilde{g} c_{\theta _1} c_{\theta _2} {A}_{\text{++}}^{11} s_{\theta _\text{w}} \right.$\\
			&$\left. -4 \tilde{g} \Theta _{{3LB}} c_{\theta _2} c_{\theta _\text{w}}+4 \tilde{g} \Theta _{{3RB}} c_{\theta _1} s_{\theta _2}+4 \tilde{g} \Theta _{{3RB}} c_{\theta _2} s_{\theta _1} s_{\theta _\text{w}}+4 \tilde{g} \Theta _{{BB}} c_{\theta _1} c_{\theta _2} s_{\theta _\text{w}}-4 \tilde{g} \Theta_{{BB}} s_{\theta _1} s_{\theta _2} \right.$\\
			&$\left. +2 \kappa _1^2 \tilde{g} c_{\theta _1} c_{\theta _2} (\mathcal{C}_{\phi D}^{\{21\}\{(R)(r)\}}+2 \mathcal{C}_{\phi D}^{(R)11(r)}) s_{\theta _\text{w}}+2 \kappa _2^2 \mathcal{C}_{\phi D}^{\{21\}\{(R)(r)\}} \tilde{g} c_{\theta _1} c_{\theta _2} s_{\theta _\text{w}}+2 \kappa _2^2 \mathcal{C}_{\phi D}^{21(R)(r)} \tilde{g} c_{\theta _1} c_{\theta _2} s_{\theta _\text{w}} \right.$\\
			&$\left. -2 \mathcal{C}_{\phi D}^{\{{Rr}\}\{(R)(r)\}} \tilde{g} v_R^2 s_{\theta _1} s_{\theta _2}-2 \kappa _2^2 \mathcal{C}_{\phi D}^{\{21\}\{(R)(r)\}} \tilde{g} s_{\theta _1} s_{\theta _2}-2 \kappa _1^2 \tilde{g} s_{\theta _1} s_{\theta _2} (\mathcal{C}_{\phi D}^{\{21\}\{(R)(r)\}}+2 \mathcal{C}_{\phi D}^{(R)11(r)})-2 \kappa _2^2 \mathcal{C}_{\phi D}^{21(R)(r)} \tilde{g} s_{\theta _1} s_{\theta _2} \right.$\\
			&$\left. -4 g c_{\theta _1} s_{\theta _2} A_{\text{++}}^{11}-4 g c_{\theta _2} s_{\theta _1} A_{\text{++}}^{11} s_{\theta _\text{w}}-g c_{\theta _2} c_{\theta _\text{w}} \left(4 \Theta_{{3L3R}}+\kappa _2^2 \mathcal{C}_{\phi D}^{R(r)1(1)}\right)+4 g \Theta_{{3R3R}} c_{\theta _1} s_{\theta _2}+4 g \Theta_{{3R3R}} c_{\theta _2} s_{\theta _1} s_{\theta _\text{w}} \right.$\\
			&$\left. +4 g \Theta_{{3RB}} c_{\theta _1} c_{\theta _2} s_{\theta _\text{w}}-4 g \Theta_{{3RB}} s_{\theta _1} s_{\theta _2}+2 g \mathcal{C}_{\phi D}^{\{\text{Rr}\}\{(R)(r)\}} c_{\theta _1} v_R^2 s_{\theta _2}+2 g \mathcal{C}_{\phi D}^{\{\text{Rr}\}\{(R)(r)\}} c_{\theta _2} v_R^2 s_{\theta _1} s_{\theta _\text{w}} \right.$\\
			&$\left. +g \kappa _2^2 c_{\theta _1} s_{\theta _2} (2 \mathcal{C}_{\phi D}^{\{21\}\{(R)(r)\}}+2 \mathcal{C}_{\phi D}^{21(R)(r)}-\mathcal{C}_{\phi D}^{R(r)1(1)})+2 g \kappa _1^2 c_{\theta _1} s_{\theta _2} (\mathcal{C}_{\phi D}^{\{21\}\{(R)(r)\}}+2 \mathcal{C}_{\phi D}^{(R)11(r)}) \right.$\\
			&$\left. +2 g \kappa _1^2 c_{\theta _2} s_{\theta _1} (\mathcal{C}_{\phi D}^{\{21\}\{(R)(r)\}}  +2 \mathcal{C}_{\phi D}^{(R)11(r)}) s_{\theta _\text{w}} +2 g \kappa _2^2 \mathcal{C}_{\phi D}^{\{21\}\{(R)(r)\}} c_{\theta _2} s_{\theta _1} s_{\theta _\text{w}}+2 g \kappa _2^2 \mathcal{C}_{\phi D}^{21(R)(r)} c_{\theta _2} s_{\theta _1} s_{\theta _\text{w}} \right.$\\
			&$\left. +4 \tilde{g} s_{\theta _1} s_{\theta _2} A_{\text{++}}^{11}+2 \mathcal{C}_{\phi D}^{\{\text{Rr}\}\{(R)(r)\}} \tilde{g} c_{\theta _1} c_{\theta _2} v_R^2 s_{\theta _\text{w}}-g \kappa _2^2 \mathcal{C}_{\phi D}^{R(r)1(1)} c_{\theta _2} s_{\theta _1} s_{\theta _\text{w}}\right)$\\
			\hline
			$\delta_{L}^{\mp \mp} \delta_{R}^{\pm \pm}Z_{1}^{\mu}$ &$\frac{1}{2 \Lambda ^2}\left(-c_{\theta _1} \left(\tilde{g} c_{\theta _2} s_{\theta _\text{w}} \left(k_{\mp \mp}^{\mu }+p_{\pm \pm}^{\mu }\right) \left(\kappa _1^2 (\mathcal{C}_{\phi D}^{(L)11(r)}+\mathcal{C}_{\phi D}^{(L)12(r)}+\mathcal{C}_{\phi D}^{(R)11(l)})-2 A_{\text{++}}^{12}\right)+g s_{\theta _2} \left(k_{\mp \mp}^{\mu } \left(\kappa _1^2 (\mathcal{C}_{\phi D}^{(L)11(r)} \right.\right.\right.\right.$\\
			&$\left.\left.\left.\left. +\mathcal{C}_{\phi D}^{(L)12(r)}+\mathcal{C}_{\phi D}^{(R)11(l)})-A_{\text{++}}^{12}\right)-A_{\text{++}}^{12} p_{\pm \pm}^{\mu }\right)\right)+\tilde{g} s_{\theta _1} s_{\theta _2} \left(k_{\mp \mp}^{\mu }+p_{\pm \pm}^{\mu }\right) \left(\kappa _1^2 (\mathcal{C}_{\phi D}^{(L)11(r)}+\mathcal{C}_{\phi D}^{(L)12(r)}+\mathcal{C}_{\phi D}^{(R)11(l)})-2 A_{\text{++}}^{12}\right)\right.$\\
			&$\left.+g c_{\theta _2} \left(c_{\theta _\text{w}} \left(p_{\pm \pm}^{\mu } \left(\kappa _1^2 (\mathcal{C}_{\phi D}^{(L)11(r)}+\mathcal{C}_{\phi D}^{(L)12(r)}+\mathcal{C}_{\phi D}^{(R)11(l)})-A_{\text{++}}^{12}\right)-k_{\mp \mp}^{\mu } A_{\text{++}}^{12}\right)-s_{\theta _1} s_{\theta _\text{w}} \left(k_{\mp \mp}^{\mu } \left(\kappa _1^2 (\mathcal{C}_{\phi D}^{(L)11(r)}+\mathcal{C}_{\phi D}^{(L)12(r)}\right.\right.\right.\right.$\\
			&$\left.\left.\left.\left.+\mathcal{C}_{\phi D}^{(R)11(l)})-A_{\text{++}}^{12}\right)-A_{\text{++}}^{12} p_{\pm \pm}^{\mu }\right)\right)\right)$\\
			\hline
		\end{tabular}%
	}
		\caption{MLRSM: Coupling of a pair of doubly charged scalars ($\delta^{\pm\pm}$) with a neutral gauge boson ($Z_1$). Here, $\mathit{s}_{\theta_i}=\sin \theta_i$ and $\mathit{c}_{\theta_i}=\cos \theta_i$. $k^{\mu}_{\pm \pm}$ and $p^{\mu}_{\pm \pm}$  are the four momenta of $\delta^{\pm \pm}_{L}$ and $\delta^{\pm \pm}_{R}$ respectively.  $\tilde{\theta}_{2}$ is the redefined angle defined in Eq.~\ref{eq:MLRSM-nGB-rot-angle-full}.}\label{tab:MLRSM-dcs-dcs-nGB1}
\end{table}

\begin{table}[h!]
	\centering
	{ \scriptsize
		\renewcommand{\arraystretch}{2.0}
		\begin{tabular}{|c|l|}
			\hline 
			{\bf Vertex} & {\bf  MLRSM: Effective vertex $( \delta^{\pm \pm}\,\delta^{\mp \mp}\,Z_2)$ factor}\\
			\hline
			$\delta_{L}^{--} \delta_{L}^{++}Z_{2}^{\mu}$&$-\left(k_{--}^{\mu }+k_{++}^{\mu }\right) \left( \tilde{g} c_{\tilde{\theta} _2} s_{\theta _1}+ \tilde{g} c_{\theta _1} s_{\tilde{\theta} _2} s_{\theta _\text{w}}- g s_{\tilde{\theta} _2} c_{\theta _\text{w}}\right)-\frac{1}{4 \Lambda ^2}\left(k_{--}^{\mu }+k_{++}^{\mu }\right) \left(-4 \tilde{g} c_{\theta _2} s_{\theta _1} A_{\text{++}}^{22}-4 \tilde{g} c_{\theta _1} s_{\theta _2} A_{\text{++}}^{22} s_{\theta _\text{w}} \right.$\\
			&$\left. +s_{\theta _1} s_{\theta _2} s_{\theta _\text{w}} \left(4 \tilde{g} \Theta_{{3RB}}+4 g \Theta_{{3L3R}}+g \kappa _2^2 (-\mathcal{C}_{\phi D}^{L(l)1(1)})\right)-4 \tilde{g} \Theta_{{3LB}} s_{\theta _2} c_{\theta _\text{w}}-4 \tilde{g} \Theta_{{3RB}} c_{\theta _1} c_{\theta _2}+4 \tilde{g} \Theta_{{BB}} c_{\theta _2} s_{\theta _1} \right.$\\
			&$\left. +4 \tilde{g} \Theta_{{BB}} c_{\theta _1} s_{\theta _2} s_{\theta _\text{w}}+2 \mathcal{C}_{\phi D}^{\{\text{Ll}\}\{(L)(l)\}} \tilde{g} c_{\theta _2} v_L^2 s_{\theta _1}+2 \mathcal{C}_{\phi D}^{\{\text{Ll}\}\{(L)(l)\}} \tilde{g} c_{\theta _1} v_L^2 s_{\theta _2} s_{\theta _\text{w}}+2 \kappa _2^2 \mathcal{C}_{\phi D}^{\{21\}\{(L)(l)\}} \tilde{g} c_{\theta _2} s_{\theta _1} \right.$\\
			&$\left. +2 \kappa _1^2 \tilde{g} c_{\theta _2} s_{\theta _1} (\mathcal{C}_{\phi D}^{\{21\}\{(L)(l)\}}+2 \mathcal{C}_{\phi D}^{(L)11(l)})+2 \kappa _1^2 \tilde{g} c_{\theta _1} s_{\theta _2} (\mathcal{C}_{\phi D}^{\{21\}\{(L)(l)\}}+2 \mathcal{C}_{\phi D}^{(L)11(l)}) s_{\theta _\text{w}}+2 \kappa _2^2 \mathcal{C}_{\phi D}^{\{21\}\{(L)(l)\}} \tilde{g} c_{\theta _1} s_{\theta _2} s_{\theta _\text{w}}\right.$\\
			&$\left.+2 \kappa _2^2 \mathcal{C}_{\phi D}^{21(L)(l)} \tilde{g} c_{\theta _2} s_{\theta _1}+2 \kappa _2^2 \mathcal{C}_{\phi D}^{21(L)(l)} \tilde{g} c_{\theta _1} s_{\theta _2} s_{\theta _\text{w}}+4 g s_{\theta _2} A_{\text{++}}^{22} c_{\theta _\text{w}}-4 g \Theta_{{3L3L}} s_{\theta _2} c_{\theta _\text{w}}+g c_{\theta _2} \left( c_{\theta _1} \left( \kappa _2^2 \mathcal{C}_{\phi D}^{L(l)1(1)} \right.\right.\right.$\\
			&$\left.\left.\left. -4 \Theta_{{3L3R}}\right)+4 \Theta_{{3LB}} s_{\theta _1}\right)+4 g \Theta_{{3LB}} c_{\theta _1} s_{\theta _2} s_{\theta _\text{w}}-2 g \mathcal{C}_{\phi D}^{\{\text{Ll}\}\{(L)(l)\}} v_L^2 s_{\theta _2} c_{\theta _\text{w}}+g \kappa _2^2 s_{\theta _2} c_{\theta _\text{w}} (-2 \mathcal{C}_{\phi D}^{\{21\}\{(L)(l)\}}-2 \mathcal{C}_{\phi D}^{21(L)(l)} \right.$\\
			&$\left. -\mathcal{C}_{\phi D}^{L(l)1(1)})-2 g \kappa _1^2 s_{\theta _2} (\mathcal{C}_{\phi D}^{\{21\}\{(L)(l)\}}+2 \mathcal{C}_{\phi D}^{(L)11(l)}) c_{\theta _\text{w}}\right)$\\			
			\hline
			$\delta_{R}^{--} \delta_{R}^{++}Z_{2}^{\mu}$&$ \left(p_{--}^{\mu }+p_{++}^{\mu }\right) \left(- \tilde{g} c_{\theta _2} s_{\theta _1}- \tilde{g} c_{\theta _1} s_{\theta _2} s_{\theta _\text{w}}+ g c_{\theta _1} c_{\theta _2}- g s_{\theta _1} s_{\theta _2} s_{\theta _\text{w}}\right)+\frac{1}{4 \Lambda ^2}\left(p_{--}^{\mu }+p_{++}^{\mu }\right) \left(4 \tilde{g} c_{\theta _2} s_{\theta _1} A_{\text{++}}^{11}+4 \tilde{g} c_{\theta _1} s_{\theta _2} A_{\text{++}}^{11} s_{\theta _\text{w}} \right. $\\
			&$\left. +s_{\theta _2} c_{\theta _\text{w}} \left(4 \tilde{g} \Theta_{{3LB}}+4 g \Theta_{{3L3R}}+g \kappa _2^2 \mathcal{C}_{\phi D}^{R(r)1(1)}\right)+4 \tilde{g} \Theta_{{3RB}} c_{\theta _1} c_{\theta _2}-4 \tilde{g} \Theta_{{3RB}} s_{\theta _1} s_{\theta _2} s_{\theta _\text{w}}-4 \tilde{g} \Theta_{{BB}} c_{\theta _2} s_{\theta _1}-4 \tilde{g} \Theta_{{BB}} c_{\theta _1} s_{\theta _2} s_{\theta _\text{w}} \right. $\\
			&$\left. -2 \mathcal{C}_{\phi D}^{\{\text{Rr}\}\{(R)(r)\}} \tilde{g} c_{\theta _2} v_R^2 s_{\theta _1}-2 \mathcal{C}_{\phi D}^{\{\text{Rr}\}\{(R)(r)\}} \tilde{g} c_{\theta _1} v_R^2 s_{\theta _2} s_{\theta _\text{w}}-2 \kappa _2^2 \mathcal{C}_{\phi D}^{\{21\}\{(R)(r)\}} \tilde{g} c_{\theta _2} s_{\theta _1}-2 \kappa _1^2 \tilde{g} c_{\theta _2} s_{\theta _1} (\mathcal{C}_{\phi D}^{\{21\}\{(R)(r)\}} \right. $\\
			&$\left. +2 \mathcal{C}_{\phi D}^{(R)11(r)})-2 \kappa _1^2 \tilde{g} c_{\theta _1} s_{\theta _2} (\mathcal{C}_{\phi D}^{\{21\}\{(R)(r)\}}+2 \mathcal{C}_{\phi D}^{(R)11(r)}) s_{\theta _\text{w}}-2 \kappa _2^2 \mathcal{C}_{\phi D}^{\{21\}\{(R)(r)\}} \tilde{g} c_{\theta _1} s_{\theta _2} s_{\theta _\text{w}}-2 \kappa _2^2 \mathcal{C}_{\phi D}^{21(R)(r)} \tilde{g} c_{\theta _2} s_{\theta _1} \right. $\\
			&$\left. -2 \kappa _2^2 \mathcal{C}_{\phi D}^{21(R)(r)} \tilde{g} c_{\theta _1} s_{\theta _2} s_{\theta _\text{w}}-4 g c_{\theta _1} c_{\theta _2} A_{\text{++}}^{11}+4 g s_{\theta _1} s_{\theta _2} A_{\text{++}}^{11} s_{\theta _\text{w}}+4 g \Theta_{{3R3R}} c_{\theta _1} c_{\theta _2}-4 g \Theta_{{3R3R}} s_{\theta _1} s_{\theta _2} s_{\theta _\text{w}} \right. $\\
			&$\left. -4 g \Theta_{{3RB}} c_{\theta _2} s_{\theta _1}-4 g \Theta_{{3RB}} c_{\theta _1} s_{\theta _2} s_{\theta _\text{w}}+2 g \mathcal{C}_{\phi D}^{\{\text{Rr}\}\{(R)(r)\}} c_{\theta _1} c_{\theta _2} v_R^2+g \kappa _2^2 c_{\theta _1} c_{\theta _2} (2 \mathcal{C}_{\phi D}^{\{21\}\{(R)(r)\}}+2 \mathcal{C}_{\phi D}^{21(R)(r)}-\mathcal{C}_{\phi D}^{R(r)1(1)})\right. $\\
			&$\left. +2 g \kappa _1^2 c_{\theta _1} c_{\theta _2} (\mathcal{C}_{\phi D}^{\{21\}\{(R)(r)\}}  +2 \mathcal{C}_{\phi D}^{(R)11(r)})-2 g \mathcal{C}_{\phi D}^{\{\text{Rr}\}\{(R)(r)\}} v_R^2 s_{\theta _1} s_{\theta _2} s_{\theta _\text{w}}+g \kappa _2^2 s_{\theta _1} s_{\theta _2} s_{\theta _\text{w}} (-2 \mathcal{C}_{\phi D}^{\{21\}\{(R)(r)\}}-2 \mathcal{C}_{\phi D}^{21(R)(r)}\right. $\\
			&$\left.+\mathcal{C}_{\phi D}^{R(r)1(1)})-2 g \kappa _1^2 s_{\theta _1} s_{\theta _2} (\mathcal{C}_{\phi D}^{\{21\}\{(R)(r)\}}+2 \mathcal{C}_{\phi D}^{(R)11(r)}) s_{\theta _\text{w}}\right)$\\
			\hline
			$\delta_{L}^{\mp \mp} \delta_{R}^{\pm \pm}Z_{2}^{\mu}$ &$\frac{1}{2 \Lambda ^2}\left(-\tilde{g} c_{\theta _2} s_{\theta _1} \left(k_{\mp \mp}^{\mu }+p_{\pm \pm}^{\mu }\right) \left(\kappa _1^2 (\mathcal{C}_{\phi D}^{(L)11(r)}+\mathcal{C}_{\phi D}^{(L)12(r)}+\mathcal{C}_{\phi D}^{(R)11(l)})-2 A_{\text{++}}^{12}\right)+c_{\theta _1} \left(g c_{\theta _2} \left(k_{\mp \mp}^{\mu } \left(\kappa _1^2 (\mathcal{C}_{\phi D}^{(L)11(r)} \right.\right.\right.\right.$\\
			&$ \left.\left.\left.\left.+\mathcal{C}_{\phi D}^{(L)12(r)}+\mathcal{C}_{\phi D}^{(R)11(l)})-A_{\text{++}}^{12}\right)-A_{\text{++}}^{12} p_{\pm \pm}^{\mu }\right)-\tilde{g} s_{\theta _2} s_{\theta _\text{w}} \left(k_{\mp \mp}^{\mu }+p_{\pm \pm}^{\mu }\right) \left(\kappa _1^2 (\mathcal{C}_{\phi D}^{(L)11(r)}+\mathcal{C}_{\phi D}^{(L)12(r)}+\mathcal{C}_{\phi D}^{(R)11(l)})-2 A_{\text{++}}^{12}\right)\right)\right. $\\
			&$\left. +g s_{\theta _2} \left(c_{\theta _\text{w}} \left(p_{\pm \pm}^{\mu } \left(\kappa _1^2 (\mathcal{C}_{\phi D}^{(L)11(r)}+\mathcal{C}_{\phi D}^{(L)12(r)}+\mathcal{C}_{\phi D}^{(R)11(l)})-A_{\text{++}}^{12}\right)-k_{\mp \mp}^{\mu } A_{\text{++}}^{12}\right)-s_{\theta _1} s_{\theta _\text{w}} \left(k_{\mp \mp}^{\mu } \left(\kappa _1^2 (\mathcal{C}_{\phi D}^{(L)11(r)}+\mathcal{C}_{\phi D}^{(L)12(r)} \right.\right.\right.\right.$\\
			&$ \left.\left.\left.\left. +\mathcal{C}_{\phi D}^{(R)11(l)})-A_{\text{++}}^{12}\right)-A_{\text{++}}^{12} p_{\pm \pm}^{\mu }\right)\right)\right)$\\
			\hline
		\end{tabular}%
	}
		\caption{MLRSM: Coupling of a pair of doubly charged scalars ($\delta^{\pm\pm}$) with a neutral gauge boson ($Z_2$). Here, $\mathit{s}_{\theta_i}=\sin \theta_i$ and $\mathit{c}_{\theta_i}=\cos \theta_i$. $k^{\mu}_{\pm \pm}$ and $p^{\mu}_{\pm \pm}$  are the four momenta of $\delta^{\pm \pm}_{L}$ and $\delta^{\pm \pm}_{R}$ respectively.  $\tilde{\theta}_{2}$ is the redefined angle defined in Eq.~\ref{eq:MLRSM-nGB-rot-angle-full}.}\label{tab:MLRSM-dcs-dcs-nGB2}
\end{table}

\begin{table}[h!]
	\centering
	{ \scriptsize
		\renewcommand{\arraystretch}{2.0}
		\begin{tabular}{|c|l|}
			\hline 
			{\bf Vertex} & {\bf  MLRSM: Effective vertex $( \delta^{\pm \pm}\,\delta^{\mp \mp}\,\gamma)$ and $( \delta^{\pm}\,\delta^{\mp}\,\gamma)$ factor}\\
			\hline
			$\delta_{L}^{--} \delta_{L}^{++}A^\mu$&$\frac{1}{4} \left(k_{\mp \mp}^{\mu}+k_{\pm \pm}^{\mu}\right) \left(4 \tilde{g} c_{\theta _1} c_{\theta_{\text{w}}}+4 g s_{\theta_{\text{w}}}\right)+\frac{1}{4 \Lambda ^2}\left(k_{\mp \mp}^{\mu}+k_{\pm \pm}^{\mu}\right) \left(-4 \tilde{g} c_{\theta _1} A_{\text{++}}^{22} c_{\theta_{\text{w}}}+4 \tilde{g} \left(\Theta_{{3LB}} s_{\theta_{\text{w}}}+\Theta_{{3RB}} s_{\theta _1} c_{\theta_{\text{w}}}\right)\right.$\\
			&$\left.+4 \tilde{g} \Theta_{{BB}} c_{\theta _1} c_{\theta_{\text{w}}}+2 \mathcal{C}_{\phi D}^{\{\text{Ll}\}\{(L)(l)\}} \tilde{g} c_{\theta _1} v_L^2 c_{\theta_{\text{w}}}+2 \kappa _2^2 \mathcal{C}_{\phi D}^{\{21\}\{(L)(l)\}} \tilde{g} c_{\theta _1} c_{\theta_{\text{w}}}+2 \kappa _2^2 \mathcal{C}_{\phi D}^{21(L)(l)} \tilde{g} c_{\theta _1} c_{\theta_{\text{w}}}\right.$\\
			&$\left.+4 \kappa _1^2 \mathcal{C}_{\phi D}^{(L)11(l)} \tilde{g} c_{\theta _1} c_{\theta_{\text{w}}}-4 g A_{\text{++}}^{22} s_{\theta_{\text{w}}}+4 g \Theta_{{3L3L}} s_{\theta_{\text{w}}}+g c_{\theta_{\text{w}}} \left(s_{\theta _1} \left(4 \Theta_{{3L3R}}-\kappa _2^2 \mathcal{C}_{\phi D}^{L(l)1(1)}\right)+4 \Theta_{{3LB}} c_{\theta _1}\right)\right.$\\
			&$\left.+2 g \mathcal{C}_{\phi D}^{\{\text{Ll}\}\{(L)(l)\}} v_L^2 s_{\theta_{\text{w}}}+2 g \kappa _+^2 \mathcal{C}_{\phi D}^{\{21\}\{(L)(l)\}} s_{\theta_{\text{w}}}+2 g \kappa _2^2 \mathcal{C}_{\phi D}^{21(L)(l)} s_{\theta_{\text{w}}}+4 g \kappa _1^2 \mathcal{C}_{\phi D}^{(L)11(l)} s_{\theta_{\text{w}}}+g \kappa _2^2 \mathcal{C}_{\phi D}^{L(l)1(1)} s_{\theta_{\text{w}}}\right)$\\			
			\hline
			$\delta_{R}^{--} \delta_{R}^{++}A^\mu$&$\frac{1}{4} \left(p_{\mp \mp}^{\mu}+p_{\pm \pm}^{\mu}\right) \left(4 \tilde{g} c_{\theta _1} c_{\theta_{\text{w}}}+4 g s_{\theta _1} c_{\theta_{\text{w}}}\right)+\frac{1}{4 \Lambda ^2}\left(p_{\mp \mp}^{\mu}+p_{\pm \pm}^{\mu}\right) \left(-4 \tilde{g} c_{\theta _1} A_{\text{++}}^{11} c_{\theta_{\text{w}}}+4 \tilde{g} \left(\Theta_{{3LB}} s_{\theta_{\text{w}}}+\Theta_{{3RB}} s_{\theta _1} c_{\theta_{\text{w}}}\right)\right.$\\
			&$\left.+4 \tilde{g} \Theta_{{BB}} c_{\theta _1} c_{\theta_{\text{w}}}+2 \mathcal{C}_{\phi D}^{\{\text{Rr}\}\{(R)(r)\}} \tilde{g} c_{\theta _1} v_R^2 c_{\theta_{\text{w}}}+2 \kappa _+^2 \mathcal{C}_{\phi D}^{\{21\}\{(R)(r)\}} \tilde{g} c_{\theta _1} c_{\theta_{\text{w}}}+2 \kappa _2^2 \mathcal{C}_{\phi D}^{21(R)(r)} \tilde{g} c_{\theta _1} c_{\theta_{\text{w}}}+4 \kappa _1^2 \mathcal{C}_{\phi D}^{(R)11(r)} \tilde{g} c_{\theta _1} c_{\theta_{\text{w}}}\right.$\\
			&$\left.-4 g s_{\theta _1} A_{\text{++}}^{11} c_{\theta_{\text{w}}}+g s_{\theta_{\text{w}}} \left(4 \Theta_{{3L3R}}+\kappa _2^2 \mathcal{C}_{\phi D}^{R(r)1(1)}\right)+4 g \Theta_{{3R3R}} s_{\theta _1} c_{\theta_{\text{w}}}+4 g \Theta_{{3RB}} c_{\theta _1} c_{\theta_{\text{w}}}+2 g \mathcal{C}_{\phi D}^{\{\text{Rr}\}\{(R)(r)\}} v_R^2 s_{\theta _1} c_{\theta_{\text{w}}}\right.$\\
			&$\left.+2 g \kappa _+^2 \mathcal{C}_{\phi D}^{\{21\}\{(R)(r)\}} s_{\theta _1} c_{\theta_{\text{w}}}+2 g \kappa _2^2 \mathcal{C}_{\phi D}^{21(R)(r)} s_{\theta _1} c_{\theta_{\text{w}}}+4 g \kappa _1^2 \mathcal{C}_{\phi D}^{(R)11(r)} s_{\theta _1} c_{\theta_{\text{w}}}-g \kappa _2^2 \mathcal{C}_{\phi D}^{R(r)1(1)} s_{\theta _1} c_{\theta_{\text{w}}}\right) $\\
			\hline
			$\delta_{L}^{\mp \mp} \delta_{R}^{\pm \pm} A^\mu$ &$-\frac{1}{2 \Lambda ^2}\tilde{g} c_{\theta _1} c_{\theta_{\text{w}}} \left(k_{\mp \mp}^{\mu}+p_{\pm \pm}^{\mu}\right) \left(2 A_{\text{++}}^{12}-\kappa _1^2 (\mathcal{C}_{\phi D}^{(L)11(r)}+\mathcal{C}_{\phi D}^{(L)12(r)}+\mathcal{C}_{\phi D}^{(R)11(l)})\right)+g A_{\text{++}}^{12} \left(k_{\mp \mp}^{\mu}+p_{\pm \pm}^{\mu}\right) \left(s_{\theta _1} c_{\theta_{\text{w}}}+s_{\theta_{\text{w}}}\right)$\\
			&$-g \kappa _1^2 (\mathcal{C}_{\phi D}^{(L)11(r)}+\mathcal{C}_{\phi D}^{(L)12(r)}+\mathcal{C}_{\phi D}^{(R)11(l)}) \left(k_{\mp \mp}^{\mu} s_{\theta _1} c_{\theta_{\text{w}}}+p_{\pm \pm}^{\mu} s_{\theta_{\text{w}}}\right)$\\
			\hline \hline
			$\delta_{L}^{\mp} \delta_{L}^{\pm}A^\mu$ &$\tilde{g} c_{\theta _1} \left(k_{\mp}^{\mu}+k_{\pm}^{\mu}\right) c_{\theta_{\text{w}}}-\frac{1}{8 \Lambda ^2}\left(k_{\mp}^{\mu}+k_{\pm}^{\mu}\right) \left(8 \tilde{g} c_{\theta _1} A_+^{44} c_{\theta_{\text{w}}}-8 \tilde{g} \left(\Theta_{{3LB}} s_{\theta_{\text{w}}}+\Theta_{{3RB}} s_{\theta _1} c_{\theta_{\text{w}}}\right)-8 \tilde{g} \Theta_{{BB}} c_{\theta _1} c_{\theta_{\text{w}}}\right. $\\
			&$\left.-4 \mathcal{C}_{\phi D}^{\{\text{Ll}\}\{(L)(l)\}} \tilde{g} c_{\theta _1} v_L^2 c_{\theta_{\text{w}}}-2 \mathcal{C}_{\phi D}^{(L)\text{rR}(l)} \tilde{g} c_{\theta _1} v_R^2 c_{\theta_{\text{w}}}-4 \kappa _+^2 \mathcal{C}_{\phi D}^{\{21\}\{(L)(l)\}} \tilde{g} c_{\theta _1} c_{\theta_{\text{w}}}-2 \kappa _+^2 \mathcal{C}_{\phi D}^{21(L)(l)} \tilde{g} c_{\theta _1} c_{\theta_{\text{w}}}\right.$\\
			&$\left.-4 \kappa _+^2 \mathcal{C}_{\phi D}^{(L)11(l)} \tilde{g} c_{\theta _1} c_{\theta_{\text{w}}}+g \left(\kappa _2^2-\kappa _1^2\right) \mathcal{C}_{\phi D}^{L(l)1(1)} s_{\theta _1} c_{\theta_{\text{w}}}+g s_{\theta_{\text{w}}} \left(8 C_{\square }^{\text{Ll}(\text{Ll})} v_L^2+\kappa _-^2 \mathcal{C}_{\phi D}^{L(l)1(1)}\right)\right)$\\
			\hline
			$\delta_{R}^{\mp} \delta_{R}^{\pm}A^\mu$ &$\tilde{g} c_{\theta _1} \left(p_{\mp}^{\mu}+p_{\pm}^{\mu}\right) c_{\theta_{\text{w}}}-\frac{1}{8 \Lambda ^2}\left(p_{\mp}^{\mu}+p_{\pm}^{\mu}\right) \left(8 \tilde{g} c_{\theta _1} A_+^{33} c_{\theta_{\text{w}}}-8 \tilde{g} \left(\Theta_{{3LB}} s_{\theta_{\text{w}}}+\Theta_{{3RB}} s_{\theta _1} c_{\theta_{\text{w}}}\right)-8 \tilde{g} \Theta_{{BB}} c_{\theta _1} c_{\theta_{\text{w}}}\right.$\\
			&$\left.-2 \mathcal{C}_{\phi D}^{(R)\text{lL}(r)} \tilde{g} c_{\theta _1} v_L^2 c_{\theta_{\text{w}}}-4 \mathcal{C}_{\phi D}^{\{\text{Rr}\}\{(R)(r)\}} \tilde{g} c_{\theta _1} v_R^2 c_{\theta_{\text{w}}}-16 \mathcal{C}_{\phi D}^{R(r)R(r)} \tilde{g} c_{\theta _1} v_R^2 c_{\theta_{\text{w}}}-4 \kappa _+^2 \mathcal{C}_{\phi D}^{\{21\}\{(R)(r)\}} \tilde{g} c_{\theta _1} c_{\theta_{\text{w}}}\right.$\\
			&$\left.-2 \kappa _+^2 \mathcal{C}_{\phi D}^{21(R)(r)} \tilde{g} c_{\theta _1} c_{\theta_{\text{w}}}-4 \kappa _+^2 \mathcal{C}_{\phi D}^{(R)11(r)} \tilde{g} c_{\theta _1} c_{\theta_{\text{w}}}+g s_{\theta _1} c_{\theta_{\text{w}}} \left(4 v_R^2 (3 \mathcal{C}_{\phi D}^{R(r)R(r)}+C_{\square }^{\text{Rr}(\text{Rr})})-\kappa _-^2 \mathcal{C}_{\phi D}^{R(r)1(1)}\right)+g \kappa _-^2 \mathcal{C}_{\phi D}^{R(r)1(1)} s_{\theta_{\text{w}}}\right)$\\	
			\hline
			$\delta_{L}^{\mp} \delta_{R}^{\pm}A^\mu$ &$-\frac{1}{4 \Lambda ^2}\tilde{g} c_{\theta _1} c_{\theta_{\text{w}}} \left(4 A_+^{34} \left(k_{\mp}^{\mu}+p_{\pm}^{\mu}\right)-k_{\mp}^{\mu} \left(v_L v_R (2 \mathcal{C}_{\phi D}^{(L)\text{rL}(r)}+\mathcal{C}_{\phi D}^{(L)\text{rR}(l)})+\kappa _+^2 (\mathcal{C}_{\phi D}^{(L)11(r)}+\mathcal{C}_{\phi D}^{(L)12(r)}+\mathcal{C}_{\phi D}^{(R)11(l)})\right)\right.$\\
			&$\left.-p_{\pm}^{\mu} \left(v_L v_R (2 \mathcal{C}_{\phi D}^{(L)\text{rL}(r)}+\mathcal{C}_{\phi D}^{(R)\text{lL}(r)})+\kappa _+^2 (\mathcal{C}_{\phi D}^{(L)11(r)}+\mathcal{C}_{\phi D}^{(L)12(r)}+\mathcal{C}_{\phi D}^{(R)11(l)})\right)\right)$\\
			&$+g v_L v_R \left(s_{\theta _1} c_{\theta_{\text{w}}} \left(\mathcal{C}_{\phi D}^{(L)\text{rL}(r)} k_{\mp}^{\mu}+\mathcal{C}_{\phi D}^{(R)\text{lL}(r)} p_{\pm}^{\mu}\right)+s_{\theta_{\text{w}}} \left(\mathcal{C}_{\phi D}^{(L)\text{rR}(l)} k_{\mp}^{\mu}+\mathcal{C}_{\phi D}^{(L)\text{rL}(r)} p_{\pm}^{\mu}\right)\right)$\\
			\hline
		\end{tabular}%
	}
	\caption{MLRSM: Coupling of a pair of doubly charged scalars ($\delta^{\pm\pm}$) and  singly charged scalars ($\delta^{\pm}$) with photon ($A^\mu$). Here, $\mathit{s}_{\theta_i}=\sin \theta_i$ and $\mathit{c}_{\theta_i}=\cos \theta_i$. $k^{\mu}_{\pm \pm}$, $p^{\mu}_{\pm \pm}$, $k^{\mu}_{\pm}$ and $p^{\mu}_{\pm}$ are the four momenta of $\delta^{\pm \pm}_{L}$, $\delta^{\pm \pm}_{R}$, $\delta_{L}^{\pm}$ and $\delta_{R}^{\pm}$ respectively. $\tilde{\theta}_{2}$ is the redefined angle defined in Eq.~\ref{eq:MLRSM-nGB-rot-angle-full}.}\label{tab:MLRSM-dcs-dcs-gamma}
\end{table}

\begin{table}[h!]
	\centering
	{ \scriptsize
		\renewcommand{\arraystretch}{2.45}
		\begin{tabular}{|l|l|}
			\hline 
			{\bf Vertex} & {\bf  MLRSM: Effective vertex $( \delta^{\mp \mp}\,\delta^{\pm \pm}\,W_3[B]\,W_3[B])$ factor}\\			
			\hline \hline
			$\delta^{--}_{L} \delta^{++}_{L} W^{\mu}_{3L} W_{3L\mu}$&$g^2+\frac{g}{8 \Lambda ^2} \left(16 \tilde{g} \Theta_{{3LB}}-8 g A_{\text{++}}^{22}+16 g \Theta_{{3L3L}}+8 g \mathcal{C}_{\phi D}^{\{\text{Ll}\}\{(L)(l)\}} v_L^2+4 g \kappa _1^2 \mathcal{C}_{\phi D}^{\{21\}\{(L)(l)\}}+4 g \kappa _2^2 \mathcal{C}_{\phi D}^{\{21\}\{(L)(l)\}} \right.$\\
			&$\left.+4 g \kappa _2^2 \mathcal{C}_{\phi D}^{21(L)(l)}+8 g \kappa _1^2 \mathcal{C}_{\phi D}^{(L)11(l)}+4 g \kappa _2^2 \mathcal{C}_{\phi D}^{L(l)1(1)}+g \kappa _1^2 \mathcal{C}_{\phi D}^{\{\text{Ll}\}\{(2)(1)\}}+g \kappa _2^2 \mathcal{C}_{\phi D}^{\{\text{Ll}\}\{(2)(1)\}}+g \kappa _2^2 \mathcal{C}_{\phi D}^{\text{Ll}(2)(1)}\right)$\\
			\hline
			$\delta^{\mp \mp}_{L} \delta^{\pm \pm}_{R} W^{\mu}_{3L} W_{3L\mu}$&$\frac{g^2}{8 \Lambda ^2} \left(\kappa _1^2 \mathcal{C}_{\phi D}^{(2)\text{rL}(1)}-4 A_{\text{++}}^{12}\right)$\\
			\hline
			$\delta^{--}_{R} \delta^{++}_{R} W^{\mu}_{3L} W_{3L\mu}$&$\frac{g^2}{8 \Lambda ^2} \left(\kappa _1^2 \mathcal{C}_{\phi D}^{\{\text{Rr}\}\{(2)(1)\}}+\kappa _2^2 (\mathcal{C}_{\phi D}^{\{\text{Rr}\}\{(2)(1)\}}+\mathcal{C}_{\phi D}^{\text{Rr}(2)(1)})\right)$\\
			\hline \hline
			$\delta^{--}_{L} \delta^{++}_{L} W^{\mu}_{3R} W_{3R\mu}$&$\frac{g^2}{8 \Lambda ^2} \left(\kappa _1^2 \mathcal{C}_{\phi D}^{\{\text{Ll}\}\{(2)(1)\}}+\kappa _2^2 (\mathcal{C}_{\phi D}^{\{\text{Ll}\}\{(2)(1)\}}+\mathcal{C}_{\phi D}^{\text{Ll}(2)(1)})\right)$\\
			\hline
			$\delta^{\mp \mp}_{L} \delta^{\pm \pm}_{R} W^{\mu}_{3R} W_{3R\mu}$&$\frac{g^2}{8 \Lambda ^2} \left(\kappa _1^2 \mathcal{C}_{\phi D}^{(2)\text{rL}(1)}-4 A_{\text{++}}^{12}\right)$\\
			\hline
			$\delta^{--}_{R} \delta^{++}_{R} W^{\mu}_{3R} W_{3R\mu}$&$g^2+\frac{g}{8 \Lambda ^2} \left(16 \tilde{g} \Theta_{{3RB}}-8 g A_{\text{++}}^{11}+16 g \Theta_{{3R3R}}+8 g \mathcal{C}_{\phi D}^{\{\text{Rr}\}\{(R)(r)\}} v_R^2+4 g \kappa _1^2 \mathcal{C}_{\phi D}^{\{21\}\{(R)(r)\}}+4 g \kappa _2^2 \mathcal{C}_{\phi D}^{\{21\}\{(R)(r)\}}\right.$\\
			&$\left.+4 g \kappa _2^2 \mathcal{C}_{\phi D}^{21(R)(r)}+8 g \kappa _1^2 \mathcal{C}_{\phi D}^{(R)11(r)}-4 g \kappa _2^2 \mathcal{C}_{\phi D}^{R(r)1(1)}+g \kappa _1^2 \mathcal{C}_{\phi D}^{\{\text{Rr}\}\{(2)(1)\}}+g \kappa _2^2 \mathcal{C}_{\phi D}^{\{\text{Rr}\}\{(2)(1)\}}+g \kappa _2^2 \mathcal{C}_{\phi D}^{\text{Rr}(2)(1)}\right)$\\
			\hline \hline
			$\delta^{--}_{L} \delta^{++}_{L} B^{\mu} B_{\mu} $&$\tilde{g}^2+\frac{\tilde{g}}{2 \Lambda ^2} \left(-2 \tilde{g} A_{\text{++}}^{22}+4 \tilde{g} \Theta_{{BB}}+2 \mathcal{C}_{\phi D}^{\{\text{Ll}\}\{(L)(l)\}} \tilde{g} v_L^2+\kappa _2^2 \tilde{g} (\mathcal{C}_{\phi D}^{\{21\}\{(L)(l)\}}+\mathcal{C}_{\phi D}^{21(L)(l)})+\kappa _1^2 \tilde{g} (\mathcal{C}_{\phi D}^{\{21\}\{(L)(l)\}}\right.$\\
			&$\left.+2 \mathcal{C}_{\phi D}^{(L)11(l)})+4 g \Theta_{{3LB}}\right)$\\
			\hline
			$\delta^{\mp \mp}_{L} \delta^{\pm \pm}_{R} B^{\mu} B_{\mu} $&$\frac{\tilde{g}^2}{2 \Lambda ^2} \left(\kappa _1^2 (\mathcal{C}_{\phi D}^{(L)11(r)}+\mathcal{C}_{\phi D}^{(L)12(r)}+\mathcal{C}_{\phi D}^{(R)11(l)})-2 A_{\text{++}}^{12}\right)$\\
			\hline
			$\delta^{--}_{R} \delta^{++}_{R} B^{\mu} B_{\mu} $&$\tilde{g}^2+\frac{\tilde{g}}{2 \Lambda ^2} \left(-2 \tilde{g} A_{\text{++}}^{11}+4 \tilde{g} \Theta_{{BB}}+2 \mathcal{C}_{\phi D}^{\{\text{Rr}\}\{(R)(r)\}} \tilde{g} v_R^2+\kappa _1^2 \mathcal{C}_{\phi D}^{\{21\}\{(R)(r)\}} \tilde{g}+\kappa _2^2 \mathcal{C}_{\phi D}^{\{21\}\{(R)(r)\}} \tilde{g}+\kappa _2^2 \mathcal{C}_{\phi D}^{21(R)(r)} \tilde{g} \right.$\\
			&$\left. +2 \kappa _1^2 \mathcal{C}_{\phi D}^{(R)11(r)} \tilde{g}+4 g \Theta_{{3RB}}\right)$\\
			\hline \hline
			$\delta^{--}_{L} \delta^{++}_{L} W^{\mu}_{3L} W_{3R\mu}$&$\frac{g}{4 \Lambda ^2} \left(8 \tilde{g} \Theta_{{3RB}}+8 g \Theta_{{3L3R}}-g \left(\kappa _2^2 (2 \mathcal{C}_{\phi D}^{L(l)1(1)}+\mathcal{C}_{\phi D}^{\{\text{Ll}\}\{(2)(1)\}}+\mathcal{C}_{\phi D}^{\text{Ll}(2)(1)})+\kappa _1^2 \mathcal{C}_{\phi D}^{\{\text{Ll}\}\{(2)(1)\}}\right)\right)$\\
			\hline
			$\delta^{\mp \mp}_{L} \delta^{\pm \pm}_{R} W^{\mu}_{3L} W_{3R\mu}$&$-\frac{g^2 \kappa _1^2}{4 \Lambda ^2} (\mathcal{C}_{\phi D}^{(2)\text{rL}(1)}-2 (\mathcal{C}_{\phi D}^{(L)11(r)}+\mathcal{C}_{\phi D}^{(L)12(r)}+\mathcal{C}_{\phi D}^{(R)11(l)}))$\\
			\hline
			$\delta^{--}_{R} \delta^{++}_{R} W^{\mu}_{3L} W_{3R\mu}$&$\frac{g}{4 \Lambda ^2} \left(8 \tilde{g} \Theta_{{3LB}}+8 g \Theta_{{3L3R}}-g \left(\kappa _2^2 (-2 \mathcal{C}_{\phi D}^{R(r)1(1)}+\mathcal{C}_{\phi D}^{\{\text{Rr}\}\{(2)(1)\}}+\mathcal{C}_{\phi D}^{\text{Rr}(2)(1)})+\kappa _1^2 \mathcal{C}_{\phi D}^{\{\text{Rr}\}\{(2)(1)\}}\right)\right)$\\
			\hline \hline
			$\delta^{--}_{L} \delta^{++}_{L} W^{\mu}_{3L} B_{\mu}$&$2 g \tilde{g}+\frac{1}{2 \Lambda ^2}\left(-4 g \tilde{g} A_{\text{++}}^{22}+4 g \tilde{g} \Theta_{{3L3L}}+4 \tilde{g}^2 \Theta_{{3LB}}+4 g \tilde{g} \Theta_{{BB}}+2 g \kappa _1^2 \mathcal{C}_{\phi D}^{\{21\}\{(L)(l)\}} \tilde{g}\right.$\\
			&$\left.+2 g \kappa _2^2 \mathcal{C}_{\phi D}^{\{21\}\{(L)(l)\}} \tilde{g}+2 g \kappa _2^2 \mathcal{C}_{\phi D}^{21(L)(l)} \tilde{g}+4 g \kappa _1^2 \mathcal{C}_{\phi D}^{(L)11(l)} \tilde{g}+g \kappa _2^2 \mathcal{C}_{\phi D}^{L(l)1(1)} \tilde{g}+4 g^2 \Theta_{{3LB}}\right)$\\
			\hline
			$\delta^{\mp \mp}_{L} \delta^{\pm \pm}_{R} W^{\mu}_{3L} B_{\mu}$&$\frac{g \tilde{g}}{2 \Lambda ^2} \left(\kappa _1^2 (\mathcal{C}_{\phi D}^{(L)11(r)}+\mathcal{C}_{\phi D}^{(L)12(r)}+\mathcal{C}_{\phi D}^{(R)11(l)})-2 A_{\text{++}}^{12}\right)$\\
			\hline
			$\delta^{--}_{R} \delta^{++}_{R} W^{\mu}_{3L} B_{\mu}$&$\frac{\tilde{g}}{2 \Lambda ^2} \left(4 \tilde{g} \Theta_{{3LB}}+4 g \Theta_{{3L3R}}+g \kappa _2^2 \mathcal{C}_{\phi D}^{R(r)1(1)}\right)$\\
			\hline \hline
			$\delta^{--}_{L} \delta^{++}_{L} W^{\mu}_{3R} B_{\mu} $&$\frac{\tilde{g}}{2 \Lambda ^2} \left(4 \tilde{g} \Theta_{{3RB}}+4 g \Theta_{{3L3R}}+g \kappa _2^2 (-\mathcal{C}_{\phi D}^{L(l)1(1)})\right)$\\
			\hline
			$\delta^{\mp \mp}_{L} \delta^{\pm \pm}_{R} W^{\mu}_{3R} B_{\mu} $&$\frac{g \tilde{g}}{2 \Lambda ^2} \left(\kappa _1^2 (\mathcal{C}_{\phi D}^{(L)11(r)}+\mathcal{C}_{\phi D}^{(L)12(r)}+\mathcal{C}_{\phi D}^{(R)11(l)})-2 A_{\text{++}}^{12}\right)$\\
			\hline
			$\delta^{--}_{R} \delta^{++}_{R} W^{\mu}_{3R} B_{\mu} $&$2 g \tilde{g}+\frac{1}{2 \Lambda ^2}\left(-4 g \tilde{g} A_{\text{++}}^{11}+4 g \tilde{g} \Theta_{{3R3R}}+4 \tilde{g}^2 \Theta_{{3RB}}+4 g \tilde{g} \Theta_{{BB}}+2 g \kappa _1^2 \mathcal{C}_{\phi D}^{\{21\}\{(R)(r)\}} \tilde{g}+2 g \kappa _2^2 \mathcal{C}_{\phi D}^{\{21\}\{(R)(r)\}} \tilde{g}\right.$\\
			&$\left.+2 g \kappa _2^2 \mathcal{C}_{\phi D}^{21(R)(r)} \tilde{g}+4 g \kappa _1^2 \mathcal{C}_{\phi D}^{(R)11(r)} \tilde{g}-g \kappa _2^2 \mathcal{C}_{\phi D}^{R(r)1(1)} \tilde{g}+4 g^2 \Theta_{{3RB}}\right)$\\
			\hline
		\end{tabular}
	}
		\caption{MLRSM: Coupling of a pair of doubly charged scalars ($\delta^{\pm\pm}$) with a pair two of neutral gauge bosons ($B, W_{3L,3R}$) in unphysical basis. The vertices can be written in terms of ($A, Z_{1,2}$) using rotation matrix $\tilde{\mathcal{R}}_{b}$.}\label{tab:MLRSM-dcs-dcs-nGB-nGB}
\end{table}

\begin{table}[h!]
	\centering
	{ \scriptsize
		\renewcommand{\arraystretch}{2.6}
		\begin{tabular}{|l|l|}
			\hline 
			{\bf Vertex} & {\bf MLRSM: Effective vertex $( \delta^{\mp\mp}\,\delta^{\pm \pm}\,W^{\mp}\,W^{\pm})$ factor}\\			
			\hline \hline
			$\delta^{\mp \mp}_{L} \delta^{\pm \pm}_{L} W_{L}^{\pm\mu} W_{L\mu}^{\mp}$&$g^2+\frac{g^2}{4 \Lambda ^2}\left(-4 A_{\text{++}}^{22}+4 v_L^2 (\mathcal{C}_{\phi D}^{\{\text{Ll}\}\{(L)(l)\}}+3 \mathcal{C}_{\phi D}^{L(l)L(l)}+\mathcal{C}_{\square }^{\text{Ll(Ll)}})+\mathcal{C}_{\phi D}^{(L)\text{rR}(l)} v_R^2+8 \Theta _{W_{\text{LL}}}+2 \kappa _+^2 \mathcal{C}_{\phi D}^{\{21\}\{(L)(l)\}}\right.$\\
			&$\left.+\kappa _+^2 \mathcal{C}_{\phi D}^{21(L)(l)}+2 \kappa _+^2 \mathcal{C}_{\phi D}^{(L)11(l)}+2 \kappa _1 \kappa _2 \mathcal{C}_{\phi D}^{L(l)1(1)}+\kappa _+^2 \mathcal{C}_{\phi D}^{\{\text{Ll}\}\{(2)(1)\}}+\kappa _2^2 \mathcal{C}_{\phi D}^{\text{Ll}(2)(1)}\right)$\\
			\hline
			$\delta^{\mp \mp}_{L} \delta^{\pm \pm}_{R} W_{L}^{\pm\mu} W_{L\mu}^{\mp}$&$\frac{g^2}{4 \Lambda ^2} \left(\kappa _2^2 \mathcal{C}_{\phi D}^{(2)\text{rL}(1)}-2 A_{\text{++}}^{12}\right)$\\
			\hline
			$\delta^{\mp \mp}_{R} \delta^{\pm \pm}_{R} W_{L}^{\pm\mu} W_{L\mu}^{\mp}$&$\frac{g^2}{4 \Lambda ^2} \left(\mathcal{C}_{\phi D}^{(L)\text{rR}(l)} v_L^2+\kappa _1^2 \mathcal{C}_{\phi D}^{\{\text{Rr}\}\{(2)(1)\}}+\kappa _2^2 (\mathcal{C}_{\phi D}^{\{\text{Rr}\}\{(2)(1)\}}+\mathcal{C}_{\phi D}^{\text{Rr}(2)(1)})\right)$\\
			\hline \hline
			$\delta^{\mp \mp}_{L} \delta^{\pm \pm}_{L} W_{R}^{\pm\mu} W_{R\mu}^{\mp}$&$\frac{g^2}{4 \Lambda ^2} \left(\mathcal{C}_{\phi D}^{(R)\text{lL}(r)} v_R^2+\kappa _2^2 \mathcal{C}_{\phi D}^{\{\text{Ll}\}\{(2)(1)\}}+\kappa _1^2 (\mathcal{C}_{\phi D}^{\{\text{Ll}\}\{(2)(1)\}}+\mathcal{C}_{\phi D}^{\text{Ll}(2)(1)})\right)$\\
			\hline
			$\delta^{\mp \mp}_{L} \delta^{\pm \pm}_{R} W_{R}^{\pm\mu} W_{R\mu}^{\mp}$&$\frac{g^2}{4 \Lambda ^2} \left(\kappa _1^2 \mathcal{C}_{\phi D}^{(2)\text{rL}(1)}-2 A_{\text{++}}^{12}\right)$\\
			\hline
			$\delta^{\mp \mp}_{R} \delta^{\pm \pm}_{R} W_{R}^{\pm\mu} W_{R\mu}^{\mp}$&$g^2+\frac{g^2}{4 \Lambda ^2} \left(-4 A_{\text{++}}^{11}+\mathcal{C}_{\phi D}^{(R)\text{lL}(r)} v_L^2+4 v_R^2 (\mathcal{C}_{\phi D}^{\{\text{Rr}\}\{(R)(r)\}}+3 \mathcal{C}_{\phi D}^{R(r)R(r)}+\mathcal{C}_{\square }^{\text{Rr(Rr)}})+8 \Theta _{W_{\text{RR}}}+2 \kappa _+^2 \mathcal{C}_{\phi D}^{\{21\}\{(R)(r)\}}\right.$\\
			&$\left.+\kappa _+^2 \mathcal{C}_{\phi D}^{21(R)(r)}+2 \kappa _+^2 \mathcal{C}_{\phi D}^{(R)11(r)}-2 \kappa _1^2 \mathcal{C}_{\phi D}^{R(r)1(1)}+\kappa _+^2 \mathcal{C}_{\phi D}^{\{\text{Rr}\}\{(2)(1)\}}+\kappa _1^2 \mathcal{C}_{\phi D}^{\text{Rr}(2)(1)}\right)$\\
			\hline \hline
			$\delta^{\mp \mp}_{L} \delta^{\pm \pm}_{L} W_{L}^{\pm\mu} W_{R\mu}^{\mp}$&$\frac{g^2}{4 \Lambda ^2} \left(4 \Theta _{W_{\text{LR}}}-\kappa _1 \left(\kappa _1 \mathcal{C}_{\phi D}^{L(l)1(1)}+\kappa _2 (2 \mathcal{C}_{\phi D}^{\{\text{Ll}\}\{(2)(1)\}}+\mathcal{C}_{\phi D}^{\text{Ll}(2)(1)})\right)\right)$\\
			\hline
			$\delta^{\mp \mp}_{L} \delta^{\pm \pm}_{R} W_{L}^{\pm\mu} W_{R\mu}^{\mp}$&$\frac{g^2}{4 \Lambda ^2} \left(\mathcal{C}_{\phi D}^{(L)\text{rL}(r)} v_L v_R-\kappa _1 \kappa _2 \mathcal{C}_{\phi D}^{(2)\text{rL}(1)}\right)$\\
			\hline
			$\delta^{\mp \mp}_{R} \delta^{\pm \pm}_{R} W_{L}^{\pm\mu} W_{R\mu}^{\mp}$&$\frac{g^2}{4 \Lambda ^2} \left(4 \Theta _{W_{\text{RL}}}+\kappa _1 \kappa _2 (\mathcal{C}_{\phi D}^{R(r)1(1)}-2 \mathcal{C}_{\phi D}^{\{\text{Rr}\}\{(2)(1)\}}-\mathcal{C}_{\phi D}^{\text{Rr}(2)(1)})\right)$\\
			\hline
		\end{tabular}
	}
		\caption{MLRSM: Coupling of a pair of doubly charged scalars ($\delta^{\pm\pm}$) with a pair of two charged gauge bosons ($W^{\pm}_{L,R}$) in unphysical basis. These can be easily translated in physical gauge boson basis using rotation matrix $\widetilde{\mathcal{R}}_{a}$.}\label{tab:MLRSM-dcs-dcs-cGB-cGB}
\end{table}

\begin{table}[h]
	\centering
	{ \scriptsize
		\renewcommand{\arraystretch}{2.3}
		\begin{tabular}{|c|l|}
			\hline 
			{\bf Vertex} & {\bf  MLRSM: Effective vertex $( \delta^{\mp}\,\delta^{\pm}\,Z_{1,2})$ factor}\\
			\hline
			$\delta_{L}^{-} \delta_{L}^{+}Z_{1}^{\mu}$&$-\frac{1}{8} \left(k^{\mu}_{-}+k^{\mu}_{+}\right) \left(8 \tilde{g} c_{\theta _1} c_{\tilde{\theta} _2} s_{\theta _\text{w}}-8 \tilde{g} s_{\theta _1} s_{\tilde{\theta} _2}\right)-\frac{1}{8 \Lambda ^2}\left(k^{\mu}_{-}+k^{\mu}_{+}\right) \left[-8 \tilde{g} c_{\theta _1} c_{\theta _2} s_{\theta _\text{w}}A_+^{44} +8 \tilde{g} s_{\theta _1} s_{\theta _2} A_+^{44}-8 \tilde{g} c_{\theta _2} c_{\theta _\text{w}}\Theta_{{3LB}}\right.$\\
			&$\left.+8 \tilde{g}  c_{\theta _1} s_{\theta _2}\Theta_{{3RB}}+8 \tilde{g} c_{\theta _2} s_{\theta _1} s_{\theta _\text{w}}\Theta_{{3RB}}+8 \tilde{g} c_{\theta _1} c_{\theta _2} s_{\theta _\text{w}}\Theta_{{BB}}-8 \tilde{g} s_{\theta _1} s_{\theta _2}\Theta_{{BB}}+4 \mathcal{C}_{\phi D}^{\{\text{Ll}\}\{(L)(l)\}} \tilde{g} c_{\theta _1} c_{\theta _2} v_L^2 s_{\theta _\text{w}}\right.$\\
			&$\left.+2 \mathcal{C}_{\phi D}^{(L)\text{rR}(l)} \tilde{g} c_{\theta _1} c_{\theta _2} v_R^2 s_{\theta _\text{w}}+4 \kappa _1^2 \mathcal{C}_{\phi D}^{\{21\}\{(L)(l)\}} \tilde{g} c_{\theta _1} c_{\theta _2} s_{\theta _\text{w}}+4 \kappa _2^2 \mathcal{C}_{\phi D}^{\{21\}\{(L)(l)\}} \tilde{g} c_{\theta _1} c_{\theta _2} s_{\theta _\text{w}}+2 \kappa _1^2 \mathcal{C}_{\phi D}^{21(L)(l)} \tilde{g} c_{\theta _1} c_{\theta _2} s_{\theta _\text{w}}\right.$\\
			&$\left.+2 \kappa _2^2 \mathcal{C}_{\phi D}^{21(L)(l)} \tilde{g} c_{\theta _1} c_{\theta _2} s_{\theta _\text{w}}+4 \kappa _1^2 \mathcal{C}_{\phi D}^{(L)11(l)} \tilde{g} c_{\theta _1} c_{\theta _2} s_{\theta _\text{w}}+4 \kappa _2^2 \mathcal{C}_{\phi D}^{(L)11(l)} \tilde{g} c_{\theta _1} c_{\theta _2} s_{\theta _\text{w}}-4 \mathcal{C}_{\phi D}^{\{\text{Ll}\}\{(L)(l)\}} \tilde{g} v_L^2 s_{\theta _1} s_{\theta _2}\right.$\\
			&$\left.-2 \mathcal{C}_{\phi D}^{(L)\text{rR}(l)} \tilde{g} v_R^2 s_{\theta _1} s_{\theta _2}-4 \kappa _1^2 \mathcal{C}_{\phi D}^{\{21\}\{(L)(l)\}} \tilde{g} s_{\theta _1} s_{\theta _2}-4 \kappa _2^2 \mathcal{C}_{\phi D}^{\{21\}\{(L)(l)\}} \tilde{g} s_{\theta _1} s_{\theta _2}-2 \kappa _1^2 \mathcal{C}_{\phi D}^{21(L)(l)} \tilde{g} s_{\theta _1} s_{\theta _2}-2 \kappa _2^2 \mathcal{C}_{\phi D}^{21(L)(l)} \tilde{g} s_{\theta _1} s_{\theta _2}\right.$\\
			&$\left.-4 \kappa _1^2 \mathcal{C}_{\phi D}^{(L)11(l)} \tilde{g} s_{\theta _1} s_{\theta _2}-4 \kappa _2^2 \mathcal{C}_{\phi D}^{(L)11(l)} \tilde{g} s_{\theta _1} s_{\theta _2}+g \left(c_{\theta _2} \left(c_{\theta _\text{w}} \left(v_L^28 \mathcal{C}_{\square }^{\text{Ll}(\text{Ll})} +\kappa _1^2 \mathcal{C}_{\phi D}^{L(l)1(1)}-\kappa _2^2 \mathcal{C}_{\phi D}^{L(l)1(1)}\right)\right.\right.\right.$\\
			&$\left.\left.\left.+\kappa _-^2 \mathcal{C}_{\phi D}^{L(l)1(1)} s_{\theta _1} s_{\theta _\text{w}}\right)+\kappa _-^2 \mathcal{C}_{\phi D}^{L(l)1(1)} c_{\theta _1} s_{\theta _2}\right)\right] $\\		
			\hline
			$\delta_{L}^{\mp} \delta_{R}^{\pm}Z_{1}^{\mu}$ &$\frac{1}{4 \Lambda ^2}\left[g v_L v_R \left(c_{\theta _2} \left(s_{\theta _1} s_{\theta _\text{w}} \left(\mathcal{C}_{\phi D}^{(L)\text{rL}(r)} k^{\mu}_{\mp}+\mathcal{C}_{\phi D}^{(R)\text{lL}(r)} p^{\mu}_{\pm}\right)-c_{\theta _\text{w}} \left(\mathcal{C}_{\phi D}^{(L)\text{rR}(l)} k^{\mu}_{\mp}+\mathcal{C}_{\phi D}^{(L)\text{rL}(r)} p^{\mu}_{\pm}\right)\right)\right.\right.$\\
			&$\left.\left.+c_{\theta _1} s_{\theta _2} \left(\mathcal{C}_{\phi D}^{(L)\text{rL}(r)} k^{\mu}_{\mp}+\mathcal{C}_{\phi D}^{(R)\text{lL}(r)} p^{\mu}_{\pm}\right)\right)-\tilde{g} \left(s_{\theta _1} s_{\theta _2}-c_{\theta _1} c_{\theta _2} s_{\theta _\text{w}}\right) \left( (k^{\mu}_{\mp}+p^{\mu}_{\pm}) \left(4A_+^{34}- \left(2 v_L v_R  \mathcal{C}_{\phi D}^{(L)\text{rL}(r)}\right.\right.\right.\right.$\\
			&$\left.\left.\left.\left.+\kappa _+^2 (\mathcal{C}_{\phi D}^{(L)11(r)}+\mathcal{C}_{\phi D}^{(L)12(r)}+\mathcal{C}_{\phi D}^{(R)11(l)})\right)\right)- v_L v_R\left(k^{\mu}_{\mp}\mathcal{C}_{\phi D}^{(L)\text{rR}(l)}+p^{\mu}_{\pm} \mathcal{C}_{\phi D}^{(R)\text{lL}(r)}\right)\right)\right]$\\
			\hline
			{\bf {\large$\vdots$}} & \hspace{7cm} {\bf{\large$\vdots$}} \\
			\hline
			$\delta_{R}^{-} \delta_{R}^{+}Z_{2}^{\mu}$&$+\frac{1}{8} \left(p^{\mu}_{-}+p^{\mu}_{+}\right) \left(-8 \tilde{g} c_{\theta _2} s_{\theta _1}-8 \tilde{g} c_{\theta _1} s_{\theta _2} s_{\theta _\text{w}}\right)+\frac{1}{8 \Lambda ^2}\left(p^{\mu}_{-}+p^{\mu}_{+}\right) \left[8 \tilde{g} c_{\theta _2} s_{\theta _1} A_+^{33}+8 \tilde{g} c_{\theta _1} s_{\theta _2}  s_{\theta _\text{w}}A_+^{33}+8 \tilde{g}  s_{\theta _2} c_{\theta _\text{w}}\Theta_{{3LB}}\right.$\\
			&$\left.+8 \tilde{g} c_{\theta _1} c_{\theta _2}\Theta_{{3RB}}-8 \tilde{g}  s_{\theta _1} s_{\theta _2} s_{\theta _\text{w}}\Theta_{{3RB}}-8 \tilde{g}  c_{\theta _2} s_{\theta _1}\Theta_{{BB}}-8 \tilde{g}  c_{\theta _1} s_{\theta _2} s_{\theta _\text{w}}\Theta_{\text{BB}}-2 \mathcal{C}_{\phi D}^{(R)\text{lL}(r)} \tilde{g} c_{\theta _2} v_L^2 s_{\theta _1}-2 \mathcal{C}_{\phi D}^{(R)\text{lL}(r)} \tilde{g} c_{\theta _1} v_L^2 s_{\theta _2} s_{\theta _\text{w}}\right.$\\
			&$\left.-4 \mathcal{C}_{\phi D}^{\{\text{Rr}\}\{(R)(r)\}} \tilde{g} c_{\theta _2} v_R^2 s_{\theta _1}-4 \mathcal{C}_{\phi D}^{\{\text{Rr}\}\{(R)(r)\}} \tilde{g} c_{\theta _1} v_R^2 s_{\theta _2} s_{\theta _\text{w}}-16 \mathcal{C}_{\phi D}^{R(r)R(r)} \tilde{g} c_{\theta _2} v_R^2 s_{\theta _1}-16 \mathcal{C}_{\phi D}^{R(r)R(r)} \tilde{g} c_{\theta _1} v_R^2 s_{\theta _2} s_{\theta _\text{w}}\right.$\\
			&$\left.-4 \kappa _1^2 \mathcal{C}_{\phi D}^{\{21\}\{(R)(r)\}} \tilde{g} c_{\theta _2} s_{\theta _1}-4 \kappa _2^2 \mathcal{C}_{\phi D}^{\{21\}\{(R)(r)\}} \tilde{g} c_{\theta _2} s_{\theta _1}-4 \kappa _1^2 \mathcal{C}_{\phi D}^{\{21\}\{(R)(r)\}} \tilde{g} c_{\theta _1} s_{\theta _2} s_{\theta _\text{w}}-4 \kappa _2^2 \mathcal{C}_{\phi D}^{\{21\}\{(R)(r)\}} \tilde{g} c_{\theta _1} s_{\theta _2} s_{\theta _\text{w}}\right.$\\
			&$\left.-2 \kappa _1^2 \mathcal{C}_{\phi D}^{21(R)(r)} \tilde{g} c_{\theta _2} s_{\theta _1}-2 \kappa _2^2 \mathcal{C}_{\phi D}^{21(R)(r)} \tilde{g} c_{\theta _2} s_{\theta _1}-2 \kappa _1^2 \mathcal{C}_{\phi D}^{21(R)(r)} \tilde{g} c_{\theta _1} s_{\theta _2} s_{\theta _\text{w}}-2 \kappa _2^2 \mathcal{C}_{\phi D}^{21(R)(r)} \tilde{g} c_{\theta _1} s_{\theta _2} s_{\theta _\text{w}}-4 \kappa _1^2 \mathcal{C}_{\phi D}^{(R)11(r)} \tilde{g} c_{\theta _2} s_{\theta _1}\right.$\\
			&$\left.-4 \kappa _2^2 \mathcal{C}_{\phi D}^{(R)11(r)} \tilde{g} c_{\theta _2} s_{\theta _1}-4 \kappa _1^2 \mathcal{C}_{\phi D}^{(R)11(r)} \tilde{g} c_{\theta _1} s_{\theta _2} s_{\theta _\text{w}}-4 \kappa _2^2 \mathcal{C}_{\phi D}^{(R)11(r)} \tilde{g} c_{\theta _1} s_{\theta _2} s_{\theta _\text{w}}+g \left(s_{\theta _2} \left(-\kappa _-^2 \mathcal{C}_{\phi D}^{R(r)1(1)} c_{\theta _\text{w}}\right.\right.\right.$\\
			&$\left.\left.\left.+s_{\theta _1} s_{\theta _\text{w}} \left(4 v_R^2 (3 \mathcal{C}_{\phi D}^{R(r)R(r)}+\mathcal{C}_{\square }^{\text{Rr}(\text{Rr})})-\kappa _-^2 \mathcal{C}_{\phi D}^{R(r)1(1)}\right)\right)-c_{\theta _1} c_{\theta _2} \left(4 v_R^2 (3 \mathcal{C}_{\phi D}^{R(r)R(r)}+\mathcal{C}_{\square }^{\text{Rr}(\text{Rr})})-\kappa _-^2 \mathcal{C}_{\phi D}^{R(r)1(1)}\right)\right)\right]$\\
			\hline
		\end{tabular}
		\caption{MLRSM: Coupling of a pair of singly charged scalars ($\delta^{\pm}$) with a neutral gauge boson ($Z_{1,2}$). Here $\mathit{s}_{\theta_{i}}=\sin\theta_{i}$ and $\mathit{c}_{\theta_{i}}=\cos\theta_{i}$. $k^{\mu}_{\pm}$ and $p^{\mu}_{\pm}$ are the four momenta of $\delta_{L}^{\pm}$ and $\delta_{R}^{\pm}$. $\tilde{\theta}_{2}$ is the redefined angle defined in Eq.~\ref{eq:MLRSM-nGB-rot-angle-full}.}
		\label{tab:MLRSM-scs-scs-nGB}
	}
\end{table}

\begin{table}[h]
	\centering
	{ \scriptsize
		\renewcommand{\arraystretch}{2.3}
		\begin{tabular}{|l|l|}
			\hline 
			{\bf Vertex} & {\bf  MLRSM: Effective vertex $( \delta^{\mp}\,\delta^{\pm}\,W_3[B]\,W_3[B])$ factor}\\			
			\hline 
			$\delta^{-}_{L} \delta^{+}_{L} W_{L}^{+\mu} W_{L\mu}^{-}$&$2 g^2+\frac{g^2}{8 \Lambda ^2} \left(-16 A_+^{44}+4 v_L^2 (3 \mathcal{C}_{\phi D}^{\{\text{Ll}\}\{(L)(l)\}}+9 \mathcal{C}_{\phi D}^{L(l)L(l)}-\mathcal{C}_{\square }^{\text{Ll}(\text{Ll})})+4 \mathcal{C}_{\phi D}^{(L)\text{rR}(l)} v_R^2+32 \Theta _{W_{\text{LL}}}+8 \kappa _1^2 \mathcal{C}_{\phi D}^{\{21\}\{(L)(l)\}}\right.$\\
			&$\left.+8 \kappa _2^2 \mathcal{C}_{\phi D}^{\{21\}\{(L)(l)\}}+4 \kappa _1^2 \mathcal{C}_{\phi D}^{21(L)(l)}+4 \kappa _2^2 \mathcal{C}_{\phi D}^{21(L)(l)}+8 \kappa _1^2 \mathcal{C}_{\phi D}^{(L)11(l)}+8 \kappa _2^2 \mathcal{C}_{\phi D}^{(L)11(l)}+8 \kappa _1 \kappa _2 \mathcal{C}_{\phi D}^{L(l)1(1)}\right.$\\
			&$\left.+2 \kappa _1^2 \mathcal{C}_{\phi D}^{\{\text{Ll}\}\{(2)(1)\}}+2 \kappa _2^2 \mathcal{C}_{\phi D}^{\{\text{Ll}\}\{(2)(1)\}}+\kappa _1^2 \mathcal{C}_{\phi D}^{\text{Ll}(2)(1)}+\kappa _2^2 \mathcal{C}_{\phi D}^{\text{Ll}(2)(1)}\right)$\\
			\hline
			$\delta^{-}_{L} \delta^{+}_{L} W^{\mu}_{3L} W_{3L\mu}$&$\frac{g^2}{16 \Lambda ^2} \left(8 v_L^2 (\mathcal{C}_{\phi D}^{\{\text{Ll}\}\{(L)(l)\}}+\mathcal{C}_{\phi D}^{L(l)L(l)}-\mathcal{C}_{\square }^{\text{Ll}(\text{Ll})})+\left(\kappa _1^2+\kappa _2^2\right) (2 \mathcal{C}_{\phi D}^{\{\text{Ll}\}\{(2)(1)\}}+\mathcal{C}_{\phi D}^{\text{Ll}(2)(1)})\right)$\\
			\hline
			$\delta^{-}_{L} \delta^{+}_{L} W^{\mu}_{3L} B_{\mu}$&$\frac{\tilde{g}}{4 \Lambda ^2} \left(8 \tilde{g} \Theta_{{3LB}}-4 g v_L^2 (\mathcal{C}_{\phi D}^{\{\text{Ll}\}\{(L)(l)\}}+4 \mathcal{C}_{\phi D}^{L(l)L(l)})+g \left(\kappa _2^2-\kappa _1^2\right) \mathcal{C}_{\phi D}^{L(l)1(1)}\right)$\\
			\hline
			$\delta^{-}_{L} \delta^{+}_{L} B^{\mu} B_{\mu} $&$\tilde{g}^2+\frac{\tilde{g}^2}{4 \Lambda ^2} \left(-4 A_+^{44}+8 \Theta_{{BB}}+4 v_L^2 (\mathcal{C}_{\phi D}^{\{\text{Ll}\}\{(L)(l)\}}+4 \mathcal{C}_{\phi D}^{L(l)L(l)})+\mathcal{C}_{\phi D}^{(L)\text{rR}(l)} v_R^2+\mathcal{C}_{\phi D}^{(R)\text{lL}(r)} v_R^2+2 \kappa _+^2 \mathcal{C}_{\phi D}^{\{21\}\{(L)(l)\}}\right.$\\
			&$\left.+\kappa _+^2 \mathcal{C}_{\phi D}^{21(L)(l)}+2 \kappa _+^2 \mathcal{C}_{\phi D}^{(L)11(l)}\right)$\\
			\hline
			\hspace{1cm} {\bf{\large$\vdots$}} & \hspace{6cm} {\bf{\large$\vdots$}} \\
			\hline
			$\delta^{-}_{R} \delta^{+}_{R} W_{R}^{+\mu} W_{R\mu}^{-}$&$2 g^2+\frac{g^2}{8 \Lambda ^2} \left(-16 A_+^{33}+4 \mathcal{C}_{\phi D}^{(R)\text{lL}(r)} v_L^2+12 \mathcal{C}_{\phi D}^{\{\text{Rr}\}\{(R)(r)\}} v_R^2+36 \mathcal{C}_{\phi D}^{R(r)R(r)} v_R^2-4 \mathcal{C}_{\square }^{\text{Rr}(\text{Rr})} v_R^2+32 \Theta _{W_{\text{RR}}}\right.$\\
			&$\left.+8 \kappa _1^2 \mathcal{C}_{\phi D}^{\{21\}\{(R)(r)\}}+8 \kappa _2^2 \mathcal{C}_{\phi D}^{\{21\}\{(R)(r)\}}+4 \kappa _1^2 \mathcal{C}_{\phi D}^{21(R)(r)}+4 \kappa _2^2 \mathcal{C}_{\phi D}^{21(R)(r)}+8 \kappa _1^2 \mathcal{C}_{\phi D}^{(R)11(r)}+8 \kappa _2^2 \mathcal{C}_{\phi D}^{(R)11(r)}\right.$\\
			&$\left.-4 \kappa _1^2 \mathcal{C}_{\phi D}^{R(r)1(1)}-4 \kappa _2^2 \mathcal{C}_{\phi D}^{R(r)1(1)}+2 \kappa _1^2 \mathcal{C}_{\phi D}^{\{\text{Rr}\}\{(2)(1)\}}+2 \kappa _2^2 \mathcal{C}_{\phi D}^{\{\text{Rr}\}\{(2)(1)\}}+\kappa _1^2 \mathcal{C}_{\phi D}^{\text{Rr}(2)(1)}+\kappa _2^2 \mathcal{C}_{\phi D}^{\text{Rr}(2)(1)}\right)$\\
			\hline
			$\delta^{-}_{R} \delta^{+}_{R} W^{\mu}_{3R} W_{3R\mu}$&$\frac{g^2}{16 \Lambda ^2} \left(8 v_R^2 (\mathcal{C}_{\phi D}^{\{\text{Rr}\}\{(R)(r)\}}+\mathcal{C}_{\phi D}^{R(r)R(r)}-\mathcal{C}_{\square }^{\text{Rr}(\text{Rr})})+\left(\kappa _1^2+\kappa _2^2\right) (2 \mathcal{C}_{\phi D}^{\{\text{Rr}\}\{(2)(1)\}}+\mathcal{C}_{\phi D}^{\text{Rr}(2)(1)})\right)$\\
			\hline
			$\delta^{-}_{R} \delta^{+}_{R} W^{\mu}_{3R} B_{\mu}$&$\frac{\tilde{g}}{4 \Lambda ^2} \left(8 \tilde{g} \Theta_{{3RB}}-4 g v_R^2 (\mathcal{C}_{\phi D}^{\{\text{Rr}\}\{(R)(r)\}}+4 \mathcal{C}_{\phi D}^{R(r)R(r)})+g \left(\kappa _1^2-\kappa _2^2\right) \mathcal{C}_{\phi D}^{R(r)1(1)}\right)$\\
			\hline
			$\delta^{-}_{R} \delta^{+}_{R} B^{\mu} B_{\mu} $&$\tilde{g}^2+\frac{\tilde{g}^2}{4 \Lambda ^2} \left(-4 A_+^{33}+8 \Theta_{{BB}}+v_L^2 (\mathcal{C}_{\phi D}^{(L)\text{rR}(l)}+\mathcal{C}_{\phi D}^{(R)\text{lL}(r)})+4 \mathcal{C}_{\phi D}^{\{\text{Rr}\}\{(R)(r)\}} v_R^2+16 \mathcal{C}_{\phi D}^{R(r)R(r)} v_R^2+2 \kappa _+^2 \mathcal{C}_{\phi D}^{\{21\}\{(R)(r)\}}\right.$\\
			&$\left.+\kappa _+^2 \mathcal{C}_{\phi D}^{21(R)(r)}+2 \kappa _+^2 \mathcal{C}_{\phi D}^{(R)11(r)}\right)$\\
			\hline
		\end{tabular}
		\caption{MLRSM: Coupling of a pair of singly charged scalars ($\delta^{\pm}$) with either a pair of charged ($W_{L,R}^{\pm}$) or neutral $(B,W_{3L,3R})$ gauge bosons in unphysical basis. These can be translated in physical gauge boson basis using rotation matrix $\widetilde{\mathcal{R}}_{a,b}$.}
		\label{tab:MLRSM-scs-scs-GB-GB}
	}
\end{table}
\clearpage




\subsection{Rare Processes}

So far, we have discussed the impact of the modified Feynman vertices on different low energy observables and also on some significant and promising collider phenomenological processes. 
Now we would like to focus on possible rare processes which may get additional contributions from the higher dimensional operators. As these processes are yet to be observed, the rate of their occurrence must be very small, which in turn puts a stringent upper limit on the corresponding Wilson Coefficients. Thus not only for the sake of completeness but also to estimate the allowed maximum values of the WCs these channels must be taken into account. As we have a complete set of operators, there will be many rare processes, but we can not include all of them. We will provide an outline to show how these effective operators can affect different rare events. 

\begin{figure}[h!]
	\centering
	{
		\includegraphics[trim={1.3cm 4cm 1.4cm 0},scale=0.3]{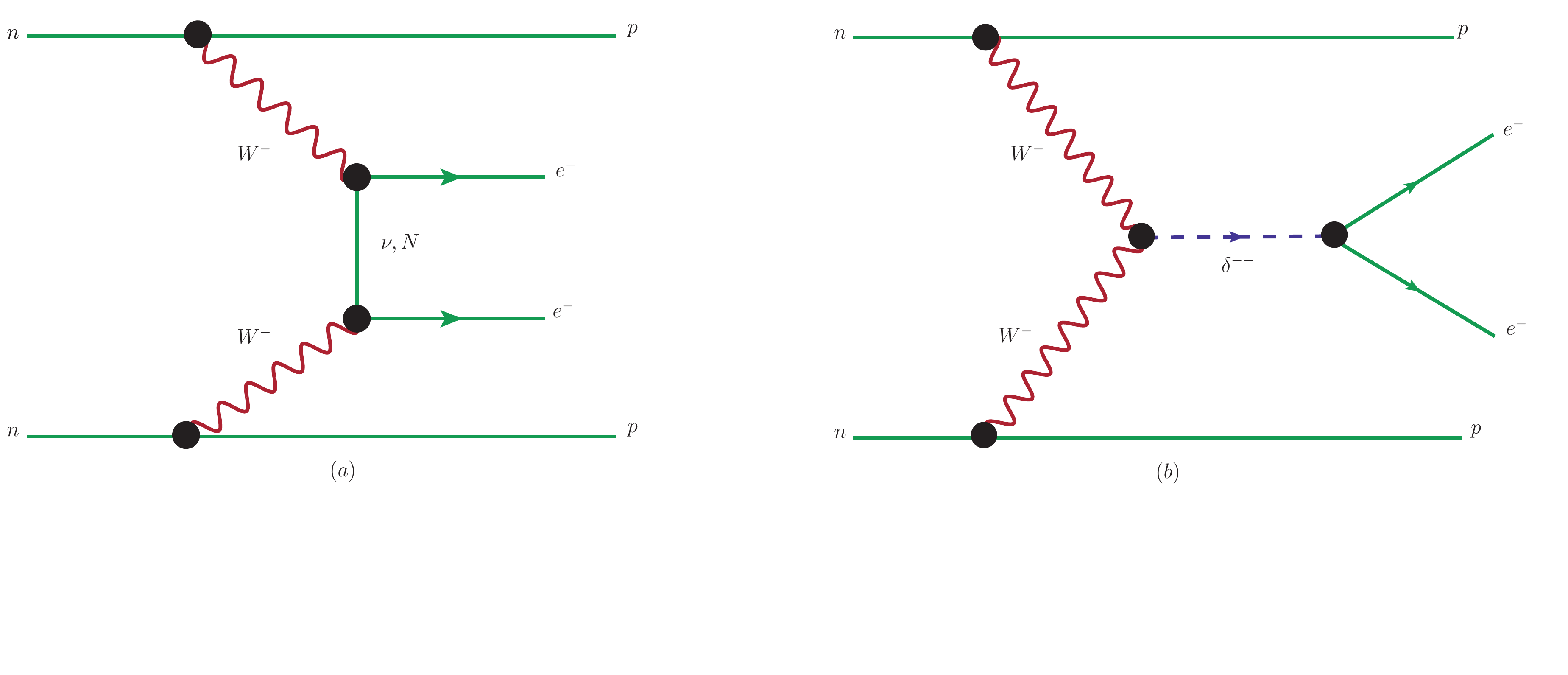}
	}\caption{Feynman diagrams representing the neutrinoless double beta decay: $2n\to 2p+2e^-$ ($0\nu2\beta$). The  [{\Large $\bullet$}] represents effective vertex.  $0\nu2\beta$ can acquire additional contributions from (a) light and heavy Majorana neutrino exchange, and (b)  from doubly charged scalars too.}
	\label{fig:nulessbetadeacy}
\end{figure}

\begin{figure}[h!]
	\centering
	{
		\includegraphics[trim={0.3cm 0 1.0cm 0},scale=0.5]{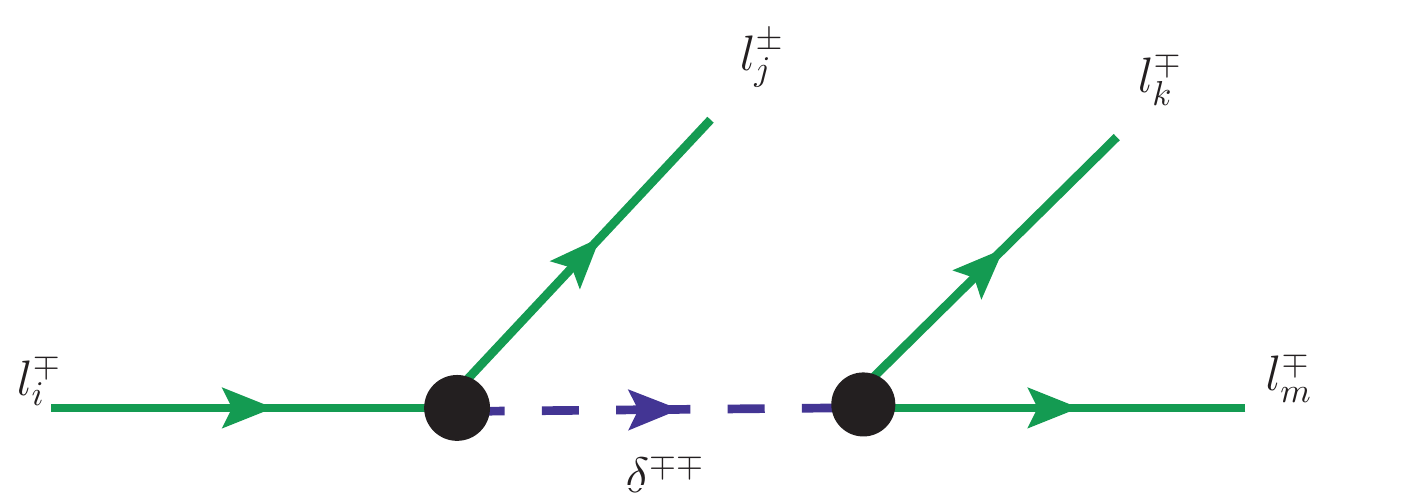}
	}\caption{Feynman diagram representing the lepton number and flavour violating three body decays of charged lepton: $l_i^\mp \to l_j^\pm l_k^\mp l_m^\mp$. The  [{\Large $\bullet$}] represents effective vertex.}
	\label{fig:3bodydecay}
\end{figure}

Looking into the VBF channel, Fig.~\ref{fig:VBF}, it is expected that the amplitude for neutrinoless double beta decay ($0\nu2\beta$), $2n\to 2p+2e^-$, Fig.~\ref{fig:nulessbetadeacy}(b), will be modified   as compared to the earlier analysis based on the renormalizable interactions  \cite{Chakrabortty:2012mh,Dev:2014xea}. The other diagram for $0\nu2\beta$ mediated by the Majorana neutrinos, Fig.~\ref{fig:nulessbetadeacy}(a),  in MLRSM will be also be affected due to the effective gauge boson-fermion vertex, see Tables~\ref{tab:MLRSM-cGB-F-F-I},~\ref{tab:MLRSM-cGB-F-F-II}.  So MLRSM-EFT scenario offers  very convoluted contributions to $0\nu2\beta$ process, and the usual constraints need to be revisited. The leptonic current for this process, and the meson decays ($M^+ \to M'^{-} l^+ l^+$) are the same \cite{Ma:2009fi,Cvetic:2010rw,Bambhaniya:2015nea,Wang:2018bgp}. Thus the prediction for meson decays through Majorana neutrinos and doubly charged scalars will also be modified. The lepton flavour as well as number violating three body decays of charged leptons, e.g., $l_i^\mp \to l_j^\pm l_k^\mp l_m^\mp$ \cite{Chakrabortty:2015zpm}, can get additional contributions from the effective operators, see Fig.~\ref{fig:3bodydecay}. The possibility of having charged lepton flavour violating radiative decays, $l_{i}\rightarrow\;l_{j}+\gamma,\; (\text{with}~ i\neq j)$ can not be ignored, see Figs.~\ref{fig:mag-moment}(a),~(b). 
The sensitivity of these rare events are expected to increase significantly with future experiments. Thus the respective WCs will be severely constrained and may play a crucial role to decide the minimum energy scale to unveil these rare events at the colliders. 

%% file: Theoretical-constraints.tex
\subsection{Theoretical Constraints in the Scalar sector}

The shape of the scalar potential is determined by the quartic couplings in the asymptotic limit. The physical masses of the scalar fields are linear functions of these quartic couplings. Thus it is evident that any constraint on these couplings can be translated appropriately to put some bounds on the scalar spectrum.  Here, we have outlined two such theoretical constraints: (i) Tree Unitarity (TU) \cite{Huffel:1980sk, Marciano:1989ns} and (ii) Vacuum Stability (VS) \cite{Kannike:2016fmd}.
Both the bounds are computed at the large value of the scalar field. Thus the relevant part of the scalar potential for us is the scalar four-point interactions. The TU sets an upper limit on the amplitudes of the scalar four-point interactions, and that is $\leq 8\pi$ \cite{Marciano:1989ns}. Now we know that the amplitude for such processes is the quartic couplings. This, in turn, sets a maximum allowed limit on the quartic couplings.  On the other side the VS criteria ensure that potential is bounded from below such that the vacuum is stable\footnote{Here, we have kept vacuum stability and boundedness criteria in the same footing. The meta-stability  is overlooked for this discussion.}. These TU and VS criteria are modified in the presence of effective operators, and thus the limits on the scalar spectrum may be altered \cite{Ghosh:2017coz,Corbett:2017qgl}.

\begin{figure}[h!]
	\centering
	{
		\includegraphics[trim={1.5cm 0 1.5cm 0},scale=0.74]{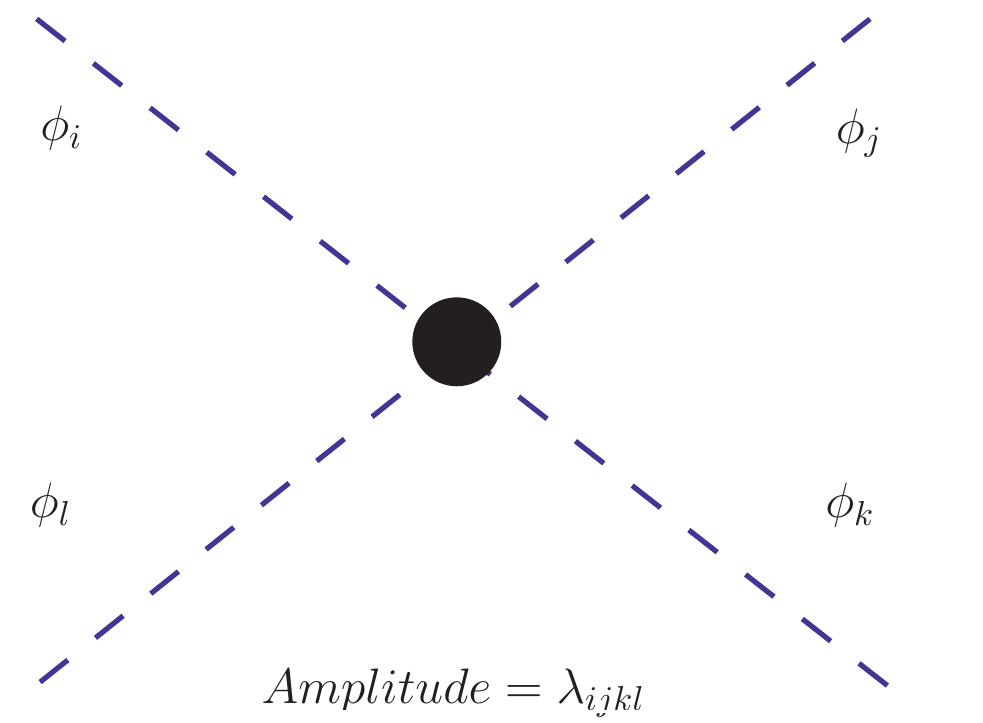}
	}
	\caption{Scalar four-point interactions: Amplitude is the effective scalar quartic coupling $\lambda_{ijkl}$. The  [{\Large $\bullet$}] represents the effective four-point scalar vertex.}\label{fig:scalar-four-point}
\end{figure}

In this paper we have computed the scalar four-point interactions, see Fig.~\ref{fig:scalar-four-point}, which also represent the amplitude for the process $\phi_i \phi_j \to \phi_k \phi_l$ including the modifications due the dimension-6 operators. The added contributions are directly from $\phi^6$ operators, and also from the $\phi^4 D^2$ operators through the redefinition of the scalar fields.

\begin{table}[h!]
	\centering
	{ \scriptsize
		\renewcommand{\arraystretch}{2.3}
		\begin{tabular}{|c|l|}
			\hline
			{\bf Vertex}&{\bf 2HDM: Effective vertex $ (\phi_{i}\,\phi_{j}\,\phi_{k}\,\phi_{l})$ factor}\\
			\hline 
			$\phi_1^- \phi_1^- \phi_1^+ \phi_1^+$  &  $\lambda _1-\frac{\lambda _6 A_{+}^{12}}{\Lambda ^2}-\frac{2 \lambda _1 A_{+}^{11}}{\Lambda ^2}+\frac{3 \mathcal{C}_{6}^{111111} v_1^2}{2 \Lambda ^2}\textcolor{purple}{+\frac{\mathcal{C}_{6}^{111112} v_1 v_2}{\Lambda ^2}}+\frac{\mathcal{C}_{6}^{111122} v_2^2}{2 \Lambda ^2}$  \\
			$\phi _1^- \phi_2^- \phi _1^+ \phi_2^+$  &  $\lambda _3+\lambda _4-\frac{\lambda _3 A_{+}^{22}}{\Lambda ^2}-\frac{\lambda _4 A_{+}^{22}}{\Lambda ^2}-\frac{2 \lambda _7 A_{+}^{12}}{\Lambda ^2}-\frac{\lambda _3 A_{+}^{11}}{\Lambda ^2}-\frac{\lambda _4 A_{+}^{11}}{\Lambda ^2}-\frac{2 \lambda _6 A_{+}^{11}}{\Lambda ^2}+\frac{\mathcal{C}_{6}^{111122} v_1^2}{\Lambda ^2}$  \\
			$\text{}$  &  $\textcolor{purple}{+\frac{\mathcal{C}_{6}^{112212} v_2 v_1}{\Lambda ^2}}+\frac{\mathcal{C}_{6}^{112222} v_2^2}{\Lambda ^2}+\frac{\mathcal{C}_{6}^{122111} v_1^2}{2 \Lambda ^2}\textcolor{purple}{+\frac{2 \mathcal{C}_{6}^{121221} v_2 v_1}{\Lambda ^2}}+\frac{\mathcal{C}_{6}^{122122} v_2^2}{2 \Lambda ^2}$  \\
			$\rho _1^2 \phi _1^- \phi _1^+$  &  $\lambda _1-\frac{\lambda _6 A_{r}^{12}}{2 \Lambda ^2}-\frac{\lambda _1 A_{r}^{11}}{\Lambda ^2}-\frac{\lambda _1 A_{+}^{11}}{\Lambda ^2}-\frac{\lambda _6 A_{+}^{12}}{2 \Lambda ^2}+\frac{9 \mathcal{C}_{6}^{111111} v_1^2}{2 \Lambda ^2}\textcolor{purple}{+\frac{3 \mathcal{C}_{6}^{111112} v_1 v_2}{\Lambda ^2}}+\frac{\mathcal{C}_{6}^{111122} v_2^2}{2 \Lambda ^2}+\frac{\mathcal{C}_{6}^{121211} v_2^2}{2 \Lambda ^2}$  \\
			$\text{}$  &  $+\frac{\mathcal{C}_{6}^{122111} v_2^2}{4 \Lambda ^2}$  \\
			$\rho _1 \rho _2 \phi _1^- \phi _1^+$  &  $\lambda _6-\frac{\lambda _6 A_{r}^{22}}{2 \Lambda ^2}-\frac{\lambda _1 A_{r}^{12}}{\Lambda ^2}-\frac{\lambda _3 A_{r}^{12}}{2 \Lambda ^2}-\frac{\lambda _6 A_{r}^{11}}{2 \Lambda ^2}-\frac{\lambda _4 A_{+}^{12}}{2 \Lambda ^2}-\frac{\lambda _5 A_{+}^{12}}{2 \Lambda ^2}-\frac{\lambda _6 A_{+}^{11}}{\Lambda ^2}\textcolor{purple}{+\frac{3 \mathcal{C}_{6}^{111112} v_1^2}{\Lambda ^2}}$  \\
			$\text{}$  &  $+\frac{2 \mathcal{C}_{6}^{111122} v_1 v_2}{\Lambda ^2}\textcolor{purple}{+\frac{3 \mathcal{C}_{6}^{112212} v_2^2}{2 \Lambda ^2}}+\frac{2 \mathcal{C}_{6}^{121211} v_1 v_2}{\Lambda ^2}+\frac{\mathcal{C}_{6}^{122111} v_1 v_2}{\Lambda ^2}$  \\
			$\rho _2^2 \phi _1^- \phi _1^+$  &  $\frac{\lambda _3}{2}-\frac{\lambda _3 A_{r}^{22}}{2 \Lambda ^2}-\frac{\lambda _6 A_{r}^{12}}{2 \Lambda ^2}-\frac{\lambda _3 A_{+}^{11}}{2 \Lambda ^2}-\frac{\lambda _7 A_{+}^{12}}{2 \Lambda ^2}+\frac{\mathcal{C}_{6}^{111122} v_1^2}{2 \Lambda ^2}\textcolor{purple}{+\frac{3 \mathcal{C}_{6}^{112212} v_1 v_2}{2 \Lambda ^2}}+\frac{\mathcal{C}_{6}^{121211} v_1^2}{2 \Lambda ^2}+\frac{\mathcal{C}_{6}^{122111} v_1^2}{4 \Lambda ^2}$  \\
			$\text{}$  &  $+\frac{3 \mathcal{C}_{6}^{112222} v_2^2}{2 \Lambda ^2}$  \\
			$\eta _1^2 \phi _1^- \phi _1^+$  &  $\lambda _1-\frac{\lambda _6 A_{i}^{12}}{2 \Lambda ^2}-\frac{\lambda _1 A_{i}^{11}}{\Lambda ^2}-\frac{\lambda _1 A_{+}^{11}}{\Lambda ^2}-\frac{\lambda _6 A_{+}^{12}}{2 \Lambda ^2}+\frac{3 \mathcal{C}_{6}^{111111} v_1^2}{2 \Lambda ^2}+\textcolor{purple}{\frac{\mathcal{C}_{6}^{111112} v_1 v_2}{\Lambda ^2}}+\frac{\mathcal{C}_{6}^{111122} v_2^2}{2 \Lambda ^2}-\frac{\mathcal{C}_{6}^{121211} v_2^2}{2 \Lambda ^2}$  \\
			$\text{}$  &  $+\frac{\mathcal{C}_{6}^{122111} v_2^2}{4 \Lambda ^2}$  \\
			$\eta _1 \eta _2 \phi _1^- \phi _1^+$  &  $\lambda _6-\frac{\lambda _6 A_{i}^{22}}{2 \Lambda ^2}-\frac{\lambda _1 A_{i}^{12}}{\Lambda ^2}-\frac{\lambda _3 A_{i}^{12}}{2 \Lambda ^2}-\frac{\lambda _6 A_{i}^{11}}{2 \Lambda ^2}-\frac{\lambda _4 A_{+}^{12}}{2 \Lambda ^2}-\frac{\lambda _5 A_{+}^{12}}{2 \Lambda ^2}-\frac{\lambda _6 A_{+}^{11}}{\Lambda ^2}\textcolor{purple}{+\frac{\mathcal{C}_{6}^{111112} v_1^2}{\Lambda ^2}}$  \\
			$\text{}$  &  $\textcolor{purple}{+\frac{\mathcal{C}_{6}^{112212} v_2^2}{2 \Lambda ^2}}+\frac{2 \mathcal{C}_{6}^{121211} v_1 v_2}{\Lambda ^2}$  \\
			$\phi_2^- \phi_2^-  \phi_2^+ \phi_2^+ $  &  $\lambda _2-\frac{2 \lambda _2 A_{+}^{22}}{\Lambda ^2}-\frac{\lambda _7 A_{+}^{11}}{\Lambda ^2}+\frac{\mathcal{C}_{6}^{112222} v_1^2}{2 \Lambda ^2}\textcolor{purple}{+\frac{\mathcal{C}_{6}^{122222} v_1 v_2}{\Lambda ^2}}+\frac{3 \mathcal{C}_{6}^{222222} v_2^2}{2 \Lambda ^2}$  \\
			$\rho _1^2 \phi_2^- \phi_2^+$  &  $\frac{\lambda _3}{2}-\frac{\lambda _3 A_{r}^{11}}{2 \Lambda ^2}-\frac{\lambda _7 A_{r}^{12}}{2 \Lambda ^2}-\frac{\lambda _3 A_{+}^{22}}{2 \Lambda ^2}-\frac{\lambda _6 A_{+}^{11}}{2 \Lambda ^2}+\frac{3 \mathcal{C}_{6}^{111122} v_1^2}{2 \Lambda ^2}\textcolor{purple}{+\frac{3 \mathcal{C}_{6}^{112212} v_1 v_2}{2 \Lambda ^2}}+\frac{\mathcal{C}_{6}^{121222} v_2^2}{2 \Lambda ^2}+\frac{\mathcal{C}_{6}^{112222} v_2^2}{2 \Lambda ^2}$  \\
			$\text{}$  &  $+\frac{\mathcal{C}_{6}^{122122} v_2^2}{4 \Lambda ^2}$  \\
			$\rho _1 \rho _2 \phi_2^- \phi_2^+$  &  $\lambda _7-\frac{\lambda _7 A_{r}^{22}}{2 \Lambda ^2}-\frac{\lambda _2 A_{r}^{12}}{\Lambda ^2}-\frac{\lambda _3 A_{r}^{12}}{2 \Lambda ^2}-\frac{\lambda _7 A_{r}^{11}}{2 \Lambda ^2}-\frac{\lambda _4 A_{+}^{11}}{2 \Lambda ^2}-\frac{\lambda _5 A_{+}^{11}}{2 \Lambda ^2}-\frac{\lambda _7 A_{+}^{22}}{\Lambda ^2}\textcolor{purple}{+\frac{3 \mathcal{C}_{6}^{112212} v_1^2}{2 \Lambda ^2}}$  \\
			$\text{}$  &  $+\frac{2 \mathcal{C}_{6}^{121222} v_1 v_2}{\Lambda ^2}+\frac{2 \mathcal{C}_{6}^{112222} v_1 v_2}{\Lambda ^2}+\frac{\mathcal{C}_{6}^{122122} v_1 v_2}{\Lambda ^2}\textcolor{purple}{+\frac{3 \mathcal{C}_{6}^{122222} v_2^2}{\Lambda ^2}}$  \\
			$\rho _2^2 \phi_2^- \phi_2^+$  &  $\lambda _2-\frac{\lambda _2 A_{r}^{22}}{\Lambda ^2}-\frac{\lambda _7 A_{r}^{12}}{2 \Lambda ^2}-\frac{\lambda _2 A_{+}^{22}}{\Lambda ^2}-\frac{\lambda _7 A_{+}^{11}}{2 \Lambda ^2}+\frac{\mathcal{C}_{6}^{121222} v_1^2}{2 \Lambda ^2}+\frac{\mathcal{C}_{6}^{112222} v_1^2}{2 \Lambda ^2}+\frac{\mathcal{C}_{6}^{122122} v_1^2}{4 \Lambda ^2}\textcolor{purple}{+\frac{3 \mathcal{C}_{6}^{122222} v_1 v_2}{\Lambda ^2}}$  \\
			$\text{}$  &  $+\frac{9 \mathcal{C}_{6}^{222222} v_2^2}{2 \Lambda ^2}$  \\
			$\eta _1^2 \phi_2^- \phi_2^+$  &  $\frac{\lambda _3}{2}-\frac{\lambda _3 A_{i}^{11}}{2 \Lambda ^2}-\frac{\lambda _7 A_{i}^{12}}{2 \Lambda ^2}-\frac{\lambda _3 A_{+}^{22}}{2 \Lambda ^2}-\frac{\lambda _6 A_{+}^{11}}{2 \Lambda ^2}+\frac{\mathcal{C}_{6}^{111122} v_1^2}{2 \Lambda ^2}\textcolor{purple}{+\frac{\mathcal{C}_{6}^{112212} v_1 v_2}{2 \Lambda ^2}}-\frac{\mathcal{C}_{6}^{121222} v_2^2}{2 \Lambda ^2}+\frac{\mathcal{C}_{6}^{112222} v_2^2}{2 \Lambda ^2}$  \\
			$\text{}$  &  $+\frac{\mathcal{C}_{6}^{122122} v_2^2}{4 \Lambda ^2}$  \\
			$\eta _1 \eta _2 \phi_2^- \phi_2^+$  &  $\lambda _7-\frac{\lambda _7 A_{i}^{22}}{2 \Lambda ^2}-\frac{\lambda _2 A_{i}^{12}}{\Lambda ^2}-\frac{\lambda _3 A_{i}^{12}}{2 \Lambda ^2}-\frac{\lambda _7 A_{i}^{11}}{2 \Lambda ^2}-\frac{\lambda _4 A_{+}^{11}}{2 \Lambda ^2}-\frac{\lambda _5 A_{+}^{11}}{2 \Lambda ^2}-\frac{\lambda _7 A_{+}^{22}}{\Lambda ^2}\textcolor{purple}{+\frac{\mathcal{C}_{6}^{112212} v_1^2}{2 \Lambda ^2}}$  \\
			$\text{}$  &  $+\frac{2 \mathcal{C}_{6}^{121222} v_1 v_2}{\Lambda ^2}\textcolor{purple}{+\frac{\mathcal{C}_{6}^{122222} v_2^2}{\Lambda ^2}}$  \\
			\hline    
		\end{tabular}
		\caption{2HDM: Scalar four-point vertices in the unphysical basis.}
		\label{tab:2HDM-scalar-4-pt-I}
	}
\end{table}
\begin{table}[h!]
	\centering
	{ \scriptsize
	\renewcommand{\arraystretch}{2.3}%
	\begin{tabular}{|c|l|}
		\hline 
		{\bf Vertex}&{\bf 2HDM: Effective vertex $ (\phi_{i}\,\phi_{j}\,\phi_{k}\,\phi_{l})$ factor}\\
		\hline
		$\rho _1^4$  &  $\frac{\lambda _1}{4}-\frac{\lambda _6 A_{r}^{12}}{4 \Lambda ^2}-\frac{\lambda _1 A_{r}^{11}}{2 \Lambda ^2}+\frac{15 \mathcal{C}_{6}^{111111} v_1^2}{8 \Lambda ^2}\textcolor{purple}{+\frac{5 \mathcal{C}_{6}^{111112} v_1 v_2}{4 \Lambda ^2}}+\frac{\mathcal{C}_{6}^{111122} v_2^2}{8 \Lambda ^2}+\frac{\mathcal{C}_{6}^{121211} v_2^2}{4 \Lambda ^2}+\frac{\mathcal{C}_{6}^{122111} v_2^2}{8 \Lambda ^2}$  \\
		$\rho _2^4$  &  $\frac{\lambda _2}{4}-\frac{\lambda _2 A_{r}^{22}}{2 \Lambda ^2}-\frac{\lambda _7 A_{r}^{12}}{4 \Lambda ^2}+\frac{\mathcal{C}_{6}^{121222} v_1^2}{4 \Lambda ^2}+\frac{\mathcal{C}_{6}^{112222} v_1^2}{8 \Lambda ^2}+\frac{\mathcal{C}_{6}^{122122} v_1^2}{8 \Lambda ^2}\textcolor{purple}{+\frac{5 \mathcal{C}_{6}^{122222} v_1 v_2}{4 \Lambda ^2}}+\frac{15 \mathcal{C}_{6}^{222222} v_2^2}{8 \Lambda ^2}$  \\
		$\rho _1^3 \rho _2$  &  $\frac{\lambda _6}{2}-\frac{\lambda _6 A_{r}^{22}}{4 \Lambda ^2}-\frac{\lambda _1 A_{r}^{12}}{2 \Lambda ^2}-\frac{\lambda _3 A_{r}^{12}}{4 \Lambda ^2}-\frac{\lambda _4 A_{r}^{12}}{4 \Lambda ^2}-\frac{\lambda _5 A_{r}^{12}}{4 \Lambda ^2}-\frac{3 \lambda _6 A_{r}^{11}}{4 \Lambda ^2}\textcolor{purple}{+\frac{5 \mathcal{C}_{6}^{111112} v_1^2}{2 \Lambda ^2}}+\frac{\mathcal{C}_{6}^{111122} v_1 v_2}{\Lambda ^2}$  \\
		$\text{}$  &  $\textcolor{purple}{+\frac{3 \mathcal{C}_{6}^{112212} v_2^2}{4 \Lambda ^2}}+\frac{2 \mathcal{C}_{6}^{121211} v_1 v_2}{\Lambda ^2}\textcolor{purple}{+\frac{3 \mathcal{C}_{6}^{121212} v_2^2}{4 \Lambda ^2}}+\frac{\mathcal{C}_{6}^{122111} v_1 v_2}{\Lambda ^2}\textcolor{purple}{+\frac{3 \mathcal{C}_{6}^{121221} v_2^2}{4 \Lambda ^2}}$  \\
		$\rho _1^2 \rho _2^2$  &  $\frac{\lambda _3}{4}+\frac{\lambda _4}{4}+\frac{\lambda _5}{4}-\frac{\lambda _3 A_{r}^{22}}{4 \Lambda ^2}-\frac{3 \lambda _6 A_{r}^{12}}{4 \Lambda ^2}-\frac{3 \lambda _7 A_{r}^{12}}{4 \Lambda ^2}-\frac{\lambda _3 A_{r}^{11}}{4 \Lambda ^2}-\frac{\lambda _4 A_{r}^{11}}{4 \Lambda ^2}-\frac{\lambda _5 A_{r}^{11}}{4 \Lambda ^2}$  \\
		$\text{}$  &  $-\frac{\lambda _4 A_{r}^{22}}{4 \Lambda ^2}-\frac{\lambda _5 A_{r}^{22}}{4 \Lambda ^2}+\frac{3 \mathcal{C}_{6}^{111122} v_1^2}{4 \Lambda ^2}\textcolor{purple}{+\frac{9 \mathcal{C}_{6}^{112212} v_1 v_2}{4 \Lambda ^2}}+\frac{3 \mathcal{C}_{6}^{121211} v_1^2}{2 \Lambda ^2}\textcolor{purple}{+\frac{9 \mathcal{C}_{6}^{121212} v_1 v_2}{4 \Lambda ^2}}+\frac{3 \mathcal{C}_{6}^{121222} v_2^2}{2 \Lambda ^2}+\frac{3 \mathcal{C}_{6}^{122111} v_1^2}{4 \Lambda ^2}$  \\
		$\text{}$  &  $\textcolor{purple}{+\frac{9 \mathcal{C}_{6}^{121221} v_1 v_2}{4 \Lambda ^2}}+\frac{3 \mathcal{C}_{6}^{112222} v_2^2}{4 \Lambda ^2}+\frac{3 \mathcal{C}_{6}^{122122} v_2^2}{4 \Lambda ^2}$  \\
		$\eta _1^2 \rho _1^2$  &  $\frac{\lambda _1}{2}-\frac{\lambda _6 A_{i}^{12}}{4 \Lambda ^2}-\frac{\lambda _1 A_{i}^{11}}{2 \Lambda ^2}-\frac{\lambda _1 A_{r}^{11}}{2 \Lambda ^2}-\frac{\lambda _6 A_{r}^{12}}{4 \Lambda ^2}+\frac{9 \mathcal{C}_{6}^{111111} v_1^2}{4 \Lambda ^2}\textcolor{purple}{+\frac{3 \mathcal{C}_{6}^{111112} v_1 v_2}{2 \Lambda ^2}}+\frac{\mathcal{C}_{6}^{111122} v_2^2}{4 \Lambda ^2}+\frac{\mathcal{C}_{6}^{122111} v_2^2}{4 \Lambda ^2}$  \\
		$\eta _1 \eta _2 \rho _1^2$  &  $\frac{\lambda _6}{2}-\frac{\lambda _6 A_{i}^{22}}{4 \Lambda ^2}-\frac{\lambda _1 A_{i}^{12}}{2 \Lambda ^2}-\frac{\lambda _3 A_{i}^{12}}{4 \Lambda ^2}-\frac{\lambda _4 A_{i}^{12}}{4 \Lambda ^2}+\frac{\lambda _5 A_{i}^{12}}{4 \Lambda ^2}-\frac{\lambda _6 A_{i}^{11}}{4 \Lambda ^2}-\frac{\lambda _5 A_{r}^{12}}{2 \Lambda ^2}-\frac{\lambda _6 A_{r}^{11}}{2 \Lambda ^2}$  \\
		$\text{}$  &  $\textcolor{purple}{+\frac{3 \mathcal{C}_{6}^{111112} v_1^2}{2 \Lambda ^2}}\textcolor{purple}{+\frac{\mathcal{C}_{6}^{112212} v_2^2}{4 \Lambda ^2}}+\frac{3 \mathcal{C}_{6}^{121211} v_2 v_1}{\Lambda ^2}\textcolor{purple}{+\frac{9 \mathcal{C}_{6}^{121212} v_2^2}{4 \Lambda ^2}+\frac{\mathcal{C}_{6}^{121221} v_2^2}{4 \Lambda ^2}}$  \\
		$\eta _1^2 \rho _1 \rho _2$  &  $\frac{\lambda _6}{2}-\frac{\lambda _5 A_{i}^{12}}{2 \Lambda ^2}-\frac{\lambda _6 A_{i}^{11}}{2 \Lambda ^2}-\frac{\lambda _6 A_{r}^{22}}{4 \Lambda ^2}-\frac{\lambda _1 A_{r}^{12}}{2 \Lambda ^2}-\frac{\lambda _3 A_{r}^{12}}{4 \Lambda ^2}-\frac{\lambda _4 A_{r}^{12}}{4 \Lambda ^2}+\frac{\lambda _5 A_{r}^{12}}{4 \Lambda ^2}-\frac{\lambda _6 A_{r}^{11}}{4 \Lambda ^2}$  \\
		$\text{}$  &  $\textcolor{purple}{+\frac{3 \mathcal{C}_{6}^{111112} v_1^2}{2 \Lambda ^2}}+\frac{\mathcal{C}_{6}^{111122} v_2 v_1}{\Lambda ^2}\textcolor{purple}{+\frac{3 \mathcal{C}_{6}^{112212} v_2^2}{4 \Lambda ^2}}\textcolor{purple}{-\frac{9 \mathcal{C}_{6}^{121212} v_2^2}{4 \Lambda ^2}}+\frac{\mathcal{C}_{6}^{122111} v_2 v_1}{\Lambda ^2}\textcolor{purple}{+\frac{3 \mathcal{C}_{6}^{121221} v_2^2}{4 \Lambda ^2}}$  \\
		$\eta _1 \eta _2 \rho _1 \rho _2$  &  $\lambda _5-\frac{\lambda _5 A_{i}^{22}}{2 \Lambda ^2}-\frac{\lambda _6 A_{i}^{12}}{2 \Lambda ^2}-\frac{\lambda _7 A_{i}^{12}}{2 \Lambda ^2}-\frac{\lambda _5 A_{i}^{11}}{2 \Lambda ^2}-\frac{\lambda _5 A_{r}^{22}}{2 \Lambda ^2}-\frac{\lambda _5 A_{r}^{11}}{2 \Lambda ^2}-\frac{\lambda _6 A_{r}^{12}}{2 \Lambda ^2}-\frac{\lambda _7 A_{r}^{12}}{2 \Lambda ^2}$  \\
		$\text{}$  &  $\textcolor{purple}{+\frac{\mathcal{C}_{6}^{112212} v_2 v_1}{\Lambda ^2}}+\frac{3 \mathcal{C}_{6}^{121211} v_1^2}{\Lambda ^2}\textcolor{purple}{+\frac{9 \mathcal{C}_{6}^{121212} v_2 v_1}{\Lambda ^2}}+\frac{3 \mathcal{C}_{6}^{121222} v_2^2}{\Lambda ^2}\textcolor{purple}{+\frac{\mathcal{C}_{6}^{121221} v_2 v_1}{\Lambda ^2}}$  \\
		$\rho _1^2 \rho _2^2$  &  $\frac{\lambda _3}{4}+\frac{\lambda _4}{4}+\frac{\lambda _5}{4}-\frac{\lambda _3 A_{r}^{22}}{4 \Lambda ^2}-\frac{3 \lambda _6 A_{r}^{12}}{4 \Lambda ^2}-\frac{3 \lambda _7 A_{r}^{12}}{4 \Lambda ^2}-\frac{\lambda _3 A_{r}^{11}}{4 \Lambda ^2}-\frac{\lambda _4 A_{r}^{11}}{4 \Lambda ^2}-\frac{\lambda _5 A_{r}^{11}}{4 \Lambda ^2}$  \\
		$\text{}$  &  $-\frac{\lambda _4 A_{r}^{22}}{4 \Lambda ^2}-\frac{\lambda _5 A_{r}^{22}}{4 \Lambda ^2}+\frac{3 \mathcal{C}_{6}^{111122} v_1^2}{4 \Lambda ^2}\textcolor{purple}{+\frac{9 \mathcal{C}_{6}^{112212} v_1 v_2}{4 \Lambda ^2}}+\frac{3 \mathcal{C}_{6}^{121211} v_1^2}{2 \Lambda ^2}\textcolor{purple}{+\frac{9 \mathcal{C}_{6}^{121212} v_1 v_2}{4 \Lambda ^2}}+\frac{3 \mathcal{C}_{6}^{121222} v_2^2}{2 \Lambda ^2}+\frac{3 \mathcal{C}_{6}^{122111} v_1^2}{4 \Lambda ^2}$  \\
		$\text{}$  &  $\textcolor{purple}{+\frac{9 \mathcal{C}_{6}^{121221} v_1 v_2}{4 \Lambda ^2}}+\frac{3 \mathcal{C}_{6}^{112222} v_2^2}{4 \Lambda ^2}+\frac{3 \mathcal{C}_{6}^{122122} v_2^2}{4 \Lambda ^2}$  \\
		$\rho _1 \rho _2^3$  &  $\frac{\lambda _7}{2}-\frac{3 \lambda _7 A_{r}^{22}}{4 \Lambda ^2}-\frac{\lambda _2 A_{r}^{12}}{2 \Lambda ^2}-\frac{\lambda _3 A_{r}^{12}}{4 \Lambda ^2}-\frac{\lambda _4 A_{r}^{12}}{4 \Lambda ^2}-\frac{\lambda _5 A_{r}^{12}}{4 \Lambda ^2}-\frac{\lambda _7 A_{r}^{11}}{4 \Lambda ^2}\textcolor{purple}{+\frac{3 \mathcal{C}_{6}^{112212} v_1^2}{4 \Lambda ^2}+\frac{3 \mathcal{C}_{6}^{121212} v_1^2}{4 \Lambda ^2}}$  \\
		$\text{}$  &  $+\frac{2 \mathcal{C}_{6}^{121222} v_2 v_1}{\Lambda ^2}+\frac{\mathcal{C}_{6}^{112222} v_2 v_1}{\Lambda ^2}\textcolor{purple}{+\frac{3 \mathcal{C}_{6}^{121221} v_1^2}{4 \Lambda ^2}}+\frac{\mathcal{C}_{6}^{122122} v_2 v_1}{\Lambda ^2}\textcolor{purple}{+\frac{5 \mathcal{C}_{6}^{122222} v_2^2}{2 \Lambda ^2}}$  \\
		$\eta _1^2 \rho _1 \rho _2$  &  $\frac{\lambda _6}{2}-\frac{\lambda _5 A_{i}^{12}}{2 \Lambda ^2}-\frac{\lambda _6 A_{i}^{11}}{2 \Lambda ^2}-\frac{\lambda _6 A_{r}^{22}}{4 \Lambda ^2}-\frac{\lambda _1 A_{r}^{12}}{2 \Lambda ^2}-\frac{\lambda _3 A_{r}^{12}}{4 \Lambda ^2}-\frac{\lambda _4 A_{r}^{12}}{4 \Lambda ^2}+\frac{\lambda _5 A_{r}^{12}}{4 \Lambda ^2}-\frac{\lambda _6 A_{r}^{11}}{4 \Lambda ^2}$  \\
		$\text{}$  &  $\textcolor{purple}{+\frac{3 \mathcal{C}_{6}^{111112} v_1^2}{2 \Lambda ^2}}+\frac{\mathcal{C}_{6}^{111122} v_2 v_1}{\Lambda ^2}\textcolor{purple}{+\frac{3 \mathcal{C}_{6}^{112212} v_2^2}{4 \Lambda ^2}}\textcolor{purple}{-\frac{9 \mathcal{C}_{6}^{121212} v_2^2}{4 \Lambda ^2}}+\frac{\mathcal{C}_{6}^{122111} v_2 v_1}{\Lambda ^2}\textcolor{purple}{+\frac{3 \mathcal{C}_{6}^{121221} v_2^2}{4 \Lambda ^2}}$  \\
		$\eta _1 \eta _2 \rho _1 \rho _2$  &  $\lambda _5-\frac{\lambda _5 A_{i}^{22}}{2 \Lambda ^2}-\frac{\lambda _6 A_{i}^{12}}{2 \Lambda ^2}-\frac{\lambda _7 A_{i}^{12}}{2 \Lambda ^2}-\frac{\lambda _5 A_{i}^{11}}{2 \Lambda ^2}-\frac{\lambda _5 A_{r}^{22}}{2 \Lambda ^2}-\frac{\lambda _5 A_{r}^{11}}{2 \Lambda ^2}-\frac{\lambda _6 A_{r}^{12}}{2 \Lambda ^2}-\frac{\lambda _7 A_{r}^{12}}{2 \Lambda ^2}$  \\
		$\text{}$  &  $\textcolor{purple}{+\frac{\mathcal{C}_{6}^{112212} v_2 v_1}{\Lambda ^2}}+\frac{3 \mathcal{C}_{6}^{121211} v_1^2}{\Lambda ^2}\textcolor{purple}{+\frac{9 \mathcal{C}_{6}^{121212} v_2 v_1}{\Lambda ^2}}+\frac{3 \mathcal{C}_{6}^{121222} v_2^2}{\Lambda ^2}\textcolor{purple}{+\frac{\mathcal{C}_{6}^{121221} v_2 v_1}{\Lambda ^2}}$  \\
		$\eta _1^2 \rho _1^2$  &  $\frac{\lambda _1}{2}-\frac{\lambda _6 A_{i}^{12}}{4 \Lambda ^2}-\frac{\lambda _1 A_{i}^{11}}{2 \Lambda ^2}-\frac{\lambda _1 A_{r}^{11}}{2 \Lambda ^2}-\frac{\lambda _6 A_{r}^{12}}{4 \Lambda ^2}+\frac{9 \mathcal{C}_{6}^{111111} v_1^2}{4 \Lambda ^2}\textcolor{purple}{+\frac{3 \mathcal{C}_{6}^{111112} v_1 v_2}{2 \Lambda ^2}}+\frac{\mathcal{C}_{6}^{111122} v_2^2}{4 \Lambda ^2}+\frac{\mathcal{C}_{6}^{122111} v_2^2}{4 \Lambda ^2}$  \\
		$\text{}$  &  $\textcolor{purple}{+\frac{\mathcal{C}_{6}^{112212} v_1^2}{4 \Lambda ^2}}\textcolor{purple}{+\frac{9 \mathcal{C}_{6}^{121212} v_1^2}{4 \Lambda ^2}}+\frac{3 \mathcal{C}_{6}^{121222} v_2 v_1}{\Lambda ^2}\textcolor{purple}{+\frac{\mathcal{C}_{6}^{121221} v_1^2}{4 \Lambda ^2}+\frac{3 \mathcal{C}_{6}^{122222} v_2^2}{2 \Lambda ^2}}$  \\
		\hline
	\end{tabular}
\caption{2HDM: Scalar four-point vertices in the unphysical basis.}
\label{tab:2HDM-scalar-4-pt-II}
}
\end{table}

\begin{table}[h!]
	\centering
	{ \scriptsize
	\renewcommand{\arraystretch}{2.3}
	\begin{tabular}{|c|l|}
		\hline
		{\bf Vertex}&{\bf 2HDM: Effective vertex  $ (\phi_{i}\,\phi_{j}\,\phi_{k}\,\phi_{l})$ factor}\\
		\hline 
		$\eta _1 \eta _2 \rho _1^2$  &  $\frac{\lambda _6}{2}-\frac{\lambda _6 A_{i}^{22}}{4 \Lambda ^2}-\frac{\lambda _1 A_{i}^{12}}{2 \Lambda ^2}-\frac{\lambda _3 A_{i}^{12}}{4 \Lambda ^2}-\frac{\lambda _4 A_{i}^{12}}{4 \Lambda ^2}+\frac{\lambda _5 A_{i}^{12}}{4 \Lambda ^2}-\frac{\lambda _6 A_{i}^{11}}{4 \Lambda ^2}-\frac{\lambda _5 A_{r}^{12}}{2 \Lambda ^2}-\frac{\lambda _6 A_{r}^{11}}{2 \Lambda ^2}$  \\
		$\text{}$  &  $\textcolor{purple}{+\frac{3 \mathcal{C}_{6}^{111112} v_1^2}{2 \Lambda ^2}+\frac{\mathcal{C}_{6}^{112212} v_2^2}{4 \Lambda ^2}}+\frac{3 \mathcal{C}_{6}^{121211} v_2 v_1}{\Lambda ^2}\textcolor{purple}{+\frac{9 \mathcal{C}_{6}^{121212} v_2^2}{4 \Lambda ^2}+\frac{\mathcal{C}_{6}^{121221} v_2^2}{4 \Lambda ^2}}$  \\
		$\eta _2^2 \rho _1^2$  &  $\frac{\lambda _3}{4}+\frac{\lambda _4}{4}-\frac{\lambda _5}{4}-\frac{\lambda _3 A_{i}^{22}}{4 \Lambda ^2}-\frac{\lambda _4 A_{i}^{22}}{4 \Lambda ^2}+\frac{\lambda _5 A_{i}^{22}}{4 \Lambda ^2}-\frac{\lambda _6 A_{i}^{12}}{4 \Lambda ^2}-\frac{\lambda _3 A_{r}^{11}}{4 \Lambda ^2}-\frac{\lambda _4 A_{r}^{11}}{4 \Lambda ^2}$  \\
		$\text{}$  &  $-\frac{\lambda _7 A_{r}^{12}}{4 \Lambda ^2}+\frac{\lambda _5 A_{r}^{11}}{4 \Lambda ^2}+\frac{3 \mathcal{C}_{6}^{111122} v_1^2}{4 \Lambda ^2}\textcolor{purple}{+\frac{3 \mathcal{C}_{6}^{112212} v_1 v_2}{4 \Lambda ^2}}-\frac{3 \mathcal{C}_{6}^{121211} v_1^2}{2 \Lambda ^2}\textcolor{purple}{-\frac{9 \mathcal{C}_{6}^{121212} v_1 v_2}{4 \Lambda ^2}}+\frac{\mathcal{C}_{6}^{112222} v_2^2}{4 \Lambda ^2}+\frac{3 \mathcal{C}_{6}^{122111} v_1^2}{4 \Lambda ^2}$  \\
		$\text{}$  &  $\textcolor{purple}{+\frac{3 \mathcal{C}_{6}^{121221} v_1 v_2}{4 \Lambda ^2}}+\frac{\mathcal{C}_{6}^{122122} v_2^2}{4 \Lambda ^2}$  \\
		$\eta _1 \eta _2 \rho _1 \rho _2$  &  $\lambda _5-\frac{\lambda _5 A_{i}^{22}}{2 \Lambda ^2}-\frac{\lambda _6 A_{i}^{12}}{2 \Lambda ^2}-\frac{\lambda _7 A_{i}^{12}}{2 \Lambda ^2}-\frac{\lambda _5 A_{i}^{11}}{2 \Lambda ^2}-\frac{\lambda _5 A_{r}^{22}}{2 \Lambda ^2}-\frac{\lambda _5 A_{r}^{11}}{2 \Lambda ^2}-\frac{\lambda _6 A_{r}^{12}}{2 \Lambda ^2}-\frac{\lambda _7 A_{r}^{12}}{2 \Lambda ^2}$  \\
		$\text{}$  &  $\textcolor{purple}{+\frac{\mathcal{C}_{6}^{112212} v_2 v_1}{\Lambda ^2}}+\frac{3 \mathcal{C}_{6}^{121211} v_1^2}{\Lambda ^2}\textcolor{purple}{+\frac{9 \mathcal{C}_{6}^{121212} v_2 v_1}{\Lambda ^2}}+\frac{3 \mathcal{C}_{6}^{121222} v_2^2}{\Lambda ^2}\textcolor{purple}{+\frac{\mathcal{C}_{6}^{121221} v_2 v_1}{\Lambda ^2}}$  \\
		$\eta _2^2 \rho _1 \rho _2$  &  $-\frac{\lambda _7 A_{i}^{22}}{2 \Lambda ^2}-\frac{\lambda _5 A_{i}^{12}}{2 \Lambda ^2}-\frac{\lambda _2 A_{r}^{12}}{2 \Lambda ^2}-\frac{\lambda _3 A_{r}^{12}}{4 \Lambda ^2}-\frac{\lambda _4 A_{r}^{12}}{4 \Lambda ^2}+\frac{\lambda _5 A_{r}^{12}}{4 \Lambda ^2}-\frac{\lambda _7 A_{r}^{22}}{4 \Lambda ^2}-\frac{\lambda _7 A_{r}^{11}}{4 \Lambda ^2}+\frac{\lambda _7}{2}$  \\
		$\text{}$  &  $\textcolor{purple}{+\frac{3 \mathcal{C}_{6}^{112212} v_1^2}{4 \Lambda ^2}-\frac{9 \mathcal{C}_{6}^{121212} v_1^2}{4 \Lambda ^2}}+\frac{\mathcal{C}_{6}^{112222} v_2 v_1}{\Lambda ^2}\textcolor{purple}{+\frac{3 \mathcal{C}_{6}^{121221} v_1^2}{4 \Lambda ^2}}+\frac{\mathcal{C}_{6}^{122122} v_2 v_1}{\Lambda ^2}\textcolor{purple}{+\frac{3 \mathcal{C}_{6}^{122222} v_2^2}{2 \Lambda ^2}}$  \\
		$\eta _1^2 \rho _2^2$  &  $\frac{\lambda _3}{4}+\frac{\lambda _4}{4}-\frac{\lambda _5}{4}-\frac{\lambda _3 A_{i}^{11}}{4 \Lambda ^2}-\frac{\lambda _4 A_{i}^{11}}{4 \Lambda ^2}+\frac{\lambda _5 A_{i}^{11}}{4 \Lambda ^2}-\frac{\lambda _7 A_{i}^{12}}{4 \Lambda ^2}-\frac{\lambda _3 A_{r}^{22}}{4 \Lambda ^2}-\frac{\lambda _6 A_{r}^{12}}{4 \Lambda ^2}$  \\
		$\text{}$  &  $-\frac{\lambda _4 A_{r}^{22}}{4 \Lambda ^2}+\frac{\lambda _5 A_{r}^{22}}{4 \Lambda ^2}+\frac{\mathcal{C}_{6}^{111122} v_1^2}{4 \Lambda ^2}+\textcolor{purple}{\frac{3 \mathcal{C}_{6}^{112212} v_1 v_2}{4 \Lambda ^2}-\frac{9 \mathcal{C}_{6}^{121212} v_1 v_2}{4 \Lambda ^2}}-\frac{3 \mathcal{C}_{6}^{121222} v_2^2}{2 \Lambda ^2}+\frac{3 \mathcal{C}_{6}^{112222} v_2^2}{4 \Lambda ^2}+\frac{\mathcal{C}_{6}^{122111} v_1^2}{4 \Lambda ^2}$  \\
		$\text{}$  &  $\textcolor{purple}{+\frac{3 \mathcal{C}_{6}^{121221} v_1 v_2}{4 \Lambda ^2}}+\frac{3 \mathcal{C}_{6}^{122122} v_2^2}{4 \Lambda ^2}$  \\
		$\eta _1 \eta _2 \rho _2^2$  &  $\frac{\lambda _7}{2}-\frac{\lambda _7 A_{i}^{22}}{4 \Lambda ^2}-\frac{\lambda _2 A_{i}^{12}}{2 \Lambda ^2}-\frac{\lambda _3 A_{i}^{12}}{4 \Lambda ^2}-\frac{\lambda _4 A_{i}^{12}}{4 \Lambda ^2}+\frac{\lambda _5 A_{i}^{12}}{4 \Lambda ^2}-\frac{\lambda _7 A_{i}^{11}}{4 \Lambda ^2}-\frac{\lambda _5 A_{r}^{12}}{2 \Lambda ^2}-\frac{\lambda _7 A_{r}^{22}}{2 \Lambda ^2}$  \\
		$\text{}$  &  $\textcolor{purple}{+\frac{\mathcal{C}_{6}^{112212} v_1^2}{4 \Lambda ^2}+\frac{9 \mathcal{C}_{6}^{121212} v_1^2}{4 \Lambda ^2}}+\frac{3 \mathcal{C}_{6}^{121222} v_2 v_1}{\Lambda ^2}\textcolor{purple}{+\frac{\mathcal{C}_{6}^{121221} v_1^2}{4 \Lambda ^2}+\frac{3 \mathcal{C}_{6}^{122222} v_2^2}{2 \Lambda ^2}}$  \\
		$\eta _1^2 \rho _2^2$  &  $\frac{\lambda _3}{4}+\frac{\lambda _4}{4}-\frac{\lambda _5}{4}-\frac{\lambda _3 A_{i}^{11}}{4 \Lambda ^2}-\frac{\lambda _4 A_{i}^{11}}{4 \Lambda ^2}+\frac{\lambda _5 A_{i}^{11}}{4 \Lambda ^2}-\frac{\lambda _7 A_{i}^{12}}{4 \Lambda ^2}-\frac{\lambda _3 A_{r}^{22}}{4 \Lambda ^2}-\frac{\lambda _6 A_{r}^{12}}{4 \Lambda ^2}$  \\
		$\text{}$  &  $-\frac{\lambda _4 A_{r}^{22}}{4 \Lambda ^2}+\frac{\lambda _5 A_{r}^{22}}{4 \Lambda ^2}+\frac{\mathcal{C}_{6}^{111122} v_1^2}{4 \Lambda ^2}\textcolor{purple}{+\frac{3 \mathcal{C}_{6}^{112212} v_1 v_2}{4 \Lambda ^2}-\frac{9 \mathcal{C}_{6}^{121212} v_1 v_2}{4 \Lambda ^2}}-\frac{3 \mathcal{C}_{6}^{121222} v_2^2}{2 \Lambda ^2}+\frac{3 \mathcal{C}_{6}^{112222} v_2^2}{4 \Lambda ^2}+\frac{\mathcal{C}_{6}^{122111} v_1^2}{4 \Lambda ^2}$  \\
		$\text{}$  &  $\textcolor{purple}{+\frac{3 \mathcal{C}_{6}^{121221} v_1 v_2}{4 \Lambda ^2}}+\frac{3 \mathcal{C}_{6}^{122122} v_2^2}{4 \Lambda ^2}$  \\
		$\eta _1 \eta _2 \rho _2^2$  &  $\frac{\lambda _7}{2}-\frac{\lambda _7 A_{i}^{22}}{4 \Lambda ^2}-\frac{\lambda _2 A_{i}^{12}}{2 \Lambda ^2}-\frac{\lambda _3 A_{i}^{12}}{4 \Lambda ^2}-\frac{\lambda _4 A_{i}^{12}}{4 \Lambda ^2}+\frac{\lambda _5 A_{i}^{12}}{4 \Lambda ^2}-\frac{\lambda _7 A_{i}^{11}}{4 \Lambda ^2}-\frac{\lambda _5 A_{r}^{12}}{2 \Lambda ^2}-\frac{\lambda _7 A_{r}^{22}}{2 \Lambda ^2}$  \\
		$\eta _2^2 \rho _2^2$  &  $\frac{\lambda _2}{2}-\frac{\lambda _2 A_{i}^{22}}{2 \Lambda ^2}-\frac{\lambda _7 A_{i}^{12}}{4 \Lambda ^2}-\frac{\lambda _2 A_{r}^{22}}{2 \Lambda ^2}-\frac{\lambda _7 A_{r}^{12}}{4 \Lambda ^2}+\frac{\mathcal{C}_{6}^{112222} v_1^2}{4 \Lambda ^2}+\frac{\mathcal{C}_{6}^{122122} v_1^2}{4 \Lambda ^2}\textcolor{purple}{+\frac{3 \mathcal{C}_{6}^{122222} v_1 v_2}{2 \Lambda ^2}}+\frac{9 \mathcal{C}_{6}^{222222} v_2^2}{4 \Lambda ^2}$  \\
		$\eta _1^4$  &  $\frac{\lambda _1}{4}-\frac{\lambda _6 A_{i}^{12}}{4 \Lambda ^2}-\frac{\lambda _1 A_{i}^{11}}{2 \Lambda ^2}+\frac{3 \mathcal{C}_{6}^{111111} v_1^2}{8 \Lambda ^2}\textcolor{purple}{+\frac{\mathcal{C}_{6}^{111112} v_1 v_2}{4 \Lambda ^2}}+\frac{\mathcal{C}_{6}^{111122} v_2^2}{8 \Lambda ^2}-\frac{\mathcal{C}_{6}^{121211} v_2^2}{4 \Lambda ^2}+\frac{\mathcal{C}_{6}^{122111} v_2^2}{8 \Lambda ^2}$  \\
		$\eta _1^3 \eta _2$  &  $\frac{\lambda _6}{2}-\frac{\lambda _6 A_{i}^{22}}{4 \Lambda ^2}-\frac{\lambda _1 A_{i}^{12}}{2 \Lambda ^2}-\frac{\lambda _3 A_{i}^{12}}{4 \Lambda ^2}-\frac{\lambda _4 A_{i}^{12}}{4 \Lambda ^2}-\frac{\lambda _5 A_{i}^{12}}{4 \Lambda ^2}-\frac{3 \lambda _6 A_{i}^{11}}{4 \Lambda ^2}\textcolor{purple}{+\frac{\mathcal{C}_{6}^{111112} v_1^2}{2 \Lambda ^2}}+\frac{\mathcal{C}_{6}^{121211} v_1 v_2}{\Lambda ^2}$  \\
		$\text{}$  &  $\textcolor{purple}{+\frac{\mathcal{C}_{6}^{112212} v_2^2}{4 \Lambda ^2}-\frac{3 \mathcal{C}_{6}^{121212} v_2^2}{4 \Lambda ^2}+\frac{\mathcal{C}_{6}^{121221} v_2^2}{4 \Lambda ^2}}$  \\
		$\eta _1^2 \eta _2^2$  &  $\frac{\lambda _3}{4}+\frac{\lambda _4}{4}+\frac{\lambda _5}{4}-\frac{\lambda _3 A_{i}^{22}}{4 \Lambda ^2}-\frac{3 \lambda _6 A_{i}^{12}}{4 \Lambda ^2}-\frac{3 \lambda _7 A_{i}^{12}}{4 \Lambda ^2}-\frac{\lambda _3 A_{i}^{11}}{4 \Lambda ^2}-\frac{\lambda _4 A_{i}^{11}}{4 \Lambda ^2}-\frac{\lambda _5 A_{i}^{11}}{4 \Lambda ^2}$  \\
		$\text{}$  &  $-\frac{\lambda _4 A_{i}^{22}}{4 \Lambda ^2}-\frac{\lambda _5 A_{i}^{22}}{4 \Lambda ^2}+\frac{\mathcal{C}_{6}^{111122} v_1^2}{4 \Lambda ^2}\textcolor{purple}{+\frac{\mathcal{C}_{6}^{112212} v_1 v_2}{4 \Lambda ^2}+\frac{9 \mathcal{C}_{6}^{121212} v_1 v_2}{4 \Lambda ^2}}+\frac{\mathcal{C}_{6}^{112222} v_2^2}{4 \Lambda ^2}+\frac{\mathcal{C}_{6}^{122111} v_1^2}{4 \Lambda ^2}$  \\
		$\text{}$  &  $\textcolor{purple}{+\frac{\mathcal{C}_{6}^{121221} v_1 v_2}{4 \Lambda ^2}}+\frac{\mathcal{C}_{6}^{122122} v_2^2}{4 \Lambda ^2}$  \\
		\hline
	\end{tabular}
\caption{2HDM: Scalar four-point vertices in the unphysical basis.}
\label{tab:2HDM-scalar-4-pt-III}
}
\end{table}


\begin{center}
	\underline{\bf {2HDM: Scalar-four point interactions}}  
\end{center}

The TU bounds have been computed for 2HDM  in \cite{Huffel:1980sk,Kanemura:1993hm,Akeroyd:2000wc,Horejsi:2005da,Chakraborty:2014oma,Chakrabarty:2015kmt,
	Chakrabarty:2016smc,DeCurtis:2016scv,Maalampi:1991fb,Horejsi:2005da,Kikuta:2012tf} but with the only renormalizable scalar potential.  We expect these bounds to be altered in the presence of these extra contributions from the effective operators. Thus the conclusion embracing the validity of 2HDM scenario including the renormalization group evolutions (RGEs) of the couplings is destined to change. As, after the  inclusion of these effective operators, the anomalous dimension matrices need to be included to perform the RGEs of the Wilson coefficients. This will lead to an involved computation and it will be difficult to pass any conclusive remarks on the scale up to which this theory will be valid based on these criteria.
In this section, we have summarised the complete set of four-point interactions in the unphysical basis of scalar fields in Tables~\ref{tab:2HDM-scalar-4-pt-I}-\ref{tab:2HDM-scalar-4-pt-III}.
\clearpage


\begin{center}
	\underline{\bf {MLRSM: Scalar four-point interactions}}  
\end{center}

Similar to the 2HDM case, the scalar quartic couplings in this model  are also constrained by the TU criteria. For the renormalizable potential these criteria are derived and discussed in \cite{Gunion:1989in,Mondal:2015fja,Chakrabortty:2016wkl}. These scalar four-point vertex factors, i.e., quartic couplings must satisfy the vacuum stability criteria such that the effective potential is bounded from below. This should be further analyzed using the same method mentioned in \cite{Chakrabortty:2013mha,Chakrabortty:2013zja,Kannike:2016fmd}. 
As the scalar sector of the MLRSM scenario is quite extended, the number of four-point interaction vertices are very large. Thus instead of providing all of them, unlike the 2HDM case, here, we have listed a few selective vertices, see Table~\ref{tab:MLRSM-scalar-four-pt}. Interested readers can be provided with the Mathematica file containing all such vertices for MLRSM scenario.

\begin{table}[h!]
	\centering
	{ \scriptsize
		\renewcommand{\arraystretch}{2.3}
		\begin{tabular}{|*{2}{l|}}
			\hline 
			{\bf Vertex}&{\bf MLRSM: Effective vertex $ (\phi_{i}\,\phi_{j}\,\phi_{k}\,\phi_{l})$ factor}\\
			\hline 
			$\left( \phi _1^{0r} \right)^4 $  &  $\frac{\lambda _1}{4}+\frac{\lambda _2}{2}+\frac{\lambda _3}{4}+\frac{\lambda _4}{2}+\frac{\mathcal{C}_6^{2121{Rr}} v_R^2}{32 \Lambda ^2}+\frac{15 \kappa ^2 \mathcal{C}_6^{\{21\}\{2121\}}}{8 \Lambda ^2}+\frac{15 \kappa ^2 \mathcal{C}_6^{212121}}{16 \Lambda ^2}+\frac{15 \kappa ^2 \mathcal{C}_6^{412121}}{8 \Lambda ^2}+\frac{15 \kappa ^2 \mathcal{C}_6^{414121}}{8 \Lambda ^2}$  \\
			&  $+\frac{15 \kappa ^2 \mathcal{C}_6^{414123}}{8 \Lambda ^2}+\frac{15 \kappa ^2 \mathcal{C}_6^{414141}}{8 \Lambda ^2}+\frac{\mathcal{C}_6^{2141{Rr}} v_R^2}{16 \Lambda ^2}+\frac{\mathcal{C}_6^{2211{Rr}} v_R^2}{32 \Lambda ^2}+\frac{\mathcal{C}_6^{23{Rr41}} v_R^2}{32 \Lambda ^2}+\frac{\mathcal{C}_6^{2{R21r1}} v_R^2}{32 \Lambda ^2}+\frac{\mathcal{C}_6^{4141{Rr}} v_R^2}{16 \Lambda ^2}$  \\
			&$+\frac{\mathcal{C}_6^{41{R2r1}} v_R^2}{16 \Lambda ^2}-\frac{\lambda _1 \tilde{{A}}_r^{11}}{2 \Lambda ^2}-\frac{\lambda _2 \tilde{{A}}_r^{11}}{\Lambda ^2}-\frac{\lambda _3 \tilde{{A}}_r^{11}}{2 \Lambda ^2}-\frac{\lambda _4 \tilde{{A}}_r^{11}}{\Lambda ^2}$\\
			$\left( \phi _1^{0r} \right)^3 \phi _1^{0i}$  &  $-\frac{\mathcal{C}_6^{41{R2r1}} v_R^2}{8 \Lambda ^2}$  \\
			$\left( \phi _1^{0r} \right)^3 \phi _2^{0r}$  &  $-\frac{\lambda _1 \tilde{{A}}_r^{12}}{\Lambda ^2}-\frac{\lambda _4 \tilde{{A}}_r^{12}}{\Lambda ^2}-\frac{\mathcal{C}_6^{2121{Rr}} v_R^2}{8 \Lambda ^2}-\frac{\mathcal{C}_6^{2141{Rr}} v_R^2}{8 \Lambda ^2}-\frac{\mathcal{C}_6^{2211{Rr}} v_R^2}{8 \Lambda ^2}$  \\
			$\left( \phi _1^{0r} \right)^3 \phi _2^{0i}$  &  $\frac{\mathcal{C}_6^{2141{Rr}} v_R^2}{8 \Lambda ^2}+\frac{\mathcal{C}_6^{4141{Rr}} v_R^2}{4 \Lambda ^2}+\frac{\mathcal{C}_6^{41{R2r1}} v_R^2}{8 \Lambda ^2}+\frac{5 \kappa ^2 \mathcal{C}_6^{412121}}{2 \Lambda ^2}+\frac{5 \kappa ^2 \mathcal{C}_6^{414121}}{\Lambda ^2}+\frac{5 \kappa ^2 \mathcal{C}_6^{414123}}{2 \Lambda ^2}+\frac{15 \kappa ^2 \mathcal{C}_6^{414141}}{2 \Lambda ^2}$  \\
			$\left(\phi _1^{0r}\right)^3 \delta _L^{0r}$  &  $\frac{\kappa  \mathcal{C}_6^{21{L2r1}} v_R}{2 \sqrt{2} \Lambda ^2}+\frac{\kappa  \mathcal{C}_6^{21{L4r1}} v_R}{2 \sqrt{2} \Lambda ^2}+\frac{\kappa  \mathcal{C}_6^{21{R4l1}} v_R}{2 \sqrt{2} \Lambda ^2}+\frac{\kappa  \mathcal{C}_6^{2{L41r3}} v_R}{2 \sqrt{2} \Lambda ^2}+\frac{\kappa  \mathcal{C}_6^{41{L2r3}} v_R}{2 \sqrt{2} \Lambda ^2}+\frac{\kappa  \mathcal{C}_6^{41{L41r}} v_R}{2 \sqrt{2} \Lambda ^2}+\frac{\kappa  \mathcal{C}_6^{41{R2l1}} v_R}{2 \sqrt{2} \Lambda ^2}$  \\
			${}$  &  $-\frac{\alpha _1 \tilde{{A}}_r^{14}}{4 \Lambda ^2}-\frac{\alpha _2 \tilde{{A}}_r^{14}}{2 \Lambda ^2}-\frac{\alpha _3 \tilde{{A}}_r^{14}}{8 \Lambda ^2}-\frac{\lambda _1 \tilde{{A}}_r^{14}}{2 \Lambda ^2}-\frac{\lambda _2 \tilde{{A}}_r^{14}}{\Lambda ^2}-\frac{\lambda _3 \tilde{{A}}_r^{14}}{2 \Lambda ^2}+\frac{\kappa  \mathcal{C}_6^{41{R41l}} v_R}{2 \sqrt{2} \Lambda ^2}-\frac{\lambda _4 \tilde{{A}}_r^{14}}{\Lambda ^2}$  \\
			$\left(\phi _1^{0r}\right)^3 \delta _L^{0i}$  &  $\frac{\kappa  \mathcal{C}_6^{21{L2r1}} v_R}{2 \sqrt{2} \Lambda ^2}+\frac{\kappa  \mathcal{C}_6^{21{L4r1}} v_R}{2 \sqrt{2} \Lambda ^2}-\frac{\kappa  \mathcal{C}_6^{21{R4l1}} v_R}{2 \sqrt{2} \Lambda ^2}+\frac{\kappa  \mathcal{C}_6^{2{L41r3}} v_R}{2 \sqrt{2} \Lambda ^2}+\frac{\kappa  \mathcal{C}_6^{41{L2r3}} v_R}{2 \sqrt{2} \Lambda ^2}+\frac{\kappa  \mathcal{C}_6^{41{L41r}} v_R}{2 \sqrt{2} \Lambda ^2}-\frac{\kappa  \mathcal{C}_6^{41{R2l1}} v_R}{2 \sqrt{2} \Lambda ^2}$  \\
			&$-\frac{\kappa  \mathcal{C}_6^{41{R41l}} v_R}{2 \sqrt{2} \Lambda ^2}$\\
			\hline
			\hspace{0.5cm} {\bf{\large$\vdots$}} & \hspace{6cm} {\bf{\large$\vdots$}} \\
			\hline
			$\left(\phi _1^{0r}\right)^3 \delta _R^{0r}$  &  $\frac{\kappa  \mathcal{C}_6^{2121{Rr}} v_R}{2 \sqrt{2} \Lambda ^2}+\frac{\kappa  \mathcal{C}_6^{2141{Rr}} v_R}{\sqrt{2} \Lambda ^2}+\frac{\kappa  \mathcal{C}_6^{2211{Rr}} v_R}{2 \sqrt{2} \Lambda ^2}+\frac{\kappa  \mathcal{C}_6^{23{Rr41}} v_R}{2 \sqrt{2} \Lambda ^2}+\frac{\kappa  \mathcal{C}_6^{2{R21r1}} v_R}{2 \sqrt{2} \Lambda ^2}+\frac{\kappa  \mathcal{C}_6^{4141{Rr}} v_R}{\sqrt{2} \Lambda ^2}+\frac{\kappa  \mathcal{C}_6^{41{R2r1}} v_R}{\sqrt{2} \Lambda ^2}$  \\
			${}$  &  $-\frac{\alpha _1 \tilde{{A}}_r^{13}}{4 \Lambda ^2}-\frac{\alpha _2 \tilde{{A}}_r^{13}}{2 \Lambda ^2}-\frac{\alpha _3 \tilde{{A}}_r^{13}}{8 \Lambda ^2}-\frac{\lambda _1 \tilde{{A}}_r^{13}}{2 \Lambda ^2}-\frac{\lambda _2 \tilde{{A}}_r^{13}}{\Lambda ^2}-\frac{\lambda _3 \tilde{{A}}_r^{13}}{2 \Lambda ^2}-\frac{\lambda _4 \tilde{{A}}_r^{13}}{\Lambda ^2}$  \\
			$\left(\phi _1^{0r}\right)^3 \delta _R^{0i}$  &  $0$\\
			\hline
		\end{tabular}
		\caption{  MLRSM: Scalar four-point vertices in the unphysical basis. As the number of such kind of vertices are very large, we have not mentioned all of them. These vertex factors are computed with the following assumptions: $\kappa_{1}=\kappa_{2}=\kappa, v_{L}=0$. In these limits we have redefined  $A_{r}^{ij}, A_{i}^{ij}$ as $\tilde{A}_{r}^{ij}, \tilde{A}_{i}^{ij}$ respectively.} \label{tab:MLRSM-scalar-four-pt}
	}
\end{table}
\clearpage